\documentclass{cernrep}
\usepackage{amsmath,axodraw,bbold}
\usepackage{epsfig}
\usepackage{color}
\usepackage{chngcntr}

\counterwithin{figure}{section}
\counterwithin{equation}{section}
\counterwithin{table}{section}
\def\ga{\mathrel{\rlap{\raise.6ex\hbox{$>$}}{\lower.6ex\hbox{$\sim$}}}}
\def\la{\mathrel{\rlap{\raise.6ex\hbox{$<$}}{\lower.6ex\hbox{$\sim$}}}}
\def\mec{\mbox{$m_{\rm EC}$}\xspace}

\providecommand{\PYTHIA}{{\sc{pythia}}\xspace}

\def\babar{Babar}

\def\epem 		{\ensuremath{e^+e^-}\xspace}
\def\tautau 		{\ensuremath{\tau^+\tau^-}\xspace}
\def\nb 		{\ensuremath{{\rm \,nb}}\xspace}

\newcommand{\ev}	{\ensuremath{\mathrm{\,e\kern -0.1em V}}\xspace}
\newcommand{\gev}	{\ensuremath{\mathrm{\,Ge\kern -0.1em V}}\xspace}

\def\electron 		{\ensuremath{e}\xspace}

\def\pzero		{\ensuremath{P^0}\xspace}
\def\tautolpz 		{\ensuremath{\tau^\pm \to \ell^\pm \pzero}\xspace}

\def\tautoepiz 		{\ensuremath{\tau^\pm \to \electron^\pm \pi^0}\xspace}
\def\tautompiz 		{\ensuremath{\tau^\pm \to \mu^\pm \pi^0}\xspace}

\def\etogg 		{\ensuremath{\eta \to \gamma\gamma}\xspace}
\def\etoppp 		{\ensuremath{\eta \to \pi^+\pi^-\pi^0}\xspace}
\def\eptoppe 		{\ensuremath{\eta^{\prime} \to \pi^+\pi^-\eta}\xspace}
\def\eptorg 		{\ensuremath{\eta^{\prime} \to \rho^0\gamma}\xspace}

\def\tautoleta 		{\ensuremath{\tau^\pm \to \ell^\pm \eta}\xspace}
\def\tautoeeta 		{\ensuremath{\tau^\pm \to \electron^\pm \eta}\xspace}
\def\tautometa 		{\ensuremath{\tau^\pm \to \mu^\pm \eta}\xspace}

\def\BRtaumeta 		{\ensuremath{\BR(\tau^\pm \to \mu^\pm \eta)}\xspace}

\def\tautoletap 	{\ensuremath{\tau^\pm \to \ell^\pm \eta^{\prime}\xspace}}
\def\tautoeetap 	{\ensuremath{\tau^\pm \to \electron^\pm \eta^{\prime}}\xspace}
\def\tautometap 	{\ensuremath{\tau^\pm \to \mu^\pm \eta^{\prime}}\xspace}

\newcommand{\taumg} 	{\ensuremath{\tau^{\pm} \to \mu^{\pm} \gamma}\xspace}
\newcommand{\taueg} 	{\ensuremath{\tau^{\pm} \to e^{\pm} \gamma}\xspace}
\newcommand{\taulg} 	{\ensuremath{\tau^{\pm} \to \ell^{\pm} \gamma}\xspace}

\newcommand{\BRtaumg} 	{\ensuremath{\BR(\taumg)}\xspace}

\newcommand{\taulll} 	{\ensuremath{\tau^\pm \to \ell^\pm \ell^\mp \ell^\pm}\xspace}

\newcommand{\taummm} 	{\ensuremath{\tau^\pm \to \mu^\pm \mu^\mp \mu^\pm}\xspace}
\newcommand{\BRtaummm} 	{\ensuremath{\BR(\taummm)}\xspace}

\def\invfb 		{\ensuremath{\mbox{\,fb}^{-1}}\xspace}

\def\BRul 		{{\cal{B}}_{\mathrm{UL}}}
\def\BRulninety 	{{\cal{B}}_{\mathrm{UL}}^{90}}

\newcommand{\ten}[1]{\times 10^{#1}}

\newcommand{\micron}{\ensuremath{\mathrm{\;\mu m}}\xspace}
\newcommand{\cm}{\ensuremath{\mathrm{\; cm}}\xspace}
\newcommand{\mm}{\ensuremath{\mathrm{\; mm}}\xspace}

\newcommand{\keV}{\ensuremath{\mathrm{\; keV}}\xspace}
\newcommand{\MeV}{\ensuremath{\mathrm{\; MeV}}\xspace}
\newcommand{\GeV}{\ensuremath{\mathrm{\; GeV}}\xspace}
\newcommand{\TeV}{\ensuremath{\mathrm{\; TeV}}\xspace}

\newcommand{\fbinv} {\mbox{\ensuremath{\mathrm{\; fb}^{-1}}}\xspace}
\newcommand{\percms}{\ensuremath{\mathrm{\; cm^{-2}\,s^{-1}}}\xspace}
\newcommand{\lumi}{\ensuremath{{\cal{L}}}\xspace}
\newcommand{\lowlumi}{\ensuremath{{\cal{L}}=2\times 10^{33}\; \mathrm{cm^{-2}\,s^{-1}}}\xspace}

\newcommand{\hilumi} {\ensuremath{{\cal{L}}=10^{34}\; \mathrm{cm^{-2}\,s^{-1}}}\xspace}

\newcommand{\bbbar}{\ensuremath{{b\overline{b}}}\xspace}

\newcommand{\pt}{\ensuremath{p_{\mathrm{T}}}\xspace}

\newcommand{\tanb}{\ensuremath{\tan\beta}\xspace}

\newcommand{\Hz}{\ensuremath{H^{0}}\xspace}

\newcommand{\rpv}{\ensuremath{\rlap{\kern.2em/}R}\xspace}

\DeclareMathAlphabet\EuScript{U}{eus}{m}{n}
\SetMathAlphabet\EuScript{bold}{U}{eus}{b}{n}
\newcommand{\Esca}{\EuScript{E}}

\begin{document}

\title{ Flavour physics of leptons and dipole moments
\footnote{Report of Working Group 3 of the CERN Workshop ``Flavour in
  the era of the LHC'', Geneva, Switzerland, November 2005 -- March 2007. }}
\author{
M.~Raidal$^{1}$$^{,2}$, A.~van der Schaaf$^{1}$$^{,3}$, I.~Bigi$^{1}$$^{,4}$, M.L.~Mangano$^{1}$$^{,5}$, Y.~Semertzidis$^{1}$$^{,6}$, S.~Abel$^{7}$,
S.~Albino$^{8}$, S.~Antusch$^{9}$, E.~Arganda$^{10}$, B.~Bajc$^{11}$, S.~Banerjee$^{12}$, C.~Biggio$^{9}$, M.~Blanke$^{9}$$^{,13}$, W.~Bonivento$^{14}$,
G.C.~Branco$^{15}$$^{,5}$, D.~Bryman$^{16}$, A.J.~Buras$^{13}$, L.~Calibbi$^{17}$$^{,18}$$^{,19}$, A.~Ceccucci$^{5}$, P.H.~Chankowski$^{20}$, S.~Davidson$^{21}$,
A.~Deandrea$^{21}$, D.P.~DeMille$^{22}$, F.~Deppisch$^{23}$, M.~Diaz$^{24}$, B.~Duling$^{13}$, M.~Felcini$^{5}$, W.~Fetscher$^{25}$, D.K.~Ghosh$^{26}$,
M.~Giffels$^{27}$, G.~Giudice$^{5}$, E.~Goudzovskij$^{28}$, T.~Han$^{29}$, P.G.~Harris$^{30}$, M.J.~Herrero$^{10}$, J.~Hisano$^{31}$, R.J.~Holt$^{32}$,
K.~Huitu$^{33}$, A.~Ibarra$^{34}$, O.~Igonkina$^{35}$$^{,36}$, A.~Ilakovac$^{37}$, J.~Imazato$^{38}$, G.~Isidori$^{28}$$^{,39}$, F.R.~Joaquim$^{10}$,
M.~Kadastik$^{2}$, Y.~Kajiyama$^{2}$, S.F.~King$^{40}$, K.~Kirch$^{41}$, M.G.~Kozlov$^{42}$, M.~Krawczyk$^{20}$$^{,5}$, T.~Kress$^{27}$, O.~Lebedev$^{5}$,
A.~Lusiani$^{43}$, E.~Ma$^{44}$, G.~Marchiori$^{43}$, I.~Masina$^{5}$, G.~Moreau$^{45}$, T.~Mori$^{46}$, M.~Muntel$^{2}$, F.~Nesti$^{47}$,
C.J.G.~Onderwater$^{48}$, P.~Paradisi$^{49}$, S.T.~Petcov$^{17}$$^{,50}$, M.~Picariello$^{51}$, V.~Porretti$^{18}$, A.~Poschenrieder$^{13}$, M.~Pospelov$^{52}$,
L.~Rebane$^{2}$, M.N.~Rebelo$^{15}$$^{,5}$, A.~Ritz$^{52}$, L.~Roberts$^{53}$, A.~Romanino$^{17}$, A.~Rossi$^{19}$, R.~R\"uckl$^{54}$, G.~Senjanovic$^{55}$,
N.~Serra$^{14}$, T.~Shindou$^{34}$, Y.~Takanishi$^{17}$, C.~Tarantino$^{13}$, A.M.~Teixeira$^{45}$, E.~Torrente-Lujan$^{56}$, K.J.~Turzynski$^{57}$$^{,20}$,
T.E.J.~Underwood$^{7}$, S.K.~Vempati$^{58}$, O.~Vives$^{18}$}
 \institute{
$^{1}$ Convener \\
$^{2}$ National Institute for Chemical Physics and Biophysics, 10143 Tallinn, Estonia \\
$^{3}$ Physik-Intitut der Universit\"at Z\"urich, CH-8057 Z\"urich, Switzerland \\
$^{4}$ Physics Dept., University of Notre Dame du Lac, Notre Dame, IN 46556, USA \\
$^{5}$ Physics Dept., CERN, CH-1211 Geneva, Switzerland \\
$^{6}$ Brookhaven National Laboratory, Upton, NY 11973-5000, USA \\
$^{7}$ Institute for Particle Physics Phenomenology, Durham University, Durham DH1 3LE, UK \\
$^{8}$ II. Institute for Theoretical Physics, University of Hamburg, D-22761 Hamburg, Germany \\
$^{9}$ Max-Planck-Institut f\"ur Physik, D-80805, M\"unchen, Germany \\
$^{10}$ Departamento de Fisica Teorica and IFT/CSIC-UAM, Universidad Autonoma de Madrid, E-28049 Madrid, Spain \\
$^{11}$ J. Stefan Institute, 1000 Ljubljana, Slovenia \\
$^{12}$ Dept. of Physics, University of Victoria, Victoria, BC, V8W 3P6 , Canada \\
$^{13}$ Physics Dept., TU Munich , D-85748 Garching, Germany \\
$^{14}$ Universit\`a  degli Studi di  Cagliari and INFN Cagliari, I-09042 Monserrato, (CA), Italy \\
$^{15}$ Departamento de F{\'\i}sica and Centro de F{\'\i}sica
   Te{\'o}rica de Part{\'\i}culas (CFTP), Instituto Superior T{\'e}cnico (IST), 1049-001 Lisboa, Portugal \\
$^{16}$ Dept. of Physics and Astronomy, Univ. of British Columbia, TRIUMF, Vancouver, BC, V6T 2A3, Canada \\
$^{17}$ SISSA and INFN, Sezione di Trieste, I-34013 Trieste, Italy \\
$^{18}$ Departament de F{\'\i}sica Te{\`o}rica, Universitat de Val\`encia-CSIC, E-46100, Burjassot, Spain \\
$^{19}$ Dipartimento di Fisica `G. Galilei' and INFN, I-35131 Padova, Italy \\
$^{20}$ Institute of Theoretical Physics, University of Warsaw, 00-681 Warsaw, Poland \\
$^{21}$ IPNL, CNRS, Universit\'e Lyon-1, F-69622 Villeurbanne Cedex, France \\
$^{22}$ Physics Dept., Yale University, New Haven, CO 06520, USA \\
$^{23}$ School of Physics and Astronomy, University of Manchester, Manchester, M13 9PL, UK \\
$^{24}$ Pontificia Univ. Catolica de Chile, Facultad de Fisica, Santiago 22, Chile \\
$^{25}$ Dept. of Physics, ETH Honggerberg, CH-8093 Zurich, Switzerland \\
$^{26}$ Theoretical Physics Division, Physical Research Lab., Navrangpura, Ahmedabad 380 009, India \\
$^{27}$ III. Physikalisches Institut B, RWTH Aachen, D-52056 Aachen, Germany \\
$^{28}$ Scuola Normale Superiore, I-56100 Pisa, Italy \\
$^{29}$ Univ. of Wisconsin, Dept. of Physics, High Energy Physics, Madison, WI 53706, USA \\
$^{30}$ Dept. of Physics and Astronomy, University of Sussex, Falmer, Brighton BN1 9QH, UK \\
$^{31}$ ICRR, University of Tokyo, Japan \\
$^{32}$ Physics Division, Argonne National Laboratory, Argonne, IL  60439-4843, USA \\
$^{33}$ Dept. of Physics, Univ. of Helsinki, and Helsinki Institute of Physics, FIN-00014 Helsinki, Finland \\
$^{34}$ DESY, Theory group, D-22603 Hamburg, Germany \\
$^{35}$ Physics Dept., University of Oregon, Eugene, OR, USA \\
$^{36}$ NIKHEF, 1098 SJ Amsterdam, the Netherlands \\
$^{37}$ Dept. of Physics, University of Zagreb, HR-10002 Zagreb, Croatia \\
$^{38}$ IPNS, KEK, Ibaraki 305-0801, Japan \\
$^{39}$ INFN, Laboratori Nazionali di Frascati, I-00044 Frascati, Italy \\
$^{40}$ School of Physics and Astronomy, University of Southampton, SO17 1BJ Southampton, UK \\
$^{41}$ Paul Scherrer Institut, CH-5232 Villigen, Switzerland \\
$^{42}$ Petersburg Nuclear Physics Institute, Gatchina 188300, Russia \\
$^{43}$ INFN and Dipartimento di Fisica, Universita di Pisa, I-56127 Pisa, Italy \\
$^{44}$ Dept. of Physics and Astronomy, University of California, Riverside, California 92521, USA \\
$^{45}$ Laboratoire de Physique Th\'eorique, UMR 8627 Universit\'e de Paris-Sud XI, F-91405 Orsay Cedex, France \\
$^{46}$ ICEPP, University of Tokyo, Tokyo, 113-0033, Japan \\
$^{47}$ Universit\`a dell'Aquila and INFN LNGS, I-67010, L'Aquila,  Italy \\
$^{48}$ Kernfysisch Versneller Instituut (KVI), NL 9747 AA Groningen, the Netherlands \\
$^{49}$ University of Rome ``Tor Vergata'' and INFN sez. RomaII, I-00133 Roma, Italy \\
$^{50}$ Institute of Nuclear Research and Nuclear Energy, Bulgarian Academy of Sciences, 1784 Sofia, Bulgaria \\
$^{51}$ INFN and Dipartimento di Fisica, Universit\`a del Salento, I-73100 Lecce, Italy \\
$^{52}$ Dept. of Physics and Astronomy, Univ. of Victoria, Victoria, BC, V8P 5C2, Canada \\
$^{53}$ Dept. of Physics, Boston University, Boston, MA 02215, USA \\
$^{54}$ Inst. for Theoretical Physics and Astrophysics, Univ. of W\"urzburg, D-97074 W\"urzburg, Germany \\
$^{55}$ International Centre for Theoretical Physics, Trieste, Italy \\
$^{56}$ Dept. of Physics, Univ. of Murcia, E-30100 Murcia, Spain \\
$^{57}$ Physics Dept., University of Michigan, Ann Arbor, MI 48109, USA \\
$^{58}$ Centre for High Energy Physics, Indian Institute of Science, Bangalore, 560012, India \\
 }
\maketitle

\begin{abstract}
This chapter of the report of the ``Flavour in the era of the LHC''
Workshop discusses the theoretical, phenomenological and experimental
issues related to flavour phenomena in the charged lepton sector and
in flavour-conserving CP-violating processes.  We review the current
experimental limits and the main theoretical models for the flavour
structure of fundamental particles. We analyze the phenomenological
consequences of the available data, setting constraints on explicit models
beyond the Standard Model, presenting benchmarks for the discovery potential of
forthcoming measurements both at the LHC and at low energy, and
exploring options for possible future experiments.
\end{abstract}

\tableofcontents


 \section{Charged leptons and fundamental dipole moments: alternative probes of the origin of flavour and CP-violation}\label{sec:mainintro}

The understanding of the flavour structure and CP-violation (CPV) of fundamental interactions has so far been dominated by the
phenomenology of the quark sector of the Standard Model (SM). More recently, the observation of neutrino masses and mixing has begun
extending this phenomenology to the lepton sector. While no experimental data available today link flavour and CP-violation in the
quark and in the neutrino sectors, theoretical prejudice strongly supports the expectation that a complete understanding should
ultimately expose their common origin. Most attempts to identify the common origin, whether through grand-unified (GUT) scenarios,
supersymmetry (SUSY), or more exotic electroweak symmetry breaking mechanisms, predict in addition testable correlations between the
flavour and CP-violation observables in the quark and neutrino sector on the one side, and new phenomena involving charged leptons and
flavour-conserving CP-odd effects on the other. This chapter of the ``Flavour in the era of the LHC'' report focuses precisely on the
phenomenology arising from these ideas, discussing flavour phenomena in the charged lepton sector and flavour-conserving CP-violating
processes.

Several theoretical arguments make the studies discussed in this chapter particularly interesting:
\begin{itemize}
\item the charged lepton sector provides unique opportunities to test scenarios tailored to explain flavour in the quark and neutrino
  sectors, for example by testing correlations between neutrino mixing and the rate for $\mu\to e\gamma$ decays, as predicted by specific
  SUSY/GUT scenarios. Charged leptons are therefore an indispensable element of the flavour puzzle, without which its clarification could be impossible.
\item The only observed source of CP-violation is so far the Cabibbo-Kobayashi-Maskawa (CKM) mixing matrix. On the other hand, it
  is by now well established that this is not enough to explain the observed baryon asymmetry of the Universe (BAU). The existence of other
  sources of CP-violation is therefore required. CP-odd phases in neutrino mixing, directly generating the BAU through leptogenesis,
  are a possibility, directly affecting the charged-lepton sector via, e.g., the appearance of electric dipole moments (EDMs).  Likewise,
  EDMs could arise via CP-violation in flavour conserving couplings, like phases of the gaugino fields or in extended Higgs sectors. In
  all cases, the observables discussed in this chapter provide an essential experimental input towards the understanding of the origin
  of CP-violation.
\item The excellent agreement of all flavour observables in the quark sector with the CKM picture of flavour and CP-violation has recently
  led to the concept of Minimal Flavour Violation (MFV). In scenarios beyond the SM (BSM) with MFV, the smallness of possible deviations
  from the SM is naturally built into the theory. While these schemes provide a natural setting for the observed lack of new physics (NP)
  signals, their consequence is often a reduced sensitivity to the underlying flavour dynamics of most observables accessible by the
  next generation of flavour experiments. Lepton flavour violation (LFV) and EDMs could therefore provide our only probe into this dynamics. 
\item Last but not least, with the exception of the magnetic dipole moments, where the SM predicts non-zero values and deviations due to new
  physics compete with the effect of higher-order SM corrections, the observation of a non-zero value for any of the observables discussed
  in this chapter would be unequivocal indication of new physics. In fact, while neutrino masses and mixing can mediate lepton flavour
  violating transitions, as well as induce CP-odd effects, their size is such that all these effects are by many orders of
  magnitude smaller than anything measurable in the foreseeable future. This implies that, contrary to many of the observables
  considered in other chapters of this report, and although the signal interpretation may be plagued by theoretical ambiguities or
  systematics, there is nevertheless no theoretical systematic uncertainty to claim a discovery once a positive signal is detected.
\end{itemize}
The observables discussed here are also very interesting from the experimental point of view. They call for a very broad approach, based
not only on the most visible tools of high-energy physics, namely the high-energy colliders, but also on a large set of smaller-scale
experiments that draw from a wide variety of techniques. The emphasis of these experiments is by and large on high rates and high
precision, a crucial role being played by the control of very large backgrounds and subtle systematics. A new generation of such
experiments is ready to start, or will start during the first part of the LHC operations. More experiments have been on the drawing board
for some time, and could become reality during the LHC era if the necessary resources were made available. The synergy between the
techniques and potential results provided by both the large- and small-scale experiments makes this field of research very rich and
exciting and gives it a strong potential to play a key role in exploring the physics landscape in the era of the LHC.

The purpose of this document is to provide a comprehensive overview of the field, from both the theoretical and the experimental
perspective. Several of the results presented are well known from the existing literature, but are nevertheless documented here to provide a
self-contained review, accessible to physicists whose expertise covers only some of the many diverse aspects of this subject.  Many
results emerged during the Workshop, including ideas on possible new experiments, further enrich this report. We present here a short
outline, and some highlights, of the contents.

Section~\ref{sec:framework} provides the general theoretical framework that allows to discuss flavour from a symmetry point of view. It
outlines the origin of the flavour puzzles and lists the mathematical settings that have been advocated to justify or predict the
hierarchies of the mixing angles in both the quark and neutrino sectors. Section~\ref{sec:observables} introduces the observables that
are sensitive to flavour in the charged-lepton sector and to flavour-conserving CP-violation, providing a unified description in
terms of effective operators and effective scales for the new physics that should be responsible for them. The existing data already provide 
rather stringent limits on the size of these operators, as shown in several tables. We collect here in Table~\ref{tab:summary-limits}
some of the most significant benchmark results (for details, we refer to the discussion in Section~\ref{sec:taudecays}). We constrain the
dimensionless coefficients $\epsilon_i$ of effective operators $O_i$ describing flavour- or CP-violating interactions. Examples of these
effective operators include:
\begin{equation}
\overline{\ell}_i \sigma^{\mu \nu} \gamma_5 \ell_{i} F_{\mu \nu}^{\rm em} \; , \quad \overline{\ell}_i \sigma^{\mu \nu} \ell_{j} F_{\mu \nu}^{\rm em} \; ,
\end{equation}
which describe a CP-violating electric dipole moment (EDM) of lepton $\ell_i$ or the flavour-violating decay $\ell_i \to \ell_j \gamma$, or the four-fermion operators:
\begin{equation}
\overline{\ell_i} \Gamma^a \ell_j \; \overline{q_k} \Gamma_{a} q_l \;, \quad \overline{\ell_i} \Gamma^a \ell_j \; \overline{\ell_k} \Gamma_{a} \ell_l \;,
\end{equation}
where the $\Gamma_a$ represent the various possible Lorentz structures. The overall normalization of the operators is chosen to
reproduce the strength of transitions mediated by weak gauge bosons, assuming flavour mixing angles and CP-violating phases of order unity. In this way, the
$\epsilon$ coefficients will scale like:
\begin{equation}
\epsilon \sim \frac{m_W^2}{m_{NP}^2} \; \frac{g_{NP}^2}{g_W^2} \; \delta_{CPV} \; \delta_{mix} \; ,
\end{equation}
\begin{table}[!bht]
\begin{minipage}{\textwidth}
\caption{Bounds on CP- or flavour-violating effective operators,  expressed as upper limits on their dimensionless 
coefficients $\epsilon$, scaled to the strength of weak interactions. For more details, in particular the overall normalization convention
for the effective operators, see Section~\ref{sec:taudecays} \label{tab:summary-limits} }
\begin{tabular*}{\textwidth}{@{\extracolsep{\fill}} l|ll  }
\hline\hline
Observable & Operator & Limit on $\epsilon$ \\ 
\hline
$e$EDM 	& $\overline{e_L} \sigma^{\mu \nu} \gamma_5 e_{R} F_{\mu \nu} $ & $\le 1.1 \times 10^{-3}$ \\
B($\mu\to e\gamma$) & $\overline{\mu} \sigma^{\mu \nu} e F_{\mu \nu} $ & $\le 1.4 \times 10^{-4}$ \\
B($\tau\to \mu\gamma$) & $\overline{\tau} \sigma^{\mu \nu} \mu F_{\mu \nu} $ & $\le 2.2 \times 10^{-2}$ \\
B($K^0_L\to \mu^\pm e^\mp$) & $(\overline{\mu} \gamma^\mu P_L e)(\overline{s} \gamma^\mu P_L d)$ & $\le 2.9 \times 10^{-7}$ \\
\hline\hline
\end{tabular*}
\end{minipage}
\end{table} 
where $m_{NP}$ ($m_W$) and $g_{NP}$ ($g_W$) are the mass scale and coupling strength of the new physics (of weak interactions), with
$g_{NP}$ absorbing the size of the (possibly suppressed) CP-violating and flavour-mixing phases.  The smallness of the constraints on
$\epsilon$ therefore reflects either the large mass scale of flavour phenomena, or the weakness of the relative interactions.

It is clear from this table that current data are already sensitive to mass scales much larger than the electroweak scale, or to very small
couplings. On the other hand, many of these constraints leave room for interesting signals coupled to the new physics at the TeV scale that can be directly
discovered at the LHC. For example, a mixing of order 1 between the supersymmetric scalar partners of the charged leptons, and a mass splitting among them
of the order of the lepton masses, is consistent with the current limits if the scalar lepton masses are just above 100 GeV, and could lead both to their
discovery at the LHC, and to observable signals at the next generation of $\ell \to \ell^\prime \gamma$ experiments.

Most of this report will be devoted to the discussion of the phenomenological consequences of limits such as those in Table~\ref{tab:summary-limits},
setting constraints on explicit BSM models, presenting benchmarks for the discovery potential of forthcoming measurements both at the LHC
and at low energy, and exploring options for future experiments aimed at increasing the reach even further.

Section~\ref{sec:observables} also introduces the phenomenological parameterizations of the quark and lepton mixing matrices that are
found in the literature, emphasizing with concrete examples the correlations among the neutrino and charged-lepton sectors that arise
in various proposed models of neutrino masses. The section is completed by a discussion of the possible role played by leptogenesis and
cosmological observables in constraining the neutrino sector.

Section~\ref{sec:organising} reviews the organizing principles for flavour physics. With a favourite dynamical theory of flavour
still missing, the extended symmetries of BSM theories can provide some insight in the nature of the flavour structures of quarks and
leptons, and give phenomenologically relevant constraints on low-energy correlations between them. In GUT theories, for example,
leptons and quarks belong to the same irreducible representations of the gauge group, and their mass matrices and mixing angles are consequently
tightly related. Extra-dimensional theories provide a possible dynamical origin for flavour, linking flavour to the geometry of the
extra dimensions. This section also discusses the implications of models adopting for the lepton sector the same concept of MFV
already explored in the case of quarks.

Section~\ref{sec:phenomenology} discusses at length the phenomenological consequences of the many existing models, and
represents the main body of this document. We cover models based on SUSY, as well as on alternative descriptions of electroweak symmetry
breaking, such as Little Higgs or extended Higgs sectors. In this section we discuss the predictions and the detection prospects of
standard observables, such as $\ell \to \ell ^\prime \gamma$ decays or EDMs, and connect the discovery potential for these observables with
the prospects for direct detection of the new massive particles at the LHC or at a future Linear Collider. 

This section underlines, as is well known, that the exploration of these processes has great discovery potential, since most BSM models
anticipate rates that are within the reach of the forthcoming experiments. From the point of view of the synergy with collider physics,
the remarkable outcome of these studies is that the sensitivities reached in the searches for rare lepton decays and dipole moments
are often quite similar to those reached in direct searches at high energy. We give here some explicit examples. In SO(10) SUSY GUT models,
where the charged-lepton mixing is induced via renormalization-group evolution of the heavy neutrinos of different generations, the observation
of $B(\mu\to e\gamma)$ at the level of 10$^{-13}$, within the range of the just-starting MEG experiment, is suggestive of the existence of
squarks and gluinos with a mass of about 1~TeV, well within the discovery reach of the LHC. Squarks and gluinos in the range of 2-2.5~TeV, at
the limit of detectability for the LHC, would push $B(\mu\to e\gamma)$ down to the level of 10$^{-16}$. While this is well beyond the MEG sensitivity,
it would well fit the ambitious goals of the next-generation $\mu\to e$ conversion experiments, strongly endorsing their plans. The decay $\mu \to e \gamma$
induced by the mixing of the scalar partners of muon and electron, and with a $B(\mu\to e\gamma)$ at the level of 10$^{-13}$, could give a
$\chi_2^0 \to \chi_1^0\mu^\pm e^\mp$ signal at the LHC, with up to 100 events after 300 fb$^{-1}$. Higher statistics and a cleaner signal would
arise at a Linear Collider. Models where neutrino masses arise not from a seesaw mechanism at the GUT scale, but from triplet Higgs fields at
the TeV scale, can be tested at the LHC, where processes like $pp\to H^{++}H^{- -}$ can be detected for $m_{H^{++}}$ up to 700~GeV, using the
remarkable signatures due to $B(H^{++}\to\tau^+\tau^+)$=$B(H^{++}\to \mu^+\mu^+)$=$B(H^{++}\to\mu^+\tau^+)$=1/3. 

Should signals of new physics be observed, alternative interpretations can be tested by exploiting different patterns of correlations that they predict
among the various observables. For example, while typical SUSY scenarios predict $B(\mu \to 3e)\sim 10^{-2}B(\mu\to e\gamma)$, these branching ratios
are of the same order in the case of Little Higgs models with T parity. Important correlations also exist in seesaw SUSY GUT models between $B(\mu\to
e\gamma)$ and $B(\tau\to \mu\gamma)$ or $B(\tau\to e\gamma)$. Furthermore, SUSY models with CP-violation in the Higgs or gaugino mass matrix, be them supergravity 
(SUGRA) inspired or of the split-SUSY type, predict the ratio of electron and neutron EDM to be in the range of $10^{-2} - 10^{-1}$. Furthermore, in SUSY GUT
models with seesaw mechanism correlations exist between the values of the neutron and deuteron EDMs and the heavy neutrino masses.

Section~\ref{sec:exp:universality} discusses studies of lepton universality. The branching ratios $\Gamma(\pi\to\mu\nu)/\Gamma(\pi\to e\nu)$ and
$\Gamma(K\to\mu\nu)/\Gamma(K\to e\nu)$, for example, are very well known theoretically within the SM. Ongoing experiments (at PSI and TRIUMF
for the pion, and at CERN and Frascati for the kaon) test the existence of flavour-dependent charged-Higgs couplings, by improving the existing
accuracies by factors of order 10.

In Section~\ref{sec:exp:CP} we consider CP-violating charged lepton decays, which offer interesting prospects as alternative probes of BSM phenomena.
SM-allowed $\tau$ decays, such as $\tau \to \nu K\pi$, can be sensitive to new CP-violating effects. The decays being allowed by the SM, the CP-odd asymmetries
are proportional to the interference of a SM amplitude with the BSM, CP-violating one. As a result, the small CP-violating amplitude
contributes linearly to the rate, rather than quadratically, enhancing the sensitivity. In the specific case of $\tau \to \nu K\pi$, and
for some models, a CP asymmetry at the level of $10^{-3}$ would correspond to  $B(\tau \to \mu \gamma)$ around $10^{-8}$. Another example is the CP-odd
transverse polarization of the muon, $P_T$, in $K\to \pi\mu\nu$ decays. The current sensitivity of the KEK experiment E246, which resulted in
$P_T<5\times 10^{-3}$ at 90\% CL, can be improved to the level of $10^{-4}$, by TREK proposed at J-PARC, probing models such as multi-Higgs or R-parity-violating SUSY.

Section~\ref{sec:exp:LFV} discusses experimental searches for charged LFV processes. Transitions between $e$, $\mu$, and $\tau$ might be found in the decay of almost
any weakly decaying particle and searches have been performed in $\mu$, $\tau$, $\pi$, $K$, $B$, $D$, $W$ and $Z$ decay. Whereas highest experimental sensitivities
were reached in dedicated $\mu$ and $K$ experiments, $\tau$ decay starts to become competitive as well. In Section~\ref{sec:exp:LFV} experimental limitations to the
sensitivities for the various decay modes are discussed in some detail, in particular for $\mu$ and $\tau$ decays, and some key experiments are presented. 
The sensitivities reached in searches for $\mu^+ \to e^+ \gamma$ are limited by accidental $e^+ \gamma$ coincidences and muon beam intensities have to be reduced now
already. Searches for $\mu - e$ conversion, on the other hand, are limited by the available beam intensities and large improvements in sensitivity may still be
achieved. Similarly, in rare $\tau$ decays some decay modes are already background limited at the present $B$-factories and future sensitivities may not scale
with the accumulated luminosities. Prospects of LFV decays at the LHC are limited to final states with charged leptons, such as $\tau \to 3\mu$ and
$B^0_{d,s}\to e^\pm\mu^\mp$, which are discussed in detail. This section finishes with the preliminary results of a feasibility study for in-flight $\mu \to \tau$
conversions using a wide beam of high-momentum muons. No working scheme emerged yet.

Section~\ref{sec:exp:dipoles} covers electric and magnetic dipole moments. The muon magnetic moment has been much discussed recently, so
we limit ourselves to a short review of the theoretical background and of the current and foreseeable experimental developments. In
the case of EDMs, we provide an extensive description of the various theoretical approaches and experimental techniques applied to test
electron and quark moments, as well as other possible sources of flavour diagonal CP-violating effects, such as the gluonic $\theta
\tilde{F}F$ coupling, or CP-odd 4-fermion interactions. While the experimental technique may differ considerably, the various systems provide
independent and complementary information. EDMs of paramagnetic atoms such as Tl are sensitive to a combination
of the fundamental electron EDM and CP-odd 4-fermion interactions between nucleons and electrons. EDMs of diamagnetic atoms such as Hg are
sensitive, in addition, to the intrinsic EDM of quarks, as well as to a non-zero QCD $\theta$ coupling. The neutron EDM more directly
probes intrinsic quark EDMs, $\theta$, and possible higher-dimension CP-odd quark couplings. EDMs of the electron, without contamination
from hadronic EDM contributions, can be tested with heavy diatomic molecules with unpaired electrons, such as YbF. 
In case of a positive signal the combination of measurements would help to disentangle the various contributions.

The experimental situation looks particularly promising, with several new experiments about to start or under construction. For example,
new ultracold-neutron setups at ILL, PSI and Oak Ridge will increase the sensitivity to a neutron EDM by more than 2 orders of magnitude,
to a level of about $10^{-28}$ $e\,$cm in 5--10 years. This sensitivity probes e.g. CP-violating SUSY phases of the order of 10$^{-4}$ or
smaller. Similar improvements are expected for the electron EDM. One of the main new ideas developed in the course of the Workshop is the use
of a storage ring to measure the deuteron EDM. The technical issues related to the design and construction of such an
experiment, which could have a statistical sensitivity of about $10^{-29}$ $e\,$cm, are discussed here in some detail. 

All the results presented in this document prove the great potential of this area of particle physics to shed light on one of the main
puzzles of the Standard Model, namely the origin and properties of flavour. Low-energy experiments are sensitive to scales of new physics
that in several cases extend beyond several TeV. The similarity with the scales directly accessible at the LHC supports the expectation of
an important synergy with the LHC collider programme, a synergy that clearly extends to future studies of the neutrino and quark sectors.
The room for improvement, shown by the projections suggested by the proposed experiments, finally underscores the importance of keeping
these lines of research at the forefront of the experimental high energy physics programme, providing the appropriate infrastructure, support and funding.

\section{Theoretical framework and flavour symmetries}\label{sec:framework}
\subsection{The flavour puzzle}\label{sec:flavourpuzzle}
The flavour puzzle in the Standard Model  is associated to the
 presence of three fermion families with identical gauge quantum
 numbers. The very origin of this replication of families constitutes
 the first element of the SM flavour puzzle. The second element has to
 do with the Yukawa interactions of those three families of
 fermions. While the gauge principle allows to determine all SM gauge
 interactions in terms of three gauge couplings only (once the SM
 gauge group and the matter gauge quantum numbers have been
 specified), we do not have a clear evidence of a guiding principle
 underlying the form of the $3\times 3$ matrices describing the SM
 Yukawa interactions. Finally, a third element of the puzzle is
 represented by the peculiar pattern of fermion masses and mixing
 originating from those couplings.

The replication of SM fermion families can be rephrased in terms of
the symmetries of the gauge part of the SM Lagrangian. The latter is
in fact symmetric under a U(3)$^5$ symmetry acting on the family
indexes of each of the 5 inequivalent SM representations forming a
single SM family ($q, u^c, d^c, l, e^c$ in Weyl notation). In other
words, the gauge couplings and interactions do not depend on the
(canonical) basis we choose in the flavour space of each of the 5 sets
of fields $q_i, u_i^c, d_i^c, l, e_i^c$, $i=1,2,3$.

This U(3)$^5$ symmetry is explicitly broken in the Yukawa sector by
the fermion Yukawa matrices. It is because of this breaking that the
degeneracy of the three families is broken and the fields
corresponding to the physical mass eigenstates, as well as their
mixing, are defined. An additional source of breaking is provided by
neutrino masses. The smallness of neutrino masses is presumably due to
the breaking of the accidental lepton symmetry of the SM at a scale
much larger than the electroweak, in which case neutrino masses and
mixing can be accounted for in the SM effective Lagrangian in terms
of a dimension-five operator breaking the U(3)$^5$ symmetry in the
lepton doublet sector.

As mentioned, the special pattern of masses and mixing originating
from the U(3)$^5$ breaking is an important element of the flavour
puzzle. This pattern is quite peculiar. It suffices to mention
the smallness of neutrino masses; the hierarchy of charged fermion
masses and the milder or absent hierarchy between the two heavier
neutrinos; the smallness of Cabibbo-Kobayashi-Maskawa mixing in
the quark sector and the two large mixing angles in
Pontecorvo-Maki-Nakagawa-Sakata (PMNS) matrix in the lepton sector;
the mass hierarchy in the up quark sector, more pronounced than in the
down quark and charged lepton sectors; the presence of a large
CP-violating phase in the quark sector and the need of additional
CP-violation to account for baryogenesis; the approximate equality of
bottom and tau masses at the scale at which the gauge couplings
unify\footnote{Needless to say, precise unification requires an
extension of the SM, with supersymmetry doing best from this
point of view.} and the approximate factor of 3 between the strange
and muon masses, both pointing at a grand unified picture at high
energy.

The origin of family replication and of the peculiar pattern of
fermion masses and mixing are among the most interesting open
questions in the SM, which a theory of flavour, discussed in
Section~\ref{sec:framework}, should address.  As seen in
Section~\ref{sec:observables}, experiment is ahead of theory in this
field.  All the physical parameters describing the SM flavour
structure in the quark sector have been measured with good accuracy.
In the lepton sector crucial information on lepton mixing and neutrino
masses has been gathered and a rich experimental program is under way
to complete the picture.

Several tools are used to attack the flavour problem. Grand Unified
Theories  allow to relate quark and lepton masses at the GUT
scale and provide an appealing framework to study neutrino masses,
leptogenesis, flavour models, etc. Note that in a grand unified
context the U(3)$^5$ symmetry of the gauge sector is reduced (to U(3)
in the case in which all fermions in a family are unified in a single
representation, as in SO(10)). Extra-dimensions introduce new ways to
account for the hierarchy of charged fermion masses (and in some cases
for the smallness of neutrino masses) through the mechanism of
localization in extra-dimensions and by providing a new framework for
the study of flavour symmetries.  The ideology of minimal flavour
violation may also provide a framework for addressing flavour.  Impact
of those organizing principles on flavour physics is discussed in
detail in Section~\ref{sec:organising}.

From experimental point of view, however, additional handles are
needed to gain a firmer understanding on the origin of
flavour. Essentially this requires a discovery of new physics beyond
the SM.  New physics at the TeV scale may in fact be associated with
an additional flavour structure, whose origin might well be related to
the origin of the Yukawa couplings. Some of the presents attempts to
understand the pattern of fermion masses and mixing do link the
flavour structure of the SM and that of the new physics sectors. In
which case, the search for indirect effects at low energy and for
direct effects at colliders may play a primary role in clarifying our
understanding of flavour. And conversely, the attempts to understand
the pattern of fermion masses and mixing might lead to the prediction
of new flavour physics effects. Those issues are addressed in
Section~\ref{sec:phenomenology}.

Finally, lepton flavour physics is not just related to the lepton
flavour violation  or CP- violation in the lepton sector
but also to understanding the unitarity and universality in the lepton
sector. Possible deviations from those are discussed in
Section~\ref{sec:unitarity}.

\subsection{Flavour symmetries}\label{sec:flavoursymmetries}

The SM Lagrangian is $U(3)^5$ invariant in the limit in which the
Yukawa couplings vanish. This might suggest that the Yukawa couplings,
or at least some of them, arise from the spontaneous breaking of a
subgroup of $U(3)^5$. Needless to say, the use of (spontaneously
broken) symmetries as organizing principles to understand physical
phenomena has been largely demonstrated in the past (chiral symmetry
breaking, electroweak, etc). In the following, we discuss the
possibility of using such an approach to address the origin of the
pattern of fermion masses and mixing, the constraints on the flavour
structure of new physics, and to put forward expectations for flavour
observables.

The spontaneously broken ``flavour'' or ``family'' symmetry can be
local or global. Many (most) of the consequences of flavour symmetries
are independent of this. The flavour breaking scale must be
sufficiently high in such a way to suppress potentially dangerous
effects associated with the new fields and interactions, in particular
with the new gauge interactions (in the local case) or the unavoidable
pseudo-Goldstone bosons (in the global case).  In the context of an
analysis in terms of effective operators of higher dimensions, a
generic bound of about $10^3$ TeV on the flavour scale from flavour
changing neutral currents (FCNC) processes would be
obtained. Nevertheless, a certain evidence for $b$-$\tau$ unification
and the appeal of the see-saw mechanism for neutrino masses seem to
suggest that these Yukawa couplings are already present near the GUT
scale. This is indeed what most flavour models assume and we will also
assume in the following.

The SM matter fields belong to specific representations of the flavour
group, such that, in the unbroken limit the Yukawa couplings have a
particularly simple form. Typically some or all Yukawa couplings (with
the possible exception of third generation ones) are not allowed. The
spontaneous symmetry breaking of the flavour symmetry is provided by
the vacuum expectation value (VEV) of fields often called ``flavons''. As the breaking presumably
arises at a scale much higher than the electroweak scale, such flavons
are SM singlets (or contain a SM singlet in the case of SM extensions)
and typically they are only charged under the flavour symmetry.
Flavour breaking is communicated dynamically to the SM fields by some
physics (possibly renormalizable, often not specified) living at a
scale $\Lambda_f$ not smaller than the scale of the flavour symmetry
breaking. A typical example for these physics that communicate the
breaking is the exchange of heavy fermions whose mass terms respect
the flavour symmetry. In that case the scale $\Lambda_f$ would
correspond to this fermion mass $M_f$.  Many consequences of the
flavour symmetry are actually independent of the mediator physics. It
is therefore useful to consider an effective field theory approach
below the scale $\Lambda_f$ in which the flavour messengers have been
integrated out. Once the flavon fields have acquired their VEVs, the
structure of the Yukawa matrices (and other flavour parameters) can be
obtained from an expansion in non-renormalizable operators involving
the flavon fields and respecting the different symmetries (flavour and
other symmetries) of the theory.

There are several possibilities for the flavour symmetry, local,
global, accidental, continuous or discrete, Abelian or
non-Abelian. Many examples are available in the literature for each of
those possibilities. Some of them will be discussed in next
subsections in relation to the implications considered in this study.

\subsubsection{Flavour symmetries - continuous examples }
\label{sec:contflav}
In order to provide an explicit example, we shortly discuss here one
of the simplest possibilities, which goes back to the pioneering work
of Froggatt-Nielsen \cite{Froggatt:1978nt}.  In this model we have a
$U(1)$ flavour symmetry under which the three generation of SM fields
have different charges. In the simplest version we assign positive
integer charges to the SM fermionic fields, the Higgs field is neutral
and we have a single flavon field $\theta$ of charge $-1$. The VEV of the
flavon field is somewhat smaller than the mass of the heavy mediator
fields $M_f$, so that the ratio, $\epsilon = v/M_f \ll 1$. In this
way the different entries in the Yukawa matrices are determined by
epsilon to the power of the sum of the fermion charges with an
undetermined order one coefficient. This mechanism explains nicely the
hierarchy of fermion masses and mixing angles.

This idea is the basis for most flavour symmetries. It can be
implemented in a great variety of different models. For the sake of
definiteness, we show here how it works using as a concrete example a
supersymmetric GUT model. Its superpotential is of the form: 
\begin{equation}
W_{\rm Yukawa} = c^d_{ij} ~\epsilon^{q_i+d^c_j}~Q_i D^c_j H_1 +
c^u_{ij}~\epsilon^{q_i+u^c_j}~ Q_i U^c_j H_2 +
c^e_{ij}~\epsilon^{l_i+e^c_j} ~L_i E^c_j H_1 + c^\nu_{ij}~
\epsilon^{l_i+l_j}~ L_i L_j \frac{H_2 H_2}{\bar M}
\label{effspot}
\end{equation}
where the $c$'s are $O(1)$ coefficients and $\bar M$ is the scale
associated to $B-L$ breaking.  The last term in this equation is an
effective operator, giving Majorana masses to neutrinos, which can be
generated, e.g., through a seesaw mechanism.  Notice that the power of
$\epsilon$ in each Yukawa coupling is proportional to the sum of the
fermion charges: $Y^u_{ij}=c^u_{ij} \epsilon^{q_i+u^c_j}$,
$Y^d_{ij}=c^d_{ij} \epsilon^{q_i+d^c_j}$, etc.  Hence, this mechanism
explains the hierarchy of fermion masses and mixing angles through a
convenient choice of charges. The value of these charges and the
expansion parameter $\epsilon$ are constrained by the observed masses
and angles.  A convenient set of charges for example is given in
Table~\ref{tab:charges2}.
\begin{table}[t!] 
\caption{Transformation of the matter superfields under the family
symmetries. The i-th generation SM fermion fields are grouped into the
representation $\bar 5_i=(D^c,L)_i$, $10_i=(Q,U^c,E^c)_i$,
$1_i=(N^c)_i$.\label{tab:charges2}}
\begin{tabular*}{1.00\textwidth}{  @{\extracolsep{\fill}} c|c c c c c c c c  c c}
\hline\hline Field &$10_3$ &$10_2$ &$10_1$ &$\bar 5_3$ &$\bar 5_2$
&$\bar 5_1$ &$1_3$ &$1_2$ &$1_1$ &$\theta$ \\ \hline $U(1)$ &$0$ &2 &3
&$0$ &$0$ &1 &$n^c_3$ &$n^c_2$ &$n^c_1$ &${-1}$ \\ \hline\hline
\end{tabular*}
\end{table}
It turns out that this set of charges is the only one compatible with
minimal $SU(5)$ unification. By introducing three right-handed
neutrinos with positive charges it is also possible to successfully
realize the seesaw mechanism.

These charges give rise to the following Dirac Yukawa couplings for
charged fermions at the GUT scale
\begin{equation}
\label{chargedYuk}
Y_u= \left( \begin{matrix} \epsilon^6&\epsilon^5&\epsilon^3 \cr
\epsilon^5&\epsilon^4&\epsilon^2 \cr \epsilon^3&\epsilon^2&1
\end{matrix} \right), \qquad \qquad 
\left( \begin{matrix} \epsilon^4&\epsilon^3&\epsilon^3 \cr
\epsilon^3&\epsilon^2&\epsilon^2 \cr \epsilon&1&1 \end{matrix}
\right), 
\end{equation}
where $O(1)$ coefficients in each entry are understood here and in the
following.  With $\epsilon =O(\lambda_c)$ (the Cabibbo angle), the
observed features of charged fermion masses and mixing are
qualitatively well reproduced. It is known that the high energy
relation $Y_e^T = Y_d$ is not satisfactory for the lighter families
and should be relaxed by means of some mechanism
\cite{Georgi:1979df,Ellis:1979fg,Anderson:1993fe}.  The Dirac neutrino
Yukawa couplings and the Majorana mass matrix of right handed
neutrinos are
\begin{equation}
Y_\nu = \left( \begin{matrix}
\epsilon^{n^c_1+1}&\epsilon^{n^c_2+1}&\epsilon^{n^c_3+1} \cr
\epsilon^{n^c_1}&\epsilon^{n^c_2}&\epsilon^{n^c_3} \cr
\epsilon^{n^c_1}&\epsilon^{n^c_2}&\epsilon^{n^c_3} \end{matrix}
\right) ,~~~~ M_R=\left( \begin{matrix} \epsilon^{2n^c_1} &
\epsilon^{n^c_1+n^c_2} & \epsilon^{n^c_1 + n^c_3} \cr \epsilon^{n_1^c
+ n_2^c}& \epsilon^{2 n^c_2}& \epsilon^{n^c_2 + n^c_3} \cr
\epsilon^{n_1^c + n_3^c}& \epsilon^{n_2^c + n_3^c}& \epsilon^{2 n_3^c}
\end{matrix} \right) \bar M .
\end{equation}      
Applying the seesaw mechanism to obtain the effective light neutrino
mass matrix $M_\nu$ in the basis of diagonal charged lepton Yukawa
couplings\footnote{Notice, going to the basis of diagonal charged
leptons will only change the $O(1)$ coefficients, but not the power in
$\epsilon$ of the different entries.}, it is well known
\cite{Irges:1998ax,Elwood:1998kf}, if all right-handed neutrino masses
are positive, that the dependence on the right-handed charges
disappears:
\begin{equation}
U_{PMNS}^*~ m_\nu^{diag}~ U_{PMNS}^\dagger=m_\nu = \left( \begin{matrix} \epsilon^2&\epsilon&\epsilon \cr \epsilon&1&1 \cr \epsilon&1&1 \end{matrix}  \right)  \frac{v_2^2}{\bar M}.
\end{equation}
Experiments require $\bar M\sim 5\times 10^{14}$ GeV.  The features of
neutrino masses and mixing are quite satisfactorily reproduced -- the
weak point being the tuning in the 23-determinant
\cite{Irges:1998ax,Elwood:1998kf} that has to be imposed.  For later
application, it is useful to introduce the unitary matrices which
diaginalize $Y_\nu$ in the basis where both $Y_e$ and $M_R$ are
diagonal: $V_L Y_\nu V_R =
Y_\nu^{diag}\approx$diag$(\epsilon^{n^c_1},\epsilon^{n^c_2},\epsilon^{n^c_3})$.
Notice that, as a consequence of the equal charges of the lepton
doublets $L_2$ and $L_3$, the model predicts that $V_L$ has a large
mixing, although not necessarily maximal, in the 2--3 sector as
observed in $U_{PMNS}$.

The literature is very rich of models based on flavour symmetries, some references are
\cite{Froggatt:1978nt,Leurer:1992wg,Dine:1993np,Nir:1993mx,Leurer:1993gy,Kaplan:1993ej,Carone:1995xw,Pomarol:1995xc,Barbieri:1995uv,Nir:1996am,Binetruy:1996xk,
Binetruy:1996cs,Dudas:1996fe,Barbieri:1996ww,Barbieri:1997tg,Carone:1997qg,Barbieri:1997tu,Irges:1998ax,Elwood:1998kf,Barbieri:1998em,Hall:1998cu,Barbieri:1998qs,
Aranda:1999kc,Barbieri:1999pe,Berezhiani:2000cg,Aranda:2000tm,Aranda:2001rd,Chen:2000fp,King:2001uz,Ma:2001dn,Chkareuli:2001dq,Lavignac:2001vp,Roberts:2001zy},
for more recent attempts the interested reader is referred for instance to \cite{Babu:2002dz,Ross:2002fb,Nir:2002ah,Dreiner:2003yr,King:2003rf,Ross:2004qn,
Grimus:2004rj,Grimus:2005mu, Vives:2005ze, Berezhiani:2005tp,Altarelli:2005yx,deMedeirosVarzielas:2005ax,Kane:2005va,Chankowski:2005qp,Abel:2000hn,Diaz-Cruz:2005qz, 
Masina:2006ad, Masina:2006pe, Ferretti:2006df,Appelquist:2006ag,Hagedorn:2006ug,deMedeirosVarzielas:2006fc,King:2006np,Feruglio:2007uu}.

\subsubsection{Flavour symmetries - discrete examples}
\subsubsubsection{Finite groups} 

Discrete flavour symmetries have gained popularity because they seem
to be appropriate to address the large mixing angles observed in
neutrino oscillations.  To obtain a non-Abelian discrete symmetry, a
simple heuristic way is to choose two specific non-commuting matrices
and form all possible products.  As a first example, consider the two
$2 \times 2$ matrices:
\begin{equation}
\label{eq:ma0}
A = \begin{pmatrix}  0 & 1 \\ 1 & 0
\end{pmatrix}, 
~~~ B = \begin{pmatrix} \omega & 0 \\ 0 & 
\omega^{-1}
\end{pmatrix},
\end{equation}
where $\omega^n = 1$, i.e. $\omega = \exp(2\pi i/n)$.  Since $A^2=1$ and 
$B^n=1$, this group contains $Z_2$ and $Z_n$.  For $n=1,2$, we obtain 
$Z_2$ and $Z_2 \times Z_2$ respectively, which are Abelian.  For $n=3$, 
the group generated has 6 elements and is in fact the smallest non-Abelian 
finite group $S_3$, the permutation group of 3 objects.  This particular 
representation is not the one found in text books, but is related to it 
by a unitary transformation \cite{Ma:2004pt}, and was first used in 1990 for 
a model of quark mass matrices \cite{Ma:1990qh,Deshpande:1991zh}. For $n=4$, the group 
generated has 8 elements which are in fact $\pm 1$, $\pm i\sigma_{1,2,3}$, 
where $\sigma_{1,2,3}$ are the usual Pauli spin matrices.  This is the 
group of quaternions $Q$, which has also been used \cite{Frigerio:2004jg} 
for quark and lepton mass matrices.  In general, the groups generated by 
Eq.~(\ref{eq:ma0}) have $2n$ elements and may be denoted as $\Delta(2n)$.

Consider next the two $3 \times 3$ matrices:
\begin{equation}
\label{eq:ma00}
A = \begin{pmatrix}0 & 1 & 0 \\ 0 & 0 & 1 \\ 1 & 0 & 0\end{pmatrix} ~~~ 
B = \begin{pmatrix}\omega & 0 & 0 \\ 0 & \omega^2 & 0 \\ 0 & 0 & \omega^{-3}\end{pmatrix}. 
\end{equation}
Since $A^3=1$ and $B^n=1$, this group contains $Z_3$ and $Z_n$.  For $n=1$, 
we obtain $Z_3$.  For $n=2$, the group generated has 12 elements and is 
$A_4$, the even permutation group of 4 objects, which was first used in 
2001 in a model of lepton mass matrices \cite{Ma:2001dn,Babu:2002dz}.  It is also 
the symmetry group of the tetrahedron, one of five perfect geometric 
solids, identified by Plato with the element ``fire'' \cite{Ma:2002ge}.  In 
general, the groups generated by Eq.~(\ref{eq:ma00}) have $3n^2$ elements and may 
be denoted as $\Delta(3n^2)$ \cite{Luhn:2007uq}.  They are in fact subgroups of 
$SU(3)$. In particular, $\Delta(27)$ has also been used \cite{Ma:2006ip,
deMedeirosVarzielas:2006fc}. 
Generalizing to $k \times k$ matrices, we then have the series 
$\Delta(kn^{k-1})$. However, since there are presumably only 3 families, 
$k>3$ is probably not of much interest.

Going back to $k=2$, but using instead the following two matrices:
\begin{equation}
A = \begin{pmatrix}0 & 1 \\ 1 & 0\end{pmatrix} ~~~ B =
\begin{pmatrix}\omega & 0 \\ 0 & 1\end{pmatrix}.
\end{equation}
Now again $A^2=1$ and $B^n=1$, but the group generated will have
$2n^2$ elements.  Call it $\Sigma(2n^2)$.  For $n=1$, it is just
$Z_2$.  For $n=2$, it is $D_4$, i.e. the symmetry group of the square,
which was first used in 2003 \cite{Grimus:2003kq,Grimus:2004rj}.  For
$k=3$, consider
\begin{equation}
A = \begin{pmatrix}0 & 1 & 0 \\ 0 & 0 & 1 \\ 1 & 0 & 0\end{pmatrix}, ~~~ 
B = \begin{pmatrix}\omega & 0 & 0 \\ 0 & 1 & 0 \\ 0 & 0 & 1\end{pmatrix}, 
\end{equation}
then the groups generated have $3n^3$ elements and may be denoted as
$\Sigma(3n^3)$. They are in fact subgroups of $U(3)$.  For $n=1$, it
is just $Z_3$.  For $n=2$, it is $A_4 \times Z_2$.  For $n=3$, the
group $\Sigma(81)$ has been used \cite{Ma:2006ht} to understand the
Koide formula \cite{Koide:1982si} as well as lepton mass matrices
\cite{Ma:2007ku}.  In general, we have the series $\Sigma(kn^k)$.

\subsubsubsection{Model recipe} 

\begin{enumerate}
\item Choose a group, e.g. $S_3$ or $A_4$, and write down its possible
representations.  For example $S_3$ has \underline{1},
\underline{1}$'$, \underline{2}; $A_4$ has \underline{1},
\underline{1}$'$, \underline{1}$''$, \underline{3}.  Work out all
product decompositions.  For example $\underline{2} \times
\underline{2} = \underline{1} + \underline{1}' + \underline{2}$ in
$S_3$, and $\underline{3} \times \underline{3} = \underline{1} +
\underline{1}' + \underline{1}'' + \underline{3} + \underline{3}$ in
$A_4$.

\item Assign $(\nu,l)_{1,2,3}$ and $l^c_{1,2,3}$ to the
representations of choice.  To have only
renormalizable interactions,  it is necessary to add Higgs doublets (and
perhaps also triplets and singlets) and, if so desired, neutrino singlets.

\item The Yukawa structure of the model is restricted by the choice of
particle content and their representations.  As the Higgs bosons
acquire vacuum expectation values (which may be related by some extra
or residual symmetry), the lepton mass matrices will have certain
particular forms, consistent with the known values of $m_e$, $m_\mu$,
$m_\tau$, etc.  If the number of parameters involved is less than the
number of observables, there will be one or more predictions.

\item In models with more than one Higgs doublet, flavour
non-conservation will appear at some level.  Its phenomenological
consequences need to be worked out, to ensure the consistency with
present experimental constraints.  The implications for phenomena at
the TeV scale can then be explored.

\item Insisting on using only the single SM Higgs doublet requires
 effective non-renormalizable interactions to support the discrete
 flavour symmetry.  In such models, there are no predictions beyond
 the forms of the mass matrices themselves.

\item Quarks can be considered in the same way.  The two quark mass
matrices ${m}_u$ and ${m}_d$ must be nearly aligned so that their
mixing matrix involves only small angles.  In contrast, the mass
matrices ${m}_\nu$ and ${m}_e$ should have different structures so
that large angles can be obtained.
\end{enumerate}
Some explicit examples will be now outlined.

\subsubsubsection{$S_3$}

Being the simplest, the non-Abelian discrete symmetry $S_3$ was used
already \cite{Yamaguchi:1964} in the early days of strong
interactions.  There are many recent applications
\cite{Kubo:2003iw,Kubo:2004ps,Chen:2004rr,
Grimus:2005mu,Lavoura:2005kx,Teshima:2005bk,
Koide:2005ep,Mohapatra:2006pu,Morisi:2005fy,Kaneko:2006wi}, some of
which are discussed in \cite{Ma:2006ay}.  Typically, such models often
require extra symmetries beyond $S_3$ to reduce the number of
parameters, or assumptions of how $S_3$ is spontaneously and softly
broken.  For illustration, consider the model of Kubo et
al.~\cite{Kubo:2003iw} which has recently been updated by Felix et
al.~\cite{Felix:2006pn}.  The symmetry used is actually $S_3 \times
Z_2$, with the assignments
\begin{equation}
(\nu,l), ~l^c, ~N, ~(\phi^+,\phi^0) \sim \underline{1} + \underline{2},
\end{equation}
and equal vacuum expectation values for the two Higgs doublets
transforming as \underline{2} under $S_3$.  The $Z_2$ symmetry serves
to eliminate 4 Yukawa couplings otherwise allowed by $S_3$, resulting
in an inverted ordering of neutrino masses with
\begin{equation}
\theta_{23} \simeq \pi/4, ~~ \theta_{13} \simeq 0.0034, ~~ m_{ee}
\simeq 0.05~{\rm eV},
\end{equation}
where $m_{ee}$ is the effective Majorana neutrino mass measured in
neutrinoless double beta decay.  This model relates $\theta_{13}$ to
the ratio $m_e/m_\mu$.

\subsubsubsection{$A_4$}

To understand why quarks and leptons have very different mixing
matrices, $A_4$ turns out to be very useful.  It allows the two
different quark mass matrices to be diagonalized by the same unitary
transformations, implying thus no mixing as a first approximation, but
because of the assumed Majorana nature of the neutrinos, a large
mismatch may occur in the lepton sector, thus offering the possibility
of obtaining the so-called tri-bi-maximal mixing matrix
\cite{Harrison:2002er,He:2003rm}, which is a good approximation to
present data. One way of doing this is to consider the decomposition
\begin{equation}
\label{ma1}
U_{PMNS} = \begin{pmatrix}\sqrt{2/3} & 1/\sqrt{3} & 0 \\ -1/\sqrt{6} & 1/\sqrt{3} 
& -1/\sqrt{2} \\ -1/\sqrt{6} & 1/\sqrt{3} & 1/\sqrt{2}\end{pmatrix} = 
\frac{1}{\sqrt{3}} \begin{pmatrix}1 & 1 & 1 \\ 1 & \omega & \omega^2 \\ 1 & 
\omega^2 & \omega\end{pmatrix} \begin{pmatrix}0 & 1 & 0 \\ 1/\sqrt{2} & 0 & -i/\sqrt{2} \\ 
1/\sqrt{2} & 0 & i/\sqrt{2}\end{pmatrix},
\end{equation}
where $U_{PMNS}$ is the observed neutrino mixing matrix and
 $\omega = \exp(2 \pi i/3) = -1/2 + i\sqrt{3}/2$.  The matrix involving 
$\omega$ has equal moduli for all its entries and was conjectured already 
in 1978 \cite{Cabibbo:1977nk,Wolfenstein:1978uw} to be a possible candidate for the $3 \times 3$ 
neutrino mixing matrix.

Since $U_{PMNS} = V_e^\dagger V_\nu$, where $V_e$, $V_\nu$ diagonalize 
matrices ${m}_e {m}_e^\dagger$, ${m}_\nu {m}_\nu^\dagger$ 
respectively, Eq.~(\ref{ma1}) may be obtained if we have
\begin{equation}
\label{ma2}
V_e^\dagger =  \frac{1}{\sqrt{3}} \begin{pmatrix}1 & 1 & 1 \\ 1 & \omega & 
\omega^2 \\ 1 & \omega^2 & \omega\end{pmatrix}
\end{equation}
and
\begin{equation}
\label{ma3}
{m}_\nu = \begin{pmatrix}a+2b & 0 & 0 
\\ 0 & a-b & d \\ 0 & d & a-b\end{pmatrix}\nonumber
\end{equation}
\begin{equation} 
= \begin{pmatrix}0 & 1 & 0 \\ 1/\sqrt{2} & 0 & -i/\sqrt{2} \\ 1/\sqrt{2} 
& 0 & i/\sqrt{2}\end{pmatrix} \begin{pmatrix}a-b+d & 0 & 0 \\ 0 & a+2b & 0 \\ 0 & 0 & 
-a+b+d\end{pmatrix} \begin{pmatrix}0 & 1/\sqrt{2} & 1/\sqrt{2} \\ 1 & 0 & 0 \\ 
0 & -i/\sqrt{2} & i/\sqrt{2}\end{pmatrix}.
\end{equation}

It was discovered in Ref.~\cite{Ma:2001dn}  that Eq.~(\ref{ma2}) is naturally obtained 
with $A_4$ if
\begin{equation}
\label{ma4}
(\nu,l)_{1,2,3} \sim \underline{3}, ~~ l^c_{1,2,3} \sim \underline{1} + 
\underline{1}' + \underline{1}'', ~~  (\phi^+,\phi^0)_{1,2,3} \sim 
\underline{3}
\end{equation}
for $\langle \phi^0_1 \rangle = \langle \phi^0_2 \rangle = \langle \phi^0_3 
\rangle$.  This assignment also allows $m_e$, $m_\mu$, $m_\tau$ to take 
on arbitrary values because there are here exactly three independent 
Yukawa couplings invariant under $A_4$.  If we use this also for quarks 
\cite{Babu:2002dz}, then $V_u^\dagger$ and $V_d^\dagger$ are also given by Eq.~(\ref{ma2}), 
resulting in $U_{CKM}=1$, i.e. no mixing.  This should be considered as a 
good first approximation because the observed mixing angles are all small. 
In the general case without any symmetry, we would have expected $V_u$ and 
$V_d$ to be very different.

It was later discovered in Ref.~\cite{Ma:2004zv}  that Eq.~(\ref{ma3}) may also be 
obtained with $A_4$, using two further assumptions.  Consider the 
most general $3 \times 3$ Majorana mass matrix in the form
\begin{equation}
{m}_\nu = \begin{pmatrix}a+b+c & f & e \\ f & a+b\omega +c\omega^2 & d \\ 
e & d & a+b\omega^2 +c\omega\end{pmatrix},
\end{equation}
where $a$ comes from \underline{1}, $b$ from \underline{1}$'$, $c$ from 
\underline{1}$''$, and $(d,e,f)$ from \underline{3} of $A_4$.  To get 
Eq.~(\ref{ma3}), we need $e=f=0$, i.e. the effective scalar $A_4$ triplet 
responsible for neutrino masses should have its vacuum expectation value 
along the (1,0,0) direction, whereas that responsible for charged-lepton 
masses should be (1,1,1) as I remarked earlier.  This misalignment is a 
technical challenge to all such models   \cite{Altarelli:2005yp,Babu:2005se,
Ma:2005sh,Zee:2005ut,Ma:2005qf,Altarelli:2005yx,
He:2006dk,Adhikary:2006wi,Adhikary:2006jx,Yin:2007rv,Altarelli:2006kg,
He:2006et}.  The other requirement is that 
$b=c$. Since they come from different representations of $A_4$, this is rather 
{\it ad hoc}.  A very clever solution \cite{Altarelli:2005yp,Altarelli:2005yx} is to eliminate both, 
i.e. $b=c=0$.  This results in a normal ordering of neutrino masses with 
the prediction 
\cite{Ma:2005sh}
\begin{equation}
|m_{\nu_e}|^2 \simeq |m_{ee}|^2 + \Delta m^2_{atm}/9.
\end{equation}
Other applications \cite{Ma:2002iq,Ma:2002yp,Babu:2002ki,Chen:2005jm,
Hirsch:2005mc,
Ma:2005qy,Ma:2005mw,Ma:2005tr,
Ma:2006sk,Ma:2006wm,Ma:2006vq,Lavoura:2006hb,deMedeirosVarzielas:2005qg,
King:2006np,Morisi:2007ft,Koide:2007kw,Hirsch:2007kh} of $A_4$ have also been 
considered.  A natural (spinorial) extension of $A_4$ is the binary 
tetrahedral group \cite{Aranda:1999kc,Aranda:2000tm} which is under active 
current discussion \cite{Carr:2007qw,Feruglio:2007uu,Chen:2007af,
Frampton:2007et}.

Other recent applications of non-Abelian discrete flavour symmetries include 
those of $D_4$ \cite{Grimus:2003kq,Grimus:2004rj,Kobayashi:2006wq}, 
$Q_4$ \cite{Frigerio:2004jg}, $D_5$ 
\cite{Ma:2004br,Hagedorn:2006ir}, $D_6$ \cite{Kajiyama:2006ww}, $Q_6$ 
\cite{Frampton:1994xm,Babu:2004tn,Kubo:2005ty}, $D_7$ 
\cite{Chen:2005jt}, $S_4$ \cite{Ma:2005pd,Hagedorn:2006ug,Cai:2006mf,
Zhang:2006fv,Koide:2007sr},  $\Delta(27)$ \cite{Ma:2006ip,
deMedeirosVarzielas:2006fc}, $\Delta(75)$ \cite{Kaplan:1993ej,
Schmaltz:1994ws}, $\Sigma(81)$ \cite{Ma:2006ht,
Ma:2007ku}, 
and $B_3 \times Z_2^3$ \cite{Grimus:2005rf,Grimus:2006wy} which has 384 
elements.

 \subsubsection{Accidental flavour symmetries}\label{sec:Accidental}

While flavour symmetries certainly represent one of the leading approaches to understanding the pattern of fermion masses and mixing, it was recently found that the hierarchical structure of charged fermion masses and many other peculiar features of the fermion spectrum in the SM (neutrinos included) do not require a flavour symmetry to be understood, nor any other special ``horizontal'' dynamics involving the family indices of the SM fermions~\cite{Barr:2001vj,Ferretti:2006df}. Surprisingly enough, those features can in fact be recovered in a model in which the couplings of the three SM families not only are not governed by any symmetry, but are essentially anarchical (uncorrelated $\mathcal{O}(1)$ numbers) at a very high scale.

The idea is based on the hypothesis that the SM Yukawa couplings all arise from the exchange of heavy degrees of freedom (messengers) at a scale not far from the unification scale. Examples of diagrams contributing to the up and down quark Yukawa matrices are shown below, where $\phi$ is a SM singlet field getting a VEV. As discussed in Section~\ref{sec:flavoursymmetries} and~\ref{sec:flavoursymmetries}.1, the same exchange mechanism is often assumed to be at work in models with flavour symmetries. Here, however, the couplings of the heavy messengers to the SM fields are not constrained by any symmetry\footnote{A discrete $\mathbf{Z}_2$ symmetry, under which \emph{all} the three SM families (and the field $\phi$)\ are odd, is used for the sole purpose of distinguishing the light SM fields from the heavy messengers.}. An hierarchy among Yukawa couplings still arises because a single set of left-handed messenger fields (heavy quark doublets $Q+\bar Q$ in the quark sector and heavy lepton doublets $L+\bar L$ in the lepton sector) dominates the exchange at the heavy scale. For example, the diagrams below represents the dominant contribution to the quark Yukawa matrices. As only one field is exchanged, the Yukawa matrices have rank one. Therefore, whatever are the $\mathcal{O}(1)$ couplings in the diagram, the top and bottom Yukawa couplings are generated (at the $\mathcal{O}(1)$ level, giving large $\tan\beta$), but the first two families' are not, which is a good starting point to obtain a hierarchy of quark masses. This mechanism is similar to a the single right-handed neutrino dominance mechanism, used in neutrino model building to obtain a hierarchical spectrum of light neutrinos~\cite{King:1998jw,King:1999cm,King:1999mb,King:2002nf}. Note that the diagonalization of the quark Yukawa matrices involves large rotations, as all the couplings are supposed to be $\mathcal{O}(1)$. However, the rotations of the up and down left-handed quarks turn out to be the same (because they have same couplings to the left-handed doublet messenger). Therefore, the two rotations cancel when combined in the CKM matrix, which ends up vanishing at this level. 
\begin{figure}[h]
\thinlines
\begin{center}
\begin{picture}(360,65)(80,90)
\put(105,102){\line(1,0){110}} 
\put(130,102){\line(0,1){40}}
\put(190,102){\line(0,1){40}}
\put(157,99){\large $\times$}
\put(87,100){$u^c_i$}
\put(222,100){$q_j$}
\put(127,150){$h$}
\put(176,150){$< \phi >$}
\put(137.5,88){$Q$}
\put(172.5,88){$\bar Q$}

\put(305,102){\line(1,0){110}} 
\put(330,102){\line(0,1){40}} 
\put(390,102){\line(0,1){40}} 
\put(357,99){\large $\times$}
\put(287,100){$d^c_i$}
\put(422,100){$q_j$}
\put(327,150){$h$}
\put(376,150){$< \phi >$}
\put(337.5,88){$Q$}
\put(372.5,88){$\bar Q$}
\end{picture}
\end{center}
\end{figure}

The Yukawa couplings of the second family, and a non-vanishing $V_{cb}$ angle, are generated by the subdominant exchange of heavier right-handed messengers $D^c$, $U^c$, $E^c$, $N^c$. Altogether, the messengers form a heavy (vectorlike) replica of a SM family, with the left-handed fields lighter than the right-handed ones. The (inter-family) hierarchy between the masses of the second and the third SM family masses arises from the (intra-family) hierarchy between left and right-handed fields in the single family of messengers. In turn, in a Pati-Salam or SO(10) unified model, the hierarchy between right-handed and left-handed fields can be easily obtained by giving mass to the messengers through a breaking of the gauge group along the $T_{3R}$ direction. This way, the hierarchy among different families is explained in terms of the breaking of a gauge group acting on single families, with no need of flavour symmetries or other dynamics acting on the family indexes of the SM fermions. 

It is also possible to describe the mechanism outlined above in terms of accidental flavour symmetries. In the effective theory below the scale of the right-handed messengers, in fact, the Yukawa couplings of the two lighter families are ``protected'' by an accidental U(2) symmetry. One can also consider the effective theory below the cut-off of the model, which is supposed to lie one or two orders of magnitude above the mass of the right-handed messengers. In the effective theory below the cut-off, the second family gets a non-vanishing Yukawa coupling, but the Yukawa of the lightest family is still ``protected'' by an accidental U(1) symmetry. 

Surprisingly enough, a number of important features of the fermion spectrum can be obtained in this simple and economical model. The relation $|V_{cb} | \sim m_s /m_b$ is a direct consequence of the principles of this approach. The stronger mass hierarchy observed in the up quark sector is accounted for without introducing a new scale (besides the left-handed and right-handed messenger ones) or making the up quark sector somehow different. In spite of the absence of small coefficients, the CKM mixing angles turn out to be small. At the same time, a large atmospheric mixing can be generated in a natural way in the neutrino sector, together with normal hierarchical neutrino masses. In fact, a see-saw mechanism dominated by the single right-handed (messenger) neutrino $N^c$ is at work. The bottom and tau mass unify at the high scale, while a $B-L$ factor 3 enters the ratios of the muon and strange masses. For a detailed illustration of the model, we refer the reader to~\cite{Ferretti:2006df}.

The study of FCNC and CPV effects in a supersymmetric context is still under way. Such effects might represent the distinctive signature of the model, due to the sizeable radiative effects one obtains in the (23) block of the ``right-handed'' sfermion mass matrices in both the squark and slepton sector.

\subsubsection{Flavour/CP symmetries and their violation from
  supersymmetry breaking}\label{sec:consequences}  

While the vast literature on flavour symmetries covers a number of
interesting aspects of the theory and phenomenology of flavour, we are
interested here in a (non exhaustive) review of only those aspects
relevant to new physics. The relevance of flavour symmetries to new
physics follows from the fact that SM extensions often contain new
flavour dependent interactions. In the following we will consider the
case of supersymmetry, in which new flavour-violating gaugino or
Higgsino interactions can be induced by possible new sources of
SU(5)$^5$ breaking in the soft supersymmetry breaking terms.

While in the SM the Yukawa matrices provide the only source of flavour
(U(3)$^5$) breaking, the supersymmetric extensions of the SM are
characterized by a potentially much richer flavour structure
associated to the soft supersymmetry breaking
Lagrangian. Unfortunately, a generic flavour structure leads to FCNC
and CPV processes that can exceed the experimental bounds by up to two
orders of magnitude --- the so-called supersymmetric flavour and CP
problem.  The solution of the latter problem can lie in the
supersymmetry breaking and mediation mechanism (this is the case for
example of gauge mediated supersymmetry breaking) or in the
constraints on the soft terms provided by flavour symmetries.

In turn, the implications of flavour symmetries on the structure of
the soft terms depends on the interplay between flavour and
supersymmetry breaking. Without entering the details of specific
models, we can distinguish two opposite situations:
\begin{itemize}
\item the soft terms are flavour universal, or at least symmetric
under the flavour symmetry, at the tree level and;
\item 
flavour symmetry breaking enters the soft terms (as for the Yukawa interactions) already at the tree level, through non-renormalizable couplings to the flavon fields. 
\end{itemize}
Let us consider them in greater detail.

The first possibility is that the supersymmetry breaking mechanism
takes care of the FCNC and CPV problems. In the simplest case, the new
sfermion masses and $A$-terms do not introduce new flavour structure
at all. This is the case if
\begin{align}
  \mathbf{m}^2_{ij} = m^2_0 \delta_{ij}, \qquad \mathbf{A}_{ij} =
  A_0\, \delta_{ij}, \notag
\end{align}
where $i,j$ are family indexes and the universal values $m^2_0$, $A_0$
can be different in the different sfermion sectors\footnote{This is
the case for example of gauge mediation. In supergravity,
supersymmetry breaking can be fully flavour blind in the case of
dilaton domination. In this case, we expect the diagonal elements of
the soft mass matrices to be exactly universal. However, this is
not always the case. Moduli domination is often encountered, in which
case fields with different modular weights receive different soft
masses.}. The breaking of the flavour symmetry is felt at the tree
level only by the Yukawa matrices.  Needless to say, the tree level
universality of the soft terms will be spoiled by
\emph{renormalization effects} associated to interactions sensitive to
Yukawa couplings \cite{Borzumati:1986qx,Hall:1985dx}.  These effects
can be enhanced by large logarithms if the scale at which the soft
terms and the Yukawa interactions appear in the observable sector is sufficiently
high.  The radiative contributions of Yukawa couplings associated with
neutrino masses (or Yukawa couplings occurring in the context of grand
unification) are particularly interesting in this context because they
offer new possibilities to test flavour physics by opening a window
for physics at very large scales.  For example, in the minimal SUSY
seesaw model only the off-diagonal elements for left-slepton soft
supersymmetry breaking mass terms are generated while in
supersymmetric GUTs also the right-handed slepton masses get
renormalization induced flavour non-diagonal contributions. In any
case, all the flavour effects induced by the soft terms can be traced
back to the Yukawa couplings, which remain the only source of flavour
breaking. Such unavoidable effects of flavour breaking on the soft
terms will be discussed in Section~\ref{sec:SUSY} and
Section~\ref{sec:SUSYGUTs}.

As we have just seen, the radiative contributions to soft masses
represent an unavoidable but indirect effect of the physics at the
origin of fermion masses and mixing. On the other hand, the mechanism
generating the soft terms might not be blind to flavour symmetry
breaking, in which case we might also expect flavour breaking to enter
the soft terms in a more direct way. If this is the case, the soft
term provide a new independent source of flavour violation. Such
model-dependent \emph{``tree level''} effects of flavour breaking on
the soft terms add to the radiative effects and will be
discussed in Section~\ref{sec:flavourTree}.  The actual presence in
the soft terms of flavour violating effects directly induced by the
physics accounting for Yukawa couplings depends on the interplay of
the supersymmetry breaking and the flavour generation mechanisms.

Theoretical and
phenomenological~\cite{Ferrara:1979wa,Fayet:1977yc,Witten:1981nf,Dimopoulos:1981zb,Weinberg:1981wj,Alvarez-Gaume:1981wy}
constraints on supersymmetry breaking parameters essentially force
supersymmetry breaking to take place in a hidden sector with no
renormalizable coupling to observable fields\footnote{The fields of
the Minimal Supersymmetric Standard Model (MSSM) or its relevant extension.}. 
The soft terms are therefore
often characterized by the scale $\Lambda_\textrm{SUSY}$ at which
supersymmetry breaking is communicated to the observable sector by
some mediation mechanism. The soft terms arise in fact from
non-renormalizable operators in the effective theory below
$\Lambda_\textrm{SUSY}$ obtained by integrating out the supersymmetry
breaking messenger fields. Analogously, in the context of a theory
addressing the origin of flavour, we can define a scale $\Lambda_f$ at
which the flavour structure arises. Let us consider for definiteness
the case of flavour symmetries. The analogy with supersymmetry
breaking is in this case even more pronounced. Above $\Lambda_f$, the
theory is flavour symmetric. By this we mean that we can at least
define conserved family numbers, perhaps part of a larger flavour
symmetry. The family numbers are then spontaneously broken by the VEV
of flavons that couple to observable fields through non-renormalizable
interactions suppressed by the scale $\Lambda_f$.

We are now in the position to discuss the presence of ``tree-level''
flavour violating effects in the soft terms. A first possibility is to
have $\Lambda_f \lesssim \Lambda_\textrm{SUSY}$, as for instance in
the case of gravity mediation, in which we expect $\Lambda_f\lesssim
M_\text{Planck} = \Lambda_\textrm{SUSY}$. The soft breaking terms are
already present below $M_{\text{Planck}}$. However, the flavour symmetry
is still exact at scales larger than $\Lambda_f$. Therefore the soft
terms must respect the family symmetries. At the lower scale
$\Lambda_f$ the effective Yukawa couplings are generated as functions
of the flavon VEVs, $\langle \theta \rangle/\Lambda_f$, and
analogously the soft breaking terms will also be functions of $\langle
\theta \rangle/\Lambda_f$.  In the $\Lambda_f \lesssim
\Lambda_\textrm{SUSY}$ case, we therefore expect new ``tree-level''
sources of flavour breaking in the soft terms on top of the effects
radiatively induced by the Yukawa couplings.

On the other hand, if $\Lambda_\textrm{SUSY}\ll \Lambda_f$, the soft
terms are not present at the scale of flavour breaking. The
prototypical example in this case is gauge mediated supersymmetry
breaking (GMSB)~(see \cite{Giudice:1998bp} and references therein). At
$\Lambda_f$ the flavour interactions are integrated and supersymmetry
is still unbroken. The only renormalizable remnant of the flavour
physics below $\Lambda_f$ are the Yukawa couplings. At the scale
$\Lambda_\textrm{SUSY}$ soft breaking terms feel flavour breaking only
through the Yukawa couplings. Strictly speaking, there could also be
non-renormalizable operators involving flavon fields suppressed by the
heavier $\Lambda_f$. The contributions of these terms to soft masses
would be proportional to $\Lambda_\textrm{SUSY}/\Lambda_f$ and
therefore negligible~\cite{Giudice:1998bp}. We are then only left with
the radiatively induced effects of Yukawa couplings.  The qualitative
arguments above show that flavour physics can provide relevant
information on the interplay between the origin of supersymmetry and
flavour breaking in the observable sector.

As we just saw, the family symmetry that accounts for the structure of
the Yukawa couplings also constrains the structure of sfermion
masses. In the limit of exact flavour symmetry, this implies family
universal, or at least diagonal, sfermion mass matrices. After the
breaking of the flavour symmetry giving rise to the Yukawa couplings,
we can have two cases:
\begin{itemize}
\item
The SUSY-breaking mediation mechanism takes place at a scale higher or equal to the flavour symmetry breaking scale and is usually sensitive to flavour. 
The flavour symmetry breaking accounts for both the structure of the Yukawa couplings and the deviations of
the soft-breaking terms from universality. This is the general expectation in gravity mediation of the supersymmetry breaking from the hidden sector.
\item
The supersymmetry breaking mediation mechanism takes place at a scale
much smaller than the flavour symmetry breaking scale. In this case
the flavour mediation mechanism, which is flavour-blind, guarantees
the universality of the soft-breaking terms. The flavour symmetry
breaking generates the Yukawa couplings but flavour breaking
corrections in the soft mass matrices are suppressed by the ratio of
the two scales. This is the case of gauge-mediation models of
supersymmetry breaking \cite{Giudice:1998bp}.
\end{itemize}
We begin discussing the first case.

\subsubsubsection{``Tree level'' effects of flavour symmetries in
supersymmetry breaking terms}\label{sec:flavourTree}

After the breaking of the flavour symmetry responsible for the
structure of the Yukawa couplings, we can expect to have non-universal
contributions to the soft breaking terms at {\it tree level}. Under
certain conditions, mainly related to the SUSY-breaking mediation
mechanism, these tree-level contributions can be sizeable and have
important phenomenological effects. The main example among these
models where the tree level non-universality in the soft breaking
terms is relevant is provided by models of supergravity mediation
\cite{Cremmer:1982en,Soni:1983rm,Kaplunovsky:1993rd,Brignole:1993dj,Chankowski:2005jh}
(for a nice introduction see the appendix in \cite{Martin:1997ns}).

The structure of the scalar mass matrices when SUSY breaking is
mediated by supergravity interactions is determined by the K\"ahler
potential. We are not going to discuss here the supergravity
Lagrangian, we refer the interested reader to Refs.
\cite{Cremmer:1982en,Soni:1983rm,Kaplunovsky:1993rd,Brignole:1993dj,Martin:1997ns}. For
our purposes, we only need to know that the K\"ahler potential is a
non-renormalizable, real, and obviously gauge-invariant, function of
the chiral superfields with dimensions of mass squared. This
non-renormalizable function includes couplings with the hidden sector
fields suppressed by different powers of $M_{\rm Planck}$, $\phi
\phi^* (1 + X X^*/M_{\rm Planck}^2 + \dots)$ with $\phi$ visible
sector fields and $X$ hidden sector fields.  This K\"ahler potential
gives rise to SUSY breaking scalar masses once a certain field of the
hidden sector gets a non-vanishing F-term.  The important point here is
that these couplings with hidden sector fields that will eventually
give rise to the soft masses are present in the theory at any scale
below $M_{\rm Planck}$. Below this scale, we can basically consider
the hidden sector as frozen and renormalize these couplings only with
visible sector interactions.

Therefore, in the following, to simplify the discussion, we
concentrate only on the soft masses and treat them as couplings
present at all energies below $M_{\rm Planck}$. The structure of the
soft mass matrices is easily understood in terms of the present
symmetries. At high energies, our flavour symmetry is still an exact
symmetry of the Lagrangian and therefore the soft breaking terms have
to respect this symmetry \cite{Ross:2004qn}.  At some stage, this
symmetry is broken generating the Yukawa couplings in the
superpotential.  In the same way, the scalar masses will also receive
new contributions after flavour symmetry breaking from the flavon
field VEVs suppressed by mediator masses.

First we must notice that a mass term $\phi _{i}^{\dagger }\phi _{i}$
is clearly invariant under gauge, flavour and global symmetries and
hence gives rise to a flavour diagonal contribution to the soft masses
even before the family symmetry breaking\footnote{As we will discuss in
the following, these allowed contributions may be universal, the same
for the different generations, as in the case of non-Abelian flavour
symmetries, or they can be different for the three generations in some
cases with Abelian flavour symmetries.}. Then, after flavour symmetry
breaking, any invariant combination of flavon fields (VEVs) with a
pair of sfermion fields, $\phi_{i}^{\dagger }\phi _{j}$, can also
contribute to the sfermion mass matrix and will break the universality
of the soft masses.

 An explicit example with a continuous Abelian $U(1)$ flavour
symmetry~\cite{Froggatt:1978nt,Leurer:1992wg,Nir:1993mx,Pomarol:1995xc,Binetruy:1996xk,Dudas:1996fe,Dreiner:2003yr,Kane:2005va,Chankowski:2005qp}
was given above in Section~\ref{sec:contflav}.

We turn now to the structure of the scalar mass matrices concentrating
mainly on the slepton mass matrix
\cite{Nir:1993mx,Leurer:1993gy,Pomarol:1995xc,Nir:2002ah}. In this
case, even before the breaking of the flavour symmetry, we have three
different fields with different charges corresponding to each of the
three generations. As we have seen, diagonal scalar masses are allowed
by the symmetry, but being different fields, there is no reason a
priori for these diagonal masses to be the same, and in general, we
have
\begin{equation}
\label{eq:Lsymm}
{\cal L}^{\rm symm}_{m^2} = m_1^2 ~\phi_1^* \phi_1 + m_2^2~ \phi_2^* \phi_2 + m_3^2~ \phi_3^* \phi_3 \, .
\end{equation}
Notice, however, that this situation is very dangerous, especially in
the case of squarks, given that the rotation to the basis of diagonal
Yukawa couplings from Eq.~(\ref{chargedYuk}) will generate too large
off-diagonal entries \cite{Nir:2002ah}.  In some cases, like dilaton
domination, these allowed masses can be equal avoiding this
problem. In the following we assume $
m_1^2=m_2^2=m_3^2=m_0^2$. However, even in this case, after the
breaking of the flavour symmetry we obtain new contributions
proportional to the flavon VEVs that break this universality. All we
have to do is to write all possible combinations of two MSSM scalar
fields $\phi_i$ and an arbitrary number of flavon VEVs invariant under
the symmetry:
\begin{equation}
{\cal L}_{m^2} = m_0^2 (\phi_1^* \phi_1 + \phi_2^* \phi_2 + \phi_3^* \phi_3 +
\left(\frac{\langle\theta\rangle}{M_{\rm fl}}\right)^{q_2-q_1} \phi_1^* \phi_2 +
\left(\frac{\langle\theta\rangle}{M_{\rm fl}}\right)^{q_3-q_1}~\phi_1^* \phi_3 +
\left(\frac{\langle\theta\rangle}{M_{\rm fl}}\right)^{q_3-q_2}~\phi_2^* \phi_3 + {\rm h.c.}) .
\end{equation}
Therefore, the structure of the charged slepton mass matrix we would
have in this model at the scale of flavour symmetry breaking would be
(suppressing $O(1)$ coefficients):
\begin{eqnarray}
m^2_{\tilde{L}} \simeq \left( \begin{array}{ccc} 1 & {\epsilon} &
{\epsilon} \\ {\epsilon} & 1 & 1\\ {\epsilon} & 1 & 1 \end{array}
\right) m_0^2 \, .
\end{eqnarray}
 This structure has serious problems with the phenomenological bounds
coming from $\mu \to e \gamma$, etc. There are other $U(1)$ examples
that manage to alleviate, in part, these problems \cite{Nir:2002ah}.
However, large LFV effects are a generic problem of these models due
to the required charge assignments to reproduce the observed masses
and mixing angles.

These FCNC problems in the sfermion mass matrices of Abelian
symmetries were one of the main reasons for the introduction of
non-Abelian flavour symmetries
\cite{Barbieri:1995uv,Barbieri:1996ww}. The mechanism used in
non-Abelian flavour models to generate the Yukawa couplings is again a
variation of the Froggatt-Nielsen mechanism, very similar to the
mechanism we have just seen for Abelian symmetries. The main
difference is that in this case the left handed fermions are grouped
in larger representations of the symmetry group.  For instance, in a
$SU(3)$ symmetry all three generations are unified in a triplet. In a
$SO(3)$ flavour symmetry we can assign the three generations to a
triplet or to three singlets.  In a $U(2)$ flavour symmetry the third
generation is a singlet and the two light generations are grouped in a
doublet. Then we do not have to assign different charges to the
different generations, but in exchange, we need several stages of
symmetry breaking by different flavon fields with specially aligned
VEVs.

We begin analyzing a non-Abelian $U(2)$ flavour symmetry.  As stressed
above, if the sfermions mass matrices are only constrained by a U(1)
flavour symmetry there is no reason why $m^2_1$ should be close to
$m^2_2$ in Eq.~(\ref{eq:Lsymm}). Unless an alignment mechanism between
fermions and sfermions is available, the family symmetry should then
suppress $(\tilde m^2_1-\tilde m^2_2)/\tilde m^2$. At the same time,
in the fermion sector, the family symmetry must suppress the Yukawa
coupling of the first two families, $m_1, m_2\ll m_3$. If the small
breaking of a flavour symmetry is responsible for the smallness of
$(\tilde m^2_1-\tilde m^2_2)/\tilde m^2$ on one hand and of
$m_1/m_3,m_2/m_3$ on the other, the symmetric limit should correspond
to $\tilde m^2_1 = \tilde m^2_2$ and to $m_1 = m_2 = 0$. Interestingly
enough, the largest family symmetry compatible with SO(10) unification
that forces $m_1 = m_2 = 0$ automatically also forces $\tilde m^2_1 =
\tilde m^2_2$. This is a U(2) symmetry under which the first two
families transform as a doublet and the third one, as well as the
Higgs, as a
singlet~\cite{Pomarol:1995xc,Barbieri:1995uv,Barbieri:1997tg,Barbieri:1996ww,Barbieri:1997tu}.
\[\psi=\psi_a\oplus\psi_3.\]
The same conclusion can be obtained by using discrete subgroups~\cite{Aranda:1999kc,Feruglio:2007uu}. In the limit of unbroken U(2), only the third generation
of fermions can acquire a mass, whereas the first two generations of scalars are exactly degenerate. While the first property is not a bad approximation of the
fermion spectrum, the second one is what is needed to keep FCNC and CP-violating effects under control. This observation can actually be considered as a hint
that the flavour structure of the mass matrices of the fermions and of the scalars are related to each other by a symmetry principle. The same physics
responsible for the peculiar pattern of fermion masses also accounts for the structure of sfermion masses. 

The rank 2 of U(2) allows a two step breaking pattern
\begin{equation}
\mbox{U(2)} \stackrel{\epsilon}{\rightarrow} \mbox{U(1)}\stackrel{\epsilon'}{\rightarrow} 0, 
\label{u2br}
\end{equation}
controlled by two small parameters $\epsilon$ and $\epsilon' < \epsilon$, to be at the origin of the generation mass hierarchies $m_3\gg m_2\gg m_1$ in the
fermion spectrum. Although it is natural to view U(2) as a subgroup of U(3), the maximal flavour group in the case of full intra-family gauge unification,
U(3) will be anyhow strongly broken to U(2) by the large top Yukawa coupling.

A nice aspect of the U(2) setting is that there is little arbitrariness in the way the symmetry breaking fields couple to the SM fermions. This is unlike
what happens e.g.\ with the choice of fermion charges in the cases of U(1) symmetries. The Yukawa interactions transform as: $(\psi_3 \psi_3)$, $(\psi_3 \psi_a)$,
$(\psi_a \psi_b)$ ($a,b,c\ldots = 1,2$).  Hence the only relevant U(2) representations for the fermion mass matrices are $1$, $\phi^a$, $S^{ab}$ and $A^{ab}$,
where $S$ and $A$ are symmetric and antisymmetric tensors, and the upper indices denote a U(1) charge opposite to that of $\psi_a$. While $\phi^a$ and $A^{ab}$
are both necessary, models with~\cite{Barbieri:1996ww,Barbieri:1997tu} or without~\cite{Barbieri:1997tg} $S^{ab}$ are both possible. 

Let us first consider the case with $S^{ab}$. At leading order, the flavons couple to SM fermions through D=5 operators suppressed by a flavour scale $\Lambda$.
Normalizing the flavons to $\Lambda$, it is convenient to choose a basis in which $\phi^2={\cal O}(\epsilon)$ and $\phi^1=0$, while
$A^{12}=-A^{21}={\cal O}(\epsilon')$. If $S$ is present, it turns out to be automatically aligned with $\phi$~\cite{Barbieri:1998em}, in such a way that in the
limit $\epsilon'\rightarrow 0$ a U(1) subgroup is unbroken. More precisely, $S^{22}={\cal O}(\epsilon)$ and all other components essentially vanish. We are then
led to Yukawa matrices of the form:
\begin{equation}
\left( \begin{array}{ccc} 0 & \epsilon' & 0 \\ -\epsilon' & \epsilon & \epsilon \\ 0 & \epsilon & 1 \end{array} \right).
\label{lambda}
\end{equation}
All non vanishing entries have unknown coefficients of order unity, while still keeping $\lambda_{12}=-\lambda_{21}$. In the context of SU(5) or SO(10)
unification, the mass relations $m_\tau\approx m_b$, $m_\mu\approx 3m_s$, $3m_e\approx m_d$ are accounted for by the choice of the transformations of  $A^{ab}$,
$S^{ab}$ under the unified group. The stronger mass hierarchy in the up quark sector, a peculiar feature of the fermion spectrum, is then predicted, due to the
interplay of the U(2) and the unified gauge symmetry. 

The texture in Eq.~(\ref{lambda}) leads to the predictions
\begin{equation}
\label{eq:u2pred}
\left|\frac{V_{td}}{V_{ts}}\right| = \sqrt{\frac{m_d}{m_s}} ,\qquad \left|\frac{V_{ub}}{V_{cb}}\right| = \sqrt{\frac{m_u}{m_c}} .
\end{equation}
While the experimental determination of $|V_{td}/V_{ts}|$ based on 1-loop observables might be affected by new physics, the tree-level determination of
$|V_{ub}/V_{cb}|$ is less likely to be affected and at present is significantly away from the prediction in Eq.~(\ref{eq:u2pred})~\cite{Barbieri:1998qs,Roberts:2001zy}.
A better agreement can be obtained by i) relaxing the condition $\lambda_{12}=-\lambda_{21}$, ii) allowing for small contributions to the 11, 13, 31 entries in
Eq.~(\ref{lambda}) or by iii) allowing for asymmetric textures~\cite{Roberts:2001zy}. The latter possibility is realized in models in which the $S^{ab}$ flavon is not
present~\cite{Barbieri:1996ww}. 

While the model building degrees of freedom in the quark and charged lepton sector are limited, a virtue of the U(2) symmetry, the neutrino sector is less
constrained. This is due, in the see-saw context, to the several possible choices involved in the modelization of  the singlet neutrino mass matrix. This is
reflected for example in the possibility to get both small and large mixing angles~\cite{Carone:1997qg,Hall:1998cu,Barbieri:1999pe,Aranda:2000tm,Aranda:2001rd}. 

In the case of an $SU(3)$ flavour symmetry, all three generations are grouped in a single triplet representation, $\psi _{i}$. In addition we have 
several new scalar fields (flavons) which are either triplets, $\overline{\theta}_{3}$, $\overline{\theta}_{23}$ and $\overline{\theta}_{2}$, or 
anti-triplets, $\theta_{3}$ and $\theta_{23}$. $SU(3)_{fl}$ is broken in two steps: the first step occurs when $\theta_3$ and $\bar \theta_{3}$ get
a large VEV breaking $SU(3)$ to $SU(2)$, and defining the direction of the third generation. Subsequently a smaller VEV of
$\theta_{23}$ and $\bar \theta_{23}$ breaks the remaining symmetry and defines the second generation direction.
To reproduce the Yukawa textures the large third generation Yukawa couplings require  a $\theta_3$ (and $\bar \theta_{3}$) VEV of the order of the
mediator scale, $M_{\rm fl}$, while $\theta_{23}/M_{\rm fl}$ (and $\bar \theta_{23}/M_{\rm fl}$) have small VEVs\footnote{In fact, 
in realistic models reproducing the CKM mixing matrix, there are two different mediator scales and expansion
parameters,  $\varepsilon$ in the up-quark and $\bar \varepsilon$ in the down-quark sector \cite{Ross:2004qn,King:2001uz,King:2003rf}.}  of order $\varepsilon$. 
After this breaking chain we obtain the effective Yukawa couplings at low energies through the Froggatt-Nielsen mechanism
\cite{Froggatt:1978nt} integrating out heavy fields. The resulting superpotential invariant under $SU(3)$ would be:
\begin{eqnarray}
W_{Y} &=&H\psi _{i}\psi _{j}^{c}\left[ \theta _{3}^{i}\theta_{3}^{j}\,+\,\theta _{23}^{i}\theta _{23}^{j} \,+\,\epsilon ^{ikl}
\overline{\theta }_{23,k}\overline{\theta }_{3,l}\theta _{23}^{j}\left(\theta _{23}\overline{\theta _{3}}\right) \,+\,\right.   \nonumber \\
&&\left. \epsilon ^{ijk}\overline{\theta }_{23,k}\left( \theta _{23} \overline{\theta _{3}}\right) ^{2}\,+\,\epsilon ^{ijk}\overline{\theta }
_{3,k}\left( \theta _{23}\overline{\theta _{3}}\right) \left( \theta _{23} \overline{\theta _{23}}\right) \,+\,\dots \right]\, .   \label{YukawaW0}
\end{eqnarray}
In this equation we can see that each of the $SU(3)$ indices of the external MSSM particles (triplets) are either saturated individually with an 
anti-triplet flavon index (a ``meson'' in QCD notation) or in an antisymmetric couplings with other two triplet indices (a ``baryon'').
The presence of other singlets in the different term is due to the presence of additional global symmetries necessaries to ensure the correct hierarchy in
the different Yukawa elements \cite{Ross:2004qn,King:2001uz,King:2003rf}. This structure is quite general for the different $SU(3)$ models we can
build. Here we are not specially concerned with additional details and we refer to \cite{Ross:2004qn,King:2001uz,King:2003rf}
for more complete examples. The Yukawa texture we obtain with this superpotential is the following:
\begin{equation}
\label{thematrix}
{Y^{f}}=\left(\begin{array}{ccc} 0& \alpha ~{\varepsilon^3} & \beta~{\varepsilon^3} \\
\alpha~{\varepsilon^3} &\frac{\varepsilon^2}{a^2} & \gamma~ \frac{\varepsilon^2}{a^2} \\
\beta~{\varepsilon^3}& \gamma~\frac{\varepsilon^2}{a^2} &1 \end{array} \right) a^2,
\end{equation}
with $a=\frac{\langle\theta_3\rangle}{M}$, and $\alpha,~ \beta,~\gamma$ unknown coefficients $O(1)$.

Let us now analyze the structure of scalar soft masses. 
In analogy with the Abelian case, in the unbroken limit diagonal soft masses
are allowed. However, the three generations belong to the same representation of the flavour symmetry and now this implies the mass is the same for the 
whole triplet. After the  breaking of $SU(3)$ symmetry the scalar soft masses deviate from  exact universality 
\cite{Ross:2004qn,Ross:2002mr,King:2004tx,Antusch:2007re}.
Any invariant combination of flavon fields can also contribute to the sfermion masses, although flavour symmetry indices can be contracted with
fermion fields. Including these corrections the leading contributions to the sfermion mass matrices are given by
\begin{eqnarray}
(m^2_{\tilde f})^{ij}&=& m_0^2 (\delta ^{ij} +\frac{\displaystyle{1}}{\displaystyle{M_f^{2}}}[\theta _{3}^{i\dagger
}\theta _{3}^{j} +\theta _{23}^{i\dagger }\theta_{23}^{j}]
\label{kahler}
+\frac{1}{M_f^4}(\epsilon ^{ikl}\overline{\theta }_{3,k} \overline{\theta }_{23,l})^{\dagger }(\epsilon ^{jmn}
\overline{\theta }_{3,m}\overline{\theta }_{23,n})).
\end{eqnarray}
Notice that each term inside the parenthesis is trivially neutral under the symmetry because it contains always a field together with its own complex
conjugate field. However, as the flavour indices of the flavon fields are contracted with the external matter fields this gives a non-trivial
contribution to the sfermion mass matrices. Therefore in this model, suppressing  factors of order 1 we have, 
\begin{equation}
m_{\tilde{f}}^{2}\simeq \left( 
\begin{array}{ccc} 1 &  &  \\  & 1 &  \\  &  & 1 \end{array}
\right) m_{0}^{2} +\left( 
\begin{array}{ccc} \varepsilon^2  & 0 & 0 \\ 0 & \frac{\varepsilon^2}{a^2} & \frac{\varepsilon^2}{a^2} \\ 0 & \frac{\varepsilon^2}{a^2} & 1 \end{array}
\right) a^2 m_{0}^{2},
\label{soft1}
\end{equation}
with $a = \langle\theta_{3}\rangle /M_{\rm fl}$ which is still $O(1)$.
In the model \cite{Ross:2004qn,King:2001uz,King:2003rf}, the expansion parameter for right-handed down quarks and charged leptons is $\bar \varepsilon = 0.15$.
Using Eq.~(\ref{thematrix}) and Eq.~(\ref{soft1}) we can obtain the slepton mass matrix in the basis of diagonal charged lepton Yukawa couplings: 
\begin{eqnarray}
m^2_{\tilde{e}_R} \simeq \left(\begin{array}{ccc} 1 + \bar \varepsilon^2 & - \bar \varepsilon^3 & - \bar \varepsilon^3  \\ 
- \bar \varepsilon^3 & 1 + \bar \varepsilon^2 & \bar \varepsilon^2  \\
- \bar \varepsilon^3 & \bar \varepsilon^2 & 1 \end{array} \right)  m_0^2,
\label{softm2}
\end{eqnarray}
where we have used $a_3 \simeq {\cal{O}}(M_{\rm fl})$. Therefore  that generates the order $\bar \varepsilon^3$ entry in the $(1,2)$ element.
The modulo of this entry is order $3 \times 10^{-3}$ at $M_{GUT}$. These estimates at $M_{GUT}$ are slightly reduced through renormalization group evolution to the 
electroweak scale and is order $1 \times 10^{-3}$ at $M_W$. 
This value implies that supersymmetric contribution to 
$\mu \to e \gamma$ is very big and can even exceed the present bounds for light slepton masses and large $\tan \beta$ if we are not in the
cancellation region\cite{Masina:2002mv,Masiero:2002jn,Ciuchini:2007ha}. This makes this process perhaps the most 
promising one to find deviations from universality in flavour models. The presence of the $SU(3)$
flavour symmetry controls the structure of the sfermion mass matrices and the supersymmetric flavour problem can be nicely solved. However,
interesting signals of the supersymmetric flavour structure can be found in the near future LFV experiments.

\section{Observables and their parameterization}\label{sec:observables}

\subsection{Effective operators and low scale observables }

In spite of the clear success of the SM in reproducing all the known
phenomenology up to energies of the order of the electroweak scale,
nobody would doubt the need of a more complete theory beyond
it. There remain many fundamental problems such as the experimental
evidence for Dark Matter (DM) and neutrino masses, as well as the
theoretical puzzles posed by the origin of flavour, the three
generations, etc, that a complete theory should address. Therefore, we
can consider the SM as the low-energy effective theory of some more
complete model that explains all these puzzles. Furthermore, we have
strong reasons (gauge hierarchy problem, unification of couplings,
dark matter candidate, etc.) to expect the appearance of new physics
close to the electroweak scale. Suppose that these new particles
from the more complete theory are to be found at the LHC. Experiments
at lower energies $ E < m_{\rm NP}$ are also sensitive to this
NP. Indeed the exchange of new particles can induce:
\begin{itemize}
\item
corrections to the SM observables (such as S,T and U),
\item
the appearance of {\it new} observables or new ($d>4$) operators, ({\it e.g.} the flavour violating dipole operators).
\end{itemize}
Note that both effects can be parameterized by $SU(3) \times SU(2)
\times U(1)$--invariant operators of mass dimension $d>4$. We refer to
these non-renormalizable operators as {\it effective} operators.  Any
NP proposed to explain new phenomena at the LHC must satisfy the
experimental constraints on the effective operators it generates.

\subsubsection{Effective Lagrangian approach: ${\cal L}_{eff}$}

Considering the SM as an effective theory below the scale of NP,
$m_{\rm NP}$, where the heavy fields have been integrated out, we can describe the
physics through an effective Lagrangian, ${\cal L}_{eff}$. This effective
Lagrangian contains all possible terms invariant under the SM
gauge group and built with the SM fields.
Besides the usual SM fields, we could introduce new light singlet
fermions with renormalizable Yukawa couplings to the lepton
doublets (and possibly small Majorana masses) to accommodate the observed 
neutrino masses. In this case we would have more operators
allowed in the effective Lagrangian of the SM + extra light
sterile states.  On the assumption that the light sterile particles are weakly 
interacting, if present, and therefore not relevant to the LHC, we focus
on the effective Lagrangian that can be constructed only from the known SM 
fields. Then, the effective Lagrangian at energies $ E\ll m_{\rm NP}$ can be 
written as an expansion in $1/m_{\rm NP}$ as,
\begin{eqnarray}
{\cal L}^{SM}_{\rm eff}& =& {\cal L}_{0} + \frac{1}{m_{\rm NP}}~ {\cal L}_{1} +
\frac{1}{m_{\rm NP}^2}~ {\cal L}_{2} + \frac{1}{m_{\rm NP}^3}~ {\cal L}_{3} + 
\dots \label{eff.exp},
\end{eqnarray}
where $ {\cal L}_{0}$ is the renormalizable SM Lagrangian containing the
kinetic terms of the $ U(1)$, $ SU(2)$ and $SU(3)$ gauge bosons $ A_\mu$,
the gauge interactions and kinetic terms of the SM fermions, $\{f\}$, and
Higgs, and the Yukawa couplings of the Higgs and SM fermions.  
In order to fix the notation, we list the SM fermions as 
\begin{eqnarray}
q_{i} = \left( \begin{array}{c} u_{Li} \\ d_{Li} \end{array} \right),~
\ell_{i} = \left( \begin{array}{c} \nu_{Li} \\ e_{Li} \end{array} \right),~
u_{Ri},~d_{Ri},~e_{Ri}, 
\end{eqnarray}
where $i$ is a flavour/family/generation index. Note that, in the following
we use always four-component Dirac spinors in the different Lagrangians. 
Explicit expressions, for ${\cal L}_{0}$  in similar notation, can be found 
in \cite{Buchmuller:1985jz}.

The different ${\cal L}_{n}$ are Lagrangians of dimension $d=4+n$ invariant
under $SU(3) \times SU(2) \times U(1)$ and can be schematically written
\begin{eqnarray}
{\cal L}_{n} = \sum_a C_a\cdot {\cal O}_a(H,\{f\}, \{A_\mu\}) + h.c. 
\label{Lieff}
\end{eqnarray} 
The local operators ${\cal O}_a$ are gauge invariant combinations of SM fields of
dimension $4+n$. Their coefficient, that in the full Lagrangian has mass 
dimension  $-n$, is unknown in bottom-up effective field theory, but 
calculable in NP models. We write this coefficient as a dimensionless $C_a$
divided by the n-th power of the mass scale of the NP mediator, 
$m_{\rm NP}^n$, which for new physics relevant at LHC energies would be 
$m_{\rm NP}\sim \sqrt{s_{LHC}}$. We will later 
normalize to $G_F$ (see Eq.~(\ref{S12})).

We are mainly interested in dimension 5 and dimension 6 operators. We 
assume that any particles created at the LHC could generate dimension 6 
operators, and then we can neglect higher dimension operators 
contributing to the same physical processes. Operators of 
dimension 7 include the lepton number violating operator
$\epsilon_{ab} \epsilon_{cd} H^a \ell^b_{[i} \sigma^{\mu \nu} H^c \ell^d_{j]} 
F_{\mu \nu}$ which gives neutrino transition moments (flavour-changing dipole 
moments) after electroweak symmetry breaking (EWSB). 
At dimension 8 are two-Higgs-four-fermion operators,
which can give 4-fermion operators after EWSB, with a different flavour
structure from the dimension 6 terms.  We will not analyze these operators
here, but they are studied in the context of non-standard neutrino interactions
\cite{Berezhiani:2001rs}. 
Therefore, in the following, we restrict our analysis to ${\cal L}_{1}$ and 
${\cal L}_{2}$.

The unique operator allowed with the Standard Model fields and symmetries 
at dimension 5 is 
${\cal O}_{\ell\ell}^{ij} = \epsilon_{ab} \epsilon_{cd} H^a \overline{\ell^{c}}^{b}_i H^c \ell^d_j$ 
($a,b,c,d$ are SU(2)
indices). Thus we have, 
\begin{eqnarray}
{\cal L}_{1}~ =~ 
\frac{1}{4}~ \kappa_{\nu \ell\ell}^{\,ij} \cdot \epsilon_{ab}
\epsilon_{cd} H^a \overline{\ell^{c}}^{b}_i H^c 
\ell^d_j~ +~ h.c.~,
\label{L1eff}
\end{eqnarray}
where $\ell^{c}$ is the charge conjugate of the lepton doublet.
After electroweak symmetry breaking,
this  gives rise to a Majorana mass matrix
$\frac{1}{4}~ \kappa_{\ell\ell}^{ij} \langle H^0 \rangle^2
 \overline{\nu^{c}}_i
\nu_j~ +~ h.c.~$. In the neutrino mass eigenstate basis,
the masses are $\kappa_{\ell\ell}^{ii} \langle H^0 \rangle^2/2$. The coefficient
$\kappa_{\ell\ell}^{ij} =2 Y_{ki} M_k^{-1} Y_{kj}$
is generated for instance after
integrating out  heavy right-handed neutrinos of mass $M_k$
in a seesaw mechanism with Yukawa coupling $Y$.

${\cal L}_{2}$ is constructed with dimension 6 operators  which give 
interactions among 3 or 4  ``light'' external legs. 
We can classify the possible operators according to the external legs as:
\begin{itemize}
\item
operators with a pair of leptons and an (on-shell) photon:
\begin{eqnarray} \label{OeB}
{\cal O}^{ij}_{eB}  = \overline{\ell}_i \sigma^{\mu \nu}e_{R j} H B_{\mu \nu}, \qquad
&{\cal O}^{ij}_{eW}  = \overline{\ell}_i \sigma^{\mu \nu} \tau^I e_{R j} H
W^I_{\mu \nu} .\qquad
\end{eqnarray} 
\item
four-lepton operators, with Lorenz structure $\overline{L}L\overline{L}L$, $\overline{R}R\overline{R}R$ or $\overline{L}R\overline{R}L$,
 singlet or triplet SU(2) gauge contractions (described in the operator subscript), and all possible
inequivalent flavour index combinations (see Section~\ref{sec:taudecays}). The $SU(2) \times U(1)$ invariant
operators, with flavour indices in the superscript, are: 
\begin{eqnarray}\label{OLR}
{\cal O}_{(1) \ell \ell}^{ijkl} =
(\overline{\ell}_i \gamma^\mu \ell_j)(\overline{\ell}_k \gamma_\mu \ell_l),
\qquad\qquad~~~& {\cal O}_{(3) \ell \ell}^{ijkl} =
(\overline{\ell}_i \tau^I \gamma^\mu \ell_j)(\overline{\ell}_k\tau^I
\gamma_\mu \ell_l),\quad  \nonumber \\
{\cal O}_{ee}^{ijkl}  = 
 (\overline{e}_i \gamma^\mu P_R e_j)(\overline{e}_k \gamma_\mu P_R e_l),\qquad&
 {\cal O}_{ \ell e}^{ijkl} =  (\overline{\ell}_i e_j)(\overline{e}_k \ell_l) .~~\qquad\qquad
\end{eqnarray}
\end{itemize}
Therefore the Lagrangian ${\cal L}_{2}$ involving leptons is\footnote{Note
  that we do not include here 2 quark--2 lepton operators and, in the
  following, we will only consider the photon component of the dipole operators.}, 
\begin{eqnarray}
{\cal L}_{2}& =&  ~C_{eB}^{ij} \cdot {\cal O}^{ij}_{eB}~ +~
 ~C_{eW}^{ij} \cdot {\cal O}^{ij}_{eW}~ +~ \frac{1}{1+\delta} \left(
~C_{(1)\ell\ell}^{ijkl} \cdot {\cal O}^{ijkl}_{(1)\ell\ell}~
+~ ~C_{(3)\ell\ell}^{ijkl} \cdot {\cal
  O}^{ijkl}_{(3)\ell\ell}~ +~\nonumber\right. \\&&\left. ~C_{ee}^{ijkl} \cdot {\cal
  O}^{ijkl}_{ee}~ +~ 2~C_{\ell e}^{ijkl} \cdot {\cal
  O}^{ijkl}_{\ell e}.~\right)~+~ h.c.~,
\label{L2eff} 
\end{eqnarray}
where  we introduce the parameter $\delta$ to cancel possible factors of 2 
that can arise from the $+~h.c.$: it is 1 for
$ {\cal O}_{\dots}^{ij\dots} = [{\cal O}_{\dots}^{ij\dots}]^\dagger$,
otherwise it is $ 0$. The sums over  
$ i,j,k,l$  run over inequivalent  operators,
taking an operator to be inequivalent if neither it, nor its h.c., are already 
in the list. 
The factor of 2 in the definition of ${\cal O}_{ \ell e}$ is included to 
compensate the 1/2 in the Fiertz rearrangement below (second line of 
Eq.~(\ref{Olr}))\footnote{Note there will sometimes be other 2s for
  identical  fermions.}. The effective operators 
whose coefficients we constrain in the next section are related to those of 
Eq.~(\ref{L2eff}) through an expansion in terms of the $SU(2)$ 
components of the fields and taking into account the electroweak symmetry 
breaking :

\begin{eqnarray}
 \label{OeB2}
{\cal O}^{ij}_{eB} &\!\!\!= \overline{\ell}_i \sigma^{\mu \nu}e_{R j} H B_{\mu \nu} \qquad &\!\!\!= \cos \theta_W ~~  \langle H \rangle~
\overline{e}_i \sigma^{\mu \nu} P_R e_j 
F^{em}_{\mu \nu},\\
\label{OeW}
{\cal O}^{ij}_{eW} &\!\!\!=   \overline{\ell}_i \sigma^{\mu \nu} \tau^I e_{R j} H W^I_{\mu \nu}\quad &\!\!\!= - \sin \theta_W 
~\langle H \rangle~ 
 \overline{e}_i \sigma^{\mu \nu}P_R e_j 
F^{em}_{\mu \nu},\\
{\cal O}_{(1) \ell \ell}^{ijkl}  &\!\!\!=  
(\overline{\ell}_i \gamma^\mu \ell_j)(\overline{\ell}_k \gamma_\mu \ell_l)~~\quad &\!\!\!=
(\overline{\nu}_i \gamma^\mu P_L \nu_j + \overline{e}_{i} \gamma^\mu P_Le_{j}) 
(\overline{\nu}_k \gamma_\mu P_L \nu_l + \overline{e}_{k} \gamma_\mu P_L e_{l}) ,\label{Oll1} \\
{\cal O}_{(3) \ell \ell}^{ijkl} &\!\!\!=
(\overline{\ell}_i \tau^I \gamma^\mu \ell_j)(\overline{\ell}_k\tau^I \gamma_\mu \ell_l)&\!\!\!= 2~ (\overline{\nu}_i \gamma^\mu P_L e_{j})(\overline{e}_k \gamma_\mu P_L\nu_l) + 2 
(\overline{e}_{i} \gamma^\mu P_L \nu_{j}) (\overline{\nu}_k \gamma_\mu P_L
e_{l}) \nonumber \\
&&~~ + \left[(\overline{\nu}_i \gamma^\mu P_L \nu_j)(\overline{\nu}_k \gamma_\mu P_L \nu_l) +
(\overline{e}_i \gamma^\mu P_L e_{j})(\overline{e}_k \gamma_\mu P_L e_{l}) \right. \nonumber \\
&&~~\left. - (\overline{\nu}_i \gamma^\mu P_L \nu_j)(\overline{e}_k \gamma_\mu P_L e_{l})
-(\overline{e}_i \gamma^\mu P_L e_{j})(\overline{\nu}_k \gamma_\mu P_L
\nu_l)\right] ,\label{Oll3} \\
{\cal O}_{ \ell e}^{ijkl}  &\!\!\!= \quad 2~ (\overline{\ell}_i e_j)(\overline{e}_k \ell_l)~\quad
&\!\!\!= 2~ \left[ (\overline{\nu}_i P_R e_j)(\overline{e}_k P_L \nu_l)+ (\overline{e}_i P_R e_j)(\overline{e}_k P_L e_l)\right]  \nonumber \\
&&\!\!\!=- \left[ (\overline{\nu}_i \gamma^\mu P_L \nu_l) (\overline{e}_k \gamma_\mu P_R e_j)+ (\overline{e}_i \gamma^\mu P_L e_l)
(\overline{e}_k \gamma_\mu P_R e_j) \right] . \label{Olr}
\end{eqnarray}

All these operators, together with $O_{ee}^{ijkl}$, induce dipole moments and four-charged-lepton 
(4CL) vertices, as appear to the right-hand side (RHS) in the above equations.
Constraints on the coefficients of the 4CL operators 
\begin{eqnarray}
{\cal O}^{ijkl}_{PP} = \frac{1}{1+ \delta} (\overline{e}_i \gamma^\mu P e_j)(\overline{e}_k \gamma_\mu P e_l), ~~~~
{\cal O}^{ijkl}_{RL} = \frac{1}{1+ \delta} (\overline{e}_i \gamma^\mu P_R e_j)(\overline{e}_k \gamma_\mu P_L e_l) ,~~~~ \label{4ftobound}
\end{eqnarray} 
where $P = P_R$ or $P_L$, are listed in Tables~\ref{tab:4lfc}, \ref{tab:4lDL1}, \ref{tab:4lDL2} and \ref{tab:4lDL21}. 

After electroweak symmetry breaking, the operators ${\cal O}^{ij}_{eB}$ and 
${\cal O}^{ij}_{eW}$ become the chirality-flipping dipole moments as written 
in Eqs.~(\ref{OeB2},\ref{OeW}) (where we did not include the
$Z$--lepton--lepton operators \cite{Brignole:2004ah}). These dipole which can
be flavour conserving or transition dipole moments. The flavour diagonal
operators are specially interesting because they correspond to the anomalous
magnetic moments and the electric dipole moments of the different fermions. 
Taking $C^{ij}_{e\gamma} (q^2) = C^{ij}_{eB} (q^2) \cos \theta_W -
C^{ij}_{eW} (q^2) \sin \theta_W$ as the Wilson coefficient with 
momentum transfer equal to $q^2$, we have for $q^2=0$,
\begin{eqnarray}\label{DipoleMoments}
\frac{C^{ii}_{e\gamma} (q^2=0)}{m_{\rm NP}^{2}}~\langle H \rangle~ 
\overline{e}_i \sigma^{\mu \nu} P_R e_i~ 
F^{em}_{\mu \nu} &+& h.c.~~ = \nonumber \\ 
\frac{{\rm Re} \{C^{ii}_{e\gamma} (q^2=0)\}}{m_{\rm NP}^{2}}~\langle H \rangle~ 
 \overline{e}_i \sigma^{\mu \nu} e_i ~F^{em}_{\mu \nu} &+& \frac{ {\rm Im}  
\{C^{ij}_{e\gamma} (q^2=0)\}}{m_{\rm NP}^{2}} ~\langle H \rangle~
 ~ i~ \overline{e}_i \sigma^{\mu \nu} \gamma_5 e_i~ F^{em}_{\mu \nu}~ = 
\nonumber \\
e~ \frac{a_{e_i}}{4 m_{e_i}}~ 
 \overline{e}_i \sigma^{\mu \nu} e_i ~F^{em}_{\mu \nu} &+&~ \frac{i}{2}~~ 
d_{e_i}~~ \overline{e}_i \sigma^{\mu \nu} \gamma_5 e_i~ F^{em}_{\mu \nu},
\end{eqnarray} 
with $a_{e_i} =(g_{e_i}-2)/2$ the anomalous magnetic moment and $d_{e_i}$ the
electric dipole moment  of the lepton $e_i$ that can be found in 
\cite{Yao:2006px}.

In a given model, the coefficients of the effective operators can be obtained
by matching the effective theory of Eq.~(\ref{eff.exp}) onto the model, at
some matching scale (for instance, the mass scale of new
particles). However, in particular models there can appear various pitfalls in
constraining the generic coefficients $C^{ijkl}_{\dots}$. This is illustrated, 
for example, in
the model of \cite{Bilenky:1993bt} which corresponds to adding a singlet 
slepton $\tilde{E}^c$ of flavour $k$, in R-parity violating (RPV) SUSY. In this case, after
integrating out the heavy slepton we obtain the following effective operator:
\begin{eqnarray}
\label{RPVex}
\frac{\lambda_{[ij]}^k \lambda_{[mn]}^{*k}}{M^2} {\Big (}\overline{(\nu_L)^c}_i e_{Lj} {\Big )}
{\Big (} \overline{(e_L)}_n (\nu_{L})^c_m {\Big )} = \frac{\lambda_{[ij]}^k \lambda_{[mn]}^{*k}}{2M^2}
(\overline{e}_{n} \gamma^\mu P_Le_{j}) (\overline{\nu}_m \gamma_\mu P_L \nu_i ),
\end{eqnarray}
where $\lambda_{[ij]}^k$ is anti-symmetric in $i,j$ because the SU(2) contraction of $\ell_i \ell_j$ is antisymmetric. This is an example of operator ${\cal O}_{\ell \ell (1)}$,
but since it is induced by singlet scalar exchange, there is no four-charged-lepton operator (compare to Eq.~(\ref{Oll1})).
This illustrates that the bounds obtained here, by assuming that $C^{ijkl}_{\dots} \neq 0$ for one choice of $ijkl$ at a time, are not generic.
Each process receives contributions from a sum of operators, and that sum could contain cancellations in a particular model. 

Many models of new physics introduce new TeV-scale particles carrying a conserved quantum number ($e.g.$ R-parity, T-parity...). Such
particles appear in pairs at vertices, so contribute via boxes and penguins to the four-fermion and dipole moment operators
considered here. Generic formulae for the one-loop contribution to a dipole
moment can be found in \cite{Lavoura:2003xp}, and for boxes in
\cite{Inami:1980fz}. Extra Higgses \cite{Brignole:2003iv,Paradisi:2006jp} 
would contribute to the same operators constructed from SM fields, so are 
constrained by the experimental limits on the coefficients of such operators.  

\subsubsection{Constraints on low scale observables }\label{sec:taudecays}

In this section we present the low-energy constraints on the different Wilson
coefficients introduced before. Any NP found at LHC will necessarily respect 
the bounds presented here.

\subsubsubsection{Dipole transitions}

After electroweak symmetry breaking, the operators of
Eqs.~(\ref{OeB2}), (\ref{OeW}) generate magnetic and electric dipole
moments for the charged leptons. Flavour-diagonal operators give rise
to anomalous magnetic moments and electric dipole moments as shown in
Eq.~(\ref{DipoleMoments}).  The anomalous magnetic moment of the
electron $a_e = (g-2)_e/2$ is used to determine $\alpha_{\rm em}$. The
current measurement of the muon anomalous moment $a_\mu=(g-2)_\mu/2$
deviates from the (uncertain) SM expectation by 3.2 $\sigma$ using
$e^+ e^-$--data \cite{Jegerlehner:2007xe}, and can be taken as a
constraint, or indication on the presence of New Physics.  Currently
there is only an upper bound on the magnetic moment of the $\tau$ from
the analysis of $e^+ e^- \to \tau^+ \tau^-$
\cite{Gonzalez-Sprinberg:2000mk,Yao:2006px}.  Electric dipole moments
have not yet been observed, although we have very constraining bounds
specially on the electron dipole moment. In Table~\ref{tab:dipoles} we
present the bounds of flavour-diagonal dipole moments. The EDMs are
discussed in detail in Section~\ref{sec:phenomenology}.

The bounds on off-diagonal dipole transitions are presented in Table~\ref{tab:dipoles}. It is convenient to normalize these coefficients,
 $C^{ij}_{e\gamma}= C^{ij}_{eB} 
\cos \theta_W - C^{ij}_{eW} \sin \theta_W$, 
to the Fermi interactions given our ignorance on the scale of new 
physics $m_{\rm NP}$ :
\begin{eqnarray} 
\frac{C_{e \gamma}^{ij}}{m_{\rm NP}^2} = \frac{4G_F}{\sqrt{2}} \epsilon_{e \gamma}^{ij} ~~~~,\label{D-eps}
\end{eqnarray}
In the literature, it is customary to use the left and right form-factors for
lepton flavour violating transitions defined as,
\begin{equation}
\label{dipformfac1}
\Delta {\cal L}_2=m_{l_i} A_{\mu}\overline{e}_j [i \sigma^{\mu\nu}q_\nu
 (A_{L}^{ij}P_{L}+A_{R}^{ij}P_{R})]e_i ~+~h.c.
\end{equation}
The radiative decay $f_i \rightarrow f_j + \gamma$ proceeds at the rate
$\Gamma = m_i^5 e^2/(16 \pi) (|A_L^{ij}|^2 + |A_R^{ij}|^2)$ \cite{Hisano:1995cp}.
Bounds on the dimensionless coefficients $C^{ij}_{e\gamma}$ and $\epsilon_{e
  \gamma}^{ij}$ can be  obtained by translating from $A_L^{ij}$ and $A_R^{ij}$ 
as: 
\begin{eqnarray}
\label{dipformfac2}
\frac{C^{ij}_{e \gamma}}{m_{\rm NP}^2} \langle H \rangle = \frac{m_i}{2} A_R^{ij},~~ ~~~~~~
\frac{{C^{ji}}^*_{e \gamma}}{m_{\rm NP}^2} \langle H \rangle = \frac{m_i}{2} A_L^{ij} .
\end{eqnarray}
The experimental bounds on radiative lepton decays can be used to set
bounds on these off-diagonal Wilson coefficients. The current experimental
bounds are B$(\mu \rightarrow e \gamma) < 1.2 \times 10^{-11}$\cite{Brooks:1999pu}, 
B$(\tau \rightarrow \mu \gamma) < 4.5 \times 10^{-8}$  \cite{Hayasaka:2007vc}, and  
B$(\tau \rightarrow e \gamma) < 1.1 \times 10^{-7}$\cite{Aubert:2005wa}.

For the off-shell photon, $q^2\neq 0,$ there exist additional form factors,
\begin{equation}
\label{dipformfacnew}
\Delta {\cal L}=m_{l_i} A_{\mu}
\overline{e}_j  \left[ \left( g_{\mu\nu} -\frac{q_\mu q_\nu}{q^2}\right) \gamma_\nu \,
 (B_{L}^{ij}P_{L}+B_{R}^{ij}P_{R})\right]e_i  ~+~h.c.,
\end{equation}
which induce contributions to the four-fermion operators to be discussed in the 
next subsections. These form factors may be enhanced by a large factor compared
to the on-shell photon form factors \cite{Raidal:1997hq},
$\ln (m_{NP}/m_{l_i}),$ depending on the nature of new physics.
Therefore those operators become relevant for constraining new physics in
$R$-parity violating SUSY \cite{Huitu:1997bi} and in low-scale type-II seesaw 
models \cite{Raidal:1997hq}.

\begin{table}[bht]
\begin{minipage}{\textwidth}
\caption{Bounds on the different dipole coefficients. Flavour diagonal dipole coefficients are given in terms of the corresponding 
anomalous magnetic moment, $a_{e_i}$, and the dipole moment,  $d_{e_i}$. Bounds on transition moments are given in terms of the dimensionless 
coefficients $|\epsilon^{ij}_{e \gamma}|$ (defined in Eq.~(\ref{D-eps})) from the bounds on the branching ratios given in the last column. 
These bounds apply also both to $|\epsilon^{ij}_{e \gamma}|$ and $|\epsilon^{ji}_{e \gamma}|$. See Section~\ref{sec:taudecays} for details. \label{tab:dipoles} }
\begin{tabular*}{\textwidth}{@{\extracolsep{\fill}} l|ccl  }
\hline\hline
($ij$) 			&$ a_i = \frac{g_i-2}{2}$ 			&edm$_i$ (e~cm) 			&Ref. 				\\
\hline
$\overline{e}e$ 	&0.0011596521859(38) 				&$d_e  \le 1.6 \times 10^{-27}$  	&PDG \cite{Yao:2006px},\cite{Regan:2002ta}	\\
$\overline{\mu}\mu$ 	&$11 659 208.0(5.4)(3.3) \times 10^{-10} $ 	&$d_{\mu} \le 2.8\times 10^{-19}$ 	&Muon g-2 Coll. \cite{Bennett:2006fi,McNabb:2004tj}  \\
$\overline{\tau}\tau$ 	&$-0.052 <a_{\tau}<0.013$			&$(-2.2 <d_{\tau}<4.5)\times 10^{-17}$	&LEP2 \cite{Abdallah:2003xd}, BELLE \cite{Inami:2002ah}\\
\hline
&&&\\[-3mm]
($ij$) 			&$\overline{\ell}_i \sigma^{\mu \nu}e_{R j} F_{\mu \nu}^{\rm em} $ &&Ref.\\
\hline
$\overline{e}\mu$ 	&$\le 1.1 \times 10^{-10}$ 			&& MEGA Coll. \cite{Brooks:1999pu}\\
$\overline{e}\tau$ 	&$\le 4.3 \times 10^{-7}$  			&& BABAR \cite{Aubert:2005wa}\\
$\overline{\mu}\tau$ 	&$\le 2.8 \times 10^{-7}$			&& Belle, BABAR \cite{Hayasaka:2007vc,Aubert:2005ye}\\
\hline\hline
\end{tabular*}
\end{minipage}
\end{table}

\subsubsubsection{Four-charged-lepton operators}

As before, to present the bounds on the dimensionless four-charged-fermion coefficients
in Eq.~(\ref{4ftobound}), we normalize them to the Fermi interactions :
\begin{eqnarray} 
\frac{C_{(n)\ell \ell}^{ijkl}}{m_{\rm NP}^2} = - \frac{4G_F}{\sqrt{2}} \epsilon_{(n)\ell \ell}^{ijkl} ~~~~, ~~~ \frac{C_{ee}^{ijkl}}{m_{\rm NP}^2}
=  - \frac{4G_F}{\sqrt{2}} \epsilon_{ee}^{ijkl} ~~~, ~~~ \frac{C_{\ell e}^{ilkj}}{m_{\rm NP}^2} = 
\frac{4G_F}{\sqrt{2}} \epsilon_{\ell e}^{ijkl} .~~~~ \label{S12}
\end{eqnarray}
The current low-energy constraints on the dimensionless $\epsilon$'s
are shown in Tables~\ref{tab:4lfc}, \ref{tab:4lDL1}, \ref{tab:4lDL2} and
\ref{tab:4lDL21}.  The rows of the tables are labeled by the flavour
combination, and the column by the Lorentz structure.  The numbers
given in this tables correspond to the best current experimental bound
on the coefficient of each operator, assuming it is the only non-zero
coefficient present.  The last column in the table lists the
experiment setting the bound.  The compositeness search limits
$\Lambda @ $ LEP are at 95~$\%$ C.L., the decay rate bounds at 90~$\%$
C.L.

Regarding the definition of the different coefficients we have to make some
comments. First, note the flavour index permutation between 
$C_{\ell e}$ and $\epsilon_{\ell e}$ as,
\begin{equation}
C_{\ell e}^{ilkj} (\overline{\ell}_i e_l)(\overline{e}_k \ell_j)=  - \frac{1}{2}\epsilon_{\ell e}^{ijkl}
(\overline{\ell}_i \gamma^\mu \ell_j)(\overline{e}_k \gamma_\mu e_l ).
\end{equation}
There are relations between the  flavour indices of the different
operators. For $ {\cal O}_{LL} = (\overline{e} \gamma^\mu P_L e) (\overline{e}
\gamma_\mu P_L e)$ and ${\cal O}_{RR} = (\overline{e} \gamma^\mu P_R
e)(\overline{e} \gamma_\mu P_R e) $ we have:
\begin{eqnarray}
{\cal O}_{PP}^{ijkl} = {\cal O}_{PP}^{klij} , ~~~ {\cal O}_{PP}^{ijkl} = {\cal O}_{PP}^{*jilk} , ~~~ {\cal O}_{PP}^{ijkl} = {\cal O}_{PP}^{ilkj},
\label{equalities}
\end{eqnarray}
by symmetry, Hermitian conjugation and Fiertz rearrangement, respectively. Therefore the constraints on  $ \overline{e} e \overline{\mu} \tau$  in the first two
columns of Tables~\ref{tab:4lfc} to \ref{tab:4lDL21} apply to
$\epsilon_{(n)xx}^{ e e \mu \tau}$, 
$\epsilon_{(n)xx}^{ \mu \tau e e}$, 
$\epsilon_{(n)xx}^{* e e \tau \mu}$, 
$\epsilon_{(n)xx}^{* \tau \mu e e}$, 
$\epsilon_{(n)xx}^{ e \tau \mu e }$, 
$\epsilon_{(n)xx}^{ \mu e e \tau }$, 
$\epsilon_{(n)xx}^{* \tau e e \mu}$, and 
$\epsilon_{(n)xx}^{*e \mu \tau e }$ with $(n)xx$ equal to $(3) \ell \ell, 
(1) \ell \ell$, or $(1) ee$. Note, however that it
is calculated assuming only one of these $\epsilon$ is non-zero.
 Similarly, the operator ${\cal O}^{ijkl}_{LR} = (\overline{e}_i \gamma_\mu
 P_L e_j) (\overline{e}_k \gamma^\mu P_R e_l)$, with coefficient  
$\epsilon_{\ell e}^{ijkl}$, is 
related by Hermitian conjugation:
\begin{eqnarray}
{\cal O}_{LR}^{ijkl} = {\cal O}_{LR}^{*jilk} ,
\end{eqnarray}
so again the bounds on $\epsilon_{\ell e}^{ijkl}$ apply to 
$\epsilon_{\ell e}^{*jilk}$. 
We can usually  apply also  these bounds 
 to $\epsilon_{\ell e}^{klij}$ because the chirality of the fermion legs does
 not affect the matrix element squared, but $\epsilon_{\ell e}^{ilkj }$ is 
bounded  separately in the Tables.

The bounds from $Z$ decays in Tables~\ref{tab:4lfc} and \ref{tab:4lDL1} are 
estimated from the one-loop penguin diagram obtained closing two of the legs
of the four fermion operator and coupling it with the $Z$
\cite{Davidson:2003ha}. These bounds would be more correctly included by
renormalization group mixing between the four fermion operators and the Z
-fermion-fermion operators discussed in \cite{Brignole:2004ah}. They are
listed in the tables to indicate the existence of a constraint.
The  bound can be applied to $\epsilon_{\ell e}^{iikl}$ and 
$\epsilon_{\ell e}^{ijkk}$ but it does not apply to $\epsilon_{\ell e}^{ilki}$.

Contact interaction bounds are usually quoted on the scale $\Lambda$, where
\begin{equation}
\epsilon_{a b}^{ijkl} \frac{4 G_F}{\sqrt{2}} = \pm \frac{1}{1+ \delta} \frac{4 \pi}{\Lambda^2},
\end{equation}
and $\delta = 1$ for the operators ${\cal O}_{LL}^{eeee}$ and ${\cal
  O}_{RR}^{eeee}$ of Eq.~(\ref{4ftobound}), 0 otherwise.  Since our
  normalization does not have this factor of 2, we have a Feynman rule
  $\epsilon 8 G_F/\sqrt{2}$ for these operators, and correspondingly stricter
  bounds on the $\epsilon$'s.  The bounds are the same for  
$\epsilon_{\ell e}^{ikki}$ and $\epsilon_{\ell e}^{kiik}$.
However, contact interaction bounds are not quoted
on operators of the form 
$(\overline{e}_i \gamma^\mu P_L e_j)(\overline{e}_j \gamma_\mu P_R e_i)$,
corresponding to $\epsilon_{\ell e}^{iijj}$.  Such operators
are generated by sneutrino exchange in R-parity violating SUSY, 
so we estimate  the  bound $ \lambda^2/m_{\tilde{\nu}}^2 < 4/(9$~TeV$^2$) 
from the plotted constraints in \cite{Acciarri:2000uh}, and
impose $ 4 | \epsilon_{a b}^{ijkl} | G_F/\sqrt{2} < \lambda^2/(2m_{\tilde{\nu}}^2) $.

\begin{table}[bht]
\begin{minipage}{\textwidth}
\caption{ Bounds on coefficients of flavour conserving 4-lepton operators, from four-charged-lepton processes.
The number is the upper bound on the dimensionless operator coefficient $ \epsilon^{ijkl}$ (defined in Eq.~(\ref{S12})), arising from the measurement
in the last column. The bound applies also to $ \epsilon^{klij}$. The second column is the bounds on
$ \epsilon_{(3) \ell \ell}^{ijkl}$,  and  $ \epsilon_{(1) \ell \ell}^{ijkl}$ [except in the case
of the bracketed limits, which are the upper bound on  $ \epsilon_{(1) \ell \ell}^{ijkl}$ and  $2 \epsilon_{(1) \ell \ell}^{ijkl}$]. 
 The third  column is the bound on  $ \epsilon_{(1) ee}^{ijkl}$. The bounds in these two columns
apply also when the flavour indices are permuted to   ${jilk}$ and ${ilkj}$. The fourth column is the bound on $  \epsilon_{ \ell e}^{ijkl}$ (which does not apply
to the flavour permutation $ilkj$, so this is listed with a line of its own). 
The constraints in [brackets] apply to the 2-charged-lepton-2-neutrino operator of the same flavour structure, and arise from lepton universality in $\tau$ decays. 
See Section~\ref{sec:taudecays} for details.  . \label{tab:4lfc}}
\begin{tabular*}{\textwidth}{@{\extracolsep{\fill}}l|rrrll  }
\hline\hline
($ijkl$) 	&\multicolumn{3}{c}{$(\overline{e} \gamma^\mu P_L e)(\overline{e} \gamma_\mu P_L e)\;\; (\overline{e}\gamma^\mu P_R e)(\overline{e} \gamma_\mu P_R e)\;\;
(\overline{e} \gamma_\mu P_L e)(\overline{e} \gamma^\mu P_R e)$}
		&expt. limit 		&Ref.		\\
\hline
$\overline{e}e\overline{e}e$ 		&(-1.8 $-$ +2.8) $\cdot 10^{-3}$  &(-1.8 $-$ +2.8) $\cdot 10^{-3}$ 	&(-2.4 $-$ +4.9) $\cdot 10^{-3}$ 
		&$\Lambda @$LEP2 	&\hspace*{-3mm}\cite{Bourilkov:2001pe}				\\	
$\overline{e}e\overline{\mu}\mu$ 	&(-7.2 $-$ +5.2) $\cdot 10^{-3}$  &(-7.8 $-$ +5.8) $\cdot 10^{-3}$ 	&(-9.0 $-$ +9.6) $\cdot 10^{-3}$ 
		&$\Lambda @$LEP2	&\hspace*{-3mm}\cite{Abbiendi:2003dh,Acciarri:2000uh} 		\\
$\overline{e}\mu\overline{\mu}e $	&(-7.2 $-$ +5,2) $\cdot 10^{-3}$  &(-7.8 $-$ +5.8) $\cdot 10^{-3}$ 	&1.3 $\cdot 10^{-2}$ 
		&$\Lambda, RPV@$LEP2	&\hspace*{-3mm}\cite{Abbiendi:2003dh,Acciarri:2000uh} 		\\
$\overline{e}e\overline{\tau} \tau$ 	&(-7.3 $-$ +13) $\cdot 10^{-3}$   &(-8.0 $-$ +15) $\cdot 10^{-3}$ 	&(-1.2 $-$ +1.8) $\cdot 10^{-2}$ 
		&$\Lambda @$LEP2	&\hspace*{-3mm}\cite{Abbiendi:2003dh,Acciarri:2000uh}			\\
$\overline{\tau}e\overline{e} \tau$  	&(-7.3 $-$ +13) $\cdot 10^{-3}$   &(-8.0 $-$ +15) $\cdot 10^{-3}$ 	&1.3 $\cdot 10^{-2}$ 
		&$\Lambda,RPV @$LEP2 	&\hspace*{-3mm}\cite{Abbiendi:2003dh,Acciarri:2000uh}			\\
$\overline{\mu}\mu \overline{\mu}\mu$ 	&$\sim 1$ 			  &$\sim 1$ 				& $\sim 1$
		&$B(Z \to \mu \bar{\mu})$ 							\\
$\overline{\mu}\mu \overline{\tau}\tau$ &$\sim 1\;\; [0.0014]$ 		  &$\sim 1 				$ &$\sim 1 \;\; [0.01]$
		&$B(Z \to \mu \bar{\mu})$							\\
$\overline{\mu}\tau \overline{\tau}\mu$ &$\sim 1\;\; [0.0014]$   	  &$\sim 1$ 				&
		&$B(Z \to \mu \bar{\mu})$							\\
$\overline{\tau}\tau \overline{\tau}\tau$&$\sim 1$			  &$\sim 1$				&$\sim 1$
		&$B(Z \to \tau \bar{\tau})$							\\
\hline\hline
\end{tabular*}
\end{minipage}
\end{table}

\begin{table}[bht]
\begin{minipage}{\textwidth}
\caption{Bounds on coefficients of 4-lepton operators with  $\Delta L_\alpha = -\Delta L_\beta = 1$. 
They apply also to flavour index permutations $klij$ and $ilkj$, except in the case of $\tau \tau e \mu$, where the
bound on $\tau \mu e \tau$ in the fourth column is from $\mu$ decay and is listed separately.
See the caption of Table~\ref{tab:4lfc} and Section~\ref{sec:taudecays} for  further details.\label{tab:4lDL1} }
\begin{tabular*}{\textwidth}{@{\extracolsep{\fill}}l|cccl}
\hline\hline
($ijkl$) 				&$(\overline{e} \gamma^\mu P_L e)(\overline{e} \gamma_\mu P_L e)$ &$(\overline{e} \gamma^\mu P_R e)(\overline{e} \gamma_\mu P_R e)$ 
					&$(\overline{e} \gamma_\mu P_L e)(\overline{e} \gamma^\mu P_R e)$ &expt. limit 				\\
\hline
$\overline{e}e\overline{e}\mu$ 		&$7.1\cdot 10^{-7}$ 						&$7.1\cdot 10^{-7}$
					&$7.1\cdot 10^{-7}$ 						&B$(\mu \to e\overline{e}e) < 10^{-12}$ \\
$\overline{e}e\overline{e} \tau$ 	&$7.8\cdot 10^{-4}$ 						&$7.8\cdot 10^{-4}$
					&$7.8\cdot 10^{-4}$						&B$(\tau \to e\overline{e}e) < 2 \cdot 10 ^{-7}$ \\
$\overline{e}e\overline{\mu} \tau$ 	&$1.1\cdot 10^{-3}$ 						&$1.1\cdot 10^{-3}$
					&$1.1\cdot 10^{-3}$						&B$(\tau \to \overline{e}e \mu) < 1.9 \cdot 10^{-7} $\\
$\overline{\mu}\mu\overline{e}\mu$ 	&$\sim 1 $ 							&$\sim 1 $
					&$\sim 1 $ 							&B$(Z \to e \bar{\mu}) < 1.7 \cdot 10^{-6}$ \\
$\overline{\mu}\mu\overline{e} \tau$ 	&$1.1\cdot 10^{-3}$ 						&$1.1\cdot 10^{-3}$
					&$1.1\cdot 10^{-3}$						&B$(\tau \to \overline{\mu}e \mu) < 2.0 \cdot 10^{-7} $ \\
$\overline{\mu}\mu\overline{\mu} \tau$ 	&$7.8\cdot 10^{-4}$ 						&$7.8\cdot 10^{-4}$
					&$7.8\cdot 10^{-4}$						&B$(\tau \to 3\mu) < 1.9 \cdot 10^{-7}$ \\
$\overline{\tau}\tau\overline{e}\mu$ 	&$\sim 1\;\; [0.05]$ 						&$\sim 1 $
					&$\sim 1\;\; [0.05]$ 						&B$(Z \to e \bar{\mu}) < 1.7 \cdot 10^{-6}$ \\
$\overline{\tau}\mu\overline{e}\tau$ 	&$\sim 1\;\; [0.05]$ 						&$\sim 1 $
					&$  [0.05]$ 							&B$(Z \to e \bar{\mu}) < 1.7 \cdot 10^{-6}$ \\
$\overline{\tau}\tau\overline{e}\tau$ 	&$\sim 3\;\;[0.05]$ 						&$\sim 3$
					&$\sim 3\;\;[0.05]$ 						&B$(Z \to e \bar{\tau}) < 9.8 \cdot 10^{-6}$ \\
$\overline{\tau}\tau\overline{\tau}\mu$ &$\sim 3\;\;[0.05]$ 						&$\sim 3$
					&$\sim 3\;\;[0.05]$ 						&B$(Z \to \tau \bar{\mu}) < 1.2 \cdot 10^{-5}$\\
\hline\hline
\end{tabular*}
\end{minipage}
\end{table}

\begin{table}[bht]
\begin{minipage}{\textwidth}
\caption{Bounds on coefficients of  4-lepton operators with $\Delta L_\alpha =\Delta L_\beta = 2$. 
See the caption of Table~\ref{tab:4lfc} and Section~\ref{sec:taudecays} for details. \label{tab:4lDL2} }
\begin{tabular*}{\textwidth}{@{\extracolsep{\fill}}c|cccc}
\hline\hline
($ijkl$) 				&$(\overline{e}\gamma^\mu P_L e)(\overline{e}\gamma_\mu P_L e)$ &$(\overline{e}\gamma^\mu P_R e)(\overline{e}\gamma_\mu P_R e) $ 
					&$(\overline{e}\gamma_\mu P_L e)(\overline{e}\gamma^\mu P_R e)$ & expt. limit 		\\
\hline
$\overline{e} \mu\overline{e} \mu$ &	$3.0\cdot 10^{-3}$						& $3.0\cdot 10^{-3}$
					&$2.0\cdot 10^{-3}$ 						& $(\bar{\mu}e) \leftrightarrow (\bar{e} \mu)$ \\
$\overline{e} \tau\overline{e} \tau$ 	&$[0.05]$ 							&
					&$[0.05]$ 							&						\\
$\overline{\mu}\tau\overline{\mu} \tau$ &$[0.05]$ 							&
					&$[0.05]$ 							&						\\ 
\hline\hline
\end{tabular*}
\end{minipage}
\end{table}

\begin{table}[bht]
\begin{minipage}{\textwidth}
\caption{Bounds on coefficients of  4-lepton operators with $\Delta L_\alpha = \Delta L_\beta = -\frac{1}{2}\Delta L_\rho$.
See the caption of Table~\ref{tab:4lfc} and Section~\ref{sec:taudecays} for details.  . \label{tab:4lDL21} }
\begin{tabular*}{\textwidth}{@{\extracolsep{\fill}}l|c c c c}
\hline\hline
($ijkl$) 				&$(\overline{e}\gamma^\mu P_L e)(\overline{e}\gamma_\mu P_L e)$ &$(\overline{e}\gamma^\mu P_R e)(\overline{e}\gamma_\mu P_R e)$ 
					&$(\overline{e}\gamma^\mu P_L e)(\overline{e}\gamma^\mu P_R e)$	& expt. limit 						\\
\hline
$\overline{e}\mu\overline{e}\tau$	&$2.3\cdot 10^{-4}$ 						&$2.3\cdot 10^{-4}$
					&$2.3\cdot 10^{-4}$						&$B(\tau \to \overline{\mu}e e) < 1.1 \cdot 10^{-7} $ 	\\
$\overline{\mu}e\overline{\mu}\tau$ 	&$2.6\cdot 10^{-4}$						&$2.6\cdot 10^{-4}$
					&$2.6\cdot 10^{-4}$						&$B(\tau \to\overline{e}\mu\mu) < 1.3 \cdot 10^{-7} $	\\
$\overline{\tau} e\overline{\tau} \mu$  &$[0.05]$							&
					&$[0.05]$							&							\\
\hline\hline
\end{tabular*}
\end{minipage}
\end{table}

Many of the 4CL operators involving two $\tau$'s are poorly constrained. In
some  cases, see Eqs.~(\ref{Oll1}, \ref{Oll3}), new physics that generates 4CL 
operators also induces $(\overline{e}_i \gamma^\lambda P e_j)(\overline{\nu}_k
\gamma_\lambda L \nu_l)$. The coefficients of operators of the form 
$(\overline{\mu} \gamma^\lambda P e)(\overline{\nu}_k
\gamma_\lambda L \nu_l)$, $(\overline{\mu} \gamma^\lambda P
\tau)(\overline{\nu}_k \gamma_\lambda L \nu_l)$ or $(\overline{e}
\gamma^\lambda P \tau)(\overline{\nu}_k \gamma_\lambda L \nu_l)$,  
are constrained from lepton universality measurements in $\mu$ and  $\tau$ 
decays \cite{Pich:1995vj}. The decay rate  
$\tau \to e_i \nu_k \bar{\nu}_l $ in the presence of the operators of Eq.~(\ref{4ftobound}), divided by the SM prediction for 
$\tau \to e_i \nu_\tau \overline{\nu}_i$, is
\begin{eqnarray}
( 1 - 2 \delta_{k \tau} \delta_{ il} {\rm Re} \{ \epsilon_{(1) \ell \ell}^{ \tau \tau i i} + 2\epsilon_{(3) \ell \ell}^{\tau \tau i i} \}
+ \frac{4 m_i}{ m_\tau} \delta_{k \tau} \delta_{ il} {\rm Re} \{ \epsilon_{\ell e}^{\tau \tau i i} \}  +
 |\epsilon_{(1) \ell \ell}^{i \tau kl}|^2 + 4|\epsilon_{(3) \ell \ell}^{i \tau kl}|^2 
 + |\epsilon_{\ell e}^{i \tau kl}|^2 ). 
\end{eqnarray}
Within the experimental accuracy, the weak $\tau$ and $\mu$ decays verify lepton universality and agree with LEP precision measurements
of $m_W$. Rough bounds on the $\epsilon$'s can therefore be obtained by requiring the new physics contribution to the decay rates to be
less than the errors $\frac{\Delta B}{B} (\tau \to e \nu \overline{\nu})= 0.05/17.84$, $\frac{\Delta B}{B} 
(\tau \to \mu \nu \overline{\nu})= 0.05/17.36$. These are listed in the tables in [brackets]. The bracketed  limit
in the second column applies to $\epsilon^{ijkl}_{(1)\ell ell}$; the bound on $\epsilon^{ijkl}_{(3)\ell ell}$ is 1/2
the quoted number. The limit on $\epsilon_{\ell e}^{\tau  e \tau \mu}$ is from its contribution to $\mu \to e \nu_\tau \bar{\nu}_\tau$.

Finally, we would like to remind the reader the various caveats to these 4-fermion vertex bounds.
\begin{list}{-}{}
\item The constraints are calculated ``one operator at a time''. 
This is unrealistic; new physics is likely to induce many non-renormalizable 
operators. In some cases, see Eq.~(\ref{RPVex}), a symmetry
in the new physics can cause cancellations such that it does not contribute to
certain observables.
\item The coefficients of the 4CL operators, and two-$\nu$- two-charged-lepton (2$\nu$2CL) operators 
 may differ by a factor of few, because they are induced by the exchange of different members
of a multiplet, whose masses differ \cite{Bergmann:1999pk}. 
\item The list of operators is incomplete.  Perhaps some of the neglected
operators give relevant constraints on New Physics. For instance, bounds
from lepton universality on  the $(H^* \overline{\ell}) \gamma^\mu \partial_\mu (H \ell)$
operator \cite{Gonzalez-Garcia:1991be} are relevant to extra-dimensional scenarios 
\cite{DeGouvea:2001mz}.
\item Operators of dimension $>6$ are neglected. If  the mass scale of the
New Physics is $\sim $ TeV, then higher dimension operators with
Higgs VEVs\cite{Berezhiani:2001rt}, such as $HH \bar{\psi}\psi \bar{\psi}\psi$ are not
significantly suppressed. 
\end{list}

\subsubsubsection{Two lepton--two quark operators}

Once more, we normalize the coefficients of the two lepton-two quark operators
in Eq.~(\ref{OLR}) to the Fermi interactions:
\begin{equation}
\begin{array}{cccc} 
\frac{C_{(n)\ell q}^{ijkl}}{m_{\rm NP}^2} = - \frac{4G_F}{\sqrt{2}} \epsilon_{(n)\ell q}^{ijkl} ,&
\frac{C_{ed}^{ijkl}}{m_{\rm NP}^2}        = - \frac{4G_F}{\sqrt{2}} \epsilon_{ed}^{ijkl} ,&
\frac{C_{\ell d}^{ijkl}}{m_{\rm NP}^2}    =   \frac{4G_F}{\sqrt{2}} \epsilon_{\ell d}^{ijkl} ,&\\[2mm]
\frac{C_{e u}^{ijkl}}{m_{\rm NP}^2}       = - \frac{4G_F}{\sqrt{2}} \epsilon_{e u}^{ijkl} ,&
\frac{C_{\ell u}^{ijkl}}{m_{\rm NP}^2}    = - \frac{4G_F}{\sqrt{2}} \epsilon_{\ell u}^{ijkl} ,&
\frac{C_{\ell q S}^{ijkl}}{m_{\rm NP}^2}  = - \frac{4G_F}{\sqrt{2}} \epsilon_{\ell q S}^{ijkl} ,&
\frac{C_{q d e}^{ijkl}}{m_{\rm NP}^2}     = - \frac{4G_F}{\sqrt{2}} \epsilon_{q d e}^{ijkl} .
\label{S2Q12}
\end{array}
\end{equation}
The main bounds on the dimensionless $\epsilon$s
are given in Tables~\ref{tab:2qlv} and \ref{tab:2qlvs}. These numbers
correspond to the best current experimental bound
on the coefficient of each operator, assuming it is the only non-zero
coefficient present. The bounds on $\epsilon_{\ell q}$ in
Table~\ref{tab:2qlv} apply both to $\epsilon_{(1)\ell q}$ and
$\epsilon_{(3)\ell q}$. These bounds have been obtained from the corresponding
bounds on leptoquark couplings in references
\cite{Davidson:1993qk,Herz:2002gq} that can be checked for further details.

\begin{table}[bht]
\begin{minipage}{\textwidth}
\caption{ Bounds on coefficients of the left-handed vector 2 quark-2 lepton 
operators.
Bound is the upper bound on the dimensionless operator coefficient $ \epsilon^{ijkl}$ (defined in Eq.~(\ref{S2Q12})), arising from the
experimental determination of the observable in the next column. Bounds with a $*)$ are also valid under the
exchange of the lepton indices. \label{tab:2qlv}}
\begin{tabular*}{\textwidth}{@{\extracolsep{\fill}}l|cl|l|cl}
\hline\hline
\multicolumn{6}{c}{$(\overline{e} \gamma^\mu P_L e)(\overline{q} \gamma_\mu  P_L q)$ }	\\
\hline
($ijkl$)     &Bound on $\epsilon_{\ell q}^{ijkl}$ &observable                    &
($ijkl$)     &Bound on $\epsilon_{\ell q}^{ijkl}$ &observable                    \\
\hline
11 11        &5.1 $\cdot 10^{-3}$          &$R_\pi$                              &
22 11        &5.1 $\cdot 10^{-3}$          &$R_\pi$                              \\
12 11        &8.5 $\cdot 10^{-7}$          &$\mu - e$ conversion on Ti           &
12 12$^{*)}$ &2.9 $\cdot 10^{-7}$          &$B(K^0_L \to \overline{\mu}e)$       \\
$ij$ 12      &4.5$\cdot 10^{-6}$           &$\frac{B(K^+ \to \pi^+ \overline{\nu} \nu)}{B(K^+ \to \pi^0 e^+ \nu_e)}$ &
$ij$ 22      &1.0                          &$V_{cs}$                             \\
$ij$ 13      &3.6 $\cdot 10^{-3}$          &$V_{ub}$                             &
$ij$ 23      &4.2 $\cdot 10^{-2}$          &$V_{cb}$                             \\
11 23        &6.6 $\cdot 10^{-5}$          &$B(B^+ \to e^+ e^-  K^+)$            &
11 13        &9.3 $\cdot 10^{-4}$          &$B(B^+ \to e^+ e^- \pi^+)$           \\
22 23        &5.4 $\cdot 10^{-5}$          &$B(B^+ \to \mu^+ \mu^- K^+)$         &
22 13        &1.4 $\cdot 10^{-3}$          &$B(B^+ \to \mu^+ \mu^- \pi^+)$       \\
21 23$^{*)}$ &4.5 $\cdot 10^{-3}$          &$B(B^+ \to e^+ \mu^- K^+)$           &
21 13$^{*)}$ &3.9 $\cdot 10^{-5}$          &$B(B^+ \to e^+ \mu^- \pi^+)$         \\
12 23$^{*)}$ &1.2 $\cdot 10^{-2}$          &$B(B^0_s \to \mu^+ e^-)$             &
33 12        &6.6 $\cdot 10^{-2}$          &$K$--$\overline K$                   \\
22 22        &6.0 $\cdot 10^{-2}$          &$\frac{B( D_s^+ \to \mu^+ \nu_\mu)}{B( D_s^+ \to \tau^+ \nu_\tau)}$ &
33 22        &6.0 $\cdot 10^{-2}$          &$\frac{B( D_s^+ \to \mu^+ \nu_\mu)}{B( D_s^+ \to \tau^+ \nu_\tau)}$ \\
32 23$^{*)}$ &1.2 $\cdot 10^{-3}$          &$B(B^+ \to \mu^+ \tau^- X^+)$        &
33 23        &9.3 $\cdot 10^{-3}$          &$B( B^+ \to \tau^+ \tau^- X^+)$     \\
\hline\hline
\end{tabular*}
\end{minipage}
\end{table}

\begin{table}[bht]
\begin{minipage}{\textwidth}
\caption{ Bounds on coefficients of the right-handed vector and scalar 
2 quark-2 lepton operators.
Bound is the upper bound on the dimensionless operator coefficient $ \epsilon^{ijkl}$ (defined in Eq.~(\ref{S2Q12})), arising from the
experimental determination of the observable in the next column. Bounds with a $*)$ are also valid under the
exchange of the lepton indices. \label{tab:2qlvs}}
\begin{tabular*}{\textwidth}{@{\extracolsep{\fill}}l|cl||l|cl  }
\hline\hline
\multicolumn{6}{c}{$(\overline{e} \gamma^\mu P_R e)(\overline{q} \gamma_\mu
  P_R q)$ }	\\ \hline
($ijkl$) &Bound on $\epsilon_{e u}^{ijkl}$  &observable         &
($ijkl$) &Bound on $\epsilon_{e u}^{ijkl}$  &observable	        \\
\hline
11 12 &1.7 $\cdot 10^{-2}$      &$\frac{B(D^+ \to \pi^+ e^+ e^-)}{B(D^0 \to \pi^- e^+ \nu_e)}$         &
21 12$^{*)}$ &1.3 $\cdot 10^{-2}$&$\frac{B(D^+ \to \pi^+ \mu^- e^+)}{B(D^0 \to \pi^- e^+ \nu_e)}$       \\
22 12 &9.0 $\cdot 10^{-3}$      &$\frac{B(D^+ \to \pi^+ \mu^+ \mu^-)}{B(D^0 \to \pi^- e^+ \nu_e)}$     &
33 12 &0.19                     &$B(D^0$-- $\overline{D}^0$) 			                       \\
\hline
\multicolumn{6}{c}{$(\overline{\ell} P_R e)(\overline{d} P_L q)$ }	                               \\
\hline
($ijkl$) &Bound on $\epsilon_{q d e}^{ijkl}$ &observable                                               &
($ijkl$) &Bound on $\epsilon_{q d e}^{ijkl}$ &observable                                               \\
\hline
11 11 &1.5 $\cdot 10^{-7}$      &$R_\pi$                                                                        &
22 11 &3.0 $\cdot 10^{-4}$      &$R_\pi$			                                                \\
12 11 &5.1 $\cdot 10^{-3}$      &$B(\pi^+ \to \mu^+ \nu_ e$)                                                    &
12 12$^{*)}$ &2.1 $\cdot 10^{-8}$&$B(K^0_L \to \mu^+ e^-$)		                                       	\\
11 12 &2.7 $\cdot 10^{-8}$      &$B(K^0_L \to e^+ e^-$)                                                         &
22 12 &8.4 $\cdot 10^{-7}$      &$B(K^0_L \to \mu^+\mu^-$)		                                      	\\
22 21 &1.3 $\cdot 10^{-2}$      &$B(D^+ \to \mu^+ \nu_\mu$)                                                     &
22 22 &1.2 $\cdot 10^{-2}$      &$\frac{B(D^+_s \to \mu^+ \nu_\mu)}{B(D^+_s \to \tau^+ \nu_\tau)}$		\\
33 22 &0.2                      &$\frac{B(D^+_s \to \mu^+ \nu_\mu)}{B(D^+_s \to \tau^+ \nu_\tau)}$              &
33 13 &2.5 $\cdot 10^{-3}$      &$B(B^+ \to \tau^+ \nu_\tau$)		                                        \\
11 13 &9.0 $\cdot 10^{-5}$      &$B(B^0 \to e^+ e^-$)                                                           &
12 13$^{*)}$ &1.2 $\cdot 10^{-4}$&$B(B^0 \to \mu^+ e^-$)                                               		\\
13 13$^{*)}$ &2.5 $\cdot 10^{-3}$&$B(B^0 \to \tau^+ e^-$)                                                        &
23 13$^{*)}$ &3.3 $\cdot 10^{-3}$&$B(B^0 \to \tau^+ \mu^-$)	                                                \\
22 13 &7.5 $\cdot 10^{-5}$      &$B(B^0 \to \mu^+ \mu^-$)                                                       &
11 23 &6.0 $\cdot 10^{-4}$      &$B(B^0_s \to e^+ e^-$)		                                                \\
12 23$^{*)}$ &2.1 $\cdot 10^{-4}$&$B(B^0_s \to \mu^+ e^-$)                                                       &
22 23 &1.2 $\cdot 10^{-4}$      &$B(B^0_s \to \mu^+ \mu^-$)                                        		\\
\hline\hline
\end{tabular*}
\end{minipage}
\end{table}

\subsection{Phenomenological parameterizations of quark and lepton Yukawa couplings}

\subsubsection{Quark sector}

The quark Yukawa sector is described by the following Lagrangian
\begin{equation}
{\cal L}_{quark}={u_{R}^{c}}_iY_{ij}^u Q_j\overline{H}+{d_{R}^{c}}_iY_{ij}^d Q_j H +{\rm h.c.} ,\label{eq:Lag-quark}
\end{equation}
where $i,j=1,2,3$ are generation indices, $Q_i=({d_L}_i, {u_L}_i)$ are
the left-handed quark doublets, $u_{R}^{c}$ and $d_{R}^{c}$ are the
right-handed up and down quark singlets respectively, and $H$ is the
Higgs field. On the other hand, $Y^{u}$ and $Y^{d}$ are complex
$3\times 3$ matrices, which can be cast by means of a singular value
decomposition as
\begin{eqnarray}
Y^{u}&=&V^{u}_R D^u_Y {V^u_L}^{\dagger} \;,\nonumber \\
Y^{d}&=&V^{d}_R D^d_Y {V^d_L}^{\dagger} \;.
 \label{eq:sing-value-quark}
\end{eqnarray} 
Here, $D^{u}_Y={\rm diag}(y^{u}_1,y^{u}_2,y^{u}_3)$ is a diagonal matrix whose entries can be chosen real and positive with $y^{u}_1< y^{u}_2< y^{u}_3$,
and similarly for $D^{d}_Y$. $V^{u,d}_R$ and $V^{u,d}_L$ are $3\times 3$ unitary matrices that depend on three real parameters and six phases.
The unitary matrices $V^{u,d}_R$ can be absorbed in the definition of the right-handed fields without any physical effect. In neutral currents the left rotations cancel out via the
Glashow-Iliopoulos-Maiani (GIM) mechanism \cite{Glashow:1970gm}.  
On the other hand, the redefinition of the left-handed fields produces flavour mixing in the charged currents.
In the physical basis where both the up and down Yukawa couplings are simultaneously diagonal, the charged current reads:
\begin{equation}
J^{\mu}_{cc}= u^c_L \frac{\gamma^\mu(1-\gamma_5)}{2}  ({V^u_L}^{\dagger}V^d_L) d_L\;. \label{eq:charged-current-quark}
\end{equation}
The matrix ${V^u_L}^{\dagger}V^d_L$ can be generically written as ${V^u_L}^{\dagger}V^d_L=\Phi_1 U_{CKM} \Phi_2$,
where $\Phi_{1,2}$ are diagonal unitary matrices (thus, containing only phases) that can be absorbed by appropriate redefinitions
of the left handed fields. Finally, $U_{CKM}$ depends on three angles and one phase that cannot be removed by field redefinitions
and accounts for the physical mixing between quark generations and the CP-violation
\cite{Cabibbo:1963yz,Kobayashi:1973fv}. 
It is usually parameterized as:
\begin{equation}
U_{CKM}= \begin{pmatrix}
c_{13}c_{12} & c_{13}s_{12} & s_{13}e^{-i\delta}\\
-c_{23}s_{12}-s_{23}s_{13}c_{12}e^{i\delta} & c_{23}c_{12}-s_{23}s_{13}s_{12}e^{i\delta} & s_{23}c_{13}\\
s_{23}s_{12}-c_{23}s_{13}c_{12}e^{i\delta} & -s_{23}c_{12}-c_{23}s_{13}s_{12}e^{i\delta} &
c_{23}c_{13}
\end{pmatrix},
\label{eq:CKM}
\end{equation}
where $s_{ij}=\sin\theta_{ij}$, $c_{ij}=\cos\theta_{ij}$ and $\delta$ is the CP-violating phase. Experiments show a hierarchical structure in the off-diagonal
entries of the CKM matrix: $|V_{ub}|\ll V_{cb}\ll V_{us}$, that can be well described by the following phenomenological parameterization of the CKM matrix,
proposed by Wolfenstein \cite{Wolfenstein:1983yz}. It reads:
\begin{equation}
U_{CKM}= \begin{pmatrix} 1-\frac{\lambda^2}{2}  & \lambda & A\lambda^3(\rho-i\eta)\\
-\lambda & 1- \frac{\lambda^2}{2} & A\lambda^2 \\
A\lambda^3(1-\rho-i\eta) & -A\lambda^2 & 1
\end{pmatrix} 
+ {\cal O}(\lambda^4) , \label{eq:Wolfenstein}
\end{equation}
where $\lambda$ is determined with a very good precision in semileptonic $K$ decays, giving $\lambda \simeq 0.23$, and $A$ is measured
in semileptonic $B$ decays, giving $A\simeq 0.82$. The parameters $\rho$ and $\eta$ are more poorly measured, although a rough
estimate is $\rho\simeq 0.1$, $\eta\simeq 0.3$ \cite{Battaglia:2003in} 

\subsubsection{Leptonic sector with Dirac neutrinos}

A Dirac mass term for the neutrinos requires the existence of three right-handed neutrinos, which are singlets under the Standard Model gauge 
group. In consequence, the leptonic Lagrangian would contain in general a Majorana mass term for the right-handed neutrinos, that has
to be forbidden by imposing exact lepton number conservation. Then, the leptonic Lagrangian reads
\begin{equation}
{\cal L}_{lep}={e_{R}^{c}}_iY_{ij}^e L_j\overline{H}+{\nu_{R}^{c}}_iY_{ij}^\nu L_j H +{\rm h.c.},
 \label{eq:Lag-Dirac}
\end{equation}
where $L_i=({\nu_L}_i, {e_L}_i)$ are the left-handed lepton doublets and $e_{R}^{c}$ and $\nu_{R}^{c}$ are respectively 
the right-handed charged lepton and neutrino singlets. Analogously to the quark sector, the Yukawa couplings can be decomposed as:
\begin{eqnarray}
Y^{e}=V^{e}_R D^e_Y {V^e_L}^{\dagger}\;, \\
Y^{\nu}=V^{\nu}_R D^\nu_Y {V^\nu_L}^{\dagger} \;, \label{eq:sing-value-lepton}
\end{eqnarray}
where $V^{e,\nu}_R$ do not have any physical effect, whereas $V^{e,\nu}_L$ have an effect in the charged current, that in the
basis where the charged lepton and neutrino Yukawa couplings are simultaneously diagonal reads:
\begin{equation}
J^{\mu}_{cc}= e^c_L \frac{\gamma^\mu(1-\gamma_5)}{2} ({V^e_L}^{\dagger}V^\nu_L) \nu_L\;. \label{eq:charged-current-lep}
\end{equation}
As in the case of the quark sector, the matrix ${V^e_L}^{\dagger}V^\nu_L$ depends on three angles and six phases and can be expressed as
${V^e_L}^{\dagger}V^\nu_L=\Phi_1 U_{PMNS} \Phi_2$. The matrices $\Phi_1$ and $\Phi_2$ can be absorbed by appropriate redefinitions of the left-handed fields,  
yielding a physical mixing matrix $U_{PMNS}$ \cite{Pontecorvo:1957qd,Maki:1962mu}
that depends on three angles and one phase, and that can be parameterized by the same structure as for the
quark sector, Eq.~(\ref{eq:CKM}). However, the values for the angles differ substantially from the quark sector. The experimental values that 
result from the global fit are $\sin^2\theta_{12}=0.26-0.36$, $\sin^2\theta_{23}=0.38-0.63$ and
$\sin^2\theta_{13}\leq 0.025$ at 2$\sigma$ \cite{Maltoni:2004ei}. On the other hand, the CP-violating phase $\delta$ is
completely unconstrained by present experiments.

\subsubsection{Leptonic sector with Majorana neutrinos}
Neutrino masses can also be accommodated in the Standard Model without extending the particle content, just by adding
a dimension five operator to the leptonic Lagrangian \cite{Weinberg:1979sa}:
\begin{equation}
{\cal L}_{lep}={e_{R}^{c}}_iY_{ij}^e L_j\overline{H}+\frac{1}{4} \kappa_{ij}(L_i H)(L_jH) +{\rm h.c.} 
\label{eq:Lag-Majorana}
\end{equation}
with $\kappa$ a $3\times 3$ complex symmetric matrix that breaks explicitly lepton number and that has dimensions of mass$^{-1}$. 
Then, after the electroweak symmetry breaking, a Majorana mass term for neutrinos is generated:
\begin{equation}
{ m}_{\nu}=\frac{1}{2}\kappa \langle H^0\rangle^2\;. 
\label{eq:majorana-mass}
\end{equation}
This term can be diagonalized as ${m}_{\nu}={V^\nu_L}^*D_{{m}_{\nu}} {V^\nu_L}^{\dagger}$,
so that the charged current reads as in Eq.~({\ref{eq:charged-current-lep}), with ${V^e_L}^{\dagger}V^\nu_L=\Phi_1 U \Phi_2$,
where the matrix $U$ has the form of the CKM matrix, Eq.~(\ref{eq:CKM}). The matrix $\Phi_1$ containing three phases can be removed by a redefinition
of the left-handed charged lepton fields. However, due to the Majorana nature of the neutrinos, the matrix $\Phi_2$ cannot
be removed and is physical, yielding a leptonic mixing matrix  \cite{Pontecorvo:1957qd,Maki:1962mu}
$U_{PMNS}=U \Phi_2$ that is defined by three angles and three phases, 
one associated to $U$, the "Dirac phase", and two associated to $\Phi_2$, the "Majorana phases".

In the leptonic Lagrangian given by Eq.~(\ref{eq:Lag-Majorana}) the origin of the dimension five operator remains open. In the rest of this
Section, we will review the heavy Majorana singlet (right-handed) neutrino mass mechanism 
(type I seesaw)
 \cite{Minkowski:1977sc,Yanagida:1979as,Gell-Mann:1980vs,Glashow:1979nm,Mohapatra:1979ia}
and the triplet Higgs mass mechanism (type II seesaw) 
\cite{Magg:1980ut, Schechter:1980gr,Lazarides:1980nt,Mohapatra:1980yp,Gelmini:1980re}  as the possible origins of this effective operator.
The third \cite{Ma:1998dn} tree level realization of the operator Eq.~(\ref{eq:Lag-Majorana}) 
via triplet fermion (type III seesaw) \cite{Foot:1988aq} is discussed in Section~\ref{sec:GUTs}.

\subsubsubsection{Type I seesaw}\label{sec:seesawI}

In the presence of singlet right-handed neutrinos, the most general Lagrangian compatible with the Standard Model gauge symmetry reads
\begin{equation}
{\cal L}_{lep}={e_{R}^{c}}_iY_{ij}^e L_j\overline{H}+{\nu_{R}^{c}}_i Y_{ij}^{\nu} L_jH
 -\frac{1}{2} {\nu_{Ri}^{cT}}  M^{}_{ij} {\nu_{R}^{c}}_j  +{\rm h.c.},
\label{eq:Lag-typeI}
\end{equation}
where lepton number is explicitly broken by the Majorana mass term for the singlet right-handed neutrinos\footnote{Here we explicitly assume three generations of singlet neutrinos. For the phenomenology of a  large number of singlets, as predicted by string theories, 
see \cite{Buchmuller:2007zd,Ellis:2007wz}.}. 
The seesaw mechanism is implemented when 
${\rm eig}(M^{})\gg\langle H^0\rangle$. If this is the case, at low energies the right-handed neutrinos are decoupled
and the theory can be well described by the effective Lagrangian for Majorana neutrinos, Eq.~(\ref{eq:Lag-Majorana}), with 
\cite{Minkowski:1977sc,Yanagida:1979as,Gell-Mann:1980vs,Glashow:1979nm,Mohapatra:1979ia}
\begin{equation}
\kappa= 2 {Y^\nu}^T {M}^{-1}{Y^\nu}\;. \label{eq:seesaw}
\end{equation}
Working in the basis where the charged lepton Yukawa matrix and the right-handed mass matrix are simultaneously diagonal,
it can be checked that the complete  Lagrangian, Eq.~(\ref{eq:Lag-typeI}), contains fifteen independent
real parameters and six complex phases \cite{Ellis:2001xt}. 
Of these, three correspond to the charged lepton masses, three to the right-handed masses, 
and the remaining nine real parameters and six phases, to the neutrino Yukawa coupling. The independent
parameters of the neutrino Yukawa coupling can be expressed in several ways. The most
straightforward parameterization uses the singular value decomposition of the neutrino Yukawa matrix:
\begin{equation}
Y_{\nu}=V^{\nu}_R D^\nu_Y {V^\nu_L}^{\dagger} \;, \label{Yukawa-singular-value}
\end{equation}
where $D^\nu_Y={\rm diag}(y^\nu_1,y^\nu_2,y^\nu_3)$, with $y^\nu_i\geq 0$ and $y^\nu_1\leq y^\nu_2\leq y^\nu_3$. On the other hand, $V^\nu_L$ 
and $V^\nu_R$ are $3\times 3$ unitary matrices, that depend in general on three real parameters and six phases. Both can be
generically written as $\Phi_1 V \Phi_2$, where $V$ has the form of the CKM matrix and $\Phi_{1,2}$ are diagonal
unitary matrices (thus, containing only phases). One can check that for $V^\nu_R$ the $\Phi_2$ matrix can be absorbed
into the definition of $V^\nu_L$, so that
\begin{equation}
V^\nu_R=\begin{pmatrix} e^{i \alpha^R_1} & & \\ 
& e^{i \alpha^R_2} & \\
 & & 1
\end{pmatrix}
\begin{pmatrix} 
c^R_2c^R_3 &   c^R_2s^R_3 &   s^R_2e^{-i\delta^R}\\
-c^R_1s^R_3-s^R_1s^R_2c^R_3e^{i\delta^R} &   c^R_1c^R_3-s^R_1s^R_2s^R_3e^{i\delta^R} & s^R_1c^R_2\\ 
s^R_1s^R_3-c^R_1s^R_2c^R_3e^{i\delta^R} &   -s^R_1c^R_3-c^R_1s^R_2s^R_3e^{i\delta^R} &   c^R_1c^R_2
\end{pmatrix}\;.
\label{eq:VRnu}
\end{equation}
Similarly, for $V_L$ the $\Phi_1$ matrix can be absorbed into the definition of $L$ and $e_R$, while keeping 
$Y_e$ diagonal and real. In consequence,
\begin{equation}
V^\nu_L= \begin{pmatrix} c^L_2c^L_3 &   c^L_2s^L_3 &   s^L_2e^{-i \delta^L}\\
-c^L_1s^L_3-s^L_1s^L_2c^L_3e^{i\delta^L} &   c^L_1c^L_3-s^L_1s^L_2s^L_3e^{i\delta^L} & s^L_1c^L_2\\
 s^L_1s^L_3-c^L_1s^L_2c^L_3e^{i\delta^L} &   -s^L_1c^L_3-c^L_1s^L_2s^L_3e^{i\delta^L} &   c^L_1c^L_2
\end{pmatrix}
\begin{pmatrix}
e^{i \alpha^L_1} & & \\
& e^{i \alpha^L_2} & \\& & 1
\end{pmatrix}\;.
\label{eq:VLnu}
\end{equation}

Therefore, in this parameterization the independent parameters in the
Yukawa coupling can be identified with the three Yukawa eigenvalues,
$y_i$, the three angles and three phases in $V_L$, and the three
angles and three phases in $V_R$
\cite{Ellis:2001xt,Pascoli:2003uh,Casas:2006hf}. The requirement that
the low energy phenomenology is successfully reproduced imposes
constraints among these parameters. To be precise, the low energy
leptonic Lagrangian depends just on the three charged lepton masses
and the six real parameters and three complex phases of the effective
neutrino mass matrix. In consequence, there are still six real
parameters and three complex phases that are not determined by low
energy neutrino data; this information about the high-energy
Lagrangian is ``lost'' in the decoupling of the three right-handed
neutrinos and cannot be recovered just from neutrino experiments.

The ambiguity in the determination of the high-energy parameters can be encoded in the three right-handed neutrino masses and
an orthogonal complex matrix $R$ defined as \cite{Casas:2001sr}
\begin{equation}
R=D_{\sqrt{M}}^{-1} {\bf Y_\nu}U_{PMNS} D_{\sqrt{m}}^{-1} \langle H^0 \rangle\;, \label{eq:def-R}
\end{equation}
so that the most general Yukawa coupling compatible with the low energy data is given by:
\begin{equation}
Y^\nu=D_{ \sqrt{M}} R D_{\sqrt{m}} U_{PMNS}^{\dagger} \langle H^0 \rangle\;. \label{eq:Yukawa-R}
\end{equation}
It is straightforward to check that this equation indeed satisfies the seesaw formula, Eq.~(\ref{eq:seesaw}).
In this expression, $D_{\sqrt{m}}$ and $D_{\sqrt{M}}$ are diagonal matrices whose entries are the square roots of the 
light neutrino and the  right-handed neutrino masses, respectively, and  $U_{PMNS}$ is the leptonic mixing matrix. It is customary to 
parameterize $R$ in terms of three complex angles, $\hat \theta_i$:
\begin{equation}
R=\begin{pmatrix} \hat c_2\hat c_3 & -\hat c_1\hat s_3-\hat s_1\hat s_2\hat
c_3  & \hat s_1\hat s_3-\hat c_1\hat s_2\hat c_3 \\ \hat c_2\hat s_3 & \hat c_1\hat c_3-\hat s_1\hat s_2\hat s_3   & -\hat s_1\hat c_3-\hat
c_1\hat s_2\hat s_3 \\ \hat s_2  & \hat s_1\hat c_2 & \hat c_1\hat c_2
\end{pmatrix}  \;, \label{eq:parametrisation-R}
\end{equation}
up to reflections., where $\hat c_i \equiv \cos \hat \theta_i$, $\hat s_i \equiv \sin \hat \theta_i$.

Whereas the physical interpretation of the right-handed masses is very transparent, the meaning of $R$ is more obscure. 
$R$ can be interpreted as a dominance matrix in the sense that \cite{Lavignac:2002gf} :
\begin{itemize}
\item $R$ is an orthogonal transformation from the basis of the left-handed leptons mass eigenstates to the one of the right-handed neutrino mass eigenstates;\\
\item if and only if an eigenvalue $m_i$ of ${m}_\nu$ is dominated - in the sense already given before - by one right-handed
neutrino eigenstate $N_j$, then $|R_{ji}| \approx 1\, ;$\\
\item if a light pseudo-Dirac pair is dominated by a heavy pseudo-Dirac pair, then the corresponding  $2 \times 2$ sector in $R$ is a boost.
\end{itemize}

An interesting limit of this dominance behaviour is the seesaw model with two right-handed neutrinos (2RHN) \cite{Frampton:2002qc,Raidal:2002xf}. 
In this limit, the parameterization Eq.~(\ref{eq:Yukawa-R}) still holds, with the substitutions $D_{\sqrt{M}}={\rm diag}(M^{-1}_1,M^{-1}_2)$
and \cite{Ibarra:2003xp,Ibarra:2003up,Chankowski:2004jc,Petcov:2005jh}
\begin{equation}
R=\left( \begin{array}{ccc} 0 & \cos \hat\theta & \xi \sin \hat\theta \\ 
0 & -\sin \hat\theta & \xi \cos \hat\theta  
\end{array} \right) (\rm{normal~hierarchy}), 
\label{eq:R-2RHN-nh}
\end{equation}
\begin{equation}
R=\left( \begin{array}{ccc} \cos \hat\theta & \xi \sin \hat\theta & 0\\  
-\sin \hat\theta & \xi \cos \hat\theta  & 0
\end{array} \right) (\rm{inverted~hierarchy}),  \label{eq:R-2RHN-ih}
\end{equation}
with $\hat\theta$ a complex parameter and $\xi=\pm 1$ a discrete parameter that accounts for a discrete indeterminacy in $R$. 

A third possible parameterization of the neutrino Yukawa coupling uses the Gram-Schmidt decomposition, in order to cast the Yukawa
coupling as a product of a unitary matrix and a lower triangular matrix \cite{Endoh:2000hc}:
\begin{equation}
Y^\nu= U_{\triangle} Y_{\triangle}= U_{\triangle} \left(\begin{array}{ccc} y_{11} & 0 & 0 \\ y_{21} & y_{22} & 0 \\ y_{31} & y_{32} & y_{33} 
\end{array}\right)\;, \label{eq:yukawa-triangular}
\end{equation} 
where the diagonal elements of $Y_{\triangle}$ are real. Three of the six phases in $U_{\triangle}$ can be absorbed
into the definition of the charged leptons. Therefore, the nine real parameters and the six phases of the neutrino
Yukawa coupling are identified with the three angles and three phases in  $U_{\triangle}$ and the six real parameters and three phases in $Y_{\triangle}$.

In the SM extended with right-handed neutrinos, the charged lepton masses and the effective neutrino
mass matrix are the only source of information about the leptonic sector. However, if supersymmetry is discovered, the structure
of the low energy slepton mass matrices would provide additional information about the leptonic sector, provided
the mechanism of supersymmetry breaking is specified. Assuming that the slepton mass matrices are proportional
to the identity at the high energy scale, quantum effects induced by the right-handed neutrinos would yield
at low energies a left-handed slepton mass matrix with a complicated structure, whose measurement would provide
additional information about the seesaw parameters \cite{Borzumati:1986qx,Hall:1985dx}. 
To be more specific, in the minimal supersymmetric seesaw model the off-diagonal elements of 
the low energy  left-handed and right-handed slepton 
mass matrices and $A$-terms read, in the leading log approximation \cite{Hisano:1995cp}
\begin{eqnarray} \label{softafterRG} 
\left(m^2_{\tilde{L} } \right)_{ij}  &\simeq&
-\frac{1}{8\pi^2}(3m_0^2 + A_0^2) {Y^\nu_{ik}}^{\dagger} Y^\nu_{kj} \log\frac{M_X}{M_k}\ , \\
\left( m^2_{\tilde{e}_R} \right)_{ij}  &\simeq&
 0\ , \\
\left( A_e \right)_{ij}  &\simeq &
-\frac{3}{8\pi^2} A_0 Y_e  {Y^\nu_{ik}}^{\dagger} Y^\nu_{kj} \log\frac{M_X}{M_k}\ ,
\end{eqnarray}
where $m_0$ and $A_0$ are the universal soft supersymmetry breaking parameters
at high scale $M_X.$ Note that the diagonal elements of those mass matrices
include the tree level soft mass matrix, the radiative corrections from gauge and charged lepton Yukawa interactions, and the
mass contributions from F- and D-terms (that are different for charged sleptons and sneutrinos). Therefore, the measurement at low energies of rare
lepton decays, electric dipole moments and slepton mass splittings would provide information about the combination 
\begin{equation}
C_{ij}\equiv \sum_k{Y^\nu_{ik}}^{\dagger} Y^\nu_{kj} \log\frac{M_X}{M_k}
\equiv 
\left(  Y_\nu^\dagger L Y_\nu   \right)_{ij}
\;, \label{eq:def-P}
\end{equation}
where $L_{ij}=\log\frac{M_X}{M_i}\delta_{ij}.$

Interestingly enough, $C$ encodes precisely the additional information needed to reconstruct the complete seesaw Lagrangian from
low energy observations \cite{Davidson:2001zk,Ellis:2002fe} (note in particular that $C$ is a Hermitian
matrix that depends on six real parameters and three phases, that together with the nine real parameters and three phases
of the neutrino mass matrix sum up to the independent fifteen real parameters and six complex phases in $Y_{\nu}$ and $M$).

To determine $Y_{\nu}$ and $M$ from the low energy observables $C$ and ${m}_{\nu}$, it is convenient to define
\begin{eqnarray}
\widetilde  Y^{\nu}&=&{\rm diag}(\sqrt{\log\frac{M_X}{M_1}}, \sqrt{\log\frac{M_X}{M_2}},\sqrt{\log\frac{M_X}{M_3}}) Y^{\nu}, \nonumber \\
\widetilde M_k &=& M_k \log\frac{M_X}{M_k}\;, \label{eq:change-of-variables}
\end{eqnarray}
so that the effective neutrino mass matrix and $C$ now read:
\begin{eqnarray}
&&{m}_\nu= \widetilde Y^{\nu t} {\rm diag}(\widetilde M^{-1}_1,\widetilde M^{-1}_2,\widetilde M^{-1}_3) \widetilde Y^{\nu} \langle H^0_u\rangle^2\nonumber, 
\nonumber \\ &&C=\widetilde Y^{\nu \dagger}\widetilde Y^{\nu}\;. \label{eq:new-mass-P}
\end{eqnarray}
where $H^0_u$ is the neutral component of the up-type Higgs doublet. Using the singular value decomposition $\widetilde Y^\nu=\widetilde V^\nu_R 
\widetilde D^\nu_Y \widetilde V^{\nu \dagger}_L$, one finds that $\widetilde V^{\nu \dagger}_L$ and $\widetilde D^\nu_Y$ could be
straightforwardly determined from $C$, since
\begin{equation}
C \equiv  \widetilde Y^{\nu \dagger} \widetilde Y^\nu = \widetilde V_L^{\dagger}\widetilde  D_Y^2 \widetilde V_L . \label{eq:step1}
\end{equation}
On the other hand, from ${m}_\nu = \widetilde  Y^{\nu t} \widetilde D^{-1}_{M} \widetilde  Y^{\nu}\langle H^0_u \rangle^2$ and
the singular value decomposition of $ \widetilde Y^{\nu}$,
\begin{equation}
\widetilde D_Y^{-1} \widetilde V_L^*  {m}_\nu \widetilde V^{\nu\dagger}_L \widetilde D_Y^{\nu -1} = 
\widetilde V_R^{\nu *}  \widetilde D_{M}^{-1} \widetilde V^{\nu\dagger}_R, \label{eq:step2}
\end{equation}
where the left hand side of this equation is known (${m}_\nu$ is one of  our inputs, and $\widetilde V^{\nu}_L$ and $\widetilde D^\nu_Y$ were obtained 
from Eq.~(\ref{eq:step1})). Therefore, $\widetilde V^\nu_R$ and  $\widetilde D_{M}$ can also be determined. This simple
procedure shows that starting from the low energy observables ${m}_\nu$ and $C$ it is possible to determine uniquely
the matrices $\widetilde D_{M}$ and $\widetilde Y^\nu=\widetilde V^\nu_R \widetilde D^\nu_Y \widetilde V^{\nu \dagger}_L$. Finally, inverting
Eq.~(\ref{eq:change-of-variables}), the actual parameters of the Lagrangian $M_k$ and $Y^{\nu}$ can be computed.

This procedure is particularly powerful in the case of the two right-handed neutrino model, as the number of
independent parameters involved (either at high energies or at low energies) is drastically reduced. The matrix $C$ defined in Eq.~(\ref{eq:def-P})
depends in general on six moduli and three phases. However, since the Yukawa coupling depends in the 2RHN model on only three unknown
moduli and one phase, so does $C$, and consequently it is possible to obtain predictions on the moduli
of three $C$-matrix elements and the phases of two $C$-matrix elements. Namely, from Eq.~(\ref{eq:Yukawa-R}) one obtains that:
\begin{equation}
U^{\dagger} C U =  U^{\dagger}  \widetilde Y^{\nu\dagger} \widetilde Y^{\nu} U = D_{\sqrt{m}} R^{\dagger}  \widetilde D_{M} R D_{\sqrt{m}}
/\langle H^0_u\rangle^2. \label{eq:master-2RHN}
\end{equation}
where we have written $U\equiv U_{PMNS}$. Since $m_1=0$ in the 2RHN model\footnote{Here we are assuming a neutrino spectrum with normal hierarchy. In the
case with inverted hierarchy, the analysis is similar, using $m_3=0$.}, it follows that $(U^{\dagger} C U)_{1i}=0$, for $i=1,2,3$,
leading to three relations among the elements in $C$. For instance, one could derive the diagonal elements in $C$ in terms of the off-diagonal elements:
\begin{eqnarray}
C_{11}&=&-\frac{C^*_{12}U^*_{21}+C^*_{13}U^*_{31}} {U^*_{11}}, \nonumber \\
C_{22}&=&-\frac{C_{12}U^*_{11}+C^*_{23}U^*_{31}}{U^*_{21}}, \nonumber \\ 
C_{33}&=&-\frac{C_{13}U^*_{11}+C_{23}U^*_{21}}{U^*_{31}}. \label{eq:diagonal-2RHN}
\end{eqnarray}
The observation of these correlations would be non-trivial tests of the 2RHN model.

The relations for the phases arise from the hermiticity of $C$, since the diagonal elements in $C$ have to be real. Taking as the independent phase the argument
of $C_{12}$, one can derive from Eq.~(\ref{eq:diagonal-2RHN}) the arguments of the remaining elements:
\begin{eqnarray}
e^{i{\rm arg}C_{13}}&=&\frac{-i~{\rm Im}(C_{12}U_{21}U^*_{11})\pm \sqrt{|C_{13}|^2|U_{11}|^2|U_{31}|^2-[{\rm Im}(C_{12}U_{21}U^*_{11})]^2}}
{ |C_{13}| U_{31}U^*_{11}}, \nonumber \\
e^{i{\rm arg}C_{23}}&=&\frac{i~{\rm Im}(C_{12}U_{21}U^*_{11})\pm \sqrt{|C_{23}|^2|U_{21}|^2|U_{31}|^2-[{\rm Im}(C_{12}U_{21}U^*_{11})]^2}}
{ |C_{23}| U_{31}U^*_{21}}, \label{eq:phases-2RHN}
\end{eqnarray}
where the $\pm$ sign has to be chosen so that the eigenvalues of $C$ are positive. We conclude then that the $C$-matrix parameters $C_{12}$, $|C_{13}|$ and
$|C_{23}|$ can be regarded as independent and can be used as an alternative parameterization of the 2RHN model \cite{Ibarra:2005qi}. Together with the five
moduli and the two phases of the neutrino mass matrix, sum up to the eight moduli and the three phases necessary to reconstruct the high-energy Lagrangian of the
2RHN model. 

\subsubsubsection{Type II seesaw}\label{sec:seesawII}

The type II seesaw mechanism \cite{Magg:1980ut, Schechter:1980gr,Lazarides:1980nt,Mohapatra:1980yp,Gelmini:1980re} 
consists on adding to the SM particle content a Higgs triplet
\begin{equation}
T=\begin{pmatrix} T^0 & -\frac{1}{\sqrt{2}}T^+ \\ -\frac{1}{\sqrt{2}}T^+ & -T^{++} \end{pmatrix}.
\label{eq:def-triplet}
\end{equation}
Then, the leptonic potential compatible with the SM gauge symmetry reads 
\begin{equation}
{\cal L}_{lep}={e_{R}^{c}}_iY_{ij}^e L_j\overline{H}+ Y^T_{ij} L_i T L_j +{\rm h.c.} \label{eq:Lag-typeII}
\end{equation}
From this Lagrangian, it is apparent that the triplet $T$ carries
lepton number $-2$.  If the neutral component of the triplet acquires
a VEV and breaks lepton number spontaneously as happens in the
Gelmini-Roncadelli model \cite{Gelmini:1980re}, the associated
massless Majoron rules out the model. Therefore phenomenology suggests
to break lepton number explicitly via the triplet coupling to the SM
Higgs boson \cite{Ma:1998dx}.  The most general scalar potential
involving one Higgs doublet and one Higgs triplet reads
\begin{equation}
V=m_H^2  H^{\dagger} H + \frac{1}{2} \lambda_1  (H^{\dagger} H)^2+ M_T^2 T^{\dagger} T + \frac{1}{2} \lambda_2 (T^{\dagger} T)^2+
\lambda_3  (H^{\dagger} H) (T^{\dagger} T) + { \mu'}  H^{\dagger}T  H^{\dagger},
 \label{eq:potential-triplet}
\end{equation}
where the term proportional to $\mu'$
 breaks lepton number explicitly. The type II seesaw mechanism is implemented when
$M_T\gg \langle H^0\rangle$. Then, the minimization of the scalar potential yields:
\begin{equation}
\langle H^0 \rangle^2  \simeq  \frac{-m_H^2}{\lambda_1-2 \mu_{L\!\!\!/}^2/M_T^2} ~~~~~~~~~ \langle T^0 \rangle \simeq
\frac{-\mu' \langle H^0 \rangle^2}{M_T^2} \label{eq:vevs-triplet}
\end{equation}
that produce Majorana masses for the neutrinos given by
\begin{equation}\label{numass-t1}
m_\nu=  Y_T \frac{-\mu' \langle H^0 \rangle^2}{M_T^2} \, .
\end{equation}
The Yukawa matrix $Y^T$ has the same flavour structure as the non-renormalizable operator $\kappa$ defined in 
Eq.~(\ref{eq:Lag-Majorana}) for the effective Lagrangian of Majorana neutrinos. Therefore, the parameterization 
of the type II seesaw model is completely identical to that case.

Supersymmetric models with low scale triplet Higgses have been extensively  considered in studies 
of collider phenomenology \cite{Huitu:1993gf,Huitu:1993uv,Raidal:1998vi}.
The model   \cite{Ma:1998dx} was first supersymmetrised  in 
\Ref~\cite{Hambye:2000ui} as a
possible scenario for leptogenesis. 
The requirement of a holomorphic superpotential implies introducing the triplets in a vector-like
$SU(2)_W\times U(1)_Y$ representation, as $T \sim (3,1)$ and $\bar{T} \sim (3,-1)$. The relevant superpotential terms are
\begin{equation}
\label{rossiL-T} 
\frac{1}{\sqrt{2}}Y^{ij}_{T} L_i T L_j  + \frac{1}{\sqrt{2}}\lambda_1 H_1 T H_1 + \frac{1}{\sqrt{2}} \lambda_2
H_2 \bar{T} H_2 +  M_T T \bar{T} + \mu H_2 H_1 \, ,
\end{equation}
where $L_i$ are the $SU(2)_W$ lepton doublets and $H_1 (H_2)$ is the Higgs doublet with hypercharge $Y=-1/2 (1/2)$. 
Decoupling the triplet  at high scale at the electroweak scale the Majorana neutrino mass matrix is given 
by ($v_2 =\langle H_2\rangle$):
\begin{equation}
\label{T-mass} 
{m}^{ij}_\nu = Y^{i j}_T \frac{v_2^2 \lambda_2}{M_T} .
\end{equation}
Note that in the supersymmetric case there is only one mass parameter, $M_T$, while the mass  parameter $\mu'$ of the
non-supersymmetric version is absent.

The couplings $Y_T$ also induce LFV in the slepton mass matrix $m^2_{\tilde{L}}$ through
renormalization group (RG) running from $M_X$ to the decoupling scale $M_T$ \cite{Rossi:2002zb}. 
In the leading-logarithm approximation those   are given by ($i\neq j$):
\begin{eqnarray}\label{log1}
(m^2_{\tilde{L}})_{ ij}& \approx& \frac{-1}{8\pi^2} (9 m^2_0 + 3 A^2_0)(Y^\dagger_T Y_T)_{ij} \log\frac{M_X}{M_T} ,  \nonumber\\
(m^2_{\tilde{e}_R})_{ ij} &\approx& 0 ,  \nonumber \\
(A_{e})_{ ij}& \approx& \frac{-9}{16\pi^2}  A_0 (Y_e Y^\dagger_T Y_T)_{ij} \log\frac{M_X}{M_T} .
\end{eqnarray}
Phenomenological implications of those relations will be presented in 
Section~\ref{sec:phenomenology}.

\subsubsubsection{Renormalization of the neutrino mass matrix}\label{sec:nurenorm}

To make connection between high scale parameters and low scale observables 
one needs to consider renormalization effects on neutrino masses and mixing.
Below the scale where the dimension five operator is generated, the running of the neutrino mass matrix is governed by the renormalization
group (RG) equation of the coupling matrix $\kappa_\nu$, given by \cite{Babu:1993qv,Chankowski:1993tx,Antusch:2001ck,Antusch:2001vn}
\begin{equation}\label{Eq:RGE_kappa}
(4\pi)^2 \frac{d}{d \ln\mu}\kappa_\nu=(4\pi)^2 A_g\,\kappa_\nu+C_{e}\left({(Y_e^{\dagger}Y_e)^{\sf T}}\kappa_\nu+\kappa_\nu \,{Y_e^{\dagger}Y_e}\right)\;,
\end{equation}
where $C_e=-3/2$ for the SM and $C_e=1$ for the MSSM.  The first term does not affect the running of the neutrino mixing angles
and CP-violation phases, however it affects of course the running of the neutrino mass eigenvalues.  The flavour universal factor $A_g$ is given by
\begin{equation}
A_g=\begin{cases}-3 \,{\alpha_2}\, (4\pi) +{\lambda}+2\,\mathrm{tr}\left(3\,{Y_u^{\dagger}Y_u}+3\, {Y_d^{\dagger}Y_d}+{Y_e^{\dagger}Y_e}\right)&\text{SM}\\
-2\, {\alpha_1} \,(4\pi) - 6\,{\alpha_2} \,(4\pi)+\mathrm{tr}\left({Y_u^{\dagger}Y_u}\right)&\text{MSSM}\end{cases}\;,
\end{equation}
where $\lambda$ denotes the Higgs self-coupling constant and
 $\alpha_i=g_i^2/(4\pi)$, where $g_1$ and $g_2$ are the $U(1)_Y$ and
SU(2) gauge coupling constants, respectively.

Due to the smallness of the tau-Yukawa coupling in the SM, the mixing angles are not affected significantly by the renormalization group running below the
generation scale of the dimension five operator. However, if the neutrino mass matrix $m_\nu = \frac{\langle\phi \rangle^2}{2} \kappa_\nu$ is realized in the
seesaw scenario (type~I), running effects above and between the seesaw scales can also lead to relevant running effects in the SM. 
Note that in the MSSM case the running of the mixing angles and CP-violation phases can be large even below the seesaw scales due to the
possible enhancement of the tau-Yukawa coupling by the factor $(1+\tan \beta^2)^{1/2}$.

In order to understand generic properties of the RG evolution and to estimate the typical size of the RG effects, it is
useful to consider RGEs for the leptonic mixing angles, CP phases and neutrino masses themselves, which can be derived from the RGE in
Eq.~(\ref{Eq:RGE_kappa}).  For example, below the seesaw scales, up to ${\mathcal{O}} (\theta_{13})$ corrections, the evolution of the mixing
angles in the MSSM is given by~\cite{Antusch:2003kp} (see also~\cite{Chankowski:1999xc,Casas:1999tg})
\begin{eqnarray}\label{Eq:RGEsMNSangles}
\label{Eq:t12}
\!\!\!\!\!\!\!\!\frac{d {\theta}_{12}}{d\ln\mu} \!\!\!&=&\!\!\! \frac{- y_\tau^2}{32\pi^2} \, \!\sin 2\theta_{12} \, s_{23}^2\,  
     \frac{ | {m_1}\, e^{i \alpha_{\scriptscriptstyle M}} \!\!+\! {m_2}|^2}{\Delta m^2_{21} }\;, \\ 
\label{Eq:t13}
\!\!\!\!\!\!\!\!\frac{d {\theta}_{13}}{d\ln\mu} \!\!\!&=&\!\!\! \frac{y_\tau^2}{32\pi^2} \, \sin 2\theta_{12} \, \sin 2\theta_{23} \,
       \frac{m_3}{\Delta m^2_{31} \left( 1+\zeta \right)} \, I(m_1, m_2, \alpha_{\scriptscriptstyle M}, \beta_{\scriptscriptstyle M}, \delta)\;, \\
\label{Eq:t23}
\!\!\!\!\!\!\!\!\frac{d {\theta}_{23}}{d\ln\mu} \!\!\!&=&\!\!\! \frac{-y_\tau^2}{32\pi^2} \,
       \frac{\sin 2\theta_{23}}{\Delta m^2_{31}}
      \left[
        c_{12}^2 \, |m_2\, e^{i \beta_{\scriptscriptstyle M}} + m_3\, e^{i \alpha_{\scriptscriptstyle M}}|^2     
      +  s_{12}^2 \, \frac{|m_1\, e^{i \beta_{\scriptscriptstyle M}}\! + m_3|^2}{1+\zeta}
       \right]\;,
\label{eq:Theta23Dot} 
\end{eqnarray}
where $I(m_1, m_2, \alpha_{\scriptscriptstyle M}, \beta_{\scriptscriptstyle M}, \delta)\equiv m_1 \cos(\beta_{\scriptscriptstyle M} -\delta) - ( 1+\zeta ) \times 
m_2 \cos(\alpha_{\scriptscriptstyle M} - \beta_{\scriptscriptstyle M} + \delta) - \zeta m_3 \cos\delta$, $s_{ij}=\sin\theta_{ij}$, $c_{ij}=\cos\theta_{ij}$,
and $\zeta = \Delta m^2_{21}/\Delta m^2_{31}$. $y_\tau$ denotes the tau-Yukawa coupling,
and one can safely neglect the contributions coming from the electron- and muon-Yukawa couplings. For the matrix $P$ containing the Majorana phases, we use
the convention $P = \mbox{diag}\, ( 1, e^{ i  \alpha_{\scriptscriptstyle M}/2}, e^{ i \beta_{\scriptscriptstyle M}/2} )$. In 
addition to the above formulae, formulae for the running of the CP phases have been derived~\cite{Antusch:2003kp}. For example, the running of the Dirac
CP-violating phase $\delta$, observable neutrino oscillation experiments, is given by
\begin{equation} \label{eq:DeltaPrimeWithNonZeroDelta}
 \frac{d \delta}{d\ln\mu} \,=\, \frac{C y_\tau^2}{32\pi^2} \frac{\delta^{(-1)}}{\theta_{13}} +\frac{C y_\tau^2}{8\pi^2}\delta^{(0)}+{\cal O}(\theta_{13})\;.
\end{equation}
The coefficients $\delta^{(-1)}$ and $\delta^{(0)}$ are omitted here and can be found 
in~\cite{Antusch:2003kp}, where
also formulae for the running of the Majorana CP phases and for the neutrino mass eigenvalues (mass squared differences) can be 
found. From Eq.~(\ref{eq:DeltaPrimeWithNonZeroDelta}), it can be seen that the Dirac CP phase generically becomes more unstable under RG
corrections for smaller $\theta_{13}$.

In the seesaw scenario (type~I), the SM or MSSM  are extended by heavy right-handed neutrinos and their superpartners,
which are SM gauge singlets. Integrating them out below their mass scales $M_R$ yields the dimension five operator for neutrino masses in
the SM or MSSM.  Above $M_R$, the neutrino Yukawa couplings are active, and the RGEs in the MSSM above the scales $M_R$ are
\begin{eqnarray}
\!\!(4\pi)^2 \frac{d\kappa_{\nu}}{d\ln\mu}\!&=&\! \left\{-\frac{6}{5} \,\alpha_1 \,(4\pi)  -    6 \,\alpha_2 \,(4\pi)
+2\,\mathrm{tr}(Y_{\nu}^{\dagger}Y_{\nu})+6\,\mathrm{tr}(Y_u^{\dagger}Y_u)\right\}\kappa_{\nu}\nonumber\\
\!&&\!+(Y_e^{\dagger}Y_e)^{\sf T}\kappa_{\nu}+\kappa_{\nu}(Y_e^{\dagger}Y_e)+(Y_{\nu}^{\dagger}Y_{\nu})^{\sf T}\kappa_{\nu}+\kappa_{\nu}(Y_{\nu}^{\dagger}Y_{\nu})]\;,\\
\!\!(4\pi)^2 \frac{dM_R}{d \ln\mu}\!&=&\!\frac{1}{8\pi^2}\left[(Y_{\nu}Y_{\nu}^{\dagger})M_R +M_R (Y_{\nu}Y_{\nu}^{\dagger})^{\sf T}\right]\;,\\
\!\!(4\pi)^2 \frac{d Y_{\nu}}{d\ln\mu}\!&=
&\!-{Y_{\nu}}\left[\frac{3}{5}\,\alpha_1\,(4\pi)+3\,\alpha_2\,(4\pi)- \mathrm{tr}(3\,Y_u^{\dagger}Y_u+Y_{\nu}^{\dagger}Y_{\nu})\right. \nonumber \\
\!&&\!\left.- 3\,Y_{\nu}^{\dagger}Y_{\nu} - Y_{e}^{\dagger}Y_{e} \right]\;.
\end{eqnarray}
For non-degenerate seesaw scales, a method for dealing with the effective theories, where the heavy singlets are partly integrated
out, can be found in~\cite{Antusch:2002rr}. Analytical formulae for the running of the neutrino parameters above the seesaw scales are
derived in~\cite{Antusch:2005gp,Mei:2005qp}.  The two loop beta functions can be found in Ref.~\cite{Antusch:2002ek}.

The running correction to the neutrino mass matrix and its effects on the related issue have been widely analyzed (see $\mbox{\it
e.g.}$~\cite{Babu:1993qv,Chankowski:1993tx,Antusch:2001ck,Antusch:2001vn,Antusch:2003kp,Haba:1998fb,Haba:1999ca,Haba:1999fk,Ellis:1999my,Casas:1999tp,Casas:1999ac,Miura:2002nz,Haba:2000tx,Antusch:2002hy,Miura:2003if,Kanemura:2005it,Shindou:2004tv,Petcov:2003zb,Ellis:2005dr,Petcov:2005yh,Nielsen:2002pc,GonzalezFelipe:2003fi,Branco:2005ye,Mei:2004rn,Chankowski:1999xc,Casas:1999tg,Antusch:2002rr,Antusch:2005gp,Mei:2005qp,Antusch:2002ek}).
We will summarize below some of the features of RG running of the neutrino mixing parameters in the MSSM (c.f.~Eq.~(\ref{Eq:t12})--(\ref{Eq:t23})).
\begin{itemize}
\item
The RG effects are enhanced for relatively large $\tan\beta$, because the tau-Yukawa coupling becomes large.
\item
The mixing angles are comparatively stable with respect to the RG
running in the case of {\it normal hierarchical} neutrino mass
spectrum, $m_1\ll m_2\ll m_3$ even when $\tan\beta$ is
large~\cite{Haba:1998fb,Haba:1999ca,Haba:1999fk,Ellis:1999my,Casas:1999tp,Casas:1999ac,Miura:2002nz}.
Nevertheless, the running effects can have important implications
facing the high precision of future neutrino oscillation experiments.
\item
For $m_1 \gtrsim 0.05$ eV and the case of $\tan\beta \gtrsim 10$, the RG running effects can be rather large and the leptonic
mixing angles can run significantly. Particularly, the RGE effects
can be very large for the solar neutrino mixing 
angle $\theta_{12}$~\cite{Haba:1998fb,Haba:1999ca,Haba:1999fk,Ellis:1999my,Casas:1999tp,Casas:1999ac,Miura:2002nz,Ellis:2005dr,Petcov:2005yh}.
\item
The solar neutrino mixing angle $\theta_{12}$ at $M_R$ depends strongly on the Majorana phase
$\alpha_{\scriptscriptstyle M}$~\cite{Miura:2002nz,Haba:2000tx,Antusch:2003kp,Petcov:2005yh}, which is
the relative phase between $m_1$ and $m_2$, and plays very important role in the predictions of the effective Majorana mass in
$(\beta\beta)_{0\nu}-$decay.  The effect of RG running for $\theta_{12}$ is smallest for the CP-conserving odd case $\alpha_{\scriptscriptstyle M} = \pm \pi$,
while it is significant for the CP-conserving even case $\alpha_{\scriptscriptstyle M} = 0$.  For $\alpha_{\scriptscriptstyle M} = 0$ and $\tan\beta \sim 50$, for
instance, we have $\tan^2\theta_{12}(M_R) \lesssim 0.5 \times \tan^2\theta_{12}(M_Z)$ for $m_1 \gtrsim 0.02$ eV.
\item
The RG running effect on $\theta_{12}$ due to the $\tau$-Yukawa coupling always makes $\theta_{12}(M_Z)$ larger than
$\theta_{12}(M_R)$~\cite{Miura:2002nz}.  This constrains the models which predict the value of solar neutrino mixing angle at
$M_R$, $\theta_{12}(M_R)>\theta_{12}(M_Z)$.  For example, the bi-maximal models are strongly restricted.  However, the running
effects due to the neutrino Yukawa couplings are free from this feature~\cite{Antusch:2002rr}.  Thus, bi-maximal models can predict
the correct value of neutrino mixing angles with the neutrino Yukawa contributions~\cite{Antusch:2002hy,Miura:2003if,Kanemura:2005it,Shindou:2004tv}.
\item
The RG corrections to neutrino mixing angles depend strongly on the deviation of 
the seesaw parameter matrix $R$ (\ref{eq:def-R}) from identity~\cite{Ellis:2005dr}. 
For  hierarchical light neutrinos, $m_1 \lesssim 0.01$ eV,  $\tan\beta \lesssim 30$ and $R$ nontrivial, 
the correction to $\theta_{23}$ and $\theta_{13}$ can be beyond their likely 
future experimental errors while $\theta_{12}$ is quite stable against the RG 
corrections~\cite{Ellis:2005dr}.
\item
The correction to $\theta_{23}$ can be large when $m_1$ and/or $\tan\beta$ are/is relatively large, \mbox{\it e.g.}, (i) when $m_1 \gtrsim 0.2$ eV 
if $\tan\beta \lesssim 10$, and (ii) for any $m_1$ and $\alpha_{\scriptscriptstyle M}$ if $\tan\beta \gtrsim 40$~\cite{Ellis:2005dr,Petcov:2005yh}.  
\item
The RG corrections to $\sin\theta_{13}$ can be relatively small, even for the large $\tan\beta$ if $m_1 \lesssim 0.05$ eV, and for
any $m_1 \gtrsim 0.30$ eV, if $\theta_{13}(M_Z) \cong 0$ and  $\alpha_{\scriptscriptstyle M} \cong 0$ 
(with $\beta_M = \delta = 0$). For $\alpha_{\scriptscriptstyle M} = \pi$ and $\tan\beta \sim 50$ one can have
$\sin\theta_{13}(M_R) \gtrsim 0.10$ for $m_1 \gtrsim 0.08$ eV even if $\sin\theta_{13}(M_Z) = 0$~\cite{Ellis:2005dr,Petcov:2005yh}.
\item
For $\tan\beta \gtrsim 30$, the value of $\Delta m^2_{21}(M_R)$ depends strongly on $m_1$ in the interval $m_1 \gtrsim 0.05$ eV, and
on $\alpha_{\scriptscriptstyle M}$, $\beta_{\scriptscriptstyle M}$, $\delta$, and $s_{13}$ for $m_1 \gtrsim 0.1$ eV.  
The dependence of $\Delta m^2_{31}(M_R)$ on $m_1$ and the CP phases is rather weak, unless $\tan\beta \gtrsim 40$,
$m_1 \gtrsim 0.10$ eV, and $s_{13}\gg 0.05$~\cite{Petcov:2005yh}.
\item
Some products of the neutrino mixing parameters, such as $s_{12}c_{12}c_{23}(m_1/m_2 $ $- e^{i \alpha_{\scriptscriptstyle M}})$
are practically stable with respect to RG running if one neglects the first and second generation charged-lepton Yukawa couplings 
and $s_{13}$~\cite{Haba:2000tx,Petcov:2003zb,Petcov:2005yh}. 
\end{itemize}

\subsubsection{Quark-lepton complementarity}

\subsubsubsection{Golden complementarity}

Quark-lepton complementarity~\cite{Raidal:2004iw,Minakata:2004xt} 
is based on  the observation 
that $\theta_{12}+\theta_C $ is numerically close to $\pi/4$. Here $\theta_{12}$ is 
the solar neutrino mixing angle and $\theta_C$ is the 
Cabibbo angle. For hierarchical light neutrino masses this result is 
relatively stable against the renormalization effects~\cite{Ellis:2005dr}.
To illustrate the idea we first review the 
model  of exact golden complementarity.

Consider  the following textures~\cite{Kajiyama:2007gx}
 for the light neutrino Majorana mass matrix $m_\nu$
and for the charged lepton Yukawa couplings $Y_e$:
\begin{equation}
\label{m1}
m_\nu = \left(
\begin{array}{ c  c c}
0 &m&0 \\
m &m&0\\
0&0&m_{\rm atm}
\end{array}
\right), \;\;\;\;\;\;
Y_e = 
\begin{pmatrix}
\lambda_e & 0 &0 \cr
0& \lambda_\mu/\sqrt{2} & 
\lambda_\tau/\sqrt{2}\cr
0& -\lambda_\mu/\sqrt{2} & 
\lambda_\tau/\sqrt{2}
\end{pmatrix}.
\end{equation}
It just assumes some texture zeroes
and some strict equalities among different entries.
The mass eigenstates of the neutrino mass matrix are given by
$m_1=-m/\varphi,$ $m_2=m\varphi$, $m_3=m_{\rm atm},$
where $\varphi = (1+\sqrt{5})/2 = 
 1 + 1/\varphi\approx 1.62$ is known as the golden ratio~\cite{Euclid}.
Thanks to its peculiar mathematical properties this  constant appears in various 
natural phenomena,  possibly including solar neutrinos. 
The three neutrino mixing angles obtained from Eq.~(\ref{m1}) are
$\theta_{\rm atm}=\pi/4$, $\theta_{13}=0$ and,
more importantly, 
\begin{equation}
\label{pred1}
\tan^2 \theta_{12}=1/\varphi^2=0.382,\qquad \hbox{i.e.}\qquad
\sin^2 2\theta_{12} = 4/5,
\end{equation}
in terms of the parameter $\sin^2 2\theta_{12}$ directly measured by 
vacuum oscillation experiments, such as KamLAND.
This prediction for $\theta_{12}$ is 1.4$\sigma$ below the experimental  best fit value.
A  positive measurement of $\theta_{13}$
might imply that the prediction for $\theta_{12}$
suffers an uncertainty up to  $\theta_{13}.$

Those properties   follow from the Z$_2\otimes {\rm Z}'_2$
symmetry of the neutrino mass matrix. Explicitly
$R m_\nu R^T=m_\nu $ where
\begin{equation}
 \label{R}
R=
\begin{pmatrix}
-1/\sqrt{5} & 2/\sqrt{5} & 0\cr
 2/\sqrt{5} & 1/\sqrt{5} & 0\cr
 0&0&1
 \end{pmatrix},
  \qquad
  R' = 
  \begin{pmatrix}
  1&0&0\cr 0&1&0\cr 0&0&-1
  \end{pmatrix},
  \end{equation}
and the rotations satisfy $\det R= -1$, $R\cdot R^T=1$ and $R\cdot
R=1$. The first Z$_2$ is a reflection along the diagonal of the golden
rectangle in the $(1,2)$ plane, see Fig.~\ref{fig:comp}. The second
Z$'_2$ is the $L_3 \to -L_3$ symmetry. Those symmetries allow
contributions proportional to the identity matrix to be added to
$m_\nu.$ This property allows to extend this type symmetries to the quark
sector.

\begin{figure}[b!]
\parbox{0.48\linewidth}{
\includegraphics*[clip, trim=0 0 0 0, width=\linewidth]{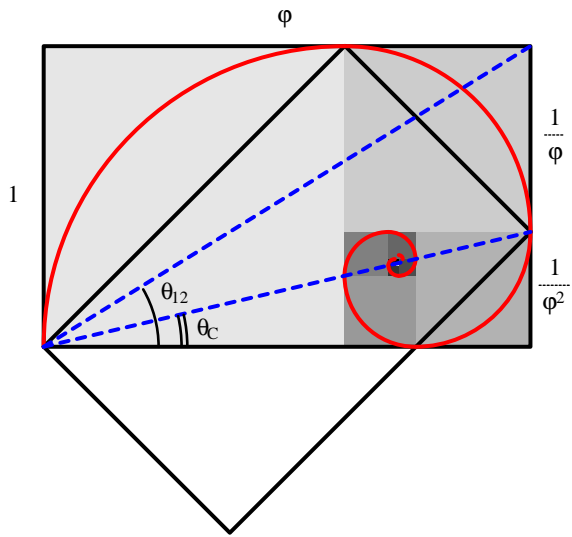}}\hspace*{\fill}
\parbox{0.48\linewidth}{
\caption{Geometrical illustration of the connection between the predictions for $\theta_{12} $ and $\theta_{C}$ and the golden rectangle.
The two dashed lines are the reflection axis of the ${\rm Z}_2$ symmetry for the neutrino mass matrix and for the up-quark mass matrix.\label{fig:comp}}}
\end{figure}

A seesaw model with singlet neutrinos satisfying the Z$_2\otimes {\rm
Z}'_2$ symmetry and giving rise to the mass matrix (\ref{m1}) is
presented in ~\cite{Kajiyama:2007gx}.

Noticing that the golden prediction (\ref{pred1}) satisfies with high accuracy the 
quark-lepton complementarity motivates one to give a golden geometric
explanation also to the Cabibbo angle.
SU(5) unification relates the down-quark Yukawa matrix $Y_d$ to $Y_e$
and suggests that the up-quark Yukawa matrix $Y_u$ is symmetric, like $m_\nu$.
One can therefore assume that $Y_d$ is diagonal in the two first generations and that
$Y_u$ is invariant under a Z$_2$ reflection described by a matrix analogous to
$R$ in Eq.~(\ref{R}), but with the factors $1\leftrightarrow 2$ exchanged.
Figure~\ref{fig:comp} illustrates the geometrical meaning of two reflection axis (dashed lines):
the up-quark reflection is along the diagonal of the golden rectangle tilted by $\pi/4$;
note also the connection with the decomposition of the golden rectangle as
an infinite sum of squares (`golden spiral').
Similarly to the neutrino case, this symmetry allows for
two independent terms that can be tuned such that $m_u \ll m_c$:
\begin{equation}
Y_u= \lambda  \left(
\begin{array}{ c  c  c}
1 &0  &0\\
0 &1 &0 \\0 &0 &1 \\
\end{array}\right)  +
\frac{\lambda}{\sqrt{5}}
\left(
\begin{array}{ c  c  c}
-2 &1 &0\\
1 & 2 &0\\
0 & 0 & c\\
\end{array}
\right).
\end{equation}
The second term fixes $\cot\theta_C=\varphi^3$, as can be geometrically seen 
from Fig.~\ref{fig:comp}.
We therefore have
\begin{equation}
\sin^2 2\theta_C = 1/5\quad
\hbox{i.e.}\quad \theta_{12}+\theta_C = \pi/4\quad\hbox{i.e.}\quad
V_{us}=\sin\theta_C = (1+\varphi^6)^{-1/2}= 0.229.
\end{equation}
This prediction is $1.9\sigma$ above the present best-fit value,
$\sin \theta_C = 0.2258\pm 0.0021$.
However, as the basic elements of flavour presented here  follow by construction 
from the 2x2 submatrices, one  naturally expects 
that the golden prediction for $V_{us}$ has an uncertainty at least comparable to
$|V_{ub}|\sim |V_{td}|\sim\hbox{few}\cdot 10^{-3}$. Thus the numerical accuracy is amazing.
Should the $1.4\sigma$ discrepancy between the golden prediction (\ref{pred1}) 
and the experimental measurement
hold after final SNO and KamLAND results, analogy with the quark sector would
allow one to predict the order of magnitude of neutrino mixing angle $\theta_{13}.$

Interestingly, similar predictions on the mixing angles are obtained 
if some suitably chosen assumptions are made on the properties of 
neutral currents of quarks and leptons \cite{Duret:2007pr}}.

\subsubsubsection{Correlation matrix from $S_3$ flavor symmetry in GUT}

On more general phenomenological ground the quark-lepton
complementarity~\cite{Raidal:2004iw,Minakata:2004xt} 
 can be described by the correlation matrix $V^M$ 
between  the CKM and the PMNS mixing matrices,
\begin{equation}\label{eq:fund}
V^M=U^{CKM}\, \Omega\,  U^{PMNS},
\end{equation}
where $\Omega={\rm diag}(e^{i\omega_i})$ is a diagonal matrix.
In the singlet seesaw mechanism the correlation matrix $V^M$ diagonalizes the 
symmetric matrix
\begin{eqnarray}
\label{eq:C}
{\cal C}&=&m_D^{diag} {V_R^\nu}^\dagger \frac{1}{M} {V_R^\nu}^\star m_D^{diag},
\end{eqnarray}
where $M$ is the heavy neutrino Majorana mass matrix and $V_R^\nu$
diagonalizes the neutrino Dirac matrix $m_D$ from the right.
In GUT models such as $SO(10)$ or $E_6$ we have intriguing relations between the Yukawa coupling of the quark sector and the one of the lepton sector.
For instance, in minimal renormalizable $SO(10)$ with Higgs in the $\bf 10$, $\bf 126$, and $\bf 120$, we have $Y_e \approx Y_d^T$.
In fact the flavor symmetry implies the structure of the Yukawa matrices: the equivalent entries of $Y_e$ and $Y_d$ are usually of the same order
of magnitude. 
In such a case one  gets 
 $$U^{PMNS} = (U^{CKM})^\dagger V^M.$$ 
As a consequence of that,  a $S_3$ flavor permutation symmetry,
softly broken into $S_2$, gives us the prediction of $V^M_{13}=0$
\cite{Chauhan:2006im}
and the correlations between CP-violating phases and the mixing 
angle $\theta_{12}$  \cite{Picariello:2006sp}.

The six generators of the $S_3$ flavor symmetry are the elements of the permutation group of three objects. The action of $S_3$ on the fields is to permute the
family label of the fields. In the following we will introduce the $S_2$ symmetry with respect the 2nd and 3rd generations. The $S_2$ group is an Abelian one
and swap the second family $\{\mu_L, (\nu_\mu)_L, s_L, c_L, \mu_R, (\nu_\mu)_R, s_R, c_R\}$ with
the third one $\{\tau_L, (\nu_\tau)_L, b_L, t_L, \tau_R, (\nu_\tau)_R, b_R, t_R\}$.

Let us assume that there is an $S_3$ flavor symmetry at high energy, which is softly broken into $S_2$ \cite{Morisi:2005fy}.
In this case, before the $S_3$ breaking all the Yukawa matrices have the following structure:
\begin{equation}\label{eq:YS3}
Y =\begin{pmatrix}  a&b&b\\ b&a&b\\ b&b&a \end{pmatrix},
\end{equation}
where $a$ and $b$ independent. The $S_3$ symmetry implies that $(1/\sqrt{3},1/\sqrt{3},1/\sqrt{3})$ is an eigenvector of our matrix in Eq.~(\ref{eq:YS3}).
Moreover these kind of matrices have two equal eigenvalues. This gives us an undetermined mixing angle in the diagonalizing mixing matrices.

When $S_3$ is softly broken into $S_2$, one gets
\begin{equation}\label{eq:YS2}
Y =\begin{pmatrix} a&b&b\\ b&c&d\\ b&d&c \end{pmatrix},
\end{equation}
with $c\approx a$ and $d\approx b$. When $S_3$ is broken the degeneracy is removed. In general the $S_2$ symmetry implies that $(0,1/\sqrt{2},-1/\sqrt{2})$
is an exact eigenvector of our matrix (\ref{eq:YS2}). The fact that $S_3$ is only softly broken into $S_2$ allows us to say that $(1/\sqrt{3},1/\sqrt{3},1/\sqrt{3})$
is still in a good approximation an eigenvector of Y in Eq.~(\ref{eq:YS2}). Then the mixing matrix that diagonalize from the right
the Yukawa mixing matrix in Eq.~(\ref{eq:YS2}) is given in good approximation by
the tri-bi-maximal mixing matrix (\ref{ma1}).

Let us now investigate the $V^M$ in this model. The mass matrix $m_D$ will have the general structure in Eq.~(\ref{eq:YS2}). To be more defined,
let us assumed that there is an extra softly broken $Z_2$ symmetry under which the 1st and the 2nd families are even, while the 3rd family is odd.
This extra softly broken $Z_2$ symmetry gives us a hierarchy between the off-diagonal and the diagonal elements of $m_D$, i.e. $b,d<< a,c$.
In fact if $Z_2$ is exact both $b$ and $d$ are zero. For simplicity, we assume also a quasi-degenerate spectrum for
the eigenvalues of the Dirac neutrino matrix as in \cite{Caravaglios:2006aq}.

The right-handed neutrino Majorana mass matrix is of the form
\begin{equation}
M =\begin{pmatrix} a&b&b^\prime\\ b&c&d\\ b^\prime&d&e \end{pmatrix}.
\end{equation}
Because $S_3$ is only softly broken into $S_2$ we have that $a\approx c \approx e$, and $b\approx b^\prime\approx d$.
In this approximation the $M$ matrix is diagonalized by a $U$ of the form in Eq.~(\ref{ma1}).
In this case we have that $m_\nu$  is near to be $S_3$ and $S_2$ symmetric, 
then it is diagonalized by a
mixing matrix $U_\nu$ near the tri-bi-maximal one given in Eq.~(\ref{ma1}). 
The ${\cal C}$ matrix is diagonalized by the mixing matrix $V_M=U_\nu U.$
We obtain that $V_M$ is a rotation in the $(1,2)$ plane, i.e. it contains a zero in the $(1,3)$ entry.
As shown in \cite{Caravaglios:2006aq}, it is possible to fit the CKM and the PMNS mixing matrix within this model.

\subsection{Leptogenesis and cosmological observables}

\subsubsection{Basic concepts and results}\label{sec:lepto}

CP-violation in the leptonic sector can have profound cosmological implications, playing a  crucial r\^ ole in the generation, via leptogenesis, 
of the observed baryon number asymmetry of the Universe \cite{Bennett:2003bz}:
\begin{equation}
\frac{n_{B}}{n_{\gamma}}= (6.1 ^{+0.3}_{-0.2}) \times 10^{-10}.
\end{equation}   
In the original framework a CP asymmetry is generated through out-of-equilibrium L-violating decays of heavy Majorana neutrinos \cite{Fukugita:1986hr} leading to a lepton
asymmetry $L \neq 0$. In the presence of sphaleron processes \cite{Kuzmin:1985mm}, which are $(B+L)$-violating and $(B-L)$-conserving, 
the lepton asymmetry is partially transformed to a baryon asymmetry.

The lepton-number asymmetry resulting from the decay of heavy Majorana neutrinos, $\varepsilon _{N_{j}}$, was computed by
several authors \cite{Liu:1993tg,Flanz:1994yx,Covi:1996wh}. The evaluation of $\varepsilon _{N_{j}}$, involves the computation of the interference between the
tree level diagram and one loop diagrams for the decay of the heavy Majorana neutrino $N_j$ into charged leptons $l_\alpha^\pm$ ($\alpha$ = e, $\mu$ , $\tau$). 
Summing the asymmetries  $\varepsilon^\alpha _{N_{j}}$ over  charged lepton flavour, one obtains:     
\begin{eqnarray}
\varepsilon _{N_{j}} = \frac{g^2}{{M_W}^2} \sum_{\alpha, k \ne j} \left[{\rm Im} \left((m_D^\dagger)_{j \alpha}( m_D)_{\alpha k} (m_D^\dagger m_D)_{jk} \right)
\frac{1}{16 \pi} \left(I(x_k)+ \frac{\sqrt{x_k}}{1-x_k} \right)\right]\frac{1}{(m_D^\dagger m_D)_{jj}}   ,
 \label{rmy}
\end{eqnarray}
where $M_k$ denote the heavy neutrino masses, the variable $x_k$ is defined as  $x_k=\frac{{M_k}^2}{{M_j}^2}$ and
$ I(x_k)=\sqrt{x_k} (1+(1+x_k) \log(\frac{x_k}{1+x_k}))$. From Eq.~(\ref{rmy}) it can be seen that, when one sums over all charged leptons, the
lepton-number asymmetry is only sensitive to the CP-violating phases appearing in $m_D^\dagger m_D$ in the basis where $M_R $ is diagonal. Note that this combination is
insensitive to rotations of the left-hand neutrinos.

If the lepton flavours are distinguishable in the final state,  it is the flavoured asymmetries  which are relevant
\cite{Barbieri:1999ma,Abada:2006fw,Nardi:2006fx,Abada:2006ea}. Below $T \sim 10^{12}$ GeV, the $\tau$ Yukawa  interactions are fast compared to
the Hubble rate, so  at least one flavour may  be distinguishable. The asymmetry in family  $\alpha$, generated from the decay 
of the $k$th heavy Majorana neutrino  depends on the combination \cite{Fujihara:2005pv} Im$( (m_D^\dagger m_D)_{k k^\prime}(m_D^*)_{\alpha k}
(m_D)_{\alpha k^\prime}) $ as well as on Im$((m_D^\dagger m_D)_{k^\prime k}(m_D^*)_{\alpha k} (m_D)_{\alpha k^\prime})$.
Summing over all leptonic flavours $\alpha $ the second term becomes real so that its imaginary part vanishes and the first term gives rise to the combination 
Im$((m_D^\dagger m_D)_{jk} (m_D^\dagger m_D)_{jk})$ that appears in Eq.~(\ref{rmy}). Clearly, when one works with 
separate flavours the matrix $U_{PMNS}$ 
does not cancel out and one is lead to the interesting possibility of having viable
leptogenesis even in the case of $R$ being a real matrix \cite{Pascoli:2006ie,Branco:2006hz,Branco:2006ce,Uhlig:2006xf}.

The simplest leptogenesis scenario corresponds to the case  of heavy hierarchical neutrinos where $M_1$ is much smaller than $M_2$ and $M_3$.
In this limit,  the asymmetries generated by $N_2$ and $N_3$ are frequently ignored, because the production of $N_2$ and $N_3$ can be suppressed by
kinematics (for instance, they are not produced thermally, if the re-heat temperature after inflation is $< M_2,M_3$), and the asymmetries from their decays
are partially washed out  \cite{Plumacher:1996kc,Barbieri:1999ma,Engelhard:2006yg}. In this hierarchical limit, the $\varepsilon^\alpha _{N_{1}} $ can be simplified  into:
\begin{equation}
\varepsilon^\alpha _{N_{1}}\simeq -\frac{3}{16\,\pi v^{2}}\,\left( I^\alpha_{12}\, \frac{M_{1}}{M_{2}}+I^\alpha_{13}\,\frac{M_{1}}{M_{3}}\right) \,,  \label{lepto3}
\end{equation}
where
\begin{equation}
I^\alpha_{1i}\equiv \frac{\mathrm{Im} \left[ (m_D^{\dagger })_{1 \alpha}(m_D)_{ \alpha i} (m_D^{\dagger }m_D)_{1i}\right] }{(m_D^{\dagger }\,m_D)_{11}}\ .  \label{lepto4}
\end{equation}
The flavour-summed CP asymmetry $\varepsilon _{N_{1}}$ can be written 
 in terms of the parameterization Eq.~(\ref{eq:Yukawa-R}) as
\begin{equation}\label{eqn:CPAsymmetry}
 \varepsilon_ {N_{1}}
 \approx
  -\frac{3}{8\pi}\frac{M_1}{v^2}\frac{\sum_i
  m_i^2 \text{Im} \left(R_{1i}^2\right)}{\sum_{i}
  m_i\left|R_{1i}\right|^2}.
\end{equation}
 In this case, obviously, leptogenesis demands non-zero imaginary parts in the \(R\) matrix. 
It has an  upper bound $|\varepsilon_ {N_{1}}|<\varepsilon_ {N_{1}}^{DI}$ 
where \cite{Davidson:2002qv}
\begin{equation}\label{eq:DI}
 \varepsilon_ {N_{1}}^{DI}=
  \frac{3}{8\pi}\frac{(m_3-m_1) M_1}{v^2},
  \end{equation}
which is  proportional to $M_1$. 
So the requirement of generating a sufficient baryon
asymmetry gives a lower bound on $M_1$ \cite{Davidson:2002qv,Hamaguchi:2001gw}. 
Depending on the cosmological scenario, the range for
minimal  $M_{1}$ varies from order $10^7$ GeV to $10^9$ 
GeV \cite{Buchmuller:2002rq,Giudice:2003jh}.  
This bound does not move much with the inclusion of flavour effects
\cite{Abada:2006fw,Blanchet:2006be,Antusch:2006gy}.
In supersymmetric world there is an upper bound $T_{RH}< 10^{8}$ GeV 
on the re-heating temperature of the
Universe from the possible overproduction of gravitinos, the so called gravitino problem
\cite{Ellis:1984eq,Ellis:1984er,Ellis:1995mr,Moroi:1995fs}.
Together with the lower bound on $M_1$ the gravitino problem puts severe constraints
on supersymmetric thermal leptogenesis scenarios.

However, the upper bound (\ref{eq:DI}) is based on 
the (natural) assumption that higher order corrections suppressed by $M_1/M_2, $  $M_1/M_3 $ 
in Eq.~(\ref{rmy}) are negligible. This may not be true as explicitly demonstrated 
in Ref. \cite{Raidal:2004vt} in which neutrino mass model is presented realizing
$\varepsilon_ {N_{1}}\gg \varepsilon_ {N_{1}}^{DI}.$ In such a case low scale 
standard thermal leptogenesis consistent with the gravitino bound 
is possible also for hierarchical heavy neutrinos.

Thermal leptogenesis is a rather involved thermodynamical non-equilibrium process and depends on additional parameters and on the proper treatment  of thermal effects \cite{Giudice:2003jh}. 
In the simplest case,  the $N_i$ are hierarchical,
and $N_1$ decays into a combination of flavours which are indistinguishable \footnote{This  can occur above  $ \sim 10^{12}$GeV, before the $\tau$ Yukawa interaction becomes fast
compared to the Hubble  rate, or in the case where the $N_1$ decay rate is faster than the charged lepton Yukawa interactions\cite{Blanchet:2006ch}.}.
In this case, the baryon asymmetry only depends on four parameters \cite{Buchmuller:2002rq,Buchmuller:2002jk,Davidson:2003fj,Davidson:2002qv}:
the mass $M_{1}$ of the lightest heavy neutrino, together with the corresponding CP asymmetry $\varepsilon _{N_{1}}$ in its decay, as well as the rescaled
$N_1$ decay rate, or effective neutrino mass  $\tilde{m}_{1}$ defined as
\begin{equation}
\tilde{m}_{1}=\sum_\alpha (m_D^{\dagger })_{1 \alpha}(m_D)_{\alpha 1}/M_{1}  \label{mtil},
\end{equation}
in the weak basis where $M_R$ is diagonal, real and positive. Finally, the baryon asymmetry depends also on the sum of all light neutrino masses squared, 
${\bar{m}}^{2}=m_{1}^{2}+m_{2}^{2}+m_{3}^{2}$, since it has been shown that this sum controls an important class of washout processes. If lepton flavours are
distinguishable, the final baryon asymmetry  depends on partial decay rates $\widetilde{m}_1^\alpha$ and CP asymmetries $\epsilon^\alpha_{1}$.

The $N_1$ decays in the early Universe at  temperatures $T \sim  M_1$,  producing asymmetries in the distinguishable final states. A particular asymmetry will survive once
washout by inverse decays go out of equilibrium. In the unflavoured calculation (where lepton flavours are indistinguishable), the fraction of the asymmetry that survives
is of order $ {\rm min }\{1, H/\Gamma \}$, where the Hubble rate $H$ and  the $N_1$ total decay rate $\Gamma$ are evaluated at $T = M_1$. This is usually
written $H/\Gamma  = m^*/\widetilde{m}_1 $, where \cite{Kolb:1990vq,Fischler:1990gn,Buchmuller:1992qc}:
\begin{equation} 
m_{*}=\frac{16\pi ^{5/2}}{3\sqrt{5}}g_{*}^{1/2}\frac{v^{2}}{M_{\text{Planck}}}\simeq 10^{-3}\ \mbox{eV}\; ,  \label{enm}
\end{equation}  
and  $M_{\text{Planck}}$ is the Planck mass ($M_{\text{Planck}}=1.2\times 10^{19}$ GeV), $v=\langle \phi ^{0}\rangle /\sqrt{2}\simeq 174\,$GeV is the weak scale 
and $g_{*}$ is the effective number of relativistic degrees of freedom in the plasma and equals 106.75 in the SM case. In a  flavoured calculation, the fraction of
a flavour  asymmetry that survives can be estimated  in the same way,  replacing  $\Gamma$ by the partial  decay rate.

\subsubsection{Implications of flavour effects}

For a long time the flavour effects in thermal leptogenesis were known \cite{Barbieri:1999ma} 
but  their phenomenological implications were considered only in specific
neutrino flavour models \cite{Raidal:2002xf}. 
As discussed,  in the single-flavour calculation, the  most important parameters 
for thermal leptogenesis from $N_1$
decays are $M_1$, 
$\tilde{m}_1$, $\epsilon_{N_1}$ and
the light neutrino mass scale. 
 Including  flavour effects gives this parameter space more dimensions
($M_1,$ $ \epsilon^{\alpha}, $ $\tilde m_1^{\alpha}$),
but it can still be  projected onto $M_1$, $\tilde{m}$ space.
For the readers convenience we summarize here some general results on the 
implications of flavoured leptogenesis.

In the unflavoured calculation, leptogenesis does not work for degenerate light neutrinos with a mass scale above $\sim  0.1$eV
\cite{Buchmuller:2003gz,Buchmuller:2004nz,Buchmuller:2004tu,Hambye:2003rt}. This bound does not survive  in the flavoured calculation,  where models with
a neutrino mass scale up to the cosmological bound, $\sum m_{\nu}<0.68$eV \cite{Spergel:2006hy}, can be tuned to work \cite{Abada:2006fw,Blanchet:2006ch}.

Considering the scale of leptogenesis, flavoured 
leptogenesis works for $M_1$ a factor
of $ \sim 3 $ smaller in the
 ``interesting'' region of $\tilde{m} < m_{atm}$.
But the lower bound on $M_1$, in the optimized
 $ \tilde{m} $ region,
remains $\sim 10^{9}$ GeV 
\cite{Blanchet:2006be,Antusch:2006gy}.
A smaller $M_1$ could be possible for
very degenerate light neutrinos\cite{Abada:2006fw}.

An important, but disappointing, observation in
single-flavour leptogenesis  was 
the lack of a  model-independent  connection between
CP-violation for leptogenesis and PMNS phases. It was shown
\cite{Branco:2001pq,Ellis:2002xg} that thermal leptogenesis can work
with no CP-violation  in $U_{PMNS}$, and conversely,
that leptogenesis can fail in spite
 of phases in $U_{PMNS}$.   In the
  ``flavoured'' leptogenesis case,
it is still true that the   baryon asymmetry  is
not sensitive  to PMNS phases \cite{Davidson:2007va,Antusch:2006cw} 
(leptogenesis can work  for any value of
the PMNS phases). However, interesting observations
can be made in classes of models 
\cite{Nardi:2006fx,Pascoli:2006ie,Branco:2006ce,Antusch:2006cw}.

\subsubsection{Other scenarios}

We have presented a brief discussion of minimal thermal leptogenesis in the 
context of type I seesaw with hierarchical heavy neutrinos. 
This scenario is  the most popular one because it is generic, supported by 
neutrino mass mechanism and, most importantly, it has predictions
for the allowed seesaw parameter space, as described above.
There are many other scenarios
in which leptogenesis may also be viable. 

Resonant leptogenesis \cite{Flanz:1994yx,Pilaftsis:1997jf} may occur when two or more
heavy neutrinos are nearly degenerate in mass and in this scenario the scale of the heavy neutrino masses can be lowered whilst still 
being compatible with thermal 
leptogenesis \cite{Pilaftsis:1997jf,Pilaftsis:1997dr,Ellis:2002eh,Pilaftsis:2003gt}. 
Heavy neutrinos of TeV scale or below could in principle be detected
at large colliders \cite{delAguila:2006dx}. 
In the seesaw context low scale heavy neutrinos may follow from 
extra symmetry 
principles \cite{Buchmuller:1992wm,Gluza:2002vs,Ellis:2002eh,Pilaftsis:2005rv}.
Also, the SM extensions with heavy neutrinos at TeV scale or below, include 
Kaluza--Klein modes in models with extra dimensions or extra matter content of Little Higgs models.

Leptogenesis from the out-of-equilibrium decays of a Higgs triplet 
\cite{Ma:1998dx,Hambye:2003ka,Hambye:2005tk} is another viable scenario
but requires the presence of at least two triplets for non-zero CP asymmetry.
Despite the presence of gauge interactions the washout effects 
in this scenario are not drastically larger than those in the singlet leptogenesis scenario
\cite{Hambye:2005tk}. Hybrid leptogenesis from type I and type II seesaw
can for instance occur in $SO(10)$ models \cite{Antusch:2004xy,Antusch:2007km}. 
In that case there 
are twelve independent CP-violating phases.

``Soft leptogenesis''
\cite{Grossman:2003jv,D'Ambrosio:2003wy}  
can work  in a  one-generational SUSY seesaw model because  CP-violation
in this scenario comes from complex supersymmetry breaking terms.
If the soft SUSY-breaking terms  are of suitable size,
there is  enough  CP-violation in  $\tilde{N} -\tilde{N}^*$ mixing
to imply the observed asymmetry.
Unlike non-supersymmetric triplet Higgs leptogenesis, 
soft leptogenesis with a triplet scalar \cite{D'Ambrosio:2004fz,Hambye:2005tk}
can also work in the minimal supersymmetric model of type II seesaw mechanism.

A very predictive supersymmetric leptogenesis scenario is obtained if
the sneutrino is playing the role of 
inflaton \cite{Murayama:1992ua,Murayama:1993em,Hamaguchi:2001gw,Ellis:2003sq,Antusch:2004hd}.
In this scenario the Universe is dominated by $\tilde N.$ Relating $\tilde N$ properties
to neutrino masses via the seesaw mechanism implies a lower bound 
$T_{RH}>10^6$ GeV on the 
re-heating temperature of the Universe \cite{Ellis:2003sq}.  A connection of this 
scenario with LFV is discussed in Section~\ref{sec:SUSY}.

Dirac leptogenesis is
another possibility considered in the literature. In this case neutrinos are of Dirac type rather than Majorana. In the original paper  \cite{Dick:1999je}       
two Higgs doublets were required and their decays create the leptonic asymmetry. Recently some authors have studied 
the connection  between leptogenesis and low energy data with two Higgs doublets \cite{Atwood:2005bf}.

Finally, let us mention that right-handed
neutrinos could have been produced non-thermally in the early Universe, by direct couplings to the inflation field. If this is the case, the constraints on neutrino
parameters from leptogenesis depend on the details of the inflationary model \cite{Asaka:1999yd,Giudice:1999fb,Endo:2006nj}.

For a recent overview of the present knowledge of neutrino masses and mixing and what can be learned about physics beyond the Standard Model 
from the various proposed neutrino experiments see Ref. \cite{Mohapatra:2005wg} and references therein.

\section{Organizing principles for flavour physics}\label{sec:organising}

\subsection{Grand Unified Theories}\label{sec:GUTs}

Grand unification is an attempt to unify all known interactions but gravity in a single simple gauge group.  It is motivated in part 
by the arbitrariness of electromagnetic charge in the standard model. One has charge quantization in a purely non-Abelian theory, without 
an U(1) factor, as in Schwinger's original idea \cite{Schwinger:1957em} of a SU(2) theory of electroweak interactions. 
The minimal gauge group which unifies weak and strong interactions, $SU(5)$  \cite{Georgi:1974sy},  automatically implies a 
quantized $U(1)$ piece too. While Dirac needed a monopole to achieve charge quantization \cite{Dirac:1948um}, grand unification in turn predicts the existence 
of magnetic monopoles \cite{'tHooft:1974qc,Polyakov:1974ek}. Since it unifies quarks and leptons \cite{Pati:1974yy}, it also predicts another remarkable phenomenon: 
the decay of the proton.
Here we are mostly interested in GUT implications on the flavour structure of Yukawa matrices.

\subsubsection{SU(5): the minimal theory}
 
The $24$ gauge bosons reduce to the $12$ ones of the SM plus a SU(2) doublet, color triplet pair $(X_\mu, Y_\mu)$ (vector leptoquarks), 
with $Y=5/6$ (charges $+4/3,+1/3$) and their antiparticles. The $15$ fermions of a single family in the SM fit in  the $\overline 5_F$ and 
$10_F$ anomaly-free representations of $SU(5)$, and the new  super-weak interactions of leptoquarks with fermions are ($\alpha$, $\beta$ and 
$\gamma$ are colour indices):
 \begin{eqnarray}
{\cal L}(X,Y) &=&\frac{g_5}{\sqrt{2}}X_\mu^{(-4/3)\alpha}\left(\bar e\gamma^\mu d^c_\alpha+\bar d_\alpha\gamma^\mu e^c-
\epsilon_{\alpha\beta\gamma}\overline{u^c}^\beta\gamma^\mu u^\gamma\right)\\
&-&\frac{g_5}{\sqrt{2}}Y_\mu^{(-1/3)\alpha}\left(\bar\nu\gamma^\mu d^c_\alpha+\bar u_\alpha\gamma^\mu e^c+
\epsilon_{\alpha\beta\gamma}\overline{u^c}^\beta\gamma^\mu d^\gamma\right)+h.c.\;,\nonumber
\end{eqnarray}
where all fermions above are explicitly left-handed and 
$\psi^c\equiv C\bar\psi^T$.

The exchange of the heavy gauge bosons leads to the effective interactions suppressed by two powers of their mass $m_X$ 
($m_X \simeq m_Y$ due to $SU(2)_L$ symmetry), which preserves $B-L$, but breaks both B and L symmetries and leads 
to ($d=6$) proton decay \cite{Weinberg:1979sa,Wilczek:1979hc}. From $\tau_P\mathrel{\rlap{\lower4pt\hbox{\hskip1pt$\sim$}}\raise1pt\hbox{$>$}} 6\times 10^{33}$ yr \cite{Kobayashi:2005pe}, $m_X \mathrel{\rlap{\lower4pt\hbox{\hskip1pt$\sim$}}\raise1pt\hbox{$>$}} 10^{15.5}$ GeV. 

The Higgs sector consists of an adjoint $24_H$ and a fundamental $5_H$, the first breaks $SU(5) \to SM$, the latter completes the symmetry 
breaking \'a l\`a Weinberg-Salam. Now, $5_H = (T, D)$, where $T$ is a color triplet and $D$ the usual Higgs $SU(2)_L$ doublet of the SM 
and so the Yukawa interactions in the matrix form
\begin{equation}
\label{star}
{\cal L}_Y=10_F\,y_u\,10_F\,5_H+\overline 5_F\,y_d\,10_F\,5_H^*,
\end{equation}
give the quark and lepton mass matrices
\begin{equation}
m_u =y_u\langle D\rangle\;,\;m_d=m_e^T=y_d \langle D\rangle\;.
\end{equation}

Note the correlation between down quarks and charged leptons \cite{Buras:1977yy}, valid at the GUT scale, and  impossible to 
be true for all three generations. Actually, in the SM it is wrong for all of them. It can be corrected by an extra Higgs, $45_H$ 
\cite{Georgi:1979df}, or higher dimensional non-renormalizable interaction \cite{Ellis:1979fg}. 

From (\ref{star}), one gets also the interactions of the triplet, which lead to proton decay and thus the triplet $T$ must 
be superheavy, $m_T \mathrel{\rlap{\lower4pt\hbox{\hskip1pt$\sim$}}\raise1pt\hbox{$>$}} 10^{12} $GeV. The enormous split between $m_T$ and $m_D\simeq m_W$ can be achieved through the large scale 
of the breaking of $SU(5)$
\begin{equation}
\langle 24_H\rangle = v_X {\rm diag}(2,2,2,-3,-3)\;,
\end{equation}
with $m_X^2 = m_Y^2 = \frac{25}{4} g_5^2 v_X^2 $. This fine-tuning is known as the doublet-triplet problem. Whatever solution one may adopt, 
the huge hierarchy can be preserved in perturbation theory only by supersymmetry with low scale breaking of order TeV. 

The consistency of grand unification requires that the gauge couplings of the SM unify at a single scale, in a tiny window 
$10^{15.5}$ GeV $\mathrel{\rlap{\lower4pt\hbox{\hskip1pt$\sim$}}\raise1pt\hbox{$<$}} M_{GUT} \mathrel{\rlap{\lower4pt\hbox{\hskip1pt$\sim$}}\raise1pt\hbox{$<$}} 10^{18}$GeV (lower limit from proton decay, upper limit from perturbativity, i.e. to stay below $M_{Pl}$). 
Here the minimal ordinary $SU(5)$ theory described above fails badly, while the version with low energy supersymmetry does great 
\cite{Dimopoulos:1981yj,Ibanez:1981yh,Einhorn:1981sx,Marciano:1981un}. Actually, one needed a heavy top quark \cite{Marciano:1981un}, 
with $m_t \simeq 200$ GeV in order for the theory to work. The same is needed in order to achieve a radiative 
symmetry breaking of the SM gauge symmetry, where only the Higgs doublet becomes tachyonic \cite{Inoue:1982pi,Alvarez-Gaume:1983gj}. One can then 
define the minimal supersymmetric $SU(5)$ GUT with the three families of fermions $10_F$ and $\overline 5_F$, and with $24_H$ and $5_H$ and 
$\overline 5_H$ supermultiplets. It predicts $m_d = m_e^T$ at $M_{GUT}$ which works  well for the 3rd generation; the first two can be corrected by 
higher dimensional operators. Although this theory  typically has a very fast $d=5$ \cite{Sakai:1981pk,Weinberg:1981wj,Hisano:1992jj,Lucas:1996bc,Goto:1998qg} 
proton decay \cite{Murayama:2001ur}, the higher dimensional operators can easily make it in accord with experiments 
\cite{Bajc:2002bv,Bajc:2002pg,Emmanuel-Costa:2003pu}. The main problem are massless neutrinos, unless one breaks 
R-parity (whose approximate or exact conservation must be assumed in supersymmetric SU(5), contrary to some supersymmetric 
SO(10)). Other ways out include adding singlets, right-handed neutrinos 
(type I see-saw  \cite{Minkowski:1977sc,Yanagida:1979as,Gell-Mann:1980vs,
Glashow:1979nm,Mohapatra:1979ia}), 
or a $15_H$ multiplet (type II see-saw \cite{Magg:1980ut,Schechter:1980gr,Lazarides:1980nt,Mohapatra:1980yp}). 
In both cases their Yukawa 
are not connected to the charged sector, so it is much more appealing to go to SO(10) theory, which unifies all fermions 
(of a single family) too, besides the interactions. 

Before we move to SO(10), what about ordinary non-supersymmetric SU(5)? In order to have $m_\nu \neq 0$ and to achieve the unification 
of gauge couplings one can add either (a) $15_H$ Higgs multiplet \cite{Dorsner:2005fq} or (b) $24_F$ fermionic multiplet \cite{Bajc:2006ia}. 
The latter one is particularly interesting, since it leads to the mixing of the type I and type III see-saw \cite{Foot:1988aq,Ma:1998dn}, 
with the remarkable 
prediction of a light $SU(2)$ fermionic triplet below TeV and $M_{GUT} \leq 10^{16}$ GeV, which offers hope both for the observable 
see-saw at LHC and a detectable proton decay in a future generation of experiments now planned \cite{Wilkes:2005rg}.

These fermionic triplets $T_F$ would be produced in pairs
through a Drell-Yan process. The production cross section for
the sum of all three possible final states, $T_F^+ T_F^-$, $T_F^+ T_F^0$
and $T_F^- T_F^0$, can be read from Fig.2 of Ref. \cite{Cheung:2005ba}:
it is approximately 20 pb for 100 GeV triplet mass, and around
40 fb for 500 GeV triplets. The triplets then decay into 
$W$ or $Z$ and a light lepton through the
same Yukawa couplings that enter into the seesaw. 

The clearest signature would be 
the three charged lepton decay of the charged triplet, but it
has only a $3\%$ branching ratio. A more promising situation is
the decay into two jets with SM gauge boson invariant mass
plus a charged lepton: this happens in
approximately 23\% of all decays. 
The signatures in this case is two same charge leptons plus two pairs
of jets having the W or Z mass and peaks in the lepton-dijet mass.
From the above estimates the cross section for such events is
around 1pb (2fb) for 100 (500) GeV triplet mass. 
Such signatures were suggested originally in
L-R symmetric theories 
\cite{Keung:1983uu}, but are quite generic of the seesaw mechanism.

\subsubsection{SO(10): the minimal theory of matter and gauge coupling unification}
 
There are a number of features that make SO(10) special:
\begin{list}{-}{}
\item 
a family of fermions is unified in a 16-dimensional spinorial representation; this in turn predicts the existence of
right-handed neutrinos, making the implementation of the see-saw mechanism almost automatic;
\item 
$L-R$ symmetry \cite{Pati:1974yy,Mohapatra:1974gc,Senjanovic:1975rk, Senjanovic:1978ev} is a finite gauge transformation in the form of 
charge conjugation. This is a consequence of both left-handed fermions $f_{L}$ and its charged conjugated counterparts 
$(f^{c})_{L} \equiv C \overline{f}_{R}^{T}$ residing in the same representation $16_{F}$; 
\item 
in the supersymmetric version, the matter parity $M= (-1)^{3(B-L)}$, equivalent to the R-parity $R= M (-1)^{2 S}$, is a gauge 
transformation \cite{Mohapatra:1986su,Font:1989ai,Martin:1992mq}, a part of the center $Z_{4}$ of SO(10). In the 
renormalizable version of the theory it remains exact at all energies \cite{Aulakh:1997ba,Aulakh:1998nn,Aulakh:1999cd}. 
The lightest supersymmetric partner (LSP) is then stable and is a natural candidate for the dark matter of the universe; 
\item 
its other maximal subgroup, besides $SU(5)\times U(1)$, is 
$G_{PS}= SU(2)_{L}\times SU(2)_{R}\times SU(4)_C$ 
quark-lepton symmetry of Pati and Salam, which plays an important role in relating quark and lepton masses and mixings;
\item 
the unification of gauge couplings can be achieved even without supersymmetry (for a recent and complete work and references 
therein see \cite{Deshpande:1992au,Deshpande:1992em}).
\end{list}

Fermions belong to the spinor representation $16_{F}$ (for useful reviews on spinors and SO(2N) group theory in general see \cite{Mohapatra:1979nn,Wilczek:1981iz,
Slansky:1981yr,Nath:2001yj,Aulakh:2002zr}). From
\begin{equation}
16 \times 16 = 10 + 120 + 126\;,
\end{equation}
the most general Yukawa sector in general contains $10_{H}$, $120_{H}$ and $\overline{126}_{H}$, respectively the fundamental vector 
representation, the three-index antisymmetric representation and the five-index antisymmetric and anti-self-dual representation. 
$\overline{126}_{H}$ is necessarily complex, supersymmetric or not; $10_{H}$ and $\overline{126}_{H}$ Yukawa matrices are symmetric 
in generation space, while the $120_{H}$ one is antisymmetric.

The decomposition of the relevant representations under $G_{PS}$ gives
\begin{eqnarray}
{\bf 16}&=&(2,1,4)+(1,2,\bar 4) ,\nonumber \\
{\bf 10}&=&(2,2,1)+(1,1,6) ,\nonumber \\
{\bf 120}&=&(2,2,1)+(3,1,6)+(1,3,6)+(2,2,15)+(1,1,10)+(1,1,\overline{10}),\nonumber \\
{\bf \overline{126}}&=&(3,1,\overline{10})+(1,3,10)+(2,2,15)+(1,1,6).
\end{eqnarray}

The see-saw mechanism, whether type I or II, requires $\overline{126}$: it contains both $(1,3,10)$ whose VEV gives a 
mass to $\nu_{R}$ (type I), and $(3,1,\overline{10})$, which contains a color singlet, $B-L= 2$ field $\Delta_{L}$, that can give directly 
a small mass to $\nu_{L}$ (type II). In SU(5) language this is seen from the decomposition 
\begin{equation}
{\bf \overline{126}} = 1 + 5 + 15 + \overline{45} + 50.
\end{equation}
The $1$ of SU(5) belongs to the  $(1,3,10)$ of $G_{PS}$ and gives a mass for $\nu_{R}$, while $15$ corresponds to the $ (3,1,\overline{10})$ 
and gives the direct mass to $\nu_{L}$. 

$ \overline{126}$ can be a fundamental field, or a composite of two $\overline{16}_{H}$ fields (for some realistic examples see 
for example \cite{Albright:2000sz,Babu:1998wi,Blazek:1999ue}), or can even be induced as a two-loop effective representation 
built out of a $10_{H}$ and two gauge $45$-dim representations \cite{Witten:1979nr,Bajc:2004hr,Bajc:2005aq}. 

Normally the light Higgs is chosen to be the smallest one, $10_{H}$. Since $\langle 10_{H}\rangle = \langle (2,2,1)\rangle$ is a 
$SU(4)_C$ singlet, $m_{d} = m_{e}$ follows immediately, independently of the number of $10_{H}$. Thus we must add either $120_{H}$ or 
$\overline{126}_{H}$ or both in order to correct the bad mass relations. Both of these fields contain $(2,2,15)$, which 
VEV alone gives the relation $m_{e}= -3 m_{d}^T$.

As $\overline{126}_{H}$ is needed anyway for the see-saw, it is natural to take this first. The crucial point here is that in general 
$(2,2,1)$ and $(2,2,15)$ mix through $\langle (1,3,10)\rangle$ \cite{Lazarides:1980nt,Babu:1992ia} and thus the light Higgs is a 
mixture of the two. In other words, $\langle (2,2,15)\rangle$ in $\overline{126}_{H}$ is in general non-vanishing (in supersymmetry 
this is not automatic, but depends on the Higgs superfields needed to break SO(10) at $M_{GUT}$ or on the presence of higher dimensional 
operators).

If one considers all the operators allowed by SO(10) for the Yukawa couplings, there are too many model parameters, and so no prediction
is really possible. One option is to assume that the minimal number of parameters must be employed. It has been shown that 4 (3 of them
non-renormalizable) operators are enough in models with $10$ and $45$ Higgs representations only \cite{Anderson:1993fe}.  Although this is
an important piece of information and it has been the starting point of a lot of model building, it is difficult to see a reason for some
operators (of different dimensions) to be present and other not, without using some sort of flavour
symmetry, so these type of models will not be considered in this subsection.  On the other hand, a self consistent way of truncating
the large number of SO(10) allowed operators without relying on extra symmetries is to consider only the renormalizable ones.  This is
exactly what we will assume.

In this case there are just two ways of giving mass to $\nu_R$: by a nonzero VEV 
of the Higgs $\overline{126}$, or generate an effective non-renormalizable operator radiatively \cite{Witten:1979nr}. We will consider in turn both of them.

\subsubsubsection{Elementary $\overline{126}_H$}

It is rather appealing that $10_{H}$ and $ \overline{126}_{H}$ may be sufficient for all the fermion masses, with only two sets of symmetric Yukawa coupling matrices. 
The mass matrices at $M_{GUT}$ are 

\begin{eqnarray}
\label{md}
m_d&=&v_{10}^dY_{10}+v_{126}^dY_{126}\;,\\
\label{mu}
m_u&=&v_{10}^uY_{10}+v_{126}^uY_{126}\;,\\
\label{me}
m_e&=&v_{10}^dY_{10}-3v_{126}^dY_{126}\;,\\
\label{mn}
m_\nu&=&-m_{D}M_{R}^{-1}m_{D}+m_{\nu_L}\;,
\end{eqnarray}
where
\begin{eqnarray}
\label{mnd}
m_{D}&=&v_{10}^uY_{10}-3v_{126}^uY_{126}\;,\\
\label{mnr}
M_{R}&=&v_RY_{126}\;,\\
\label{mnl}
m_{\nu_L}&=&v_LY_{126}\;.
\end{eqnarray}
These relations are valid at $M_{GUT}$, so it is there that their validity must be tested. The analysis done so far used the results 
of renormalization group running from $M_Z$ to $M_{GUT}$ from \cite{Fusaoka:1998vc,Das:2000uk}. 

The first attempts in fitting the mass matrices assumed the domination of the type I seesaw. It was pioneered by 
treating CP-violation perturbatively in a non-supersymmetric framework \cite{Babu:1992ia}, and later improved with a 
more detailed treatment of complex parameters and supersymmetric low-energy effective theory 
\cite{Matsuda:2000zp,Matsuda:2001bg,Fukuyama:2002ch}. Nevertheless, these fits had problems to reproduce correctly the PMNS matrix parameters.

A new impetus to the whole program was given by the observation that in case type II seesaw dominates (a way to enforce it is 
to use a $54$ dimensional Higgs representation \cite{Goh:2004fy}) the neutrino mass, an 
interesting relation in these type of models between $b-\tau$ unification and large atmospheric mixing angle can be found 
\cite{Bajc:2001fe,Bajc:2002iw,Bajc:2004fj}. The argument is very simple and it can be traced to the relation \cite{Brahmachari:1997cq} 
\begin{equation}
m_\nu \propto m_d-m_e\;,
\end{equation}
which follows directly from (\ref{md}), (\ref{me}) and (\ref{mnl}), if only the second term (type II) in (\ref{mn}) is considered. 
Considering only the heaviest two generations as an example and taking the usually good approximation of small second generation 
masses and small mixing angles, one finds all the elements of the right-hand-side small except the $22$ element, which is 
proportional to the difference of two big numbers, $m_b-m_\tau$. Thus, a large neutrino atmospheric mixing angle is linked to 
the smallness of this $22$ matrix element, and so to $b-\tau$ unification. Note that in these types of models 
$b-\tau$ unification is no more automatic due to the presence of the $\overline{126}$, which breaks SU(4)$_C$. It is however quite a 
good prediction of the RGE running in the case of low-energy supersymmetry. 

The numerical fitting was able to reproduce also a large solar mixing angle both in case of type II \cite{Goh:2003sy,Goh:2003hf} 
or mixed seesaw \cite{Dutta:2004wv}, predicting also a quite large $|U_{e3}| \approx 0.16$ mixing element, close to the experimental 
upper bound. The difficulty in fitting the CKM CP-violating phase in the first quadrant was overcome by new solutions found in 
\cite{Bertolini:2005qb,Babu:2005ia}, maintaining the prediction of large $|U_{e3}| \ge 0.1$ matrix element. 

All these fittings were done assuming no constraints coming from the Higgs sector. Regarding it, it was found that the minimal
supersymmetric model \cite{Clark:1982ai,Aulakh:1982sw,Lee:1994je} has only 26 model parameters
\cite{Aulakh:2003kg}, on top of the usual supersymmetry breaking soft terms, as in the MSSM.  When one considers this minimal model, the
VEVs in the mass formulae (\ref{md})-(\ref{mnl}) are not completely arbitrary, but are connected by the restrictions of the Higgs
sector. This has been first noticed in \cite{Aulakh:2005bd,Bajc:2005qe,Aulakh:2005mw} showing a possible
clash with the positive results of the unconstrained Yukawa sector studied in \cite{Bertolini:2005qb,Babu:2005ia}. The issue has been
pursued in \cite{Bertolini:2006pe}, showing that in the region of parameter space where the fermion mass fitting is successful, there
are necessarily intermediate scale thresholds which spoil perturbativity of the RGE evolution of the gauge couplings.

To definitely settle the issue two further checks should be done: a) the $\chi^2$ analysis used in the fitting procedure should be
implemented at $M_Z$, not at $M_{GUT}$. The point is in fact that while the errors at $M_Z$ are uncorrelated, they become strongly
correlated after running to $M_{GUT}$, due to the large Yukawa coupling of top and possibly also of bottom, tau and neutrino.
 b) Another issue is to consider also the
effect of the possible increased gauge couplings on the Yukawas.
 Only after these two checks
will be done, this minimal model could be ruled out.

A further important point is that in the case of VEVs constrained by the Higgs sector one finds from the charged fermion masses that the model
predicts large $\tan \beta\simeq 40$, as confirmed by the last fits in~\cite{Bertolini:2006pe}.  In this regime there may be sizeable
corrections to the ``down'' fermion mass matrices from the soft SUSY breaking parameters~\cite{Hall:1993gn}; this brings into the game also
the soft SUSY breaking sector, lowering somewhat the predictivity but relaxing the difficulty in fitting the experimental data. In this
scenario predictions on masses would become predictions on the soft sector.

Some topics have to be still mentioned in connection with the above: the important calculation of the mass spectrum and Clebsch-Gordan
coefficients in SO(10) \cite{He:1990jw,Lee:1993rr,Aulakh:2002zr,Bajc:2004xe,Aulakh:2004hm,Fukuyama:2004xs,Fukuyama:2004ps,
Fukuyama:2004ti,Aulakh:2005sq, Aulakh:2005ic}, the doublet-triplet splitting problem \cite{Lee:1993jw,Babu:2006nf}, the Higgs doublet mass
matrix \cite{Bajc:2004xe,Aulakh:2002zr}, the running of the gauge couplings at two loops together with threshold corrections
\cite{Aulakh:2004hm}, and the study of proton decay \cite{Goh:2003nv,Fukuyama:2004xs,Fukuyama:2004pb}.

What if this model turns out to be wrong? There are other models on the market. The easiest idea is to add a $120$ dimensional Higgs, that
may also appear as a natural choice, being the last of the three allowed representations that couple with fermions.  There are three
different ways of doing it considered in the literature: a) take $120$ as a small, non-leading, contribution, i.e. a perturbation to the previous formulae 
\cite{Bertolini:2004eq,Yang:2004xt,Dutta:2004hp}; b) consider $120$ on an equal footing as $10$ and $\overline{126}$,
but assume some extra discrete symmetry or real parameters in the superpotential, breaking CP spontaneously 
\cite{Dutta:2004zh,Dutta:2005ni,Grimus:2006bb,Grimus:2006rk} (and suppressing in the first two references the dangerous $d=5$ proton decay modes); 
c) assume small $\overline{126}$ contributions to the charged fermion masses \cite{Aulakh:2006vi,Lavoura:2006dv,Aulakh:2006vj,Aulakh:2006hs}.

Another limit is to forget the $10_H$ altogether, as has been proposed for non-supersymmetric theories \cite{Bajc:2005zf}. The two generation
study predicts a too small ratio $m_b/m_\tau\approx 0.3$, instead of the value $0.6$ that one gets by straight running. The idea is that
this could get large corrections due to Dirac neutrino Yukawas \cite{Vissani:1994fy} and the effect of finite second generation
masses, as well as the inclusion of the first generation and CP-violating phases. This is worth pursuing for it provides an 
alternative minimal version of SO(10), and after all, supersymmetry may not be there.

\subsubsubsection{Radiative $\overline{126}_H$}

The original idea \cite{Witten:1979nr} is that there is no $\overline{126}_H$ representation in the theory, but the same operator is generated by loop 
corrections. The representation that breaks the rank of SO(10) is now $16_H$, which VEV let us call $M_\Lambda$. Generically there is a contribution to the 
righthanded neutrino mass at two loops:
\begin{equation}
M_{R}\approx \left(\frac{\alpha}{4\pi}\right)^2\frac{M_\Lambda^2}{M_{GUT}}
\frac{M_{SUSY}}{M_{GUT}}Y_{10}\;,
\end{equation}
which is too small in low-energy supersymmetry (low breaking scale $M_{SUSY}$) as well as non-supersymmetric theories ($M_{SUSY}=M_{GUT}$, 
but low intermediate scale $M_\Lambda$ required by gauge coupling unification). The only exception, proposed in \cite{Bajc:2004hr}, could be split
supersymmetry \cite{Arkani-Hamed:2004fb,Giudice:2004tc}.

In the absence of $\overline{126}_H$, the charged fermion masses must be given by only $10_H$ and $120_H$ \cite{Bajc:2004hr}, together 
with radiative corrections. The simplest analysis of the tree order two generation case gives three interesting predictions-relations 
\cite{Bajc:2005aq,Bajc:2006pa}: 1) almost exact $b-\tau$ unification; 2) large atmospheric mixing angle related to the small quark 
$\theta_{bc}$ mixing angle; 3) somewhat  degenerate neutrinos. For a serious numerical analysis one needs to use the RGE for the 
case of split supersymmetry, taking a very small $\tan{\beta}<1$ to get an approximate $b-\tau$ unification \cite{Giudice:2004tc}. 
One needs also some fine-tuning of the parameters to account for the small ratio $M_{SUSY}/M_{GUT}\le 10^{-(3-4)}$ required in 
realistic models to have gluinos decay fast enough \cite{Arvanitaki:2005fa}.

\subsection{Higher-dimensional approaches}
\label{sec:extra-th}

Recently, in the context of theories with extra spatial dimensions, some new approaches toward the question of SM fermion mass hierarchy and flavour structure have
arisen \cite{Yoshioka:1999ds,Bando:2000jd,Neronov:2001qv,Arkani-Hamed:1999yy,Dienes:1998vh,Dienes:1998vg,Arkani-Hamed:1999dc,Gherghetta:2000qt}.  
 For instance, the SM fermion mass spectrum can be generated naturally by permitting the quark/lepton
masses to evolve with a power-law dependence on the mass scale 
 \cite{Dienes:1998vh,Dienes:1998vg}.
 The most studied and probably most attractive idea for generating a non-trivial
flavour structure is the displacement of various SM fermions along extra dimension(s). This approach is totally different from the one discussed in
Section~\ref{sec:framework} as it is purely geometrical and thus does not rely on the existence of any novel symmetry in the short-distance theory. The displacement
idea applies to the scenarios with large flat \cite{Arkani-Hamed:1999dc} 
or small warped \cite{Gherghetta:2000qt} 
extra dimension(s), as we develop in the following subsections. 

\subsubsection{Large extra dimensions}\label{subsec:LEDI} 
In order to address the gauge hierarchy problem, a scenario with large flat extra dimensions has been proposed by Arkani-Hamed, Dimopoulos and Dvali (ADD) 
\cite{Arkani-Hamed:1998rs,Antoniadis:1998ig,Arkani-Hamed:1998nn}, 
 based on a reduction of the fundamental gravity scale down to the TeV scale. In this scenario, gravity propagates in the bulk whereas SM fields
live on a 3-brane. One could assume that this 3-brane has a certain thickness $L$ along an extra dimension  (as for example in \cite{Barenboim:2001wy}). 
Then SM fields would feel an extra dimension of size $L$, 
exactly as in a Universal Extra Dimension (UED) model \cite{Appelquist:2000nn}  
(where SM fields propagate in the bulk) with one extra dimension of size $L$
\footnote{The constraint from electroweak precision measurements is $R^{-1} \gtrsim 2-5$ TeV, the one from direct search at LEP collider is $L^{-1} \gtrsim 5$ TeV and the
expected LHC sensitivity is about $L^{-1} \sim 10$ TeV. }. 

In such a framework, the SM fermions can be localized at different positions along  
this extra dimension $L$. Then the relative displacements of quark/lepton wave function  
peaks produce suppression factors in the effective 4-dimensional Yukawa couplings.  
These suppression factors being determined by the overlaps of fermion  
wave functions (getting smaller as the distance between wave function 
peaks increases), they can vary with the fermion flavours and thus induce a mass  
hierarchy. This mechanism was first suggested in \cite{Arkani-Hamed:1999dc} and 
its variations have been  studied  in 
\cite{Libanov:2000uf,Frere:2000dc,Frere:2001ug,Libanov:2002tm,Dvali:2000ha,Hung:2002qp,Kaplan:2000av,Kaplan:2001ga,Kakizaki:2001en,Chang:2003sx,Nussinov:2001ps}.

Let us describe this mechanism more precisely. The fermion
localization can be achieved through either non-perturbative effects
in string/M theory or field-theoretical methods. One field-theoretical
possibility is to couple the SM fermion fields $\Psi_i(x_\mu,x_5)$
[$i=1,...,3$ being the family index and $\mu=1,...,4$ the usual
coordinate indexes] to 5-dimensional scalar fields with VEV
$\Phi_i(x_5)$ depending on the extra dimension (parameterized by
$x_5$)
\footnote{Although we concentrate here on the case with only one extra dimension, for simplicity, the mechanism can be directly extended to more extra dimensions.}.
Indeed, chiral fermions are confined in solitonic backgrounds \cite{Jackiw:1975fn}.
 If the scalar field profile behaves as a linear function of the form
$\Phi_i(x_5)=2\mu^2x_5-m_i$ around its zero-crossing point $x^0_i=m_i/2\mu^2$, the zero-mode of 5-dimensional fermion acquires a Gaussian wave function of typical
width $\mu^{-1}$ and centered at $x^0_i$ along the $x_5$ direction: $\Psi^{(0)}_i(x_\mu,x_5) = Ae^{-\mu^2(x_5-x^0_i)^2}\psi_i(x_\mu)$,
$\psi_i(x_\mu)$ being the 4-dimensional fermion field and $A=(2\mu^2/\pi)^{1/4}$ a normalization factor. Then the 4-dimensional Yukawa couplings between the
5-dimensional SM Higgs boson $H$ and zero-mode fermions, obtained by integration on $x_5$ over the wall width $L$ \footnote{Here, the factor $\sqrt L$ compensates
with the Higgs component along $x_5$, since the Higgs boson is not localized.}:
\begin{eqnarray}
{\cal S}_{\textrm Yukawa}  =  \int d^5x \sqrt L \kappa H(x_\mu,x_5) \bar \Psi^{(0)}_i(x_\mu,x_5) \Psi^{(0)}_j(x_\mu,x_5)  =
\int d^4x Y_{ij} h(x_\mu) \bar \psi_i(x_\mu) \psi_j(x_\mu), \label{eq:Yukawa}
\end{eqnarray}
are modulated by the following effective coupling constants,
\begin{equation}
Y_{ij}=\int dx_5 \kappa A^2 e^{-\mu^2(x_5-x^0_i)^2} e^{-\mu^2(x_5-x^0_j)^2}=
\kappa e^{-\frac{\mu^2}{ 2}(x^0_i-x^0_j)^2}.
\label{eq:lambdaij}
\end{equation}
It can be considered as natural to have a 5-dimensional Yukawa coupling constant equal to $\sqrt L \kappa$, where the dimensionless parameter $\kappa$ is universal
(in flavour and nature of fermions) and of order unity, so that the flavour structure is mainly generated by the field localization effect through the exponential
suppression factor in Eq.~(\ref{eq:lambdaij}). The remarkable feature is that, due to this exponential factor, large hierarchies can be created among the physical
fermion masses, even for all fundamental parameters $m_i$ of order of the same energy scale $\mu$.

This mechanism can effectively accommodate all the data on quark and 
charged lepton masses and mixings \cite{Mirabelli:1999ks,Branco:2000rb,Hung:2001hw}. In case where 
right-handed neutrinos are added to the SM so that neutrinos acquire  
Dirac masses (as those originating from Yukawa couplings (\ref{eq:Yukawa})),  
neutrino oscillation experiment results can also be reproduced 
\cite{Barenboim:2001wy}. The fine-tuning, arising there on relative $x^0_i$ parameters,   
turns out to be improved when neutrinos get Majorana masses instead  
\cite{Frere:2003hn} (see also \cite{Raidal:2002xf,Klapdor-Kleingrothaus:2002ir}).

\subsubsection{Small extra dimensions}\label{sec:SED1} 
An other type of higher-dimensional scenario solving the gauge hierarchy problem   
was suggested by Randall and Sundrum (RS) \cite{Gogberashvili:1998vx,Randall:1999ee}. There, the unique extra   
dimension is warped and has a size of order $M_{Pl}^{-1}$ ($M_{Pl}$ being the reduced   
Planck mass: $M_{Pl}=2.44\ 10^{18}\mbox{GeV}$)   
leading to an effective gravity scale around the TeV.  
In the initial version, gravity propagates in the bulk and SM particles are all  
stuck on the TeV-brane.  
An extension of the original RS model was progressively proposed
 \cite{Grossman:1999ra}-\cite{Bajc:1999mh},   
motivated by its interesting features with respect to the gauge coupling unification   
\cite{Pomarol:2000hp}-\cite{Agashe:2002pr} and dark matter problem
\cite{Agashe:2004ci,Agashe:2004bm}. 
This new set-up is  
characterized by the presence of SM fields, except the Higgs boson (to ensure that the gauge  
hierarchy problem does not re-emerge), in the bulk.

In this RS scenario with bulk matter, a displacement of SM fermions along the extra dimension is also possible \cite{Gherghetta:2000qt}:  the effect is that the effective
4-dimensional Yukawa couplings are affected by exponential suppression factors, originating from the wave function overlaps between bulk fermions and Higgs boson
(confined on our TeV-brane). If the fermion localization depends on the flavour and nature of fermions, then the whole structure in flavour space can be generated
by these wave function overlaps. In particular, if the top quark is located closer to the TeV-brane than the up quark, then its overlap with the Higgs boson, and
thus its mass after electroweak symmetry breaking, is larger relatively to the up quark (for identical 5-dimensional Yukawa coupling constants). 

More precisely, the fermions can acquire different localizations if each field $\Psi_i(x_\mu,x_5)$ is coupled to a distinct 5-dimensional mass $m_{i}$: 
$\int d^{4}x \int dx_5 \ \sqrt{G} \ m_{i} \bar{\Psi}_{i} \Psi _{i}$, $G$ being the determinant of the RS metric. To modify the location 
of fermions, the masses $m_{i}$ must have a non-trivial dependence on $x_5$, like $m_{i}= sign(x_5) c_{i} k$,   
where $c_{i}$ are dimensionless parameters and $1/k$ is the curvature radius of Anti-de-Sitter space.
Then the fields decompose as, $\Psi _{i}(x^{\mu },x_5)= \sum_{n=0}^{\infty }\psi_{i}^{(n)}(x^{\mu })f_{n}^{i}(x_5)$
[$n$ labeling the tower of Kaluza-Klein (KK) excitations], admitting the following solution for the zero-mode wave function,
$f_{0}^{i}(x_5)=e^{(2-c_{i})k|x_5|}/N_{0}^{i}$, where $N_{0}^{i}$ is a normalization factor.

The Yukawa interactions with the Higgs boson $H$ read here as,
\begin{equation} 
{\cal S}_{\textrm Yukawa} = \int d^5x \ \sqrt{G} \ \bigg ( Y_{ij}^{(5)} \ H \bar \Psi_{+ i} \Psi_{- j} + h.c. \bigg ) = \int d^4x \ M_{ij} \ \bar \psi_{L i}^{(0)} 
\psi_{R j}^{(0)} + h.c. + \dots  \label{eq:Yuk} 
\end{equation} 
The $Y_{ij}^{(5)}$ are the 5-dimensional Yukawa coupling constants and the dots stand for KK mass terms. The fermion mass matrix is obtained after integrating: 
\begin{equation} 
M_{ij} = \int dx_5 \ \sqrt{G} \ Y_{ij}^{(5)} \ H f_0^i(x_5) f_0^j(x_5).  \label{eq:MassMatrix} 
\end{equation} 
The $Y_{ij}^{(5)}$ can be chosen almost universal so that the quark/lepton mass hierarchies are mainly governed by the overlap mechanism. Large fermion mass
hierarchies can be produced for fundamental mass parameters $m_i$ all of order of the unique scale of the theory $k \sim M_{Pl}$. 

With this mechanism, the quark masses and CKM mixing angles can be effectively  
accommodated \cite{Huber:2000ie,Huber:2003tu,Chang:2005ya}, as well as the lepton masses and PMNS  
mixing angles in both cases where neutrinos acquire Majorana masses (via either  
dimension five operators \cite{Huber:2002gp} or the see-saw mechanism \cite{Huber:2003sf})  
and Dirac masses (see \cite{Huber:2001ug}, and,
\cite{Moreau:2005kz,Moreau:2006np} for order 
unity Yukawa couplings leading to mass hierarchies 
essentially generated by the geometrical mechanism).


\subsubsection{Sources of FCNC in extra dimension scenarios}
GIM-violating FCNC effects in extra dimension scenarios may appear
 both from tree level and from loop effects.

At tree level FCNC processes can be induced by exchanges of KK
excitations of neutral gauge bosons.  The neutral current
action of the effective 4-dimensional coupling, between SM fermions
$\psi_{i}^{(0)}(x^{\mu})$ and KK excitations of any neutral gauge
boson $A_{\mu}^{(n)}(x^{\mu})$, reads in the interaction basis as,
\begin{equation} 
S_{\rm NC}=g_L^{SM}\int d^4x \sum_{n=1}^{\infty}\bar\psi_{Li}^{(0)}\
\gamma^{\mu}\ {\cal C}_{Lij}^{(n)}\ \psi_{Lj}^{(0)}\ A_{\mu}^{(n)}\ +\
\{L\leftrightarrow R\},
\label{Sweak} 
\end{equation} 
Therefore, FCNC interactions can be induced by
the non-universality of the effective coupling constants
$g_{L/R}^{SM} \times C_0^{i \ (n)}$ between KK modes of the gauge
fields and the three SM fermion families (which have different
locations along $x_5$).

At the loop level, KK fermion excitations may invalidate the GIM
cancellation, as discussed e.g. in \cite{Huber:2001ug,Ilakovac:1994kj}  for
$\ell^\pm_{\alpha} \to \ell^\pm_{\beta} \gamma$. Indeed, these
excitations have KK masses which are not negligible (and thus not
quasi-degenerate in family space) compared to $m_{W^{\pm }}$. The GIM
mechanism is also invalidated by the loop contributions of the KK
$W^{\pm (n)}$ modes which couple (KK level by level), e.g. to leptons
in the 4-dimensional theory, via an effective mixing matrix of type
$V_{MNS}^{eff}=U_{L}^{l\dagger} \mathcal{C}_{L}^{(n)} U_{L}^{\nu}$
being non-unitary due to the non-universality of
\begin{equation} 
\mathcal{C}_{L}^{(n)} \equiv diag(C_{m}^{1 \ (n)},C_{m}^{2 \ (n)},C_{m}^{3 \ (n)}).
\label{eq:GeomFact}
\end{equation} 
In this diagonal matrix, $C_{m}^{i \ (n)}$ quantifies the wave
function overlap along the extra dimension between the $W^{\pm (n)}$
[$n \geq 1$] and exchanged (m-th level KK) fermion $f_{m}^{i}(x_5)$
[$i=\{1,2,3\}$ being the generation index] (see later for more
details).

The GIM mechanism for leptons can be clearly restored if the 3
coefficients $C_{m}^{i \ (n)}$ as well as the 3 KK fermion masses
$m_{KK}^{i \ (m)}$ are equal to each other, i.e are universal with
respect to $i=\{1,2,3\}$ (KK level by level)  \cite{Kim:2002kk}.  Within the
quark sector, on the other hand, 
the top quark mass
cannot be totally neglected relatively to the KK up-type quark
excitation scales, leading to a mass shift of the KK
top quark mode from the rest of the KK up-type quark modes and
removing the degeneracy among 3 family masses of the up quark
excitations at fixed KK level (with regard to $m_{W^{\pm
(n)}}$). Moreover, this means that the Yukawa interaction with the
Higgs boson induces a substantial mixing of the top quark KK tower
members among themselves \cite{delAguila:2000aa,Kaplan:2001ga}. 

For example, the data on $b\to s\gamma$ (receiving a contribution from
the exchange of a $W^{\pm (n)}$ [$n=0,1\dots $] gauge field and an up
quark, or its KK excitations, at one loop-level) can be accommodated
in the RS model with $m(W^{\pm (1)}) \simeq 1$ TeV, as shown in
\cite{Kim:2002kk}  using numerical methods for the diagonalization of a large
dimensional mass matrix and taking into account the top quark mass
effects described previously.

\subsubsection{Mass bounds on Kaluza-Klein excitations}
In this subsection we develop constraints on the KK gauge boson 
masses derived from the tree level FCNC effect described above. Our purpose 
is to determine whether these constraints still allow the KK gauge bosons to be  
sufficiently light to imply potentially visible signatures at LHC.

\subsubsubsection{Large extra dimensions} 
 Let us consider the generic framework of a flat extra dimension, with a large size $L$, 
along which gravity as well as gauge bosons propagate. The SM fermions are located at 
different points of the fifth dimension, so that their mass hierarchy can be interpreted 
in term of the geometrical mechanism described in details in Section \ref{subsec:LEDI}. 
In such a framework the exchange of the KK excitations of the gluon can bring  
important contributions to the $K^0-\bar K^0$ mixing ($\Delta F=2$) at tree level. Indeed, the  
KK gluon can couple the d quark with the s quark, if these light down-quarks are displaced along  
the extra dimension. The obtained KK contribution to the mass splitting 
$\Delta m_K$ in the kaon system depends on the KK gluon coupling between the s and d quarks  
(which is fixed by quark locations) and mainly on the mass of the first KK gluon  
$M_{KK}^{(1)}$. Assuming that the s,d quark locations are such that the $m_s$, $m_d$ mass  
values are reproduced, the obtained $\Delta m_K$ and also $| \varepsilon_K |$ are smaller  
than the associated experimental values for, respectively, 
\begin{equation}  
M_{KK}^{(1)} \gtrsim 25 \mbox{TeV}, \ \ \mbox{and} \ \ \ 
M_{KK}^{(1)} \gtrsim 300 \mbox{TeV},  
\label{eq:ADD25TeVlimit}  
\end{equation}    
as found by the authors of \cite{Delgado:1999sv}. The same bound coming from the $D^0$ meson system is weaker. 

In the lepton sector  the experimental upper limit on the branching ratio 
$B(\mu \to e e e)$ imposes typically the constraint \cite{Delgado:1999sv}
\begin{equation}  
M_{KK}^{(1)} \gtrsim 30 \mbox{TeV},  
\label{eq:ADD30TeVlimit}  
\end{equation}    
since the exchange of the KK excitations of the electroweak neutral gauge bosons contributes  
to the decay $\mu \to e e e$. 
 
 To conclude, we stress that if the extra dimensions treat families in a non-universal way 
(which could explain the fermion mass hierarchy), the indirect bounds from FCNC physics like  
the ones in Eq.(\ref{eq:ADD25TeVlimit})-(\ref{eq:ADD30TeVlimit}) force the mass of the KK gauge  
bosons to be far from the collider reach. As a matter of fact, the LHC will be able to probe  
the KK excitations of gauge bosons only up to $6-7$ TeV 
\cite{Rizzo:1999br,Antoniadis:1999bq,Nath:1999mw,Rizzo:1999en} in the present context.

\subsubsubsection{Small extra dimensions} 
 In the context of the RS model with SM fields in the bulk, described in Section  
\ref{sec:SED1}, the exchange of KK excitations of neutral gauge bosons (like e.g.  
the first $Z^0$ excitation: $Z^{(1)}$) also  
contributes to FCNC processes at tree level 
\cite{Gherghetta:2000qt,Huber:2003tu,Burdman:2003nt,Agashe:2004ay,Agashe:2004cp,Agashe:2006iy,Agashe:2006wa}
since these KK states   
possess FC couplings if the different families of fermions are displaced along the  
warped extra dimension.  
 There exist some configurations of fermion locations, 
pointed out in \cite{Moreau:2006np}, which simultaneously reproduce all quark/lepton masses 
and mixing angles via the wave function effects 
{\it and} lead to amplitudes of FCNC reactions [$l_\alpha \to l_\beta l_\gamma l_\gamma$, 
$Z^0 \to l_\alpha l_\beta$, $P^0-\bar P^0$ mixing of a generic meson $P$, $\mu-e$ conversion, 
$K^0 \to l_\alpha l_\beta$ and $K^+ \to \pi^+ \nu \nu$] compatible with the corresponding 
experimental constraints even for light neutral KK gauge bosons:   
\begin{equation}  
M_{KK}^{(1)} \gtrsim 1 \mbox{TeV}.  
\label{eq:RSTeVlimit}  
\end{equation}  
The explanation of this result is the following.  
If the SM fermions with different locations 
are localized typically close to the Planck-brane, they 
have quasi-universal couplings $C_0^{i \ (n)}$ [{\it c.f.} Eq.(\ref{Sweak})] with  
the KK gauge bosons which have a wave function almost constant along the fifth dimension  
near the Planck-brane. Therefore, small FC couplings are generated in the physical basis 
for these fermions  leading to the weak bound (\ref{eq:RSTeVlimit}).  
The fermions from the third family, associated to  
heavy flavours, cannot be localized extremely close to the Planck-brane since their
wave function overlap with the Higgs boson [confined on the TeV-brane] must be large  
in order to generate high effective Yukawa couplings.
Nevertheless, this is compensated by the fact that phenomenological FCNC constraints  
are usually less severe in the third generation sector.  
 
As a result, the order of lower limits on $M_{KK}^{(1)}$   
coming from the considerations on both fermion mass data and FCNC processes can be   
as low as  TeV. From the purely theoretical point of view, the favored order of magnitude for 
$M_{KK}^{(1)}$ is ${\cal O}(1)$TeV which corresponds to a satisfactory solution for  
the gauge hierarchy problem.
From the model building point of view one has to rely on an appropriate 
extension of the RS model insuring  that, for light KK masses, 
the deviations of the electroweak precision observables do not conflict  
with the experimental results.  The existing RS extensions,  
like the scenarios with brane-localized kinetic terms for fermions \cite{delAguila:2003bh}  
and gauge bosons \cite{Carena:2002me} (see \cite{Carena:2002dz,Carena:2003fx} for 
the localized gauge boson kinetic terms   and \cite{Carena:2004zn} for the fermion ones), or the scenarios with an extended gauge symmetry 
(see \cite{Agashe:2003zs}, \cite{Agashe:2006at} and \cite{Djouadi:2006rk} for different fermion charges under this  
broken symmetry), allow $M_{KK}^{(1)}$ to be as low as $\sim 3$ TeV.     
In such a case, one can expect  
a direct detection of the KK excited gauge bosons at LHC.

\subsection{Minimal Flavour Violation in the lepton sector}
\subsubsection{Motivations and basic idea}\label{sec:MFVmfv}
Within the SM
 the dynamics of flavour-changing transitions is controlled by the structure of fermion mass matrices. In the quark sector,
up and down quarks have mass eigenvalues which are up to $10^5$ times smaller than the electroweak scale, and mass matrices which are
approximately aligned. This results in the effective CKM and GIM suppressions of charged and neutral flavour-violating interactions,
respectively. Forcing this connection between the low-energy fermion mass matrices and the flavour-changing 
couplings to be valid also beyond the SM, leads to new-physics scenarios with a high level of predictivity (in the flavour sector) 
and a natural suppression of flavour-changing transitions. The latter achievement is a key ingredient to maintain a 
good agreement with experiments in models where flavoured degrees of freedom are expected around the TeV scale.

This is precisely the idea behind the Minimal Flavour Violation
principle~\cite{Chivukula:1987py,Hall:1990ac,D'Ambrosio:2002ex}. 
It is a fairly general hypothesis that can be implemented in strongly-interacting theories~\cite{Chivukula:1987py}, low-energy
supersymmetry~\cite{Hall:1990ac,D'Ambrosio:2002ex}, multi Higgs~\cite{D'Ambrosio:2002ex,Manohar:2006ga} and
GUT~\cite{Grinstein:2006cg} models. In a model-independent formulation, the MFV construction consists in identifying the
flavour symmetry and symmetry-breaking structure of the SM and enforce it in a more general effective theory
(written in terms of SM fields and valid above the electroweak scale). In the quark sector this procedure is unambiguous:
the largest group of flavour-changing field transformations commuting with the gauge group is 
${\cal G}_q = SU(3)_{Q_L}\times SU(3)_{u_R}\times SU(3)_{d_R}$, and this group is broken only by the Yukawa couplings. 
The invariance of the SM Lagrangian under ${\cal G}_q$ can be formally recovered elevating the Yukawa matrices to
spurion fields with appropriate transformation properties under ${\cal G}_q$. The hypothesis of MFV states that these are the only 
spurions breaking ${\cal G}_q$ also beyond the SM. Within the effective theory formulation, this implies that all the higher dimensional 
operators constructed from SM and Yukawa fields  must be (formally) invariant under ${\cal G}_q$. The consequences of this hypothesis 
in the quark sector have been extensively analyzed in the literature (see e.g.~Refs.~\cite{Buras:2000dm,Bona:2005eu}). 
Without entering into the details, we can state that the MFV hypothesis provides a plausible explanation of why no new-physics effects have been 
observed so far in the quark sector. 

Apart from arguments based on the analogy with quarks, and despite the scarce experimental information, the definition of a Minimal Lepton Flavour 
Violation (MLFV) principle~\cite{Cirigliano:2005ck} is demanded by a severe fine-tuning problem in LFV decays of charged leptons.
Within a generic effective theory approach, the radiative decays $l_i\rightarrow l_j \gamma$ proceed through the following gauge-invariant operator
\begin{equation}  
\frac{\delta^{RL}_{ij}}{\Lambda_{\textrm{LFV}}^2}\, H^\dagger \bar e^i_R \sigma^{\sigma\rho} L^j_L  F_{\sigma\rho},
\end{equation}
where $\delta^{RL}_{ij}$ are the generic flavour-changing couplings and  $\Lambda_\textrm{LFV}$ denotes the cut-off of the  effective theory.  
In absence of a specific flavour structure, it is natural to expect $\delta^{RL}_{ij}=\mathcal{O}(1)$. In this case the experimental limit for
$\mu\rightarrow e\gamma$ implies $\Lambda_{\textrm{LFV}} > 10^5$ TeV, in clear tension with the expectation of new degrees of freedom 
close to the TeV scale in order to stabilize the Higgs sector of the SM.

The implementation of a MFV principle in the lepton sector is not as simple as in the quark sector. The problem is
that the neutrino mass matrix itself cannot be accommodated within the renormalizable part of the SM Lagrangian. 
The most natural way to describe neutrino masses, explaining their strong suppression, is to assume they are Majorana mass 
terms suppressed by the heavy scale of lepton number violation (LNV). In other words, neutrino masses
are described by a non-renormalizable interaction of the type Eq.~(\ref{L1eff})
suppressed by the scale    
$\Lambda_{\rm LNV} \gg v= |\langle H \rangle|$. 
This implies that we have to face a two scale problem
(presumably with the hierarchy $\Lambda_{\textrm{LNV}} \gg \Lambda_{\textrm{LFV}}$) and that we need some additional hypothesis to identify the irreducible 
flavour-symmetry breaking structures. As we will illustrate in the following, we can choose whether to extend or not the field content of the SM. 
The construction of the effective theory based on one of these realizations of the MLFV hypothesis can be viewed as a general tool to exploit
the observable consequences of a specific (minimalistic) hypothesis about the irreducible sources of lepton-flavour  symmetry breaking. 

\subsubsection{MLFV with minimal field content}\label{sec:MFVmfc}
The lepton field content is the SM one: three left-handed doublets $L_L^i$ and three right-handed charged lepton singlets $e_R^i$. The
flavour symmetry group is ${\cal G}_l = SU(3)_{L_L} \times SU(3)_{e_R}$ and we assume the following 
flavour-symmetry-breaking Lagrangian
\begin{eqnarray} 
\mathcal{L}_{\textrm{Sym.Br.}} &=&  - Y_e^{ij} \,\bar e^i_R(H^\dagger L^j_L) -\frac1{2 \Lambda_\textrm{LNV}}\,\kappa_\nu^{ij}(\bar
L^{ci}_L\tau_2 H)(H^T\tau_2L^j_L)+ \textrm{h.c.}\nonumber\\
&\to& - v Y_e^{ij} \,\bar e^i_Re^j_L - \frac{v^2}{2\Lambda_\textrm{LNV}}\,\kappa_\nu^{ij}\,\bar\nu^{ci}_L\nu^j_L+\textrm{h.c.}
\end{eqnarray} 
Here the two irreducible sources of LFV are the coefficient of dimension-five LNV operator ($\kappa_\nu^{ij}$) and the charged-lepton Yukawa coupling ($Y_e$), 
transforming respectively as $(6,1)$ and $(\bar 3, 3)$ under ${\cal G}_l$. An explicit realization of this scenario is provided by the so-called triplet 
see-saw mechanism (or see-saw of type II). This approach has the advantage of being highly predictive, but it differs in an essential way from the 
MFV hypothesis in the quark sector since one of the basic spurion originates from a non-renormalizable coupling. 

\begin{figure}[t]
\begin{center}
\includegraphics[width=7cm]{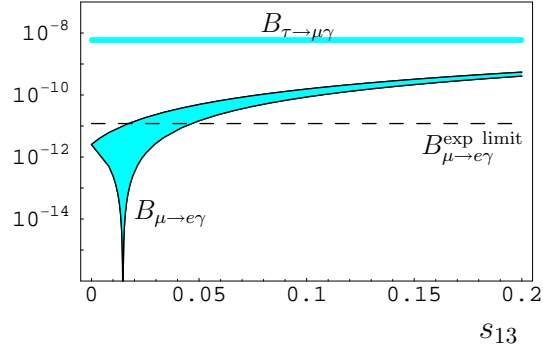}
\caption{$B_{l_i \to l_j \gamma} = \mathcal{B}(l_i \to l_j \gamma)/\mathcal{B}(l_i \to l_j \nu_i 
\bar \nu_j)$ for $\mu  \to  e \gamma$ and  $\tau \to \mu \gamma$ as a function of $\sin\theta_{13}$ in the MLFV 
framework with minimal field content~\cite{Cirigliano:2005ck}. The normalization of the vertical axis corresponds to 
$\Lambda_{\textrm{LNV}}/\Lambda_{\textrm{LFV}} = 10^{10}$. The shading is due to different values of the phase $\delta$ and the 
normal/inverted spectrum. \label{fig:MFVmeg}}
\end{center} 
\end{figure}

Having identified the irreducible sources of flavour symmetry breaking and their transformation properties, we can classify 
the non-renormalizable operators suppressed by inverse powers of $\Lambda_{\textrm{LFV}}$ which contribute to flavour-violating processes. 
These operators must be invariant combinations of SM fields and the spurions $Y_e$ and $\kappa_\nu$. The complete list of the 
leading operators contributing to LFV decays of charged leptons is given in Refs.~\cite{Cirigliano:2005ck,Cirigliano:2006su}. 
The case of the radiative decays $l_i\rightarrow l_j\gamma$ is particularly simple since there are only two 
dimension-six operators (operators with a structure as in Eq.~(\ref{L1eff}), with $F_{\sigma\rho}$ replaced by the
stress tensors of the $U(1)_Y$ and $SU(2)_L$ gauge groups, respectively). The MLFV hypothesis forces the flavour-changing couplings of these 
operators to be a spurion combination transforming as $(\bar 3,3)$ under ${\mathcal G}_l$: 
\begin{equation}
\left(\delta^{RL}_{\rm min} \right)_{ij} ~ \propto ~ \left(Y_e \kappa_\nu^\dagger \kappa_\nu\right)_{ij} +\ldots
\end{equation}
where the dots denote terms with higher powers of $Y_e$ or $\kappa_\nu$. Up to the overall normalization, this combination 
can be completely determined in terms of the neutrino mass eigenvalues and the PMNS matrix. In the basis where $Y_e$ is diagonal we can write, 
\begin{eqnarray}
&& \left(Y_e \kappa_\nu^\dagger \kappa_\nu\right)_{i\not=j} ~=~ \frac{m_{l_i}}{v}\left(\frac{\Lambda^2_{\textrm{LNV}}}{v^4}\,U_{\textrm{PMNS}}
\,m_\nu^2\,U^\dagger_{\textrm{PMNS}}\right)_{i\not=j}\, \nonumber \\
&& \qquad \to \frac{m_{l_i}}{v}\,\frac{\Lambda_{\textrm {LNV}}^2}{v^4} \left[ (U_\textrm{PMNS})_{i2}(U_\textrm{PMNS})_{j2}^* \, 
\Delta m^2_\textrm{sol} \pm (U_\textrm {PMNS})_{i3} (U_\textrm{PMNS})_{j3}^* \,  \Delta m^2_\textrm{atm} \right]~,
\label{eq:spur}
\end{eqnarray}
where $\Delta m^2_{\textrm atm}$ and $\Delta m^2_{\textrm sol}$ denote the squared mass differences deduced from atmospheric- and
solar-neutrino data, and $+/-$ correspond to normal/inverted hierarchy, respectively. The overall factor $\Lambda_{\textrm{LNV}}^2/v^2$ implies 
that the absolute normalization of LFV rates suffers of a large uncertainty. Nonetheless, a few interesting conclusions can still be drawn~\cite{Cirigliano:2005ck}:
\begin{itemize}
\item
The LFV decay rates are proportional to $\Lambda_{\textrm{LNV}}^4/\Lambda_{\textrm{LFV}}^4$ and could be detected only in presence of a large
hierarchy between these two scales. In particular, ${\mathcal B}(\mu \to e \gamma) > 10^{-13}$ only if $\Lambda_{\textrm{LNV}} > 10^{9} \Lambda_{\textrm{LFV}}$.
\item
Ratios of similar LFV decay rates, such as $\mathcal{B}(\mu\to e\gamma)/\mathcal{B}(\tau\to \mu\gamma)$, are free from the normalization ambiguity and can be  
predicted in terms of neutrino masses and PMNS angles: violations of these predictions would unambiguously signal the presence of additional sources of lepton-flavour 
symmetry breaking. One of these prediction is the $10^{-2}$--$10^{-3}$ enhancement of $\mathcal{B}(\tau\to \mu\gamma)$ vs $\mathcal{B}(\mu \to e\gamma)$ 
shown in Fig.~\ref{fig:MFVmeg}. Given the present and near-future experimental prospects on these modes, this modest enhancement implies that the 
$\mu \to e\gamma$ search is much more promising within this framework.
\item
Ratios of LFV transitions among the same two families (such as $\mu \to e \gamma$ vs $\mu \to 3 e$ or $\tau \to \mu \gamma$ vs
$\tau \to 3 \mu$ and $\tau \to \mu e \bar{e}$)  are determined by known phase space factors and ratios of various Wilson coefficients.
As data will become available on different lepton flavor violating processes, if the flavour patter is consistent with the MLFV hypothesis,
from these ratios it will be possible to disentangle the contributions of different operators.
\item  
A definite prediction of the MLFV hypothesis is that the rates for decays involving light hadrons ($\pi^0\rightarrow\mu e$,
$K_L \to \mu e$, $\tau\to\mu\pi^0\ldots$) are exceedingly small.
\end{itemize}

\subsubsection{MLFV with extended field content}\label{sec:MFVefc}
In this scenario we assume three heavy right-handed Majorana neutrinos in addition to the SM fields. As a consequence, 
the maximal flavour group becomes ${\mathcal G}_l \times SU(3)_{\nu_R}$. In order to minimize the number of free parameters (or to maximize the 
predictivity of the model), we assume that the Majorana mass term for the right-handed neutrinos is proportional to the identity
matrix in flavour space: $(M_R)_{ij} = M_R \times \delta_{ij}$. This mass term breaks $SU(3)_{\nu_R}$ to $O(3)_{\nu_R}$ and is assumed 
to be the only source of LNV  ($M_R \leftrightarrow \Lambda_{\textrm{LNV}}$). 

Once the field content of model is extended, there are in principle many alternative options to define the irreducible sources of lepton flavour 
symmetry breaking (see e.g.~Ref.\cite{Davidson:2006bd} for an extensive discussion). However, this specific choice has two important advantages:
it is predictive and closely resemble the MFV hypothesis in the quark sector. The $\nu_R$'s are the counterpart of right-handed up quarks and, similarly to 
the quark sector, the symmetry-breaking sources are two Yukawa couplings of Eq.~(\ref{eq:Lag-typeI}). 
An explicit example of MLFV with extended field content is the Minimal Supersymmetric Standard Model with degenerate right-handed neutrinos.

The classification of the higher-dimensional operators in the
effective theory proceeds as in the minimal field content case. The
only difference is that the basic spurions are now $Y_\nu$ and $Y_e$,
transforming as $(\bar 3, 1, 3)$ and $(\bar 3, 3, 1)$ under ${\mathcal
G}_l \times O(3)_{\nu_R}$, respectively.  The determination of the
spurion structures in terms of observable quantities is more involved
than in the minimal field content case. In general, inverting the
see-saw relation allows us to express $Y_\nu$ in terms of neutrino
masses, PMNS angles and an arbitrary complex-orthogonal matrix $R$ of
Eq.~(\ref{eq:def-R})~\cite{Casas:2001sr}.  Exploiting the
$O(3)_{\nu_R}$ symmetry of the MLFV Lagrangian, the real orthogonal
part of $R$ can be rotated away. We are then left with a
Hermitian-orthogonal matrix $H$~\cite{Pascoli:2003rq} which can be
parameterized in terms of three real parameters ($\phi_i$) which
control the amount of CP-violation in the right-handed sector:
\begin{equation} 
Y_\nu = \frac{M_R^{1/2}}{v}\,H(\phi_i)~ m_{\rm diag}^{1/2} ~U^\dagger_{\textrm{PMNS}}\,.
\end{equation}  
With this parameterization for $Y_\nu$ the flavour changing coupling relevant to $l_i\rightarrow l_j\gamma$ decays reads
\begin{eqnarray}
\label{eq:MVFDRLext}
\delta^{RL}_{\textrm{ext}} ~ \propto ~ Y_e\,\left(Y_\nu^\dagger Y_\nu\right)
 \to  \frac{m_e}{v}\,\left(\frac{M_R}{v^2}U_{\textrm{PMNS}}\, m_{\rm diag}^{1/2} H^2 m_{\textrm diag}^{1/2}U^\dagger_{\textrm{PMNS}}\right).
\end{eqnarray} 
In the CP-conserving limit $H \rightarrow I$ and the phenomenological predictions turns out to be quite similar to the minimal
field content scenario \cite{Cirigliano:2005ck}. In particular, all the general observations listed in the previous section remain valid. 
In the general case, i.e.~for $ H\neq I$, the predictivity of the model is substantially weakened. However, in principle some information about 
the matrix $H$ can be extracted by studying baryogenesis through leptogenesis in the MLFV framework~\cite{Cirigliano:2006nu}.

\subsubsection{Leptogenesis}\label{sec:MFVlep}
On general grounds, we expect that the tree-level degeneracy of heavy
 neutrinos is lifted by radiative corrections. This allows the
 generation of a lepton asymmetry in the interference between
 tree-level and one-loop decays of right-handed neutrinos. Following
 the standard leptogenesis scenario, we assume that this lepton
 asymmetry is later communicated to the baryon sector through
 sphaleron effects and that saturates the observed value of the baryon
 asymmetry of the universe.

\begin{figure}[t]
\begin{center}
\includegraphics[width=6cm,angle=270]{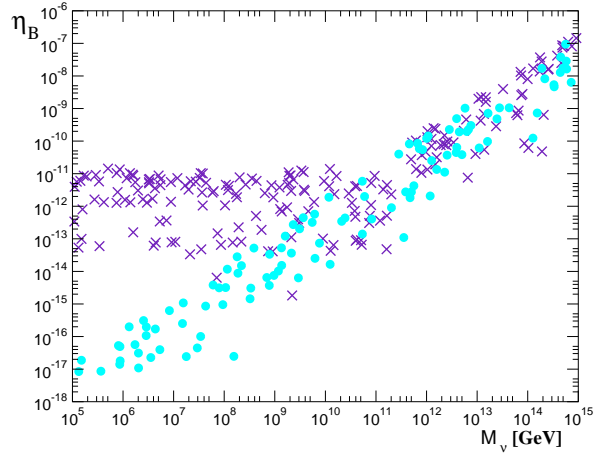}
\caption{Baryon asymmetry ($\eta_B$) as a function of the right-handed neutrino mass scale ($M_R$) for $c_{\nu l} = 0$ (dots) and
$c_{\nu l} \neq 0$ (crosses) in the MLFV framework with extended field content~\cite{Cirigliano:2006nu}.}\label{fig:MFVeta}
\end{center}
\end{figure}

The most general form of the  $\nu_R$ mass-splittings allowed within the MLFV framework has the following form:
\begin{eqnarray} 
&&
\frac{\Delta M_R}{M_R} = c_\nu \left[ Y_\nu Y_\nu^\dagger + (Y_\nu Y_\nu^\dagger)^T  \right] \nonumber + c^{(1)}_{\nu\nu} \left[
Y_\nu Y_\nu^\dagger Y_\nu Y_\nu^\dagger + (Y_\nu Y_\nu^\dagger Y_\nu Y_\nu^\dagger )^T \right] \nonumber  \\
&&\quad  +c^{(2)}_{\nu\nu} \left[ Y_\nu Y_\nu^\dagger (Y_\nu Y_\nu^\dagger)^T \right]
+ c^{(3)}_{\nu\nu}  \left[ (Y_\nu Y_\nu^\dagger)^T Y_\nu Y_\nu^\dagger \right]\nonumber  + c_{\nu l} \left[
Y_\nu Y_e^\dagger Y_e Y_\nu^\dagger + (Y_\nu Y_e^\dagger Y_e Y_\nu^\dagger )^T \right] + \ldots
\end{eqnarray}
Even without specifying the value of the $c_i$, this form allows us to derive a few general conclusions \cite{Cirigliano:2006nu}:
\begin{itemize}
\item
The term proportional to $c_\nu$ does not generate a CPV asymmetry, but sets the scale for the mass splittings: these are of the order of 
magnitude of the decay widths, realizing in a natural way the condition of resonant leptogenesis. 
\item
The right amount of leptogenesis can be generated even with $Y_e = 0$, if all the $\phi_{i}$ are non vanishing. 
However, since $Y_\nu\sim \sqrt{M_R}$, for low values of $M_R$ ($\lesssim 10^{12}$ GeV) the asymmetry generated by the $c_{\nu l}$ term
dominates. In this case $\eta_B$ is typically too small to match the observed value and has a flat dependence on $M_R$. At $M_R \gtrsim 10^{12}$ GeV 
the quadratic terms $c^{(i)}_{\nu\nu}$ dominate, determining an approximate linear growth of  $\eta_B$ with $M_R$. These two regimes 
are illustrated in Fig.~\ref{fig:MFVeta}.
\end{itemize}
As demonstrated in Ref.~\cite{Cirigliano:2006nu}, baryogenesis through leptogenesis is viable in MLFV models.
In particular, assuming a loop hierarchy between the $c_i$ (as expected in a perturbative scenario) and neglecting 
flavour-dependent effects in  the Boltzmann equations (one-flavour approximation of Ref.\cite{Blanchet:2006dq}), 
the right size of $\eta_B$ is naturally reached for $M_R \gtrsim 10^{12}$ GeV. As discussed in Ref.~\cite{Branco:2006hz} (see also \cite{Uhlig:2006xf}), 
this lower bound can be weakened by the inclusion of flavour-dependent effects in the Boltzmann equations and/or by the $\tan\beta$-enhancement of 
$Y_e$ occurring in two-Higgs doublet models.

From the phenomenological point of view, an important difference with respect to the CP-con\-ser\-ving case is the fact that non-vanishing $\phi_i$ change
the predictions of the LFV decays, typically producing an enhancement of the  $\mathcal{B}(\mu\to e\gamma)/\mathcal{B}(\tau\to \mu\gamma)$ ratio. 
For $M_R\gg 10^{12}$ GeV  their effect is moderate and the CP-conserving predictions are recovered. The other important information following 
from the leptogenesis analysis is the fact that the large $M_R$ regime is favored. Assuming $\Lambda_{\textrm{LFV}}$ to be close to the TeV scale,
the $M_R$ regime favored by leptogenesis favors a  $\mu\rightarrow e\gamma$ rate within the reach of the MEG experiment~\cite{Grassi:2005ac}.

\subsubsection{GUT implementation}\label{sec:MFVgut}
Once we accept the idea that flavour dynamics obeys a MFV principle, both in the quark and in the lepton sector, it is
interesting to ask if and how this is compatible with a grand-unified theory (GUT), where quarks and leptons sit in the same
representations of a unified gauge group. This question has recently been addressed in Ref.~\cite{Grinstein:2006cg}, 
considering the exemplifying case of $SU(5)_{\rm gauge}$.

Within $SU(5)_{\rm gauge}$, the down-type singlet quarks ($d^c_{iR}$) and the lepton doublets 
($L_{iL}$) belong to the $\bar {\bf 5}$ representation; the quark doublet ($Q_{iL}$), the up-type ($u^c_{iR}$) and lepton singlets ($e_{iR}^c$) 
belong to the ${\bf 10}$ representation, and finally the right-handed neutrinos ($\nu_{iR}$) are singlet. In this framework the largest 
group of flavour transformation commuting with the gauge group is ${\mathcal G}_{GUT} = SU(3)_{\bar 5} \times SU(3)_{10}\times SU(3)_1$, 
which is smaller than the direct product of the quark and lepton groups discussed before (${\mathcal G}_q \times {\mathcal G}_l$).  
We should therefore expect some violations of the MFV+MLFV predictions either in the quark or in the lepton sector or in both. 

A phenomenologically acceptable description of the low-energy fermion mass matrices requires the introduction of at least four irreducible 
sources of ${\mathcal G}_{GUT}$ breaking. From this point of view the situation is apparently similar to the non-unified case: the four 
 ${\mathcal G}_{GUT}$ spurions can be put in one-to-one correspondence with the low-energy spurions $Y_u$,$Y_d$, 
$Y_e$, and $Y_\nu$. However, the smaller flavour group does not allow the diagonalization of $Y_d$ and
$Y_e$ (which transform in the same way under ${\mathcal G}_{GUT}$) in the same basis. As a result, two additional mixing matrices 
can appear in the expressions for flavour changing rates: $C = V_{e_{R}}^T V_{d_{L}}$ and $G = V_{e_{L}}^T V_{d_{R}}$. 
The hierarchical texture of the new mixing matrices is known since they reduce to the identity matrix in the limit 
$Y_e^T = Y_d$. Taking into account this fact, and analyzing the structure of the allowed higher-dimensional operators, 
a number of reasonably  firm phenomenological consequences can be deduced \cite{Grinstein:2006cg}: 
\begin{itemize}
\item 
There is a well defined limit in which the standard MFV scenario for the quark  sector is  fully recovered: $M_R \ll 10^{12}$ GeV
and small $\tan \beta$ (in a two-Higgs doublet case). For $M_R \sim  10^{12}$ GeV and small $\tan \beta$, deviations from the standard MFV pattern 
can be expected in rare $K$ decays but  not in $B$ physics. Ignoring fine-tuned scenarios, $M_R \gg  10^{12}$~GeV is excluded by the present constraints 
on quark FCNC transitions. Independently from the value of $M_R$, deviations from the standard MFV pattern can appear both in $K$ and in $B$ physics
for $\tan\beta \gtrsim m_t/m_b$. 
\item 
Contrary to the non-GUT MFV framework, the rate for $\mu \to e \gamma$ (and other LFV decays) cannot be 
arbitrarily suppressed by lowering the  average mass $M_R$ of the heavy  $\nu_R$. This fact can easily be understood by looking at the flavour structure 
of the relevant effective couplings, which now assume the following form:
\begin{equation}
\label{eq:MFVgut}
\delta^\textrm{RL}_\textrm{GUT} = ~ c_{1}~Y_e Y_\nu^\dagger Y_\nu ~+~c_{2}~Y_u Y_u^\dagger Y_e ~+~c_{3}~Y_u
Y_u^\dagger Y_d^T\, +~\ldots
\end{equation}
In addition to the terms involving $Y_\nu\sim \sqrt{M_R}$ already present in the non-unified case, the GUT group allows also $M_R$-independent 
terms involving the quark Yukawa couplings. The latter become competitive for $M_R \lesssim 10^{12}$ GeV and their contribution is such that for 
$\Lambda_{\rm LFV} \lesssim 10$ TeV  the  $\mu \to e \gamma$ rate is above $10^{-13}$ (i.e.~within the reach of  MEG~\cite{Grassi:2005ac}).
\item 
Improved experimental information on $\tau \to \mu \gamma$ and $\tau \to e \gamma$ would be a powerful  tool in discriminating  the relative size of the 
standard  MFV contributions versus  the characteristic GUT-MFV contributions due to the different hierarchy 
pattern among $\tau \to \mu$, $\tau \to e$, and $\mu \to e$ transitions. 
\end{itemize}

\section{Phenomenology of theories beyond the Standard Model}
\label{sec:phenomenology}

\subsection{Flavour violation in non-SUSY models directly testable at LHC}\label{sec:nonSUSY}

\subsubsection{Multi-Higgs doublet models}

The arbitrariness of quark masses, mixing and CP-violation
in the Standard Model stems from the fact that
gauge invariance does not constrain the flavour structure
of Yukawa interactions. In the SM neutrinos are strictly
massless. No neutrino Dirac mass term can be introduced, due
to the absence of right-handed neutrinos and no Majorana mass terms can be
generated, due to exact B-L conservation. Since
neutrinos are massless, there is no leptonic mixing in the SM,
which in turn leads to separate lepton number conservation.
Therefore, the recent observation of neutrino oscillations is
evidence for physics beyond the SM. Fermion masses, mixing
and CP-violation are closely related to each other and also to
the Higgs sector of the theory.

It has been shown that gauge theories with fermions, but without
scalar fields, do not break CP symmetry \cite{Grimus:1995zi}. A
scalar (Higgs) doublet is used in the SM to break both the gauge
symmetry and generate gauge boson masses as well as fermion masses
through Yukawa  interactions. This is known as the Higgs mechanism,
which was proposed by several authors \cite{Englert:1964et},
\cite{Guralnik:1964eu}, \cite{Higgs:1964ia,Higgs:1966ev}. It
predicts the existence of one neutral scalar Higgs particle - the Higgs
boson. In the SM where a single Higgs doublet is introduced, it is
not possible to have spontaneous CP-violation since any phase in the
vacuum expectation value
can be eliminated by rephasing the Higgs field. Furthermore, in the
SM it is also not possible to violate CP explicitly in the Higgs
sector since gauge invariance together with renormalizability
restrict the Higgs potential to have only quadratic and quartic
terms and hermiticity constrains both of these to be real. Thus, CP
violation in the SM requires the introduction of complex Yukawa
couplings.

The scenario of spontaneous CP and T violation
has the nice feature of putting the breakdown of discrete symmetries
on the same footing as the breaking of the gauge symmetry, which is
also spontaneous in order to preserve renormalizability.
A simple extension of the Higgs sector that may give rise to spontaneous
CP-violation requires the presence of at least two Higgs doublets,
and was introduced by Lee \cite{Lee:1973iz}.

If one introduces two Higgs doublets, it is possible to
have either explicit or spontaneous CP breaking.
Explicit CP-violation in the Higgs sector
arises due to the fact that in this case there are gauge invariant terms
in the Lagrangian which can have complex coefficients. Note however that
the presence of complex coefficients does not always lead to
explicit CP breaking.

Extensions of the SM with extra Higgs doublets are very natural
since they keep the $\rho$ parameter at tree level equal to one
\cite{Bernreuther:1998rx}. In multi-Higgs systems there are in
general, additional sources of CP-violation in the Higgs sector
\cite{Branco:1999fs}. The most general renormalizable polynomial
consistent with  the $SU(2)\times U(1)\times SU(3)_{c}$ model with
$n_{d}$ Higgs doublets, $\phi_i $, may be written as:
\begin{equation}
{\mathcal{L}}_{\phi }=Y_{ab}\ \phi _{a}^{\dagger }\phi _{b}+Z_{abcd}\
\left( \phi _{a}^{\dagger }\phi _{b}\right) \left( \phi _{c}^{\dagger }\phi
_{d}\right),  \label{laghiggs}
\end{equation}
where repeated indices are summed. Hermiticity of
${\mathcal{L}}_{\phi }$ implies:
\begin{equation}
\begin{array}{ccc}
Y_{ab}^{\ast }=Y_{ba} & \quad ;\quad & Z_{abcd}^{\ast }=Z_{badc} \ .
\end{array}
\label{hermzy}
\end{equation}
Furthermore, by construction it is obvious that:
\begin{equation}
Z_{abcd}=Z_{cdab} \ .  \label{symz}
\end{equation}

In models with more than one Higgs doublet, one has the freedom to make
Higgs-basis transformations (HBT) that do not change the physical
content of the model, but do change both the quadratic and the quartic
coefficients. Coefficients that are complex in one Higgs basis
may become real in another basis.
Furthermore, a given model may have complex quartic
coefficients in one Higgs basis, while they may all become real
in another basis, with only the quadratic coefficients
now complex, thus indicating that in that particular model CP is only softly
broken. Such Higgs-basis transformations
leave the Higgs kinetic energy term invariant and are of the form:
\begin{equation}
\begin{array}{ccc}
\phi _{a}\stackrel{\mathrm{HBT}}{\longrightarrow }
\phi _{a}^{\prime }=V_{ai}\ \phi _{i}
& \quad ,\quad & \phi _{a}^{\dagger }
\stackrel{\mathrm{HBT}}{\longrightarrow }\left(
\phi ^{\prime }\right) _{a}^{\dagger }=V_{ai}^{\ast} \
\left( \phi ^{\prime }\right)
_{i}^{\dagger }\ ,
\end{array}
\label{trafowb}
\end{equation}
where $V$ is an $n_{d}\times n_{d}$ unitary matrix acting in the
space of Higgs doublets. In  \cite{Branco:2005em} conditions  for a
given Higgs potential to violate CP at the Lagrangian level,
expressed in terms of CP-odd Higgs-basis invariants, were derived.
These conditions are expressed in terms of couplings of the unbroken
Lagrangian, therefore they are relevant even at high energies, where
the $SU(2)\times U(1)$ symmetry is restored. This feature renders
them potentially useful for the study of baryogenesis. The
derivation of these conditions follows the general method proposed
in \cite{Bernabeu:1986fc} and already mentioned in previous sections.
The method consists of imposing invariance of the Lagrangian under
the most general CP transformation of the Higgs doublets, which is a
combination of a simple CP transformation for each Higgs field with
a Higgs-basis transformation:
\begin{equation}
\begin{array}{ccc}
\phi _{a}\stackrel{\mathrm{CP}}{\longrightarrow }
W_{ai}\ \phi _{i}^{\ast } & \quad
;\quad & \phi _{a}^{\dagger }\stackrel{\mathrm{CP}}{\longrightarrow }
W_{ai}^{\ast }\
\phi _{i}^{T}
\end{array}
\label{trafocp}
\end{equation}
here $W$ is an $n_{d}\times n_{d}$ unitary matrix operating in Higgs
doublets space.

A set of necessary and
sufficient conditions for CP invariance in the case of two Higgs doublets
have been derived  \cite{Branco:2005em}:
\begin{equation}
\begin{array}{l}
I_{1}\equiv \mathrm{Tr}[Y\ Z_{Y}\ \widehat{Z}-\widehat{Z}\ Z_{Y}\ Y]=0 \\
\\
I_{2}\equiv \mathrm{Tr}[Y\ Z_{2}\ \widetilde{Z}-\widetilde{Z}\ Z_{2}\ Y]=0 \ ,
\end{array}
\label{inv}
\end{equation}
where all matrices inside the parenthesis are
$2 \times 2$ matrices. In the general case these are $n_d \times n_d$
matrices, and are defined by:
\begin{equation}
\left( Z_{Y}\right) _{ij}\equiv Z_{ijmn}Y_{mn}; \qquad
\widehat{Z}_{ij}\equiv Z_{ijmm}; \qquad
\left( Z_{2}\right) _{ij}\equiv Z_{ipnm}Z_{mnpj}; \qquad
\widetilde{Z}_{ij}\equiv Z_{immj}
\label{ndnd}
\end{equation}
CP-odd HBT invariants are also useful \cite{Branco:2005em} to find
out whether, in a given model, there is hard or soft CP breaking.
One may also construct CP-odd weak basis invariants, involving
$v_i \equiv <0|\phi^0_i |0>$, i.e., after spontaneous
gauge symmetry breaking has occurred \cite{Lavoura:1994fv},
\cite{Botella:1994cs}.
Further discussions on  Higgs-basis independent methods for the
two-Higgs-doublet model can be found in \cite{Davidson:2005cw},
\cite{Gunion:2005ja}, \cite{Ivanov:2005hg}, \cite{Haber:2006ue}.

So far, we have considered CP-violation at the Lagrangian level in
models with  multi-Higgs doublets, i.e., explicit CP-violation. It
is also possible to derive criteria  \cite{Branco:1983tn} to verify
whether CP and T in a given model are spontaneously broken. Under T
the Higgs fields $\phi _{j}$ transform as
\begin{equation}
T \ \phi _{j} \ T^{-1} = U_{jk} \phi _{k},
\label{c1}
\end{equation}
where $U$ is a unitary matrix which may mix the scalar doublets. If
no extra symmetries beyond $SU(2) \times U(1)$ are
present in the Lagrangian, $U$ reduces to a diagonal matrix
possibly with phases.
Invariance of the vacuum under T  leads
to the following condition:
\begin{equation}
<0|\phi^0_j |0> =  U_{jk}^\ast <0|\phi^0_k |0>^\ast .
\label{c2}
\end{equation}
Therefore, a set of vacua lead to spontaneous T, CP-violation
if there is no unitary matrix $U$ satisfying Eqs.(\ref{c1}) and
(\ref{c2}) simultaneously. \\

Most of the previous discussion dealt with the general case of n-Higgs
doublets. We analyze now the case of two Higgs doublets, where the
most general gauge
invariant Higgs potential can be explicitly written as:
\begin{equation}
\begin{array}{ll}
V_{H_{2}}=&m_{1}\ \phi _{1}^{\dagger }\phi _{1}+p\ e^{i\varphi }\ \phi _{1}^{\dagger }\phi _{2}+p\ e^{-i\varphi }\
 \ \phi _{2}^{\dagger }\phi _{1}+m_{2}\ \phi _{2}^{\dagger }\phi _{2}+ \\
& +a_{1}\ \left( \phi _{1}^{\dagger }\phi _{1}\right) ^{2}+a_{2}\ \left( \phi_{2}^{\dagger }\phi _{2}\right) ^{2}+b\ \left( \phi _{1}^{\dagger }\phi
_{1}\right) \left( \phi _{2}^{\dagger }\phi _{2}\right) +b^{\prime }\ \left(
\phi _{1}^{\dagger }\phi _{2}\right) \left( \phi _{2}^{\dagger }\phi
_{1}\right) + \\
&+c_{1}\ e^{i\theta _{1}}\ \left( \phi _{1}^{\dagger }\phi _{1}\right) \left(
\phi _{2}^{\dagger }\phi _{1}\right) +c_{1}\ e^{-i\theta _{1}}\ \left( \phi
_{1}^{\dagger }\phi _{1}\right) \left( \phi _{1}^{\dagger }\phi _{2}\right)
+c_{2}\ e^{i\theta _{2}}\ \left( \phi _{2}^{\dagger }\phi _{2}\right) \left(
\phi _{2}^{\dagger }\phi _{1}\right) + \\
&+ c_{2}\ e^{-i\theta _{2}}\ \left( \phi
_{2}^{\dagger }\phi _{2}\right) \left( \phi _{1}^{\dagger }\phi _{2}\right)
+d\ e^{i\delta }\ \left( \phi _{1}^{\dagger }\phi _{2}\right) ^{2}+
d\ e^{-i\delta }\ \left( \phi _{2}^{\dagger }\phi _{1}\right) ^{2},
\end{array}
\label{higgs2}
\end{equation}
where $m_i$, $p$, $a_i$ , $b$, $b^{\prime}$, $c_i$, and $d$ 
are real and all
phases are explicitly displayed. It is clear that this
potential contains an excess of parameters. With the appropriate
choice of Higgs basis some of these may be eliminated, without loss
of generality, leaving eleven independent parameters
\cite{Lavoura:1994yu,Ginzburg:2004vp,Gunion:2002zf}. The Higgs
sector contains  five spinless particles: three neutral and a pair
of charged ones, usually denoted by $h,H$ (CP-even), $A$  (CP-odd)
(or if CP is violated $h_{1,2,3}$) and $H^\pm$.

In general, models with two Higgs doublets have tree level
Higgs-mediated flavour changing neutral currents (FCNC).
This is a problem in
view of the present stringent experimental limits on FCNC.
In order to solve this problem the concept of natural flavour
conservation (NFC) was introduced by imposing extra symmetries on the
Lagrangian. These symmetries constrain the Yukawa couplings of the neutral
scalars in such a way that the resulting neutral currents are diagonal.
Glashow and Weinberg
\cite{Glashow:1976nt} and Paschos \cite{Paschos:1976ay} have
shown that the only way to achieve NFC is to ensure that only one
Higgs doublet gives mass to quarks of a given charge.

In the case of two Higgs doublets the simplest solution to avoid FCNC is
to require invariance of the Lagrangian under the following
transformation of the $Z_2$ type:
\begin{equation}
\phi _{1} \longrightarrow \phi _{1} \qquad \phi _{2} \longrightarrow
- \phi _{2} \qquad d_R \longrightarrow d_R \qquad  u_R
\longrightarrow -u_R, \label{des}
\end{equation}
where $d_R$ ($u_R$) denote the right-handed down (up) quarks; all
other fields remain unchanged.

It is clear from Eq.~(\ref{higgs2})  that this symmetry eliminates explicit
CP-violation in the Higgs sector, since the only term of the
Higgs potential with a phase that survives
is the one with coefficient $d$, moreover a HBT of the form
$\phi _{1} \longrightarrow e^{i \delta /2} \phi _{1}$,
$\phi _{2}\longrightarrow \phi _{2}$, eliminates the phase from
the Higgs potential.
Furthermore, it can  be shown that
this symmetry also eliminates the possibility of having  spontaneous CP
violation.

In conclusion, models with two Higgs doublets and exact NFC cannot
give rise to spontaneous CP-violation. Explicit CP-violation in this
case requires complex Yukawa couplings leading to the
Kobayashi-Maskawa mechanism with no additional source of CP
violation through neutral scalar Higgs boson exchange. An
interesting alternative scenario in the case of two Higgs doublets
was considered in \cite{Branco:1985aq} with no NFC. Here CP
violating Higgs FCNC are naturally suppressed through a permutation
symmetry which is softly broken, still allowing for spontaneous CP
violation.

Three Higgs doublet  models have been considered in an attempt to
introduce CP-violation in an extension of the SM with NFC
\cite{Glashow:1976nt} in the Higgs sector. It was shown that indeed,
in such models  it is possible to violate CP in the Higgs sector
either at the Lagrangian level \cite{Weinberg:1976hu} or
spontaneously \cite{Branco:1979pv,Branco:1980sz,Branco:1985pf}.

It is also possible to generate spontaneous CP-violation with only
one additional Higgs singlet \cite{Bento:1991ez}, but in this case
at least one isosinglet vectorial quark is required in order to
generate a non trivial phase at low energies in the
Cabibbo-Kobayashi-Maskawa matrix. Such models may provide a solution
to the strong CP problem of the type proposed by Nelson
\cite{Nelson:1983zb}, \cite{Nelson:1984hg} and Barr
\cite{Barr:1984qx} as well as a common origin to all CP-violations
\cite{Branco:2003rt} , \cite{Achiman:2004qf} including the
generation of the observed baryon asymmetry of the Universe. The fact 
that the SM cannot provide the observed baryon
asymmetry \cite{Gavela:1994dt},
\cite{Huet:1994jb}, \cite{Anderson:1991zb}, \cite{Buchmuller:1993bq},
\cite{Kajantie:1995kf}, \cite{Fromme:2006cm}, provides yet another
reason to study an enlarged Higgs sector.

\vskip 0.5cm

A lot of work has been done by many authors
on possible extensions of the Higgs sector and their implications
both for the hadronic and the leptonic sectors at the existing and
future colliders, see e.g. \cite{Accomando:2006ga}. Among  the
simplest multi-Higgs  models are the two Higgs Doublet Models 
(2HDM) which have been analyzed in detail in many different 
realizations. 
The need to avoid potentially dangerous tree level Higgs
FCNC has led to the consideration of different variants of this model
with  a certain discrete $Z_2$ symmetry  imposed.

In the Type-I 2HDM the $Z_2$ discrete symmetry imposed on the
Lagrangian is such that only one of the Higgs doublets couples to
quarks and leptons. A very well known fermiophobic Higgs boson may
arise in such model
\cite{Brucher:1999tx,Akeroyd:2003xi,Delaere:2005jj} . Another
example is the Inert Doublet Model, with an unbroken discrete $Z_2$
symmetry which forbids one Higgs doublet to couple to fermions and
to get a non-zero VEV \cite{Deshpande:1977rw,Cao:2007rm}. Physical
particles related to such doublets are called  "inert" particles,
the lightest is stable and contributes to the Dark Matter
density. In \cite{Barbieri:2006dq}, the
naturalness problem has been addressed in the framework of an 
Inert Doublet Model with a heavy (SM-like) Higgs boson.
In this context Dark Matter may be composed of neutral
inert Higgs bosons. Predictions are given 
for multilepton events with missing transverse energy at
the LHC, and for the direct detection of dark matter.

The Type-II 2HDM allows one of the Higgs doublet to couple only to
the rigthanded up quarks while the other Higgs doublet can only
couple to right-handed down-type quarks and charged leptons. This is
achieved by the introduction of an appropriate  $Z_2$ symmetry,
analogous to the one in  Eq.~(\ref{des}). The Higgs sector of the
MSSM model can be viewed as a particular realization of Type-II
models but with additional constraints required by supersymmetry.
Various scenarios are possible for these models - with and without
decoupling of heavy Higgs particles
\cite{Gunion:2002zf,Ginzburg:2004vp,Ginzburg:2001wj}.

Type-III 2HDM are models where, unlike in models of Type-I and II,
NFC is not imposed on the Yukawa interactions.  This class of models
has in general scalar mediated FCNC at tree level. Various schemes
have been proposed to suppress these currents, including the ad-hoc
assumption that FCNC couplings are approximately given by the
geometric mean of the Yukawa couplings of the two generations
\cite{Cheng:1987rs}. A very interesting alternative
\cite{Branco:1996bq} is to have an exact symmetry of the Lagrangian
which constrains FCNC couplings to be related in an exact way to the
elements of the CKM matrix in such a way that FCNC are non-vanishing
but naturally suppressed by the smallness of CKM mixing. Another
example of Type III 2HDM is the Top Two Higgs Doublet Model which
was first proposed in Ref.~\cite{Das:1995df}, and recently analyzed
in detail in Ref.~\cite{Lunghi:2007ak}. In this framework a discrete
symmetry is imposed allowing only the top quark to have Yukawa
couplings to one of the doublets while all other quarks and leptons
have Yukawa couplings to the other doublet.

Lepton flavour violation is a feature common to many possible
extensions of the SM. It can occur both through charged and neutral
currents. The possibility of having lepton flavour violation
in extensions of the SM, has been considered long before the
discovery of neutrino masses \cite{Cheng:1976uq}. For example,
in the case of multi-Higgs doublet models, it has been pointed
out that even for massless neutrinos  lepton flavour can be
violated
\cite{Bjorken:1977vt}, \cite{Branco:1977rt}.
In the context of the minimal extension of
the SM, necessary to accommodate neutrino masses, where only
right-handed neutrinos are included LFV effects are extremely small.
It is well known that the effects of LFV can be large in
supersymmetry.

CLEO submitted recently a paper \cite{Besson:2006gj}
where the ratio of the tauonic
and muonic branching fractions is examined for the three
$\Upsilon(1S, 2S, 3S)$ states. Agreement with expectations
from lepton universality is found. The conclusion is that lepton
universality is respected within the current experimental accuracy
which is roughly $10 \% $. However there is tendency
for the tauonic branching fraction to turn out systematically larger than
the muonic at a few per cent level.

\subsubsection{Low scale singlet neutrino scenarios}

In the pre-LHC era neutrino oscillations have provided some of the
most robust evidence for physics beyond the SM. Many open questions
still remain in this field; why is the absolute mass scale for the
neutrinos so small with respect to the other SM particles?  what is
this mass scale? why is the pattern of mixing so different from the
quark sector?  If nature has chosen the singlet seesaw scenario
\cite{Minkowski:1977sc,Yanagida:1979as,Gell-Mann:1980vs,Glashow:1979nm,
Mohapatra:1979ia} as an answer to those questions we face the prospect
of never being able to produce the heavy neutrinos at a collider.
Nevertheless, several extensions of this minimal see-saw scenario
contain heavy neutrinos at or around the TeV scale, these include
models based around the group $E_6$ \cite{Mohapatra:1986bd,Nandi:1985uh} 
and also in SO(10)
models \cite{Witten:1979nr}.

Furthermore, even within the usual see-saw scenario, the observed
nearly maximal mixing pattern of the light neutrinos requires further
explanation.  Flavour symmetries are often invoked as possible reasons
for the almost tri-bi-maximal structure of the PMNS mixing matrix
\cite{Harrison:1999cf}. It is also possible that the small magnitude
of the light neutrino masses is due to an approximate symmetry,
allowing the right-handed neutrinos to be as light as
$\mathcal{O}$(200~GeV) \cite{Pilaftsis:2005rv}.

TeV scale right-handed neutrinos can also arise in radiative
mechanisms of neutrino mass generation. Generically, in these models a
tree-level neutrino mass is forbidden or suppressed by a symmetry but
small neutrino masses may arise through loops sensitive to symmetry
breaking effects \cite{Ma:1998dn,Pilaftsis:1991ug}.  Indeed, several supersymmetric
realizations of radiative mechanisms contain TeV scale right-handed
neutrinos linked to the scale of supersymmetry breaking
\cite{Arkani-Hamed:2000bq,Borzumati:2000mc}.

\subsubsubsection{Heavy neutrinos accessible to the LHC}

A low, electroweak-scale mass is not sufficient to imply that heavy
neutrinos could be produced and detected at the LHC. They must have a
large enough coupling (mixing) with other SM fields so that
experiments will be able to distinguish their production and decay
from SM background processes.  In this review we concentrate on the
case where heavy neutrino production and decay occurs through mixing
with SM fields only.  Quantitatively, we can consider a generalization
of the Langacker-London parameters, $\Omega_{ll'}$, defined as
\begin{equation}
\Omega_{ll'} = \delta_{ll'} - \sum^3_{i=1} B_{li} B^{*}_{l'i} = \sum^{(3 + n_R)}_{i=4} B_{li} B^{*}_{l'i}\,\,,
\end{equation}
where $l,l'=e,\mu,\tau$ and $B_{li}$ is the full $3 \times (3+n_R)$ neutrino mixing matrix taking into account all (3 light and $n_R$ heavy) neutrinos. The
3$\times$3 matrix $B_{li}$ where $i=1\ldots3$ is a good approximation to the usual PMNS matrix and $\Omega_{ll'}$ essentially measures the deviation from
unitarity of the PMNS matrix.

The $\Omega_{ll'}$ are constrained by precision electroweak data \cite{Bergmann:1998rg} and the following upper limits have been set at 90\% C.L.
\begin{equation}
\Omega_{ee} \le 0.012\,,\qquad \Omega_{\mu\mu} \le 0.0096\,,\qquad \Omega_{\tau\tau} \le 0.016\,.
\end{equation}
In addition, the off-diagonal elements of $\Omega_{ll'}$ are constrained by limits on lepton flavour violating processes such as $\tau,\mu \to e \gamma$
and $\tau,\mu \to eee$ and $\mu \to e$ conversion in nuclei \cite{Ilakovac:1994kj,Ioannisian:1999cw}. 
These limits are rather model dependent but for $M_R \gg M_W$ and $m_D \ll M_W$ (where $m_D$ is
the Dirac component of the neutrino mass matrix), the present upper bounds are \cite{Aubert:2005wa}
\begin{equation}
|\Omega_{e\mu}| \le 0.0001\,,\qquad |\Omega_{e\tau}| \le 0.02\,,\qquad |\Omega_{\mu\tau}| \le 0.02\,.
\label{const}
\end{equation}

It has been pointed out that a heavy Majorana neutrino ($N$) may be produced via a DY type of mechanism at hadron colliders
\cite{Dicus:1991fk,Pilaftsis:1991ug,Datta:1993nm,Almeida:2000pz,Panella:2001wq,Han:2006ip}, $pp \to W^{+*} \to \ell^+ N$, where $N\to \ell^+ W^-$, leading to
lepton number violation by 2. Most of the previous studies were concentrated on the $ee$ mode, which would result in a too week signal to be  appreciable due to
the recent very stringent bound $|V_{eN}|^2/m_N<5\times 10^{-8}$ GeV$^{-1}$, from the absence of the neutrinoless double beta decay. It has been
recently proposed to search for the unique and clean signal, $\mu^\pm\mu^\pm+$2 jets at the LHC \cite{Han:2006ip}. It was concluded
that a search at the LHC with an integrated luminosity of 100 fb$^{-1}$ can be sensitive to a mass range of $m_N\sim 10-400$ GeV at a 2$\sigma$
level, and up to 250 GeV at a 5$\sigma$ level. If this type of signal could be established, it would be even
feasible to consider the search for CP-violation in the heavy Majorana sector \cite{Bray:2007ru}.

A recent analysis \cite{Aguila:2007em} studied more background processes including some fast detector simulations. In particular,
the authors claimed a large background due to the faked leptons $b\bar b\to \mu^+\mu^+$. The search sensitivity is thus reduced to
175 GeV at a 5$\sigma$ level. However, the background estimate for processes such as $b\bar b+$n-jet has large uncertainties due to
QCD perturbative calculations and kinematical acceptance. More studies remain to be done for a definitive conclusion. 

\subsubsubsection{Low scale model with successful baryogenesis}

As a more detailed example satisfying the constraints of Eq.(\ref{const}) we consider a model potentially accessible to colliders, where $M_R \simeq 250$~GeV which
has been shown to successfully explain the baryon asymmetry of the Universe \cite{Pilaftsis:2005rv}.

Leptogenesis has been discussed in Section~\ref{sec:lepto}. Low scale
leptogenesis scenario would be possible with nearly degenerate heavy neutrinos, where self-energy effects on the leptonic asymmetries become
relevant \cite{Flanz:1994yx,Covi:1996wh}. In this case the CP asymmetry in the heavy neutrino decays can be resonantly enhanced \cite{Pilaftsis:1997jf}, to the extent that the observed
baryon asymmetry can be explained with heavy neutrinos as light as the electroweak scale \cite{Pilaftsis:2003gt,Pilaftsis:2005rv}.

We will consider a model with right-handed neutrinos which transform under an SO(3) flavour symmetry. Ignoring effects from the neutrino Yukawa couplings
this symmetry is assumed to be exact at some high scale, e.g. the GUT scale $M_{\rm GUT}$. This restricts the form of the heavy Majorana neutrino mass
matrix at $M_{\rm GUT}$
\begin{equation}
M_R = \mathbf{1}\,m_N + \delta M_S\,,
\end{equation}
where $\delta M_S=0$ at $M_{\rm GUT}$. This form has also been considered in a class of ``minimal flavour violating''  models of the lepton sector
\cite{Cirigliano:2005ck} 
and naturally provides nearly degenerate heavy neutrinos compatible with resonant leptogenesis. 

All other fields are singlets under this SO(3) flavour symmetry and so the neutrino Yukawa couplings will break SO(3) explicitly. 
We can still choose heavy neutrino Yukawa couplings 
$Y^{\nu}$ so that a subgroup of the SO(3)$\times$U(1)$_{L_e}\times$ 
U(1)$_{L_\mu}\times$U(1)$_{L_\tau}$ flavour
symmetry present without the neutrino Yukawa couplings remains unbroken. In this case a particular flavour direction can be singled out leaving
SO(2)$\,\simeq\,$U(1) unbroken. This residual U(1) symmetry acts to prevent the light Majorana neutrinos from acquiring a mass.  The form of the neutrino
Yukawa couplings can be written
\begin{equation}
\label{hmatrix}
Y^{\nu T} \ =\ \left(\! \begin{array}{ccc}   0  & a\, e^{-i\pi/4}  & a\, e^{i\pi/4} \\
   0  & b\, e^{-i\pi/4}  & b\, e^{i\pi/4} \\
   0  & c\, e^{-i\pi/4}  & c\, e^{i\pi/4} \end{array} \!\right)\ +\ \delta Y^{\nu}\;.
\end{equation}
The residual U(1) symmetry is broken both by small SO(3) breaking effects in the heavy Majorana mass matrix, $\delta M_S$, and by small effects
parameterized by $\delta Y^{\nu}$ in the Yukawa couplings. Although we will not consider the specific origin of these effects, $\delta M_S$ could arise
through renormalization group running for example.

In \cite{Pilaftsis:2005rv}, a specific model was considered where $m_N = 250$~GeV and which successfully explained the baryon asymmetry of the Universe. One of
either $a$, $b$ or $c$ was constrained to be small to allow a single lepton flavour asymmetry (and subsequently a baryon asymmetry) to be generated at
$T\sim 250$~GeV. The other two parameters could be as large as ${\cal O}(10^{-2})$. This scenario has the features necessary for a model to be
visible at the LHC; heavy neutrinos with masses around ${\cal O}$(1~TeV) and sufficient mixing between these neutrinos and the light neutrinos to allow
them to be produced from a vector boson. Specifically
\begin{equation}
\Omega_{ee} = \frac{|a|^2\,v^2}{m^2_N}\,,\qquad \Omega_{\mu\mu} = \frac{|b|^2\,v^2}{m^2_N}\,,\qquad \Omega_{\tau\tau} = \frac{|c|^2\,v^2}{m^2_N}\,,
\end{equation}
where $v = 246$~GeV is the vacuum expectation value of the Higgs field.

It should be noted that in this model the heavy neutrinos produced at the LHC would be linked indirectly with the mechanism providing light neutrinos with
small masses. The light neutrinos acquire masses directly through the mechanism responsible for breaking the flavour symmetries. However, studying the
properties of the heavy neutrinos accessible to the LHC would allow us to better understand the underlying symmetry protecting light neutrinos from
large masses and may give us insight into the observed pattern of large mixing. In addition, further knowledge of heavy neutrinos seen at the LHC, for
example small couplings with one or more lepton flavours or large, resonantly enhanced CP-violation, would provide us with further information
on possible explanations for the baryon asymmetry of the Universe.


\subsubsection{Lepton flavour violation from the mirror leptons in Little Higgs models}
Little Higgs models \cite{Arkani-Hamed:2001ca,Cheng:2001vd,Arkani-Hamed:2001nc,Schmaltz:2005ky,Perelstein:2005ka} offer an alternative route
to the solution of the little hierarchy problem. One of the most attractive models of
this class is the Littlest Higgs model \cite{Arkani-Hamed:2002qy} with T-parity (LHT) \cite{Chang:2003zn,Cheng:2003ju,Cheng:2004yc},
where the discrete symmetry forbids tree-level corrections to electroweak
  observables, thus weakening the electroweak precision constraints \cite{Hubisz:2005tx}. {Under this new symmetry the particles have distinct transformation properties, that is, they are either T-even or T-odd. The model is based on a two-stage spontaneous symmetry breaking occurring at the scale $f$ and the electroweak scale $v$. Here the scale $f$ is taken to be larger than about 500 GeV, which allows to expand expressions in the small parameter $v/f$. The additionally introduced}
gauge bosons, fermions and scalars are sufficiently light to be 
discovered at LHC and there is a dark matter candidate \cite{Hubisz:2004ft}. Moreover, 
the flavour structure of the LHT model is richer than the one of the 
SM, mainly due to the presence of three doublets 
of mirror quarks and three doublets of mirror leptons and their 
weak interactions with the ordinary quarks and leptons, as discussed in {\cite{Low:2004xc,Hubisz:2005bd,Blanke:2006xr}}.

Now, it is well known that in the SM the FCNC processes in the lepton sector, like $\ell_i\to\ell_j\gamma$ and {$\mu\to eee$},
are very strongly suppressed due to tiny neutrino masses. In particular, the
branching ratio for $\mu\to e \gamma$ in the SM amounts to at most $10^{-54}$,
to be compared with the present experimental upper bound, $1.2\cdot 10^{-11}$ \cite{Brooks:1999pu},
and with the one that will be available within the next two years,
$\sim 10^{-13}$ \cite{Yamada:2005tg,megexp}. 
Results close to the SM predictions are expected within the LH model without T-parity, where the lepton sector is identical to the one of the SM and the additional ${\cal O}(v^2/f^2)$ corrections have only minor impact on this result. {Similarly the new effects on $(g-2)_\mu$ turn out to be small \cite{Park:2003sq,Casalbuoni:2003ft}.}

A very different situation is to be expected in the LHT model, where
the presence of new flavour violating interactions and of mirror
leptons with masses of order $1\TeV$ can change the SM expectations by
up to 45 orders of magnitude, bringing the relevant branching ratios
for lepton flavour violating (LFV) processes close to the bounds
available presently or in the near future.

\subsubsubsection{The model}

A detailed description of the LHT model can be found
in~{\cite{Blanke:2006eb}, where also a complete set of Feynman rules
{has been derived}.} Here we just want to state briefly the
ingredients needed for the {analysis of LFV decays}.

The T-odd gauge boson sector consists of three heavy ``partners'' of the SM gauge bosons
\begin{equation}\label{2.3}
W_H^\pm\,,\qquad Z_H\,,\qquad A_H\,,
\end{equation}
with masses given to lowest order in $v/f$ by
\begin{equation}\label{2.4}
M_{W_H}=gf\,,\qquad M_{Z_H}=gf\,,\qquad
M_{A_H}=\frac{g'f}{\sqrt{5}}\,.
\end{equation}

The T-even fermion sector contains, in addition to the SM fermions, the heavy top partner $T_+$. 
On the other hand, the T-odd fermion sector~\cite{Low:2004xc} consists of three generations of mirror quarks and leptons with vectorial
couplings under $SU(2)_L\times U(1)_Y$, that are denoted by
\begin{equation}
{\begin{pmatrix}
u_H^i\\
d_H^i
\end{pmatrix}\,,\qquad 
\begin{pmatrix}\nu_H^i\\
\ell_H^i
\end{pmatrix}}
\qquad(i=1,2,3)\,.
\end{equation}
To first order in $v/f$ the masses of up- and down-type mirror fermions are equal. Naturally, their masses are of order $f$.
In the {analysis of LFV decays}, except for $K_{L,S}\to\mu e$, $K_{L,S}\to\pi^0\mu e$, $B_{d,s}\to\ell_i\ell_j$ and
$\tau\to\ell\pi,\ell\eta,\ell\eta'$, only  mirror leptons are relevant.

As discussed in detail in~\cite{Hubisz:2005bd}, one of the important ingredients of the mirror sector is the existence of four CKM-like unitary
mixing matrices, {two for mirror quarks $(V_{Hu},V_{Hd})$ and two for mirror leptons {$(V_{H\nu},V_{H\ell})$}, that are {related} via
\begin{equation}\label{2.10}
V_{Hu}^\dagger V_{Hd} = V_\text{CKM}\,,\qquad V_{H\nu}^\dagger V_{H\ell} = V_\text{PMNS}^\dagger\,.
\end{equation}
An explicit parameterization of $V_{Hd}$ and $V_{H\ell}$ in terms of three mixing angles and three complex (non-Majorana) phases can
be found in \cite{Blanke:2006xr}.}

The mirror mixing matrices  parameterize flavour violating interactions between SM fermions and mirror fermions that are mediated by the
heavy gauge bosons $W_H^\pm$, $Z_H$ and $A_H$. The {matrix notation}  indicates which of the light fermions of a given electric charge
participates in the interaction.

In the course of {the} analysis of charged LFV decays it {is} useful to introduce the following quantities ($i=1,2,3$) {\cite{Blanke:2007db}}:
\begin{equation}\label{eq:chi}
\chi_i^{(\mu e)}=V^{*ie}_{H\ell}V^{i\mu}_{H\ell}\,,\qquad
\chi_i^{(\tau e)}=V^{*ie}_{H\ell}V^{i\tau}_{H\ell}\,,\qquad
\chi_i^{(\tau\mu)}=V^{*i\mu}_{H\ell}V^{i\tau}_{H\ell}\,,
\end{equation}
that govern $\mu\to e$, $\tau\to e$ and $\tau\to\mu$ transitions, respectively.
Analogous quantities in the mirror quark sector $(i=1,2,3)$ {\cite{Blanke:2006eb,Blanke:2006sb}},
\begin{equation}
\xi_i^{(K)}=V^{*is}_{Hd}V^{id}_{Hd}\,,\qquad
\xi_i^{(d)}=V^{*ib}_{Hd}V^{id}_{Hd}\,,\qquad
\xi_i^{(s)}=V^{*ib}_{Hd}V^{is}_{Hd}\,,
\end{equation}
{are} needed for the analysis of the decays $K_{L,S}\to\mu e$, $K_{L,S}\to\pi^0\mu e$  and $B_{d,s}\to\ell_i\ell_j$. 

As an example, the branching ratio for the $\mu\to e\gamma$ decay contains the $\chi^{(\mu e)}_i$ {factors} introduced in \eqref{eq:chi}
via the short distance function~\cite{Blanke:2007db}
\begin{equation}
\label{eq:Dpoddme}
\bar D^{\prime\,\mu e}_\text{odd}=
\frac{1}{4}\frac{v^2}{f^2}\sum_i\left(\chi_i^{(\mu e)}(D'_0(y_i)-\frac{7}{6}E'_0(y_i)-\frac{1}{10}E'_0(y'_i))\right)\,,
\end{equation}
where $y_i=(m^\ell_{Hi}/M_{W_H})^2,\ y'_i = a y_i$ with $a=5/\tan^2\theta_W$, and explicit expressions for the functions
$D'_0,E'_0$ can be found in \cite{Buras:1998ra}.

The new parameters of the LHT model, relevant for the study {of LFV decays}, are
\begin{equation}\label{2.16}
f\,,\quad  m^\ell_{H1}\,,
\quad m^\ell_{H2}\,,
\quad m^\ell_{H3}\,,
\quad \theta_{12}^\ell\,,
\quad \theta_{13}^\ell\,,
\quad \theta_{23}^\ell\,,
\quad \delta_{12}^\ell\,,
\quad \delta_{13}^\ell\,,
\quad \delta_{23}^\ell
\end{equation}
and the ones in the mirror quark sector that can be probed by FCNC processes in $K$ and $B$ meson systems, as discussed in detail in
\cite{Blanke:2006sb, Blanke:2006eb}. Once the new heavy gauge bosons and mirror fermions will be discovered and their masses measured at the LHC,
the only free parameters of the LHT model will be the mixing angles $\theta^\ell_{ij}$ and the complex phases $\delta^\ell_{ij}$ of the matrix
$V_{H\ell}$, that can be determined with the help of LFV processes. Analogous comments apply to the {determination of $V_{Hd}$  parameters} in
the quark sector (see \cite{Blanke:2006eb,Blanke:2006sb} for details on $K$ and $B$ physics in the LHT model).

\subsubsubsection{Results}
LFV processes in the LHT model have for the first time been discussed in \cite{Choudhury:2006sq}, where the decays {$\ell_i\to\ell_j\gamma$}
have been considered. Further, the new contributions to $(g-2)_\mu$ in the LHT model have been calculated by these authors. In \cite{Blanke:2007db} the analysis of LFV in
the LHT model has been considerably extended, and includes the decays $\ell_i\to\ell_j\gamma$, $\mu\to eee$, the six three body leptonic decays
$\tau^-\to\ell_i^-\ell_j^+\ell_k^-$, the semi-leptonic decays $\tau\to\ell\pi,\ell\eta,\ell\eta'$ and the decays $K_{L,S}\to\mu e$, $K_{L,S}\to\pi^0\mu e$ and
$B_{d,s}\to\ell_i\ell_j$ that are flavour violating both in the {quark and lepton sector}. Moreover, $\mu-e$ conversion in nuclei and the flavour conserving $(g-2)_\mu$
have been studied. Furthermore, a detailed phenomenological analysis has been performed in that paper, paying particular attention to various ratios of LFV branching ratios
that will be useful for a clear distinction of the LHT model from the MSSM.

In contrast to $K$ and $B$ physics in the LHT model, where the SM {contributions constitute} a sizable and often the  dominant part, the T-even contributions to LFV
observables are completely negligible due to the smallness of neutrino masses and the LFV decays considered are entirely governed by mirror fermion contributions. 

\begin{figure}
\includegraphics[width=0.5\linewidth]{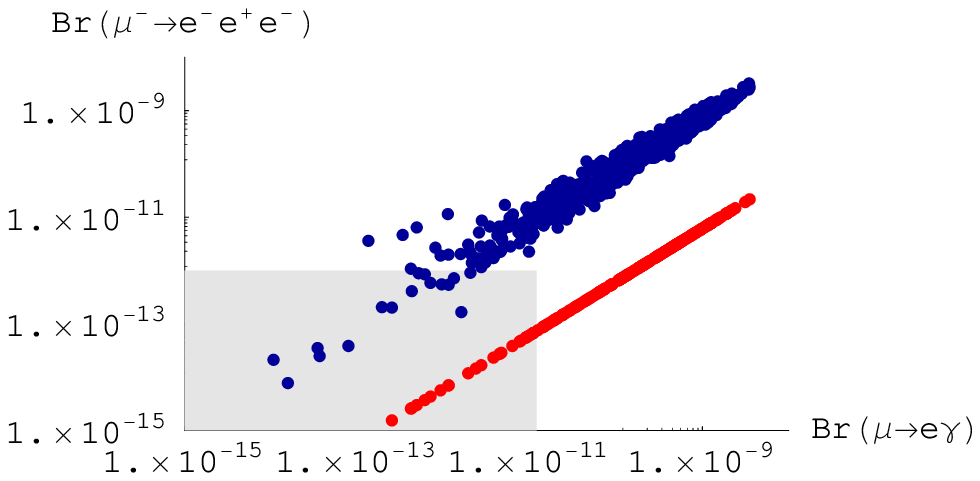}
\includegraphics[width=0.5\linewidth]{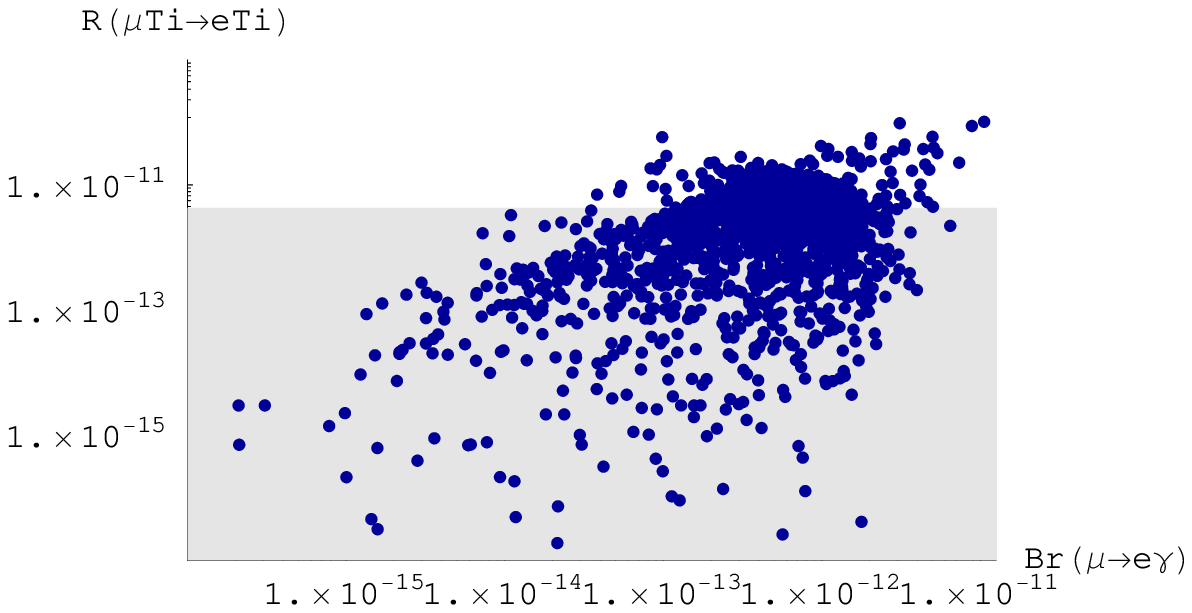}
\parbox[t]{0.48\linewidth}{
\caption{Correlation between $B(\mu\to e\gamma)$ and $B(\mu\to eee)$ in the LHT model (upper dots) \cite{Blanke:2007db}. The lower dots
represent the dipole contribution to $\mu\to eee$ separately, which, unlike in the LHT model, is the dominant contribution in the MSSM.
The grey region is allowed by the present experimental bounds.\label{fig:meg-m3ea}}}\hspace*{\fill}
\parbox[t]{0.48\linewidth}{
\caption{$R(\mu\text{Ti}\to e\text{Ti})$ as a function of $B(\mu\to e\gamma)$, after imposing the existing constraints on $\mu\to e\gamma$
and $\mu\to eee$ \cite{Blanke:2007db}. The grey region is allowed by the present experimental bounds.\label{fig:meg-m3eb}}}
\end{figure}

In order to see how large these contributions can possibly be, it is useful to consider first those decays for which the strongest constraints exist. Therefore
Fig.~\ref{fig:meg-m3ea} shows $B(\mu\to eee)$ as a function of $B(\mu\to e\gamma)$, obtained from a general scan over the mirror lepton parameter space, with $f=1\TeV$.
It is found that in order to fulfill the present bounds, either the mirror lepton spectrum has to be quasi-degenerate or the $V_{H\ell}$ matrix must be very hierarchical.
Moreover, as shown in Fig.~\ref{fig:meg-m3eb}, even after imposing the constraints on $\mu\to e\gamma$ and $\mu\to eee$, the $\mu-e$ conversion rate in Ti is very likely
to be found close to its current bound, and for some regions of the mirror lepton parameter space even violates this bound.

\begin{table}[bht]
\begin{minipage}{\textwidth}
\caption{Upper bounds on LFV $\tau$ decay branching ratios in the LHT model, for two different values of the scale $f$, after imposing the constraints on $\mu\to e\gamma$
and $\mu\to eee$ \cite{Blanke:2007db}. For $f=500\, {\rm GeV}$, also the bounds on $\tau\to\mu\pi,e\pi$ have been included. The current experimental upper bounds are also
given. The bounds in \cite{Banerjee:2007rj} have been obtained by combining Belle \cite{Enari:2005gc,Abe:2006sf} and BaBar \cite{Aubert:2006cz,Aubert:2005wa} results.
\label{tab:bounds}}
\begin{tabular*}{\textwidth}{@{\extracolsep{\fill}}l|ccr}
\hline\hline
decay 				& $f=1000\, {\rm GeV}$ 	& $f=500 \, {\rm GeV}$ 	& exp.~upper bound 				\\
\hline
$\tau\to e\gamma$ 		& $8\cdot 10^{-10}$  	&$1\cdot 10^{-8}$  	&$9.4\cdot 10^{-8}$ \cite{Banerjee:2007rj} 	\\
$\tau\to \mu\gamma$ 		& $8\cdot 10^{-10}$ 	&$2\cdot 10^{-8}$  	&$1.6\cdot 10^{-8}$ \cite{Banerjee:2007rj}	\\
$\tau^-\to e^-e^+e^-$ 		& $7\cdot 10^{-10}$ 	&$2\cdot 10^{-8}$ 	&$2.0\cdot 10^{-7}$ \cite{Aubert:2003pc}	\\
$\tau^-\to \mu^-\mu^+\mu^-$ 	& $7\cdot 10^{-10}$ 	&$3\cdot 10^{-8}$   	&$1.9\cdot 10^{-7}$ \cite{Aubert:2003pc} 	\\
$\tau^-\to e^-\mu^+\mu^-$ 	& $5\cdot 10^{-10}$ 	&$2\cdot 10^{-8}$   	&$2.0\cdot 10^{-7}$ \cite{Yusa:2004gm}		\\
$\tau^-\to \mu^-e^+e^-$ 	& $5\cdot 10^{-10}$ 	&$2\cdot 10^{-8}$  	&$1.9\cdot 10^{-7}$ \cite{Yusa:2004gm} 		\\
$\tau^-\to \mu^-e^+\mu^-$ 	& $5\cdot 10^{-14}$  	&$2\cdot 10^{-14}$ 	&$1.3\cdot 10^{-7}$ \cite{Aubert:2003pc}	\\
$\tau^-\to e^-\mu^+e^-$ 	& $5\cdot 10^{-14}$  	&$2\cdot 10^{-14}$   	&$1.1\cdot 10^{-7}$ \cite{Aubert:2003pc} 	\\
$\tau\to\mu\pi$ 		& $2\cdot 10^{-9} $  	&$5.8\cdot 10^{-8}$ 	&$5.8\cdot 10^{-8}$ \cite{Banerjee:2007rj}	\\
$\tau\to e\pi$ 			& $2\cdot 10^{-9} $ 	&$4.4\cdot 10^{-8}$	&$4.4\cdot 10^{-8}$ \cite{Banerjee:2007rj}	\\
$\tau\to\mu\eta$ 		& $6\cdot 10^{-10}$ 	&$2\cdot 10^{-8}$  	&$5.1\cdot 10^{-8}$ \cite{Banerjee:2007rj}	\\
$\tau\to e\eta$ 		& $6\cdot 10^{-10}$  	&$2\cdot 10^{-8}$  	&$4.5\cdot 10^{-8}$ \cite{Banerjee:2007rj}	\\
$\tau\to \mu\eta'$ 		& $7\cdot 10^{-10}$ 	&$3\cdot 10^{-8}$ 	&$5.3\cdot 10^{-8}$ \cite{Banerjee:2007rj}	\\
$\tau\to e\eta'$ 		& $7\cdot 10^{-10}$ 	&$3\cdot 10^{-8}$  	&$9.0\cdot 10^{-8}$ \cite{Banerjee:2007rj}	\\
\hline\hline
\end{tabular*}
\end{minipage}
\end{table}

The existing constraints on LFV $\tau$ decays are still relatively weak, so that they presently do not provide a useful constraint on the LHT parameter space. However,
as seen in Table~\ref{tab:bounds}, most branching ratios in the LHT model can reach the present experimental upper bounds, in particular for low values of $f$, and are
very interesting in view of new experiments taking place in this and the coming decade. 

The situation is different in the case of $K_L\to\mu e$, $K_L\to\pi^0\mu e$ and $B_{d,s}\to\ell_i\ell_k$, due to the double GIM suppression in the quark and lepton sectors.
E.\,g. $B(K_L\to\mu e)$ can reach values of at most $3\cdot10^{-13}$ which is still one order of magnitude below the current bound, and  $K_L\to\pi^0\mu e$ is even by two
orders of magnitude smaller. Still, measuring the rates for  $K_L\to\mu e$ and $K_L\to\pi^0\mu e$ would be desirable, as, due to their sensitivity to
$\text{Re} (\xi_i^{(K)})$ and $\text{Im}(\xi_i^{(K)})$ respectively, these decays can shed light on the complex phases present in the mirror quark sector.

While the possible huge enhancements of LFV branching ratios in the LHT model are clearly interesting, such effects are common to many other NP models, such as the MSSM,
and therefore cannot be used to distinguish these models. However, correlations between various branching ratios should allow a clear distinction of the LHT model from
the MSSM. While in the MSSM \cite{Ellis:2002fe,Arganda:2005ji,Brignole:2004ah,Paradisi:2005tk,Paradisi:2006jp} the {dominant role in decays with three leptons in the final
state and in $\mu-e$ conversion in nuclei is typically played} by the dipole operator, in \cite{Blanke:2007db} it is found that this operator is basically irrelevant in the
LHT model, where $Z^0$-penguin and box diagram contributions are {much more important}. As can be seen in Table~\ref{tab:ratios} and also in Fig.~\ref{fig:meg-m3ea} this
implies a striking difference {between} various ratios of branching ratios in the MSSM and in the LHT model and should be very useful in distinguishing these two models.
Even if for some decays this distinction is less clear when significant Higgs contributions are present \cite{Brignole:2004ah,Paradisi:2005tk,Paradisi:2006jp}, it should
be easier than through high-energy processes at LHC. 

\begin{table}[bht]
\begin{minipage}{\textwidth}
\caption{Comparison of various ratios of branching ratios in the LHT model and in the MSSM without and with significant Higgs contributions \cite{Blanke:2007db}.\label{tab:ratios}}
\begin{tabular*}{\textwidth}{@{\extracolsep{\fill}}l|ccc}
\hline\hline
ratio 							&LHT  					& MSSM (dipole) 	&MSSM (Higgs) 		\\
\hline
$B(\mu^-\to e^-e^+e^-)/B(\mu\to e\gamma)$  		&\hspace{.6cm} 0.4 -- 2.5\hspace{.6cm} &$\sim 6\cdot10^{-3}$ 	&$\sim6\cdot10^{-3}$  	\\
$B(\tau^-\to e^-e^+e^-)/B(\tau\to e\gamma)$   		&0.4 -- 2.3     			&$\sim 1\cdot10^{-2}$ 	&$\sim1\cdot10^{-2}$	\\
$B(\tau^-\to \mu^-\mu^+\mu^-)/B(\tau\to \mu\gamma)$  	&0.4 -- 2.3     			&$\sim 2\cdot10^{-3}$ 	&0.06 -- 0.1	 	\\
$B(\tau^-\to e^-\mu^+\mu^-)/B(\tau\to e\gamma)$  	&0.3 -- 1.6     			&$\sim 2\cdot10^{-3}$ 	&0.02 -- 0.04	 	\\
$B(\tau^-\to \mu^-e^+e^-)/B(\tau\to \mu\gamma)$  	&0.3 -- 1.6    			&$\sim 1\cdot10^{-2}$ 	&$\sim1\cdot10^{-2}$	\\
$B(\tau^-\to e^-e^+e^-)/B(\tau^-\to e^-\mu^+\mu^-)$     &1.3 -- 1.7   				&$\sim 5$ 		&0.3 -- 0.5		\\
$B(\tau^-\to \mu^-\mu^+\mu^-)/B(\tau^-\to \mu^-e^+e^-)$ &1.2 -- 1.6    			&$\sim 0.2$ 		&5 -- 10 		\\
$R(\mu\text{Ti}\to e\text{Ti})/B(\mu\to e\gamma)$  	&0.01 -- 100     		&$\sim 5\cdot 10^{-3}$ 	&0.08 -- 0.15	 	\\
\hline\hline
\end{tabular*}
\end{minipage}
\end{table}

Another possibility to distinguish different NP models 
through LFV processes is given by the measurement of $\mu\to e\gamma$ with polarized muons. Measuring the angular distribution of the outgoing electrons, one can determine the size of left- and right-handed contributions separately \cite{Kuno:1996kv}. In addition, detecting also the electron spin would yield information on the relative phase between these two contributions \cite{Farzan:2007us}. We recall that the LHT model is peculiar in this respect as it does not involve any right-handed contribution.

On the other hand, the contribution of mirror leptons to $(g-2)_\mu$, being a flavour conserving observable, is negligible \cite{Choudhury:2006sq,Blanke:2007db}{, so that the possible discrepancy between SM prediction and experimental data \cite{Czarnecki:2002nt} can not be cured}. {This should also be contrasted with the MSSM with large $\tan\beta$ and not too heavy scalars, where those corrections could be significant, thus allowing to solve the possible discrepancy between SM prediction and experimental data.}

\subsubsubsection{Conclusions}

{
We have seen that LFV decays open up an exciting playground for testing the LHT model. Indeed, they could offer a very clear distinction between
 this model and supersymmetry. Of particular interest are the ratios  
$B(\ell_i\to eee)/B(\ell_i\to e\gamma)$ that are ${\cal O}(1)$ in the LHT model but  strongly suppressed in  supersymmetric models even in the presence of
significant Higgs contributions. Similarly, finding the $\mu- e$ conversion rate in nuclei at the same level as $B(\mu\to e\gamma)$ would point into the direction of LHT physics rather than supersymmetry.
}

\subsubsection{Low scale triplet Higgs neutrino mass scenarios in Little Higgs models}\label{sec:lowtriplet}
%

An important open issue to address in the context of Little Higgs
models is the origin of non-zero neutrino
masses~\cite{Lee:2005mb,Han:2005nk,delAguila:2005yi,Choudhury:2005jh,Abada:2005rt}.
The neutrino mass mechanism which naturally occurs in these models is
the triplet Higgs mechanism \cite{Ma:1998dx} which employs a scalar
with the $SU(2)_L\times U(1)_Y $ quantum numbers $T\sim (3,2)$.  The
existence of such a multiplet in some versions of the Little Higgs
models is a direct consequence of global symmetry breaking which makes
the SM Higgs light. For example, in the minimal Littlest Higgs
model~\cite{Arkani-Hamed:2002qy}, the triplet Higgs with non-zero
hypercharge occurs from the breaking of global $SU(5)$ down to $SO(5)$
symmetry as one of the Goldstone bosons.  Its mass $M_T\sim g_s f,$
where $g_s<4 \pi$ is a model dependent coupling constant in the weak
coupling regime~\cite{Giudice:2007fh}, is therefore predicted to be
below the cut-off scale $\Lambda$, and could be within the mass reach
of LHC.  The present lower bound for the invariant mass of $T$ is set
by Tevatron to $M_T\ge 136$~GeV \cite{Acosta:2004uj,Abazov:2004au}.

Although the triplet mass scale is of order ${\cal O}(1)$~TeV, 
the observed neutrino masses can be obtained naturally. 
Due to the specific quantum numbers the triplet Higgs boson
couples only to the left-chiral lepton doublets $L_i\sim (2,-1)$, $i=e, \mu, \tau,$
via the Yukawa interactions of Eq.~(\ref{eq:Lag-typeII}) and to the SM Higgs
bosons via Eq.~(\ref{eq:potential-triplet}).
Those interactions 
induce lepton flavour violating decays of charged leptons which have not been
observed. The most stringent constraint on the Yukawa couplings comes from the 
upper limit on the tree-level decay $\mu\to eee$ and is\footnote{In Little Higgs models
with $T$-parity there exist additional sources of flavour violation from the mirror
fermion sector~\cite{Choudhury:2006sq,Blanke:2007db} discussed 
in the previous subsection.}
$Y_T^{ee}Y_T^{e\mu}<3\cdot 10^{-5} (M/TeV)^2$~\cite{Huitu:1996su,Yue:2007kv}.
Experimental bounds on the 
tau Yukawa couplings are much less stringent. 
The hierarchical
light neutrino masses imply $Y_T^{ee},Y_T^{e\mu}\ll Y_T^{\tau\tau}$ consistently with the 
direct experimental bounds.

Non-zero neutrino masses and mixing is presently the only experimentally verified 
signal of new physics beyond the SM. In the triplet neutrino mass 
mechanism~\cite{Ma:1998dx}
presented in Section~\ref{sec:seesawII} the neutrino masses are given by
\begin{equation}
(m_\nu)^{ij} = Y_T^{ij} v_T,
\label{mnu}
\end{equation}
where $v_T$ is the induced triplet VEV of Eq.~(\ref{eq:vevs-triplet}).
It is natural that the smallness of neutrino masses is explained by the smallness of 
$v_T.$  In the little Higgs models this can be  achieved by
requiring the Higgs mixing parameter $\mu\ll M_T$, which can be explained,
for example, via shining of explicit lepton number violation 
from extra dimensions as shown in Ref.~\cite{Ma:2000wp,Ma:2000xh},
or if the triplet is related to the Dark Energy of the Universe~\cite{Ma:2006mr,Sahu:2007uh}.
Models with additional (approximate) $T$-parity~\cite{Chang:2003zn} make
the smallness of $v_T$ technically natural (if the $T$-parity
is exact, $v_T$ must vanish). 
In that case $Y_T v_T\sim {\cal O}(0.1)$~eV while the Yukawa couplings $Y$ can be
of order charged lepton Yukawa couplings of the SM.
As a result, the branching ratio of the decay $T\to WW$ is negligible.
We also remind that $v_T$ contributes to the SM oblique corrections, 
and the precision data fit $\hat T<2\cdot 10^{-4}$~\cite{Marandella:2005wd} 
sets an upper bound $v_T \leq 1.2$~GeV on that parameter.

Notice the particularly simple connection between the flavour structure of light neutrinos
and the Yukawa couplings of the triplet via Eq.~(\ref{mnu}). Therefore, independently of the overall size 
of the Yukawa couplings, one can predict the leptonic branching ratios of the
triplet  from neutrino oscillations. 
For the normally hierarchical light neutrino masses neutrino data implies negligible
$T$ branching fractions to electrons and  
$B(T^{++} \to \mu^+ \mu^+)\approx B(T^{++} \to \tau^+ \tau^+)
\approx B(T^{++} \to \mu^+ \tau^+)\approx 1/3.$ 
Those are the final state signatures predicted by the triplet neutrino mass mechanism 
for collider experiments.

At LHC $T^{++}$ can be produced singly and in pairs.
The cross section of the single $T^{++}$ production via the $WW$
fusion process~\cite{Huitu:1996su} $qq\to q'q' T^{++}$ scales as
$\sim v_T^2.$ In the context of the littlest Higgs model
this process, followed by the decays $T^{++}\to W^+W^+,$
was studied  in Refs.~\cite{Han:2003wu,Han:2005ru,Azuelos:2004dm}.
The detailed ATLAS simulation of this channel shows~\cite{Azuelos:2004dm}
that in order to observe an $1$~TeV $T^{++},$ 
one must have $v_T>29$~GeV. This is in conflict with the 
precision physics  bound 
$v_T \leq 1.2$~GeV as well as with the neutrino 
data. Therefore the $WW$ fusion channel is not experimentally promising for
the discovery of  doubly charged Higgs.

On the other hand,  the Drell-Yan pair production 
process~\cite{Gunion:1996pq,Gunion:1989in,Huitu:1996su,Muhlleitner:2003me,
Akeroyd:2005gt,Rommerskirchen:2007jv,Hektor:2007uu,Han:2007bk}
$$pp\to T^{++}T^{--}$$
 is not suppressed by any small coupling and
 its cross section is known up to next to leading order~\cite{Muhlleitner:2003me} (possible 
 additional contributions from new physics such as $Z_H$ are strongly suppressed
 and we neglect those effects here). 
Followed by the lepton number violating decays $T^{\pm\pm} \to \ell^\pm\ell^\pm$, 
this process allows to reconstruct  $T^{\pm\pm}$ invariant mass
from the same charged
leptons rendering the SM background to be very small in the signal region.
If one also assumes  that neutrino masses come from 
the triplet Higgs interactions, one fixes the $T^{\pm\pm}$ leptonic
branching ratios. This allows to test the triplet neutrino mass model at LHC.
The pure Monte Carlo study of this scenario shows \cite{Hektor:2007uu}
that $T^{++}$ up to the mass 300 GeV is 
reachable in the first year of LHC ($L=1$ fb$^{-1}$) and  $T^{++}$ up to 
the mass 800 GeV is reachable for the luminosity $L=30$ fb$^{-1}.$ 
Including the Gaussian measurement errors to the Monte Carlo the corresponding 
mass reaches become \cite{Hektor:2007uu} 
250 GeV and 700 GeV, respectively. The errors of those estimates 
of the required luminosity for discovery depend strongly 
on the size of statistical Monte Carlo sample of the background processes.


\subsection{Flavour and CP-violation in SUSY extensions of the SM}\label{sec:SUSY} 

Supersymmetric models provide the richest spectrum of lepton flavour
and CP-violating observables among all models. They are also among the
best studied scenarios of new physics beyond the Standard Model. In
this Section we review phenomenologically most interesting aspects of
some of the supersymmetric scenarios.

\subsubsection{Mass insertion approximation and phenomenology}

In the low energy supersymmetric extensions of the SM the flavour and
CP-violating interactions would originate from the misalignment
between fermion and sfermion mass eigenstates. Understanding why all
these processes are strongly suppressed is one of the major problems
of low energy supersymmetry, the {\it supersymmetric flavour and CP
problem}.  The absence of deviations from the SM predictions in LFV
and CPV (and other flavour changing processes in the quark sector)
experiments suggests the presence of a quite small amount of
fermion-sfermion misalignment. From the phenomenological point of view
those effects are most easily described by the mass insertion
approximation.

The relevant one-loop amplitudes can be exactly written in terms of the general mass matrix of charginos and neutralinos, resulting in quite involved
expressions. To obtain simple approximate expressions, it is convenient to use the so-called mass insertion method \cite{Hall:1985dx, Gabbiani:1988rb}.
This is a particularly convenient method since, in a model independent way, the tolerated deviation from alignment is quantified by the upper limits on the mass
insertion $\delta$'s, defined as the small off-diagonal elements in terms of which sfermion propagators are expanded, normalized with an average
sfermion mass, $\delta_{i j} = \Delta_{i j}/m^2_{\tilde f}$.  They are of four types: $\delta^{LL}$, $\delta^{RR}$, $\delta^{RL}$ and $\delta^{LR}$,
according to the chiralities of the corresponding partner fermions. We shall adopt here the usual convention for the slepton mass matrix in the basis where the
lepton mass matrix $m_{\ell}$ is diagonal:
\begin{equation}
\left( \begin{array}{cc} {\tilde \ell}_L^\dagger & {\tilde \ell}_R^\dagger \end{array} \right) \ \
\left( \begin{array}{cc} m_L^2 (1 + \delta ^{LL} ) & (A^* - \mu \tan\beta) m_{\ell} + m_L m_R\delta ^{LR} \cr
(A - \mu^* \tan\beta) m_{\ell} + m_L m_R\delta ^{LR~\dagger} & m_R^2 ( 1+ \delta ^{RR} )
\end{array} \right) \ \ \left( \begin{array}{c} {\tilde \ell}_L  \cr {\tilde \ell}_R 
\end{array} \right) \nonumber
\end{equation} 
\noindent where $m_L ~, m_R ~,$ are respectively the average real masses of the left-handed and right-handed sleptons and $A$ contains only 
the diagonal entries the trilinear matrices at the electroweak scale. Notice that these flavour diagonal left--right mixing are always present in
any MSSM and play a very important role in LFV processes. In this way, our $\delta ^{LR}$ contain only the off-diagonal elements of the trilinear
matrices. This definition is then slightly different from the original definition in Refs \cite{Hall:1985dx,Gabbiani:1996hi}.
The deviations from universality are then all gathered in the different $\delta$ matrices.

Each element in these $\delta$ matrices can be tested by experiment. Searches for the decay $\ell_i \to \ell_j \gamma$ provide 
bounds on the absolute values of the off-diagonal (flavour violating) $|\delta^{LL}_{ij}|$, $|\delta^{RR}_{ij}|$, $|\delta^{LR}_{ij}|$ and
$|\delta^{RL}_{ij}|$, while measurements of the lepton EDM (MDM), 
parameters and their CP-violating phases, also provide limits on the imaginary (real) part of combinations of flavour
violating $\delta$'s, $ \delta_{ij}^{LL}\delta_{ji}^{LR}$, $\delta_{ij}^{LR}\delta_{ji}^{RR}$, $\delta_{ij}^{LL}\delta_{ji}^{RR}$
and $ \delta_{ij}^{LR}\delta_{ji}^{LR}$. Many authors have addressed the issue of the bounds on these misalignment parameters and phases in the sleptonic sector 
\cite{Gabbiani:1996hi}. Following \cite{Masina:2002mv} we present the current limits on $\mu \to e \gamma$
and we analyze the impact of the planned experimental improvements on $\tau \to \mu \gamma$. 
In the basis where $Y_{\ell}$ is diagonal, and in the mass insertion approximation, 
the branching ratio of the process reads
\begin{equation}
\label{BR}
B(\ell_i \to \ell_j \gamma) =  10^{-5} ~B(\ell_i \to \ell_j \bar \nu_j \nu_i)
~\frac{M^4_W}{\bar m^4_L} \tan^2 \beta |\delta^{LL}_{ij}|^2 F_{\rm SUSY},
\end{equation}
\noindent where $F_{\rm SUSY}=O(1)$ is a function of supersymmetric masses including both chargino and neutralino exchange (see \eg \cite{Masina:2002mv}, and
references therein). 
We focus for definiteness on the mSUGRA scenario, also assuming gaugino
and scalar universality at the gauge coupling unification scale and fixing $\mu$ as required by the radiative electroweak symmetry breaking. 

\begin{figure}
\includegraphics[clip, trim= 0 300 0 0, width=0.66\linewidth]{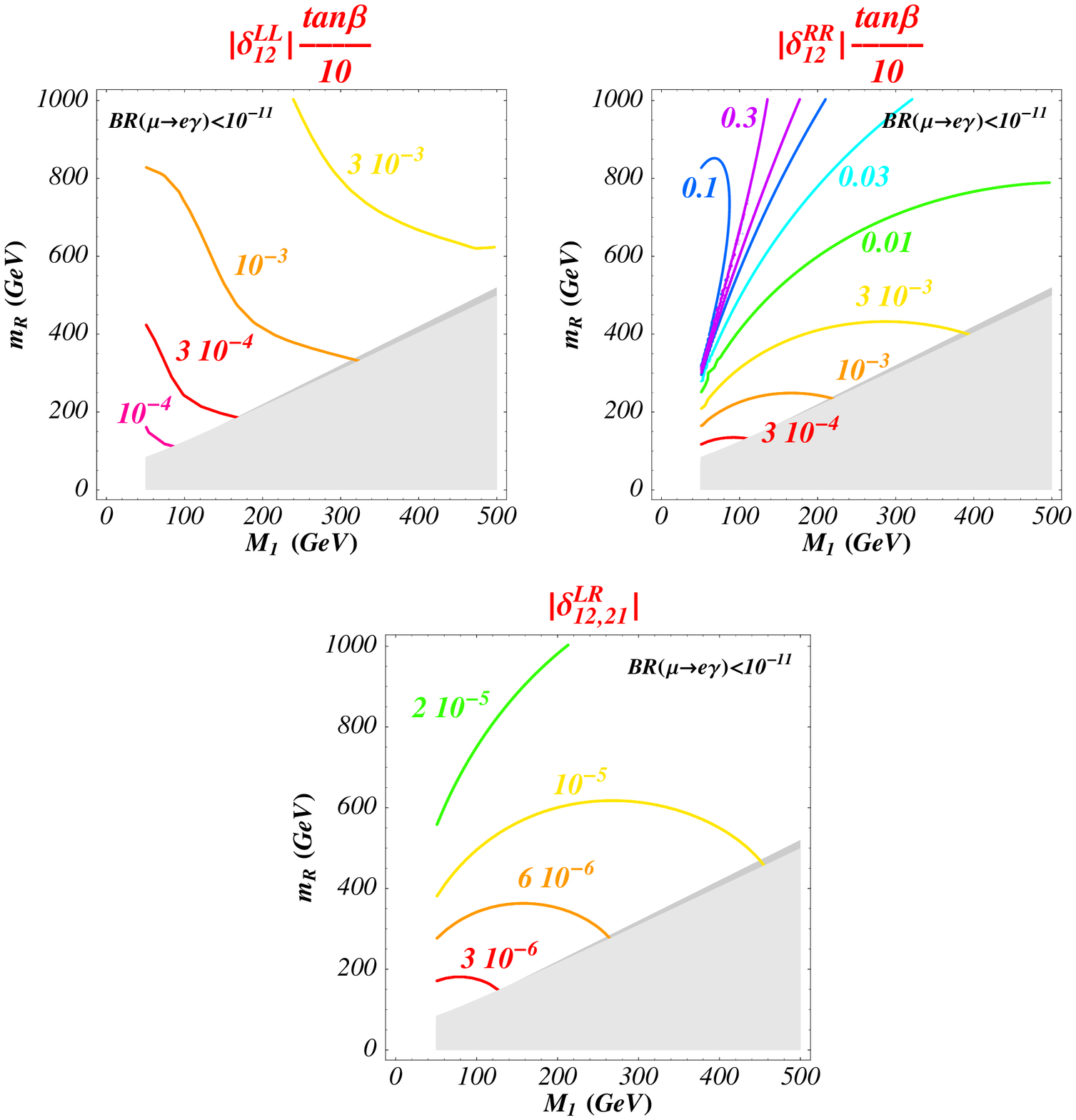}
\includegraphics[clip, trim= 160 0 125 310, width=0.33\linewidth]{figures/section52/Flfv_12rev.ps}
\caption{Upper limits on $\delta_{12}$'s in mSUGRA. Here $M_1$ and $m_{R}$
are the bino and right-slepton masses, respectively.\label{fig12}}
\includegraphics[clip, trim= 0 290 0 0, width=0.66\linewidth]{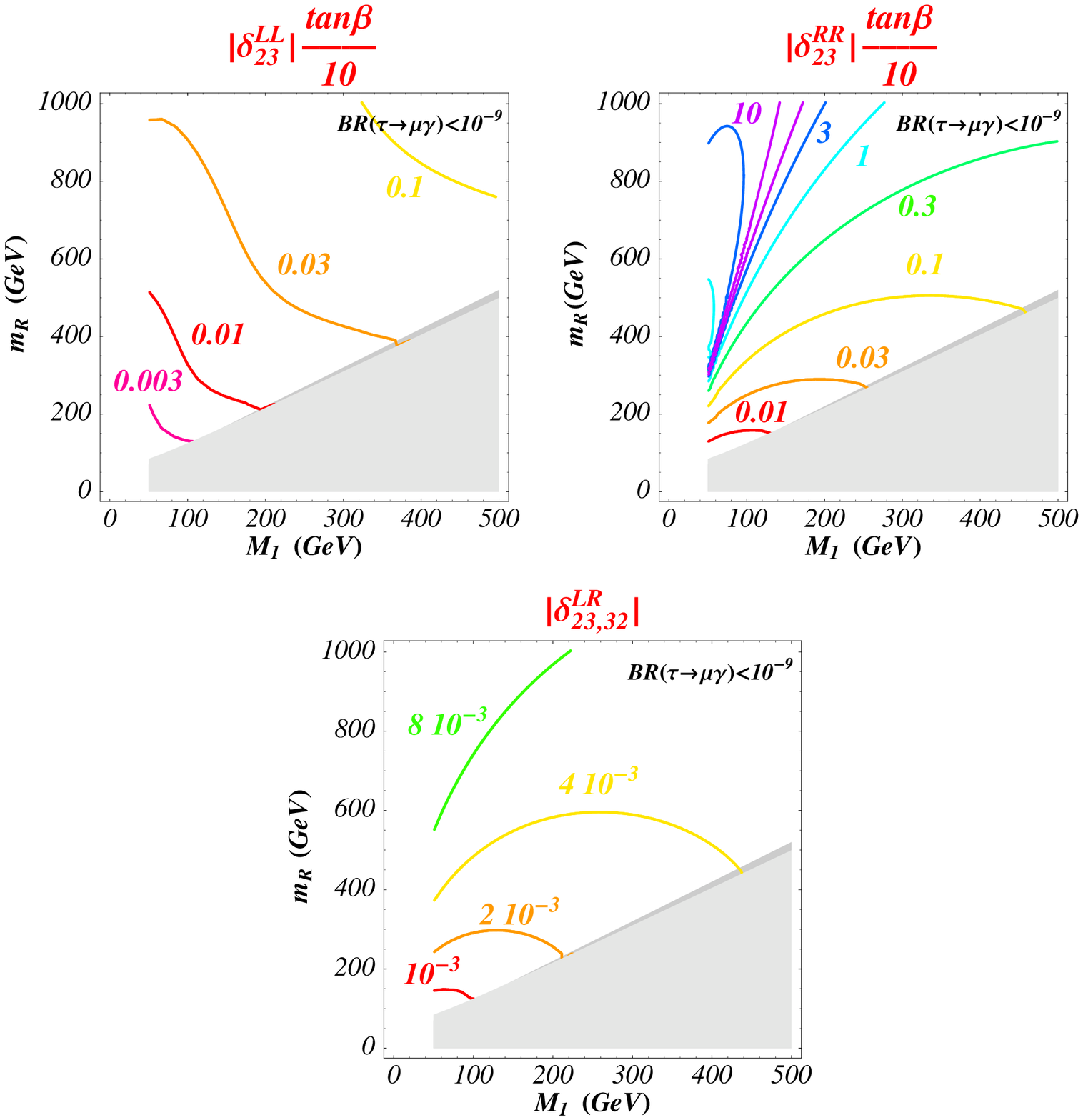}
\includegraphics[clip, trim= 155 0 135 310, width=0.33\linewidth]{figures/section52/Flfv_23.ps}
\caption{Upper limits on $\delta_{23}$'s in mSUGRA. Here $M_1$ and $m_{R}$
are the bino and right-slepton masses, respectively. \label{fig23}}
\end{figure}

As for LFV, Figs.~\ref{fig12} and~\ref{fig23} display the upper bounds on the $|\delta|$'s in the $(M_1,m_{R})$ plane, where $M_1$ and $m_{R}$
are the bino and right-slepton masses, respectively. Deviations from the mSUGRA assumptions can be estimated by means of relatively simple analytical expressions. 
In  Figs.~\ref{fig12} and \ref{fig23} we can see that the bounds on $\delta^{RR}_{ji}$ depend strongly and are practically absent for some values 
of $M_1$ and $m_R$. This fact is due to a destructive interference between the bino and bino-Higgsino amplitudes \cite{Masina:2002mv}. 
On the contrary, the limits on $\delta^{LL}_{ji}$ are robust because of a constructive interference between the chargino and bino amplitudes. 
A weaker bound on $\delta^{RR}_{12}$ on the cancellation regions can be obtained combining the experimental information from the decays $\mu \to e \gamma$,
$\mu \to e e e$ and $\mu$--e conversion in nuclei \cite{Paradisi:2005fk,Ciuchini:2007ha}. The present limits on $\mu \to e \gamma$ 
provide interesting constraints on the related $\delta$'s. As will be discussed in the following, the present sensitivity already 
allow to test these $\delta$'s at the level of the radiative effects. Such a sensitivity could hopefully be reached also in future experiments 
on $\tau \to \mu \gamma$.

Another issue is the origin of the CP-violating phases in the leptonic EDMs. Unless the sparticle masses are increased above several TeVs, the phases
in the flavour-diagonal elements of the slepton left--right mass matrices (in the lepton flavour basis), in the parameters $\mu$ and $A_i$ of supersymmetric
models, have to be quite small, and this constitutes the so-called supersymmetric CP problem. 
For the bounds on the sources of CPV also associated to FV, like e.g. $Im(\delta_{ij}^{LL}\delta_{ji}^{RR})_{ee}$ 
and so on, we refer to the plots in Ref. \cite{Masina:2002mv}.

\subsubsection{Lepton flavour violation from RGE effects in SUSY seesaw model}\label{sec:flavourRGE}

\subsubsubsection{Predictions from flavour models}

Consider first the possibility that flavour and CP are exact symmetries of the soft supersymmetry breaking sector defined at the appropriate
cutoff scale $\Lambda$ (to be identified with the Planck scale for supergravity, the messenger mass for gauge mediation, etc).
If below this scale there are flavour and CP-violating Yukawa interactions, it is well-known that in the running down to $m_{\rm SUSY}$ 
they will induce a small amount of flavour and CP-violation in sparticle masses. 

The Yukawa interactions associated to the fermion masses and mixing of the SM clearly violate {any flavour and CP symmetries.}
However, {with the exception of the third generation Yukawa couplings, all the entries in the Yukawa matrices} are very small
and the radiatively induced misalignment in the sfermion mass matrices turns out to be negligible. 
The Yukawa interactions of heavy states beyond the SM coupling to the SM fermions induce misalignments proportional to a proper combination of their 
Yukawa couplings times $\ln m_F/\Lambda$, where $m_F$ represents the heavy state mass scale. This is the case for the seesaw interactions of the 
right-handed neutrinos \cite{Borzumati:1986qx,Hall:1985dx} 
and/or the GUT interactions of the heavy colored triplets \cite{Barbieri:1994pv,Barbieri:1995tw} 
(those eventually exchanged in diagrams inducing proton decay). Notice that the observation of large mixing in light neutrino masses, may 
suggest the possibility that also the seesaw interactions could significantly violate flavour- and potentially also CP, in particular in view of the
mechanism of leptogenesis. Remarkably, for sparticle masses not exceeding the TeV, the seesaw and colored-triplet induced radiative contributions to the 
LFV decays and lepton EDM might be close to or even exceed the present or planned experimental limits.
Clearly, these processes constitute an important constraint on seesaw and/or GUT models.

\begin{figure}
\begin{center}
\includegraphics[width=13cm]{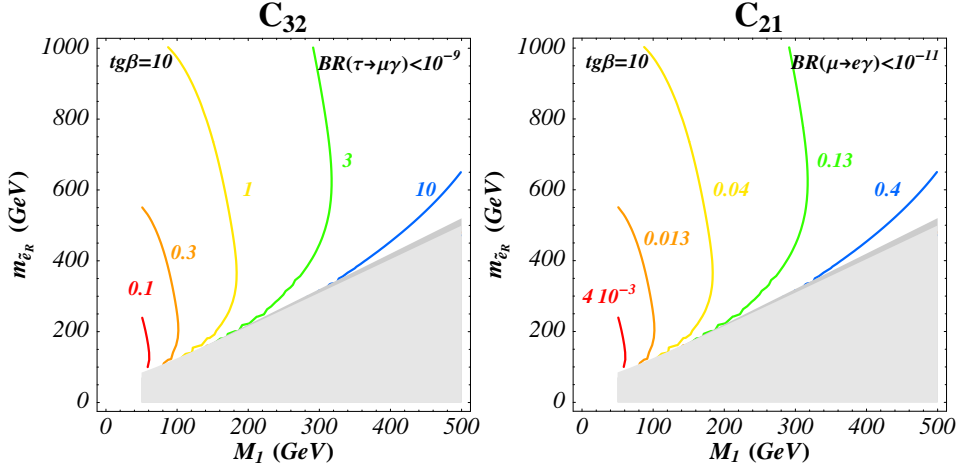}
\end{center}
\caption{Upper limit on $C_{32}$ and $C_{21}$ for the experimental sensitivities displayed \cite{Lavignac:2001vp}.}
\label{figC}
\end{figure}

For instance, in a type I seesaw model in the low-energy basis where charged leptons are diagonal, {the $ij$ element of the left-handed slepton mass
matrix provides the dominant contribution} in the decay $\ell_i \to \ell_j \gamma$. 
Assuming, for the sake of simplicity, an mSUGRA spectrum at $\Lambda= M_{\rm Pl}$, one obtains at the leading log \cite{Hisano:1995cp}: 
\begin{equation}
\delta^{LL}_{ij}= \frac{\left(m^2_{ij}\right)_{\rm LL}}{m^2_L}= -\frac{1}{8 \pi^2} \frac{3 m_0^2 + A_0^2}{m^2_L}~  C_{ij} ~~~, 
~~~~~~~~  C_{ij} \equiv \sum_k  ~ {Y_\nu}^*_{ki} ~{Y_\nu}_{kj}~  \ln\frac{M_{\rm Pl}}{M_k}  ~,
\label{eqC}
\end{equation} 
where $m_0$ and $A_0$ are respectively the universal scalar masses and trilinear couplings at $M_{\rm Pl}$, $m_L^2$ is {an average left-handed 
slepton mass and $M_k$ the mass of the right-handed neutrino with k=1,2,3}. An experimental limit on $B(\ell_i \to \ell_j \gamma)$ corresponds to 
an upper bound on $|C_{ij}|$ \cite{Lavignac:2001vp,Lavignac:2002gf}. For $\mu \to e \gamma$ and $\tau \to \mu \gamma$
this bound is shown in Fig.~\ref{figC} as a function of the right-handed selectron mass.

The seesaw model dependence resides in $C_{ij}$. Notice that in the {\it fundamental} theory at high energy, { the size of $C_{ij}$ is
determined both by the Yukawa eigenvalues and the largeness} of the mixing angles of $V_{R}, V_{L}$, the unitary matrices which diagonalize $Y_\nu$
(in the basis where $M_R$ and $Y_e$ are diagonal): $V_R Y_\nu V_L = Y_\nu^{(diag)}$. {The left-handed misalignment between neutrino and charged-lepton
Yukawa's is given by $V_L$} and, due to the mild effect of the logarithm in $C_{ij}$, in first approximation $V_L$ itself diagonalizes $C_{ij}$. 
If we consider hierarchical $Y_\nu$ eigenvalues, $Y_3>Y_2>Y_1$, the contributions from $k=1,2$ in Eq.~(\ref{eqC}) can in first approximation be neglected
with respect to the contribution from the heaviest {eigenvalue} ($k=3$):
\begin{equation}
|C_{ij}| \approx |{V_L}_{i3} {V_L}_{j3}|~ Y_3^2~~  \log (M_{\rm Pl} /M_3)
\label{lead}\end{equation}
Taking supersymmetric particle masses around the TeV scale, it has been shown that many seesaw models predict $|C_{\mu e}|$ and/or 
$|C_{\tau \mu}|$ close to the experimentally accessible range. Let us consider the predictions for the seesaw-RGE induced contribution 
to $\tau \to \mu \gamma$ and $\mu \to e \gamma$ in the flavour models discussed previously.

The present experimental bound on $\tau \to \mu \gamma$ is not very strong but nevertheless promising.
In models with ``lopsided'' $Y_\nu$, one has ${V_L}_{32} {V_L}_{32} \approx 1/2$, hence $|C_{\tau \mu}| = O( 4 \times Y_3^2 )$, for $M_3 \simeq 4 \times
10^{15}$ GeV. This is precisely the case for the $U(1)$ flavor model discussed in 
Section~\ref{sec:flavoursymmetries}, where $Y_3 \approx \epsilon^{n_3^c}$
with $\epsilon \approx 0.22$ (the Cabibbo angle). For this model, planned  $\tau \to \mu \gamma$ searches could thus be successful 
if the heaviest right handed neutrino has null charge, $n^c_3=0$. On the contrary, in models with small ${V_L}_{23}$ mixing, like in the non-abelian models
discussed previously, the seesaw-RGE induced effect is below the experimental sensitivity.

The present experimental bound on $\mu \to e \gamma$ is already very severe in constraining $|C_{\mu e}|$. For instance, if $V_{L} \approx V_{CKM}$,
one obtains $C_{\mu e} = O (10^{-3}\times Y_3^2)$. As can be seen from Fig.~\ref{figC}, $V_L$ could in future be tested at a CKM-level if $Y_3 = O(1)$
\cite{Masiero:2002jn}. The predictions for $\mu \to e \gamma$ are however very model dependent. For the simple $U(1)$ flavour model of Section~\ref{sec:flavoursymmetries}, the
mixings of $V_L$ are of the same order of magnitude as those of $U_{PMNS}$ and one expects $|C_{\mu e}| = O(8 \times \epsilon^{2 n_3^c +1} )$:
if $n_3^c=0$ the prediction exceeds the experimental limit, which is respected only with $n_3^c\ge1$ \cite{Masina:2005am}.
On the contrary, the non-abelian models discussed previously have $Y_3\sim 1$, but the $V_L$-mixings are sufficiently small 
to suppress the seesaw-RGE induced effect below the { present experimental level} \cite{Masiero:2002jn}.


\subsubsubsection{Parameter dependence for degenerate heavy neutrinos}

Eq.~(\ref{eqC}) indicates that LFV in the minimal
supersymmetric seesaw model depends on soft supersymmetry breaking
masses as well as on the seesaw parameters. The latter can be
parameterized  via the heavy and light neutrino masses, the light
neutrino mixing matrix and  the orthogonal matrix $R$ of
Eq.~(\ref{eq:def-R}). The three complex mixing angles
parameterizing $R$ can be written as $\hat \theta_j=x_j+ i y_j,$
$j=1,2,3.$ For the following numerical examples we use the mSUGRA
point SPS1a \cite{Allanach:2002nj} for SUSY breaking masses.

\begin{figure}[h!]
\centering
\includegraphics[clip,width=0.50\textwidth]{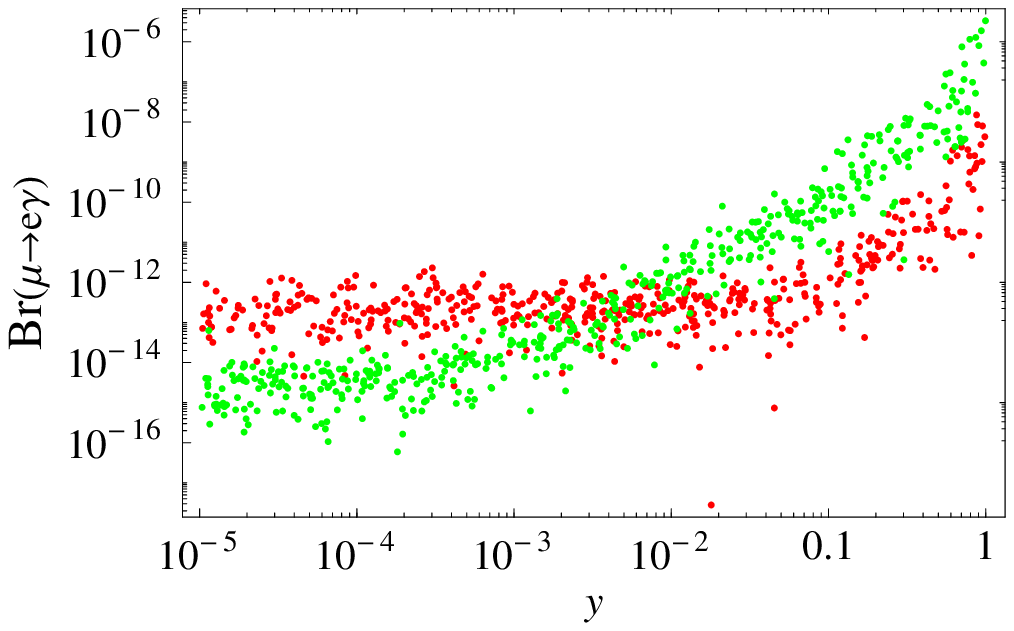}
\includegraphics[clip,width=0.49\textwidth]{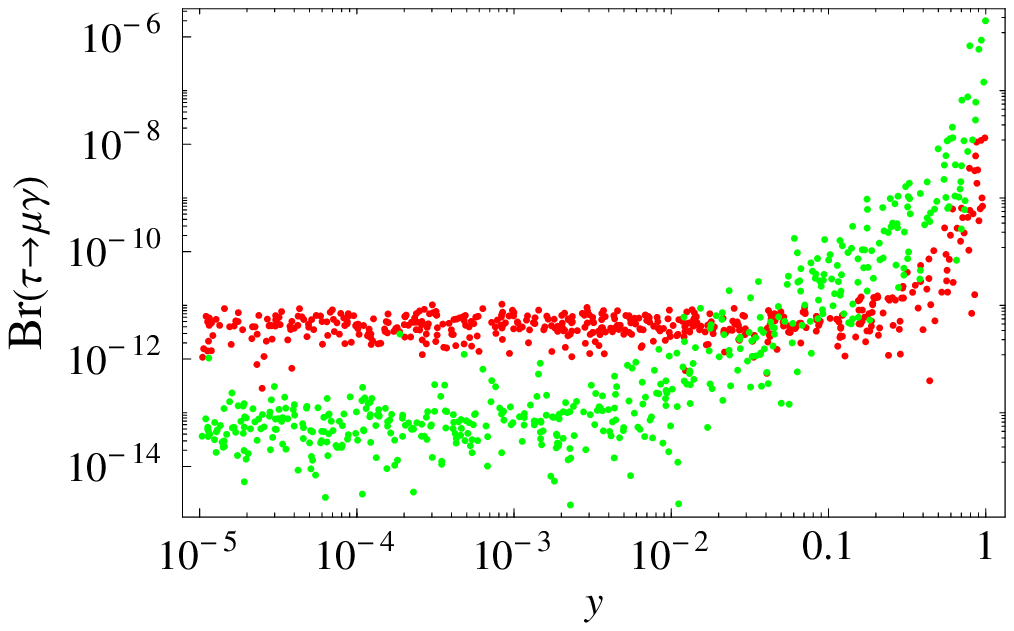}
\caption{Degenerate heavy neutrinos: LFV branching ratio versus
$|y_i|=y$ for fixed $M_i=M_R=10^{12}$ GeV in mSUGRA scenario SPS1a
for hierarchical (dark red) and degenerate (light green) light
neutrino masses. The $x_i$ are scattered over $0<x_i <2\pi$. }
\label{brijvsydegRfixMRscatallSPS1a}
\end{figure}

In the case of degenerate heavy neutrino masses, \(M_i=M_R\)
(\(i=1,2,3\)), and real \(R\), the \(R\) dependence in
Eq.~(\ref{eqC})  and hence also in \(B(l_i\to l_j\gamma)\)  drops
out. However, if $R$ is complex, the LFV observables have more
freedom since the dependence on $y_i$ can be as significant as the
$M_R$ dependence, as Fig.~\ref{brijvsydegRfixMRscatallSPS1a}
shows. For small $|y_i|$, the change in $Y_\nu^\dagger Y_\nu$  is
approximately
\begin{equation}
\Delta_R (Y_\nu^\dagger Y_\nu)\approx U_{PMNS} {\rm
diag}(\sqrt{m_i})(R^\dagger R-{\mathbf 1}){\rm
diag}(\sqrt{m_i})U_{PMNS} ^\dagger, \label{changeinYYfromy}
\end{equation}
while the renormalization effects on the soft supersymmetry
breaking masses can be estimated via~\cite{Petcov:2003zb}
\begin{equation}
m_L^8\simeq 0.5~m_0^2~M_{1/2}^2~(m_0^2 + 0.6 ~M_{1/2}^2)^2\;,
\label{eq_ms}
\end{equation}
where $M_{1/2}$ is the universal gaugino mass at high scale. In
certain cases, the leading logarithmic approximation fails, as pointed out
in~\cite{Petcov:2003zb,Chankowski:2004jc,Petcov:2005yh,Antusch:2006vw}.

Eq.~(\ref{changeinYYfromy}) implies three features seen in Fig.~\ref{brijvsydegRfixMRscatallSPS1a}:\\
(i) Compared to the case of degenerate light neutrino masses, the
$y$ dependence in the hierarchical case is weaker.
(ii) Observables like (\ref{BR}) are larger in the case of complex
\(R\) than in the case of real \(R\). For a given $M_R$, even
small values of $y$ can enhance a process by orders of magnitude.
(iii) In contrast to the real $R$ case, where \(B(l_i\to
l_j\gamma)\) for degenerate light neutrinos is always larger than
for hierarchical light neutrinos, the relative magnitude can be
reversed for complex \(R\).

\begin{figure}
\includegraphics[width=0.332\linewidth]{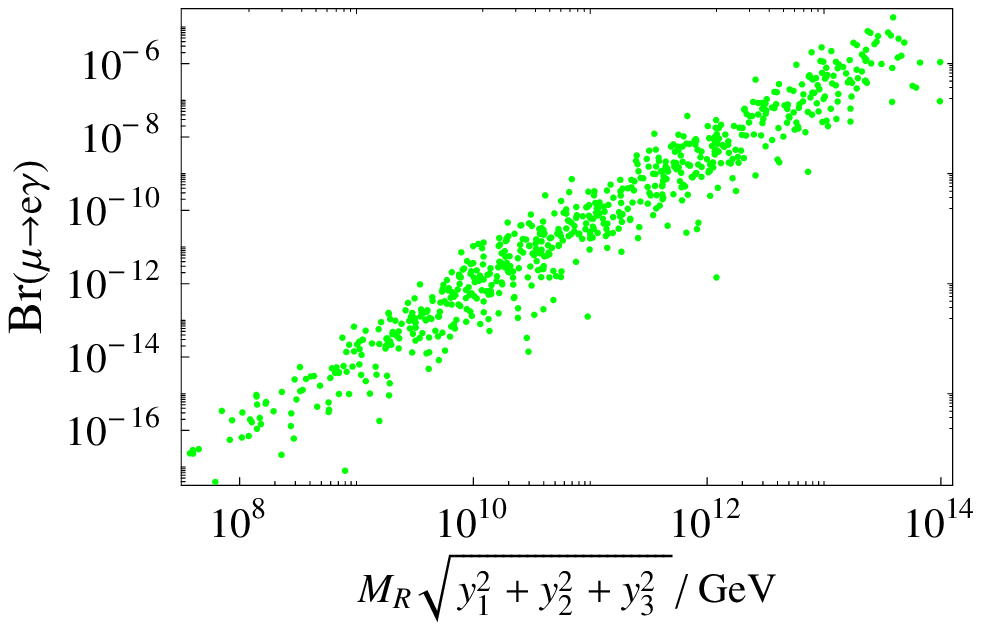}
\includegraphics[width=0.332\linewidth]{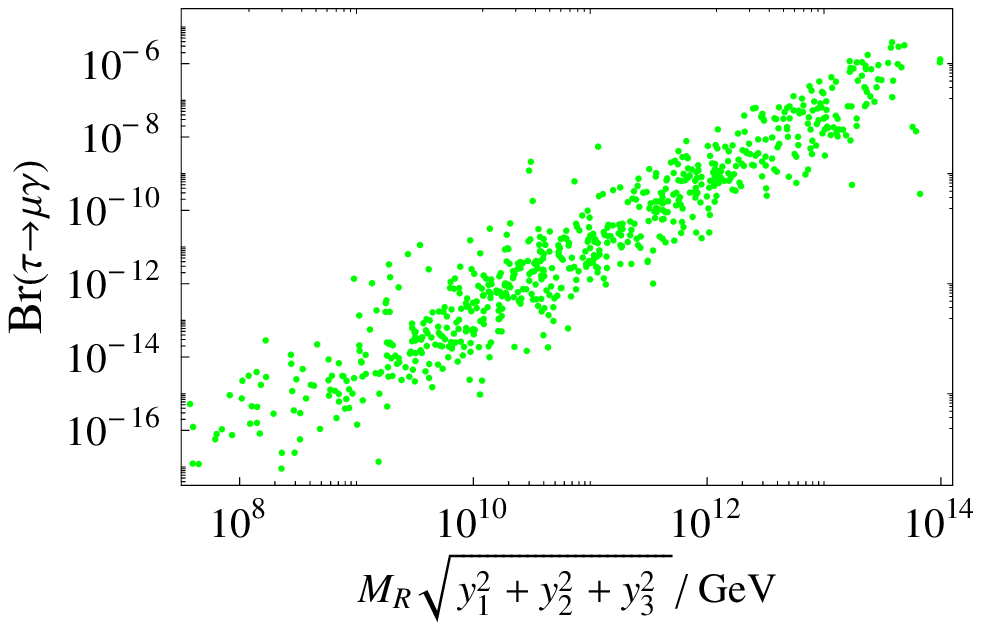}
\includegraphics[width=0.332\linewidth]{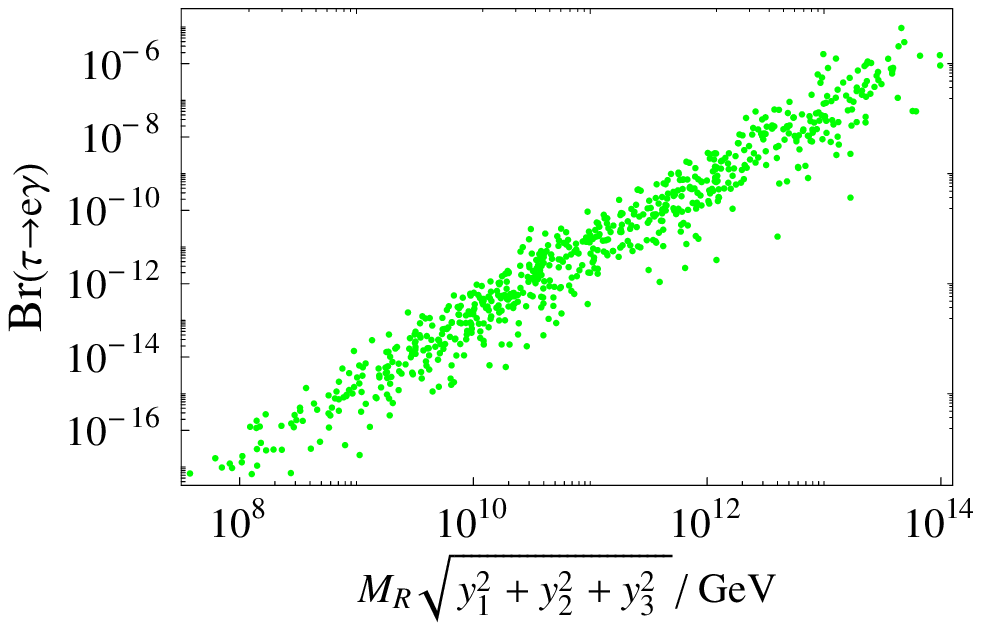}
\caption{Degenerate heavy neutrinos: LFV branching ratios versus
$M_R \sqrt{y_1^2+y_2^2+y_3^2}$, for light neutrinos. The $y_i$ are
scattered logarithmically in the range $10^{-5}<|y_i|<1$
(independently of one another) and $M_R$ is scattered
logarithmically in the range $10^{10}<M_R<M_{\rm GUT}$. The $x_i$
are scattered over $0<x_i <2\pi$. }
\label{br122331vsMRfydegRdegLscatallSPS1a}
\end{figure}

To examine the parameter dependence of rare decays at large
$|y_i|>0.1$, we extend the above analysis to the case where the
$y_i$ are independent of one another. For random values of all
parameters in their full ranges, the typical behavior
\begin{equation}
|(Y_\nu^\dagger L Y_\nu)_{jk}|^2 \propto
\begin{cases}
    M_R^2(C_1 y_1^2 +C_2 y_2^2 + C_3 y_3^2) & \text{deg. }\nu_L\\
    M_R^2                                   & \text{hier. }\nu_L
\end{cases} \qquad \qquad (j\neq k),
\end{equation}
is found, with \(C_i=O(1)\), slightly dependent on \(j,k\). This
behavior can be seen in
Fig.~\ref{br122331vsMRfydegRdegLscatallSPS1a} for degenerate light
neutrinos. Thus for large $|y_i|$ all rare decays may be of a
similar order of magnitude.
For hierarchical light neutrinos, a similar behavior is observed,
but versus \(M_R^2\) only.

\subsubsubsection{Parameter dependence for hierarchical heavy neutrinos}

 Hierarchical spectrum of heavy Majorana neutrinos, $M_1 \ll M_2 \ll M_3$,
 is well motivated by the arguments of light neutrino mass and mixing generation
  and leptogenesis. Requiring successful thermal  leptogenesis puts additional
  constraints on the seesaw parameters and constrains the LFV observables \cite{Ellis:2002xg}.
  This is the approach we take in this subsection.
 In particular, the relation
(\ref{eqn:CPAsymmetry})  implies a lower bound on \(M_1\)
\cite{Davidson:2002qv}, e.g., if \(\epsilon_1>10^{-6}\), then
\(M_1> 5\cdot 10^{9}\)~GeV. Furthermore, to allow for thermal
production of right-handed neutrinos after inflation, one has to
exclude $M_1>10^{11}$ GeV, at least in simple scenarios. Otherwise
a too high re-heating temperature would lead to an overabundance
of gravitinos, whose decays into energetic photons can spoil big
bang nucleosynthesis. Details of leptogenesis have been described
in Section~\ref{sec:lepto}.

\begin{figure}[t]
\includegraphics[clip,width=0.49\textwidth]{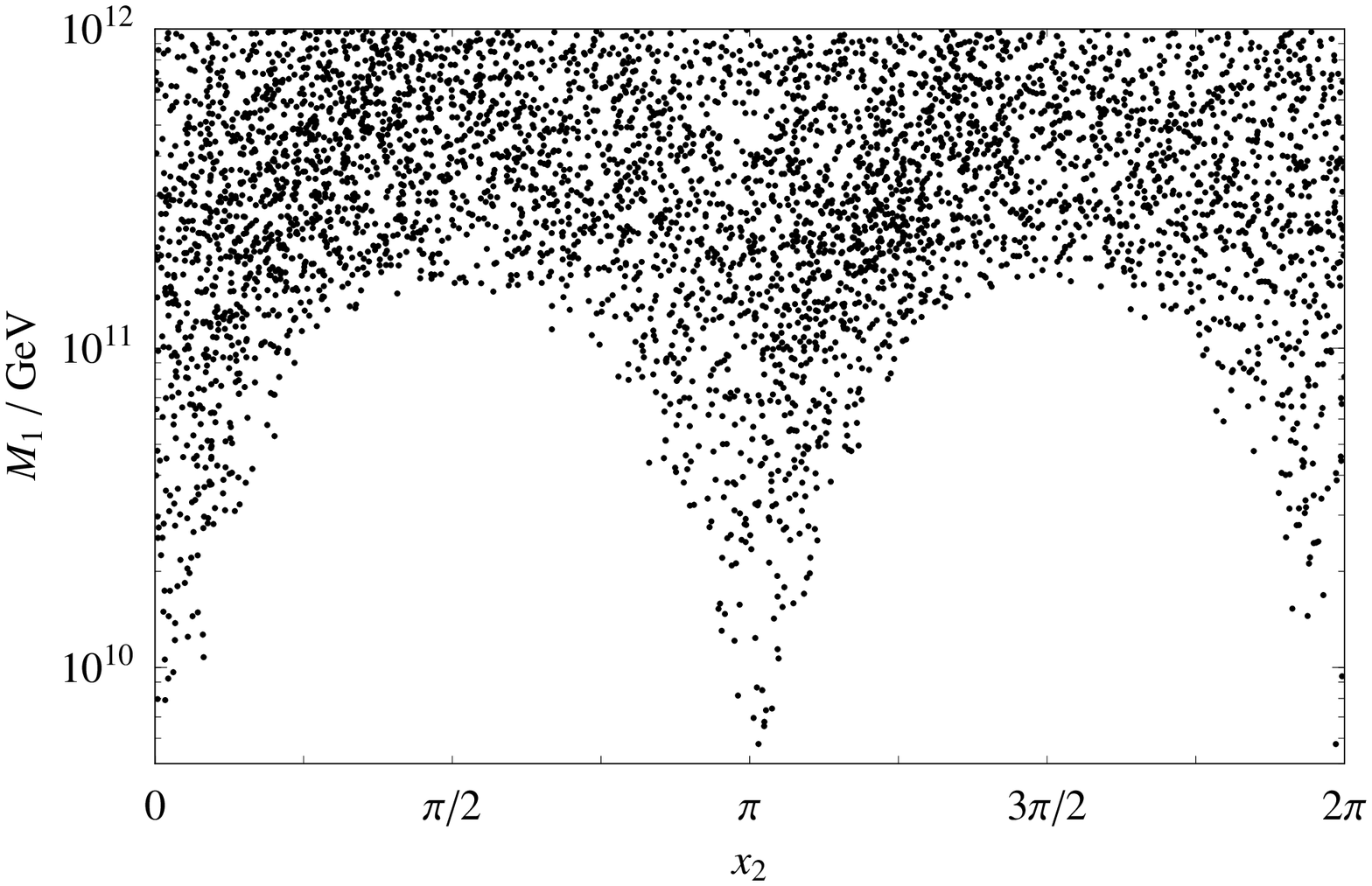}\hspace*{\fill}
\includegraphics[clip,width=0.49\textwidth]{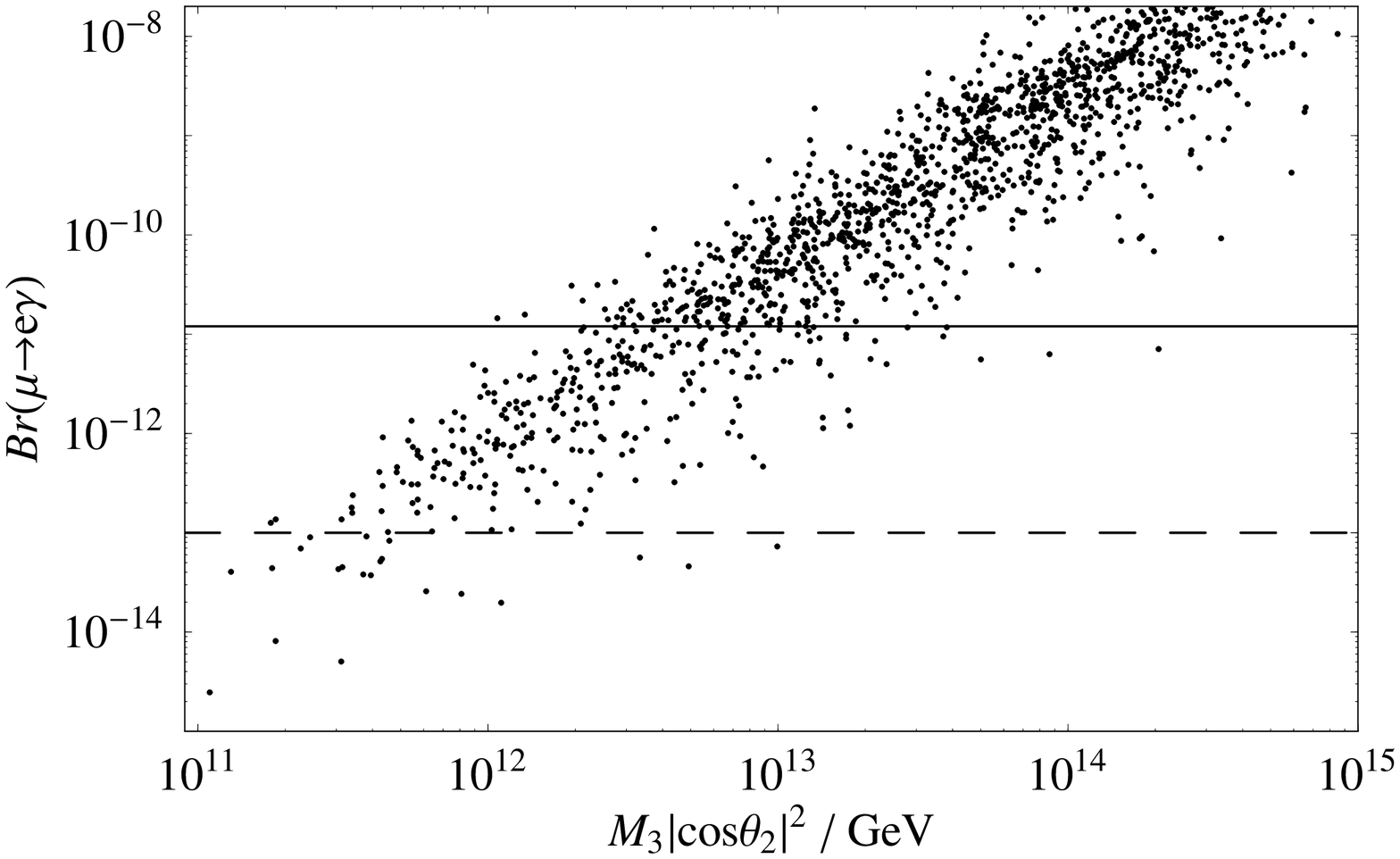}
\caption{Hierarchical heavy neutrinos: Region in the plane $(x_2,
M_1)$ consistent with the generation of the baryon asymmetry
\(\eta_B=(6.3\pm 0.3)\cdot 10^{-10}\) via leptogenesis (left). [b]
\(B(\mu\to e\gamma)\) vs $M_3|\cos^2\theta_2|$ in mSUGRA scenario
SPS1a, for \(M_1=10^{10}\)~GeV and \(x_2\approx x_3\approx
n\cdot\pi\). All other seesaw parameters are scattered in their
allowed ranges for hierarchical light and heavy neutrinos. The
solid (dashed) line indicates the present (expected future)
experimental sensitivity.} \label{fig:torboegen}
\end{figure}
Assuming hierarchical light neutrinos with \(\sqrt{\Delta
m^2_{sol}}<m_3<\sqrt{\Delta m^2_{atm}}\), the condition to
reproduce the experimental baryon asymmetry, $\eta_B=(6.3\pm
0.3)\cdot 10^{-10}$, puts constraints on \(M_1\) and the \(R\)
matrix~\cite{Deppisch:2005rv}. This is illustrated in
Fig.~\ref{fig:torboegen} in the \(M_1-x_2\) plane. For
\(M_1<10^{11}\)~GeV, \(x_2\) has to approach the values
\(0,\pi,2\pi\). A similar behavior is observed in the \(M_1-x_3\)
plane.

Taking \(M_1=10^{10}\)~GeV and \(x_2\approx x_3\approx
n\cdot\pi\), experimental bounds on \(B(\mu\to e\gamma)\) can be
used to constrain the heavy neutrino scale, here represented by
the heaviest right handed neutrino mass \(M_3\), as shown in the
right plot of Fig.~\ref{fig:torboegen}. Quantitatively, the
present bound on \(B(\mu\to e\gamma)\) already constrains $M_3$ to
be smaller than $\approx 10^{13}$~GeV, while the MEG experiment at
PSI is sensitive to \(M_3\leq O(10^{12})\)~GeV. If no signal is
observed it will be difficult to test the type I seesaw model
considered here at future colliders.

\subsubsubsection{Effects of renormalization of light neutrino masses on LFV}

The RG running of the neutrino parameters can have an important
impact on lepton flavour violating processes in MSSM extended by
right-handed neutrinos. In this example we assume universal soft
SUSY breaking terms at GUT scale and degenerate heavy neutrinos
with mass $M_R.$
\begin{figure}[t]
\includegraphics[width=0.48\linewidth]{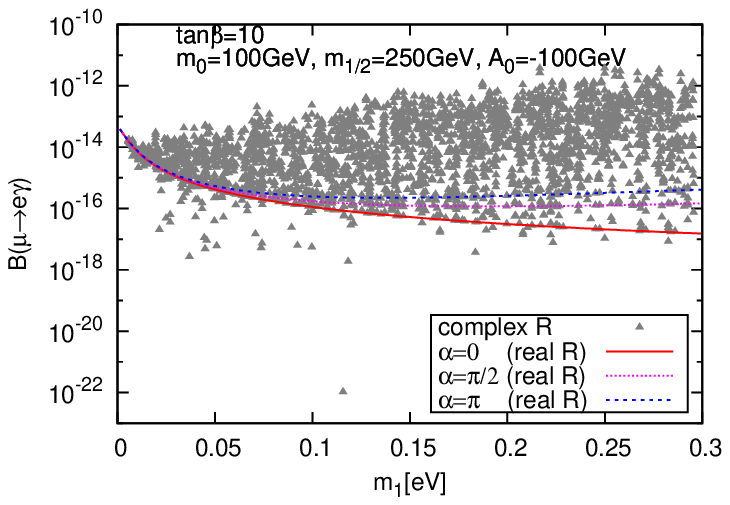}\hspace*{\fill}
\includegraphics[width=0.48\linewidth]{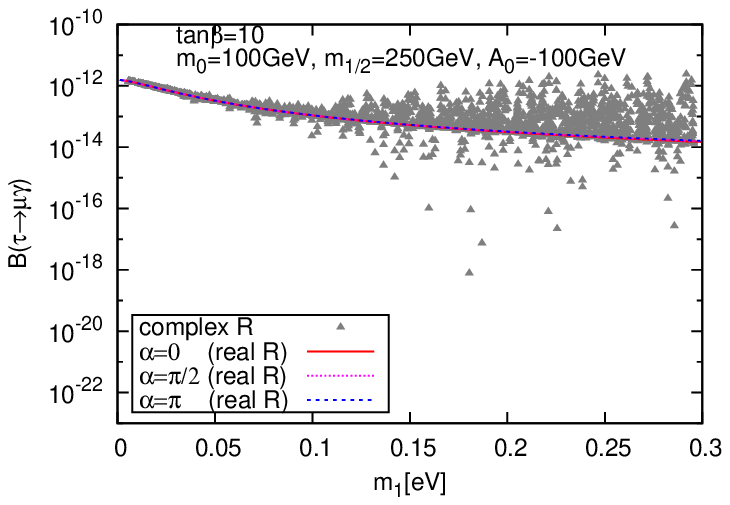}
\caption{The branching ratios of the LFV decays $\mu\to e +
\gamma$ and $ \tau \to \mu + \gamma$ versus $m_1$ in the cases of
complex and real matrix ${R}$ with $\alpha = 0;~\pi/2;~ \pi$. The
three parameters describing
 the matrix ${R}$
\cite{Pascoli:2003rq,Petcov:2005yh} are generated randomly. The
SUSY parameters are $\tan\beta=10$, $m_0=100$ GeV, $m_{1/2}=250$
GeV, $A_0=-100$ GeV, and the neutrino mixing parameters are
$\Delta m_{\odot}^2 = 8.0\times 10^{-5}~\text{eV}^2$, $\Delta
m_{\rm atm}^2 = 2.2\times 10^{-3}~\text{eV}^2$,
$\tan^2\theta_{\odot}=0.4$, $\tan^2\theta_{\rm atm} = 1$, and
$\sin\theta_{13} = 0.0$.  The neutrino mass spectrum at $M_Z$ is
assumed to be with normal hierarchy, $m_1(M_Z) < m_2(M_Z) <
m_3(M_Z)$.  The right-handed neutrino mass spectrum is taken to be
degenerate as $M_1=M_2=M_3=2\times
10^{13}$~GeV~\cite{Petcov:2005yh}.} \label{fig:neuRGE}
\end{figure}
The running effects below $M_R$ are relatively small when
$\tan\beta$ is smaller than 10 and/or $m_1$ is much smaller than
$0.05$ eV.  Because the combination $s_{12}c_{12}c_{23}(m_1 -
m_2e^{i \alpha_{\scriptscriptstyle M}})$, where we use the
notation of Section~\ref{sec:nurenorm}, is practically stable
against the RG running, and this combination is the dominant term
of $(Y_{\nu}^{\dagger}Y_{\nu})_{21}$ when
$\alpha_{\scriptscriptstyle M}=0$, $\theta_{13}=0$ and $R^*=R$ are
satisfied, the running effect on LFV can be neglected in this
case~\cite{Petcov:2005yh}.  In general,
$(Y_{\nu}^{\dagger}Y_{\nu})_{21}$ and $B(\mu \to e +
\gamma)$ can depend strongly on $\theta_{13}$ and RG running has
to be taken into account~\cite{Antusch:2006vw,Calibbi:2006ne}.
Note that due to RG running, the value of $\theta_{13}$ at $M_R$
differs from $0$, even if $\theta_{13}=0$ is assumed at low
energy~\cite{Petcov:2005yh}.

In many cases, the running of the neutrino parameters can
significantly affect the prediction of the LFV branching ratios.
In particular, for $ 0.05 \lesssim m_1 \lesssim 0.30$ eV, $30
\lesssim \tan\beta \lesssim 50$, the predicted $ \mu \to e +
\gamma$ and $\tau \to e + \gamma$ decay branching ratios,
$B(\mu \to e + \gamma)$ and $B(\tau \to e +
\gamma)$, can be enhanced by the effects of the RG running of
$\theta_{ij}$ and $m_j$ by 1 to 3 orders of magnitude if $\pi/4
\lesssim \alpha_{\scriptscriptstyle M} \lesssim \pi$, while
$B(\tau\to \mu + \gamma)$ can be enhanced by up to a factor
of 10~\cite{Petcov:2005yh}. The effects of the running of the
neutrino mixing parameters of $B(\mu \to e + \gamma)$ and
$B(\tau \to e + \gamma)$ are illustrated in
Fig.~\ref{fig:neuRGE}.

\subsubsection{Correlations between LFV observables and collider physics}

\subsubsubsection{Correlations of LFV rare decays}

\begin{figure}[t]
\center
\includegraphics[bb=88 10 338 187,height=7cm]{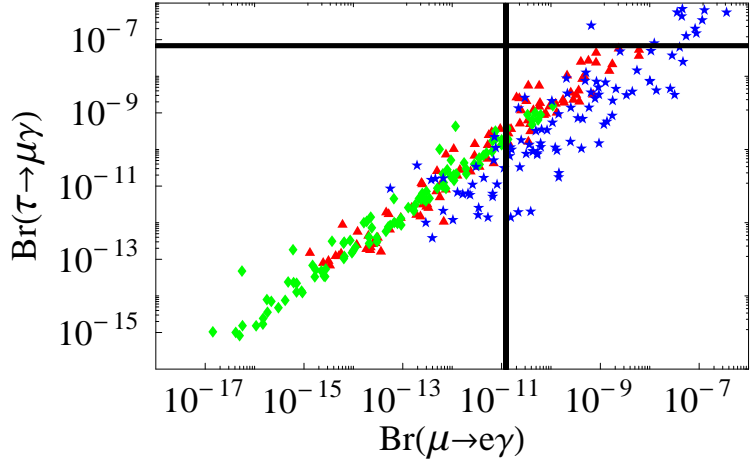}
\caption{$B(\tau \to \mu \gamma)$ versus $B(\mu \to e\gamma)$, in mSUGRA scenario SPS1a with neutrino parameters scattered within
their experimentally allowed ranges \cite{Maltoni:2003sr}. For quasi-degenerate heavy neutrino masses, both hierarchical (triangles)
and quasi-degenerate (diamonds) light neutrino masses are considered with real $R$ and $10^{11}\ {\rm GeV}<M_R <
10^{14.5}\ {\rm GeV}$. In the case of hierarchical heavy and light neutrino masses (stars), the $x_i$ are scattered over their
full ranges $0<x_i <2\pi$ and the $y_i$ and $M_i$ are scattered within the bounds demanded by leptogenesis and perturbativity.
Also indicated are the present experimental bounds $B(\mu \to e \gamma)< 1.2\times 10^{-11}$ and $B(\tau \to \mu \gamma)<6.8\times
10^{-8}$~\protect\cite{Eidelman:2004wt,Aubert:2005ye}.}\label{br12v23SPS1aallRallL2}
\end{figure}

Eq.~(\ref{BR}) and Eq.~(\ref{eqC})  imply correlations
between different LFV observables. In addition to the correlations
between different classes of LFV observables in the same flavour
mixing channels, the assumed LFV mechanism induces also
correlations among the $|(m_L)^2_{ij}|^2$ and hence among
observables of different flavour mixing channels. In this
framework, the ratios of the branching ratios are
approximately independent of SUSY parameters:
\begin{equation}
\frac{B(\tau\to\mu\gamma)} {B(\mu\to e\gamma)}
\propto\frac{|(Y_{\nu}^{\dagger}LY_{\nu})_{23}|^2}
{|(Y_{\nu}^{\dagger}LY_{\nu})_{12}|^2}\;. \label{ratio_to_ratio}
\end{equation}
Thus the measurement of the ratio between the decay rates of the different LFV channels can provide unique information on the
flavour structure of the lepton sector. The ratios of interest, such as
 Eq.~(\ref{ratio_to_ratio}), can exhibit, for instance, strong
dependence on CP-violating parameters in neutrino Yukawa couplings\cite{Petcov:2006pc} especially in the case of
quasi-degenerate heavy RH neutrinos. As a consequence such correlations have been widely studied (see, {\em e.g.},
\cite{Hisano:1995cp,Lavignac:2001vp,Ellis:2002fe,Ellis:2002xg,Ellis:2002eh,Rossi:2002zb,Deppisch:2002vz,Chankowski:2004jc,Arganda:2005ji,Petcov:2006pc}
and the references quoted therein). Consequently, bounds on one
LFV decay channel (process) will limit the parameter space of the
LFV mechanism and thus lead to bounds on the other LFV decay
channels (processes). In Fig.~\ref{br12v23SPS1aallRallL2}, the
correlation induced by the type~I seesaw mechanism between $B(\mu
\to e \gamma)$ and $B(\tau \to \mu \gamma)$ is shown, and the
bounds induced by the former on the latter can be easily read off.
Interestingly, these bounds do not depend on whether hierarchical
or quasi-degenerate heavy and light neutrinos are assumed. The
present and future prospective bounds are summarized in
Table~\ref{Br23fromBr12seesawforpresentfuture}. Note that the
present upper bound on $B(\mu \to e \gamma)$ implies a stronger
constraint on $B(\tau \to \mu \gamma)$ than its expected future bound.

\begin{table}[bht]
\begin{minipage}{\textwidth}
\caption{Present and expected future bounds on $B(\mu \to e
\gamma)$ from experiment, and bounds on $B(\tau \to \mu \gamma)$
from (i) experiment (ii) the bound on $B(\mu \to e \gamma)$
together with correlations from the SUSY type~I seesaw mechanism.
\label{Br23fromBr12seesawforpresentfuture}}
\begin{tabular*}{\textwidth}{@{\extracolsep{\fill}}l|ccc}
\hline\hline
        &$B(\mu \to e \gamma)$ (exp.)   &$B(\tau \to \mu \gamma)$ (exp.)    &$B(\tau \to \mu \gamma)$
\footnote{from $B(\mu\to e\gamma)$ (exp.) and SUSY seesaw}      \\
\hline
Present     &$1.2 \times 10^{-11}$      &$6.8 \times 10^{-8}$           &$10^{-9}$          \\
Future      &$10^{-14}$             &$10^{-9}$              & $10^{-12}$            \\
\hline\hline
\end{tabular*}
\end{minipage}
\end{table}

The above results were derived in the simplifying case of a real
\(R\) matrix. For complex \(R\) with \(|y_i|<1\) there is no
significant change with respect to the results in
Table~\ref{Br23fromBr12seesawforpresentfuture} in the case of
hierarchical heavy and hierarchical light neutrinos due to the
weak $R$ dependence of $B(\mu\to e\gamma)$ and
$B(\tau\to\mu\gamma)$. However, for quasi-degenerate light
neutrinos, $B(\tau\to \mu \gamma)$ is lowered by roughly one order
of magnitude, somewhat spoiling the overlap of all scenarios
observed in Fig.~\ref{br12v23SPS1aallRallL2}.

\begin{figure}[t]
\includegraphics[width=0.49\linewidth]{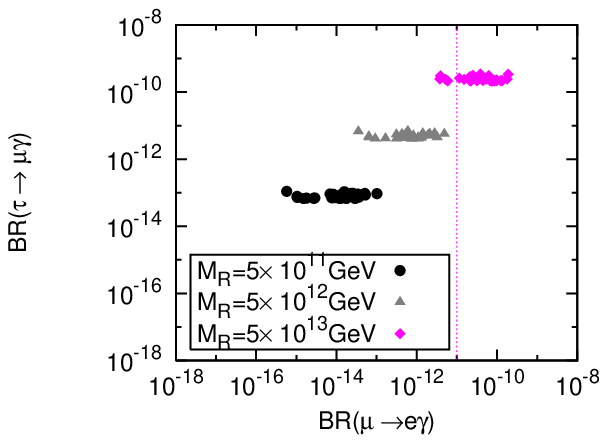}
\includegraphics[width=0.49\linewidth]{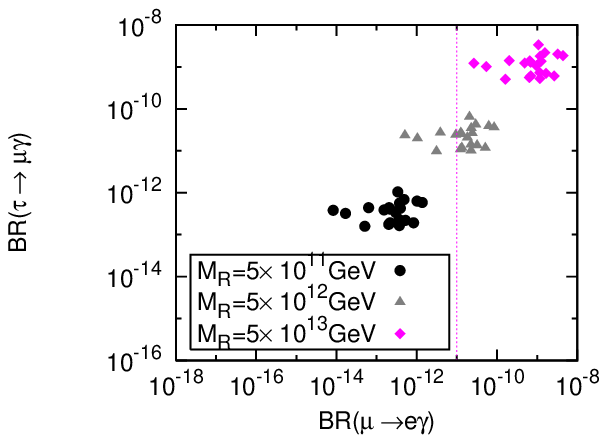}
\caption{The correlation between $B(\mu\to e\gamma)$ and
$B(\tau\to \mu \gamma)$ for quasi-degenerate heavy
neutrinos and light neutrino mass spectrum of normal hierarchical
(left panel) and inverted hierarchical (right panel) type. }
\label{br12and23fordegRandNHIHLcompR}
\end{figure}

In Fig.~\ref{br12and23fordegRandNHIHLcompR}, we display the
correlation between $B(\mu\to e\gamma)$ and
$B(\tau\to \mu\gamma)$ for complex $R$ and some fixed values
of $M_R$ in the case of quasi-degenerate RH neutrino masses and a
normal and inverted hierarchical light neutrino mass spectrum. We
note that, as Fig.~\ref{br12and23fordegRandNHIHLcompR} suggests,
$B(\tau\to \mu\gamma)$ is almost independent of the CP
violating parameters and phases respectively in $R$ and $U$, while
the dependence of $B(\mu\to e\gamma)$ on the CP-violating
quantities is much stronger. This is reflected, in particular, in
the fact that for a fixed $M_R$, $B(\tau\to \mu\gamma)$ is
practically constant while $B(\mu\to \mu\gamma)$ can change
by 2-3 orders of magnitude.

If the $\mu\to e\gamma$ and $\tau\to\mu\gamma$ decays will be
observed, the ratio of interest can give unique information on the
origin of the lepton flavour violation.

\subsubsubsection{LFV rare decays and linear collider processes}\label{sec:LC}

\begin{figure}[t]
\includegraphics[width=0.49\linewidth]{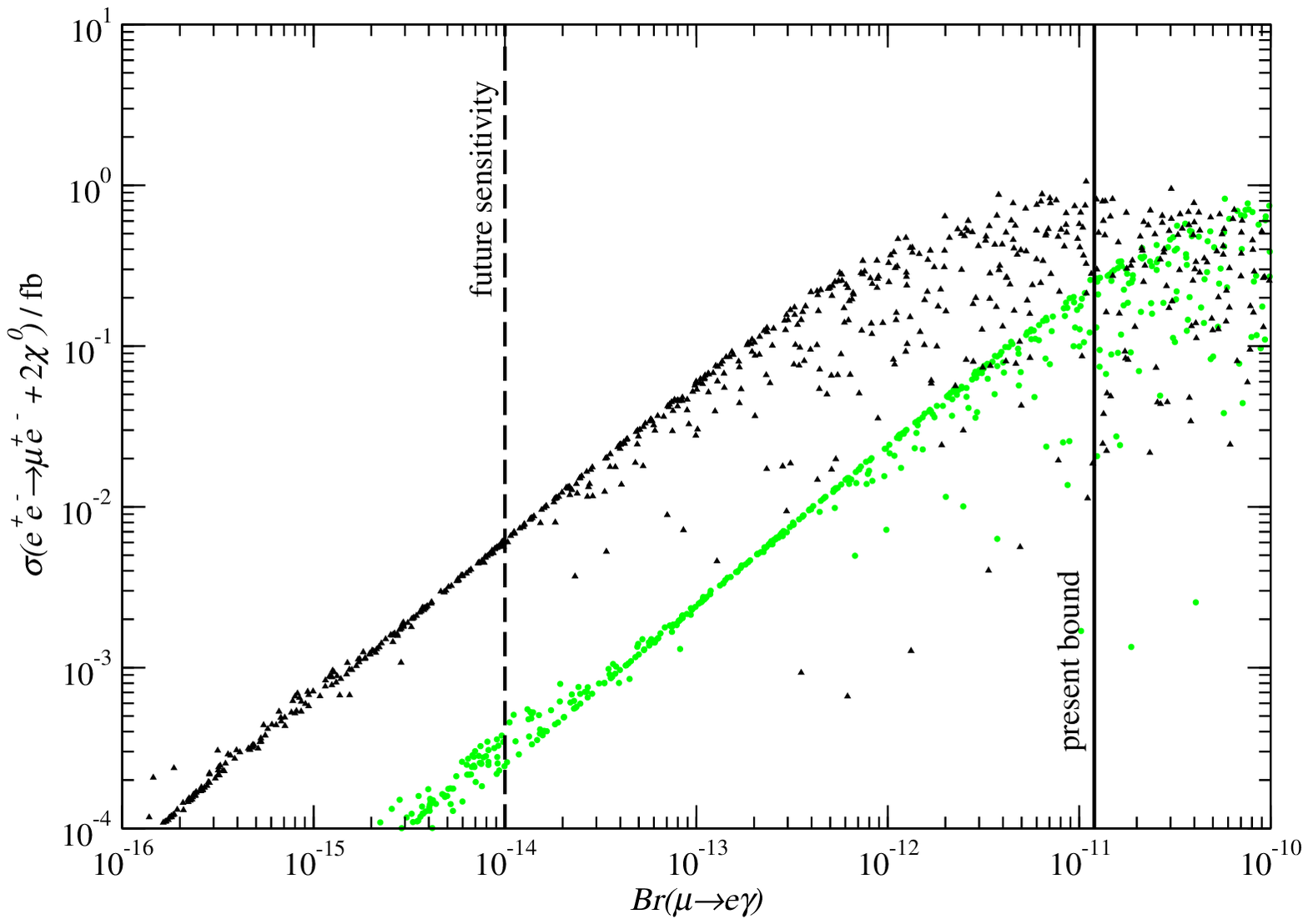}
\hspace*{\fill}\includegraphics[width=0.49\linewidth]{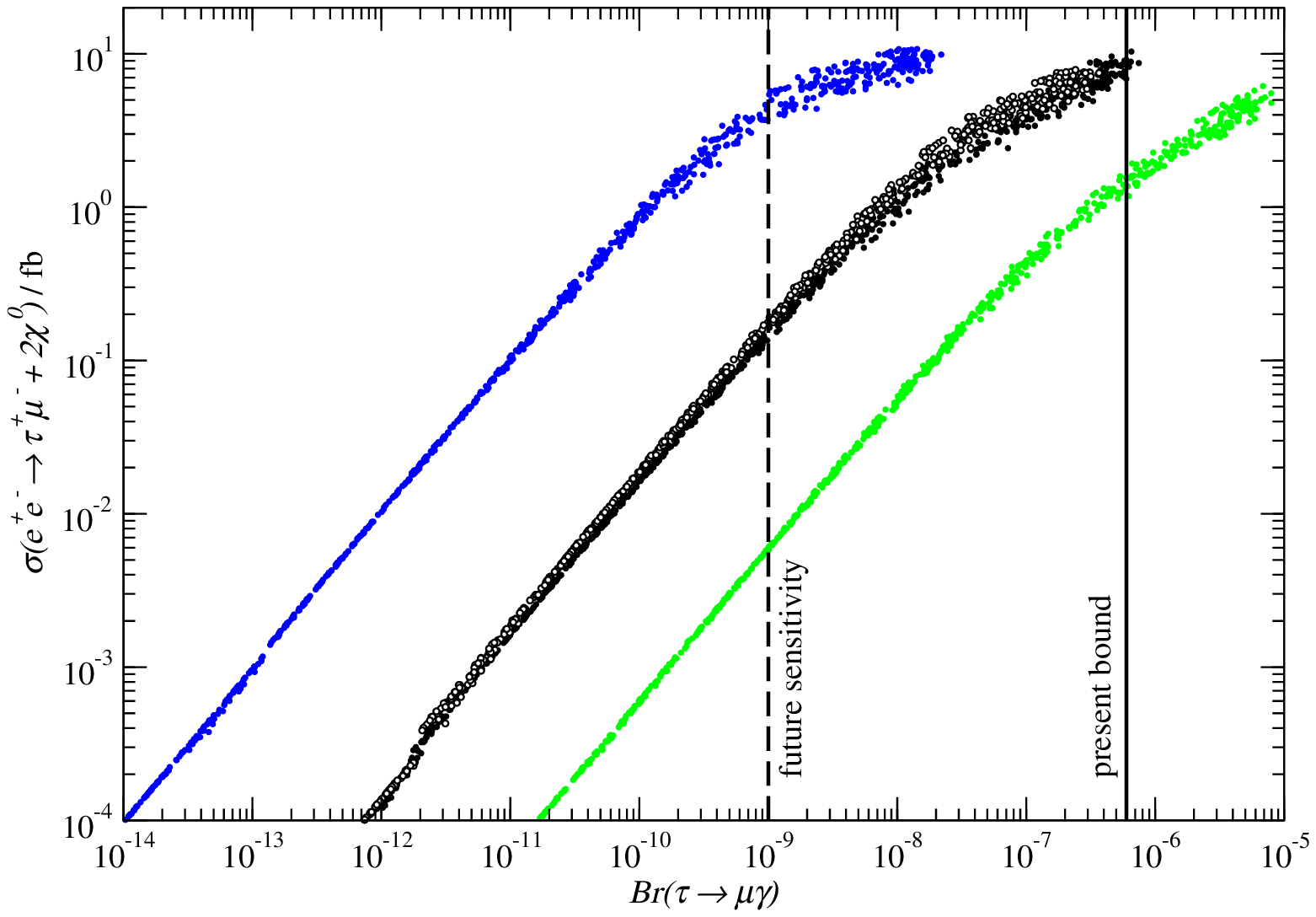}
\caption{ Correlation of LFV LC processes and rare decays in the
\(e\mu\)-channel (left) and the \(\mu\tau\)-channel (right). The
seesaw parameters are scattered as in
Fig.~\ref{br12v23SPS1aallRallL2}. The mSUGRA scenarios used are
(from left to right): SPS1a, G' (\(e\mu\)) and C', B', SPS1a, I'
(\(\mu\tau\)). }\label{br12vsMRdegRSPS1a}
\end{figure}

At high energies, feasible tests of LFV are provided by the
processes $e^+e^- \to \tilde{l}_a^-\tilde{l}^+_b\to l_i^-l^+_j +
2\tilde{\chi}^0_1$. Analogously to (\ref{BR}), one can derive the
approximate expression~\cite{Deppisch:2003wt}
\begin{equation}\label{eqn:HighEnergyApproximation}
 \sigma(e^+e^- \to l_i^- l_j^+ +2\tilde\chi^0_1)
 \approx
 \frac{|(\delta m_L)^2_{ij}|^2}{m^2_{\tilde l}
 \Gamma^2_{\tilde l}}
 \sigma(e^+e^- \to l_i^- l_i^+
 +2\tilde\chi^0_1),
\end{equation}
for the production cross section in the limit of small slepton
mass corrections. By comparing Eq.~(\ref{BR}) with
Eq.~(\ref{eqn:HighEnergyApproximation}), it is immediately
apparent that the LC processes are flavor-correlated with the rare
decays considered previously. These correlations are shown in
Fig.~\ref{br12vsMRdegRSPS1a} for the two most important channels.

This observation implies that once the SUSY parameters are known,
a measurement of, e.g., \ $B(\mu \to e \gamma)$ will lead to a
prediction for $\sigma(e^+e^- \to \mu e +2\tilde\chi^0_1)$. Quite
obviously, this prediction will be independent of the specific LFV
mechanism (seesaw or other). Figure~\ref{br12vsMRdegRSPS1a} also
demonstrates that the uncertainties in the neutrino parameters
nicely drop out except at large cross sections and branching
ratios.

\begin{figure}[t]
\centering
\includegraphics[height=8cm]{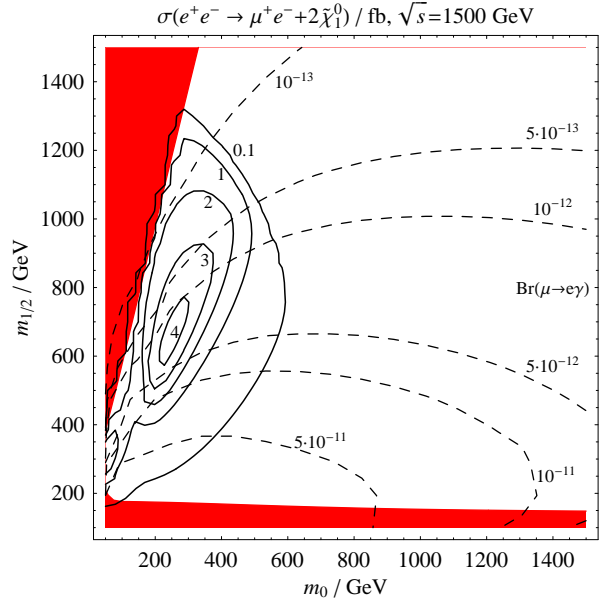}
\caption{Contours of the polarized cross section
\(\sigma(e^+e^-\to\mu^+e^- +2\tilde\chi_1^0)\) (solid) and \({\rm
B}(\mu\to e\gamma)\) (dashed) in the \(m_0-m_{1/2}\) plane. The
remaining mSUGRA parameters are \(A_0=0\)~GeV, \(\tan\beta=5\),
\(\text{sign}(\mu)=+\). The energy and beam polarizations are:
\(\sqrt{s}_{ee}=1.5\)~TeV, $P_{e^-} = +0.9, P_{e^+} = +0.7$. The
neutrino oscillation parameters are fixed at their central values
as given in \cite{Maltoni:2003sr}, the lightest neutrino mass
\(m_1=0\) and all complex phases are set to zero, and the
degenerate right handed neutrino mass scale is
\(M_R=10^{14}\)~GeV. The shaded (red) areas are already excluded
by mass bounds from various experimental sparticle
searches.}\label{fig:scan_ep1}
\end{figure}

In the previous results we have assumed a specific choice
of the as yet unknown mSUGRA parameters. The results of a more
systematic study of the model dependence are visualized in
Fig.~\ref{fig:scan_ep1} by contour plots for
\(\sigma(e^+e^-\to\mu^+e^- +2\tilde\chi_1^0)\) and \(B(\mu\to
e\gamma)\) in the \(m_0-m_{1/2}\) plane with the remaining mSUGRA
parameters fixed.

\subsubsubsection{LFV rare decays and LHC processes}
At the LHC, a
feasible test of LFV is provided by squark and
gluino production, followed by cascade decays of squarks and gluinos via
neutralinos and sleptons \cite{Agashe:1999bm,Andreev:2006sd}:
\begin{eqnarray}\label{eqn:LHCProcesses}
pp &\to& \tilde q_a \tilde q_b, \tilde g \tilde q_a, \tilde g
\tilde g,\nonumber\\
\tilde q_a(\tilde g)&\to& \tilde\chi^0_2 q_a(g),\nonumber\\
\tilde\chi^0_2 &\to& \tilde l_\alpha l_\beta, \nonumber\\
\tilde l_\alpha &\to& \tilde\chi^0_1 l_\beta,
\end{eqnarray} 
where \(a,b\) run over all squark mass eigenstates, including
antiparticles, and $\alpha,\beta$ are slepton (lepton) mass (flavour)
eigenstates, including antiparticles. LFV can occur in the decay
of the second lightest neutralino and/or the slepton, resulting in different lepton
flavors, \(\alpha\neq\beta\). The total cross section for the
signature \(l^+_\alpha l^-_\beta + X\) can then be written as
\begin{eqnarray}\label{eqn:LHCProcess}
 \sigma(pp\to l^+_\alpha l^-_\beta+X) &=&
 [\sum_{a,b}\sigma(pp\to\tilde q_a\tilde q_b)\times B(\tilde q_a\to\tilde\chi^0_2 q_a)\nonumber\\
 &\quad&+\sum_{a}\sigma(pp\to\tilde q_a\tilde g) \times(B(\tilde q_a\to\tilde\chi^0_2 q_a)+B(\tilde g\to\tilde\chi^0_2 g))\nonumber\\
 &\quad&+ \sigma(pp\to\tilde g\tilde g)\times B(\tilde g\to\tilde\chi^0_2 g)]\nonumber\\
 &\times& B(\tilde\chi^0_2\to l_\alpha^+
 l_\beta^-\tilde\chi^0_1),
\end{eqnarray}
where \(X\) can involve jets, leptons and LSPs produced by lepton
flavor conserving decays of squarks and gluinos, as well as low
energy proton remnants. The LFV branching ratio
\(B(\tilde\chi^0_2\to l_\alpha^+l_\beta^-\tilde\chi^0_1)\) is for
example calculated in \cite{Bartl:2005yy} in the framework of
model-independent MSSM slepton mixing. In general, it involves a
coherent summation over all intermediate slepton states.

\begin{figure}[t]
\includegraphics[clip,width=0.49\textwidth]{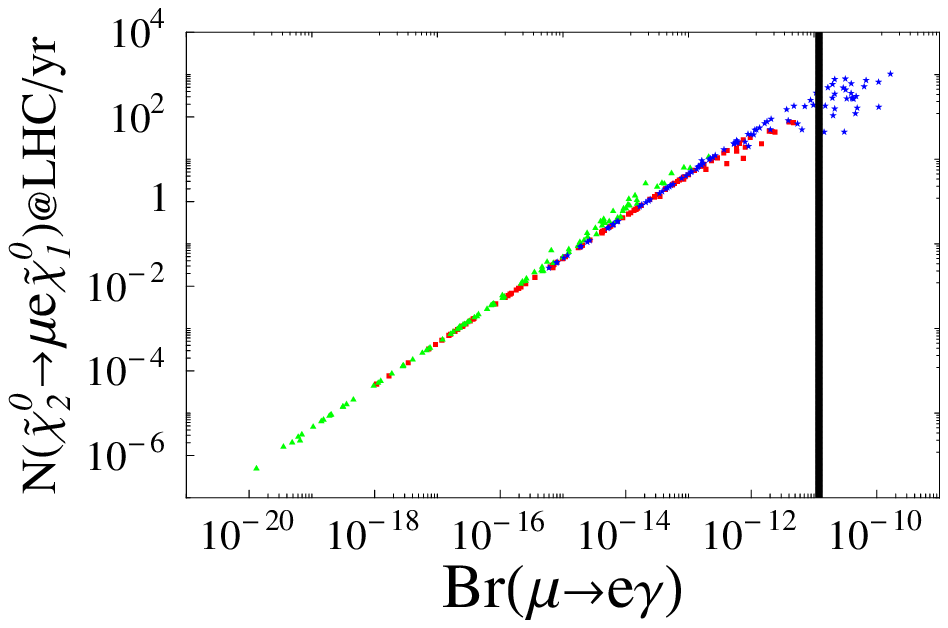}\hspace*{\fill}
\includegraphics[clip,width=0.49\textwidth]{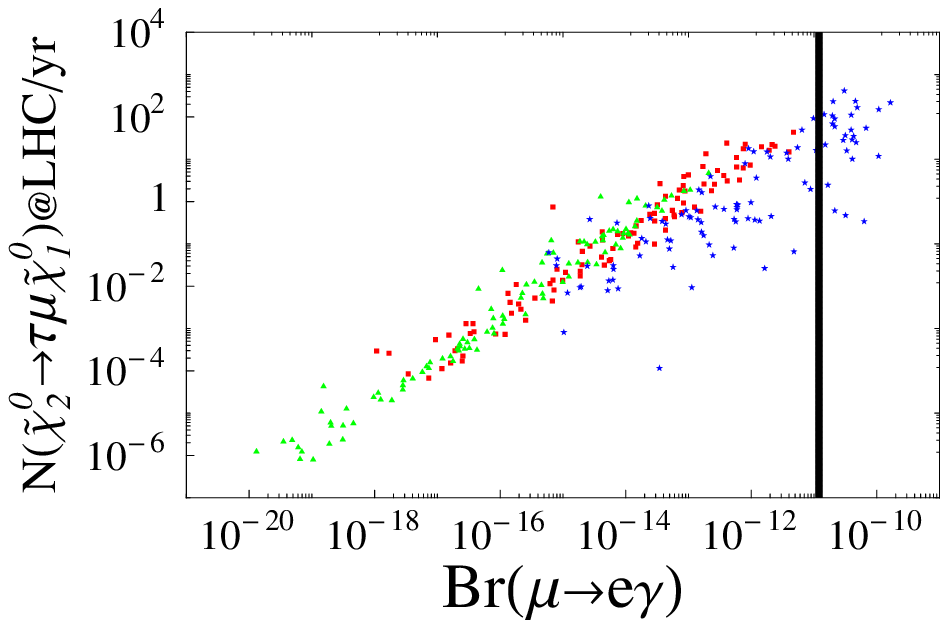}
\caption{Correlation of the number of
\(\tilde\chi_2^0\to\mu^+e^-\tilde\chi_1^0\) events per year at the
LHC and \(B(\mu\to e\gamma)\) in mSUGRA scenario C'
(\(m_0=85\)~GeV, \(m_{1/2}=400\)~GeV, \(A_0=0\)~GeV,
\(\tan\beta=10\)~GeV, \(\text{sign}\mu=+\)) for the case of hier.\
$\nu_{R/L}$ (blue stars), deg.\ $\nu_R$/hier.\ $\nu_L$ (red boxes)
and deg.\ $\nu_{R/L}$ (green triangles). The respective neutrino
parameter scattering ranges are as in
Fig.~\ref{br12v23SPS1aallRallL2}. An integrated LHC luminosity of
\(100\text{fb}^{-1}\) is assumed. The current limit on \(B(\mu\to
e\gamma)\) is displayed by the vertical line.}
\label{br12vsN2Cprime}
\end{figure}

Just as for the linear collider discussed in the previous section,
we can correlate the expected LFV event rates at the LHC with LFV
rare decays. This is shown in Fig.~\ref{br12vsN2Cprime} for the
event rates \(N(\tilde\chi_2^0\to\mu^+e^-\tilde\chi_1^0)\) and
\(N(\tilde\chi_2^0\to\tau^+\mu^-\tilde\chi_1^0)\), respectively,
originating from the cascade reactions (\ref{eqn:LHCProcesses}).
Both are correlated with \(B(\mu\to e\gamma)\), yielding maximum
rates of around \(10^{2-3}\) per year for an integrated luminosity
of (\(100\text{fb}^{-1}\)) in the mSUGRA scenario C', consistent
with the current limit on \(B(\mu\to e\gamma)\).

\begin{figure}[t]
\centering
\includegraphics[height=8cm]{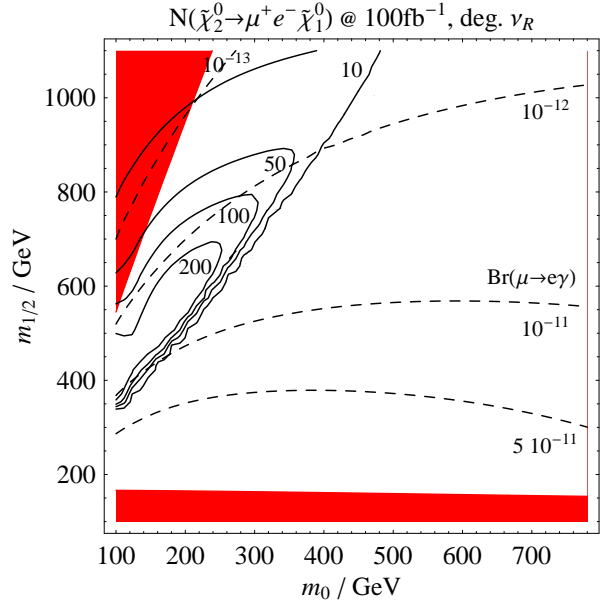}
\caption{ Contours of the number of
\(\tilde\chi_2^0\to\mu^+e^-\tilde\chi_1^0\) events at the LHC with
an integrated luminosity of \(100\text{fb}^{-1}\) (solid) and of
\(B(\mu\to e\gamma)\) in the \(m_0-m_{1/2}\) plane . The remaining
mSUGRA and neutrino oscillation parameters are as in
Fig.~\ref{fig:scan_ep1}. The shaded (red) areas are already
excluded by mass bounds from various experimental sparticle
searches.} \label{fig:scan_lhc_seesaw}
\end{figure}

As in the linear collider case, the correlation is approximately
independent of the neutrino parameters, but highly dependent on
the mSUGRA parameters. This is contemplated further in
Fig.~\ref{fig:scan_lhc_seesaw}, comparing the sensitivity of the
signature \(N(\tilde\chi_2^0\to\mu^+e^-\tilde\chi_1^0)\) at the
LHC with \(B(\mu\to e\gamma)\) in the \(m_0-m_{1/2}\) plane.
As for the linear collider, LHC searches can be
competitive with the rare decay experiments for small
\(m_0\approx200\)~GeV. Tests in the large-\(m_0\) region are again
severely limited by collider kinematics.

Up to now we have considered LFV in the class of type I SUSY
seesaw model described in Section~\ref{sec:seesawI}, which is
representative of models of flavor mixing in the left-handed
slepton sector only. However, it is instructive to analyze general
mixing in the left- and right-handed slepton sector, independent
of any underlying model for slepton flavor violation. The easiest
way to achieve this is by assuming mixing between two flavors
only, which can be parameterized by a mixing angle
\(\theta_{L/R}\) and a mass difference \((\Delta m)_{L/R}\)
between the sleptons, in the case of left-/right-handed slepton
mixing, respectively \footnote{Note that this is different to the
approach in \cite{Bartl:2005yy}, where the slepton mass matrix
elements are scattered randomly.}. In particular, the
left-/right-handed selectron and smuon sector is then diagonalized
by
\begin{equation}\label{eqn:TwoFlavorDiag}
 \begin{pmatrix}
 \tilde l_1 \\
 \tilde l_2 \\
 \end{pmatrix}
 = U\cdot \begin{pmatrix}
 \tilde e_{L/R} \\
 \tilde \mu_{L/R} \\
 \end{pmatrix},\text{ with }
 U=\begin{pmatrix}
 \cos\theta_{L/R} & \sin\theta_{L/R} \\
 -\sin\theta_{L/R} & \cos\theta_{L/R} \\
 \end{pmatrix},
\end{equation}
and a mass difference \(m_{\tilde l_2}-m_{\tilde l_1}=(\Delta
m)_{L/R}\) between the slepton mass eigenvalues\footnote{In case
of left-handed mixing, the mixing angle \(\theta_L\) and the mass
difference \((\Delta m)_L\) are also used to describe the
sneutrino sector.}. The LFV branching ratio
\(B(\tilde\chi_2^0\to\mu^+e^-\tilde\chi_1^0)\) can then be written
in terms of the mixing parameters and the flavor conserving
branching ratio \(B(\tilde\chi_2^0\to e^+e^-\tilde\chi_1^0)\) as
\begin{equation}\label{eqn:TwoFlavorMixing}
 B(\tilde\chi_2^0\to\mu^+e^-\tilde\chi_1^0)=
 2\sin^2\theta_{L/R}\cos^2\theta_{L/R}
 \frac{(\Delta m)^2_{L/R}}{(\Delta m)^2_{L/R}+\Gamma^2_{\tilde l}}
 B(\tilde\chi_2^0\to e^+e^-\tilde\chi_1^0),
\end{equation}
where  \(\Gamma_{\tilde l}\)  is the average width of the two sleptons
involved. Maximal LFV is thus achieved by choosing
\(\theta_{L/R}=\pi/4\) and \((\Delta m)_{L/R}\gg\Gamma_{\tilde
l}\). For definiteness, we use \((\Delta m)_{L/R}=0.5\)~GeV. The
results of this calculation can be seen in Fig.~\ref{fig:scan_lhc
maxmix}, which shows contour plots of
\(N(\tilde\chi_2^0\to\mu^+e^-\tilde\chi_1^0)\) in the
\(m_0-m_{1/2}\) plane for maximal left- and right-handed slepton
mixing, respectively. Also displayed are the corresponding
contours of \(B(\mu\to e\gamma)\). We see that the present bound
\(B(\mu\to e\gamma)=10^{-11}\) still permits sizeable LFV signal
rates at the LHC. However, \(B(\mu\to e\gamma)<10^{-14}\) would
exclude the observation of such an LFV signal at the LHC.

\begin{figure}[t!]
\centering
\includegraphics[clip, width=0.40\textwidth]{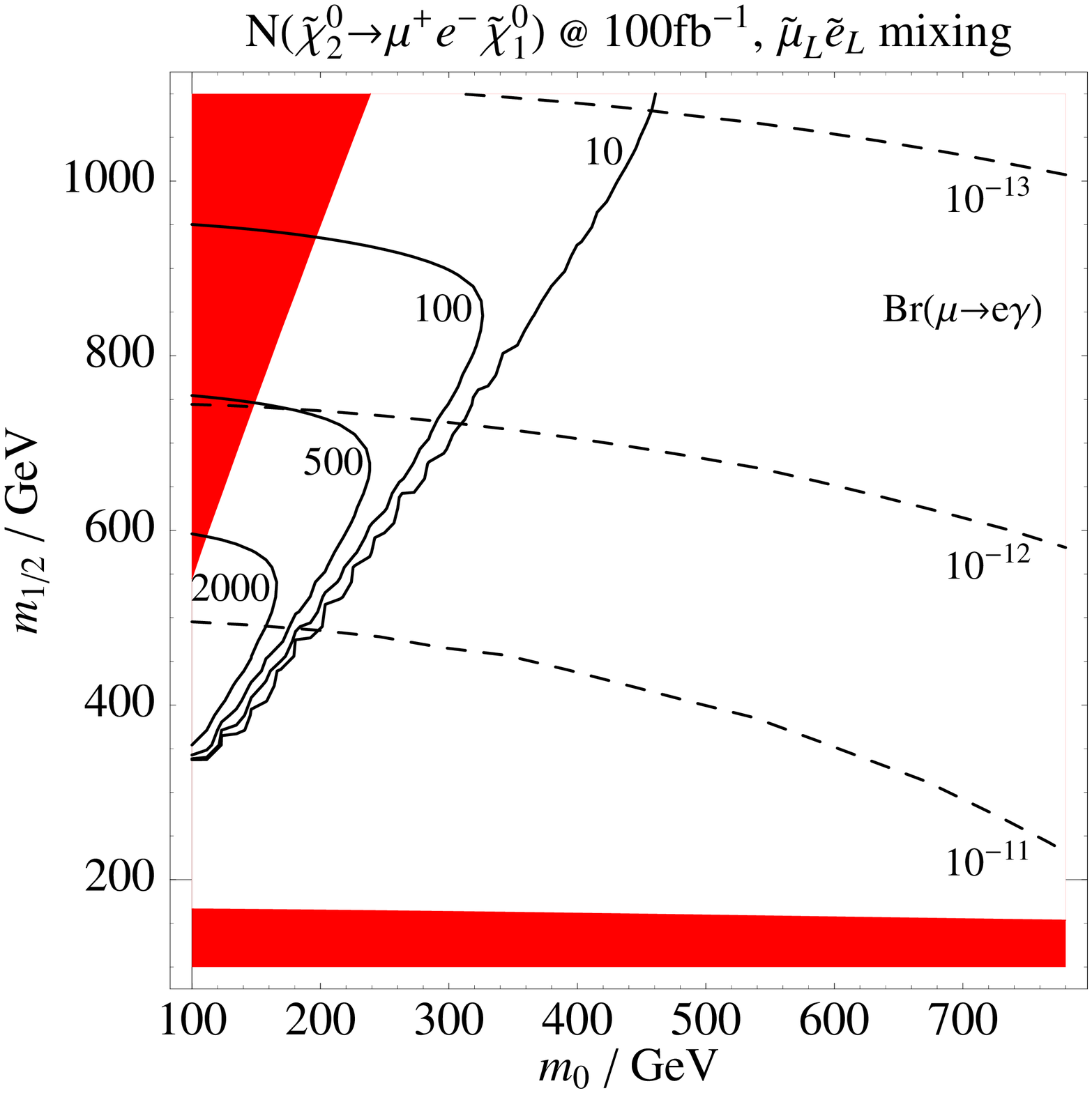}\(\qquad\)
\includegraphics[clip, width=0.40\textwidth]{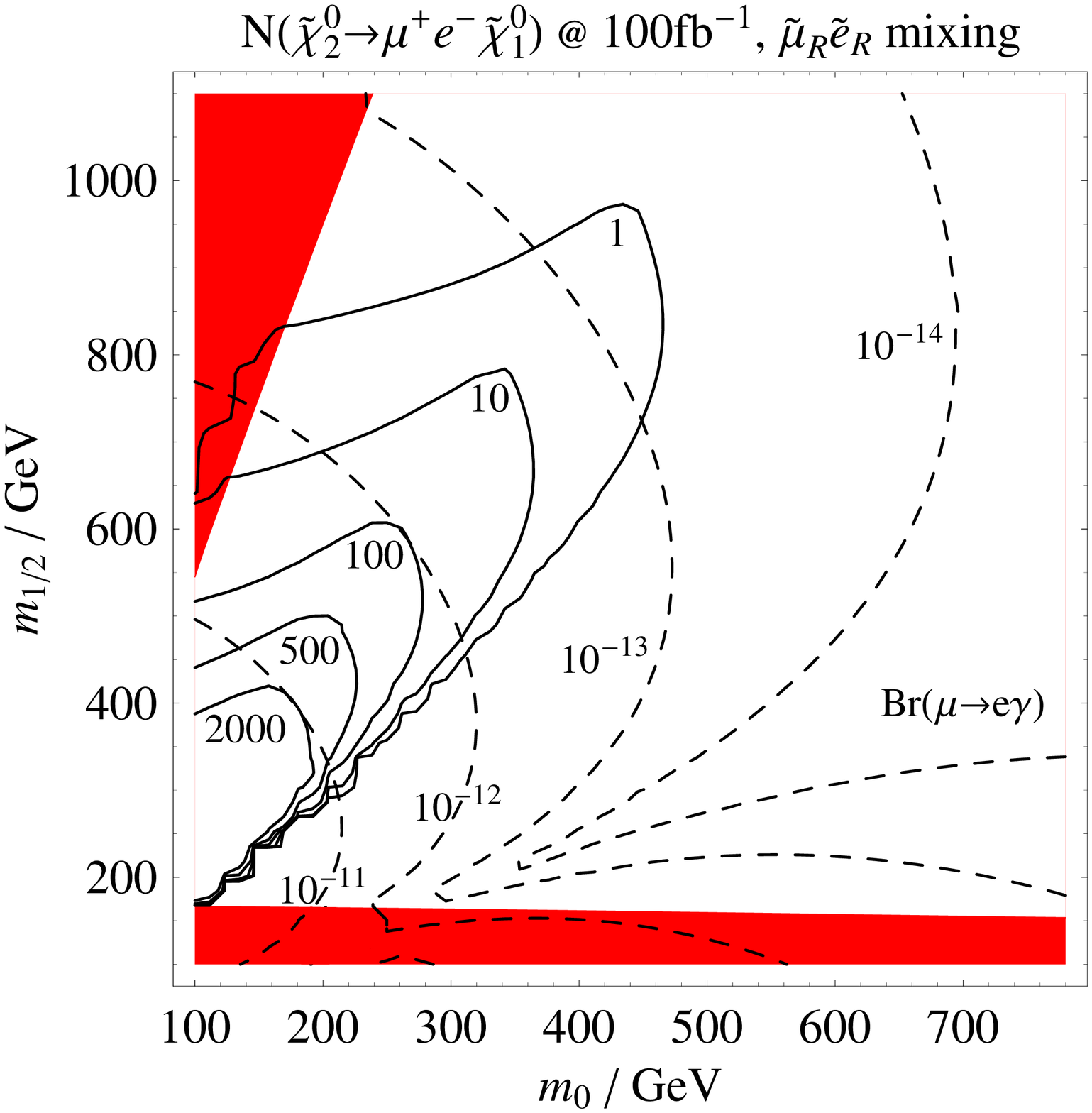}
\caption{Contours of the events per year
\(N(\tilde\chi_2^0\to\mu^+e^-\tilde\chi_1^0)\) at the LHC with an
integrated luminosity of \(100\text{fb}^{-1}\) in the
\(m_0-m_{1/2}\) plane (solid lines). The remaining mSUGRA
parameters are: \(A_0=-100\)~GeV, \(\tan\beta=10\),
\(\text{sign}(\mu)=+\). The left and right panels are for maximal
\(\tilde e_L\tilde\mu_L\) and \(\tilde e_R\tilde\mu_R\) mixing
(\(\theta=\pi/4\), \(\Delta m=1\)~GeV), respectively. For
comparison, \(B(\mu\to e\gamma)\) is shown by dashed lines. The
shaded (red) areas are forbidden by mass bounds from various
experimental sparticle searches.} \label{fig:scan_lhc maxmix}
\end{figure}

\subsubsection{Impact of $\theta_{13}$ on LFV in SUSY seesaw}

In this subsection we present the results of the LFV tau and muon decays within
the SUSY singlet-seesaw context. Specifically, we consider the Constrained 
Minimal Supersymmetric Standard Model (CMSSM) extended by 
three right handed neutrinos, $\nu_{R_i}$ and their corresponding 
SUSY partners, ${\tilde \nu}_{R_i}$, ($i=1,2,3$),
and use the seesaw mechanism for the neutrino mass generation. We include   
the predictions for the branching ratios (BRs) of two types of LFV channels, 
$l_j \to l_i \gamma$
and $l_j \to 3 l_i$, and compare them with the present bounds and future experimental sensitivities. We first analyze the
dependence of the BRs with the most relevant SUSY-seesaw parameters, and we then focus 
on the particular sensitivity to $\theta_{13}$, which we find specially interesting on the light of its potential future measurement.
We further study the constraints from the requirement of successfully producing the
Baryon Asymmetry of the Universe via thermal leptogenesis, which is another 
appealing feature of the SUSY-seesaw scenario.
We conclude with the impact that a potential measurement of the leptonic mixing angle $\theta_{13}$ can have on LFV physics.    

Regarding the technical aspects of the computation of the branching ratios, the most relevant points are (for details, see \cite{Arganda:2005ji,Antusch:2006vw}: 
\begin{itemize}
\item It is a full one-loop computation of BRs, i.e., we include all
contributing one-loop diagrams with the SUSY particles flowing in the loops. For the case 
of $l_j \to l_i \gamma$ the analytical formulas can be found in~\cite{Hisano:1995cp,Arganda:2005ji,Hisano:1995cp}. For the
case $l_j \to 3 l_i$ the complete set of diagrams (including photon-penguin, Z-penguin,
Higgs-penguin and box diagrams) and formulae are given in~\cite{Arganda:2005ji}.   
\item The computation is performed in the physical basis for all SUSY
particles entering in the loops. In other words, we do not
use the Mass Insertion Approximation (MIA). 
\item The running of the CMSSM-seesaw parameters from the
universal scale $M_X$ down to the electroweak scale is performed by
numerically solving the full one-loop Renormalization Group Equations (RGEs) (including the extended neutrino sector) 
and by means of the public Fortran Code
SPheno2.2.2.~\cite{Porod:2003um}. More concretely, we do not use the Leading Log Approximation (LLog).
\item The light neutrino sector parameters that are used in 
$m_D = \sqrt {m_N^{\rm diag}} R \sqrt {m_\nu^{\rm diag}}U^{\dagger}_{MNS}$
are those evaluated at the seesaw scale $m_R$. That is, we start with their low energy values 
(taken from data) and then apply the RGEs to run them up to $m_R$. 
\item We have added to the SPheno code extra subroutines that compute the LFV
rates for all the $l_j \to l_i \gamma$ and $l_j \to 3 l_i$ channels. 
We have also included additional subroutines to: Implement 
the requirement of successful baryogenesis (which we define as having  
$n_B/n_\gamma \in [10^{-10},10^{-9}]$) via thermal leptogenesis in the presence of upper bounds on the reheat temperature; Implement the requirement of compatibility with present bounds on lepton
electric dipole moments: {$\mbox{EDM}_{e \mu \tau}$} {$\lesssim (6.9
\times 10^{-28}, 3.7 \times 10^{-19}, 4.5 \times 10^{-17}) \, \mbox{e.cm}$}  
\end{itemize}

In what follows we present the main results for the case of hierarchical heavy neutrinos. We also include a comparison with present bounds on LFV rates~\cite{Brooks:1999pu,Aubert:2005wa,Aubert:2005ye,Bellgardt:1987du,Aubert:2003pc} and their future sensitivities~\cite{mue:Ritt,Akeroyd:2004mj,Iijima,Aysto:2001zs,PRIME,Kuno:2005mm}.
For hierarchical heavy neutrinos, the BRs are mostly sensitive to the heaviest
mass $m_{N_3}$, $\tan\beta$, $\theta_1$ and $\theta_2$ (using the $R$ parameterization of \cite{Casas:2001sr}). The other input seesaw
parameters $m_{N_1}$, $m_{N_2}$ and $\theta_3$ play a secondary role since the BRs do
not strongly depend on them. The dependence on $m_{N_1}$ and $\theta_3$ appears only 
indirectly, once the requirement of a successful BAU is imposed. 
We will comment more on this later.

\begin{figure}[t]
\parbox{0.51\linewidth}{\includegraphics[height=\linewidth, angle=270]{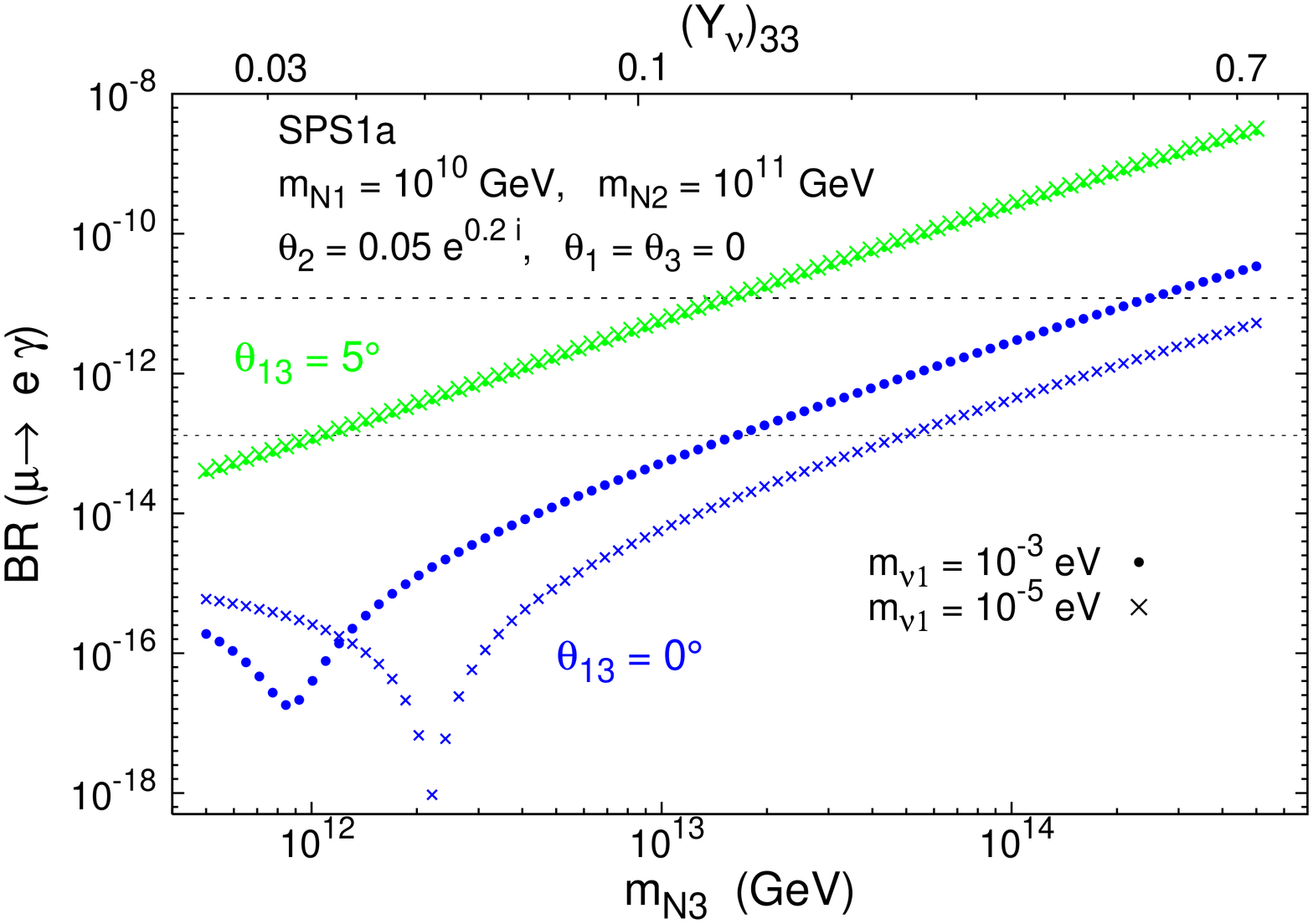}}\hspace*{\fill}
\parbox{0.45\linewidth}{\includegraphics[height=\linewidth, angle=270]{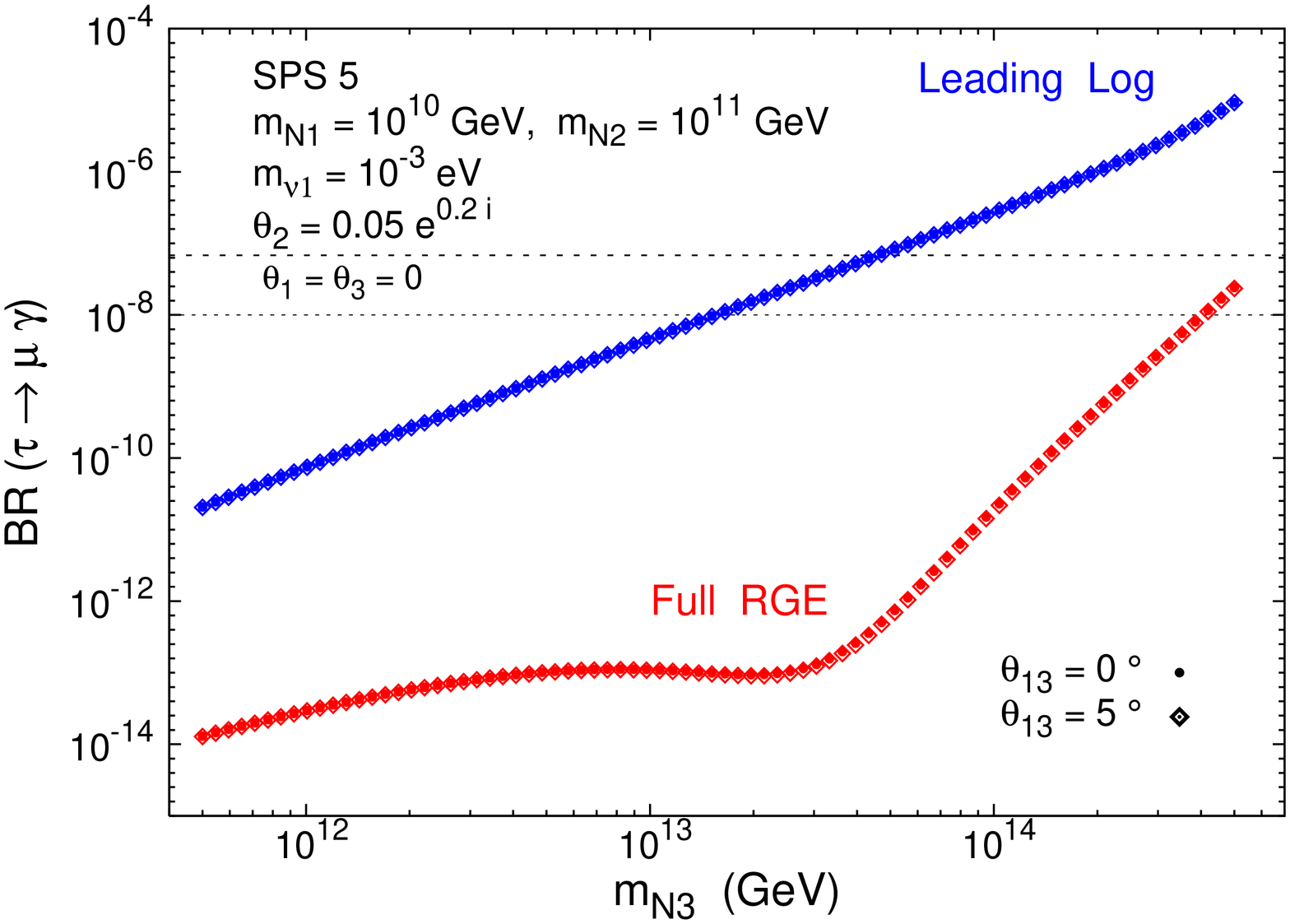}}
\caption{On the left, $B(\mu \to e\, \gamma)$ as a function of $m_{N_3}$ for SPS 1a, with
$m_{\nu_1}\,=\,10^{-5}$~eV and $m_{\nu_1}\,=\,10^{-3}$~eV (times, dots, respectively), and
$\theta_{13}=0^\circ,\,5^\circ$ (blue/darker, green/lighter lines). Baryogenesis is enabled by the choice 
$\theta_2\,=0.05\,e^{0.2\,i}$ ($\theta_1=\theta_3=0$). On the upper horizontal axis we display the associated value of
$(Y_\nu)_{33}$. A dashed (dotted) horizontal line denotes the present experimental bound (future sensitivity).
On the right, $B(\tau \to \mu \, \gamma)$ as a function of $m_{N_3}$ for SPS5, with
$m_{\nu_1}\,=\,10^{-3}$~eV and $\theta_2\,=0.05\,e^{0.2\,i}$ ($\theta_1 = \theta_3 = 0^\circ$).
The predictions for $\theta_{13} = 0^\circ, 5^\circ$ are superimposed one on the top of the other. The upper curve is obtained using the LLog
approximation and the lower one is the full RGE prediction. The dashed (dotted) horizontal line denotes the present experimental bound (future
sensitivity).\label{fig:MN3:MN2:SPS1a}}
\end{figure}

We display in Fig.~\ref{fig:MN3:MN2:SPS1a} the predictions for $B(\mu \to e \gamma)$ and $B(\tau \to \mu
\gamma)$ as a function of $m_{N_3}$, for a specific choice of the other
input parameters. This figure clearly shows the strong sensitivity of the BRs 
to $m_{N_3}$. In fact, the BRs vary by as much as six orders of magnitude in the explored range
of $5 \times 10^{11} \, {\rm GeV} \leq  m_{N_3} \leq 5 \times 10^{14} \, {\rm GeV}$.
Notice also that for the largest values of $m_{N_3}$ considered, 
the predicted rates for $\mu \to e \gamma$ enter into the present experimental reach 
and only into the future experimental sensitivity for $\tau \to \mu \gamma$. 
It is also worth mentioning 
that by comparing our full results with the LLog predictions, we find that the LLog
approximation dramatically fails in some cases. In particular, for the SPS5 point, the 
LLog predictions overestimate the BRs by about four orders of magnitude. For
the other points
SPS4, SPS1a,b and SPS2 the LLog estimate is very similar to the full result, whereas 
for SPS3 it underestimates the full computation by a factor of three. 
In general, the divergence 
of the LLog and the full computation occurs for low $M_0$ and large $M_{1/2}$~\cite{Petcov:2003zb,Chankowski:2004jc} and/or large
$A_0$ values~\cite{Antusch:2006vw}. The failure of the LLog is more dramatic for SUSY scenarios with
 large $A_0$. 
Fig.~\ref{fig:MN3:MN2:SPS1a}
also shows that while in some cases (as for instance SPS1a)
the behaviour of the BR with $m_{N_3}$ does 
follow the expected LLog approximation (BR $\sim (m_{N_3}\log m_{N_3})^2$), there
are other scenarios where this is not the case. A good example is SPS5.
It is also worth commenting on the deep minima of $B(\mu \to e \gamma)$ appearing in
Fig.~\ref{fig:MN3:MN2:SPS1a} for the lines
associated with $\theta_{13}=0^\circ$.  These minima
are induced by the effect of the running of $\theta_{13}$, shifting it
from zero to a negative value (or equivalently $\theta_{13} > 0$ and $\delta = \pi$).
In the LLog approximation, they  can be understood  as
a cancellation occurring in the relevant quantity $Y_\nu^{\dagger} L Y_\nu$, with $L_{ij}=
\log (M_X/m_{N_i})\delta_{ij}$. Most explicitly, the cancellation occurs between the terms proportional to $m_{N_3}\,L_{33}$ and 
$m_{N_2}\,L_{22}$ in the limit $\theta_{13}(m_R) \to 0^-$ 
(with $\theta_1=\theta_3=0$).
The depth of these minima is larger for smaller $m_{\nu_1}$,
as is visible in Fig.~\ref{fig:MN3:MN2:SPS1a}.

Regarding the $\tan \beta$ dependence of the BRs we obtain that, similar to
what was found for the degenerate case, the BR grow as $\tan^2 \beta$. The
hierarchy of the BR predictions for the several SPS points is dictated by the
corresponding $\tan \beta$ value, with a secondary role being played by the
given SUSY spectra. We find again the following generic hierarchy:
$B_{\rm SPS4}$~$>B_{\rm SPS1b}$~$\gtrsim B_{\rm SPS1a}$~
$>B_{\rm SPS3}$~$\gtrsim B_{\rm SPS2}$~$> B_{\rm SPS5}$.

In what concerns to the $\theta_i$ dependence of the BRs, we have found that they are 
mostly sensitive to $\theta_1$ and $\theta_2$. 
The BRs are nearly constant with $\theta_3$. 
As has been shown in \cite{Arganda:2005ji}, 
the predictions for $B(\mu \to e \gamma)$, $B(\mu \to 3 e)$, $B(\tau \to \mu
\gamma)$ and $B(\tau \to e \gamma)$ are above their corresponding experimental
bound for specific values of $\theta_1$. Particularly, the LFV muon decay rates are 
well above their present experimental bounds for most of the $\theta_1$
explored values. Notice also for SPS4 that the predicted $B(\tau \to \mu
\gamma)$ rates are very close to the present experimental reach even at
$\theta_1=0$ (that is, $R=1$). We have also explored the dependence with $\theta_2$ and
found similar results (not shown here), with the appearance of pronounced dips
at particular real values of
$\theta_2$ with the $B(\mu \to e \gamma)$, $B(\mu \to 3 e)$ and $B(\tau \to \mu
\gamma)$ predictions being above the experimental bounds for some $\theta_2$ values.

We next address the sensitivity of the LFV BRs to $\theta_{13}$. We first present the 
results for the simplest $R=1$ case and then discuss how this sensitivity
changes when
moving from this case towards the more general case of complex $R$, taking
into account additional constraints from the requirement of a successful BAU. 
 
\begin{figure}[t]
\parbox{0.48\linewidth}{\includegraphics[height=\linewidth, angle=270]{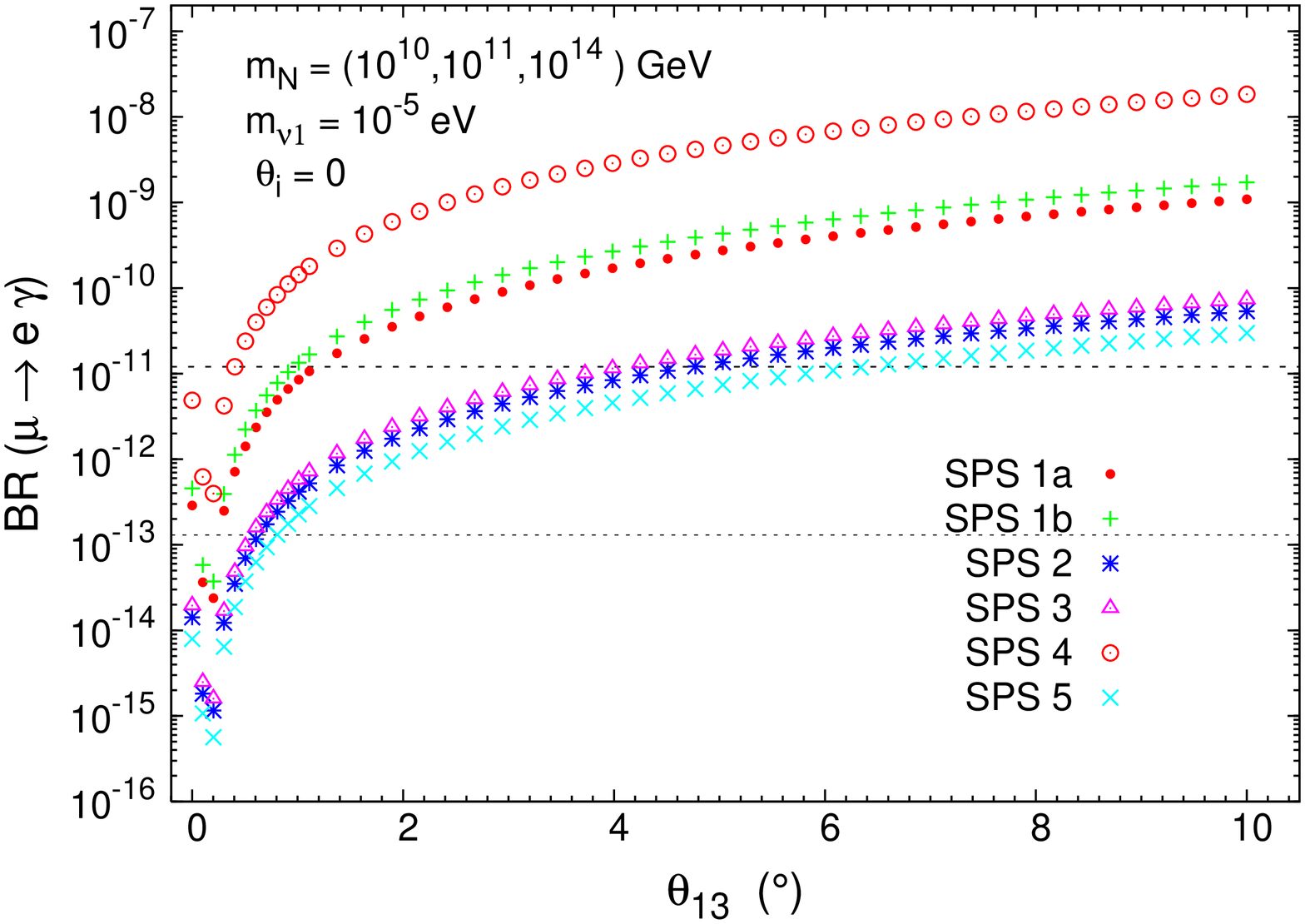}}\hspace*{\fill}
\parbox{0.48\linewidth}{\includegraphics[height=\linewidth, angle=270]{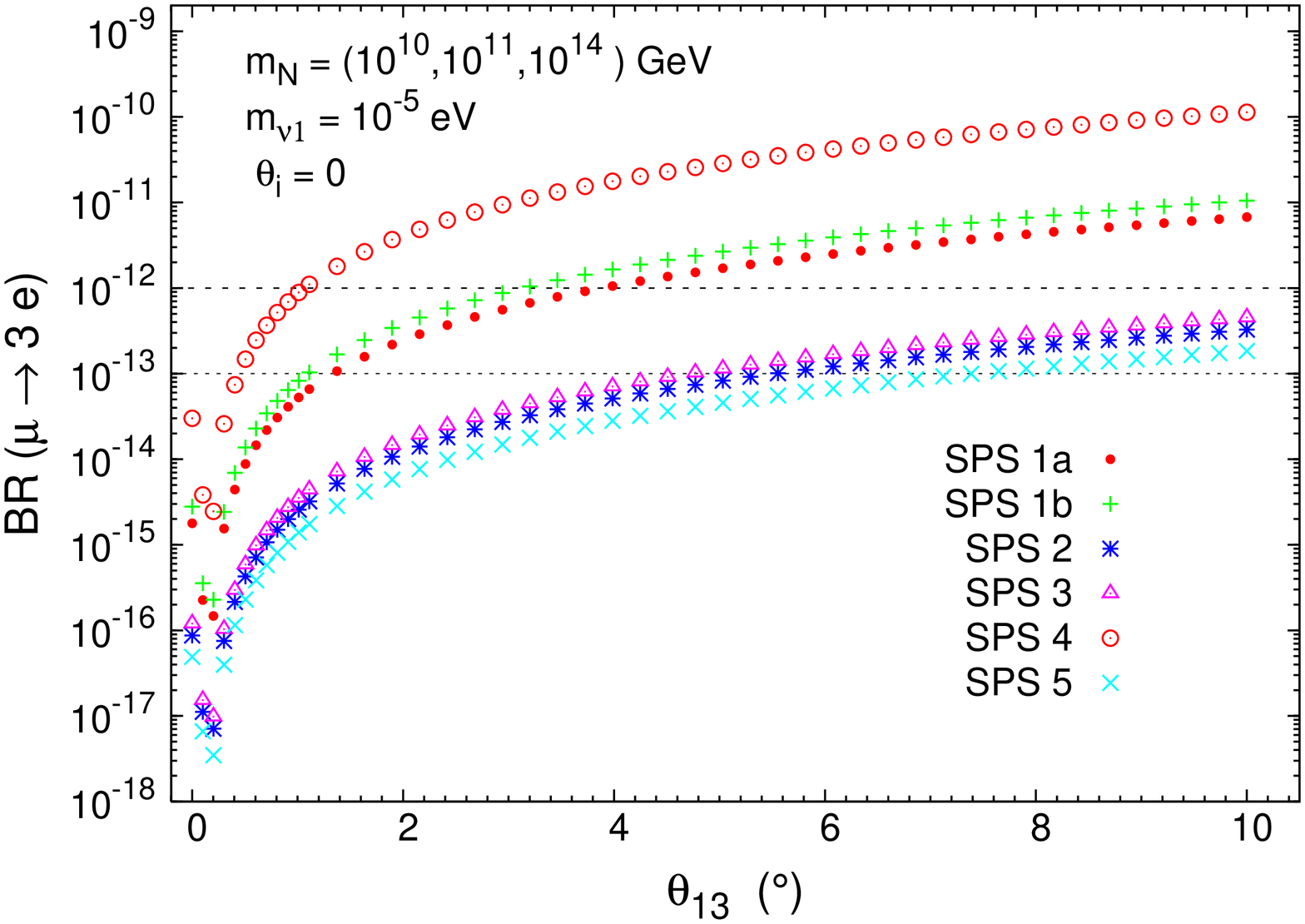}}
\parbox{0.48\linewidth}{\includegraphics[height=\linewidth, angle=270]{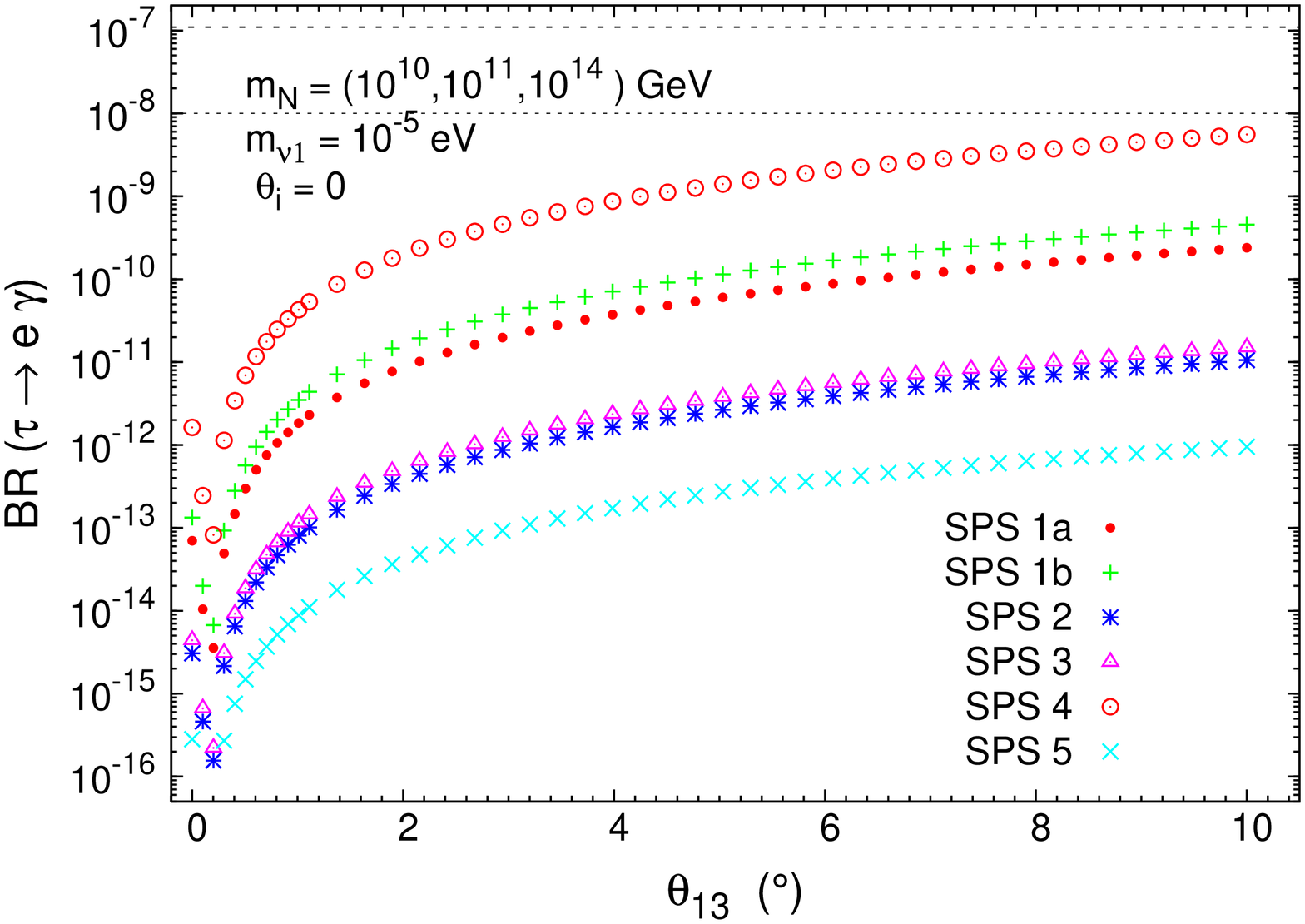}}\hspace*{\fill}
\parbox{0.48\linewidth}{\includegraphics[height=\linewidth, angle=270]{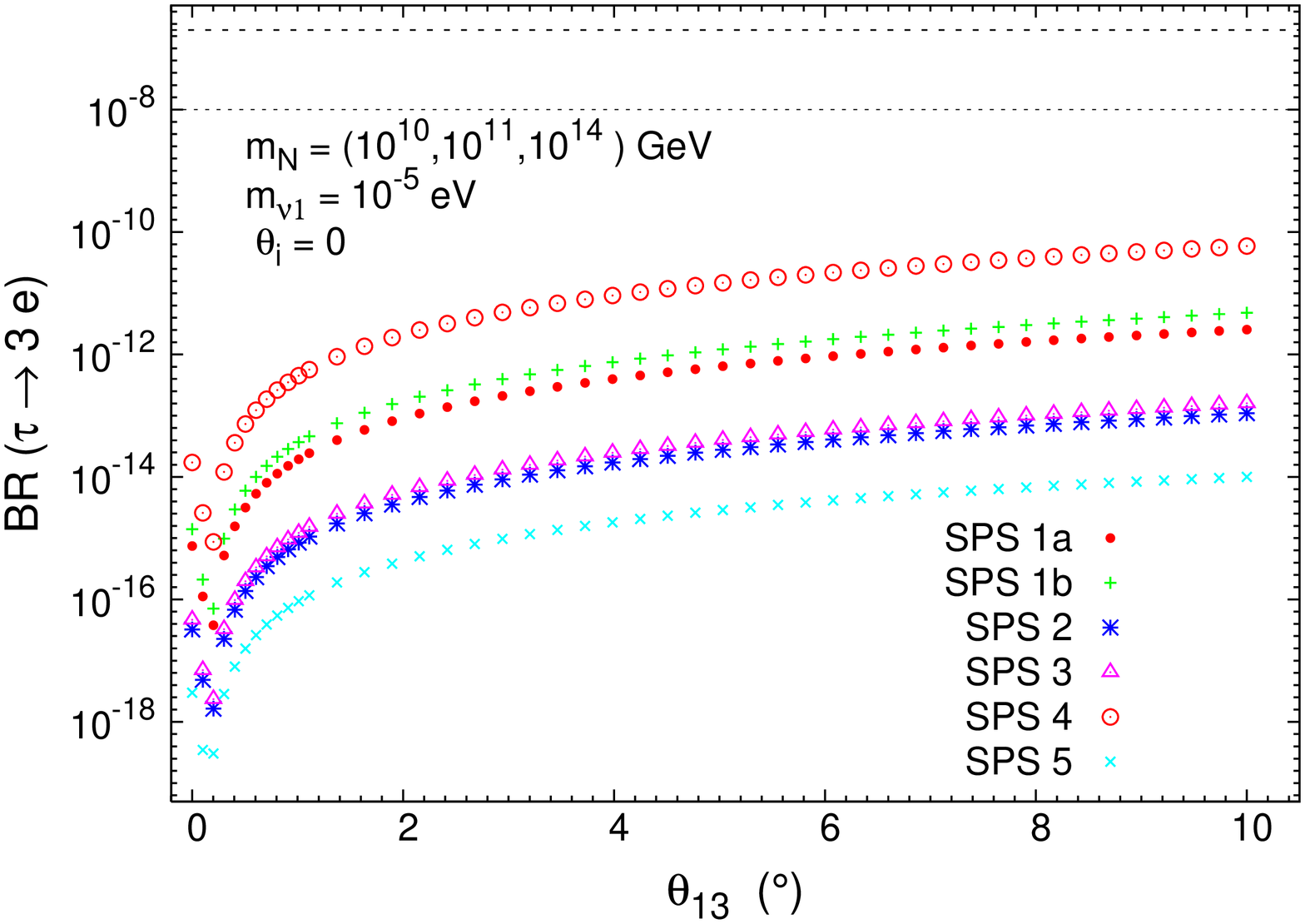}}
\caption{$B(\mu \to e\, \gamma)$ and $B(\mu \to 3\,e)$ as a function of $\theta_{13}$ (in degrees), for SPS 1a
(dots), 1b (crosses), 2 (asterisks), 3 (triangles), 4 (circles) and 5 (times). A dashed (dotted) horizontal line denotes the present experimental bound
(future sensitivity).\label{fig:SPS:t13:ad}}
\end{figure}

For $R = 1$, the predictions of the BRs as functions of 
$\theta_{13}$ in the experimentally allowed range of $\theta_{13}$, $0^\circ \leq
\theta_{13} \leq 10^\circ$ are illustrated in Fig.~\ref{fig:SPS:t13:ad}. 
In this figure we also include the present and
future experimental sensitivities for all channels. We clearly see 
that the BRs of  $\mu \to e \gamma$, $\mu \to 3 e$,
$\tau \to e \gamma$ and $\tau \to 3 e$ are extremely sensitive to $\theta_{13}$, with
their predicted rates varying many orders of magnitude along the explored $\theta_{13}$ 
interval. In the  case of $\mu \to e \gamma$ this strong sensitivity was previously 
pointed out in Ref.~\cite{Masiero:2004js}. The other LFV channels, $\tau \to \mu \gamma$ and 
$\tau \to 3 \mu$ (not displayed here), are nearly insensitive to this
parameter. The most important conclusion from Fig.~\ref{fig:SPS:t13:ad} is
that, for this choice of parameters, the predicted BRs for both 
muon decay
channels, $\mu \to e \gamma$ and
$\mu \to 3 e$, are clearly within the present 
experimental reach for
several  of the studied SPS points. The most stringent channel is manifestly $\mu \to e \gamma$ where the
predicted BRs for all the SPS points are clearly above the present experimental bound 
for $\theta_{13} \gtrsim 5^\circ$. With the expected improvement in the experimental
sensitivity to this channel, this would happen for $\theta_{13} \gtrsim 1^\circ$.

In addition to the small neutrino mass generation, the seesaw
mechanism offers the interesting possibility of 
baryogenesis via leptogenesis~\cite{Fukugita:1986hr}. Thermal leptogenesis 
is an attractive and minimal mechanism to produce a successful BAU with
rates which are compatible with present data, $n_\mathrm{B} /n_\gamma 
\,\approx\, (6.10\,\pm\,0.21)\,\times\,10^{-10}$ \cite{Spergel:2006hy}. 
In the supersymmetric version of the seesaw mechanism, it can
be successfully implemented if provided that the 
following conditions can be satisfied. Firstly, Big Bang Nucleosynthesis gravitino 
problems have to be avoided, which is possible, for instance, for 
sufficiently heavy gravitinos. Since we consider the gravitino mass 
as a free parameter, this condition can be easily achieved. In any case, further 
bounds on the reheat temperature $T_\mathrm{RH}$ still arise from 
decays of gravitinos into Lightest Supersymmetric Particles (LSPs). In the
case of heavy gravitinos and neutralino LSPs masses into the range 100-150
GeV (which is the case of the present work), one obtains 
$T_\mathrm{RH} \lesssim 2 \times 10^{10}$ GeV.
In the presence of these constraints on 
$T_\mathrm{RH}$, the favoured region by thermal leptogenesis 
corresponds to small (but non-vanishing) complex $R$-matrix angles 
$\theta_i$. For vanishing $U_{MNS}$ CP phases the constraints on $R$ are
basically $|\theta_2|,|\theta_3|  
\lesssim 1 \, \mbox{rad}$ (mod $\pi$). 
Thermal leptogenesis also
constrains $m_{N_1}$ to be roughly in the range $[10^9\:
\mbox{GeV},10 \times T_\mathrm{RH}]$ (see also \cite{Giudice:2003jh,Antusch:2006gy}). In the present work we have explicitly calculated the produced BAU in the presence of upper bounds on the reheat temperature $T_\mathrm{RH}$. We have furthermore set as ``favoured
BAU values'' those that are within the interval $[10^{-10},10^{-9}]$, which contains
the WMAP value, and choose the value of $m_{N_1} =
10^{10}$ GeV in some of our plots.
Similar studies of the constraints from leptogenesis on LFV rates have been
done in~\cite{Petcov:2005jh}.

Concerning the EDMs, which are clearly non-vanishing in the presence
of complex $\theta_i$, we have checked that all the predicted values for 
the electron, muon
and tau EDMs are well below the experimental bounds.  
In the following we therefore focus on complex but small $\theta_2$ values,  leading to
favourable BAU,  and study its effects
on the sensitivity to $\theta_{13}$. Similar results are obtained for
$\theta_3$, but for shortness are not shown here. 

\begin{figure}[t]
\parbox{0.48\linewidth}{\includegraphics[height=\linewidth, angle=270]{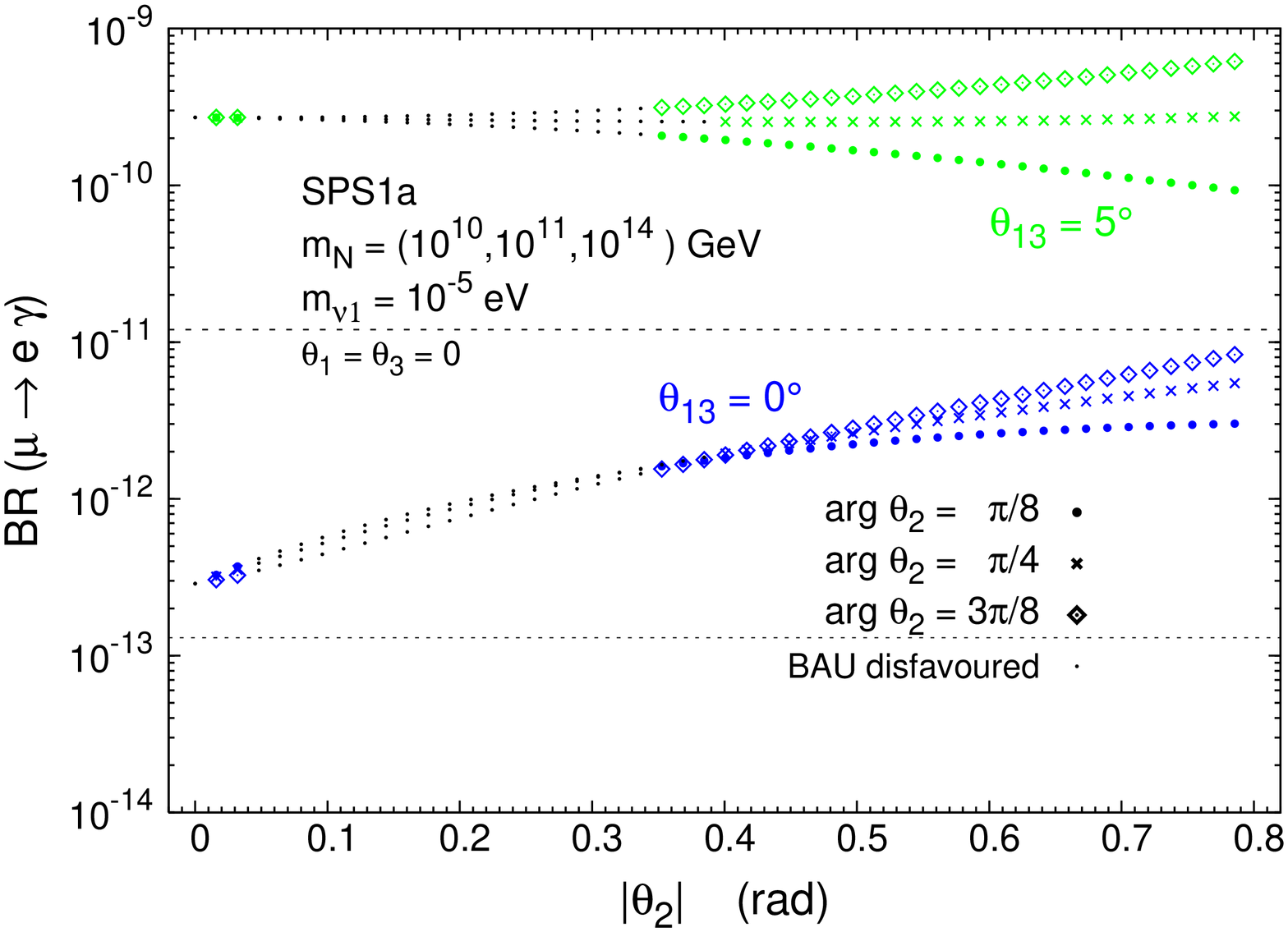}}\hspace*{\fill}
\parbox{0.48\linewidth}{\includegraphics[height=\linewidth, angle=270]{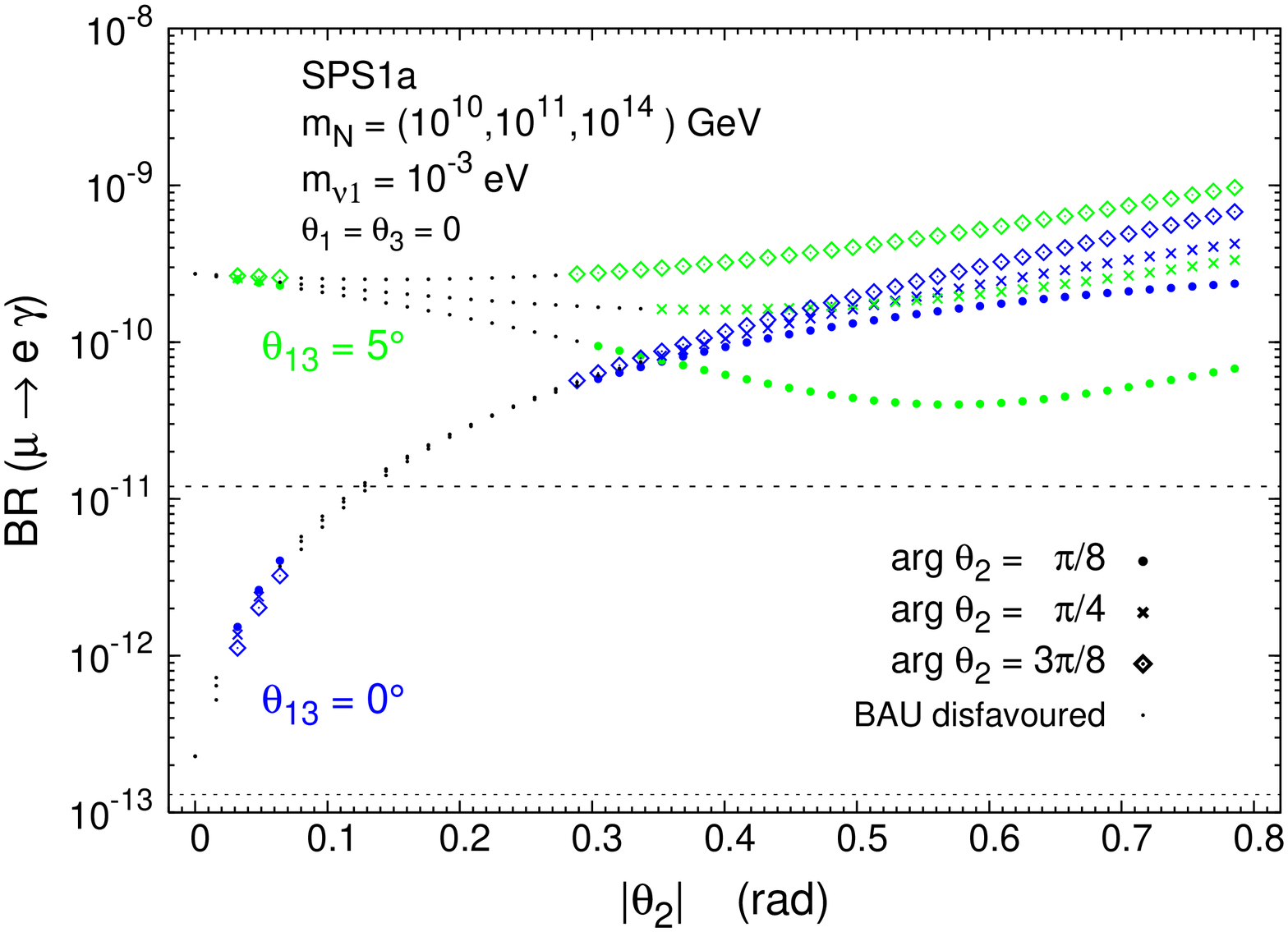}}
\caption{$B(\mu \to e\, \gamma)$ as a function of $|\theta_2|$, for $\arg \theta_2\,=\,\{\pi/8\,,\,\pi/4\,,\,3\pi/8\}$ (dots, times,
diamonds, respectively) and $\theta_{13}=0^\circ$, 5$^\circ$ (blue/darker, green/lighter lines). We take $m_{\nu_1}\,=\,10^{-5}$
($10^{-3}$) eV, on the left (right) panel. In all cases black dots represent points associated with a disfavoured BAU scenario and 
a dashed (dotted) horizontal line denotes the present experimental bound (future sensitivity).\label{fig:modt2:argt2:1214}}
\end{figure}

Fig.~\ref{fig:modt2:argt2:1214} shows the dependence of 
the most sensitive BR to $\theta_{13}$, $B(\mu \to e\, \gamma)$, on
$|\theta_2|$. We consider two
particular values of $\theta_{13}$,
$\theta_{13}=0^\circ\,,5^\circ$ and choose SPS 1a. Motivated from the thermal leptogenesis favoured $\theta_2$-regions \cite{Antusch:2006vw}, we take $0 \,\lesssim \, |\theta_2|
\,\lesssim \, \pi/4$, with $\arg \theta_2 \,=\,\{\pi/8\,,\,\pi/4\,,\,3\pi/8\}$. 
We display the numerical results,
considering $m_{\nu_1}\,=\,10^{-5}$ eV and $m_{\nu_1}\,=\,10^{-3}$~eV, 
while for the heavy neutrino masses we 
take $m_{N}\, =\, (10^{10},\,10^{11},\,10^{14})$~GeV.
There are several important conclusions to be drawn from
Fig.~\ref{fig:modt2:argt2:1214}. Let us first discuss the case 
$m_{\nu_1}\,=\,10^{-5}$~eV. We note that one can obtain a baryon asymmetry 
in the range $10^{-10}$ to $10^{-9}$
for a considerable region of the analyzed $|\theta_2|$ range. 
Notice also that there is a clear separation between the predictions
of $\theta_{13}=0^\circ$ and $\theta_{13}=5^\circ$, with the latter
well above the present experimental bound. This would imply an
experimental impact of $\theta_{13}$, in the sense that the BR predictions 
become potentially
detectable for this non-vanishing $\theta_{13}$ value. 
With the planned
MEG sensitivity~\cite{mue:Ritt}, both cases would be within experimental reach.
However, this statement is strongly dependent on the
assumed parameters, in particular $m_{\nu_1}$.
For instance, a larger value of $m_{\nu_1}=10^{-3}$~eV, illustrated on
the right panel of Fig.~\ref{fig:modt2:argt2:1214}, leads to a very
distinct situation regarding the sensitivity to $\theta_{13}$. 
While for smaller values of $|\theta_2|$
the branching ratio displays a clear sensitivity to having
$\theta_{13}$ equal or different from zero (a separation larger than two orders of
magnitude for $|\theta_2| \lesssim 0.05$), the effect of $\theta_{13}$ is
diluted for increasing values of $|\theta_2|$. For $|\theta_2| \gtrsim 0.3$ 
the $B(\mu \to e\, \gamma)$
associated with $\theta_{13}\,=\,5^\circ$ can be even smaller than for 
$\theta_{13}\,=\,0^\circ$. This implies that in this case, 
a potential measurement of $B(\mu \to e\, \gamma)$ would not be
sensitive to $\theta_{13}$. 
Whether or not a SPS 1a scenario would be disfavoured by
current experimental data on $B(\mu \to e\, \gamma)$ requires a
careful weighting of several aspects. 
Even though Fig.~\ref{fig:modt2:argt2:1214} suggests that for this
particular choice of parameters only very small values of $\theta_2$
and $\theta_{13}$ would be in agreement with current experimental
data, a distinct choice of $m_{N_3}$
(e.g. $m_{N_3}=10^{13}$ GeV) would lead to a rescaling of the
estimated BRs by a factor of approximately $10^{-2}$.
Although we do not display 
the associated plots here, in the latter case 
nearly the entire $|\theta_2|$ range would be 
in agreement with experimental data (in fact the points which are
below the present MEGA bound on Fig.~\ref{fig:modt2:argt2:1214} would
then lie below the projected MEG sensitivity).
Regarding the other SPS points, which are not shown here, 
we find BRs for SPS 1b comparable to those of SPS 1a.
Smaller ratios are associated with SPS 2, 3 and 5, while larger (more
than one order of magnitude) BRs occur for SPS 4.

Let us now address the question of whether a joint measurement of the
BRs and $\theta_{13}$ can shed some light on experimentally unreachable
parameters, like $m_{N_3}$. 
The expected improvement in the experimental sensitivity to the LFV
ratios supports the possibility that 
a BR could be measured in the future, thus
providing the first experimental evidence for new
physics, even before its discovery at the LHC.
The prospects are especially encouraging regarding $\mu \to e\,
\gamma$, where the experimental sensitivity will improve by at least two
orders of magnitude.
Moreover, and given the impressive
effort on experimental neutrino physics, a measurement
of $\theta_{13}$ will likely also occur in the 
future~\cite{Ables:MINOS:2004,Komatsu:2002sz,Migliozzi:2003pw,Huber:2006vr,Itow:2001ee,Blondel:2006su,Huber:2006wb,Burguet-Castell:2005pa,Campagne:2006yx}.  
Given that, as previously emphasized, $\mu \to e\,\gamma$ is very
sensitive to $\theta_{13}$, whereas this is not the case for 
$B(\tau \to \mu\,\gamma)$, 
and that both BRs display the same approximate behaviour with 
$m_{N_3}$ and $\tan \beta$, we now propose to study the correlation between
these two observables. This optimizes the impact of a 
$\theta_{13}$ measurement, since it allows to minimize the uncertainty
introduced from not knowing $\tan \beta$ and $m_{N_3}$, and at the
same time offers a better illustration of the uncertainty associated
with the $R$-matrix angles.
In this case, the correlation of the BRs with respect to $m_{N_3}$
means that, for a fixed set of parameters, varying $m_{N_3}$ implies
that the predicted point 
($B(\tau \to \mu\,\gamma)$,~$B(\mu \to e \, \gamma)$)) 
moves along a line with approximately constant slope in the 
$B(\tau \to \mu\,\gamma) - B(\mu \to e \, \gamma)$ plane.
On the other hand, varying $\theta_{13}$ leads to a 
displacement of the point along the vertical axis.

\begin{figure}[t]
  \begin{center} \hspace*{-10mm}
	\psfig{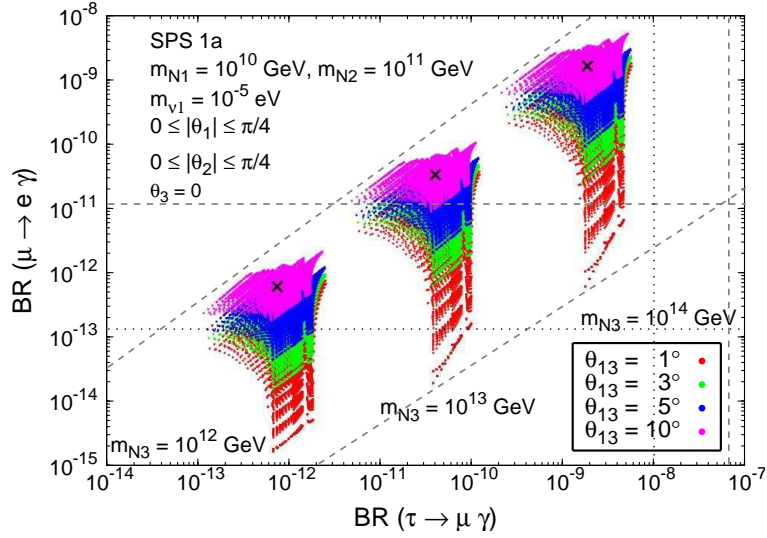} 
    \caption{Correlation between $B(\mu \to e\,\gamma)$ and 
      $B(\tau \to \mu\,\gamma)$ as a function of $m_{N_3}$, for SPS
      1a. The areas displayed represent the scan over $\theta_i$ 
      as given in Eq.~(\ref{doubleBR:input}). From bottom to top, 
      the coloured regions correspond to 
      $\theta_{13}=1^\circ$, $3^\circ$, $5^\circ$ and $10^\circ$ (red,
      green, blue and pink, respectively). Horizontal and vertical 
      dashed (dotted) lines denote the experimental bounds (future
      sensitivities). } 
    \label{fig:doubleBR}
  \end{center}
\end{figure}

In Fig.~\ref{fig:doubleBR}, we illustrate this correlation for SPS
1a, choosing distinct values of the heaviest neutrino mass, and
we scan over the BAU-enabling $R$-matrix angles (setting $\theta_3$ to
zero) as 
\begin{align}\label{doubleBR:input}
& 0\, \lesssim \,|\theta_1|\,\lesssim \, \pi/4 \,, \quad \quad
-\pi/4\, \lesssim \,\arg \theta_1\,\lesssim \, \pi/4 \,, \nonumber \\
& 0\, \lesssim \,|\theta_2|\,\lesssim \, \pi/4 \,, \quad \quad
\quad \,\,\,\,\,\,
0\, \lesssim \,\arg \theta_2\,\lesssim \, \pi/4 \,, \nonumber \\
& m_{N_3}\,=\,10^{12}\,,\,10^{13}\,,\,10^{14}\,\text{GeV}\,.
\end{align} 
 We consider the following values,
$\theta_{13}=1^\circ$, $3^\circ$, $5^\circ$ and $10^\circ$, and only
include in the plot the BR predictions which allow for a favourable BAU. 
Other SPS points have also been considered but they are not shown here for
brevity (see~\cite{Antusch:2006vw}).
We clearly observe in Fig.~\ref{fig:doubleBR} that  
for a fixed value of $m_{N_3}$, and for a given value of $\theta_{13}$, the
dispersion arising from a $\theta_1$ and $\theta_2$ variation produces a small
area rather than a point in the 
$B(\tau \to\mu\,\gamma - B(\mu \to e \, \gamma)$ plane.
The dispersion along the $B(\tau \to \mu\,\gamma)$ axis is of
approximately one order of magnitude for all $\theta_{13}$. 
In contrast, the dispersion along the $B(\mu \to e\,\gamma)$ axis
increases with decreasing $\theta_{13}$,
ranging from 
an order of magnitude for $\theta_{13}=10^\circ$,
to over three orders of magnitude for the case of small $\theta_{13}$
($1^\circ$). 
From Fig.~\ref{fig:doubleBR} 
we can also infer that other choices of $m_{N_3}$ (for $\theta_{13}
\in [1^\circ, 10^\circ]$) would lead to BR
predictions which would roughly lie within the diagonal lines depicted
in the plot. Comparing
these predictions for the shaded areas along the expected diagonal
``corridor'', with the allowed experimental region, allows to conclude
about the impact of a $\theta_{13}$ measurement on the allowed/excluded 
$m_{N_3}$ values.
The most important conclusion from Fig.~\ref{fig:doubleBR} is that for
SPS~1a, and for the parameter space defined in Eq.~(\ref{doubleBR:input}), 
an hypothetical $\theta_{13}$ measurement larger than $1^\circ$, together 
with the present experimental bound on the $B(\mu \to e\,\gamma)$,
will have the impact of excluding values of $m_{N_3} \gtrsim 10^{14}$
GeV. Moreover, with the planned MEG
sensitivity, the same $\theta_{13}$ measurement can further constrain 
$m_{N_3} \lesssim 3\times 10^{12}$~GeV.
The impact of any other $\theta_{13}$ measurement can be analogously
extracted from Fig.~\ref{fig:doubleBR}.

As a final comment let us add that, remarkably, 
within a particular SUSY scenario and scanning over specific $\theta_1$ and $\theta_2$ BAU-enabling ranges for various 
values of $\theta_{13}$, the comparison of the
theoretical predictions for $B(\mu \to e\,\gamma)$ and 
$B(\tau \to \mu\,\gamma)$ with the present experimental bounds allows 
to set $\theta_{13}$-dependent upper bounds on $m_{N_3}$. 
Together with the indirect lower bound arising from leptogenesis 
considerations, this clearly provides interesting hints on the value of the 
seesaw parameter $m_{N_3}$.
With the planned future sensitivities, these bounds would further improve by
approximately one order of magnitude.
Ultimately, a joint measurement of the LFV branching ratios, 
$\theta_{13}$ and the sparticle spectrum would be a powerful tool for 
shedding some light on otherwise unreachable SUSY seesaw parameters.
It is clear from all this study that the interplay between LFV processes and
future improvement in neutrino data is challenging for the searches of new physics.  


\subsubsection{LFV in the CMSSM with constrained sequential dominance}

Sequential Dominance (SD) \cite{King:1998jw,King:1999cm,King:1999cm,King:2002nf} 
represents classes of neutrino models where large lepton mixing angles and small hierarchical neutrino masses can be readily explained
within the seesaw mechanism. 
To understand how Sequential Dominance works, we begin by
writing the right-handed neutrino Majorana mass matrix $M_{\mathrm{RR}}$ in
a diagonal basis as $M_{RR}=\mbox{diag}(M_A,M_B,M_C)$.
We furthermore write the neutrino (Dirac) Yukawa matrix $\lambda_{\nu}$ in
terms of $(1,3)$ column vectors $A_i,$ $B_i,$ $C_i$ as
$Y_{\nu }=(A,B,C)$ using left-right convention.
The term for the light neutrino masses in the effective Lagrangian (after electroweak symmetry breaking), resulting from integrating out the massive right
handed neutrinos, is
\begin{equation}
\mathcal{L}^\nu_{eff} = \frac{(\nu_{i}^{T} A_{i})(A^{T}_{j} \nu_{j})}{M_A}+\frac{(\nu_{i}^{T} B_{i})(B^{T}_{j} \nu_{j})}{M_B}
+\frac{(\nu_{i}^{T} C_{i})(C^{T}_{j} \nu_{j})}{M_C}  
\end{equation}
where $\nu _{i}$ ($i=1,2,3$) are the left-handed neutrino fields.
Sequential dominance then corresponds to the third
term being negligible, the second term subdominant and the first term
dominant:
\begin{equation}
\frac{A_{i}A_{j}}{M_A} \gg
\frac{B_{i}B_{j}}{M_B} \gg
\frac{C_{i}C_{j}}{M_C} \, .
\end{equation}
In addition, we shall shortly see that small $\theta_{13}$ 
and almost maximal $\theta_{23}$ require that 
\begin{equation}
|A_1|\ll |A_2|\approx |A_2|.
\label{SD2}
\end{equation}
Without loss of generality, then, we shall label the dominant
right-handed neutrino and Yukawa couplings as $A$, the subdominant
ones as B, and the almost decoupled (sub-subdominant) ones as $C$. 
Note that the mass ordering of right-handed neutrinos is 
not yet specified. Again without loss of generality we shall 
order the right-handed neutrino masses as $M_1<M_2<M_3$,
and subsequently identify $M_A,M_B,M_C$ with $M_1,M_2,M_3$
in all possible ways. LFV in some of these classes of SD models has 
been analyzed in \cite{Blazek:2002wq}.
Tri-bi-maximal {\em neutrino} mixing corresponds to the choice for example
\cite{King:2005bj}, sometimes referred to as Constrained Sequential Dominance (CSD):
\begin{equation}
Y_{\nu }=
\begin{pmatrix}
0 & be^{i \beta_2} & c_1\\
-ae^{i \beta_3} & be^{i \beta_2} & c_2\\
ae^{i \beta_3} & be^{i \beta_2} & c_3
\end{pmatrix}.
\label{Y1}
\end{equation}
When dealing with LFV it is convenient to work in the 
basis where the charged lepton mass matrix is diagonal.
Let us now discuss the consequences of charged lepton corrections
with a CKM-like structure, for the neutrino Yukawa matrix with CSD.
By CKM-like structure we mean that
$V_{e_\mathrm{L}}$ is dominated by a 1-2 mixing $\theta$, i.e.\ that
its elements $(V_{e_\mathrm{L}})_{13}$, $(V_{e_\mathrm{L}})_{23}$,
$(V_{e_\mathrm{L}})_{31}$ and $(V_{e_\mathrm{L}})_{32}$ are very small
compared to $(V_{e_\mathrm{L}})_{ij}$ ($i,j = 1,2$).
After re-diagonalizing the charged lepton mass matrix, 
$Y_\nu$ in Eq.(\ref{Y1}) becomes transformed as:
$Y_\nu \rightarrow V_{e_L} \,Y_\nu $.
In the diagonal charged lepton mass basis the neutrino Yukawa
matrix therefore becomes:
\begin{equation}
Y_{\nu }=
\begin{pmatrix}
a \,s_\theta e^{- i \lambda}e^{i \beta_3} & 
b  \,(c_\theta - s_\theta e^{- i \lambda} )e^{i \beta_2} &
(c_1c_\theta-c_2 s_\theta e^{- i \lambda})\\
-a \,c_\theta e^{i \beta_3} & 
b \,(c_\theta + s_\theta e^{i \lambda})e^{i \beta_2} &  (c_1 s_\theta e^{i
  \lambda} +c_2 c_\theta )\\
ae^{i \beta_3} & 
be^{i \beta_2} & c_3
\end{pmatrix}
\label{Y2}.
\end{equation}

\begin{figure}
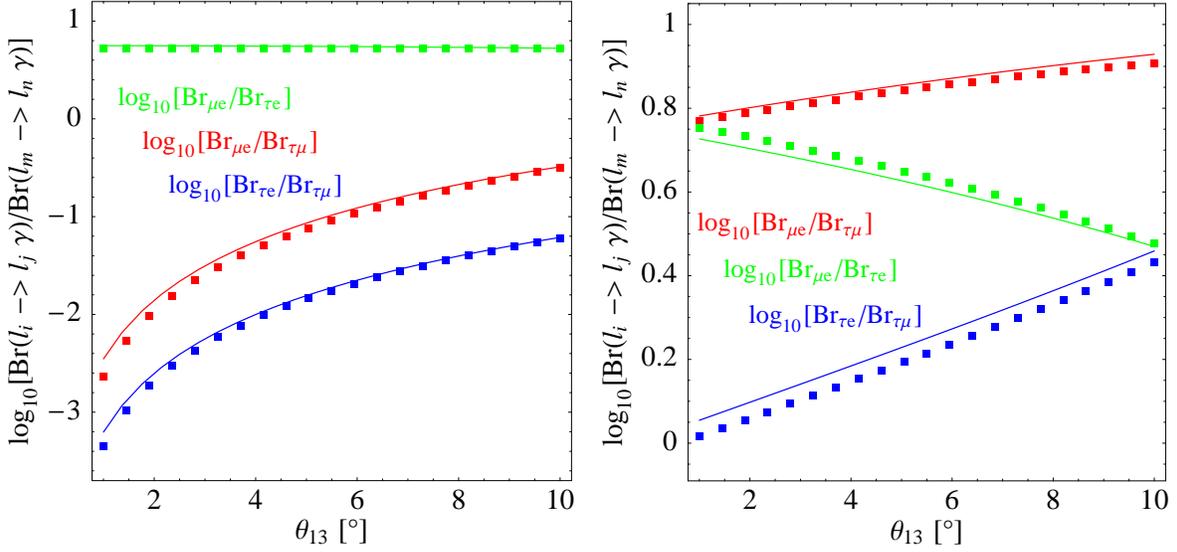

\centering
$\ensuremath{\vcenter{\hbox{\includegraphics[scale=0.9]{figures/section52/plotM3_MA_2.eps}}}}$ \;\;
$\ensuremath{\vcenter{\hbox{\includegraphics[scale=0.9]{figures/section52/plotM3_MB_2.eps}}}}$ 
 \caption{\label{fig:CSDplots} 
Ratios of branching ratios of LFV processes 
$\ell_i \to \ell_j \gamma$ in CSD for $M_3 = M_A$ 
(left panel) and $M_3 = M_B$ (right panel)
with right-handed neutrino masses $M_1=10^8$ GeV, $M_2=5\times 10^8$
GeV and $M_3=10^{14}$ GeV. The solid lines show the (naive)
prediction, from the MI and LLog approximation and with RG running
effects neglected, while the dots show the explicit numerical
computation (using SPheno2.2.2.~\cite{Porod:2003um} extended by software
packages for LFV BRs and neutrino mass matrix running \cite{Arganda:2005ji,Antusch:2006vw}) with universal CMSSM parameters chosen 
as $m_0=750$ GeV, $m_{1/2}=750$ GeV, $A_0 = 0$ GeV, $\tan \beta = 10$ and $\mbox{sign}(\mu)=+1$. 
While the ratios do not significantly depend on the choice of the SUSY model,
since the model-dependence has canceled out, they show a pronounced
dependence on $\theta_{13}$ (and $\delta$) in the case of $M_3 = M_A$ 
(and $M_3 = M_B$).}
\end{figure}

After ordering
$M_A,M_B,M_C$ according to their size, there are six possible forms of
$Y_\nu$ obtained from permuting the columns, with the convention 
always being that the dominant one is labeled by $A$, and so on.
In particular the third column of the neutrino Yukawa matrix
could be $A$, $B$ or $C$ depending on which of 
$M_A$, $M_B$ or $M_C$ is the heaviest.
If the heaviest right-handed neutrino mass is
$M_A$ then the third column of the neutrino Yukawa matrix 
will consist of the (re-ordered) first column of Eq.(\ref{Y2})
and assuming $Y^{\nu}_{33}\sim 1$
we conclude that all LFV processes will be determined approximately by
the first column of Eq.(\ref{Y2}). Similarly 
if the heaviest right-handed neutrino mass is
$M_B$ then we conclude that  
all LFV processes will be determined approximately by
the second column of Eq.(\ref{Y2}).
Note that in both cases the ratios of branching ratios
are independent of the unknown Yukawa couplings which cancel,
and only depend on the charged lepton angle $\theta$,
which in the case of tri-bi-maximal neutrino mixing is related to 
the physical reactor angle by $\theta_{13} = \theta /\sqrt{2}$
\cite{King:2005bj,Antusch:2005kw}. Also note that
$\lambda = \delta - \pi$
where $\delta$ is the Standard PDG CP-violating oscillation phase.
The results for these two cases are shown in Fig.\ref{fig:CSDplots} \cite{Antusch:2007dj}.
The third case $M_3 = M_C$ is less predictive.


\subsubsection{Decoupling of one heavy neutrino and cosmological implications}

The supersymmetric seesaw model involves 
many free parameters. 
In order to correlate the model predictions for LFV 
processes one has to resort to some supplementary hypotheses. Here we discuss
the consequences of the assumption that one of the heavy singlet neutrinos 
(not necessarily the heaviest one) decouples from the see-saw mechanism 
\cite{Chankowski:2004jc}. 

If the light neutrino masses are hierarchical, in which case the effects 
of the renormalization group (RG) running \cite{Chankowski:2001mx} of $\kappa$ are 
negligible, at least 3 arguments support this  assumption. The first one is 
the {\em naturalness of the see-saw mechanism}. Large mixing angles are not 
generic for hierarchical light neutrino masses (for a review, see \cite{Altarelli:1999gu}). They are 
natural only for special patterns of the matrix $\kappa$. One is a large 
hierarchy between one and the remaining two terms in the sum in (\ref{eq:seesaw}) 
\cite{Smirnov:1993af,King:1998jw,Lavignac:2001vp,Lavignac:2002gf}. This is what we call {\em decoupling} (one term 
hierarchically smaller) or {\em dominance} (hierarchically larger).
Seesaw with only two heavy singlet neutrinos \cite{Frampton:2002qc,Raidal:2002xf} is the limiting 
case of the decoupling of $N_3$ with $M_3\to\infty$ and ${Y}_\nu^{3A}\to0$. 
The immediate consequence of decoupling is $m_{\nu_1}\ll m_{\nu_2}$
($\kappa$ has rank 2 if there are only 2 terms in the sum in 
(\ref{eq:seesaw})). Similarly, for dominance one has $m_{\nu_2}\ll m_{\nu_3}$.

Secondly, decoupling of the lightest singlet neutrino {\em alleviates the 
gravitino problem of leptogenesis} which in the see-saw models of neutrino 
masses appears to be the most natural mechanism for producing the observed 
baryon asymmetry of the Universe\footnote{See \cite{Fukugita:1986hr};
for a review of leptogenesis, see \cite{Buchmuller:2004nz};
for a discussion of flavour effects in leptogenesis see, {\em e.g.} 
\cite{Abada:2006fw};
for recent analyses of the gravitino problem, see, {\em e.g.}, 
\cite{Pradler:2006qh}
} (BAU). As the Universe 
cools down leptonic asymmetries (subsequently converted into baryon asymmetry 
through sphaleron transitions) $Y_\alpha\equiv(n_\alpha-\bar n_\alpha)/s\neq0$ 
(where $n_\alpha$ and $\bar n_\alpha$ are the flavour $\alpha$ lepton and 
antilepton number densities, respectively and $s$ is the entropy density) 
are produced in the decays of $N_1$. The final magnitudes of $Y_\alpha$ are 
proportional to the decay asymmetries $\varepsilon_{1\alpha}$, (which in turn 
are proportional to the heavy neutrino masses) and crucially depend on the 
processes which wash out the asymmetries generated by the $N_1$ decays. The 
efficiency of these processes depends on the parameters $\tilde{m}_{1\alpha}
=\sum_A|{R}_{1A}{U}_{\alpha A}^\ast|^2m_{\nu_A}$, where
$U\equiv U_{PMNS}$  and it is the 
smallest (i.e. leptogenesis is most efficient) for $\tilde{m}_{1\alpha}$ in 
the meV range (assuming vanishing density of $N_1$ after re-heating and 
strongly hierarchical spectrum of $M_A$). If it is $N_1$ which is decoupled, 
there are essentially no lower bounds on $\tilde{m}_{1\alpha}$ and $M_1$, 
hence also the re-heating temperature $T_{\rm RH}$, already of order $10^9$ 
GeV are sufficient \cite{Davidson:2003cq,Antusch:2006gy} (see, however, {\em e.g.}, \cite{Raidal:2004vt}) to reproduce the observed BAU.\footnote{In 
contrast, for $N_2$ or $N_3$ decoupled, the washout is much stronger and $M_1$ 
has to be $\stackrel{>}{{}_\sim}10^{10}$ GeV. This requires $T_{\rm RH}$ leading to a much 
larger dangerous gravitino production \cite{Chankowski:2003rr}. 
Lower $T_{\rm RH}$ is 
in this case possible only if $N_1$ and $N_2$ are sufficiently 
degenerate \cite{Pilaftsis:1998pd}.}

Finally, one heavy singlet neutrino $N_A$ must be decoupled if its 
superpartner, $\tilde N_A$, {\em  plays the role of the inflaton field} 
\cite{Murayama:1992ua}. In such a scenario the (s)neutrino mass $M_A$ 
must be \cite{Ellis:2003sq}
$2\times10^{13}$ GeV and the re-heating temperature following inflaton decay 
is given by 
$T_{\rm RH}\sim\sqrt{\tilde m_A M_\mathrm{Pl}}(M_A/\langle H\rangle)$.
Requiring $T_{\rm RH}<10^6$ GeV (the gravitino problem) then implies
$m_{\nu_A}\leq\sum_\alpha\tilde m_{A\alpha}<10^{-17}$ eV. In this scenario, 
the leptonic asymmetries must be produced non-thermally in the inflaton decay. 
Decoupling of $N_1$ is favoured because  if it is $\tilde N_2$ or 
$\tilde N_3$ which is the inflaton the produced asymmetry may be subsequently 
washed out during the decays of $N_1$.

The assumption that $N_A$ effectively decouples from the seesaw mechanism 
or that $N_A$ effectively dominates the seesaw mechanism translates into
one of the following forms of ${R}$:
\begin{equation}
\label{rpat}
{R}_\mathrm{dec} \simeq 
{\Pi}^{(A)}
\left(
\begin{array}{ccc}
1 & 0 & 0 \\
0 & z & p \\
0 & \mp p & \pm z
\end{array}
\right)
\qquad\textrm{or}\qquad
{R}_\mathrm{dom} \simeq 
{\Pi}^{(A)}
\left(
\begin{array}{ccc}
\mp p & \pm z & 0 \\
z & p & 0 \\
0 & 0 & 1
\end{array}
\right)\, ,
\end{equation}
where $z,p$ are complex numbers satisfying $z^2+p^2=1$ and 
${\Pi}^{(A)}$ denotes permutation of the rows of 
${R}$. Both conditions can be simultaneously satisfied for 
${R}={\Pi}^{(A)}\cdot \mathbf{1}$, known as 
{\em sequential dominance} (for a review, see, e.g.~\cite{King:2003jb}).

In the framework considered, violation of the leptonic flavour is 
transmitted from the neutrino Yukawa couplings ${Y}_\nu^{}$ to the slepton mass 
matrices through the RG corrections. Branching ratios of LFV decays are well 
described by a single mass-insertion approximation via Eq.~(\ref{BR}) and
Eq.~(\ref{eqC}).
Since decoupling of $N_1$ is best motivated we discuss the results for LFV 
only in this case.\footnote{Results for $N_2$ decoupled are 
the same as for decoupled $N_1$
(including sub-leading effects if $M_1$ takes the numerical value of $M_2$).
The same is true also for $N_3$ decoupled 
(including the case with only 2 heavy singlet neutrinos)
if $M_2$ is numerically the same
as $M_3$ for decoupled $N_1$. However,  if $\tilde N_3$ is the inflaton
the LFV decays have the rates too low to be observed. 
In addition, if $N_3$ decouples due to its very 
large mass  its large Yukawa can, for 
$m_{\nu_1}/m_{\nu_3}>M_2/M_3$, still dominate the LFV 
effects which are then practically unconstrained by the oscillation data; 
some constraints can then be obtained from the limits on the electron EDM 
\cite{Joaquim:2007sm}.}

The matrix ${R}$ has then the first of the patterns 
displayed in Eq.~(\ref{rpat}) with ${\Pi}^{(1)}=\mathbf{1}$. The 
discussion simplifies if a technical assumption that 
$m_{\nu_3}M_2<m_{\nu_2}M_3$ is made. 
$\left(\tilde{{m}}_L^2\right)_{32}$ relevant for $\tau\to\mu\gamma$ 
then reads:
\begin{equation}
\label{mi32}
\left(\tilde{{m}}_L^2\right)_{32}\approx 
\frac{\kappa m_{\nu_3}M_3{U}_{33}{U}_{23}^\ast}
{\langle H\rangle^2}  \left[ (|z|^2+S|p|^2) 
+\rho\frac{{U}_{22}^\ast}{{U}_{23}^\ast}x
+\rho\frac{{U}_{32}}{{U}_{33}}x^\ast
+\rho^2\frac{{U}_{32}{U}_{22}^\ast}
{{U}_{33}{U}_{23}^\ast}(S|z|^2+|p|^2)\right]
\end{equation}
where $\rho=\sqrt{m_{\nu_2}/m_{\nu_3}}\sim 0.4$, 
$S=M_2(1+\Delta l_2/\Delta t)/M_3\sim M_2/M_3$ and $x=Sp^\ast z-z^\ast p$.
For $\left(\tilde{{m}}_L^2\right)_{A1}$ relevant for 
$\ell_A\to e\gamma$ we get:
\begin{equation}
\label{mia1}
\left(\tilde{{m}}_L^2\right)_{A1}\approx 
\frac{\kappa m_{\nu_3}M_3{U}_{A3}{U}_{12}^\ast}
{\langle H\rangle^2}\left[\frac{{U}_{13}^\ast}
{{U}_{12}^\ast}(|z|^2+S|p|^2) +\rho x
+\rho^2\frac{{U}_{A2}}{{U}_{A3}}(S|z|^2+|p|^2)\right]
\end{equation}

\begin{figure}[t]
\includegraphics*[width=0.33\linewidth]{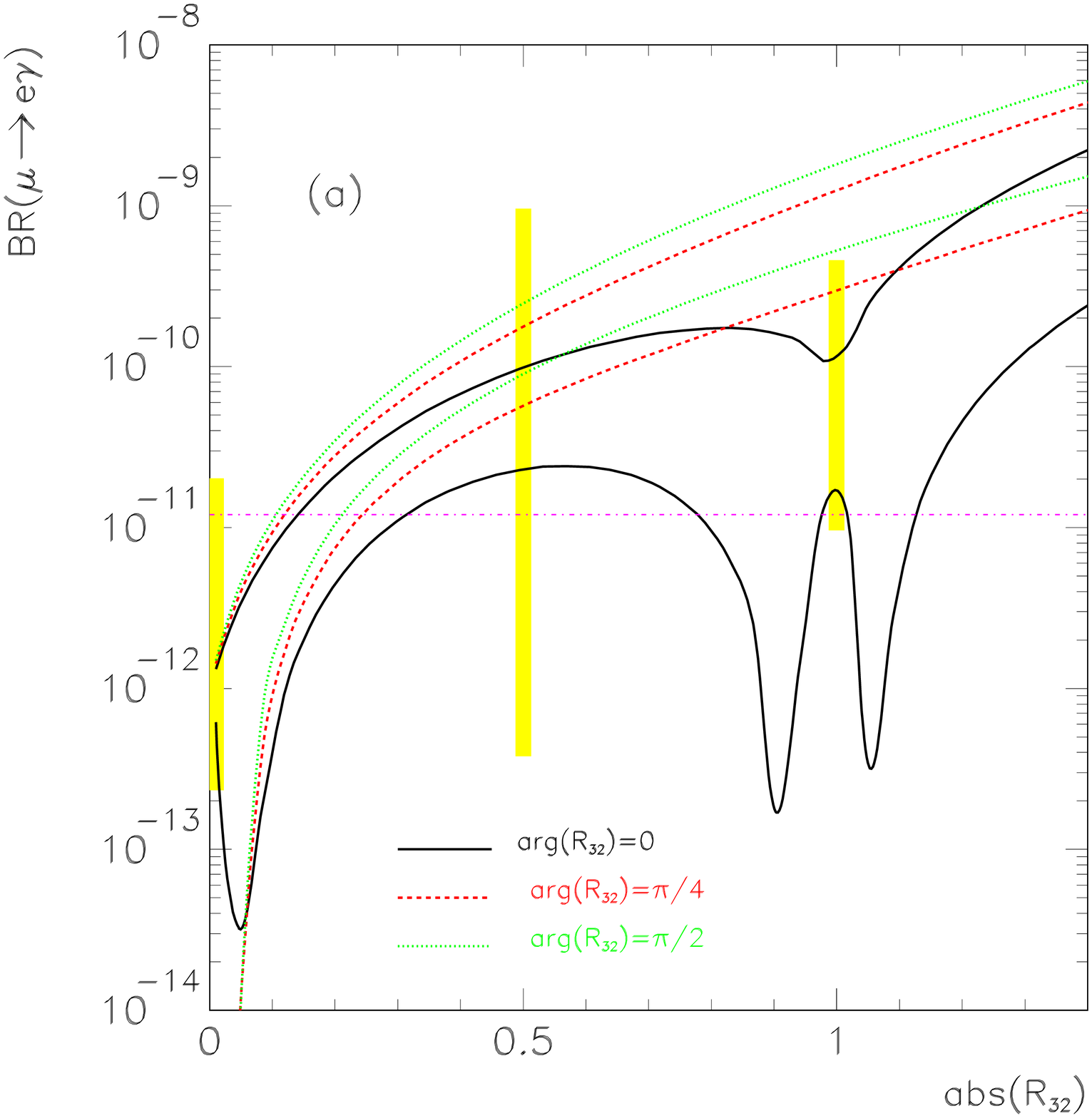}
\includegraphics*[width=0.33\linewidth]{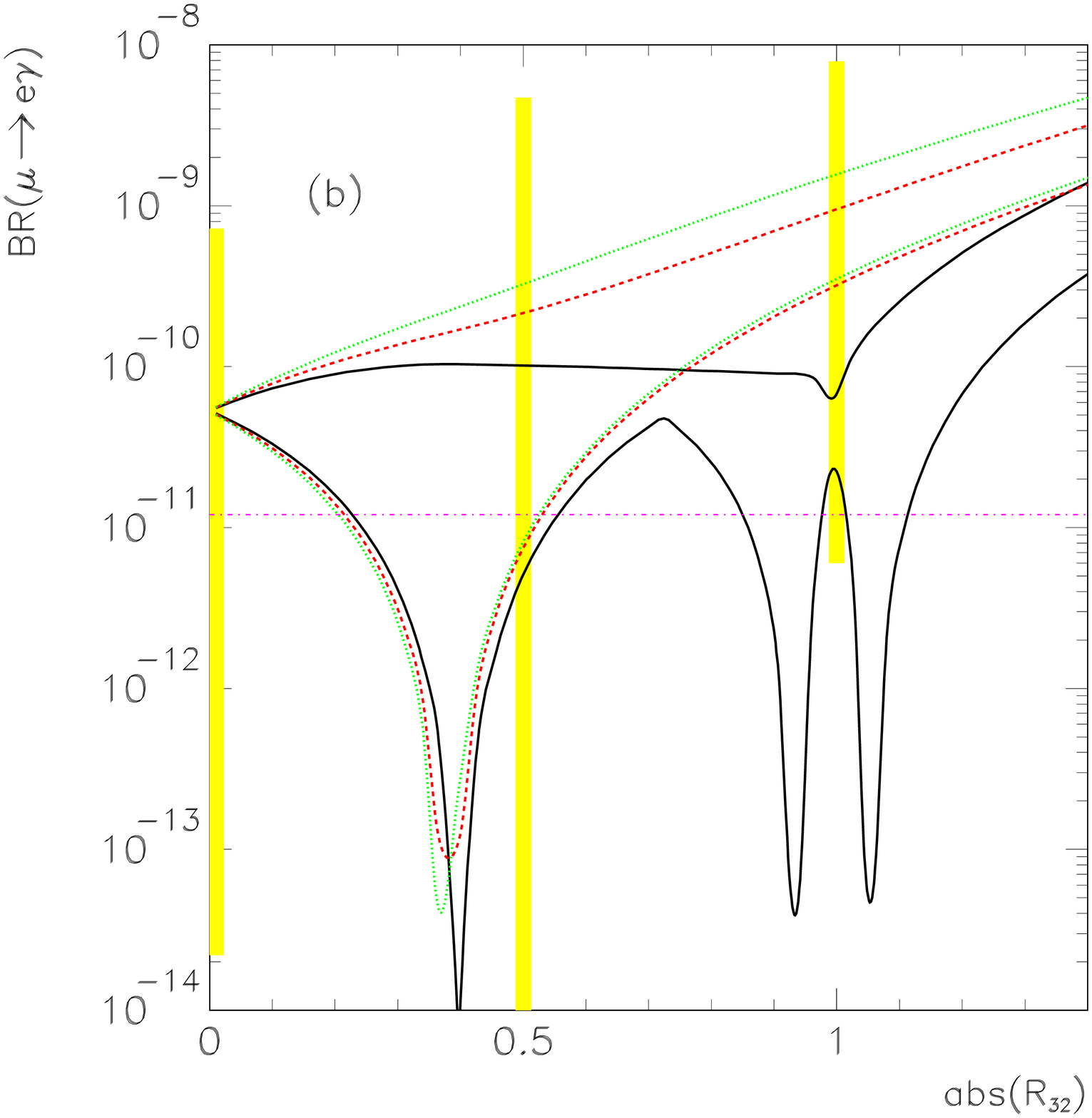}
\includegraphics*[width=0.33\linewidth]{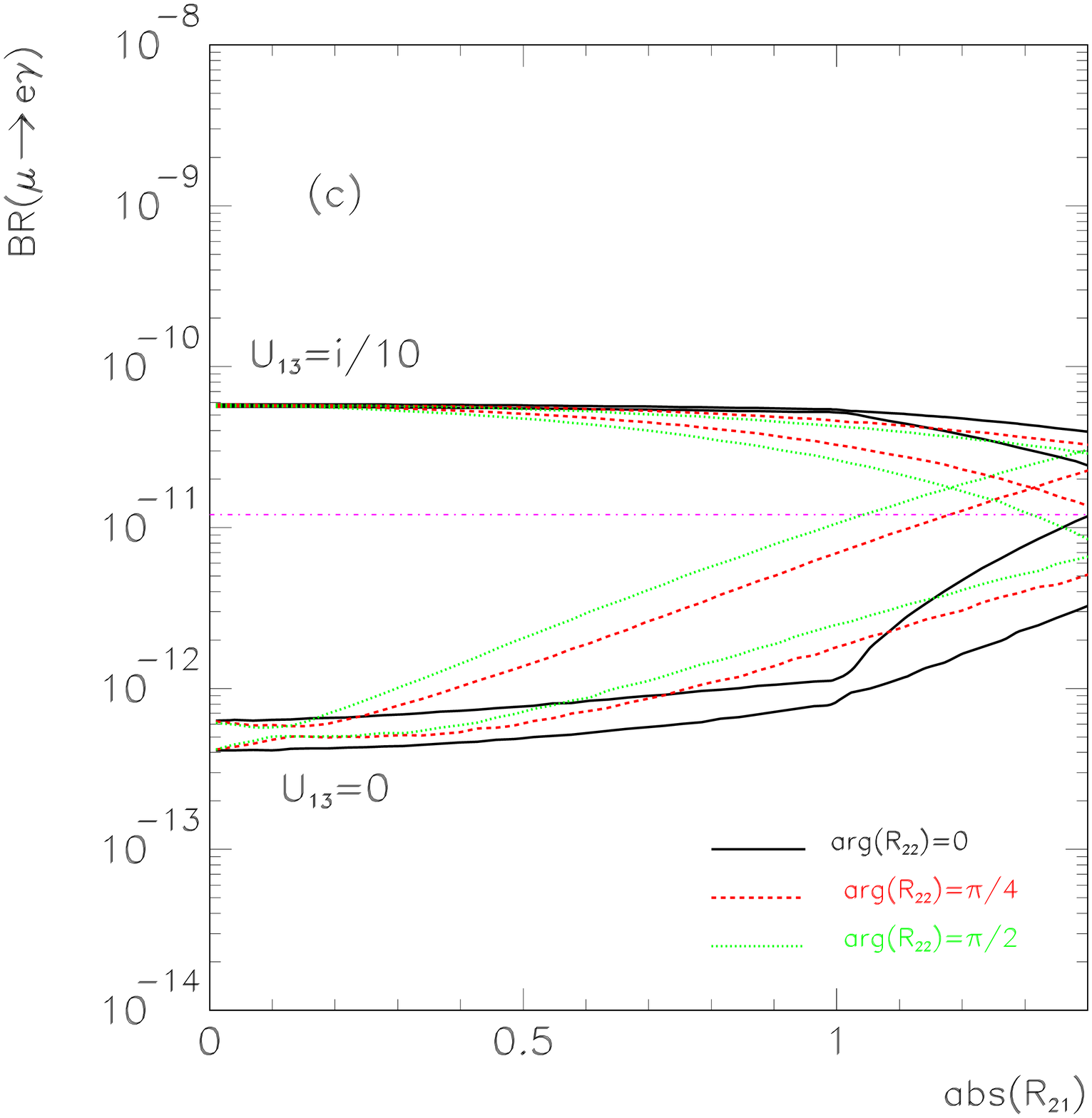}
\caption{ Predicted ranges of $B(\mu\to e\gamma)$ for 
$(M_1,M_2,M_3)=(2,3,50)\times 10^{13}$~GeV, 
$m_0=100$~GeV, $M_{1/2}=500$~GeV and $\tan\beta=10$,
for the decoupling of $N_1$ and ${U}_{13}=0$ (panel $a$) or 
${U}_{13}=0.1i$ (panel $b$). Yellow ranges show the possible 
variation for arbitrary form of ${R}$ with 
$\mathrm{arg}({R}_{32})=0$. Lower (upper) pairs of lines in the 
panel $c$ show similar ranges
for $N_3$ dominance for ${U}_{13}=0$ ($0.1i$).
The current experimental bound of 
$1.2\times 10^{-11}$ \cite{Brooks:1999pu} is also shown.  
\label{fone}}
\end{figure}

Analysis of the expressions (\ref{mi32}) and (\ref{mia1}) leads to a number 
of conclusions \cite{Chankowski:2004jc}. Firstly, the branching ratios of the LFV decays
depend (apart from the scales of soft supersymmetry breaking and the value 
of $\tan\beta$) mostly on the mass of the heaviest of the two un-decoupled 
singlet neutrinos (in this case $N_3$).
Secondly, for fixed $M_3$, they depend strongly on the magnitude and phase 
of ${R}_{32}$, mildly on the undetermined element ${U}_{13}$ 
of the light neutrino mixing matrix and, in addition, on the Majorana phases 
of ${U}$ which cannot be measured in oscillation experiments 
\cite{Pascoli:2003rq}. The latter dependence is mild for $B(\tau\to\mu\gamma)$ 
but can lead to strong destructive interference {\em either} in   
$B(\mu\to e\gamma)$ {\rm or} in
$B(\tau\to e\gamma)$ decreasing them by several orders of magnitude.
The interference effects are seen in Fig.~\ref{fone}(a) and Fig.~\ref{fone}(b)
where the predicted ranges (resulting from varying the unknown Majorana 
phases)
of $B(\mu\to e\gamma)$ are shown  
as a function of $|{R}_{32}|$ for $M_1$ appropriate for the sneutrino
inflation scenario, three different values of arg$({R}_{32})$ and for 
$m_0=100$ GeV, $M_{1/2}=500$ GeV and $\tan\beta=10$, consistent with the 
dark matter abundance \cite{Ellis:2003cw}.
Results for other values of these parameters can be obtained by 
appropriate rescalings using (\ref{BR}). 
For comparison, for selected values of ${R}_{23}$, we also indicate the 
ranges of $B(\mu\to e\gamma)$ resulting from generic form of the
matrix ${R}$ (constrained only by the conditions 
$0<{R}_{12},{R}_{13}<1.5$ and 
and $\mathrm{Re}({Y}_\nu^{AB}),\mathrm{Im}({Y}_\nu^{AB})<10$).

The bulk of the predicted values of $B(\mu\to e\gamma)$ shown in 
Fig.~\ref{fone}(a) and Fig.~\ref{fone}(b) exceed the current experimental limit. 
Since $M_{1/2}=500$ GeV leads to masses of the third generation squarks above 
1 TeV, suppressing $B(\mu\to e\gamma)$ by increasing the SUSY breaking scale 
conflicts with the stability of the electroweak scale.
Moreover, as discussed
in \cite{Chankowski:2004jc} in the scenario considered here  generically
$B(\tau\to\mu\gamma)/B(\mu\to e\gamma)\sim 0.1$. Thus the observation
$\tau\to\mu\gamma$ with $B \stackrel{>}{{}_\sim} 10^{-9}$,
accessible to future experiments would exclude this scenario.

For completeness, in Fig.~\ref{fone}(c) we also show predictions for 
$B(\mu\to e\gamma)$ in the case of $N_3$ dominance. 
$\left(\tilde{{m}}_L^2\right)_{21}$ is in this case controlled 
mainly by $|{U}_{13}|$. Moreover, 
$B(\tau\to\mu\gamma)/B(\mu\to e\gamma)\sim{\rm max}
\left(|U_{13}|^2,\rho^4S^2\right)$, while 
$B(\tau\to e\gamma)/B(\mu\to e\gamma)\sim 1$
allowing for experimental test of this scenario (cf.~\cite{Blazek:2001zm}). The limits 
${R}_{32}\to0$ in panels $a$ and $b$ and or ${R}_{21}\to0$ in 
panel $c$ correspond to pure sequential dominance. 

In conclusion, the well motivated assumption about
the decoupling/dominance of one heavy singlet neutrino significantly 
constrains the predictions for the LFV processes in supersymmetric model. 
The forthcoming experiments should be able to verify this assumption and,
in consequence, to test an interesting class of neutrino mass models.

\subsubsection{Triplet seesaw mechanism and lepton flavour violation}\label{sec:arossi-tlfv}

In this subsection we intend to discuss
the aspect of low scale LFV in rare decays 
arising in the context of the triplet seesaw mechanism of Section~\ref{sec:seesawII}.
We consider both non-SUSY and SUSY versions of it. 
The  flavour structure of the (high-energy) Yukawa matrix $Y_T$ of Eq.~(\ref{eq:Lag-typeII})
is the same as that of the (low-energy)
neutrino mass matrix $m_\nu$. Therefore, in the triplet seesaw scenario the neutrino mass matrix (containing 9 real parameters),
which can be tested in the low-energy experiments, is \emph{directly} linked to the symmetric matrix $Y_T$ (containing
also 9 real parameters), modulo the ratio $M^2_T/\mu'$, see Eq.~(\ref{numass-t1}). 
This feature has interesting implications for LFV \cite{Rossi:2002zb}. 
Collider phenomenology of the  low scale triplet was discussed in 
Section~\ref{sec:lowtriplet} 
The triplet Lagrangian also induces   LFV decays of the charged leptons through 
the one-loop exchange of the triplet states.

\begin{figure}[htb]
\begin{center}
\includegraphics[width=14.7cm]{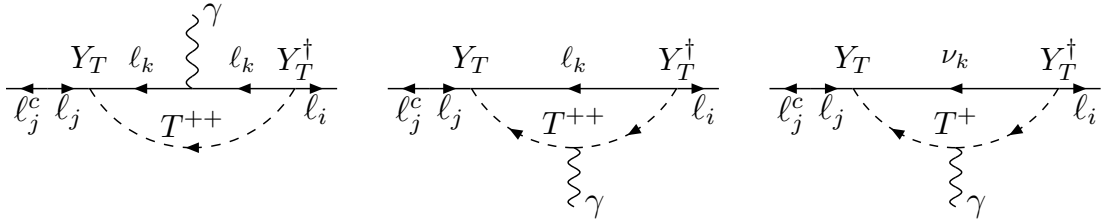}
\caption{Diagrams that contribute to the decay $\ell_j\to \ell_i \gamma$ through the exchange of the triplet scalars.}\label{fig1:arossi}
\end{center}
\end{figure}

The diagrams relevant for the LFV radiative decays $\ell_j \to \ell_i \gamma$ 
(see \eg \cite{Cuypers:1996ia,Chun:2003ej}) are depicted in  \Fref{fig1:arossi}. 
Denoting $U_{PMNS} = V \cdot {\rm diag}(1, {\rm e}^{i\phi_1},
{\rm e}^{i\phi_2})$, where,  $\phi_{1,2}$ are the Majorana phases, 
those imply the  following flavour structure:
\begin{equation}
\label{yyt} (Y^\dagger_T Y_T)_{ij} = \left(\frac{M^2_T}{\mu' v^2}\right)^2 (m^\dagger_\nu m_\nu)_{ij}= \left(\frac{M^2_T}{\mu'
v^2}\right)^2 [V (m^D_\nu)^2 V^\dagger]_{ij}\, ,
\end{equation}
where $i,j = e,\mu, \tau$ are family indices.
Therefore, the amount of LFV is \emph{ directly} and \emph{univocally} expressed in terms of the low-energy neutrino
parameters. In particular, LFV decays depend only on 7 independent neutrino parameters (there is no dependence on the Majorana phases $\phi_i$).
Notice that this simple flavour structure is peculiar of the triplet seesaw case, which represents a concrete and explicit realization of
the `minimal flavour violation' 
 hypothesis \cite{Hall:1990ac} in the lepton sector \cite{Cirigliano:2005ck}. Indeed, according to
the latter, the low-energy SM Yukawa couplings are the \emph{only source} of LFV. This is not generically the case for the seesaw
mechanism realized through the exchange of the so-called `right-handed' neutrinos, where the number of independent parameters
of the high-energy flavour structures is twice more that of the mass matrix $m_\nu$.

Finally, the parametric dependence of the dipole amplitude in \Fref{fig1:arossi} is:
\begin{equation}
\label{yy2} D_{ij} \approx \frac{(Y^\dagger_T Y_T)_{ij}}{16 \pi^2 M^2_T} = \frac{(m^\dagger_\nu m_\nu)_{ij}} {16 \pi^2
v^4}\left(\frac{M_T}{\mu'}\right)^2   \, ,
\end{equation}
From the present experimental bound on $B(\mu \to e \gamma) < 1.2\times 10^{-11}$ \cite{Brooks:1999pu}, one infers the  bound
$\mu' > 10^{-10} M_T$ [comparable limit is obtained from  $B(\tau \to \mu \gamma)$].
We can push further our discussion considering the relative size  of LFV in different family sectors:
\begin{equation}
\label{ratios1} \frac{ (Y^\dagger_T Y_T)_{\tau \mu}}  {(Y^\dagger_T Y_T)_{\mu e} } \approx \frac{\left[V (m^{D }_\nu)^2 V^\dagger\right]_{\tau \mu}} {\left[V
(m^{D }_\nu)^2 V^\dagger\right]_{\mu e}} \,~~  , \, ~~~~  \frac{ ( Y^\dagger_T Y_T)_{\tau e}}  {( Y^\dagger_T Y_T)_{\mu e} } \approx
\frac{\left[V (m^{D }_\nu)^2 V^\dagger\right]_{\tau e}} {\left[V (m^{D }_\nu)^2 V^\dagger\right]_{\mu e}} .
\end{equation}
These ratios depend only on the neutrino parameters, while do not depend on details of the model, such as the mass scales $M_T$, or
$\mu'$. By taking the present best fit values of the neutrino masses and mixing angles \cite{Fogli:2005cq}--\cite{Strumia:2006db}
provided by the analysis of the experimental data, those ratios can be explicitly expressed as:

\begin{eqnarray}\label{ratios2}
\frac{ (Y^\dagger_T Y_T)_{\tau \mu}}  {(Y^\dagger_T Y_T)_{\mu e} }  \approx \left(\frac{\Delta m^2_{A}}{\Delta m^2_{S}}\right) \frac{\sin
2\theta_{23}} {\sin 2\theta_{12} \cos\theta_{23}} \sim 40 \quad , \quad \frac{ (Y^\dagger_T Y_T)_{\tau e}} {(Y^\dagger_T Y_T)_{\mu e} } \approx
- \tan\theta_{23} \sim - 1 \, ,
\end{eqnarray}
where $\Delta m^2_{A} (\Delta m^2_{S})$ is the squared-mass difference relevant for the atmospheric (solar) neutrino oscillations. These results hold for
$\theta_{13}=0$ and for either hierarchical, quasi-degenerate or inverted hierarchical neutrino spectrum (for more details see \cite{Rossi:2002zb},
\cite{Joaquim:2006mn}). It is immediate to translate the above relations into \emph{ model-independent} predictions for  ratios of LFV processes:
\begin{eqnarray}\label{ratios3}
\frac{B(\tau \to \mu\gamma)}{B(\mu \to e\gamma)} & \approx & \left(\frac{(Y^\dagger_T Y_T)_{ \tau \mu}}{(Y^\dagger_T
Y_T)_{\mu e}}\right)^2 \frac{B(\tau \to  \mu \nu_\tau \bar{\nu}_\mu)} {B(\mu \to  e \nu_\mu \bar{\nu}_e)} \sim 300 \, , \, ~  \nonumber \\
\frac{B(\tau \to  e \gamma)}{B(\mu \to  e\gamma)}& \approx & \left(\frac{( Y^\dagger_T Y_T)_{ \tau e}}{(Y^\dagger_T
Y_T)_{\mu e}}\right)^2 \frac{B(\tau \to  e \nu_\tau \bar{\nu}_e)} {B(\mu \to e \nu_\mu \bar{\nu}_e)} \sim 0.2 \,.
\end{eqnarray}

Now we focus  upon the supersymmetric version of the triplet seesaw
 mechanism. 
 (Just recall just that in the supersymmetric case there is only one mass
 parameter, $M_T$, while the mass  parameter $\mu'$ of the
 non-supersymmetric version is absent from the superpotential and its
 role is taken by $\lambda_2 M_T$.)
 Regarding the aspect of LFV, in this case we have to consider
 besides the diagrams of \Fref{fig1:arossi} also the related ones
 with each particle in the loop replaced by its superpartner
 ($\ell_k\rightarrow \tilde{\ell}_k, T\rightarrow \tilde{T}$). Such
 additional contributions would cancel those in \Fref{fig1:arossi} in
 the limit of exact supersymmetry. In the presence of soft
 supersymmetry breaking (SSB) the cancellation is only partial and
 the overall result for the coefficient of the dipole amplitude
 behaves like
 \begin{equation}\label{dipole-susy1}
 D_{ij} \approx \frac{(Y^\dagger_T Y_T)_{ij}}{16 \pi^2 }\frac{
 \tilde{m}^2}{M^4_T} \sim \frac{(m^\dagger_\nu m_\nu)_{ij}} {16
 \pi^2(\lambda_2 v^2_2)^2} \frac{\tilde{m}^2}{M^2_T} \, ,
\end{equation}
 which is suppressed with respect to the non-supersymmetric result
 (\ref{yy2}) for $M_T > \tilde{m} \sim {\cal O} (10^2~{\rm GeV})$
 ($\tilde{m}$ denotes an average soft-breaking mass parameter).
In the supersymmetric version of the triplet seesaw mechanism flavour violation 
can also be  induced by renormalization effects via Eq.~(\ref{log1})
(the complete set of RGEs of the MSSM with the triplet states have been 
computed in \cite{Rossi:2002zb}). Thus in SUSY model the LFV processes can occur also 
in the case of very heavy triplet. In that case the relevant flavour structure
responsible for LFV is again $Y^\dagger_T Y_T$ for which we have already noticed its \emph{unambiguous} dependence on the neutrino
parameters in Eq.~(\ref{yyt}). Clearly, we find that analogous ratios as in Eq.~(\ref{ratios1}) hold also for the LFV entries of
the soft-breaking parameters, \eg
\begin{equation}\label{ratios4}
\frac{ ( m^{2 }_{\tilde{L}})_{\tau \mu}} {( m^{2 }_{\tilde{L}})_{\mu e} } \approx
\frac{\left[V (m^{D }_\nu)^2 V^\dagger\right]_{\tau \mu}} {\left[V (m^{D }_\nu)^2 V^\dagger\right]_{\mu e}} \,~~  , \, ~~~~
\frac{ ( m^{2 }_{\tilde{L}})_{\tau e}}{( m^{2 }_{\tilde{L}})_{\mu e} } \approx
\frac{\left[V (m^{D }_\nu)^2 V^\dagger\right]_{\tau e}} {\left[V (m^{D }_\nu)^2 V^\dagger\right]_{\mu e}} .
\end{equation}
Such SSB flavour-violating mass parameters induce extra contributions to the LFV processes. For example, the radiative decays $\ell_j \to \ell_i \gamma$
receive also one-loop contributions with the exchange of the charged-sleptons/neutralinos and sneutrinos/charginos, where the
slepton masses $(m^2_{\tilde{L} })_{ij}$ are the source of LFV.  The relevant dipole terms have a parametric dependence of the form
\begin{equation}
\label{dipole-SUSY2} D_{ij} \approx \frac{ g^2  }{16 \pi^2}\frac{(m^2_{\tilde{L}})_{ij}}{\tilde{m}^4}\tan\beta\approx
\frac{ g^2 }{16 \pi^2} \frac{(m^\dagger_\nu m_\nu)_{ij}}{(\lambda_2 v^2_2)^2} \frac{M^2_T}{\tilde{m}^2}\log(\frac{M_G}{M_T}) \tan\beta  \, .
\end{equation}
Notice  the inverted dependence on the ratio ${\tilde{m}}/{M_T}$ with respect to the triplet-exchange contribution. Due to this feature, the MSSM sparticle-induced  contributions (\ref{dipole-SUSY2}) tends to
dominate over the one induced by the triplet-exchange.
In this case, analogous ratios as in (\ref{ratios3}) can be derived,
\ie
\begin{eqnarray}\label{ratios6}
\frac{B(\tau \to \mu\gamma)}{B(\mu \to e\gamma)} & \approx & \left(\frac{(m^2_{\tilde{L}})_{ \tau \mu}}{(m^2_{\tilde{L}})_{\mu
e}}\right)^2 \frac{B(\tau \to  \mu \nu_\tau \bar{\nu}_\mu)} {B(\mu \to  e \nu_\mu \bar{\nu}_e)} \sim 300 \, , \, ~  \nonumber \\
\frac{B(\tau \to  e \gamma)}{B(\mu \to  e\gamma)}& \approx & \left(\frac{(m^2_{\tilde{L}})_{ \tau e}}{(m^2_{\tilde{L}})_{\mu
e}}\right)^2 \frac{B(\tau \to  e \nu_\tau \bar{\nu}_e)}{B(\mu \to e \nu_\mu \bar{\nu}_e)} \sim 0.2  .
\end{eqnarray}
(For more details see \cite{Rossi:2002zb}.)

The presence of extra $SU(2)_W$ triplet states at intermediate energy spoils the successful gauge coupling unification of the MSSM.
A simple way to recover gauge coupling unification is  to introduce more states $X$, to complete a certain representation $R$ -- such that $R = T +X$ --
of some unifying gauge group $G$, $G \supset SU(3)\times SU(2)_W\times U(1)_Y$. In general the Yukawa
couplings of the states $X$ are related to those of the triplet partners $T$. Indeed, this is generally  the case in  minimal GUT
models. In this case RG effects generates not only lepton- flavour violation but also closely correlated flavour violation  in the
quark sector (due to the $X$-couplings). An explicit scenario with $G=SU(5)$ where both lepton and quark flavour violation arise from
RG effects was discussed in \Ref~\cite{Rossi:2002zb}. In Section~\ref{sec:fjar-tgut} we review a supersymmetric $SU(5)$ model for the
triplet seesaw scenario.

\subsection{SUSY GUTs}\label{sec:SUSYGUTs}

\subsubsection{Flavour violation in the minimal supersymmetric SU(5) seesaw model}\label{sec:hisano}

In this section we review flavor- and/or CP-violating phenomena in the minimal SUSY SU(5) GUT, in which the right-handed neutrinos are
introduced to generate neutrino masses by the type-I seesaw mechanism. Here, it is assumed that the Higgs doublets in this MSSM
are embedded in {\bf 5}- and {$\bf \bar{5}$}- dimensional SU(5) multiplets. Rich flavor structure is induced even in those minimal
particle contents.  The flavor-violating SUSY breaking terms for the right-handed squarks and sleptons are generated by the GUT
interaction, while those are suppressed in the MSSM (+$\nu_R$) under the universal scalar mass hypothesis for the SUSY breaking terms.

The Yukawa interactions for quarks and leptons and the Majorana mass terms for the right-handed neutrinos in this model are given by the
following superpotential,
\begin{eqnarray}
W&=& \frac14 Y_{ij}^{u} \Psi_i \Psi_j H +\sqrt{2} Y_{ij}^{d} \Psi_i \Phi_j \overline{H}+
Y_{ij}^{\nu} \Phi_i \overline{N}_j {H}
+\frac12 M_{Nij} \overline{N}_i \overline{N}_j,\label{superp_gut_MSGUTN}
\end{eqnarray}
where $\Psi$ and $\Phi$ are for {\bf 10}- and {$\bf \bar{5}$}-dimensional multiplets, respectively, and $\overline{N}$ is for the right-handed
neutrinos.  $H$ ($\overline{H}$) is {\bf 5}- ({$\bf \bar{5}$}-) dimensional Higgs multiplets.  After removing the unphysical degrees
of freedom, the Yukawa coupling constants in Eq.~(\ref{superp_gut_MSGUTN}) are given as follows,
\begin{eqnarray}
Y^u_{ij} &=& V_{ki} Y_{u_k} {\rm e}^{i \varphi_{u_k}}V_{kj}, \nonumber\\
Y^d_{ij} &=& Y_{d_i} \delta_{ij},\nonumber\\
Y^\nu_{ij} &=& {\rm e}^{i \varphi_{d_i}} U^\star_{ij} Y_{\nu_j}. 
\label{Yukawa_MSGUTN}
\end{eqnarray}
Here, $Y_u,$ $Y_d,$ $Y_\nu$ denote diagonal Yukawa couplings,
  $\varphi_{u_i}$ and $\varphi_{d_i}$ $(i=1$--$3)$ are CP-violating phases inherent in the SUSY SU(5) GUT ($\sum_i
\varphi_{u_i} =\sum_i \varphi_{d_i} =0$).  The unitary matrix $V$ is the CKM matrix in the extension of the SM to the SUSY SU(5) GUT, and
each unitary matrices $U$ and $V$ have only a phase. When the Majorana mass matrix for the right-handed neutrinos is real and diagonal in the
basis of Eq.~(\ref{Yukawa_MSGUTN}), $U$ is the PMNS matrix measured in the neutrino oscillation experiments and the light neutrino mass
eigenvalues are given as $m_{\nu_{i}}={Y_{\nu_i}^2}\langle H_2 \rangle^2/{M_{N_i}}$, in which $M_{N_i}$ are the diagonal components.

The colored-Higgs multiplets $H_c$ and $\overline{H}_c$ are introduced in $H$ and $\overline{H}$ as SU(5) partners of the Higgs doublets
$H_f$ and $\overline{H}_f$ in the MSSM, respectively, and they have new flavor-violating interactions.  Eq.~(\ref{superp_gut_MSGUTN}) is
represented by the fields in the MSSM as follows,
\begin{eqnarray}
W&=& W_{MSSM+\overline{N}} \nonumber\\
&& + \frac12 V_{ki} Y_{u_k} {\rm e}^{i \varphi_{u_k}}V_{kj} Q_i Q_j H_c + Y_{u_i} V_{ij} {\rm e}^{i \varphi_{d_j}} \overline{U}_i \overline{E}_j H_c 
\nonumber\\
&& + Y_{d_i} {\rm e}^{-i \varphi_{d_i}} Q_i L_i \overline{H}_c + {\rm e}^{-i \varphi_{u_i}} V_{ij}^\star Y_{d_j} \overline{U}_i \overline{D}_j \overline{H}_c 
+ {\rm e}^{i \varphi_{d_i}} U_{ij}^\star Y_{\nu_j} \overline{D}_i \overline{N}_j H_c.
\end{eqnarray}
Here, the superpotential in the MSSM with the right-handed neutrinos is
\begin{eqnarray}
W_{MSSM+\overline{N}} &=& V_{ji} Y_{u_j}  Q_i \overline{U}_j H_f + Y_{d_i} Q_i \overline{D}_i \overline{H}_f + Y_{d_i} L_i \overline{E}_i \overline{H}_f \nonumber\\
&&+U^\star_{ij} Y_{\nu_j} L_i \overline{N}_j H_f +M_{ij} \overline{N}_i \overline{N}_j. \label{superp_mssm_MSGUTN}
\end{eqnarray}  
The flavor-violating interactions, which are absent in the MSSM, emerge in the SUSY SU(5) GUT due to existence of the colored-Higgs
multiplets.  The colored-Higgs interactions are also baryon-number violating \cite{Sakai:1981pk,Weinberg:1981wj}, and then proton decay
induced by the colored-Higgs exchange is a serious problem, especially in the minimal SUSY SU(5) GUT \cite{Goto:1998qg}. However, the
constraint from the proton decay depends on the detailed structure in the Higgs sector, and it is also loosened by global symmetries, such
as the Peccei-Quinn symmetry and the U(1)$_R$ symmetry. Thus, we may ignore the constraint from the proton decay while we adopt the minimal
Yukawa structure in Eq.~(\ref{superp_gut_MSGUTN}). 

The sfermion mass terms get sizable corrections by the colored-Higgs interactions, when the SUSY-breaking terms in the MSSM
are generated by dynamics above the colored-Higgs masses. In the minimal supergravity scenario the SUSY breaking terms are supposed to
be given at the reduced Planck mass scale ($M_G$). In this case, the flavor-violating SUSY breaking mass terms at low energy are induced by
the radiative correction, and they are qualitatively given in a flavor basis as
\begin{eqnarray}
(m_{{\tilde{u}_L}}^2)_{ij}  &\simeq&- V_{i3}V_{j3}^\star  \frac{Y_{b}^2}{(4\pi)^2} \;\; (3m_0^2+ A_0^2) \;\; 
(2 \log\frac{M_G^2}{M_{H_c}^2}+ \log\frac{M_{H_c}^2}{M_{SUSY^2}}),\nonumber\\
(m_{\tilde{u}_R}^2)_{ij}  &\simeq& - {\rm e}^{-i\varphi_{u_{ij}}} V_{i3}^\star V_{j3} \frac{2Y_{b}^2}{(4\pi)^2} \;\; (3m_0^2+ A_0^2) \;\;
log\frac{M_G^2}{M_{H_c}^2}, \nonumber\\
(m_{{\tilde{d}_L}}^2)_{ij}  &\simeq&-V_{3i}^\star V_{3j} \frac{Y_{t}^2}{(4\pi)^2} \;\; (3m_0^2+ A_0^2) \;\; 
(3 \log\frac{M_G^2}{M_{H_c}^2}+ \log\frac{M_{H_c}^2}{M_{SUSY}^2}),\nonumber\\
(m_{\tilde{d}_R}^2)_{ij}  &\simeq&-{\rm e}^{i\varphi_{d_{ij}}}  U^\star_{ik}U_{jk} \frac{Y_{\nu_k}^2}{(4\pi)^2} \;\; (3m_0^2+A_0^2) \;\; 
\log\frac{M_G^2}{M_{H_c}^2},\nonumber\\
(m_{\tilde{l}_L}^2)_{ij}  &\simeq&-U_{ik}U_{jk}^\star \frac{f^2_{\nu_k} }{(4\pi)^2} \;\; (3m_0^2+ A_0^2) \;\; 
\log\frac{M_G^2}{M_{N_k}^2},\nonumber\\
(m_{{\tilde{e}_R}}^2)_{ij}  &\simeq&-{\rm e}^{i\varphi_{d_{ij}}} V_{3i}V^\star_{3j} \frac{3 Y_{t}^2}{(4\pi)^2} \;\; (3m_0^2+ A_0^2)\;\; 
\log\frac{M_G^2}{M_{H_c}^2}, \label{sfermionmass_MSGUTN}
\end{eqnarray}
with $i\ne j$, where $\varphi_{u_{ij}}\equiv\varphi_{u_{i}}-\varphi_{u_{j}}$ and $\varphi_{d_{ij}}\equiv\varphi_{d_{i}}-\varphi_{d_{i}}$ and $M_{H_c}$
is the colored-Higgs mass.  Here, $M_{SUSY}$ is the SUSY-breaking scale in the MSSM, $m_0$ and $A_0$ are the universal scalar mass and
trilinear coupling, respectively, in the minimal supergravity scenario. $Y_t$ is the top quark Yukawa coupling constant while $Y_b$
is for the bottom quark.  The off-diagonal components in the right-handed squarks and slepton mass matrices are induced by the
colored-Higgs interactions, and they depend on the CP-violating phases in the SUSY SU(5) GUT with the right-handed neutrinos \cite{Moroi:2000tk}.

One of the important features of the SUSY GUTs is the correlation between the leptonic and hadronic flavor violations \cite{Hisano:2003bd,Ciuchini:2003rg}. From
Eq.~(\ref{sfermionmass_MSGUTN}), we get a relation
\begin{eqnarray}
(m_{\tilde{d}_R}^2)_{23} &\simeq& {\rm e}^{i\varphi_{d_{23}}} (m_{\tilde{l}_L}^2)^\star_{23}  
\times (\log\frac{M_G^2}{M_{H_c}^2}/\log\frac{M_G^2}{M_{N_3}^2}). \label{massrelation_MSGUTN}
\end{eqnarray}
The right-handed bottom-strange squark mixing may be tested in the $B$ factory experiments since it affects $B_s-\bar{B}_s$ mixing, CP
asymmetries in the $b-s$ penguin processes such as $B_d\to \phi K_s$, and the mixing-induced CP asymmetry in $B_d\to
M_s\gamma$. (See Chapter 2.)  The relation in Eq.~(\ref{massrelation_MSGUTN}) implies that the deviations from the
standard model predictions in the $b$-$s$ transition processes are correlated with $B(\tau\to\mu\gamma)$ in the SUSY SU(5)
GUT. We may test the model in the $B$ factories.

\begin{figure}
\begin{center}
\includegraphics[width=7cm]{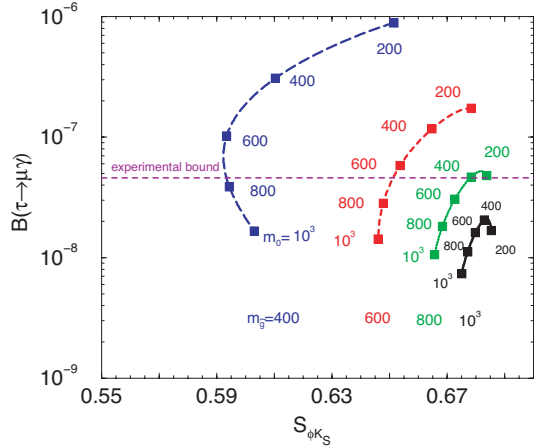} 
\caption{ $B(\tau\to\mu\gamma)$ as a function of $S_{\phi K_S}$ for fixed gluino masses $m_{\tilde{g}}=400$, 600, 800, and
 1000\,GeV. Here, $\tan\beta=10$, 200\,GeV$<m_0<$1\,TeV, $A_0=0$,  $m_{\nu_\tau}=5\times10^{-2}$\,eV, $M_{N_3}=5\times 10^{14}$\,GeV,
 and $U_{32}=1/\sqrt{2}$. $\varphi_{d_{23}}$ is taken for the  deviation of $S_{\phi K_S}$ from the SM prediction to be
 maximum. The current experimental bound on  $B(\tau\to\mu\gamma)$ 
 \cite{Abe:2006sf} is also shown in the  figure.  }
\label{fig1_MSGUTN}
\end{center}
\end{figure}

In \Fref{fig1_MSGUTN} we show the CP asymmetry in $B_d\to \phi K_s$ ($S_{\phi K_S}$) and $B(\tau\to\mu\gamma)$ as an
example of the correlation. Here, we assume the minimal supergravity hypothesis for the SUSY breaking terms. See the caption and
Ref.~\cite{Hisano:2003bd} for the input parameters and the details of the figure. It is found that $S_{\phi K_S}$ and
$B(\tau\to\mu\gamma)$ are correlated and a large deviation from the standard model prediction for $S_{\phi K_S}$ is not possible
due to the  current bound on $B(\tau\to\mu\gamma)$ in the SUSY SU(5) GUT.

In Eq.~(\ref{sfermionmass_MSGUTN}), we take the SU(5)-symmetric Yukawa interactions given in Eq.~(\ref{superp_gut_MSGUTN}), while they fails
to explain the fermion mass relations in the first and second generations. We have to extend the minimal model by introducing
non-trivial Higgs or matter contents or the higher-dimensional operators including SU(5)-breaking Higgs field. These extensions may
affect the prediction for the sfermion mass matrices. However, the relation in Eq.~(\ref{massrelation_MSGUTN}) is rather robust when the
neutrino Yukawa coupling constant of the third generation is as large as those for the top and bottom quarks and the large mixing in the
atmospheric neutrino oscillation comes from the lopside structure of the neutrino Yukawa coupling.

\begin{figure}
\includegraphics[width=7cm]{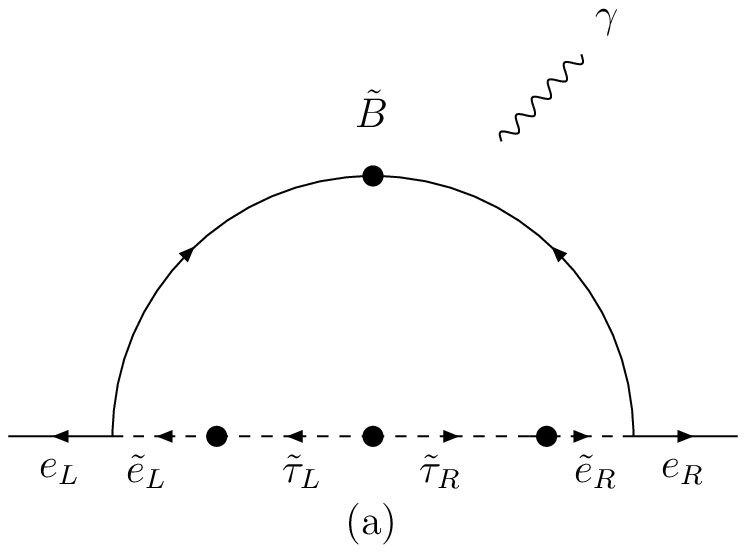} \hspace*{\fill}
\includegraphics[width=7cm]{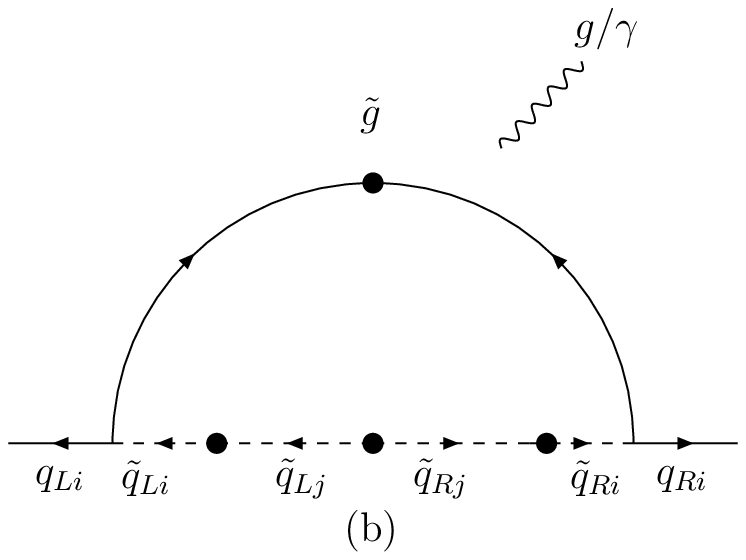} 
\caption{ (a) Diagram which generate electron EDM and (b) those which generate EDMs and CEDMs of the $i$-th quark due to flavor violation
in sfermion mass terms.  } \label{fig2_MSGUTN}
\end{figure}

Another important feature of the SUSY GUTs is that both the left- and right-handed squarks and sleptons have flavor mixing terms. In this
case, the hadronic and leptonic electric dipole moments (EDMs) are generated due to the flavor violation, and they may be large enough to
be observed in the future EDM measurements \cite{Dimopoulos:1994gj}. A diagram in \Fref{fig2_MSGUTN}(a) generates the electron EDM even at
one-loop level, when the relative phase between the left- and right-handed slepton mixing terms is non-vanishing. While this
contribution is suppressed by the flavor violation, it is compensated by a heavier fermion mass, that is, $m_\tau$. Similar diagrams in
\Fref{fig2_MSGUTN}(b) contribute to quark EDMs and chromo-electric dipole moments (CEDM), which induce the hadronic EDMs.  The EDM
measurements are important to probe the interaction of the SUSY SU(5) GUT.

\begin{figure}
\includegraphics[width=7cm]{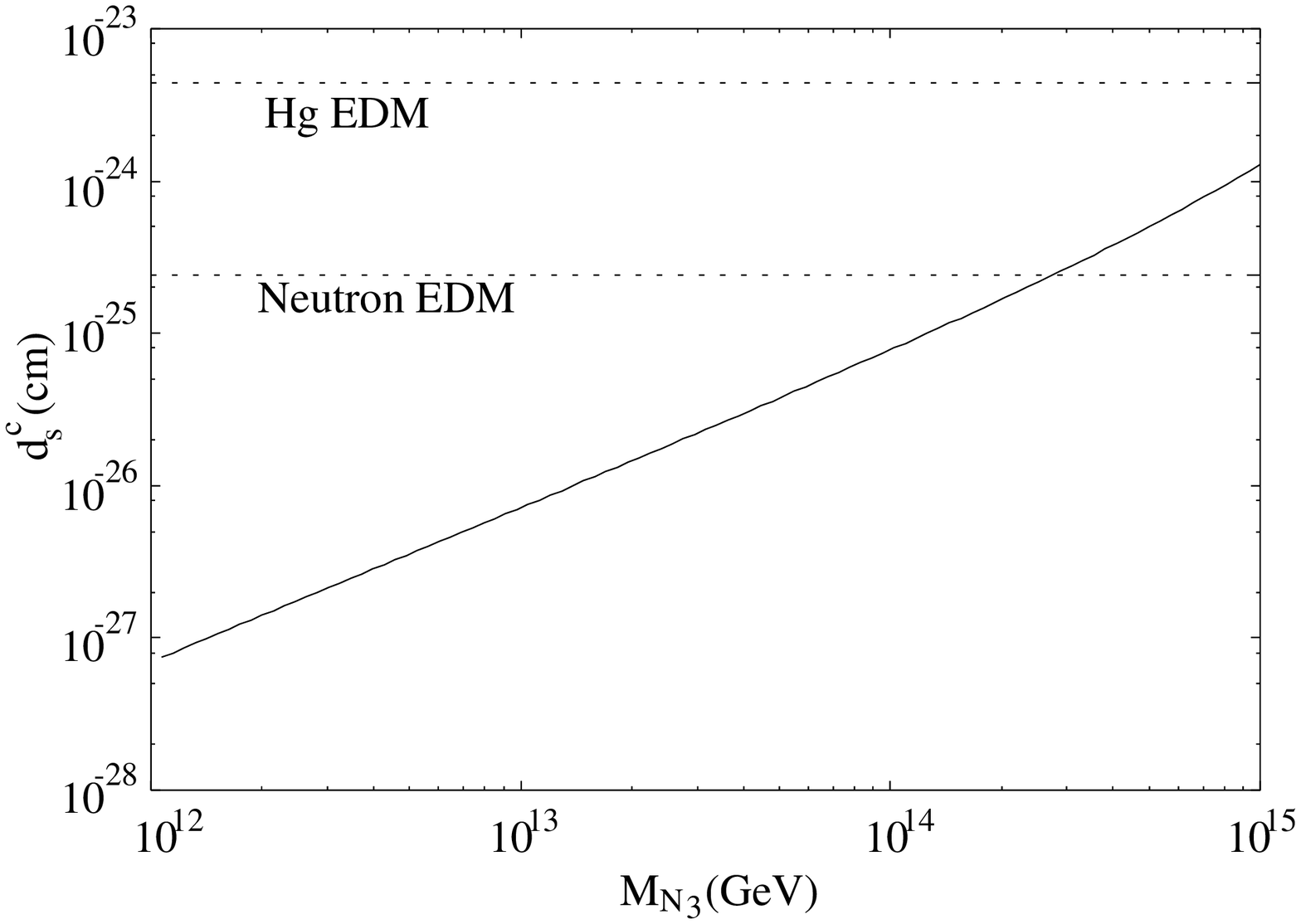} \hspace*{\fill}
\includegraphics[width=7cm]{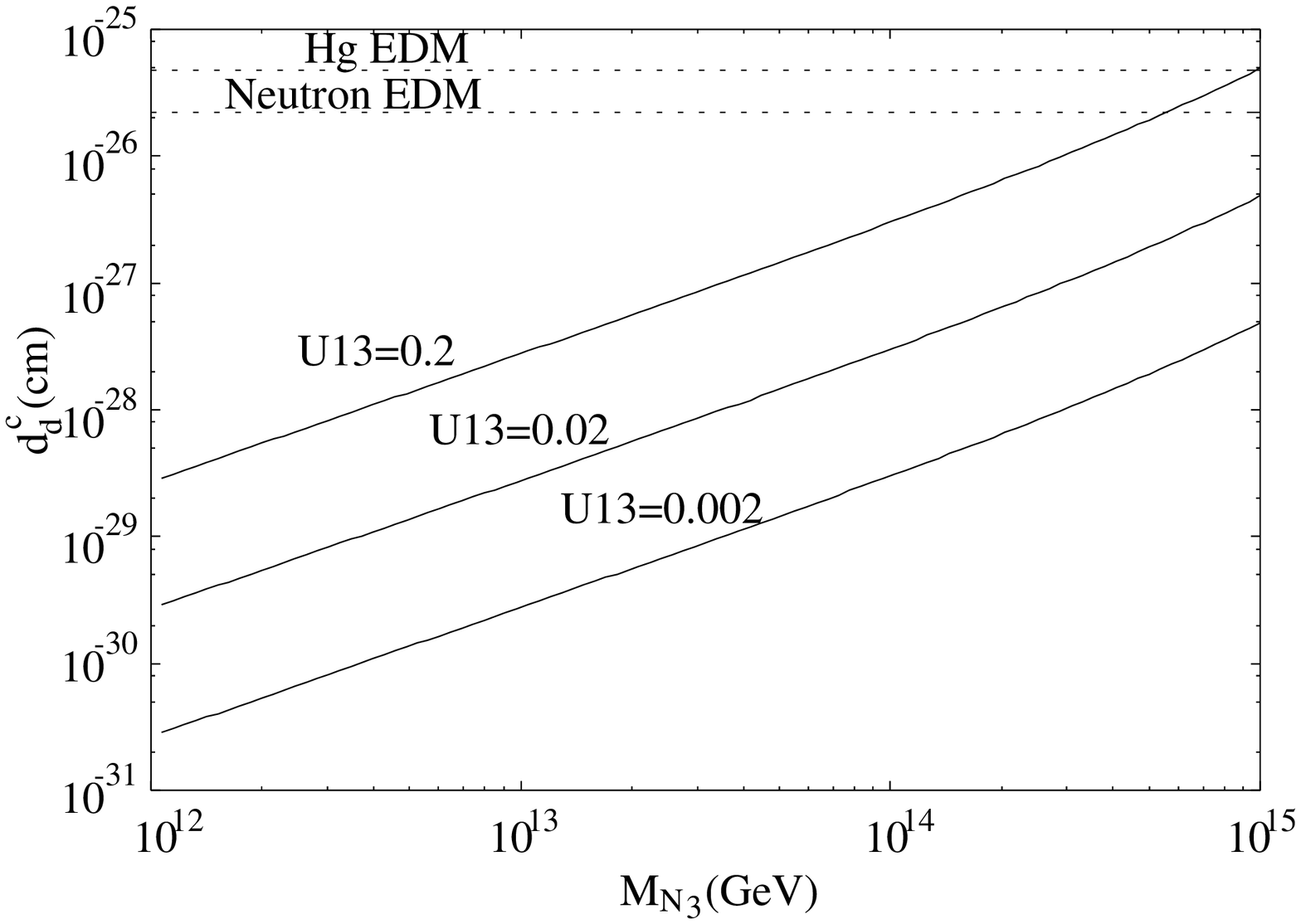} 
\caption{ CEDMs for the strange ($d_s^c$) and down quarks ($d_d^c$) as  functions of the right-handed tau neutrino mass, $M_{N_3}$.  Here,
 $M_{H_c}=2\times 10^{16}$GeV, $m_{\nu_\tau}=0.05$eV,  $U_{23}=1/\sqrt{2}$, and $U_{13}=0.2$, 0.02, and 0.002.  For the
 MSSM parameters, we take $m_0=500$GeV, $A_0=0$,  $m_{\tilde{g}}=500$GeV and $\tan\beta=10$. The CP phases
 $\varphi_{d_i}$ are taken for the CEDMs to be maximum.  The  upper bounds on the strange and down quark CEDMs from the mercury
 atom and neutron EDMs are shown in the figures.}\label{fig3_MSGUTN}
\end{figure}

In \Fref{fig3_MSGUTN} the CEDMs for strange ($d^c_s$) and down quarks ($d^c_d$) are shown as functions of the right-handed tau neutrino mass
in the SUSY SU(5) GUT with the right-handed neutrinos. See the caption and Ref.~\cite{Hisano:2004pw} for the input parameters.  The mercury
atom EDM, which is a diamagnetic atom, is sensitive to quark CEDMs via the nuclear force, while the neutron EDM depends on them in addition
to the quark EDMs.  (The evaluation of the hadronic EDMs from the effective operators at the parton level is reviewed in Section~9.1 and
also Ref.~\cite{Pospelov:2005pr}.)  The strange quark contribution to the mercury atom EDM is suppressed by the strange quark mass. On the
other hand, it is argued in Refs.~\cite{Khatsimovsky:1987bb,Zhitnitsky:1996ng} that the strange
quark component in nucleon is not negligible and the strange quark CEDM may give a sizable contribution to the neutron EDM. It implies
that we may probe the different flavor mixings by measurements of the various hadronic EDMs, though the evaluation of the hadronic EDMs
still has large uncertainties.

It is argued that the future measurements of neutron and deuteron EDMs may reach to levels of $\sim 10^{-28}e\,cm$ and $\sim 10^{-29}e\,cm$,
respectively. When the sensitivity of deuteron EDM is established, we may probe the new physics to the level of $d^c_s \sim 10^{-28}\;cm$
and $ d^c_d \sim d^c_u \sim 10^{-30}\;cm$ \cite{Hisano:2004tf}. The future measurements for the EDMs will give great impacts on the SUSY
SU(5) GUT with the right-handed neutrinos.

\subsubsection{LFV in the minimal $SU(5)$ GUT with triplet seesaw}\label{sec:fjar-tgut}

In this Section we discuss phenomenology of the minimal $SU(5)$ GUT which incorporates
the triplet seesaw mechanism, previously presented in Sections~\ref{sec:seesawII}  
and \ref{sec:arossi-tlfv}. Review of more general class of GUT models also 
including triplet Higgs has been given  in Section~\ref{sec:GUTs}.
In GUTs based on $SU(5)$  there is no natural place for incorporating singlet neutrinos, 
From this point of view $SU(5)$ presents some advantage for implementing the triplet seesaw mechanism. In particular, a very predictive scenarios can be obtained 
in the supersymmetric case \cite{Rossi:2002zb,Joaquim:2006uz,Joaquim:2006mn}. 
The triplet states $T$ ($\overline{T}$) fit into the 15
$(\overline{15})$ representation, $15 = S + T +Z$ with $S$, $T$ and $Z$ transforming as $S\sim (6,1,-\frac23), ~T \sim (1,3,1),~ Z\sim
(3,2,\frac16)$ under $SU(3) \times SU(2)_L \times U(1)_Y$ (the $\overline{15}$ decomposition is obvious).
%
We will briefly show that it is also possible to relate not only neutrino mass parameters and LFV (as shown in
Section~\ref{sec:arossi-tlfv}) but also sparticle and Higgs spectra and electroweak symmetry breakdown~\cite{Joaquim:2006uz,Joaquim:2006mn}. For this purpose,
consider that the $SU(5)$ model conserves  $B-L$, so that the relevant superpotential reads:
\begin{eqnarray}
W_{SU(5)}=  \frac{1}{\sqrt2}(Y_{15} \bar5~ 15 ~\bar5 + \lambda {5}_H ~\overline{15}~ {5}_H) + Y_5 \bar5 ~ \bar5_H 10
+ Y_{10} 10 ~10 ~5_H + M_5  5_H~\bar5_H  + \xi X 15 ~ \overline{15}\,. \nonumber \\
&&
\label{su5}
\end{eqnarray}
where the multiplets are understood as $\bar5  = (d^c, L)$, $10 =(u^c,e^c,Q)$ and the Higgs doublets fit with their coloured
partners $t$ and $\bar{t}$, like ${5}_H = (t, H_2), \bar{5}_H = (\bar{t}, H_1)$ and $X$ is a singlet superfield. The $B-L$ quantum
numbers are the combination $Q+\frac45 Y$ where $Y$ are the hypercharges and $Q_{10}=\frac15,\,  Q_{\bar{5}} = - \frac35,\,
Q_{{5}_H (\bar{5}_H )} =-\frac25 \left(\frac25\right), \, Q_{15} = \frac65, \, Q_{\overline{15}}= \frac45, \,Q_{X} = -2$. Both the
scalar $S_X$ and auxiliary $F_X$ components of the superfield $X$ are assumed to acquire a VEV through some unspecified dynamics in
the hidden sector. Namely, while $\langle S_X\rangle$ only breaks $B-L$, $\langle F_X\rangle$ breaks both SUSY and $B-L$. These
effects are parameterized  by the superpotential mass term $M_{15} 15~\overline{15}$, where $M_{15} = \xi \langle S_X\rangle$, and the
bilinear SSB term $-B_{15} M_{15}15~\overline{15}$, with $ B_{15} M_{15}= - \xi \langle F_X\rangle$. The $15$ and $\overline{15}$
states act, therefore, as {\it messengers} of both $B-L$ and SUSY breaking to the MSSM observable sector. Once $SU(5)$ is broken to
the SM group we find, below the GUT scale $M_G$,
\begin{eqnarray}
W& =& W_0 + W_T + W_{S,Z} , \nonumber \\
W_0 & = & Y_e  e^c H_1  L + Y_d d^c H_1  Q + Y_u  u^c Q  H_2  + \mu H_2 H_1 ,\nonumber \\
W_T&=& \frac{1}{\sqrt{2}}(Y_{T} L T L  + \lambda H_2 \bar{T} H_2) + M_T T \bar{T} ,\label{WT} \nonumber \\
W_{S,Z} & =& \frac{1}{\sqrt{2}}Y_S d^c S d^c + Y_Z  d^c  Z L + M_Z Z\bar{Z}+M_S S\bar{S} .
\label{su5b}
\end{eqnarray}
Here, $W_0$ denotes  the MSSM superpotential~\footnote{This should be regarded as an effective approach where the Yukawa matrices $Y_d, Y_e, Y_u$ include $SU(5)$-breaking effects 
needed to reproduce a realistic  fermion spectrum.}, the term $W_T$ is responsible for neutrino mass generation [cf.~(\ref{rossiL-T})], while
the couplings and masses of the coloured fragments $S$ and $Z$ are included in $W_{S,Z}$. It is also understood that $M_T=M_S=M_Z \equiv M_{15}$.
At the decoupling of the heavy states $S, T, Z$ we obtain at tree-level the neutrino masses, given by Eq.~(\ref{T-mass}) and at the quantum level all SSB mass parameters
of the MSSM via gauge and Yukawa interactions. At one-loop level, only the trilinear scalar couplings, the gaugino masses and the
Higgs bilinear mass term $B_H$ are generated:
\begin{eqnarray}
{A}_e & = & \frac{3 B_T}{16 \pi^2} Y_e (Y^\dagger_T Y_T + Y^\dagger_Z Y_Z)\, , ~~~ {A}_u =\frac{3B_T}{16 \pi^2} |\lambda|^2
Y_u  \,, ~~~ {A}_d =\frac{ 2 B_T}{16 \pi^2} (Y_Z Y^\dagger_Z  + 2 Y_SY^\dagger_S) Y_d \,,\nonumber \\
M_a &= &  \frac{7 B_T }{16 \pi^2}\,g_a^2\,, ~~~~~ B_H  = \frac{3 B_T }{16 \pi^2}\, |\lambda|^2\,.
\label{jr:soft1}
\end{eqnarray}
The scalar mass matrices instead are generated at the two-loop level and receive both gauge-mediated contributions proportional to $C_a^f
g^4_a$ ($C_a^f$ is the quadratic Casimir of the $f$-particle)  and Yukawa-mediated ones of the form $Y^\dagger_{p} Y_{p} (p=S,T,Z)$.
The former piece is the flavour blind contribution, which is proper of the gauge-mediated
scenarios~\cite{Dine:1981za,Dimopoulos:1981au,Alvarez-Gaume:1981wy,Nappi:1982hm,Dimopoulos:1982gm,Dine:1993yw,Dine:1994vc,Giudice:1998bp}, while the
latter ones constitute the flavour violating contributions transmitted to the SSB terms by the Yukawa's $Y_{S,T,Z}$. These
contributions are mostly relevant for the mass matrices $m^2_{\tilde{L}}$ and  $m^2_{\tilde{d}^c}$. For example,
\begin{eqnarray}
\!\!\!\!m^2_{\tilde{L}}& =& \left(\frac{B_T}{16 \pi^2}\right)^2 \left[\frac{21}{10} g^4_1 + \frac{21}{2} g^4_2 - \left(\frac{27}{5} g^2_1
+21 g^2_2\right)\!Y^\dagger_T Y_T - \left(\frac{21}{15} g^2_1 +9 g^2_2 + 16 g^2_3\right)\!Y^\dagger_Z Y_Z\right. \nonumber \\
&& \! \left. + 18  (Y^\dagger_T Y_T)^2 +15 (Y^\dagger_Z Y_Z)^2+ 3{\rm Tr}(Y^\dagger_TY_T) Y^\dagger_T Y_T + 12 Y^\dagger_Z Y_S Y^\dagger_S Y_Z \right. \nonumber \\
&& \! \left. + 3 {\rm Tr}(Y^\dagger_Z Y_Z) Y^\dagger_Z Y_Z + 9 Y^\dagger_T Y^T_Z Y^*_Z Y_T +9 (Y^\dagger_T Y_T Y^\dagger_Z Y_Z +{\rm h.c.}) \right. \nonumber \\
&& \!\left. + 3Y^\dagger_T Y^T_e Y^*_e Y_T +6 Y^\dagger_Z Y_d Y^\dagger_d Y_Z\right] \, .
\end{eqnarray}
Since the flavour structure of $m^2_{\tilde{L}}$ is proportional to $Y_T$ (and to $Y_Z$ which is $SU(5)$-related to $Y_T$),
it can be expressed in terms of the neutrino parameters [cf. Eq.~(\ref{T-mass})] and so the relative size of LFV in different leptonic
families is predicted according to the results of Eq.~(\ref{ratios4}).
\begin{figure}
\parbox{0.48\linewidth}{
\includegraphics[width=\linewidth]{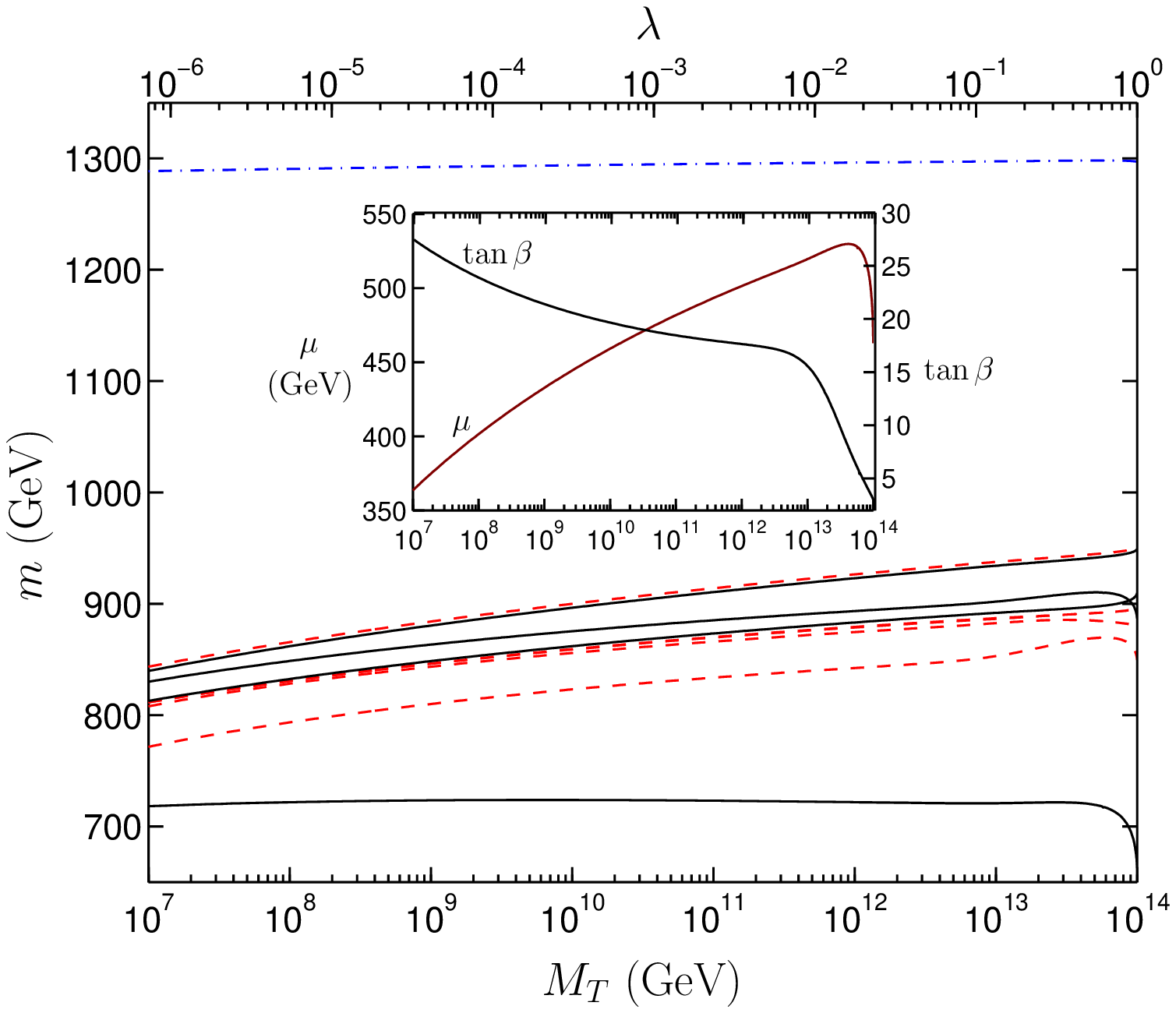}}\hspace*{\fill}
\parbox{0.48\linewidth}{
\includegraphics[width=\linewidth]{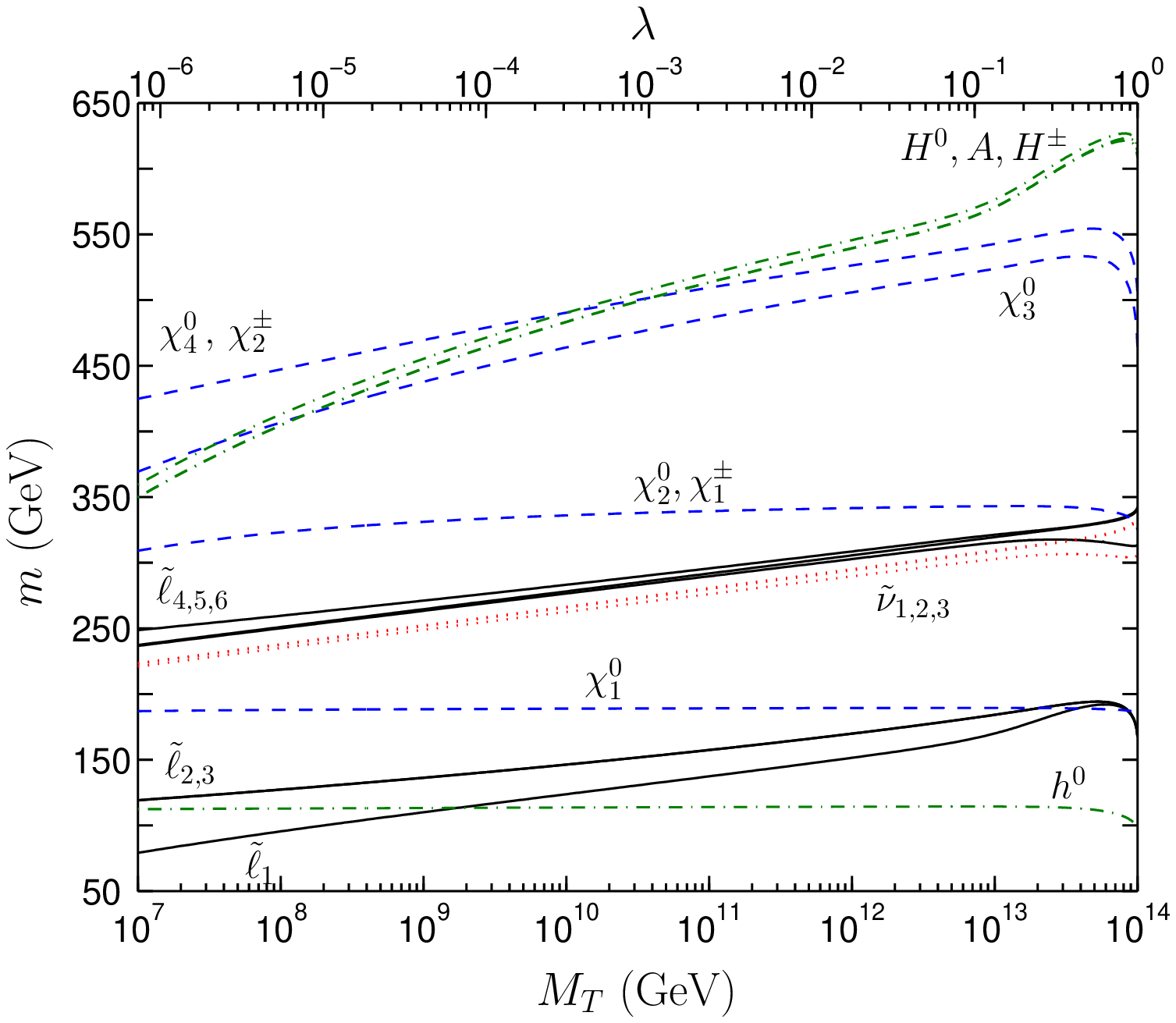}}
\caption{Sparticle and and Higgs spectrum for $B_T =20~{\rm TeV}$. Left panel: squark masses, $m_{\tilde{u}}$ (black solid line),
$m_{\tilde{d}}$(red dashed) and the gluino mass (blue dash-dotted). In the inner plot $\tan\beta$ and $\mu$ are shown as obtained
by the electroweak symmetry breaking conditions. Right panel: the masses of the charged sleptons, the sneutrinos, the charginos,
the neutralinos and the Higgs bosons as the labels indicate.}\label{fjar:f1}
\end{figure}
%
\begin{figure}
\includegraphics[width=0.5\linewidth]{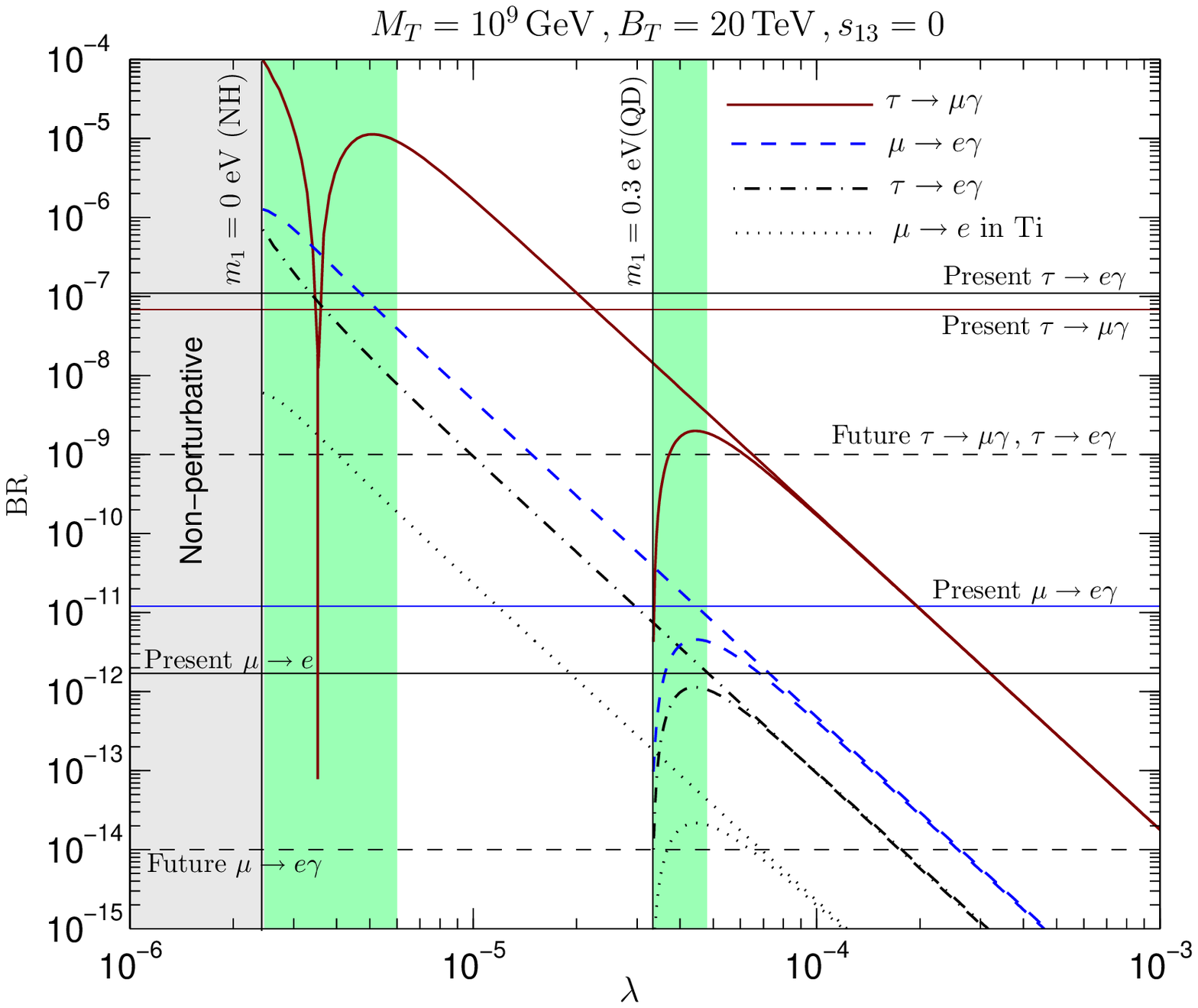}
\includegraphics[width=0.5\linewidth]{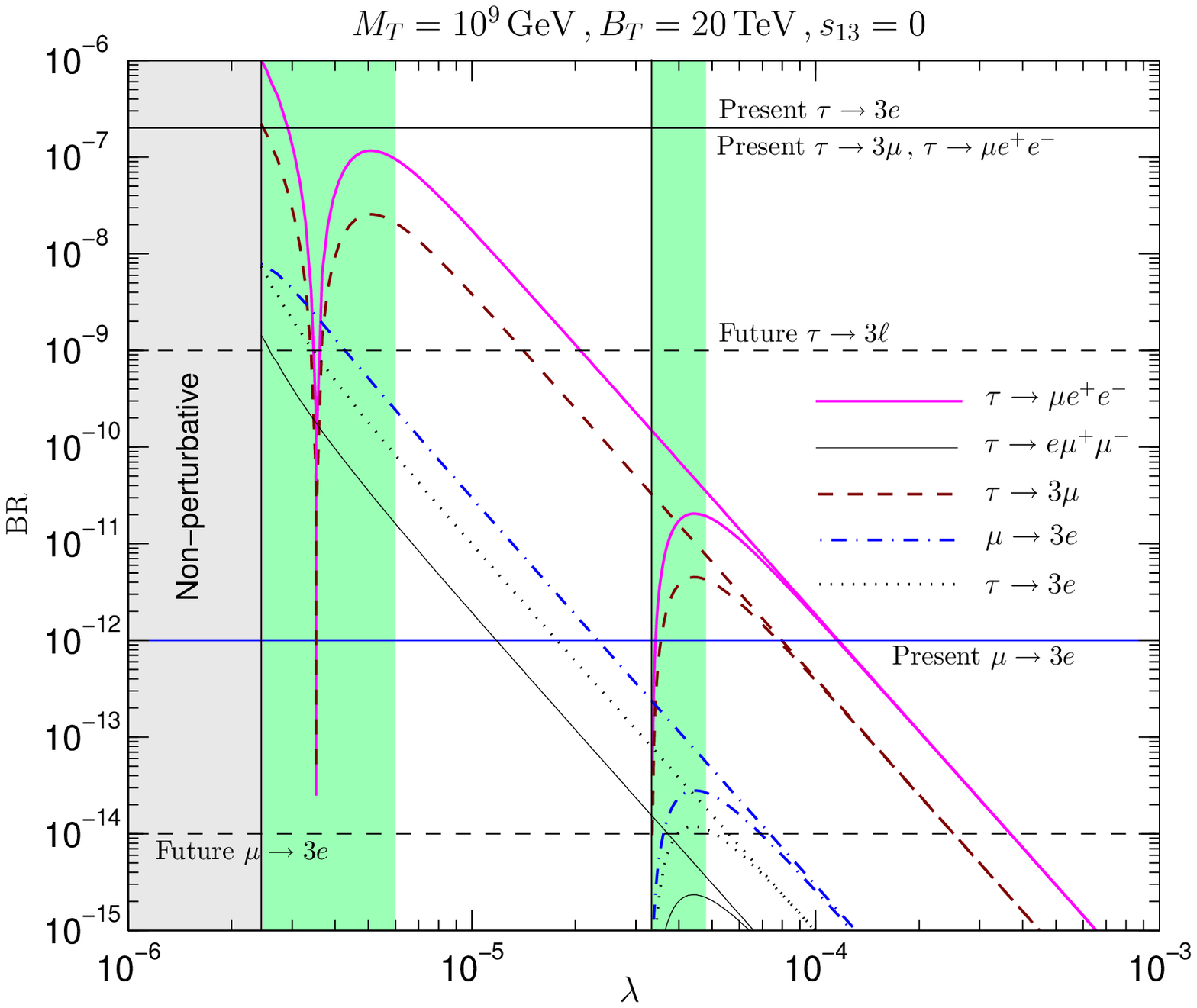}
\includegraphics[width=0.5\linewidth]{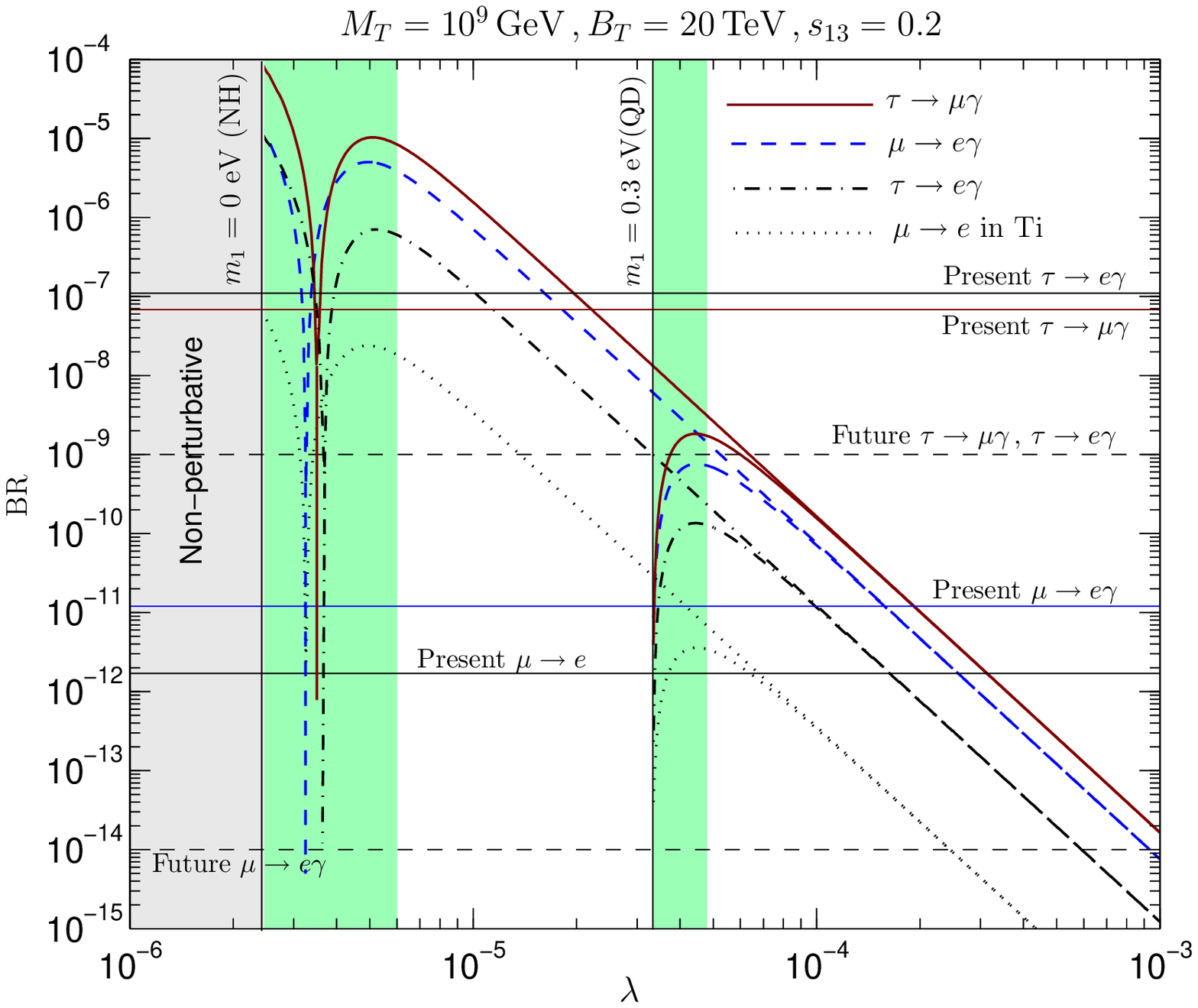}
\includegraphics[width=0.5\linewidth]{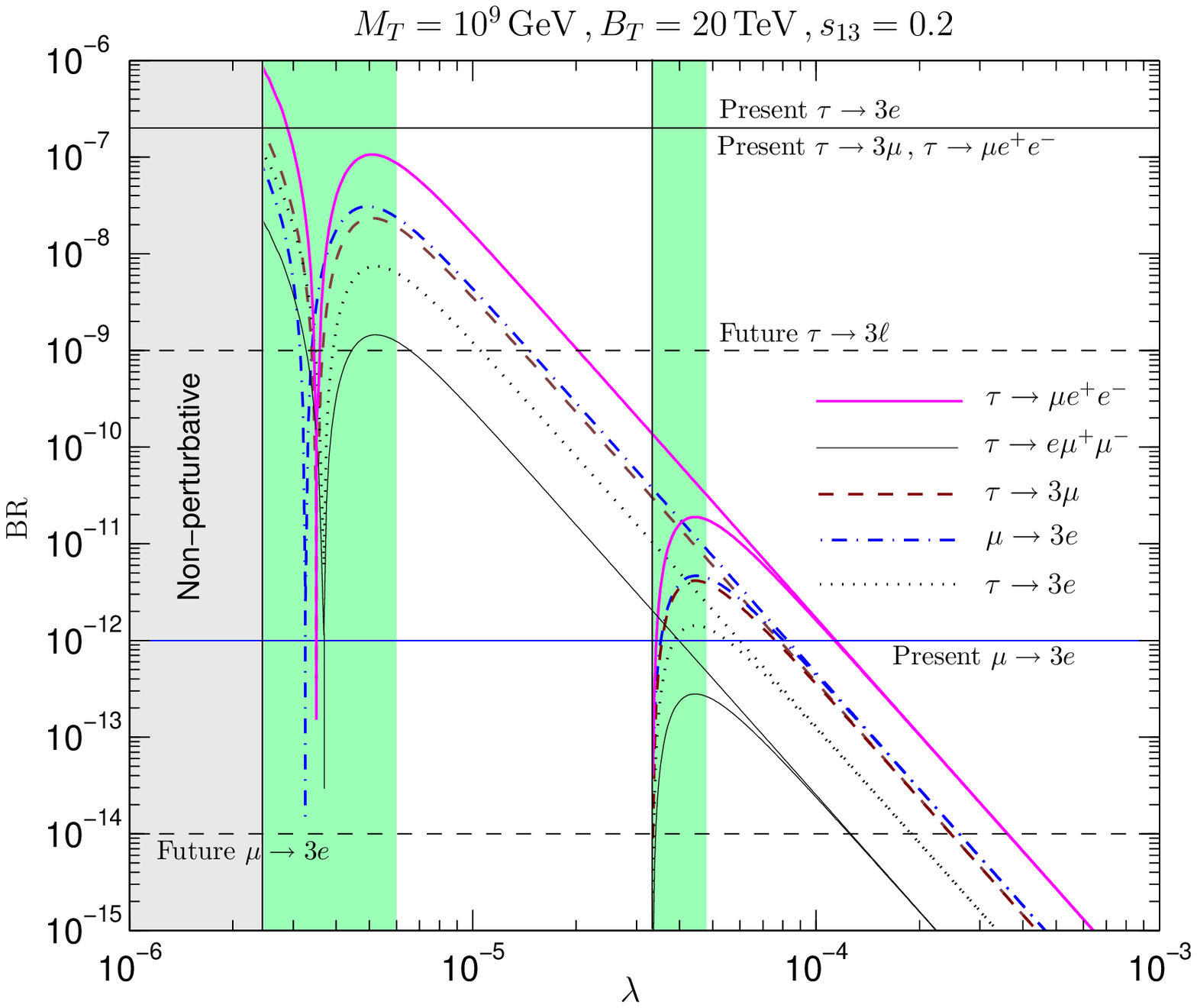}
\caption{Branching ratios of several LFV processes as a function of $\lambda$. The left (right) vertical line indicates the lower bound on
$\lambda$ imposed by requiring perturbativity of the Yukawa couplings $Y_{T,S,Z}$ when $m_1=0\,(0.3)$~eV [normal-hierarchical (quasi-degenerate) neutrino
mass spectrum]. The regions in green (grey) are excluded by the $m_{\tilde{\ell}_1} > 100$~GeV constraint (perturbativity requirement when $m_1=0$).} \label{fjar:f2}
\end{figure}

\begin{table}[bht]
\begin{minipage}{\textwidth}
\caption{Expectations for the various LFV processes assuming $B(\mu \to e\gamma) = 1.2\times 10^{-11}$. The results in
parenthesis apply to the case of the inverted-hierarchical neutrino spectrum, whenever these are different from those obtained for the
normal-hierarchical and quasi-degenerate ones.\label{fjarT}}
\begin{tabular*}{\textwidth}{@{\extracolsep{\fill}}l|ccc}
\hline\hline
					&\multicolumn{2}{c}{prediction for branching ratio}				\\		
decay mode 				&$s_{13}=0$ 			&$s_{13}=0.2$ 					\\
\hline
$\tau^- \to \mu^- \gamma$ 		&$3\times 10^{-9}$ 		&$2\,(3) \times 10^{-11}$ 	 		\\
$\tau^- \to e^- \gamma$  		&$2 \times10^{-12}$ 		&$1\,(3)\times10^{-12}$ 			\\
$\mu^-  \to e^- e^+ e^-$ 		&$6\times 10^{-14}$ 		&$6\times 10^{-14}$				\\
$\tau^- \to \mu^- \mu^+ \mu^-$		&$7\times 10^{-12}$ 		&$4\,(6)\times 10^{-14}$			\\
$\tau^- \to \mu^- e^+ e^-$		&$3\times 10^{-11}$ 		&$2\,(3)\times 10^{-13}$			\\
$\tau^- \to e^- e^+ e^-$ 		&$2\times 10^{-14}$		&$1\,(3)\times 10^{-14}$	 		\\
$\tau^- \to e^- \mu^+ \mu^-$		&$3\times 10^{-15}$ 		&$2\,(4)\times 10^{-15}$	 		\\
$\mu   \to e\,;\,{\rm Ti}$		&$6\times 10^{-14}$ 		&$6\times 10^{-14}$				\\
\hline\hline
\end{tabular*}
\end{minipage}
\end{table}

All the soft masses have the same scaling behaviour $\tilde{m}\sim B_T/(16\pi^2)$ which demands $B> 10~{\rm TeV}$ to fulfill the 
naturalness principle. This scenario appears very predictive  since it contains only three free
parameters: the triplet mass $M_T$, the effective SUSY breaking scale $B_T$ and the coupling constant $\lambda$. The parameter space
is then constrained by the experimental bounds on the Higgs boson mass, the $B(\mu \to e \gamma)$, the sfermion masses,
and the requirement of radiative electroweak symmetry breaking. The phenomenological predictions  more important and relevant for LHC,
the B-factories~\cite{Akeroyd:2004mj} the incoming MEG experiment~\cite{Grassi:2005ac}, the Super Flavour
factory~\cite{Giorgi:INFN} or the PRISM/PRIME experiment at J-PARC~\cite{Mori:Prism}, concern the sparticle and Higgs boson
spectra and the LFV decays. Regarding the spectrum, the gluino is the heaviest sparticle while, in most of the parameter space,
$\tilde{\ell}_1$ is the lightest. In the example shown in Fig.\ref{fjar:f1} the squark and slepton masses lie in the ranges 700--950~GeV
and 100 -- 300~GeV, respectively. The gluino mass is about 1.3 TeV. The
chargino masses are $m_{\tilde{\chi}^\pm_1} \sim 320$~GeV and $m_{\tilde{\chi}^\pm_2} \sim 450-550$~GeV. Moreover,
$m_{\tilde{\chi}^0_1 }\sim 190$~GeV, $m_{\tilde{\chi}^0_2} \approx m_{\tilde{\chi}^\pm_1}$ and $m_{\tilde{\chi}^0_{3,4}} \approx
m_{\tilde{\chi}^\pm_2}$. These mass ranges are within the discovery reach of the LHC.

The Higgs sector is characterized by a decoupling regime with a light SM-like Higgs boson ($h$) with mass in the range $110-120~{\rm
GeV}$ which is testable in the near future at LHC (mainly through the decay into 2 photons). The remaining three heavy states ($H,A$
and $H^\pm$) have mass $m_{H,A,H^\pm}\approx 450-550$~GeV (again, for $B_T=20$~TeV). All the spectra increase almost linearly with $B_T$.

Figure~\ref{fjar:f2}b shows instead several LFV processes: $\mu \to e X$, $\mu\to e$ conversion in nuclei, $\tau
\to e Y$ and $\tau \to  \mu Y$ $(X = \gamma, ee$, $Y= \gamma, e e, \mu \mu)$. One observe that \eg the behaviour of
the radiative-decay branching ratios is in agreement with the estimates given in Eq.~(\ref{ratios6}) for $\theta_{13}=0$. For $\theta_{13}=0.2$ one
obtains instead that $B(\tau \to \mu \gamma)/ B(\mu \to e \gamma) \sim 2$ and $B(\tau \to e \gamma)/
B(\mu \to e \gamma) \sim 0.1$ (the full analysis can be found in \Ref~\cite{Joaquim:2006mn}). The other LFV processes shown
are also correlated to the radiative ones in a model-independent way \cite{Joaquim:2006uz,Joaquim:2006mn}. The analysis shows that the
future experimental sensitivity will allow to measure at most $B(\mu \to\gamma)$, $B(\mu \to 3e)$, $B(\tau
\to\mu\gamma)$ and CR($\mu \to e\,$~Ti) for tiny $\theta_{13}$. In particular, being $B(\tau \to \mu\gamma)/
B(\mu\to e\gamma)\sim 300$, $B(\tau \to \mu\gamma)$ is expected not to exceed $3\times 10^{-9}$,
irrespective of the type of neutrino spectrum. Therefore $\tau \to \mu \gamma$  falls into the LHC capability. All the
decays $\tau \to \ell_i\ell_k\ell_k$ would have $B < \mathcal{O}(10^{-11})$.
The predictions for the LFV branching ratios in the present scenario are summarized in Table~\ref{fjarT}.

Finally, such supersymmetric $SU(5)$  framework with a 15,$\overline{15}$ pair may be realized in contexts based on string inspired
constructions~\cite{Cvetic:2002pj}--\cite{Cvetic:2006by}.

\subsubsection{LFV from a generic SO(10) framework}

The spinorial representation of the $SO(10)$, given by a 16-dimensional spinor, can accommodate all the
SM model particles as well as the right handed neutrino.  
As discussed in Section~\ref{sec:GUTs}, the product of two \textbf{16} matter representations can only 
couple to \textbf{10}, \textbf{120} or \textbf{126} representations, which can be
formed by either a single Higgs field or a non-renormalizable product of representations of several Higgs fields.
In either case, the  Yukawa matrices resulting from the couplings to \textbf{10} and \textbf{126} are complex-symmetric, whereas they are
antisymmetric when the couplings are to the \textbf{120}. Thus, the most general $SO(10)$ superpotential relevant to
fermion masses can be written as
\begin{equation}
W_{SO(10)} = Y^{10}_{ij} {\bf 16_i}~ {\bf 16_j}~ {\bf 10} + Y^{126}_{ij} {\bf 16_i}~ {\bf 16_j}~ {\bf 126} + Y^{120}_{ij} {\bf 16_i}~{\bf 16_j}~ {\bf 120},
\end{equation}
where $i,j$ refer to the generation indices. In terms of the SM fields, the Yukawa couplings relevant for fermion masses are given by 
\cite{Barbieri:1979ag,Langacker:1980js}:
\begin{eqnarray}
{\bf 16}\ {\bf 16}\ {\bf 10}\ &\supset&{\bf 5}\ ( u u^c + \nu \nu^c) + {\bf\bar 5}\ (d d^c + e e^c), \\
{\bf 16}\ {\bf 16}\ {\bf 126}\ &\supset&{\bf 1}\ \nu^c \nu^c + 15\ \nu \nu + {\bf 5}\ ( u u^c -3~ \nu \nu^c) + {\bf\bar{45}}\ (d d^c -3~ e e^c), \nonumber\\
{\bf 16}\ {\bf 16}\ {\bf 120}\ &\supset&{\bf 5}\ \nu \nu^c +{\bf 45}\ u u^c + {\bf\bar 5}\ ( d d^c + e e^c) +{\bf \bar{45}}\ (d d^c -3~ e e^c),\nonumber
\label{su5content}
\end{eqnarray}
where we have specified the corresponding $SU(5)$  Higgs representations for each  of the couplings and all the fermions are left handed fields.
The resulting up-type quarks and neutrinos' Dirac mass matrices can be written as:
\begin{eqnarray}
\label{upmats}
m^u &= & M^5_{10} + M^5_{126} + M^{45}_{120}, \\ \label{numats}
m^\nu_{D} &= & M^5_{10} - 3~ M^5_{126} + M^{5}_{120}.\label{downmats}
\end{eqnarray}

A simple analysis of the fermion mass matrices in the $SO(10)$ model, as detailed in the Eq.~(\ref{numats}) leads us to the following
result:  \textit{At least one of the Yukawa couplings in $Y^\nu~ =~ v_u^{-1}~m^\nu_{D}$ has to be as large as the top Yukawa coupling} \cite{Masiero:2002jn}. 
This result holds true in general, independently of the choice of the Higgses responsible for the masses in
Eqs.~(\ref{upmats}), (\ref{numats}), provided that no accidental fine-tuned cancellations of the different contributions in Eq.~(\ref{numats}) are
present. If contributions from the \textbf{10}'s solely dominate, $Y^\nu$ and $Y^u$ would be equal. If this occurs for the \textbf{126}'s, then $Y^\nu =- 3~ Y^u$ 
\cite{Mohapatra:1979nn}. In case both of them have dominant entries, barring a rather precisely 
fine-tuned cancellation between $M^5_{10}$ and $M^5_{126}$ in Eq.~(\ref{numats}), we expect at least one large entry to be present 
in $Y^\nu$. A dominant antisymmetric contribution to top quark mass due to the {\bf 120} Higgs is phenomenologically excluded, since it would
lead to at least a pair of heavy degenerate up quarks. Apart from sharing the property that at least one eigenvalue of both $m^u$
and $m^\nu_{D}$ has to be large, for the rest it is clear from Eqs.~(\ref{upmats}) and  (\ref{numats}) that these two matrices are not aligned
in general, and hence we may expect different mixing angles appearing from their diagonalization. This freedom is removed if one sticks to
particularly simple choices of the Higgses responsible for up quark and neutrino masses. 

Therefore, we see that the $SO(10)$ model with only two ten-plets would inevitably lead to small mixing in $Y^\nu$. In fact, with two Higgs fields
in symmetric representations, giving masses to the up-sector and the down-sector separately, it would be difficult to avoid the small CKM-like
mixing in $Y^\nu$. We will call this case the CKM case. From here, the following mass relations hold between the quark and leptonic mass matrices at the GUT 
scale\footnote{Clearly this relation cannot hold for the first two generations of down quarks and charged leptons.  One expects, small 
corrections due to  non-renormalizable operators or suppressed renormalizable operators \cite{Georgi:1979df} to be invoked.}: 
\begin{equation}
\label{massrelations} Y^u  = Y^\nu \;\;\;;\;\;\; Y^d  = Y^e . 
\end{equation}
In the basis where charged lepton masses are diagonal, we have
\begin{equation}
\label{hnumg} Y^\nu = V_{\rm CKM}^T~ Y^u_{Diag}~ V_{\rm CKM}. 
\end{equation} 
The large couplings in $Y^\nu \sim {\mathcal O}(Y_t)$ induce significant off-diagonal entries in $m_{\tilde L}^2$ through the RG
evolution between $M_{\rm GUT}$ and the scale of the right-handed Majorana neutrinos, $M_{R_i}$. The induced off-diagonal entries
relevant to $l_j \to l_i, \gamma$ are of the order of:  
\begin{equation}
\label{wcmi} (m_{\tilde L}^2)_{i\neq j}\approx -\frac{3 m_0^2+A_0^2 }{8 \pi^2}~ Y_t^2 V_{ti} V_{tj} \ln{\frac{M_{\rm GUT}}{M_{R_3}}} + \mathcal{O}(Y_c^2).
\end{equation}
where $V_{ij}$ are elements of $V_{\rm CKM}$, and $i$, $j$ flavour indices.
In this expression, the CKM angles are small but one would expect the presence of the large top Yukawa coupling to compensate such a 
suppression. The required right-handed neutrino Majorana  mass matrix, consistent with both the observed low energy neutrino masses 
and mixing as well as with CKM-like mixing in $Y^\nu$ is easily determined from the seesaw formula defined at the scale of right-handed neutrinos.

\begin{figure}[t]
\begin{center}
\includegraphics[width=0.58\textwidth]{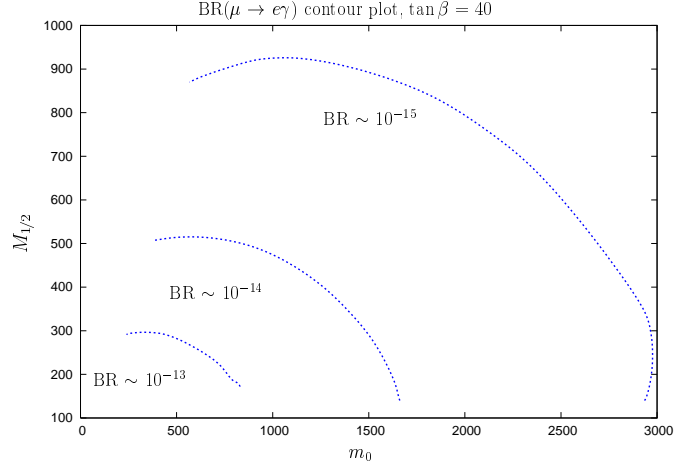}
\caption{Contour plot of $B(\mu \to e,\gamma)$ in the $(m_0, M_{1/2})$ plane, at $A_0=0$ in a CKM high $\tan\beta$
case. Note that while the plane is presently unconstrained, the planned MEG experiment sensitivity of $\mathcal{O}(10^{-13}-10^{-14})$
will be able to probe it in the $(m_0, m_{\tilde{g}}) \la 1$ TeV region.}\label{fig:cont_ckm_40}
\end{center}
\end{figure}

The $B(l_i \to l_j \gamma)$ are now predictable in this case. Considering mSUGRA boundary conditions and taking $\tan \beta = 40$, we obtain that 
reaching a sensitivity of ${\mathcal O }(10^{-13}-10^{-14})$, as planned by the MEG experiment at PSI, for $B(\mu \to e \gamma)$ would allow us to 
probe the SUSY spectrum  completely up to $m_0 = 1200\,\mathrm{GeV},\, M_{1/2} = 400\, \mathrm{GeV}$   
(notice that this corresponds to gluino and squark masses of order 1 TeV). This clearly appears from Fig.~\ref{fig:cont_ckm_40}, which shows the
$B(\mu \to e \gamma)$ contour plot in the $(m_0,M_{1/2})$ plane. Thus, in summary, though the present limits on $B(\mu \to e, \gamma)$ 
would not induce any significant constraints on the supersymmetry-breaking parameter space, an improvement in the limit to $\sim {\mathcal O }(10^{-13}-
10^{-14})$, as foreseen, would start imposing non-trivial constraints especially for the large $\tan \beta$ region. 

To obtain mixing angles larger than CKM angles, asymmetric mass matrices have to be considered. In general, it is sufficient to introduce
asymmetric textures either in the up-sector or in the down-sector. In the present case, we assume that the down-sector couples to a combination of 
Higgs representations (symmetric and antisymmetric)\footnote{The couplings of the Higgs fields in the superpotential can be either renormalizable or 
non-renormalizable. See \cite{Chang:2002mq} for a non-renormalizable example.} $\Phi$, leading to an asymmetric mass matrix in the basis where the 
up-sector is diagonal. As we will see below, this would also require that the right-handed Majorana mass matrix be diagonal in this basis. We have :
\begin{equation}
\label{mnsso10}
W_{SO(10)} = \frac{1 }{2}~  Y^{u,\nu}_{ii}~ {\bf 16_i} ~{\bf 16_i} {\bf 10^u} + \frac{1 }{2} 
~ Y^{d,e}_{ij}~ {\bf 16_i} ~{\bf 16_j} \Phi \nonumber \\
 +  \frac{1 }{2}~ Y^R_{ii}~ {\bf 16_i}~ {\bf 16_i} {\bf 126}~,
\end{equation}
where the \textbf{126}, as before, generates only the right-handed neutrino mass matrix. To study the consequences
of these assumptions, we see that at the level of $SU(5)$, we have
\begin{equation}
W_{SU(5)} =  \frac{1 }{2}~ Y^u_{ii}~ {\bf 10_i} ~{\bf 10_i} ~{\bf 5_u} + Y^\nu_{ii} ~{\bf \bar{5}_i}~ {\bf 1_i}~ {\bf 5_u} \nonumber \\
+  Y^d_{ij}~ {\bf 10_i} ~{\bf \bar{5}_j}~ {\bf \bar{5}_d} +   \frac{1 }{2}~M^R_{ii}~ {\bf 1_i} {\bf 1_i},
\end{equation}
where we have decomposed the ${\bf 16}$ into ${\bf 10} + {\bf \bar{5}} +{\bf  1}$ and ${\bf 5_u}$ and
${\bf \bar{5}_d}$ are components of ${\bf 10_u}$ and $\Phi$ respectively. To have large mixing $\sim~ U_{\rm PMNS}$  in $Y^\nu$ we see that the asymmetric matrix $Y^d$
should now give rise to both the CKM mixing as well as PMNS mixing. This is possible if
\begin{equation}
V_{\rm CKM}^T~ Y^d~ U_{\rm PMNS}^T = Y^d_{Diag}.
\end{equation}

\begin{figure}[t]
\includegraphics[width=0.48\textwidth]{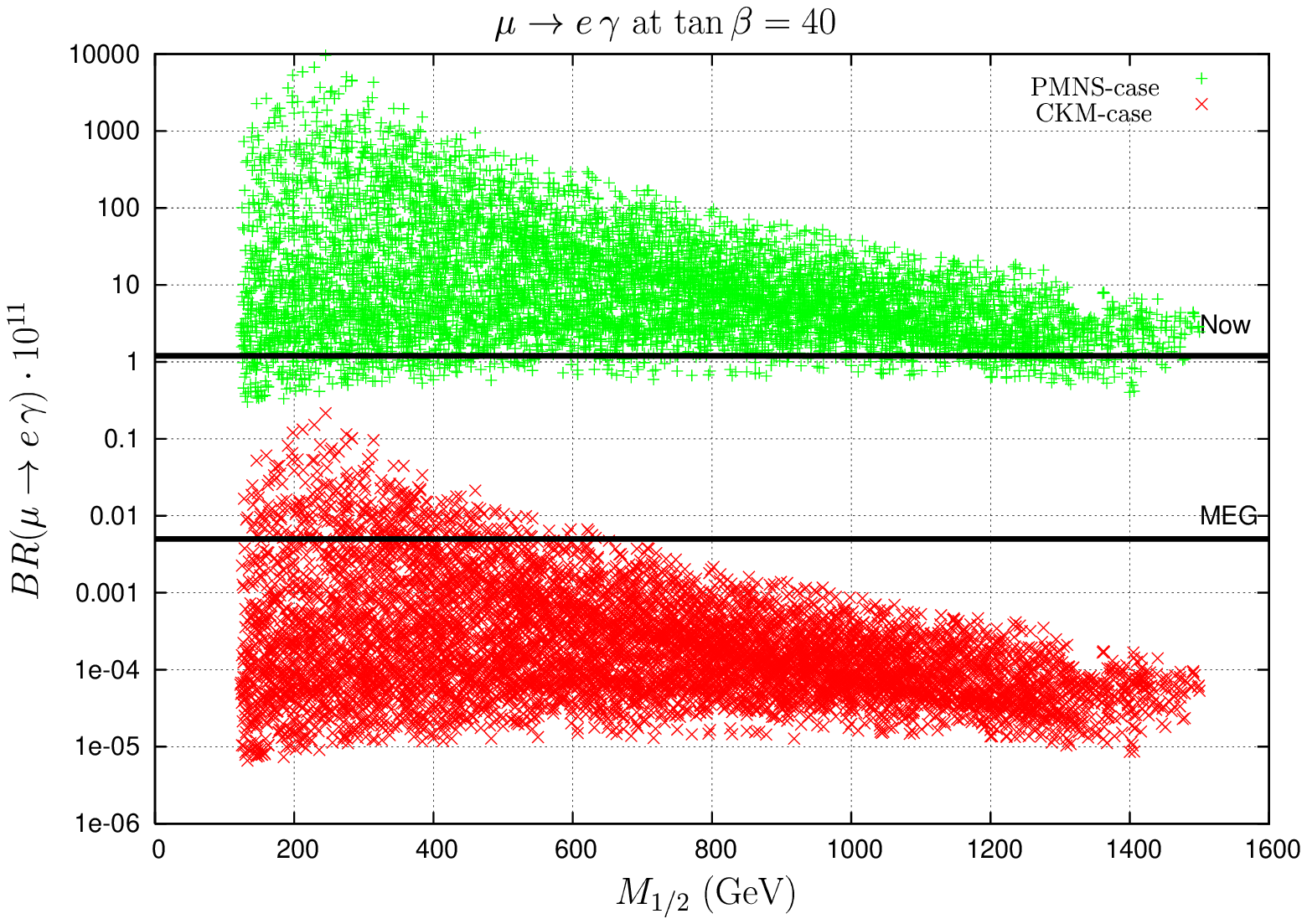}\hspace*{\fill}
\includegraphics[width=0.48\textwidth]{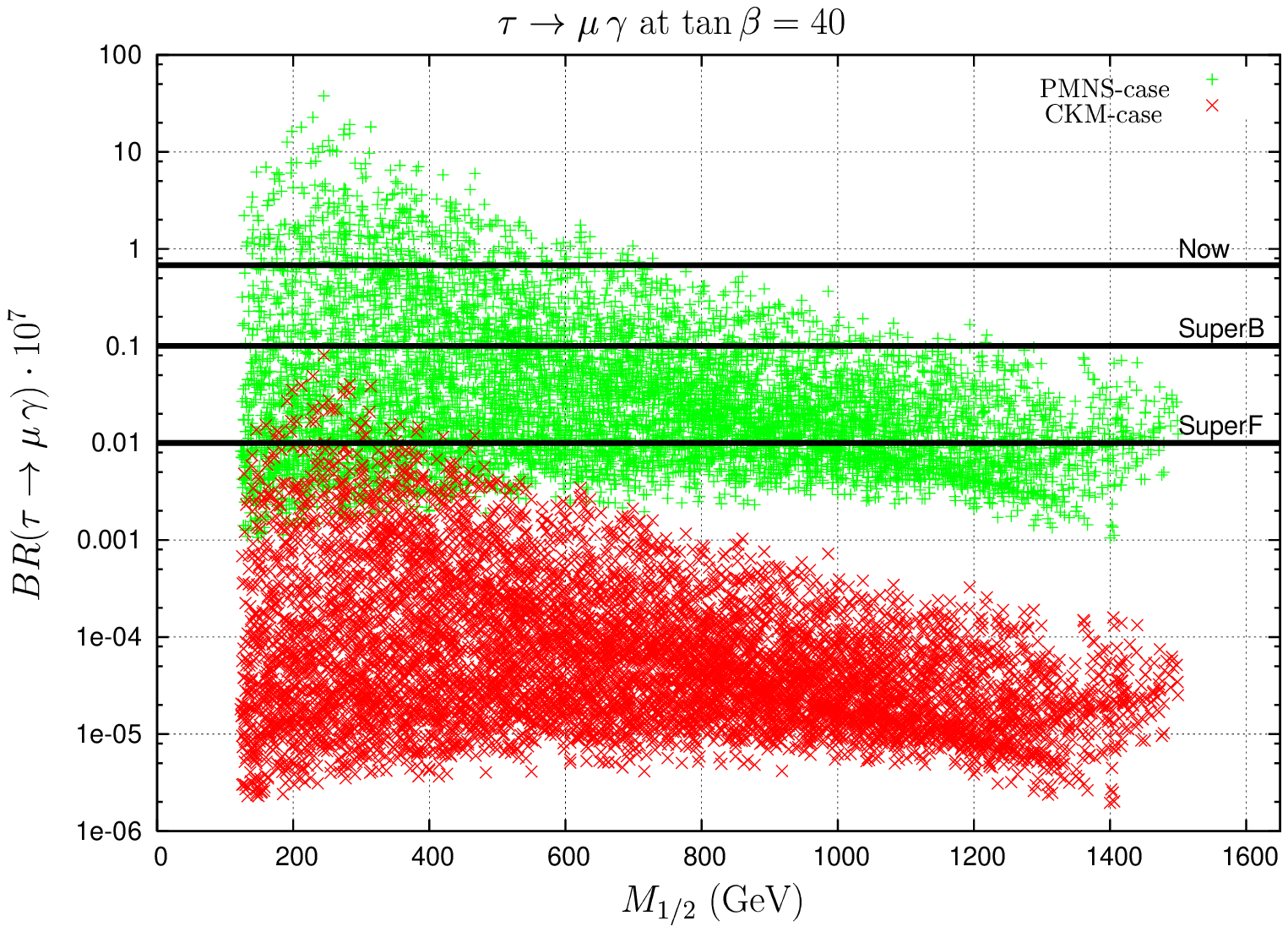}
\caption{Scatter plots of $B(\mu \to e, \gamma)$ (left) and  $B(\tau \to \mu \gamma)$ (right) versus $M_{1/2}$ for $\tan\beta = 40$, both
for the (maximal) PMNS case with $|U_{e3}|=0.07$ and the (minimal) CKM case. The plots were obtained by scanning the SUSY parameter space in the LHC accessible
region (see the text, for details).\label{fig:scatter_40}}
\end{figure}

Therefore the ${\bf 10}$ that contains the left-handed down-quarks would be rotated by the CKM matrix whereas the ${\bf \bar{5}}$ that contains the 
left-handed charged leptons would be rotated by the $U_{\rm PMNS}$ matrix to go into their respective mass bases \cite{Moroi:2000mr,Moroi:2000tk,
Akama:2001em,Chang:2002mq}. Thus we have the following relations in the basis where charged leptons and down quarks are diagonal:
\begin{eqnarray}
Y^u &=& V_{\rm CKM}~ Y^u_{Diag}~ V_{\rm CKM}^T~ ,  \\ \label{hnumns} Y^\nu &=& U_{\rm PMNS}~ Y^u_{Diag}.
\end{eqnarray}
Using the seesaw formula of Eqs.~(\ref{eq:seesaw}) and (\ref{hnumns}), we have
\begin{equation}
M_{R} = Diag\{ \frac{m_u^2 }{m_{\nu_1}},~\frac{m_c^2 }{ m_{\nu_2}},~\frac{m_t^2 }{ m_{\nu_3}} \}.
\end{equation}
We now turn our attention to  lepton flavour violation in this case. The branching ratio, $B(\mu \to e, \gamma)$ would now depend on
\begin{equation}
\label{hnusqmns} [Y^\nu Y^{\nu~T}]_{21} = Y_t^2~ U_{\mu 3}~ U_{e 3} + Y_c^2~ U_{\mu 2}~ U_{e 2} + \mathcal{O}(h_u^2).
\end{equation}
It is clear from the above that in contrast to the CKM case, the dominant contribution to the off-diagonal entries depends on the
unknown magnitude of the element $U_{e3}$ \cite{Sato:2000zh}. If $U_{e3}$ is close to its present limit $\sim~0.14$ \cite{Apollonio:1999ae}
(or at least larger than $(Y_c^2/Y_t^2)~U_{e2}~ \sim 4 \times 10^{-5}$), the first term on the RHS of the Eq.~(\ref{hnusqmns})
would dominate. Moreover, this would lead to large contributions to the off-diagonal entries in the slepton masses with $U_{\mu 3}$ of
${\mathcal O}(1)$. Thus, we have 
\begin{equation}
\label{bcmi1} (m_{\tilde L}^2)_{21} \approx -\frac{3 m_0^2+A_0^2}{8 \pi^2}~ Y_t^2 U_{e 3} U_{\mu 3} \ln\frac{M_{\rm GUT}}{M_{R_3}} + \mathcal{O}(Y_c^2).
\end{equation}
This contribution is larger than the CKM case by a factor of $(U_{\mu 3} U_{e3})/ (V_{td} V_{ts}) \sim \mathcal{O}(10^2)$.
This would mean about a factor $10^4$ times larger than the CKM case in B($\mu \to e, \gamma)$. Such enhancement with respect to the CKM case 
is clearly shown in the scatter plots of Fig.~\ref{fig:scatter_40}, where the CKM case is compared with the PMNS case with $U_{e3}=0.07$. The aim of the 
figure is to show the capability of MEG to probe the region of mSUGRA parameter space accessible to the LHC. In fact,
the plots show the value of B($\mu \to e, \gamma)$ obtained by scanning the parameter space in the large region ($0 < m_0 < 5$ TeV,
$0 < M_{1/2} < 1.5$ TeV, $-3 m_0 < A_0 < +3 m_0$, sign($\mu$)), and then keeping the points which give at least one squark lighter than 2.5 TeV 
(so roughly accessible to the LHC). 
We see that in this ``LHC accessible'' region the maximal case (with $U_{e3}=0.07$) is already excluded by the MEGA limit 
($B(\mu \to e, \gamma) < 1.2 \cdot 10^{-11}$), and therefore MEG will constrain the parameter space far beyond the LHC sensitivity for this case.
If $U_{e3}$ is very small, \textit{i.e} either zero or $\la~ (Y_c^2/Y_t^2)~U_{e2}~ \sim 4 \times 10^{-5}$, the second term $\propto ~Y_c^2$   
in Eq.~(\ref{hnusqmns}) would dominate, thus giving a strong suppression to the branching ratio. This could be not true, once RG effects on $U_{e3}$ itself 
\cite{Chankowski:2001mx,Antusch:2005gp} are taken into account. The point is that the PMNS boundary condition 
(\ref{hnumns}) is valid at high scale. Thus, it is necessary to evolve the neutrino masses and mixing
 from the low-energy scale, where measurements
are performed, up to high energy. Such effect turns out to be not negligible in case of low-energy $U_{e3}\la 10^{-3}$, giving a high-energy constant 
enhancement of order $\mathcal{O}(10^{-3})$ \cite{Calibbi:2006ne}. The consequence is that the term in Eq.~(\ref{hnusqmns}) $\propto ~Y_t^2$
always dominates, giving a contribution to the branching ratio larger than the CKM case (which turns out to be really a ``minimal'' case) 
and bringing the most of the parameter space in the realm of MEG even for very small low-energy values of $U_{e3}$ \cite{Calibbi:2006ne,Calibbi:2006nq}.

The $\tau \to \mu$ transitions are instead $U_{e3}$-independent probes of SUSY, whose importance was first pointed out in Ref.~\cite{Blazek:2001zm}. The 
off-diagonal entry in this case is given by :
\begin{equation}
\label{bcmi2} (m_{\tilde L}^2)_{32} \approx  -\frac{3 m_0^2+A_0^2}{8 \pi^2}~ Y_t^2 U_{\mu 3} U_{\tau 3} \ln\frac{M_{\rm GUT}}{M_{R_3}} + \mathcal{O}(Y_c^2).
\end{equation}
In the $\tau \to \mu \gamma$ decay the situation is at the moment similarly constrained with respect to $\mu \to e \gamma$, if $U_{e3}$
happens to be very small. The main difference is that B($\tau \to \mu \gamma$) does not depend on the value of $U_{e3}$,
so that $\tau \to \mu \gamma$ will be a promising complementary channel with respect to $\mu \to e \gamma$. 
As far as Beauty factories \cite{Abe:2003sx, Aubert:2003pc, Aubert:2005ye} are concerned, 
we see from Fig.~\ref{fig:scatter_40}, that even with the present bound
it is possible to rule out part of LHC accessible region in the PMNS high $\tan\beta$ regime; the planned accuracy of the SuperKEKB \cite{Akeroyd:2004mj} machine 
$\sim \mathcal{O}(10^{-8})$  will allow to test much of high $\tan\beta$ region  and will start probing the low $\tan\beta$
PMNS case, with a sensitivity to soft masses as high as $(m_0, m_{\tilde{g}}) \la 900$ GeV. The situation changes dramatically if one takes into account the 
possibility of a Super Flavour factory: taking the sensitivity of the most promising $\tau \to \mu \, \gamma$ process to $\sim\mathcal{O}(10^{-9})$, the
PMNS case will be nearly ruled out in the high $\tan\beta$ regime and severely constrained in the low $\tan\beta$ one; as for
the CKM case we would enter the region of interest. 

Let's finish with some remarks. Suppose that the LHC does find signals of low-energy supersymmetry, then
grand unification becomes a very appealing scenario, because of the successful unification of gauge couplings driven by the SUSY partners.
Among SUSY-GUT models, an $SO(10)$ framework is much favored as it is the `minimal' GUT to host all the fermions in a single representation and
it accounts for the smallness of the observed neutrino masses by naturally including the see-saw mechanism. In the above we
have addressed the issue by a generic benchmark analysis, within the ansatz that there is no fine-tuning in the neutrino Yukawa sector.
We can state that LFV experiments should be able to tell us much about the structure of such a SUSY-GUT scenario. If they detect
LFV processes, by their rate and exploiting the interplay between different experiments, we would be able to get hints of the structure of the
unknown neutrinos' Yukawa's. On the contrary, in the case that both MEG and a future Super Flavour factory happen not to see any LFV process, only two possibilities
should be left: (i)  the minimal mixing, low $\tan\beta$ scenario; (ii) mSUGRA $SO(10)$ see-saw without fine-tuned $Y_\nu$ couplings is not a viable
framework of physics beyond the standard model. 

Actually one should remark that LFV experiments will be able to falsify some of above scenarios even in regions
of the mSUGRA parameter space that are beyond the reach of LHC experiments. In this sense, the power of LFV experiments of testing/discriminating among
different SUSY-GUTs models results very interesting and highly complementary to the direct searches at the LHC.

\subsubsection{LFV, QFV and CPV observables in GUTs and their correlations}

In a SUSY Grand Unified Theory (GUT), quarks and leptons sit in same multiplets and are transformed ones into the others through GUT 
symmetry transformations. If the energy scale where the SUSY breaking terms are transmitted to the visible sector is larger then the GUT scale, as
in the case of gravity mediation, such breaking terms, and in particular the sfermion mass matrices, will have to respect the underlying GUT symmetry.
Hence, as already discussed in Section~\ref{sec:hisano}, the quark-lepton unification seeps also into the SUSY breaking soft sector. If the soft SUSY 
breaking terms respect boundary conditions which are subject to the GUT symmetry to start with, we generally expect the presence of relations among the
(bilinear and trilinear) scalar terms in the hadronic and leptonic sectors \cite{Ciuchini:2003rg,Ciuchini:2007ha}. Such relations hold 
true at the (superlarge) energy scale where the correct symmetry of the theory is the GUT symmetry. After its breaking, the mentioned relations 
will undergo corrections which are computable through the appropriate RGE's which are related to the specific structure of the theory between
 the GUT and the electroweak scale (for instance, new Yukawa couplings due to the presence of RH neutrinos acting down to the RH 
neutrino mass scale, presence of a symmetry breaking chain with the appearance of new symmetries at intermediate scales, etc.). As a result 
of such a computable running, we can infer the correlations between the softly SUSY breaking  hadronic and leptonic MIs at the low scale 
where FCNC tests are performed. Moreover, given that a common SUSY soft-breaking scalar term of $\mathcal{L}_{soft}$ at 
scales close to $M_{\rm Planck}$ can give rise to RG-induced  $\delta^q$'s and $\delta^l$'s at the weak scale, one may envisage the possibility to make use 
of the FCNC constraints on such low-energy $\delta$'s to infer bounds on the soft breaking parameters of the original supergravity  Lagrangian 
($\mathcal{L}_{sugra}$). Indeed, for each scalar soft parameter of $\mathcal{L}_{sugra}$ one can ascertain whether the hadronic or the 
leptonic corresponding bound at the weak scale yields the strongest constraint at the large scale \cite{Ciuchini:2007ha}.   

Let us consider the scalar soft breaking sector of the MSSM: 
\begin{eqnarray} 
\label{smsoft}
- {\cal L}_{soft} &=& m_{Q_{ii}}^2 \tilde{Q}_i^\dagger \tilde{Q}_i + m_{u^c_{ii}}^2 \tilde{u^c}_i^\star \tilde{u^c}_i + m^2_{e^c_{ii}} 
\tilde{e^c}_i^\star \tilde{e^c}_i  +   m^2_{d^c_{ii}} \tilde{d^c}^\star_i \tilde{d^c}_i + m_{L_{ii}}^2 \tilde{L}_i^\dagger \tilde{L}_i  \nonumber \\
 &+& m^2_{H_1} H^\dagger_1 H_1 +  m^2_{H_2} H_2^\dagger H_2 + A^u_{ij}~\tilde{Q}_i \tilde{u^c}_j H_2 + A^d_{ij}~ \tilde{Q}_i \tilde{d^c}_j H_1 + A^e_{ij}~
\tilde{L}_i \tilde{e^c}_j H_1 \nonumber \\
 &+&  (\Delta^l_{ij})_{LL} \tilde{L}_i^\dagger \tilde{L}_j  + (\Delta^e_{ij})_{RR} \tilde{e^c}_i^\star \tilde{e^c}_j  
+ (\Delta^q_{ij})_{LL} \tilde{Q}_i^\dagger \tilde{Q}_j  + (\Delta^u_{ij})_{RR} \tilde{u^c}_i^\star \tilde{u^c}_j \nonumber \\
 &+& 
(\Delta^d_{ij})_{RR} \tilde{d^c}_i^\star \tilde{d^c}_j  + (\Delta^e_{ij})_{LR} \tilde{e_L}_i^\star \tilde{e^c}_j  
+ (\Delta^u_{ij})_{LR} \tilde{u_L}_i^\star \tilde{u^c}_j + (\Delta^d_{ij})_{LR} \tilde{d_L}_i^\star \tilde{d^c}_j + \ldots 
\end{eqnarray}
where we have explicitly written down the various off-diagonal entries of the soft SUSY breaking matrices.
Consider now that $SU(5)$ is the relevant symmetry at the scale where the above soft terms firstly  show up.  Then, taking into account that matter 
is organized into the $SU(5)$ representations ${\bf 10}~ =~(q,u^c,e^c)$ and ${\bf\overline 5}~ = ~(l,d^c)$, one obtains the following relations
\begin{eqnarray}
m^2_{Q} = m^2_{\tilde{e^c}} = m^2_{\tilde{u^c}} = m^2_{\bf 10} ,\nonumber\\
m^2_{\tilde{d^c}} = m^2_{L} = m^2_{\bar{\bf \overline 5}} ,\nonumber\\
A^e_{ij} = A^d_{ji}\, .
\end{eqnarray}
These equations for matrices in flavour space lead to relations between the slepton and squark 
flavour violating off-diagonal entries $\Delta_{ij}$. These are: 
\begin{eqnarray}\label{cdeltas1}
(\Delta^u_{ij})_{LL} = (\Delta^u_{ij})_{RR} = (\Delta^d_{ij})_{LL} = (\Delta^l_{ij})_{RR}, \\\label{cdeltas3}
(\Delta^d_{ij})_{RR} = (\Delta^l_{ij})_{LL} ,\\\label{cdeltas4}
(\Delta^d_{ij})_{LR} = (\Delta^l_{ji})_{LR} = (\Delta^l_{ij})_{RL}^\star .
\end{eqnarray}
These GUT correlations among hadronic and leptonic scalar soft terms are summarized in the second column of Table~\ref{tbsu5}. Assuming that
no new sources of flavour structure are present from the $SU(5)$ scale down to the electroweak scale, apart from the usual SM CKM one, one
infers the relations in the first column of Table~\ref{tbsu5} at low scale. Here we have taken into account that due to their different gauge
couplings ``average'' (diagonal) squark and slepton masses acquire different values at the electroweak scale.

\begin{table}[bht]
\begin{minipage}{\textwidth}
\caption{Links between various transitions between up-type, down-type quarks and charged leptons for $SU(5)$. $m_{\tilde f}^2$ refers to the average
mass for the sfermion $f$, $m_{\tilde Q_{\rm avg}}^2= \sqrt{m_{\tilde Q}^2 m_{\tilde d^c}^2}$ and $m_{\tilde L_{\rm avg}}^2= 
\sqrt{m_{\tilde L}^2 m_{\tilde e^c}^2}$.\label{tbsu5}}
\begin{tabular*}{\textwidth}{@{\extracolsep{\fill}}ll}
\hline\hline
Relations at weak-scale 									&Boundary conditions at $M_{GUT}$ 			\\
\hline
$(\delta^u_{ij})_{RR} \approx (m_{\tilde e^c}^2/ m_{\tilde u^c}^2) (\delta^l_{ij})_{RR}$ 	&$m^2_{\tilde u^c}(0) = m^2_{\tilde e^c}(0)$ 		\\
$(\delta^q_{ij})_{LL} \approx (m_{\tilde e^c}^2/ m_{\tilde Q}^2) (\delta^l_{ij})_{RR}$ 		&$m^2_{\tilde Q}(0) = m^2_{\tilde e^c}(0)$ 		\\
$(\delta^d_{ij})_{RR} \approx (m_{\tilde L}^2/ m_{\tilde d^c}^2) (\delta^l_{ij})_{LL}$ 		&$m^2_{\tilde d^c}(0) = m^2_{\tilde L}(0)$ 		\\
$(\delta^d_{ij})_{LR} \approx (m_{\tilde L_{\rm avg}}^2 / m_{\tilde Q_{\rm avg}}^2) (m_b/ m_\tau) (\delta^l_{ij})_{LR}^\star$  & $A^e_{ij} = A^d_{ji}$	\\
\hline\hline
\end{tabular*}
\end{minipage}
\end{table}

Two comments are in order when looking at Table~\ref{tbsu5}. First, the boundary conditions on the sfermion masses at the GUT scale (last column in 
Table~\ref{tbsu5}) imply that the squark masses are \textit{always} going to be larger at the weak scale compared to the slepton masses due to the
participation of the QCD coupling in the RGEs. As a second remark, notice that the relations between hadronic and leptonic
$\delta$ MI in Table~\ref{tbsu5} always exhibit opposite ``chiralities", i.e. LL insertions are related to RR ones and vice-versa.  This stems from 
the arrangement of the different fermion chiralities in  $SU(5)$ five- and ten-plets (as it clearly appears from the final column in 
Table~\ref{tbsu5}). This restriction can easily be overcome if we move from $SU(5)$ to left-right symmetric unified models 
like  $SO(10)$ or  the  Pati-Salam (PS) case (we exhibit the corresponding  GUT boundary conditions and $\delta$ MI at the electroweak scale in Table~\ref{tbso10}). 

\begin{table}[bht]
\begin{minipage}{\textwidth}
\caption{Links between various transitions between up-type, down-type quarks and charged leptons for PS/$SO(10)$ type models.\label{tbso10}}
\begin{tabular*}{\textwidth}{@{\extracolsep{\fill}}ll}
\hline\hline
Relations at weak-scale & Boundary conditions at $M_{GUT}$ \\
\hline
$(\delta^u_{ij})_{RR} \approx (m_{\tilde e^c}^2/ m_{\tilde u^c}^2) (\delta^l_{ij})_{RR}$ 	& $m^2_{\tilde u^c}(0) = m^2_{\tilde e^c}(0)$ 	\\
$(\delta^q_{ij})_{LL} \approx (m_{\tilde L}^2/ m_{\tilde Q}^2) (\delta^l_{ij})_{LL}$ 		& $m^2_{\tilde Q}(0) = m^2_{\tilde L}(0)$ 	\\ 
\hline\hline
\end{tabular*}
\end{minipage}
\end{table}

So far we have confined the discussion within the simple $SU(5)$ model, without the presence of any extra particles like right handed (RH) neutrinos. In the
presence of RH neutrinos, one can envisage of two scenarios \cite{Masiero:2002jn}: (a) with either very small neutrino Dirac Yukawa couplings and/or very small
mixing present in the neutrino Dirac Yukawa matrix, (b) Large Yukawa and large
mixing in the neutrino sector. In the latter case, Eqs.~(\ref{cdeltas1} -- \ref{cdeltas4}) are not valid at all scales in general, as large RGE effects 
can significantly modify the sleptonic flavour structure while keeping the squark sector essentially unmodified; thus essentially breaking the GUT 
symmetric relations. In the former case where the neutrino Dirac Yukawa couplings are tiny 
and do not significantly modify the sleptonic flavour structure, the GUT symmetric relations are expected to be valid at the weak scale. 
However, in both cases it is possible to say that there exists a bound on the hadronic $\delta$ parameters of the form \cite{Ciuchini:2003rg}:
\begin{equation}
|(\delta^d_{ij})_{RR}| ~~~\geq~~~ \frac{m_{\tilde{L}}^2}{m_{\tilde{d}^c}^2} |(\delta^l_{ij})_{LL}|. 
\end{equation}
The situation is different if we try to translate the bound from quark to lepton MIs. An hadronic MI bound at low energy leads, after RGE
evolution, to a bound on the corresponding grand-unified MI at $M_{\rm GUT}$, applying both to slepton and squark mass matrices.
However, if the neutrino Yukawa couplings have sizable off-diagonal entries, the RGE running from $M_{\rm GUT}$ to $M_W$ could still generate a new
contribution to the slepton MI that exceeds this GUT bound. Therefore hadronic bounds cannot be translated to leptons unless we make some
additional assumptions on the neutrino Yukawa matrices. On general grounds, given that SM contributions in the lepton sector
are absent and that the branching ratios of leptonic processes constrain only the modulus of the MIs, it turns out that all the MI
bounds arising from the lepton sector are circles in the $\rm{Re}(\delta^d_{ij})_{AB}$--$\rm{Im}(\delta^d_{ij})_{AB}$
plane and are centered at the origin.
\begin{figure}
\includegraphics[width=0.49\linewidth]{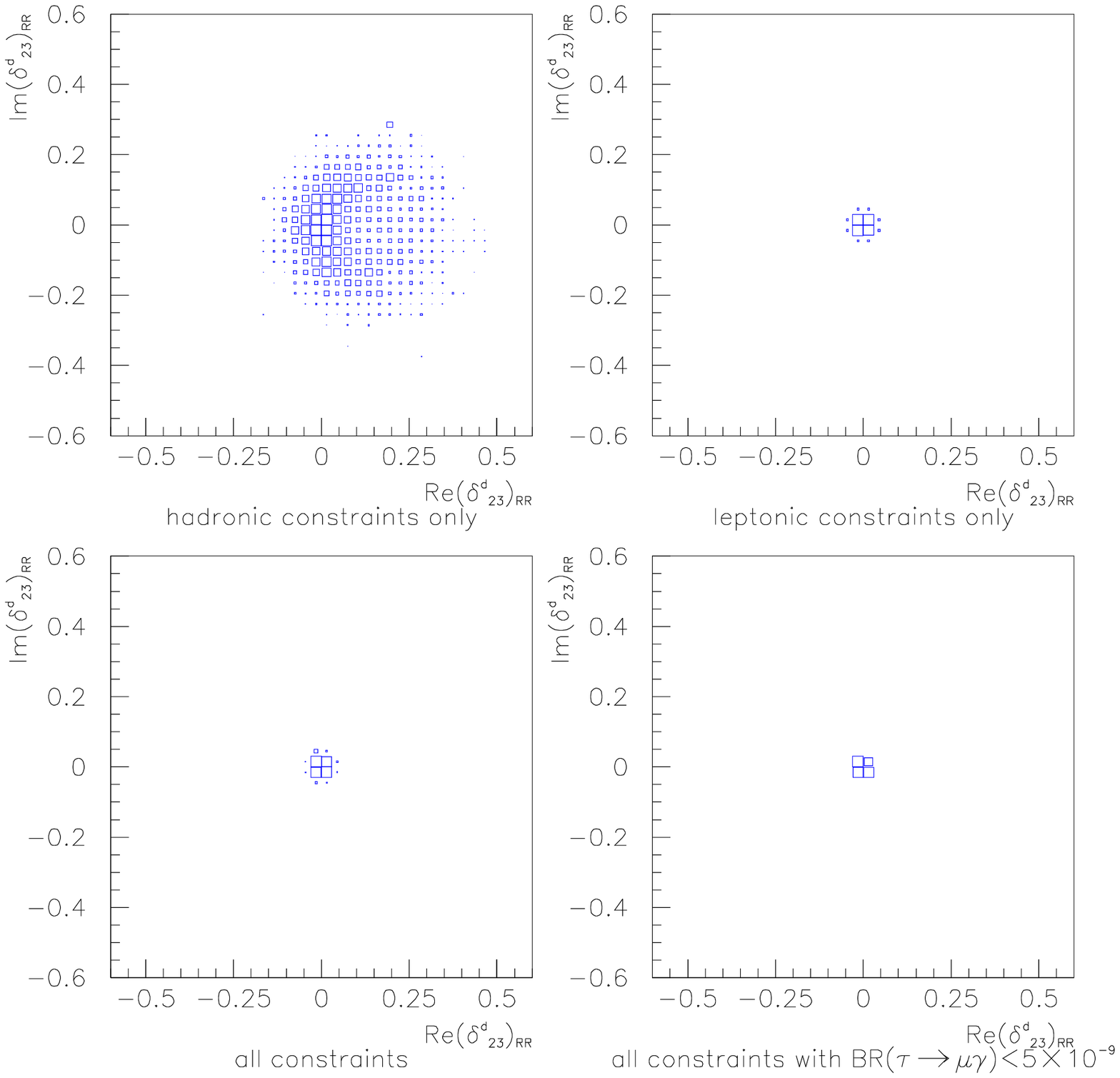}
\includegraphics[width=0.49\linewidth]{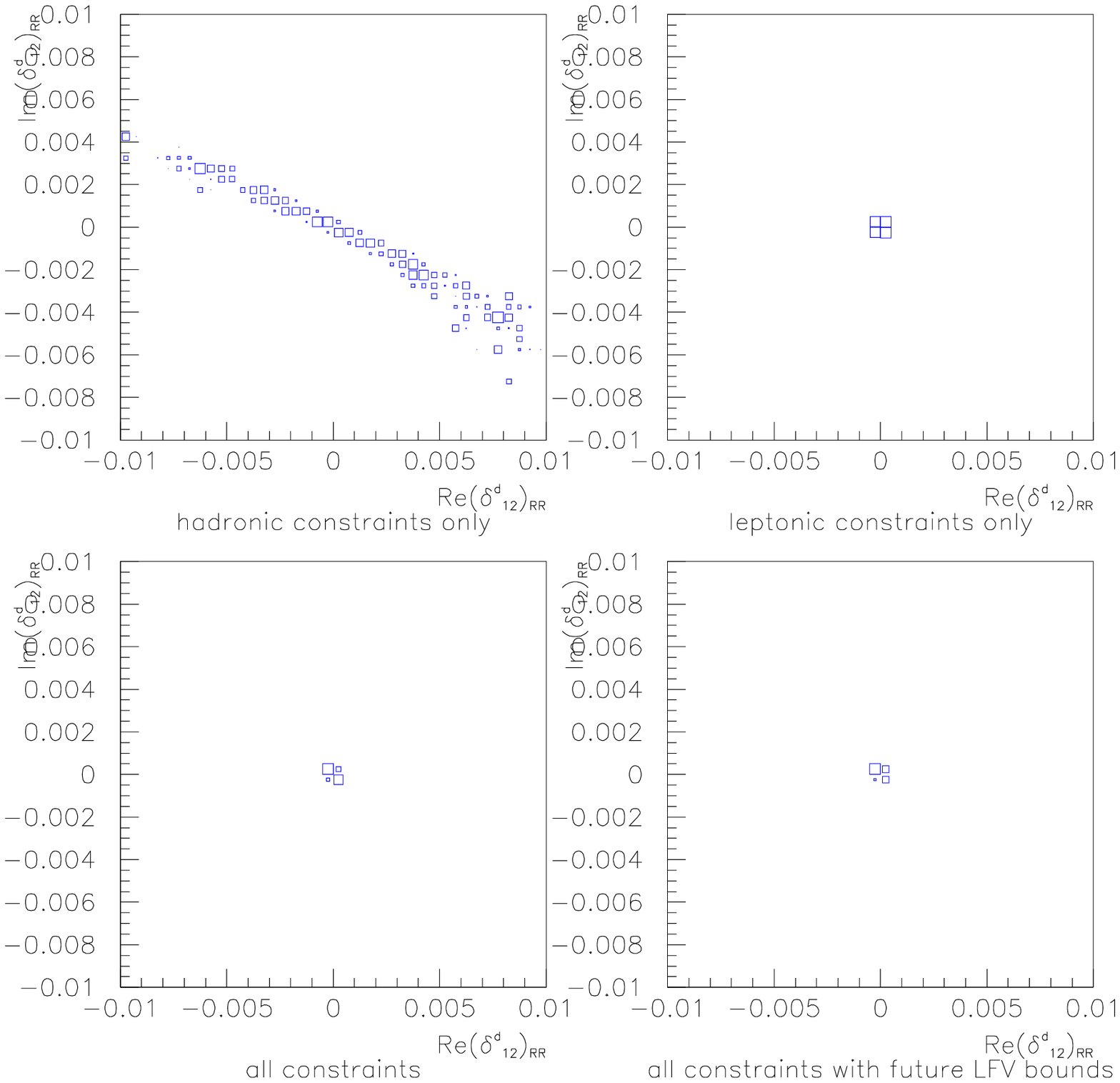}
\caption{Left four panels: allowed region for $(\delta^d_{23})_{RR}$ using constraints as indicated. Right for panels: the same
for $(\delta^d_{12})_{RR}$. For the parameter space considered, please see the text.\label{fig:RR12}}
\end{figure}

In the following the effect of leptonic bounds on the quark mass insertions are reviewed, following the results presented in 
\cite{Ciuchini:2007ha}, where constraints on $\delta$s were studied scanning
the mSUGRA parameter space in the ranges: $M_{1/2} \le 160$~GeV, $m_{0} \le 380$~GeV, $|A_{0}| \le 3 m_{0}$ and $5<\tan\beta<15$.
For instance, in presence of a $(\Delta^{d}_{23})_{LR}$ at the GUT scale, this would have effects both in the $\tau \to \mu
\gamma$ and $b \to s \gamma$ decays. Using $(\delta^{d}_{23})_{LR} \la (m_b/m_\tau)~(m_{\tilde l}^2/m_{\tilde q}^2)
(\delta^{l}_{23})_{RL}$, a bound on $(\delta^{l}_{23})_{RL}$ from the $\tau \to \mu \gamma$
decay translates into a bound on $(\Delta^{d}_{23})_{LR}$ (neglecting the effects of neutrino Yukawa's the inequality transforms into equality).
Thus, leptonic processes set a bound on the SUSY contributions to $B\left(B\to X_s\gamma\right)$.
However, it turns out that the present leptonic bounds have no effect on the $(\delta^d_{23})_{LR}$
couplings. This is due both to the existence of strong hadronic bounds from $b \to s \gamma$ and CP asymmetries and to the relatively weak
leptonic bounds here.

Similarly, in presence of a $(\Delta^{d}_{23})_{RR}$ at the GUT scale,
the corresponding MIs at the electroweak scale are
$(\delta^{d}_{23})_{RR}$ and $(\delta^{l}_{23})_{LL}$ that contribute
to $\Delta M_{B_s}$ and $\tau \to \mu \gamma$ respectively (the impact
of $(\Delta^{d}_{23})_{RR}$ on $b\to s\gamma$ and $b\to s
\ell^+\ell^-$ is not relevant because of the absence of interference
between SUSY and SM contributions).  In Fig.~\ref{fig:RR12} the
allowed values of ${\rm Re}(\delta^d_{23})_{RR}$ and ${\rm
Im}(\delta^d_{23})_{RR}$ with the different constraints are shown.
The leptonic constraints are quite effective as the bound on the
$B(\tau\to\mu\gamma)$ from B-factories is already very stringent,
while the recent measurement of $\Delta M_{B_s}$ is less
constraining. The plots correspond to $5<\tan\beta<15$, thus, the
absolute bound on $(\delta^l_{23} )_{LL}$ is set by $\tan\beta =5$ and
it scales with $\tan\beta$ as $(\delta^l_{23})_{LL}\sim
(5/\tan\beta)$\footnote{ Sizable SUSY contributions to $\Delta
M_{B_s}$ are still possible from the Higgs sector in the large $\tan
\beta$ regime both within
\cite{Buras:2002vd,Isidori:2006pk,Lunghi:2006uf} and also beyond
\cite{Foster:2006ze} the Minimal Flavor Violating
framework. However, for the considered parameter space, the above
effects are completely negligible.}.

As in the LR sector, in the LL one, there is no appreciable
improvement from the inclusion of leptonic constraints. In fact,
$\tau\to\mu\gamma$ is not effective to constrain
$(\delta^{l}_{23})_{RR}$, i.e. the leptonic MI related to
$(\delta^d_{23} )_{LL}$ in this SUSY-GUTs scheme, in large portions of
the parameter space because of strong cancellations among
amplitudes. The analysis of the constraints on the different
$(\delta^d_{13})$ MIs gives similar results to that of the
$(\delta^d_{23})$ MIs. In this case, the hadronic constraints come
mainly from $\Delta M_{B_d}$ and the different CP asymmetries measured
at B-factories, while the leptonic bounds are due to the decay $\tau
\to e \gamma$.

Coming to the 1-2 sector, let's see, as an example, the allowed values
of ${\rm Re} (\delta^d_{12})_{RR}$ and ${\rm Im}
(\delta^d_{12})_{RR}$.  In this case, as it appears from
Fig.~\ref{fig:RR12}, leptonic constraints, already using the present
limit on $B(\mu\to e\gamma)$, are competitive and constrain the
direction in which the constraint from $\varepsilon_K$ is not
effective (upper left plot). Similarly in the LR sector, even if the
hadronic bounds coming from $\varepsilon^\prime/\varepsilon$ are quite
stringent, the bounds from $\mu \to e \gamma$ are even more effective,
while the LL sector results less constrained by leptonic processes, as
an effect of the cancellations that $\mu \to e \gamma$ decay suffers
in the RR leptonic sector.

\subsection{R-parity violation}

\subsubsection{Introduction}
In supersymmetric extensions of the Standard Model (SM), baryon and lepton 
numbers are no longer automatically protected. This is the main reason 
for introducing $R$-parity. $R$-parity is associated with a $\,Z_2\,$ subgroup 
of the group of continuous $U(1)$ transformations acting on the gauge 
superfields and the two chiral doublet Higgs superfields $H_d$ and $H_u$, with 
their definition extended to quark and lepton superfields so that quarks and 
leptons carry $R\,=\,0\,$ and squarks and sleptons $R=\pm \,1\,$. 
One can express $R$-parity in terms of spin $S$, baryon $B$ and lepton $L$ 
number \cite{Farrar:1978xj}:
\begin{equation}
R\hbox{-parity} = (-1)\,^{2S} \ (-1)\,^{3(B-L)} \, .    
\end{equation} 
Taking into account the important phenomenological differences between models 
with and without $R$-parity, it is worth studying if and how $R$-parity can 
be broken. One of the main reasons to introduce R-parity is avoiding proton 
decay. However there are in principle other discrete or continuous symmetries 
that can protect proton decay while allowing for some $R$-parity violating 
couplings. In the absence of $R$-parity, $R$-parity odd terms allowed 
by renormalizability and gauge invariance 
\cite{Weinberg:1981wj}  must be included in 
the superpotential of the Minimal Supersymmetric Standard Model,
\begin{equation} 
 W_{Rp}\ =\ \mu_i\, H_u L_i\  +\ \frac{1}{2}\, \lambda_{ijk}\, L_i L_j E^c_k\
  +\ \lambda'_{ijk}\, L_i Q_j D^c_k\
  +\ \frac{1}{2}\, \lambda''_{ijk}\, U^c_i D^c_j D^c_k ,\,
\label{eq:Rpodd}
\end{equation}
where there is summation over the 
generation indices $i,j,k = 1,2,3$, and summation over gauge indices is 
understood. One has for example $L_i L_j E^c_k \equiv
(\epsilon_{ab} L^a_i L^b_j) E^c_k = (N_i E_j - E_i N_j) E^c_k$ and
$U^c_i D^c_j D^c_k \equiv
\epsilon_{\alpha \beta \gamma} U^{\alpha c}_i D^{\beta c}_j D^{\gamma c}_k$,
where $a,b = 1,2$ are $SU(2)_L$\index{$SU(2)$} indices, 
$\alpha, \beta, \gamma = 1,2,3$ are 
$SU(3)_C$\index{$SU(3)$} indices, and $\epsilon_{ab}$ and 
$\epsilon_{\alpha \beta \gamma}$ 
are totally antisymmetric tensors (with $\epsilon_{12} = \epsilon_{123}
= +1$). Gauge invariance enforces
antisymmetry of the $\lambda_{ijk}$ couplings in their first two 
indices and antisymmetry of the $\lambda''_{ijk}$ couplings in their last two 
indices,
\begin{eqnarray}
\lambda_{ijk} &=& - \lambda_{jik}\, ,\\
\lambda''_{ijk} &=& - \lambda''_{ikj}\ .
\end{eqnarray}
The bilinear terms $\mu_i H_u L_i$ in Eq.~(\ref{eq:Rpodd}) can be rotated 
away from the superpotential upon suitably redefining the lepton and Higgs 
superfields. However, in the presence of generic soft supersymmetry breaking 
terms of dimension two, bilinear $R$-parity violation will reappear. The fact 
that one can make $\mu_i = 0$ in Eq.~(\ref{eq:Rpodd}) does not mean that the 
Higgs-lepton mixing associated with bilinear $R$-parity breaking is unphysical, 
but rather that there is not a unique way of parameterizing it.
If R-parity is violated in the leptonic sector, no quantum numbers
differentiate between lepton and Higgs superfields, and they
consequently mix with each other \cite{Hall:1983id}.
The R-parity violation in the baryonic sector does not imply lepton
flavour violation, and we do not consider such option here.

A general consequence of $R$-parity violation is that unless the
relevant couplings are negligibly small, the supersymmetric model 
does not have a dark matter candidate.
Thus experimental studies on dark matter will also shed light on 
$R$-parity violation.

\subsubsection{Limits on couplings}
Limits on $R$-parity violating couplings can be obtained by direct searches at 
colliders or requiring that the $R$-parity violating contribution to
a given observable does not exceed the limit imposed by the precision of
the experimental measurement.

On the collider side $R$-parity violation implies the possibility of the 
creation, decay or exchange of single sparticles, thus allowing new decay 
channels. For example, even for relatively small $R$-parity violating 
interactions, the decay of the lightest supersymmetric particle will lead to 
collider events departing considerably from the characteristic missing momentum 
signal of $R$-parity conserving theories. 
In absence of definite theoretical predictions for the values 
of the 45 independent trilinear Yukawa couplings $\Lambda$
($\lambda_{ijk}$, $\lambda'_{ijk}$ and $\lambda''_{ijk}$), it is necessary in 
practice to assume a strong hierarchy among the couplings.
A simplifying assumption widely used for the search at
colliders is to postulate the existence of a single dominant 
$R$-parity violating coupling.
When discussing specific bounds, it is necessary
to choose a definite basis for quark and lepton superfields.
Often it is understood that the single coupling dominance
hypothesis applies in the mass eigenstate basis. It can be more
natural to apply this hypothesis in the weak eigenstate basis 
when dealing with models in which the hierarchy among couplings 
originates from a flavour theory. In this case, a single
process allows to constrain several couplings, provided one has some
knowledge of the rotations linking the weak eigenstate and mass eigenstate
bases.
Indirect bounds from loop processes typically lead to bounds on
the products of two most important $R$-parity violating couplings,
or on the sum of products of two couplings.
The limits on single dominant couplings, and on products of
couplings, as well as a more complete list of references, are collected in 
\cite{Barbier:2004ez}.

\subsubsection{Spontaneous $R$-parity breaking}
The spontaneous breaking of $R$-parity is
characterized by an $R$-parity invariant Lagrangian leading to non-vanishing
VEVs for some $R$-parity odd scalar field, which in turn generates $R$-parity 
violating terms. Such a spontaneous breakdown of $R$-parity generally also 
entails the breaking of the global $U(1)$ lepton number symmetry $L$ 
which implies the existence of a massless Nambu-Goldstone real pseudoscalar 
boson $J$, the Majoron. Another light scalar particle, denoted $\rho$,
generally accompanies the Majoron in the supersymmetric models.  If the
$U(1)$ symmetry is also explicitly broken by interaction terms in the
Lagrangian, both of these particles acquire finite masses.
The most severe constraints  on the models  with a
spontaneous  $R$-parity breaking, arise in the cases
where  the Majoron  carries electroweak gauge charges 
and hence is coupled to the $Z$ bosons and
to quarks and leptons.  The non-singlet components contribute 
to the $Z$ boson invisible width by an amount of one-half that a single light 
neutrino, $ \delta \Gamma^Z _{inv}/6 \simeq 83 $ MeV.  To 
suppress the  non-singlet  components one must allow either for 
sufficiently small sneutrino VEVs, $v_{L} /M_Z <<1 $, or for some large 
hierarchy of scales between $v_L$ and the VEV parameters associated with
additional electroweak singlet scalar fields \cite{Comelli:1993nt}.

However, it is not necessary that models with spontaneous $R$-parity 
violation have a Majoron.
Models without a Majoron include a class of models with triplet 
Higgses, where $B-L$ is a gauge symmetry, which is necessarily 
spontaneously broken unless effects of non-renormalizable terms
or some additional new fields are included 
\cite{Kuchimanchi:1993jg}.
An interesting experimental signal in these models may be a
relatively light doubly charged scalar, which decays dominantly
to same charge leptons (not necessarily of the same generation)
\cite{Huitu:1994zm}.
Another possibility for a model without a Majoron is a model where 
the lepton 
number is broken by two units explicitly, in which case the 
spontaneous breaking by one unit (which breaks the $R$-parity) does not 
lead to a Majoron \cite{Kitano:1999qb}.
The interactions in spontaneously $R$-parity breaking models through
the lepton number violation closely resemble explicitly $R$-parity 
breaking models with only bilinear $R$-parity violation.
In the case of spontaneous breaking, the parameters which are free
in the model with only bilinear couplings are related to each other 
via the sneutrino vacuum expectation value (VEV).
Thus a constraint from one process affects availability of the other
processes.
Example bounds for such a model can be found in \cite{Frank:2001tr}.

It is worth emphasizing that choosing single coupling dominance in the
case of spontaneous breaking is not possible and in this sense, the
models with spontaneous breaking are more predictive than those
without.

\subsubsection{Neutrino sector}
The presence of non-zero couplings
$\lambda_{ijk}$, $\lambda'_{ijk}$ or bilinear $R$-parity violating parameters 
implies the generation of neutrino masses and mixing
\cite{Ellis:1984gi}. 
This is  
an interesting feature of $R$-parity violating models, but it can also be a 
problem, since the contribution of $R$-parity violating couplings may exceed by 
orders of magnitude the experimental bounds.
Two types of contributions can be distinguished: tree-level or loop 
contributions.

The tree-level contributions are due to bilinear $R$-parity violation 
terms which induce a mixing between neutrinos and neutralinos
\cite{Davidson:2000ne}. This gives a 
massive neutrino state at tree level. When quantum corrections are included, all 
three neutrinos acquire a mass. The tree-level contribution arising 
from the neutrino-neutralino mixing can be understood, in the limit of small 
neutrino-neutralino mixing, as a sort of seesaw mechanism, in which the neutral 
gauginos and Higgsinos play the r\^ole of the right-handed neutrinos.

The loop contributions are induced by the trilinear $R$-parity violating 
couplings $\lambda_{ijk}$ and $\lambda'_{ijk}$ and by bilinear $R$-parity 
violating parameters \cite{Babu:1989px}. 
If bilinear $R$-parity violation is 
strongly suppressed one can concentrate on the diagrams involving
trilinear $R$-parity violating couplings only. 
The trilinear couplings $\lambda_{ijk}$ and $\lambda'_{ijk}$
contribute to each entry of the neutrino mass matrix through the
lepton-slepton and quark-squark loops. The neutrino mass matrix depends 
therefore on a large number of trilinear $R$-parity violating couplings. In 
order to obtain a predictive model, one has to make assumptions on the structure 
of the trilinear couplings.
In general, however, the bilinear $R$-parity violation contribution cannot be 
neglected. The presence of bilinear terms drastically modifies the 
calculation of one-loop neutrino masses. The neutrino mass matrix 
receives contributions already at tree level, as discussed above, and moreover 
in addition to the lepton-slepton and quark-squark loops, 
one-loop diagrams involving insertions of bilinear $R$-parity violating masses 
or slepton VEVs must be considered. One should note that the bilinear $R$-parity 
violating terms, if not suppressed, give too large loop contributions to 
neutrino masses.

The scenario known as bilinear R-Parity violation (BRpV) corresponds to
the
explicit introduction of the three mass parameters $\mu_i$ in the first
term
in Eq.~(\ref{eq:Rpodd}), without referring to their origin, and assuming
that
all the trilinear parameters are zero. The $\mu_i$ terms introduce
tree-level
mixing between the Higgs and lepton superfields. Therefore, they violate
R-Parity and lepton number, and contribute to the breaking of the
$SU(2)$
symmetry by the induction of sneutrino vacuum expectation values $v_i$.
As it was mentioned before, the mixing between neutralinos and neutrinos
leads to an effective tree-level neutrino mass matrix of the form,
\begin{equation}
m_{\nu}^{0ij}=\frac{M_1g^2+M_2g'^2}{4\det M_{\chi^0}}\Lambda_i\Lambda_j,
\end{equation}
where the parameters $\Lambda_i=\mu v_i+\mu_iv_d$ are proportional to
the
sneutrino vacuum expectation values in the basis where the $\mu_i$ terms
are
removed from the superpotential. Due to the symmetry of this mass
matrix, only
one neutrino acquires a mass. Once quantum corrections are included,
this
symmetry is broken, and the effective neutrino mass matrix takes the
form
\cite{Hirsch:2000ef},
\begin{equation}
m_{\nu}^{ij}=A\Lambda_i\Lambda_j+B(\Lambda_i\epsilon_j+\Lambda_j\epsilon_i)
+C\epsilon_i\epsilon_j .
\end{equation}
If the tree-level contribution dominates, as for example in SUGRA models
with
low values of $\tan\beta$, the atmospheric mass scale is given at tree
level,
and the solar mass scale is generated at one loop, explaining the
hierarchy between them. Most of the time, the dominant loop in SUGRA is
the
one formed with bottom quarks and squarks, followed in importance by
loops
with charginos and neutralinos. In the tree-level dominance case the
atmospheric mixing angle is well approximated by
$\tan^2\theta_{atm}=\Lambda_2^2/\Lambda_3^2$, and the reactor angle by
$\tan^2\theta_{13}=\Lambda_1^2/(\Lambda_2^2+\Lambda_3^2)$. In this case,
the
smallness of the reactor angle is achieved with a small value of
$\Lambda_1$,
and the maximal mixing in the atmospheric sector with a similar value
for
$\Lambda_2$ and $\Lambda_3$. Supergravity scenarios where tree-level
contribution does not dominate can also be found \cite{Diaz:2004fu}, in
which case the previous approximations for the angles are not valid.

\subsubsection{Lepton flavour violating processes at low energies}

Many processes, which are either rare or forbidden in the R-parity 
conserving model, become possible when interactions following from
the superpotential $W_{Rp}$ in (\ref{eq:Rpodd}) are available.
These interactions include tree-level couplings between different
lepton or quark generations, as well as tree-level couplings between
leptons and quarks, or leptons and Higgses.

In addition to the trilinear couplings 
$\lambda$ and $\lambda'$, bilinear couplings or spontaneous 
R-parity breaking contribute to the lepton flavour violating
processes mentioned below through mixing.

For references about this section, see Ref. \cite{Barbier:2004ez}.

\begin{itemize}
\item $l_i\to l_j\gamma$, $l_i\to l_jl_kl_m$, and $\mu - e$ -conversion, 
and semileptonic decays of $\tau$-leptons\\
The rare decays of leptons to lighter leptons are excellent probes of
new physics, because they do not involve any hadronic uncertainties.
Both the lepton flavour violating trilinear $\lambda$- and $\lambda'$
-type couplings give rise to LFV decays $l_i\to l_j\gamma$
(loop level process with $\tilde\nu - l$, $\nu - \tilde l$, or
$\tilde q - q'$ in the loop),
$l_i\to l_jl_kl_m$ (tree-level process via $\tilde\nu$ or 
$\tilde l$), as well as for $\mu - e$ -conversion. 
In these processes, two non-vanishing $\Lambda$ couplings are needed
and usual approach is to assume a dominant product of two couplings,
when determining bounds on couplings.
In the $\mu - e$ -conversion, certain pairs of couplings can be probed 
only in the loop-level process, mediated by virtual $\gamma $ or $Z$,
which are logarithmically enhanced compared to $\mu \to e \gamma$
\cite{Huitu:1997bi}.
The hadronic contributions to the $\mu -e$ -conversion in nuclei
make the theoretical error larger than in the decays without hadrons.
The relatively large mass of $\tau$ allows new semileptonic decay modes
for $\tau$.
The bounds from these processes vary between 
$\Lambda \sim {\cal{O}}(10^{-4}-10^{-1})$ for 100 GeV fermion masses,
and they scale as {\it mass}$^2$.

The experimental accuracies of the processes mentioned above 
are expected to increase considerably in the coming years.\\

\item Leptonic and semileptonic decays of hadrons and top quarks\\
$R$-parity violating couplings $\lambda'_{ijk}$ allow for 
lepton flavour violating decays of hadrons, {\it e.g.} 
$K_L\to e^\pm\mu^\mp$, $B_d\to\mu^+\tau^-$,
$K^+\to \pi^+\nu_i\nu_j$ \cite{Deandrea:2004ae}, as well as semileptonic 
LFV top decays, {\it e.g.} $t\to \tilde\tau^+b$, if
kinematically allowed.
The sensitivity on the couplings is restricted by the theoretical 
uncertainties in hadronic contributions.
For 100 GeV sfermions, the bounds are 
$\Lambda \sim {\cal{O}}(10^{-4}-10^{-1})$ .
\end{itemize}

\subsubsection{Anomalous muon magnetic moment $a_\mu$ and electron electric dipole moment} 

$\Lambda$ couplings affect leptons also through contributions to
dipole moments.  The experimental measurement of $a_\mu$ is quite
precise.  The theoretical calculation of the Standard Model
contribution to $a_\mu$ contains still uncertainty, which prevents
exact comparison with measurement.  The contribution of $R$-parity
violation on $a_\mu$ is small, and constrained by tiny neutrino
masses.

Contribution from complex $\Lambda$ to electron EDM could be large for
large phases.  The one-loop contribution involving both bilinear and
trilinear couplings is sizable for electron EDM, while one-loop terms
with only trilinear terms are suppressed by neutrino masses.

\subsubsection{Collider signatures}

The main advantage of collider studies compared to the low energy
probes is that the particles can 
be directly produced, and thus their masses and couplings can be 
experimentally measured.

A major difference between R-parity conserving and breaking models
from the detection point of view is the amount of missing energy.
If R-parity is violated, the supersymmetric particles
decay to the SM particles leaving little or no missing energy.
Decays of sparticles through $\lambda$ and $\lambda'$-type couplings
lead to multi-lepton final states, and $\lambda'$ and $\lambda''$ to
multi-jet final states.
Sparticles can decay first via the R-parity conserving couplings
to the lightest supersymmetric particle (LSP), which then decays via
R-parity violating couplings.  
If e.g. a neutralino is the LSP, 
it may be a cascade decay product of a sfermion, chargino, 
or a heavier neutralino.
Thus typically one gets a larger number of jets or leptons in 
the final state in R-parity violating than in 
the R-parity conserving decay.
The sparticles can also decay directly to the Standard Model 
fermions via $\lambda$, $\lambda'$, or $\lambda''$ couplings.
Assuming all the supersymmetric particles decay inside the
detector, a consequence of the decay of the LSP
is that the amount of missing energy when
R-parity is violated is considerably lower than in the 
R-parity conserving case, and only neutrinos carry the
missing energy.
When R-parity is violated, the LSP is not stable and
need not be neutral.
If then the coupling through which the LSP decays is suppressed,
a long lived possibly charged particle appears, leaving 
a heavily ionizing, easily detectable charged track in the detector.

A simplifying assumption for the search strategy at
colliders is to postulate the existence of a single dominant 
R-parity violating coupling. In case a non-vanishing coupling does exist 
with a magnitude leading to distinct phenomenology at colliders, a direct 
sensitivity to a long-lived LSP might be provided by the observation of 
displaced vertices in an intermediate range of coupling values up to 
${\cal{O}}(10^{-5}-10^{-4})$.
For larger $\Lambda$ values the presence of R-parity violating supersymmetry 
will become manifest through the decay of short-lived
sparticles produced by pair via gauge couplings. 
A possible search strategy in such cases consists of neglecting
R-parity violating contributions at production in 
non-resonant processes. This is valid provided that the interaction 
strength remains sufficiently small compared to electromagnetic or weak 
interaction strengths, for $\Lambda$ values typically below 
${\cal{O}}(10^{-2}-10^{-1})$. In a similar or larger range of 
couplings values, R-parity violation could show up at colliders 
via single resonant or non-resonant production of supersymmetric particles.

For bilinear or spontaneous breaking, the lightest supersymmetric
particle decays through mixing with the corresponding $R_p=+1$ particle.
If the LSP is neutralino or chargino, it decays through mixing
with neutrino or charged lepton, and if the LSP is a slepton it decays 
through mixing with the Higgs bosons, {\it e.g.} stau mixes with
charged Higgs.
Assuming that neutralino is the LSP, the dominant decay mode of stau
is to tau and neutralino.
Through mixing the charged Higgs has then a branching ratio to tau
and neutralino.
Thus the detection of $R$-parity violation includes precise 
measurement of the branching ratios of particles.

The main signature of BRpV is the decay of the neutralino, which decays
100\%
of the time into R-Parity and lepton number violating modes. If squarks
and
sleptons are heavy and the neutralino is heavier than the gauge bosons,
the
neutralino decays into on-shell gauge bosons and leptons:
$\chi^0_1\to W^\mp\ell_i^\pm, Z\nu_i$. If the
gauge bosons are produced off-shell, then the decay modes are
$\chi^0_1\to q\overline q'\ell_i^\pm, \,
\ell_j^\mp\nu_j\ell_i^\pm,
\, q\overline q\nu_i,\, \ell_j^\pm\ell_j^\mp\nu_i,\, \nu_j\nu_j\nu_i$.
When
sfermions cannot be neglected, the decay channels are the same, but
squarks
and sleptons contribute as intermediate particles \cite{Porod:2000hv}.
In this model, very useful quantities are formed with ratios of
branching
ratios, since they can be directly linked to R-Parity violating
parameters
and neutrino observables. We have for example,
\begin{equation}
\frac{B(\chi^0_1\to q\overline q'\mu)}
{B(\chi^0_1\to q\overline q'\tau)}\approx
\frac{\Lambda_2^2}{\Lambda_3^2}\approx \tan^2\theta_{atm}.
\end{equation}
where the last approximation is valid in the tree-level dominance
scenario.
In this way, collider and neutrino measurements, coming from very
different
experiments, can be contrasted.

Detection possibilities and extraction of limits depend a lot on the
specific
model and on the collider type and energy. On general grounds a
lepton-hadron
collider provides both leptonic and baryonic quantum numbers in the
initial state and is therefore suited for searches involving
$\lambda'$. In $e^+p$ collisions, the production of $\tilde{u}_L^j$
squarks of
the $j^{th}$generation via $\lambda'_{1j1}$ is especially interesting as
it
involves a valence $d$ quark of the incident proton.
In contrast, for $e^-p$ collisions where charge conjugate processes are
accessible, the $\lambda'_{11k}$ couplings become of special interest as
they
allow for the production, involving a valence $u$ quark, of
$\tilde{d}_R^k$
squarks of the $k^{th}$ generation.

The excluded regions of the parameter space for R-parity violating 
scenarios have been worked out from the data at LEP, HERA and Tevatron,
see e.g. \cite{Heister:2002jc,Abreu:2000ne,Achard:2001ek,Abbott:2000yu,
Aid:1996iw,Adloff:2001at}.
In the following we shall concentrate on the search possibilities 
at the LHC.

\subsubsection{Hadron colliders}

In hadron colliders the $\lambda$ or $\lambda'$ couplings can
provide a viable signal.
In many SUSY scenarios neutralinos and charginos are among the
lightest supersymmetric particles.
Their pair production or associated production of 
$\tilde\chi^\pm_1\tilde\chi^0$ via gauge couplings and decay
via $\lambda$ or $\lambda'$couplings may lead to a
tri-lepton signal from each particle, providing a clean signature.
One should notice that if the couplings are small, the vertex
may be displaced which makes the analysis more complicated.
With small enough couplings the lightest neutralino, if LSP,
decays outside the detector.

If kinematically possible,
gluinos and squarks are copiously produced at hadron colliders.
The NLO cross section has been calculated in \cite{Beenakker:1994an}.
For $m_{\tilde q}>m_{\tilde g}>m_{\tilde c_L}$, the production
with decay via $\lambda'_{121} \neq 0$ was studied at CDF.
Also coupling $\lambda'_{13k}$ from $\tilde t$ pair production
at CDF and $\lambda'$ couplings from $\chi^0_1$ decay at D0
have been investigated.

When R-parity is violated, the supersymmetric particles can be produced
singly, and thus they can be produced as resonances through
R-parity violating interactions.
In a hadron--hadron collider this
allows to probe for resonances in a wide mass range because of
the continuous energy distribution of the colliding partons.
This production mode requires non-negligible R-parity violating
coupling.
If a single R-parity violating coupling is dominant, the exchanged
supersymmetric particle may decay through the same
coupling involved in its production, giving a two fermion final state.
It is also possible that the decay of the resonant SUSY particle goes
through gauge interactions, giving rise to a cascade decay.

The resonant production of sneutrinos and charged sleptons 
(via $\lambda'$ couplings) has been investigated at hadron
colliders \cite{Kalinowski:1997zt,Moreau:2000bs,Dimopoulos:1988jw,
Dreiner:2000vf,Moreau:1999bt}.
The production of a charged lepton with neutralino leads to a
like-sign dilepton signature via $\lambda'$ couplings.
The production of a charged lepton with a chargino in the 
resonant sneutrino case decay leads to a tri-lepton final state
via $\lambda'$ couplings.
The $\tilde\chi^0_1, \;\tilde\chi^\pm_1, \;\tilde\nu$
masses can be reconstructed using the tri-lepton signal.

Single production is possible also in two-body processes
without resonance \cite{Deliot:2000mf}.
Sfermion production with a gauge boson has been studied in
 either via $\lambda$ or $\lambda'$-coupling. (The process
$\bar{q}_i q_j\to W^-\tilde\nu_k$ or 
$\bar{q}_i q_j\to W^+\tilde{l}_{kL}$ can get 
contribution also from resonant production, but e.g.
in SUGRA $m_{\tilde{l}}-m_{\tilde\nu}=\cos 2\beta m_W^2$
and resonance production is not kinematically viable).
Similarly via $\lambda'$ or $\lambda''$ gluino can be
produced with a lepton or quark, respectively.
Sneutrino production with two associated jets may also provide
a detectable signal \cite{Chaichian:2003kf}.

Resonant production of squarks can occur via $\lambda''$-type
couplings, leading eventually to jets in the final states.
Although the cross sections can be considerable for these
processes, the backgrounds in hadronic colliders are large, and
the processes seem difficult to study \cite{Dimopoulos:1988fr}.
In special circumstances the backgrounds can be small,
e.g. for stop production in $\bar d_i\bar d_j\to
\tilde t_1\to b\tilde\chi^+_1$, with
$\tilde\chi_i^+\to \bar l_i\nu_i \tilde\chi^0_i$
(here it is assumed $m_{\tilde t_1} > m_{\chi^+_1}
>  m_{\chi^0_1}$, $m_{top}> m_{\chi^0_1}$).
Then for $\lambda''_{3jk}$, $ m_{\chi^0_1}$ is stable
\cite{Allanach:1999bf,Berger:1999zt}.
Also single gluino production, $d_id_j\to
\tilde g \bar t$ via resonant stop production has a good signal to
background ratio for $\lambda''_{3jk}={\cal{O}}(0.1)$
\cite{Chaichian:2000ux}.

With the $t\overline t$ production cross section of the order of 
$800 \,pb$, the LHC can be considered a top quark factory, with
$\sim 10^8$ top quarks being produced per year, assuming an
integrated luminosity of $100 \,fb^{-1}$. This statistics allows 
for precise studies of top quark physics, in particular, for
measurements of rare RpV decays. A simulation of the signal 
and background using ATLFAST \cite{atlfast}, to take into account the 
experimental conditions prevailing at the ATLAS detector
\cite{AtlasTechRep}, was made for a top quark decaying through
a $\lambda'_{\ell 31}$ coupling to $t\to \tilde\chi^0\ell d$,
assuming only one slepton gives the leading contribution as an intermediate
state \cite{Belyaev:2004qp}. 

The importance of treating the top quark production
and decay simultaneously $gg\to t\tilde\chi^0\ell d$, rather
than
$\Gamma(gg\to t\overline t)B(t\to \tilde\chi^0\ell d)$,
was shown.
The latest approach can underestimate the cross section by a factor 
of a few units, depending on the slepton mass. The reason is
that the slepton forces the top
quark to be off-shell, becoming the resonance itself, as can be 
appreciated from $\tilde\chi^0\ell$ mass invariant distributions.

Two scenarios were chosen for the neutralino decay, $\tilde\chi^0\to
bd\nu_e$ and $\tilde\chi^0\to cde$, the last one assuming a large
stop-scharm mixing. The sensitivity of the LHC is presented as the
significance $S/\sqrt{B}$ as a function of $\lambda'_{131}$, for
slepton masses 150 and 200 GeV. The channel $t\to \tilde\chi^0 ed\to
cdeed$ is more promising with exclusion limits at $2\sigma$ c.l. for
$\lambda'>0.03$ and observation at $5\sigma$ c.l. for $\lambda'>0.05$,
with these values slightly increasing for heavier sleptons. The $t\to
\tilde\chi^0 ed\to bd\nu_e ed$ channel is observable only for a 150
GeV slepton mass. The significance is reduced to $\lambda'>0.08$ at
$2\sigma$ and $\lambda'>0.15$ at $5\sigma$ level.

Since a $\lambda'_{\ell 33}\sim h_b\epsilon_\ell/\mu$ trilinear term is 
generated in BRpV when the $\epsilon_\ell$ term is removed from the 
superpotential, we can see that the above exclusion limits for $\lambda'$
are not significant in BRpV, probing only values of $\epsilon_\ell$ 
parameters much larger than what is needed for neutrino oscillations.

\subsection{Higgs-mediated lepton flavour violation in supersymmetry}

If neutrinos are massive, one would expect LFV transitions in the
Higgs sector through the decay modes $H^0\to l_i l_j$ mediated at one
loop level by the exchange of the $W$ bosons and neutrinos. However,
as for the $\mu\to e\gamma$ and the $\tau\to \mu\gamma$ case, also the
$H^0\to l_i l_j$ rates are GIM suppressed. In a supersymmetric (SUSY)
framework the situation is completely different.  Besides the previous
contributions, supersymmetry provides new direct sources of flavour
violation, namely the possible presence of off-diagonal soft terms in
the slepton mass matrices and in the trilinear couplings
\cite{Borzumati:1986qx}. In practice, flavour violation would
originate from any misalignment between fermion and sfermion mass
eigenstates. LFV processes arise at one loop level through the
exchange of neutralinos (charginos) and charged sleptons
(sneutrinos). The amount of the LFV is regulated by a Super-GIM
mechanism that can be much less severe than in the non supersymmetric
case \cite{Borzumati:1986qx}. Another potential source of LFV in
models such as the minimal supersymmetric standard model (MSSM) could
be the Higgs sector, in fact, extensions of the SM containing more
than one Higgs doublet generally allow flavor-violating couplings of
the neutral Higgs bosons.  Such couplings, if unsuppressed, will lead
to large flavor-changing neutral currents in direct opposition to
experiments. The MSSM avoid these dangerous couplings at the tree
level segregating the quark and Higgs fields so that one Higgs $(H_u)$
can couple only to up-type quarks while the other $(H_d)$ couples only
to d-type.  Within unbroken supersymmetry this division is completely
natural, in fact, it is required by the holomorphy of the
superpotential.  However, after supersymmetry is broken, couplings of
the form $QU_cH_d$ and $QD_cH_u$ are generated at one loop
\cite{Hall:1993gn}.  In particular, the presence of a non zero $\mu$
term, coupled with SUSY breaking, is enough to induce non-holomorphic
Yukawa interactions for quarks and leptons.  For large $\tan\beta$
values the contributions to d-quark masses coming from non-holomorphic
operator $QD_cH_u$ can be equal in size to those coming from the usual
holomorphic operator $QD_cH_d$ despite the loop suppression suffered
by the former.  This is because the operator itself gets an additional
enhancement of $\tan\beta$.

As shown in reference \cite{Babu:1999hn} the presence of these loop-induced non-holomorphic couplings also leads to the appearance of flavor-changing couplings of
the neutral Higgs bosons. These new couplings generate a variety of flavor-changing processes such as $B^0 \to\mu^{+}\mu^{-}$, $\bar{B^0}-B^0$ etc.
\cite{D'Ambrosio:2002ex}. Higgs-mediated FCNC can have sizable effects also in the lepton sector \cite{Babu:2002et,Dedes:2002rh}: given a source of non-holomorphic couplings,
and LFV among the sleptons, Higgs-mediated LFV is unavoidable. These effects have been widely discussed in the recent literature
both in a generic 2HDM \cite{Chang:1993kw,Sher:1991km} and in supersymmetry \cite{Sher:2002ew,Dedes:2002rh} frameworks.
Through the study of many LFV processes as $\ell_i\to \ell_j\ell_k\ell_k$ \cite{Babu:2002et,Dedes:2002rh}, $\tau\to \ell_j\eta$ \cite{Sher:2002ew,Brignole:2004ah},
$\ell_i\to \ell_j\gamma$ \cite{Paradisi:2005tk,Paradisi:2006jp}, $\mu N\to e N$ 
\cite{Kitano:2003wn},
$\Phi^{0}\to \ell_j \ell_k$ \cite{Brignole:2003iv} (with $\ell_i=\tau,\mu$, $\ell_{j,k}=\mu,e$, $\Phi=h^0,H^0,A^0$) or the cross section of the $\mu N\to\tau X$ reaction
\cite{Kanemura:2004jt}.

\subsubsection{LFV in the Higgs sector}

SM extensions containing more than one Higgs doublet generally allow flavor-violating couplings of the neutral Higgs bosons with fermions.
Such couplings, if unsuppressed, will lead to large flavor-changing neutral currents in direct opposition to experiments.
The possible solution to this problem involves an assumption about the Yukawa structure of the model. A discrete symmetry can be invoked to allow a given 
fermion type to couple to a single Higgs doublet, and in such case FCNC's are absent at tree level. In particular, when a single Higgs field gives masses to both types of 
fermions the resulting model is referred as 2HDM-I. On the other hand, when each type of fermion couples to a different 
Higgs doublet the model is said 2HDM-II.

In the following, we will assume a scenario where the type-II 2HDM structure is not protected by any symmetry and is broken by loop effects 
(this occurs, for instance, in the MSSM).

Let us consider the Yukawa interactions for charged leptons,
including the radiatively induced LFV terms \cite{Babu:2002et}:
\begin{equation}
\mathcal{-L}\simeq \overline{l}_{Ri}Y_{l_{i}}H_1\overline{L_i}+ \overline{l}_{Ri}\left(Y_{l_{i}}\Delta_{L}^{ij}+Y_{l_{j}}\Delta_{R}^{ij}\right)H_2 \overline{L_j} + h.c.,
\end{equation}
where $H_1$ and $H_2$ are the scalar doublets, $l_{Ri}$ are lepton singlet for right handed fermions, $L_{k}$ denote the lepton doublets and $Y_{l_{k}}$ are 
the Yukawa couplings.

In the mass-eigenstate basis for both leptons and Higgs bosons, the effective flavor-violating interactions are described by the 
four dimension operators \cite{Babu:2002et}:
\begin{eqnarray}
\mathcal{-L}&\simeq&(2G_{F}^2)^{\frac{1}{4}}\frac{m_{l_i}} {c^2_{\beta}} \left(\Delta^{ij}_{L}\overline{l}^i_R l^j_L+\Delta^{ij}_{R}\overline{l}^i_L l^j_R \right)
\left(c_{\beta-\alpha}h^0-s_{\beta-\alpha}H^0-iA^0 \right)\nonumber\\\nonumber\\
&+& (8G_{F}^2)^{\frac{1}{4}}\frac{m_{l_i}} {c^2_{\beta}} \left(\Delta^{ij}_{L}\overline{l}^i_R \nu^j_L+\Delta^{ij}_{R}\nu^i_L\overline{l}^j_R\right) H^{\pm} + h.c.,
\end{eqnarray}
where $\alpha$ is the mixing angle between the CP-even Higgs bosons $h_0$ and $H_0$, $A_0$ is the physical CP-odd boson, $H^{\pm}$ are the physical
charged Higgs-bosons and $t_{\beta}$ is the ratio of the vacuum expectation value for the two Higgs (where we adopt the notation, $c_{x},s_{x}\!=\!\cos x,\sin x$ 
and $t_{x}\!=\!\tan x$). Irrespective to the mechanism of the high energy theories generating the LFV, we treat the $\Delta^{ij}_{L,R}$ terms in a model independent way.
In order to constrain the $\Delta^{ij}_{L,R}$ parameters, we impose that their contributions to LFV processes
do not exceed the experimental bounds \cite{Paradisi:2005tk,Paradisi:2006jp}.

On the other hand, there are several models with a specific ansatz about the flavor-changing couplings. For instance, the famous multi-Higgs-doublet models proposed
by Cheng and Sher \cite{Cheng:1987rs} predict that the LFV couplings of all the neutral Higgs bosons with the fermions have the form $Hf_if_j \sim \sqrt{m_im_j}$.

In Supersymmetry, the $\Delta^{ij}$ terms are induced at one loop
level by the exchange of gauginos and sleptons, provided a source of
slepton mixing.  In the so mass insertion (MI) approximation, the
expressions of $\Delta^{ij}_{L,R}$ are given by
\begin{eqnarray}
\label{deltal}
 \Delta^{ij}_{L} = &-&\frac{\alpha_{1}}{4\pi}\mu M_1 \delta^{ij}_{LL} m_{L}^2 \left[ I^{'} (M_1^2, m_{R}^2, m_{L}^2)+\frac{1}{2} I^{'} (M_1^2, \mu^2, m_{L}^2)
\right]+\nonumber\\\nonumber\\ &+& \frac{3}{2} \frac{\alpha_{2}}{4\pi} \mu M_2 \delta^{ij}_{LL} m_{L}^2 I^{'} (M_2^2, \mu^2, m_{L}^2)\ ,
\end{eqnarray}
\begin{equation}
\label{deltar}
\Delta^{ij}_{R}= \frac{\alpha_{1}}{4\pi}\mu M_1 m^{2}_{R} \delta^{ij}_{RR} \left[I^{'}\!(M^{2}_{1},\mu^2,m^{2}_{R})\!-\!(\mu\!\leftrightarrow\! m_{L}) \right],
\end{equation}
respectively, where $\mu$ is the Higgs mixing parameter, $M_{1,2}$ are the gaugino masses and $m^{2}_{L(R)}$ stands for the left-left 
(right-right) slepton mass matrix entry. The LFV mass insertions (MIs), i.e. $\delta^{3\ell}_{XX}\!=\!({\tilde m}^2_{\ell})^{3\ell}_{XX}/m^{2}_{X}$ $(X=L,R)$,
are the off-diagonal flavor changing entries of the slepton mass matrix. The loop function $I^{'}(x,y,z)$ is such that $I^{'}(x,y,z)= dI(x,y,z)/d z$, 
where $I(x,y,z)$ refers to the standard three point one-loop integral which has mass dimension -2
\begin{eqnarray}
 I_3 (x, y, z)=\frac{xy \log (x/y) + yz \log (y/z) + zx \log (z/x)}{(x-y)(z-y)(z-x)}\,.
\end{eqnarray}
The above expressions, i.e. the Eqs.~(\ref{deltal},\ref{deltar}), depend only on the ratio of the SUSY mass scales and they do not decouple for large $m_{SUSY}$.
As first shown in Ref.\cite{Brignole:2003iv}, both $\Delta^{ij}_{R}$ and $\Delta^{ij}_{L}$ couplings suffer from strong cancellations in certain 
regions of the parameter space due to destructive interferences among various contributions. For instance, from Eq.~(\ref{deltar}) it is clear that,
in the $\Delta^{ij}_{R}$ case, such cancellations happen if $\mu=m_L$.

In the SUSY seesaw model, in the mass insertion approximation, 
one obtains specific values for $\delta_{LL}^{ij}$ depending on the assumptions
on the flavour mixing in $Y_\nu$ \cite{Masiero:2002jn,Masiero:2004js}.
 If the latter is of CKM size, 
$\delta_{LL}^{21(31)}\simeq 3\cdot 10^{-5}$ and $\delta_{LL}^{32}\simeq 10^{-2}$, while in the case 
of the observed neutrino mixing, taking $U_{e3}=0.07$ at about half of the current CHOOZ 
bound, we get $\delta_{LL}^{21(31)}\simeq 10^{-2}$ and $\delta_{LL}^{32}\simeq 10^{-1}$.

\subsubsection{Phenomenology}

In order to constrain the $\Delta^{ij}_{L,R}$ parameters, we impose
that their contributions to LFV processes as $l_i\to l_jl_kl_k$ and
$l_i\to l_j\gamma$ do not exceed the experimental bounds. At tree
level, Higgs exchange contribute only to
$\ell_i\to\ell_j\ell_k\ell_k$, $\tau\to\ell_j\eta$ and $\mu N\to e
N$. On the other hand, a one loop Higgs exchange leads to the LFV
radiative decays $\ell_i\to \ell_j\gamma$.  In the following, we
report the expression for the branching ratios of the above processes.

\subsubsubsection{$\ell_i\to\ell_j\gamma$}
The $\ell_i\to\ell_j\gamma$ process can be generated by the one loop exchange of Higgs and leptons.
However, the dipole transition implies three chirality flips: two in the Yukawa vertices and one in the lepton propagator.
This strong suppression can be overcome at higher order level. Going to two loop level, one has to pay the typical price of $g^2/16\pi^2$
but one can replace the light fermion masses from Yukawa vertices with the heavy fermion (boson) masses circulating in the second loop.
In this case, the virtual Higgs boson couple only once to the lepton line, inducing the needed chirality flip.
As a result, the two loop amplitude can provide the major effects. Naively, the ratio between the two loop fermionic amplitude and the 
one loop amplitude is:
$$
\frac{A^{(2-loop)_{f}}_{l_i\to l_j\gamma}}{A^{1-loop}_{l_i\to l_j\gamma}}
\sim
\frac{\alpha_{em}}{4\pi}\frac{m^2_{f}}{m^2_{l_i}}\log\bigg(\frac{m^2_{f}}{m^2_{H}}\bigg),
$$ where $m_{f}= m_{b}, m_{\tau}$ is the mass of the heavy fermion
circulating in the loop. We remind that in a Model II 2HDM (as SUSY)
the Yukawa couplings between neutral Higgs bosons and quarks are
$H\bar{t} t \sim m_t/\tan\beta$ and $H\bar{b} b \sim m_b
\rm{\tan\beta}$.  Since the Higgs mediated LFV is relevant only at
large $\tan\beta\geq 30$, it is clear that the main contributions
arise from the $\tau$ and $b$ fermions and not from the top quark. So,
in this framework, $\tau\to l_j\gamma$ does not receive sizable two loop
effects by heavy fermionic loops, contrary to the $\mu\to e\gamma$
case.

However, the situation can drastically change when a $W$ boson circulates in the two loop Barr-Zee diagrams.
Bearing in mind that $H W^{+}W^{-}\sim m_W$ and that pseudoscalar bosons do not couple to a $W$ pair, it turns out that
$A^{(2-loop)_{W}}_{l_i\to l_j\gamma}/A^{(2-loop)_{f}}_{l_i\to l_j\gamma} \sim m^2_{W}/(m^2_{f} tan\beta )$ thus, two loop $W$ effects
are expected to dominate, as it is confirmed numerically \cite{Chang:1993kw,Paradisi:2005tk}.

As final result, the following approximate expression holds \cite{Paradisi:2005tk,Paradisi:2006jp}:
\begin{eqnarray}\underline{}
\frac{B(\ell_i\to \ell_j\gamma)}{B(\ell_i\to \ell_j\bar{\nu_j}\nu_{\tau})} &\simeq&
\frac{3}{2}\frac{\alpha_{el}}{\pi}\left(\frac{m^2_{\ell_i}}{m^2_A}\right)^2 t^{6}_{\beta}\Delta^2_{ij}\, \bigg\{\,
\frac{\delta m}{m_{A}}\log{\frac{m^2_{\ell_i}}{m^{2}_{A}}}+\frac{1}{6}+ \nonumber\\\nonumber\\
&+& \frac{\alpha_{el}}{\pi} \bigg[ \frac{m^{2}_W}{m^2_{\ell_i}}\frac{F(a_{W})}{t_{\beta}}-\sum_{f= b,\tau}\!\!N_f q^2_{f}
\frac{m^2_{f}}{m^2_{\ell_i}}\left(\log{\frac{m^2_{f}}{m^2_{\ell_i}}}+2\right)+\nonumber\\\nonumber\\
&-& \frac{N_c}{4} \bigg( q^{2}_{\tilde{t}}\,\frac{m_{t}\mu}{t_{\beta}m^2_{\ell_i}}\, s_{2\theta_{\tilde{t}}}\,
h(x_{\tilde{t}H})-q^{2}_{\tilde{b}}\,\frac{m_{b}A_{b}}{m^2_{\ell_i}}\, s_{2\theta_{\tilde{b}}}\,h(x_{\tilde{b}H}) \bigg)\bigg] \bigg\}^2
\nonumber\\\nonumber\\
&\simeq& \frac{3}{2}\,\frac{\alpha^3_{el}}{\pi^3}\, \Delta_{21}^{2}\,t^{4}_{\beta}\, \bigg(\frac{m^{4}_W}{M^{4}_{H}}\bigg)\,\bigg(F(a_{W})\bigg)^{2}\,,
\label{main}
\end{eqnarray}
where $\delta m=(m_{H}-m_{A})\sim\mathcal{O}(m^2_Z/m_{A^0})$. The
terms of the first row of Eq.~(\ref{main}) refer to one loop effects
and their role is non-negligible only in $\tau$ decays. It turns out
that pseudoscalar and scalar one loop amplitudes have opposite signs
so, being $m_A\simeq m_H$, they cancel each other to a very large
extent. Since these cancellations occur, two loop effects can become
important or even dominant. The two terms of the second row of
Eq.~(\ref{main}) refer to two loop Barr-Zee effects induced by $W$ and
fermionic loops, respectively, while the last row of Eq.~(\ref{main})
is relative two loop Barr-Zee effects with a squark loop in the second
loop. As regards the squark loop effects, it is very easy to realize
that they are negligible compared to $W$ effects.  In fact, it is well
known that Higgs mediated LFV can play a relevant or even a dominant
role compared to gaugino mediated LFV provided that slepton masses are
not below the TeV scale while maintaining the Higgs masses at the
electroweak scale (and assuming large $t_{\beta}$ values).  In this
context, it is natural to assume squark masses at least of the same
order as the slepton masses (at the TeV scale).  So, in the limit
where $x_{\tilde{f}H} = m_{\tilde{f}}^2/m_H^2 \gg 1$, the loop
function $h(x_{\tilde{f}H})$ is such that $(\log x_{\tilde{f}H}
+5/3)/6 x_{\tilde{f}H}$ thus, even for maximum squark mixing angles
$\theta_{\tilde{t},\tilde{b}}$, namely for
$s_{2\theta_{\tilde{t},\tilde{b}}}=\sin2\theta_{\tilde{t},\tilde{b}}\simeq
1$, and large $A_{b}$ and $\mu$ terms, two loop squark effects remain
much below the $W$ effects, as it is straightforward to check by
Eq.~(\ref{main}).

As a final result the main two loop effects are provided by the exchange of a $W$ boson, with the loop function $F(a_{W})\sim \frac{35}{16}(\log a_{W})^{2}$
for $a_{W}=m^{2}_{W}/m^{2}_{H}\ll 1$.
It is noteworthy that one and two loop amplitudes have the same signs. In addition, two loops effects dominate in large portions of the parameter space,
specially for large $m_H$ values, where the mass splitting $\delta m=m_{H}-m_{A}$ decreases to zero.

\begin{figure}[h]
\includegraphics[scale=0.35]{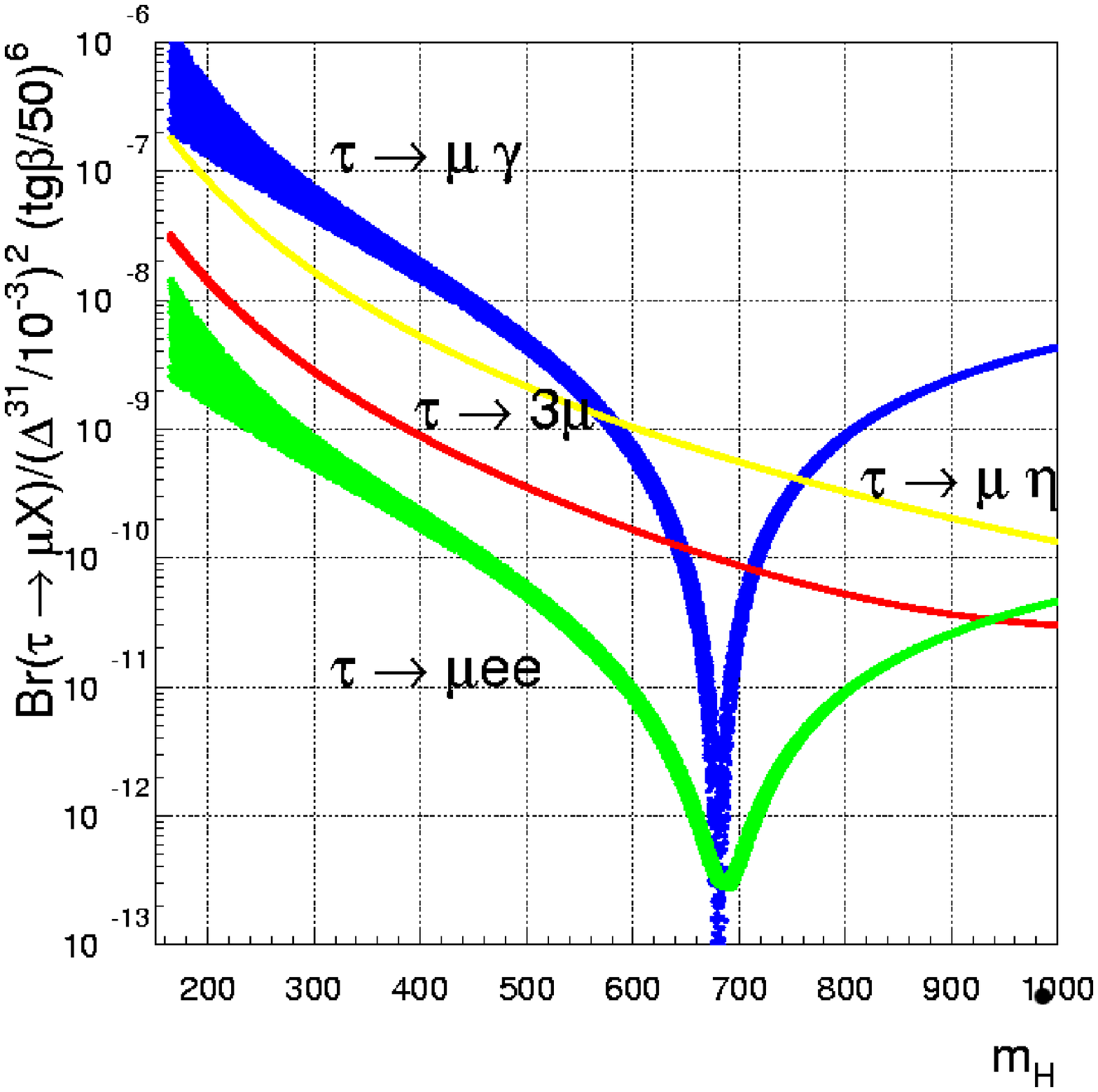} \hspace*{\fill}
\includegraphics[scale=0.35]{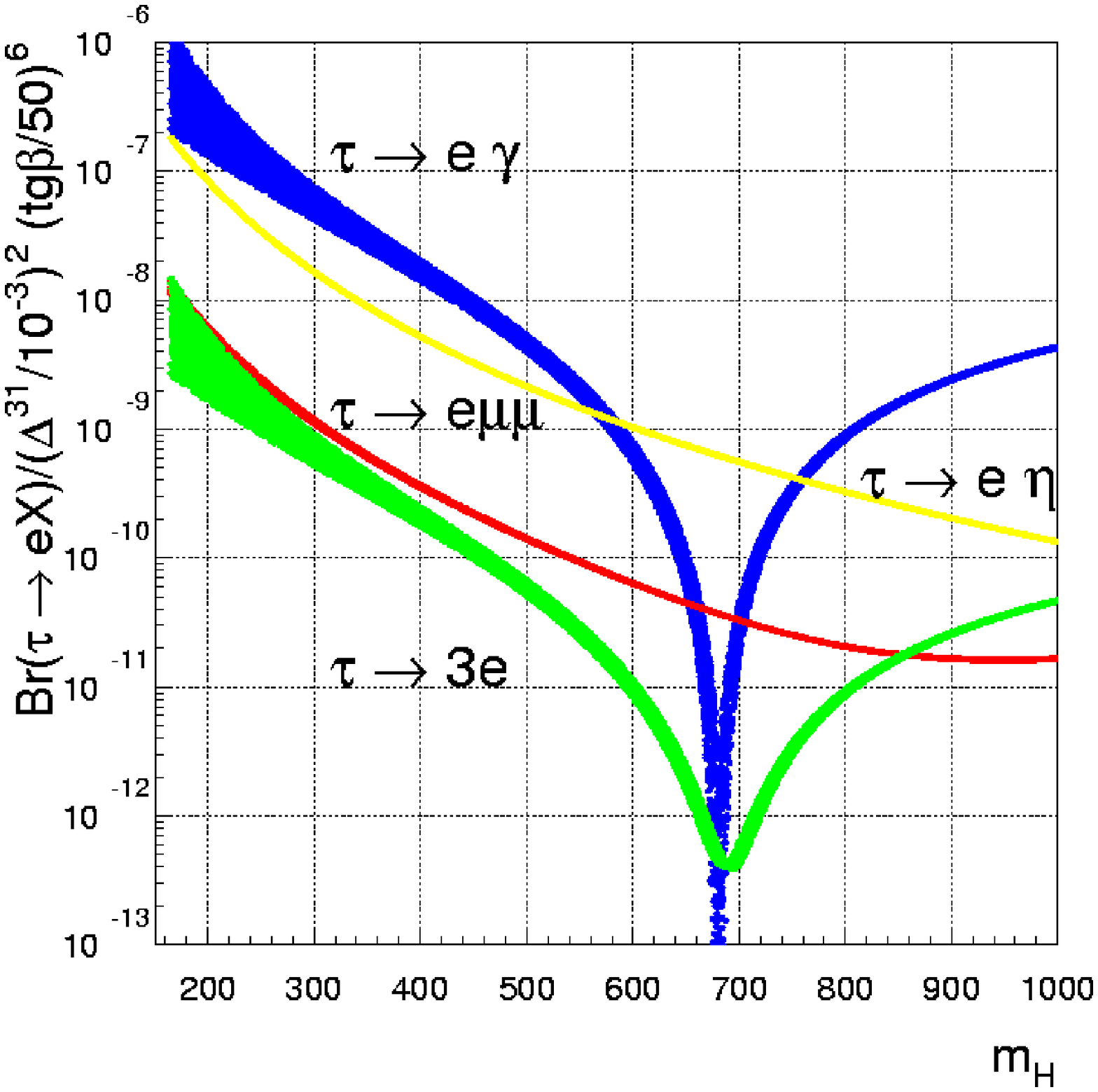}
\caption{Branching ratios of various $\tau\to\mu$ and $\tau\to e$ LFV processes vs the Higgs boson
mass $m_H$ in the decoupling limit as reported in \cite{Paradisi:2005tk}. $X= \gamma, \mu\mu, ee, \eta$.  \label{lfv1}}

\includegraphics[scale=0.35]{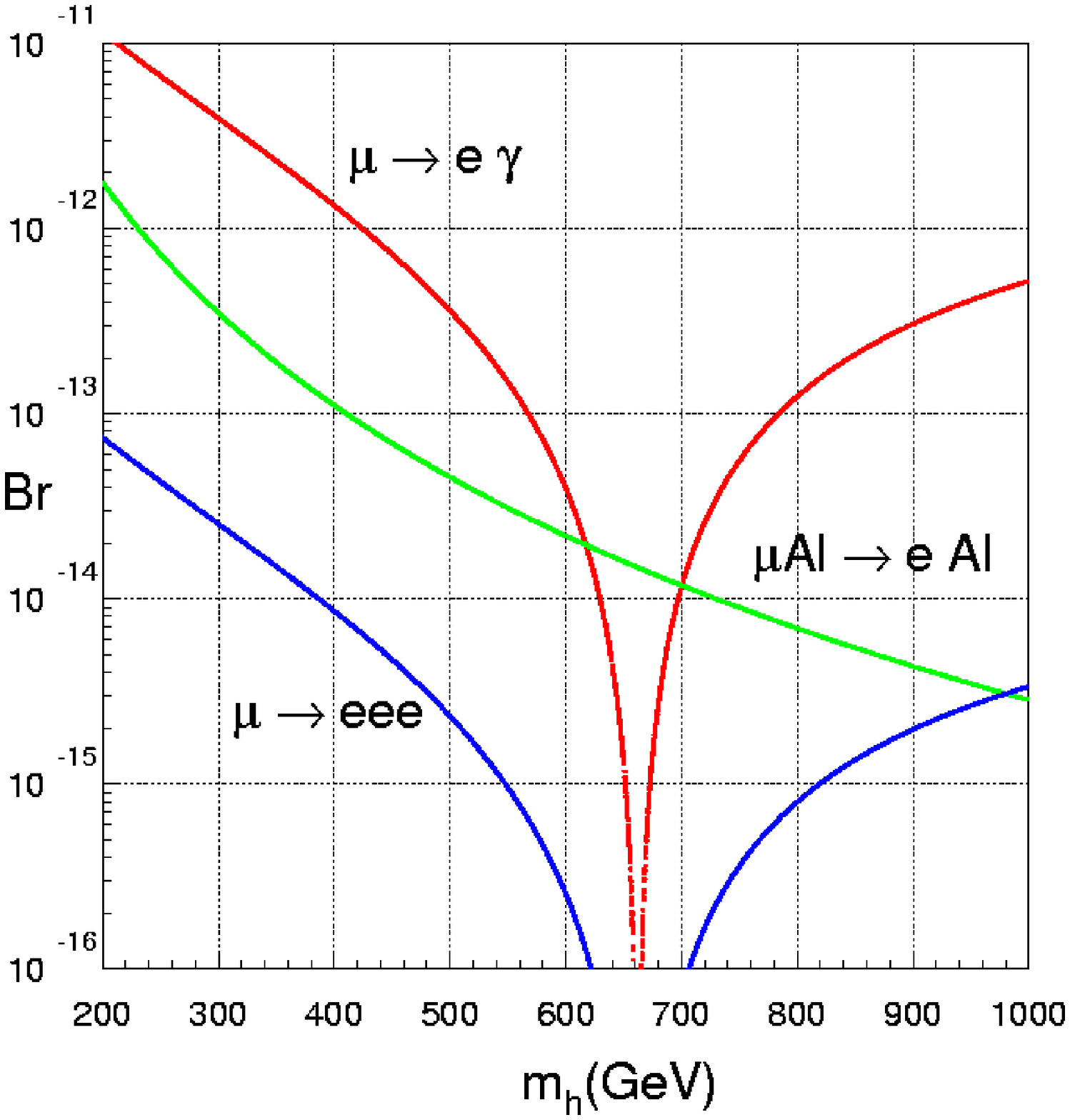}  \hspace*{\fill}
\includegraphics[scale=0.35]{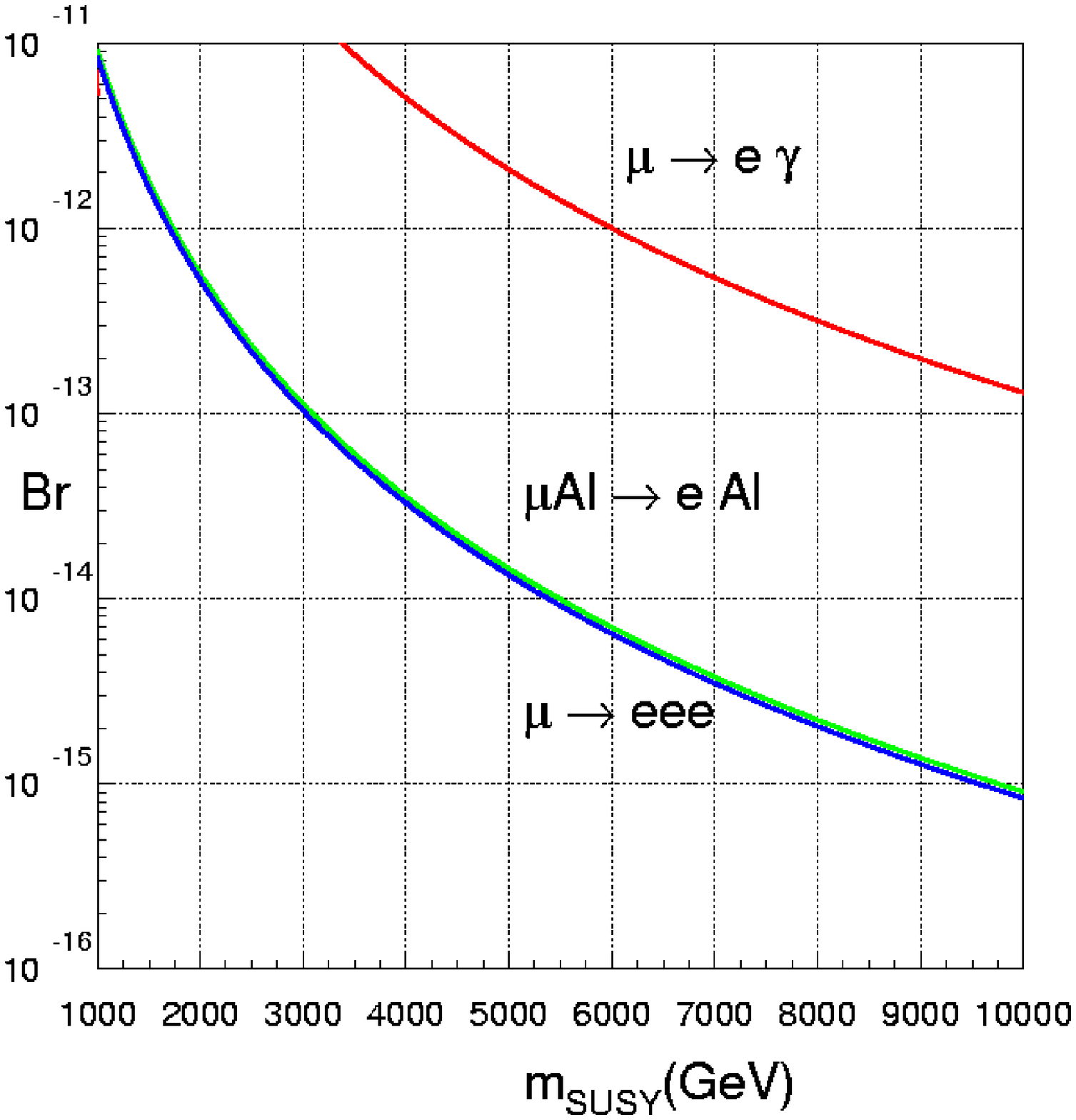}
\caption{Left: Branching ratios of $\mu\to e\gamma$, $\mu\to eee$ and $\mu Al\to e Al$ 
in the Higgs mediated LFV case vs the Higgs boson mass $m_h$ \cite{Paradisi:2006jp}. Right: Branching ratios of $\mu\to e\gamma$, 
$\mu\to eee$ and $\mu Al\to e Al$ in the gaugino mediated LFV case vs a common SUSY mass $m_{SUSY}$ \cite{Paradisi:2006jp}.
In the figure we set $t_{\beta}=50$ and $\delta^{21}_{LL}=10^{-2}$. \label{lfv2}}
\end{figure}

\subsubsubsection{ $\ell_i\to\ell_j\ell_k\ell_k$}
The $l_i\to l_jl_kl_k$ process can be mediated by a tree level Higgs exchange \cite{Babu:2002et,Dedes:2002rh}.
However, up to one loop level, $l_i\to l_jl_kl_k$ gets additional contributions induced by $l_i\to l_j\gamma^*$ amplitudes
\cite{Paradisi:2005tk,Paradisi:2006jp}. It is worth noting that the Higgs mediated monopole (chirality conserving)
and dipole (chirality violating) amplitudes have the same $\tan^3\beta$ dependence. This has to be contrasted to the non-Higgs contributions.
For instance, within SUSY, the gaugino mediated dipole amplitude is proportional to $\tan\beta$ while the monopole amplitude is $\tan\beta$ independent.
The expression for the Higgs mediated $l_i\to l_jl_kl_k$ can be approximated in the following way \cite{Paradisi:2005tk,Paradisi:2006jp}:
\begin{eqnarray}
\frac{B(\tau\to l_jl_kl_k)}{B(\tau\to l_j\bar{\nu_j}\nu_{\tau})} &\simeq& \frac{m^2_{\tau}m^2_{l_k}}{32m^4_{A}}\Delta_{\tau j}^2\tan^6\beta
\bigg[3+5\delta_{jk}\bigg]+\nonumber\\\nonumber\\
&+&\frac{\alpha_{el}}{3\pi} \bigg(\log\frac{m^2_{\tau}}{m^2_{l_k}}\!-\!3\bigg) \frac{B(\tau\to l_j\gamma)}{B(\tau\to l_j\bar{\nu_j}\nu_{\tau})}\,,
\end{eqnarray}
where we have disregarded subleading monopole effects.

\subsubsubsection{ $\mu N\to e N$}
The $\mu\to e $ conversion in Nuclei process can be generated by a scalar operator through the tree level Higgs exchange \cite{Kitano:2003wn}.
Moreover, at one loop level, additional contributions induced by $l_i\to l_j\gamma^*$ amplitudes arise \cite{Paradisi:2006jp};
however they are subleading \cite{Paradisi:2006jp}. Finally, the following expression for $B(\mu Al\to e Al)$ is derived \cite{Kitano:2003wn}:
\begin{eqnarray}
B(\mu Al\to e Al) &\simeq& \,1.8\times10^{-4}\, \frac{m^{7}_{\mu}m^{2}_{p}}{v^{4}m^{4}_{h}\omega^{Al}_{capt}}\, \Delta^{2}_{21}t_{\beta}^{6}\,,
\end{eqnarray}
where $\omega^{Al}_{capt}\simeq 0.7054\cdot 10^{6}sec^{-1}$. We observe that $B(\mu \to 3e)$ is completely dominated by
the photonic $\mu\to e\gamma^{*}$ dipole amplitude so that $B(\mu\to eee)\simeq \alpha_{em} B(\mu\to e\gamma)$.
On the other hand, tree level Higgs mediated contributions are negligible because suppressed by the electron mass through the $H(A)\bar{e}e \sim m_e$ coupling.
On the contrary, $\mu N\to e N$ is not suppressed by the light constituent quark $m_u$ and $m_d$ but only by the nucleon masses, because the Higgs-boson
coupling to the nucleon is shown to be characterized by the nucleon mass using the conformal anomaly relation \cite{Kitano:2003wn}.
In particular, the most important contribution turns out to come from the exchange of the scalar Higgs boson $H$ which couples to the strange quark \cite{Kitano:2003wn}.

In fact, the coherent $\mu-e$ conversion process, where the initial and final nuclei are in the ground state, is expected to be enhanced by a factor of $O(Z)$
(where $Z$ is the atomic number) compared to incoherent transition processes. Since the initial and final states are the same, the elements
$\langle N | \bar{p} p | N \rangle$ and $\langle N | \bar{n} n | N \rangle$ are nothing but the proton and the neutron densities in a nucleus in the non-relativistic
limit of nucleons. In this limit, the other matrix elements $\langle N | \bar{p} \gamma_5 p | N \rangle$ and $\langle N | \bar{n} \gamma_5 n | N \rangle$ vanish.
Therefore, in the coherent $\mu-e$ conversion process, the dominant contributions come from the exchange of $H$, not $A$ \cite{Kitano:2003wn}.

Moreover, we know that $\mu\to e\gamma^{*}$ (chirality conserving) monopole amplitudes are generally subdominant compared to
(chirality flipping) dipole effects \cite{Paradisi:2006jp}. Note also that, the enhancement mechanism induced by Barr-Zee type
diagrams is effective only for chirality flipping operators so, in the following, we will disregard chirality conserving one loop effects.

\subsubsubsection{ $\tau \to  \mu  P$ $(P=\pi, \eta, \eta')$}
Now we consider the implications of virtual Higgs exchange for the decays $\tau \to \mu P$, where $P$ is a neutral pseudoscalar 
meson ($P=\pi, \eta, \eta')$ \cite{Sher:2002ew,Brignole:2004ah}. Since we assume CP conservation in the Higgs sector, only 
the exchange of the $A$ Higgs boson is relevant. Moreover, in the large $\tanb$ limit, only the $A$ couplings to 
down-type quarks are important. These can be written as:
\begin{equation}
-i (\sqrt{2} G_F)^{1/2}\tanb~ A(\xi_d m_d \bar{d}_R d_L + \xi_s m_s \bar{s}_R s_L +\xi_b m_b \bar{b}_R b_L) + {\rm h.c.} 
\end{equation}
The  parameters $\xi_d, \xi_s, \xi_b$ are equal to one at tree level, but can significantly deviate from this value because of higher order 
corrections proportional to $\tanb$ \cite{Babu:1999hn,D'Ambrosio:2002ex}, generated by integrating out superpartners.
In the limit of quark flavour conservation, each $\xi_q$ $(q=d,s,b)$ has the form $\xi_q = (1 +\Delta_q \tanb)^{-1}$, where $\Delta_q$ appears in
the loop-generated term $-h_q \Delta_q H^{0 *}_2 q^c q +{\rm h.c.}$ \cite{Babu:1999hn,D'Ambrosio:2002ex}.
At energies below the bottom mass, the $b$-quark can be integrated out so
 the bilinear $-i m_b b^c b + {\rm h.c.}$ 
is effectively replaced by the gluon operator $\Omega = \frac{g^2_s}{64 \pi^2} \epsilon^{\mu \nu \rho \sigma} G^a_{\mu \nu}
G^a_{\rho \sigma}$, where $g_s$ and  $G^a_{\mu \nu}$ are the $SU(3)_C$ coupling constant and field strength, respectively \cite{Brignole:2004ah}.
In the limit in which the processes $\tau \to 3\mu$ and $\tau \to \mu \eta$ are both dominated by Higgs-exchange, these decays are related as
\cite{Brignole:2004ah}:
$$
\frac{B(\tau\to l_j\eta)}{B(\tau\to l_j\bar{\nu_j}\nu_{\tau})} \simeq 9\pi^2\!\left(\frac{f^{8}_{\eta} m^{2}_{\eta}}{m^{2}_{A}m_{\tau}}\right)^{\!2}\!
\!\left(1\!-\!\frac{m^{2}_{\eta}}{m^{2}_{\tau}}\right)^{\!2}\! \!\left[\xi_s\!+\!\frac{\xi_b}{3}\!\left(1\!+ \!\sqrt2\frac{f^{0}_{\eta}}{f^{8}_{\eta}}\right)\right]^2
\!\!\!\Delta_{3j}^2\tan^6\beta,
$$
where $m^{2}_{\eta}/m^{2}_{\tau}\simeq 9.5\times10^{-2}$ and the relevant decay constants are $f^{0}_{\eta}\sim 0.2 f_{\pi}$,
$f^{8}_{\eta}\sim 1.2 f_{\pi}$ and $f_{\pi}\sim 92$ MeV. In the above expression, both the contribution of the 
(bottom-loop induced) gluon operator $\Omega$ and the factors $\xi_q$ were included.

For $\xi_s \sim \xi_b \sim 1$, it turns out that $B(\tau^-\to\mu^-\eta)/B(\tau^-\to\mu^-\mu^+\mu^-)\simeq 5$,
but it could also be a few times larger or smaller than that, depending on the actual values of $\xi_s,\xi_b$.
Finally, let us compare $\tau \to \mu \eta'$ and  $\tau \to \mu \pi$ with $\tau \to \mu \eta$ in the limit of  Higgs-exchange domination.
Both ratios are suppressed, although for different reasons. The ratio $B(\tau \to \mu \pi)/B(\tau \to \mu \eta)$ 
is small because it is parametrically suppressed by  $m^4_\pi/m^4_\eta \sim 10^{-2}$. 
The ratio $B(\tau \to \mu \eta')/B(\tau \to \mu \eta)$, which seems to be ${\cal O}(1)$, is much smaller because the singlet and octet 
contributions to $\tau \to \mu \eta'$ tend to cancel against each other \cite{Brignole:2004ah}.

These results, combined with the present bound on $\tau \to\mu\eta$, imply that the Higgs mediated contribution to 
$B(\tau \to\mu\eta')$ and $B(\tau \to\mu\pi)$ can reach ${\cal O}(10^{-9})$ \cite{Brignole:2004ah}.

\subsubsubsection{ Higgs $\to \mu \tau $}
The LFV Higgs{\bf $\to \mu \tau $} decays and the related phenomenology have been extensively investigated in \cite{Brignole:2003iv}.
Concerning the Higgs boson decays, we have \cite{Brignole:2003iv}
\begin{equation}
{ B(A\to \mu^+\tau^-)} =  \tan^2\beta~ (|\Delta_L|^2 + |\Delta_R|^2)  { B(A\to \tau^+\tau^-)}  \, ,\label{rphi}
\end{equation}
where we have approximated $1/{\rm c}^2_\beta \simeq \tan^2\beta$ since non-negligible effects can only arise in the 
large $\tan\beta$ limit. If $A$ is replaced with $H$ [or $h$] in Eq.~(\ref{rphi}), the r.h.s. should also be multiplied by a factor 
$({\rm c}_{\beta-\alpha}/{\rm s}_\alpha)^2$ [or $({\rm s}_{\beta-\alpha}/{\rm c}_\alpha)^2$]. We recall that $B(A\to \mu \tau)$ can reach values of order 
$10^{-4}$. The same holds for the `non-standard' CP-even Higgs boson (either $H$ or $h$, depending on $m_A$).

We now  make contact with the physical observable, i.e. the $ B(\Phi^0\to \mu^+\tau^-)$, and discuss the phenomenological implications.
We outline some  general features of $B(\Phi^0\to \mu^+\tau^-)$ at large $\tanb$ and the prospects  for these decay channels at the Large Hadron Collider (LHC)
and other colliders. Let us discuss the different Higgs bosons, as reported in \cite{Brignole:2003iv},
assuming for definiteness $\tan\beta\sim 50$, $|50\Delta|^2 \sim 10^{-3}$ ($\Delta = \Delta_L$ or $\Delta_R$) and an integrated luminosity of $100 ~{\rm fb}^{-1}$ at LHC.

If $\Phi^0$ denotes one of the `non-standard' Higgs bosons, we have $C_\Phi \simeq 1$ and $ B(\Phi^0\to \tau^+\tau^-) \sim 10^{-1}$, so
$B(\Phi^0\to \mu^+\tau^-) \sim 10^{-4}$. The main production mechanisms at LHC are bottom-loop mediated gluon fusion
and associated production with $b \bar{b}$, which yield cross sections $\sigma\sim  (10^3, 10^2, 20)~{\rm pb}$ for $m_A \sim(100,200,300)~{\rm GeV}$,
respectively. The corresponding numbers of $\Phi^0\to \mu^+\tau^-$ events are about $(10^4,10^3, 2\cdot 10^2)$. These estimates do not change much if
the bottom Yukawa coupling $Y_b$ is enhanced (suppressed) by radiative corrections, since in this case the enhancement (suppression) 
of $\sigma$ would be roughly compensated by the suppression (enhancement) of $B(\Phi^0\to \mu^+\tau^-)$.

If $\Phi^0$ denotes  the other (more `Standard Model-like') Higgs boson, the factor  $C_\Phi\cdot B(\Phi^0\to \tau^+\tau^-)$  strongly depends on $m_A$, 
while the production cross section at LHC, which is dominated by top-loop mediated gluon fusion, is $\sigma \sim 30~{\rm pb}$.
For  $m_A\sim 100~{\rm GeV}$ we may have $C_\Phi\cdot B(\Phi^0\to \tau^+\tau^-)\sim 10^{-1}$ and $B(\Phi^0\to \mu^+\tau^-) \sim 10^{-4}$, which  
would imply $\sim 300$ $ \mu^+\tau^-$ events. The number of events is generically smaller for large $m_A$ since  $C_\Phi$ scales as $ 1/m_A^4$, consistently 
with the expected decoupling of LFV effects for such a Higgs boson.

The above discussion suggests  that LHC may offer good chances to detect the decays $\Phi^0\to \mu\tau$, especially in the case of non-standard Higgs bosons.
This indication should be supported by a detailed study of the background. At Tevatron the sensitivity is lower  than at LHC because 
both  the expected luminosity and  the Higgs production cross sections are smaller. The number of events would be smaller by a factor $10^2 - 10^3$. 
A few events may be expected also at $e^+ e^-$ or $\mu^+ \mu^-$ future  colliders, assuming integrated luminosities of $500~{\rm fb}^{-1}$ and $1~{\rm fb}^{-1}$, 
respectively. At a $\mu^+ \mu^-$  collider an enhancement  may occur for the non-standard Higgs bosons if radiative corrections strongly suppress $Y_b$, 
since in this case both the resonant production cross section [$\sigma \sim (4\pi/m^2_A) B( \Phi^0 \to \mu^+\mu^-)$] and the 
LFV branching ratios $B(\Phi^0\to \mu^+\tau^-)$ would be enhanced. As a result, for light $m_A$, hundreds of  $\mu^+\tau^-$ events could occur. 

\subsubsubsection{ $\mu N\to\tau X$}
Higgs mediated LFV effects can have also relevant impact on the cross section of the $\mu N\to\tau X$ reaction \cite{Kanemura:2004jt}.
The contribution of the Higgs boson mediation to the differential cross section $\mu^- N \to \tau^-  X$ is given by \cite{Kanemura:2004jt}
\begin{equation}
\frac{{\rm d}^{2} \sigma}{{\rm d} x {\rm d} y}= \sum_{q} x f_{q} (x) \left\{ \left|{\mathcal{C}_L} \right|^{2}_{q}  \left(\frac{1-\mathcal{P}_\mu}{2}\right)+
\left|{\mathcal{C}_R} \right|^{2}_{q} \left(\frac{1+\mathcal{P}_\mu}{2} \right) \right\} \frac{s}{8 \pi} y^{2}, \label{cross-Higgs}
\end{equation}
where the function $f_{q}(x)$ is the PDF for $q$-quarks, $\mathcal{P}_\mu$ is the incident muon polarization such that $\mathcal{P}_\mu=+1$ and $-1$ correspond to the 
right- and left-handed polarization, respectively, and $s$ is the center-of-mass (CM) energy. The parameters $x$ and $y$ are defined as
$x\equiv Q^{2}/2 P \cdot q$, $y\equiv 2 P \cdot q/s$, in the limit of massless tau leptons, where $P$ is the four momentum of the target, $q$ is the momentum
transfer, and $Q$ is defined as $Q^{2} \equiv -q^{2}$. As seen in Eq.~(\ref{cross-Higgs}), experimentally, the form factors of ${\mathcal{C}^{hH}_L}$ and
${\mathcal{C}^{A}_L}$ (${\mathcal{C}^{hH}_R}$ and ${\mathcal{C}^{A}_R}$) can be selectively studied by using purely left-handed (right-handed) incident muons.
In SUSY models such as the MSSM with heavy right-handed neutrinos, LFV is radiatively induced due to the left-handed 
slepton mixing, which only affects $\mathcal{C}_L^{hH}$ and $\mathcal{C}_L^{A}$. Therefore, in the following, we focus only on those $\mathcal{C}_L^{hH}$ 
and $\mathcal{C}_L^{A}$ couplings.

The magnitudes of the effective couplings are constrained by the current experimental results of searches for LFV processes of tau decays.
Therefore, both couplings are determined by the one that is more constrained, namely the pseudo-scalar coupling. It is constrained by the $\tau\to\mu\eta$
decay  ($B(\tau \to \mu \eta) < 3.4 \times 10^{-7}$). Then the constraint is given on the $s$-associated scalar and pseudo-scalar couplings by
\begin{equation}
(\left|{\mathcal{C}^{A}_L}\right|^{2})_{s}\leq 10^{-9} [\rm{GeV}^{-4}] \times \rm{B}(\tau \to \mu \eta).\label{eq:exp-limit-on-CA}
\end{equation} 
The largest values of ${\mathcal{C}^{hH}_L}$ and ${\mathcal{C}^{A}_L}$ can be realized with $m_{\rm{SUSY}} \sim \mathcal{O}(1)$ TeV and 
the Higgsino mass $\mu \sim \mathcal{O}(10)$ TeV~\cite{Sher:2002ew,Brignole:2004ah}.

\begin{figure}
\centering
\includegraphics[width=6cm]{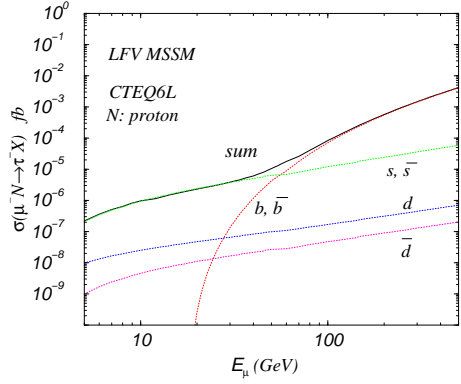}
\caption{Cross section of the $\mu^- N \to \tau^- X$ DIS process as a function of the muon energy for the Higgs mediated interaction \cite{Kanemura:2004jt}.
It is assumed that the initial muons are purely left-handed. CTEQ6L is used for the PDF.}\label{Fig:total-cross-section-vs-Emu}
\end{figure}

The cross sections of the $\mu N\to\tau X$ reaction in the DIS region is evaluated for the maximally allowed values of the effective couplings as a reference.
They are  plotted in Fig.~\ref{Fig:total-cross-section-vs-Emu} for different quark contributions as a function of the 
muon beam energy in the laboratory frame. For the PDF, CTEQ6L has been used. The target $N$ is assumed to be a proton. 
For a nucleus target, the cross section would be higher, approximately by the number of nucleons in the target.
The cross section sharply increases above $E_{\mu} \sim 50$ GeV in Fig.~\ref{Fig:total-cross-section-vs-Emu}.
This enhancement comes from  the $b$-quark contribution in addition to the $d$ and $s$-quark contributions which is enhanced by a factor of $m_{b}/m_{s}$ over the $s$-quark contribution.
The cross section is enhanced by one order of magnitude when the muon energy changes from 50 GeV to 100 GeV. Typically, 
for $E_{\mu} = 100$ GeV and $E_{\mu} = 300$ GeV, the cross section is $10^{-4}$ fb and $10^{-3}$ fb, respectively.
With the intensity of $10^{20}$ muons per year and the target mass of 100 g/cm$^2$, about $10^4$ ($10^2$) events could be
expected for $\sigma(\mu N\to \tau X)=10^{-3} ~(10^{-5})$ fb, which corresponds to $E_{\mu}=300$ $(50)$ GeV 
from Fig.~\ref{Fig:total-cross-section-vs-Emu}. This would provide good potential to improve the sensitivity by four (two)
orders of magnitude from the present limit from $\tau\to\mu\eta$ decay, respectively. 
Such a muon intensity could be available at a future muon collider and a neutrino factory.

\subsubsection{Correlations}

The numerical results shown in Fig.~\ref{lfv1} and Fig.~\ref{lfv2}
allow us to draw several observations
\cite{Paradisi:2005tk,Paradisi:2006jp}:
\begin{itemize}
\item
$\tau\to l_j\gamma$ has the largest branching ratios except for a
region around $m_H\sim$~700~GeV where strong cancellations among
two loop effects reduce their size \footnote{For a detailed discussion
about the origin of these cancellations and their connection with
non-decoupling properties of two loop $W$ amplitude, see
Ref.\cite{Chang:1993kw}.}. The following approximate relations are
found:
$$
\frac{B(\tau\to l_j\gamma)}{B(\tau\to l_j\eta)} \simeq
\left(\frac{\delta
  m}{m_A}\log\frac{m^2_{\tau}}{m^2_A}\!+\!\frac{1}{6}\!+\! 
\frac{\alpha_{el}}{\pi}\,\bigg(\frac{m^{2}_W}{m^{2}_{\tau}}\bigg)\,
\frac{F(a_{W})}{\tan\beta}\right)^2\geq 1\,,  
$$ where the last relation is easily obtained by using the
approximation for $F(z).$ If two loop effects were disregarded, then
we would obtain 
$B(\tau\to l_j\gamma)/B(\tau\to l_j\eta)\in(1/36,1)$ for $\delta m/m_A
\in(0,10\%)$. 
Two loop contributions significantly enhance $B(\tau\to l_j\gamma)$
specially for $\delta m/m_A\to 0$. 
\item
In Fig.~\ref{lfv1} non negligible mass splitting $\delta m/m_{A}$
effects can be visible at low $m_H$ regime through the bands of the  
$\tau\to l_j\gamma$ and $\tau\to l_jee$ processes. These effects tend
to vanish with increasing $m_H$ as it is correctly  
reproduced in  Fig.~\ref{lfv1} $\tau\to l_j\mu\mu$ does not receive
visible effects by $\delta m/m_{A}$ terms being dominated by the tree
level Higgs exchange. 
\item
As it is shown in Fig.~\ref{lfv1} $B(\tau\to l_j\gamma)$ is generally
larger than $B(\tau\to l_j\mu\mu)$; 
their ratio is regulated by the following approximate relation:
$$
\frac{B(\tau\to l_j\gamma)}{B(\tau\to l_j\mu\mu)} \simeq
\frac{36}{3\!+\!5\delta_{j\mu}} 
\frac{B(\tau\to l_j\gamma)}{B(\tau\to l_j\eta)}
\geq\frac{36}{3\!+\!5\delta_{j\mu}}\,, 
$$
where the last relation is valid only out of the cancellation
region. Moreover, from the above relation it turns out that: 
$$
\frac{B(\tau\to l_j\eta)}{B(\tau\to l_j\mu\mu)} \simeq
\frac{36}{3\!+\!5\delta_{j\mu}}\,. 
$$
If we relax the condition $\xi_{s,b}= 1$, $B(\tau\to l_j\eta)$ can get
values few times smaller or bigger than those in Fig.~\ref{lfv1}. 
\item
It is noteworthy that a tree level Higgs exchange predicts that $B(\tau\to l_jee)/B(\tau\to l_j\mu\mu)\sim m^2_e/m^2_{\mu}$ 
while, at two loop level, we obtain (out of the cancellation region):
$$ 
\frac{B(\tau\to l_j ee)}{B(\tau\to l_j\mu\mu)} \simeq \frac{0.4}{3\!+\!5\delta_{j\mu}}
\frac{B(\tau\to l_j\gamma)}{B(\tau\to l_j\eta)} \geq\frac{0.4}{3\!+\!5\delta_{j\mu}}\,.
$$
Let us underline that, in the cancellation region, the lower bound of $B(\tau\to l_jee)$ is given by the monopole contributions.
So, in this region, $B(\tau\to l_jee)$ is much less suppressed than $B(\tau\to l_j\gamma)$.
\item
The approximate relations among $\mu Al\to e Al$, $\mu\to e\gamma$ and $\mu\to eee$ branching ratios are
\begin{eqnarray}
\frac{B(\mu\to e \gamma)}{B(\mu Al\to e Al)} &\simeq& 10^{\,2}\,\left(\frac{F(a_W)}{\tan\beta}\right)^{2}\,\,,\,\,\,\,
\frac{B(\mu\to eee)}{B(\mu \to e \gamma)}\simeq\alpha_{el}.
\end{eqnarray}
 In the above equations we retained only dominant two loop effects
arising from $W$ exchange. The exact behavior for the examined processes is reported in Fig.~\ref{lfv2}
where we can see that $\mu\to e\gamma$ gets the largest branching ratio except for a region around $m_H\sim$~700~GeV where strong cancellations
among two loop effects sink its size.
\end{itemize}

The correlations among the rates of the above processes are an important signature of the Higgs-mediated LFV and allow us to discriminate between different SUSY scenarios.
In fact, it is well known that, in a Supersymmetric framework, besides the Higgs mediated LFV transitions, we have also LFV effects mediated by the gauginos through loops
of neutralinos (charginos)- charged sleptons (sneutrinos). On the other hand, the above contributions have different decoupling properties regulated by the mass of the
heaviest scalar mass ($m_H$) or by the heaviest mass in the slepton gaugino loops ($m_{SUSY}$). In principle, the $m_{SUSY}$ and $m_H$ masses can be unrelated so,
we can always proceed by considering only the Higgs mediated effects (assuming a relatively light $m_H$ and an heavy $m_{SUSY}$)
or only the gaugino mediated contributions (if $m_H$ is heavy). In the following, we are interested to make a comparison between Higgs and
gaugino mediated LFV effects. In order to make the comparison as simple as possible, let us consider the simple case where all the SUSY particles are degenerate. 
In this case, it turns out that
$$
\Delta^{21}_{L}\sim \frac{\alpha_{2}}{24\pi}\delta^{21}_{LL}\,,
$$
\begin{eqnarray}
\frac{B(\ell_i\to \ell_j\gamma)} {B(\ell_i\to \ell_j\bar{\nu}_j\nu_i)} \bigg|_{\rm{Gauge}} =
\frac{2\alpha_{el}}{75\pi}\bigg(1\!+\!\frac{5}{4}\tan^{2}\theta_{W}\bigg)^2 \bigg(\frac{m^{4}_{W}}{m^{4}_{SUSY}}\bigg) |\delta^{ij}_{LL}|^2 t^{2}_{\beta}\,, \nonumber
\end{eqnarray}
\begin{equation}
\frac{B(\ell_i\to \ell_j\gamma)} {B(\ell_i\to \ell_j\bar{\nu}_j\nu_i)}\bigg|_{\rm{Higgs}} \simeq 10\frac{\alpha^3_{el}}{\pi^3}
\bigg(\frac{\alpha_{2}}{24\pi}\bigg)^{2} \bigg(\frac{m^{4}_W}{M^{4}_{H}}\bigg)\, \bigg(\log\frac{m^{2}_{W}}{M^{2}_{H}}\bigg)^{4} |\delta^{ij}_{LL}|^2\,t^{4}_{\beta}\,.
\end{equation}
In Fig.~\ref{lfv2} we report the branching ratios of the examined processes as a function of the heaviest Higgs boson mass $m_H$ (in the Higgs LFV mediated case)
or of the common SUSY mass $m_{SUSY}$ (in the gaugino LFV mediated case). We set $t_{\beta}=50$ and we consider the PMNS scenario as discussed
above so that $(\delta^{21}_{LL})_{PMNS}\simeq 10^{-2}$. Sub-leading contributions proportional to $(\delta^{23}_{LL(RR)}\delta^{31}_{RR(LL)})_{PMNS}$
were neglected since, in the PMNS scenario, it turns out that $(\delta^{23}_{LL(RR)}\delta^{31}_{RR(LL)})_{PMNS}/(\delta^{21}_{LL})_{PMNS} \simeq 10^{-3}$
\cite{Masiero:2004js}. As we can see from Fig.~\ref{lfv2}, Higgs mediated effects start being competitive with the gaugino mediated ones when $m_{SUSY}$ is
roughly one order of magnitude larger then the Higgs mass $m_H$. Moreover, we stress that, both in the gaugino and in the Higgs mediated cases,
$\mu\to e\gamma$ gets the largest effects. In particular, within the PMNS scenario, it turns out that Higgs mediated $B(\mu\to e\gamma)\sim 10^{-11}$ 
when $m_{H}\sim$~200~GeV and $t_{\beta}=50$, that is just closed to the present experimental resolution.

The correlations among different processes predicted in the gaugino mediated case are different from those predicted in the Higgs mediated case.
For instance, in the gaugino mediated scenario, $B(\tau\to l_jl_kl_k)$ get the largest contributions by the dipole
amplitudes that are $\tan\beta$ enhanced with respect to all other amplitudes resulting in a precise ratio with $B(\tau\to l_j\gamma)$, namely
\begin{equation}
\frac{B(\ell_{i}\to \ell_{j}\ell_{k}\ell_{k})}{B(\ell_{i}\to \ell_{j}\gamma)}\bigg|_{Gauge}
\simeq \frac{\alpha_{el}}{3\pi}\left(\log\frac{m^2_{\tau}}{m^2_{l_{k}}}-3\right) \simeq \alpha_{el},
\end{equation}
\begin{equation}
\frac{B(\tau\to \ell_{j}ee)}{B(\tau\to \ell_{j}\mu\mu)}\bigg|_{Gauge} \simeq
\frac{\log\frac{m^2_{\tau}}{m^2_{e}}-3}{\log\frac{m^2_{\tau}}{m^2_{\mu}}-3} \simeq 5\,.
\end{equation}
Moreover, in the large $\tan\beta$ regime, one can find the simple theoretical relations
\begin{equation}
\label{relations}
\frac{B(\mu - e {\rm\ in \ Ti})}{B(\mu\!\to\!e\gamma)} \bigg|_{\rm{Gauge}} \simeq\! \alpha_{el}\,.
\end{equation}
If some ratios different from the above were discovered, then this would be clear evidence that some new process is generating the
$\ell_i\to l_j$ transition, with Higgs mediation being a leading candidate.

\subsubsection{Conclusions}

We have reviewed the allowed rates for Higgs-mediated LFV decays in a SUSY framework. In particular, we have analyzed the decay modes of the $\tau,\mu$ lepton, namely
$\ell_i\to \ell_j\ell_k\ell_k$, $\ell_i\to \ell_j\gamma$, $\tau\to l_j\eta$ and $\mu N\to e N$.
We have also discussed the LFV decay modes of the Higgs bosons $\Phi\to \ell_i\ell_j$ ($\Phi=h^{0},H^{0},A^{0}$) so as 
the impact of Higgs mediated LFV effects on the cross section of the $\mu N\to\tau X$ reaction. Analytical relations and correlations among the rates of the above
processes have been established at the two loop level in the Higgs Boson exchange. The correlations among the processes are a precise signature of the
theory. In this respect experimental improvements in all the decay channels of the $\tau$ lepton would be very welcome.
In conclusion, the Higgs-mediated contributions to LFV processes can be within the present or upcoming attained experimental resolutions and
provide an important opportunity to detect new physics beyond the Standard Model.

\subsection{Tests of unitarity and universality in the lepton sector}\label{sec:unitarity}

\subsubsection{Deviations from unitarity in the leptonic mixing matrix} 

The presence of physics beyond the SM in the leptonic sector can generate deviations from unitarity in the mixing
matrix. This is analogous to what happens in the quark sector, where the search for deviations from unitarity of the CKM matrix is
considered a sensitive way to look for new physics.

In the leptonic sector a clear example of non-unitarity is given by the seesaw 
mechanism~ \cite{Minkowski:1977sc,Yanagida:1979as,Gell-Mann:1980vs,Glashow:1979nm,Mohapatra:1979ia}.
 To generate naturally small
neutrino masses, new heavy particles -right-handed neutrinos- are added, singlet under the SM gauge group. Thus a Yukawa
coupling for neutrinos can be written, as well as Majorana masses for the new heavy fields. The mass matrix of the complete theory is now an
enlarged mass matrix ($5\times 5$ at least), whose diagonalization leads to small Majorana neutrino masses. The non-unitarity of the
$3\times 3$ leptonic mixing matrix can now be understood simply by observing that it is a sub-matrix of a bigger one which is unitary,
since the complete theory must conserve probabilities.

Another way to see this is looking at the effective theory we obtain once the heavy fields are integrated out. The unique dimension-five
operator is the well-known Weinberg operator~\cite{Weinberg:1979sa} which generates neutrino masses when the electroweak symmetry is
broken. Masses are naturally small since they are suppressed by the mass $M$ of the heavy particles which have been integrated out: $m_\nu
\sim v^2 / M$, where $v$ is the Higgs VEV. If we go on in the expansion in effective operators, we obtain only one dimension-six
operator which renormalizes the kinetic energy of neutrinos. Once we perform a field redefinition to go into a mass basis with canonical
kinetic terms, a non-unitary mixing matrix is obtained~\cite{Broncano:2002rw}. In minimal models deviations from
unitarity generated in this way are very suppressed, since the dimension-six operator is proportional to $v^2/M^2$. However, in more
sophisticated versions of this mechanism like double (or inverse) seesaw~\cite{Gonzalez-Garcia:1988rw} the suppression can be reduced
without affecting the smallness of neutrinos masses and avoiding any fine-tuning of Yukawa couplings. In terms of effective operators, this
means that it is possible to ``decouple'' the dimension-five operator from the dimension-six, permitting small neutrino masses and not so
small unitarity deviations.

Usually the elements of the leptonic mixing matrix are measured using neutrino oscillation experiments assuming unitarity. No information
can be extracted from electroweak decays on the individual matrix elements, due to the impossibility of detecting neutrino mass
eigenstates. This is quite different from the way of measuring the CKM matrix elements. Here oscillations are important too, but since quark
mass eigenstates can be tagged, direct measurements of the matrix elements can be made using electroweak decays.

The situation changes if we relax the hypothesis of unitarity of the leptonic mixing matrix. Electroweak decays acquire now an important
meaning, since they can be used to constrain deviations from unitarity. Consider as an example the decay $W \to l \, \bar{\nu}_l$.
The decay rate is modified as follows: $\Gamma = \Gamma_{SM} (NN^\dagger)_{ll}$, where $N$ is the non-unitary leptonic mixing
matrix and $\Gamma_{SM}$ is the SM decay rate. This, and other electroweak processes, can therefore be used to obtain
information on $(NN^\dagger)_{ll}$. Moreover { lepton flavour violating processes} like $\mu \to 3e$ or $\mu$-$e$ conversion in
nuclei can occur, while { rare lepton decays} like $l_i \to l_j \gamma$ can be enhanced, permitting to constrain the off-diagonal
elements of $(NN^\dagger)$. Finally, { universality violation effects} are produced, even if the couplings are universal: for
example the branching ratio of $\pi$ decay (see Section 6) is now proportional to $(NN^\dagger)_{ee}/(NN^\dagger)_{\mu\mu}$.

In Ref.~\cite{Antusch:2006vwa} all these processes have been considered, a global fit has been performed and the matrix
$|(NN^\dagger)|$ has been determined ($90\%$ C.L.):
\begin{equation}
|N N^\dagger | \approx
\left (
\begin{array}{ccc}
 0.994\pm  0.005   & < 7.0 \cdot 10^{-5}  &   < 1.6 \cdot 10^{-2}\\
 < 7.0 \cdot 10^{-5}   &  0.995 \pm  0.005 &  < 1.0 \cdot 10^{-2}  \\
 < 1.6 \cdot 10^{-2}   &  < 1.0 \cdot 10^{-2}  &  0.995 \pm 0.005
\end{array}
\right ) \, .
\end{equation}
Similar bounds can be inferred for $|N^\dagger N|$, leading to the conclusion that { deviations from unitarity in the leptonic mixing
matrix are experimentally constrained to be smaller than few percent}. Notice however that these bounds apply to a $3\times 3$
mixing matrix, i.e. they constrain deviations from unitarity induced by higher energy physics which has been integrated out\footnote{They
do not apply for instance to the case of light sterile neutrinos, where the low-energy mixing matrix is larger. Indeed in this case they
would be included in the sum over all light mass eigenstates contained inside $(NN^\dagger)_{ll}$ and unitarity would be restored.}.

However, since contrary the quark sector decays can only constrain the
elements of $|(NN^\dagger)|$, to determine the individual elements of
the leptonic mixing matrix oscillation experiments are needed. In
Ref.~\cite{Antusch:2006vwa} neutrino oscillation physics is
reconsidered in the case in which the mixing matrix is not
unitary. The main consequence of this is that the flavour basis is no
longer orthogonal, which gives rise to two physical effects:
\begin{itemize}
\item[-] ``zero-distance'' effect, i.e. flavour conversion in neutrino oscillations at $L=0$:
\\ $P_{\nu_\alpha \nu_\beta}(E, L=0) ·\propto |(NN^\dagger)_{\beta\alpha}|^2$;
\item[-]  non-diagonal matter effects.
\end{itemize}
With the new formulas for neutrino oscillations, a fit to present
oscillation experiments is performed, in order to determine the
individual matrix elements. As in the standard case, no information at
all is available on phases ($4$ or $6$, depending on the nature -Dirac
or Majorana- of neutrinos), since appearance experiments would be
needed. However the moduli of matrix elements can be determined, but
now they are all independent, so that the free parameters are $9$
instead of $3$.  The elements of the $e$-row can be constrained using
the data from CHOOZ~\cite{Apollonio:2002gd},
KamLAND~\cite{Araki:2004mb} and SNO~\cite{Aharmim:2005gt}, together
with the information on $\Delta m^2_{23}$ resulting from an analysis
of K2K~\cite{Ahn:2002up}. In contrast, less data are available for the
$\mu$-row: only those coming from K2K and
SuperKamiokande~\cite{Ashie:2005ik} on atmospheric neutrinos, and only
$|N_{\mu 3}|$ and the combination $|N_{\mu1}|^2+|N_{\mu2}|^2$ can be
determined. No information at all is available on the $\tau$-row. The
final result is the following ($3\sigma$ ranges):
\begin{eqnarray}
|N| = \left( \begin{array}{ccc} 
0.75-0.89 & 0.45-0.66 & <0.34 \\
\big[\,(|N_{\mu1}|^2+|N_{\mu2}|^2)^{1/2}= & 0.57-0.86\,\big] & 0.57-0.86 \\
? & ? & ?
\end{array} \right) \, .
\end{eqnarray}
Notice that, { without assuming unitarity, only half of the elements can be determined from oscillation experiments alone}. Adding
the information from near detectors at NOMAD~\cite{Astier:2001yj}, KARMEN~\cite{Eitel:2000by}, BUGEY~\cite{Declais:1994su} and
MINOS~\cite{Minos}, which put bounds on $|(NN^\dagger)_{\alpha\beta}|^2$ by measuring the ``zero-distance''
effect, the degeneracy in the $\mu$-row can be solved, but the $\tau$-row is still unknown.

In order to determine/constrain all the elements of the leptonic mixing matrix without assuming unitarity, data on oscillations must be
combined with data from decays. The final result is: 
\begin{eqnarray}
|N| = \left(\begin{array}{ccc} 
0.75-0.89 & 0.45-0.65 & <0.20 \\
0.19-0.55 & 0.42-0.74 & 0.57-0.82 \\
0.13-0.56 & 0.36-0.75 & 0.54-0.82
\end{array}
\right)\,,
\label{N_dec}
\end{eqnarray}
which can be compared to the one obtained with standard analysis~\cite{Gonzalez-Garcia:2004jd} where similar bounds are found.

It would be good to be able to determine the elements of the mixing matrix with oscillation experiments alone, permitting thus a
``direct'' test of unitarity. This would be for instance a way to detect light sterile neutrinos~\cite{Barger:2004db}. This could be
possible exploring the appearance channels for instance at future facilities under discussion, such as
Super-Beams~\cite{Itow:2001ee,Ayres:2004js,Gomez-Cadenas:2001eu,Campagne:2004wt}, $\beta$-Beams~\cite{Zucchelli:2002sa} and Neutrino
Factories~\cite{Geer:1997iz,DeRujula:1998hd}, where the $\tau$-row and phases could be measured. Moreover, near detectors at neutrino
factories could also improve the bounds on $(NN^\dagger)_{e\tau}$ and $(NN^\dagger)_{\mu\tau}$ by about one order of magnitude.  All this
information, coming from both decays and oscillation experiments, will be important not only to detect new physics, but even to discriminate
among different scenarios.

\subsubsection{Lepton universality}

High precision electroweak tests (HPET) represent a powerful tool to probe the SM and, hence, to constrain or obtain indirect 
hints of New Physics beyond it. A typical and relevant example of HPET is represented by the Lepton Universality (LU) breaking.
Kaon and pion physics are obvious grounds where to perform such tests, for instance in the well studied $\pi_{\ell2}$ ($\pi\rightarrow \ell\nu_{\ell}$)
and $K_{\ell2}$ ($K\rightarrow \ell\nu_{\ell}$) decays, where $l= e$ or $\mu$.

Unfortunately, the relevance of these single  decay channels in probing the SM is severely hindered by our theoretical uncertainties, 
which still remain at the percent level (in particular due to the uncertainties on non perturbative quantities like
$f_{\pi}$ and $f_{K}$). This is what prevents us from fully exploiting such decay modes in constraining new physics, 
in spite of the fact that it is possible to obtain non-SM contributions which exceed the high experimental precision which has been achieved on those modes.

On the other hand, in the ratios $R_{\pi}$ and $R_{K}$ of the electronic and muonic decay modes
$R_{\pi}\!=\!\Gamma(\pi\!\rightarrow\!e\nu)/\Gamma(\pi\!\rightarrow\! \mu\nu)$ and $R_{K}\!=\!\Gamma(K\!\rightarrow \!e\nu)/\Gamma(K\!\rightarrow\! \mu\nu)$,
the hadronic uncertainties cancel to a very large extent. As a result, the SM predictions of $R_{\pi}$ and $R_{K}$ are known 
with excellent accuracy \cite{Marciano:1993sh} and this makes it possible to fully exploit the great experimental resolutions on $R_{\pi}$ \cite{Eidelman:2004wy} 
and $R_{K}$ \cite{Eidelman:2004wy,na} to constrain new physics effects. Given our limited predictive power on  $f_{\pi}$ and $f_{K}$, deviations
from the $\mu-e$ universality represent the best hope we have at the moment to detect new physics effects in $\pi_{\ell2}$ and $K_{\ell2}$.

The most recent NA48/2 result on $R_K$:
$$
R^{exp.}_{K}=(2.416\pm 0.043_{stat.} \pm 0.024_{syst.}) \cdot 10^{-5}\,\,\,\,\,\,\rm{NA48/2},
$$
which will further improve with current analysis, significantly improves on the previous PDG value:
$$
R^{exp.}_{K}=(2.44\pm 0.11)\cdot 10^{-5}.
$$
This is to be compared with the SM prediction which reads:
$$
R^{SM}_{K}=(2.472\pm 0.001)\cdot 10^{-5}.
$$
The details of the experimental measurement of  $R_K$ are presented
in Section~\ref{sec:exp:universality:K} of this report.
Denoting by $\Delta r^{e-\mu}_{\!NP}$ the deviation from $\mu-e$ universality in $R_{K}$ due to new physics, i.e.:
\begin{equation}
\label{one}
R_{K}=R_{K}^{SM}\left(1+\Delta r^{e-\mu}_{\!NP}\right),
\end{equation}
the NA48/2 result requires (at the $2\sigma$ level): 
$$
-0.063\leq\Delta r^{e-\mu}_{\!NP}\leq 0.017 \,\,\,\,\,\,\rm{NA48/2}.
$$ 
In the following, we consider low-energy minimal SUSY extensions of the SM (MSSM) with R parity as the source of new physics to be 
tested by $R_K$ \cite{Barger:1989rk}. The question we intend to address is whether SUSY can cause deviations from $\mu-e$ universality in $K_{l2}$
at a level which can be probed with the present attained experimental sensitivity, namely at the percent level.
We will show that i) it is indeed possible for regions of the MSSM to obtain $\Delta r^{e-\mu}_{\!NP}$ of $\mathcal{O}(10^{-2})$ and
ii) such large contributions to $K_{\ell2}$ do not arise from SUSY lepton flavor conserving (LFC) effects, but, rather, from LFV ones.

At first sight, this latter statement may seem rather puzzling. The $K\!\rightarrow \!e\nu_e$ and $K\!\rightarrow\! \mu\nu_{\mu}$ decays are
LFC and one could expect that it is through LFC SUSY contributions affecting differently the  two decays that one obtains
the dominant source  of lepton flavor non-universality in SUSY. On the other hand, one can easily guess that, whenever new physics
intervenes in $K\!\rightarrow\! e\nu_e$ and $K\!\rightarrow\! \mu\nu_{\mu}$ to create a departure from the strict SM $\mu-e$ universality,
these new contributions will be proportional to the lepton masses; hence, it may happen (and, indeed, this is what occurs in the SUSY case)
that LFC contributions are suppressed with respect to the LFV ones by higher powers of the first two generations lepton masses
(it turns out that the first contributions to $\Delta r^{e-\mu}_{\!NP}$ from LFC terms arise at the cubic order in $m_{\ell}$, with $\ell=e,\mu$).
A second, important reason for such result is that among the LFV contributions to $R_K$ one can select those which involve flavor changes
from the first two lepton generations to the third one with the possibility of picking up terms proportional to the tau-Yukawa coupling
which can be large in the large $\rm{\tan\beta}$ regime (the parameter $\rm{\tan\beta}$ denotes the ratio of Higgs vacuum expectation
values responsible for the up- and down- quark masses, respectively). Moreover, the relevant one-loop induced LFV Yukawa interactions are known 
\cite{Babu:2002et} to acquire an additional $\rm{\tan\beta}$ factor with respect to the tree level LFC Yukawa terms.
Thus, the loop suppression factor can be (partially) compensated in the large $\rm{\tan\beta}$ regime.

Finally, given the NA48/2 $R_K$ central value below the SM prediction, one may wonder whether SUSY contributions could have the correct sign
to account for such an effect. Although the above mentioned  LFV terms can only add positive contributions 
to $R_K$ (since their amplitudes cannot interfere with the SM one), it turns out that there exist LFC contributions arising from double LFV
mass insertions (MI) in the scalar lepton propagators which can destructively interfere with the SM contribution.
We will show that there exist regions of the SUSY parameter space where the total $R_K$ arising from all such SM and SUSY terms is indeed lower
than $R^{SM}_K$.

Finally, we also discuss the potentiality of $\tau-\mu(e)$ universality breaking in $\tau$ decays to probe New Physics effects.

\subsubsubsection{$\mu-e$ universality in $\pi\rightarrow \ell\nu$ and $K\rightarrow \ell\nu$ decays}

Due to the V-A structure of the weak interactions, the SM contributions to $\pi_{\ell2}$ and $K_{\ell2}$ are helicity suppressed;
hence, these processes are very sensitive to non-SM effects (such as multi-Higgs effects) which might induce an effective pseudoscalar
hadronic weak current.

In particular, charged Higgs bosons ($H^\pm$) appearing in any model with two Higgs doublets (including the SUSY case) can contribute at tree level to
the above processes. The relevant four-Fermi interaction for the decay of charged mesons induced by $W^\pm$ and $H^\pm$ has the following form:
\begin{equation}
\frac{4G_F}{\sqrt{2}}V_{ud} \left[(\,\overline{u}\gamma_{\mu}P_Ld\,) (\,\overline{l}\gamma^{\mu}P_L\nu\,)
-\tan^{\!2}\!\beta\left(\frac{m_{d} m_l}{m^{2}_{H^\pm}}\right)(\,\overline{u}P_Rd\,)(\,\overline{l}P_L\nu\,)\right],
\end{equation}
where $P_{R,L}=(1\pm \gamma_5)/2$. Here we keep only the $\tan\beta$ enhanced part of the $H^\pm ud$ coupling, namely the $m_d\tan\beta$ term.
The decays $M\rightarrow l\nu$ (being $M$ the generic meson) proceed via the axial-vector part of the $W^\pm$ coupling and via the pseudoscalar part of the
$H^\pm$ coupling. Then, once we implement the PCAC's
\begin{equation}
<0|\overline{u}\gamma_{\mu}\gamma_{5}d|M^{-}>=if_M p^{\mu}_M\,\,\,\,\,\,,\,\,\,\,\,
<0|\overline{u}\gamma_{5}d|M^{-}>=-if_M \frac{m^{2}_M}{m_{d}+m_u}\,,
\end{equation}
we easily arrive at the amplitude
\begin{equation}
\mathcal{M}_{M\rightarrow l\nu}=\frac{G_F}{\sqrt{2}}V_{u(d,s)}f_M \left[m_l-m_l\,\,\tan^{2}\!\beta\bigg(\frac{m_d}{m_d\!+\!m_u}\bigg)
\frac{m^{2}_M}{m^{2}_{H^\pm}}\,\right]\overline{l}(1-\gamma_{5})\nu.
\end{equation}
We observe that the SM term is proportional to $m_l$ because of the helicity suppression while the charged Higgs term is proportional to $m_l$ because
of the Yukawa coupling. The tree level partial width is given by \cite{Babu:2002et}:
\begin{equation}
\Gamma(M^-\rightarrow l^-\overline{\nu})=\frac{G^{2}_F}{8\pi}|V_{u(d,s)}|^2f^{2}_M m_M m^{2}_l
\left(1-\frac{m^{2}_l}{m^{2}_M} \right)\times r_M,
\end{equation}
where
\begin{equation}
r_M=\left[1-\tan^2\beta\,\,
\left(\frac{m_{d,s}}{m_u\!+\!m_{d,s}}\right)\frac{m^{2}_M}{m^{2}_{H^\pm}}\right]^2\,,
\label{tree}
\end{equation}
and where $m_{u}$ is the mass of the up quark while $m_{s,d}$ stands for the down-type quark mass of the $M$ meson ($M=K, \pi$).
From Eq.~(\ref{tree}) it is evident that such tree level contributions do not introduce any lepton flavour dependent correction.
The first SUSY contributions violating the $\mu-e$ universality in $\pi\rightarrow \ell\nu$ and $K\rightarrow \ell\nu$ decays
arise at the one-loop level with various diagrams involving exchanges of (charged and neutral) Higgs scalars, charginos, neutralinos and sleptons.
For our purpose, it is relevant to divide all such contributions into two classes: i) LFC contributions where the charged meson M decays without FCNC 
in the leptonic sector, i.e. $M\rightarrow \ell\nu_{\ell}$; ii) LFV contributions $M\rightarrow \ell_i\nu_k$, with $i$ and $k$ 
referring to different generations (in particular, the interesting case will be for $i= e,\mu$, and $k=\tau$).

\subsubsubsection{The lepton flavour conserving case}

One-loop  corrections to $R_{\pi}$ and $R_{K}$ include box, wave function renormalization and vertex contributions from SUSY particle exchange.
The complete calculation of the $\mu$ decay in the MSSM \cite{Chankowski:1993eu} can be easily applied to the meson decays.

The dominant diagrams containing one loop corrections to the $lW\nu_l$ vertex have the following suppression factors (compared to the tree level graph):
\begin{itemize}
\item $\frac{g^{2}_{2}}{16\pi^2}\frac{m^{2}_{l}}{m^{2}_{W}}\tan\beta\frac{m^{2}_{W}}{m^{2}_h}$
- for loops with $h W^{\pm}l$ exchange (with $h=H^0,h^0$),
\item $\frac{g^{2}_{2}}{16\pi^2}\frac{m^{2}_{l}}{m^{2}_{W}}\tan^2\beta\frac{m^{2}_{W}}{m^{2}_h}$
- for loops with $h H^{\pm}l$ exchange (with $h=H^0,h^0$ and $A^0$),
\item $\frac{g^{2}_{2}}{16\pi^2}\frac{m^{2}_{W}}{M^{2}_{SUSY}}$
- for loops generated by charginos/neutralinos and sleptons.
\end{itemize}
For dominant box contributions we have the following estimates:
\begin{itemize}
\item $\frac{g^{2}_{2}}{16\pi^2}\frac{m_{d}m_{l}}{M^{2}}\tan^2\beta$
- for boxes with $h W^{\pm}l$ or $Z^0 H^{\pm}l$ 
exchange (where M is the heavier mass circulating in the loop),
\item $\frac{g^{2}_{2}}{16\pi^2}\left(\frac{m_{d}m_{l}}{m_{W}M_{H^{\pm}}}\right)^2 \tan^4\beta$
- for boxes with $h H^{\pm}l$ ,
\item $\frac{g^{2}_{2}}{16\pi^2}\frac{m^{2}_{W}}{M^{2}_{SUSY}}$ - for loops generated 
by charginos/neutralinos and sleptons (where $M_{SUSY}$ is the heavier mass circulating in the loop).
\end{itemize}
To get a feeling of the order of magnitude of the above contributions let us show 
the explicit expression of the dominant Higgs contributions to the $lW\nu_l$ vertex
\cite{Chankowski:1993eu}:
$$\Delta r^{e-\mu}_{SUSY}=\frac{\alpha_{2}}{32\pi}\frac{m^2_{\mu}}{M_W^{2}}\tan^2\!\beta
\left(-2+I(A^0,H^{\pm})+c^{2}_{\alpha} I(H^0,H^{\pm})+s^{2}_{\alpha} I(h^0,H^{\pm})\right),$$
where
$$
I(1,2)=\frac{1}{2}\frac{m^2_1+m^2_2}{m^2_1-m^2_2}\log\frac{m^2_1}{m^2_2},
$$
and $\alpha$ is the mixing angle in the CP-even Higgs sector. Even if we assume $\tan\beta=50$ and arbitrary relations among the Higgs boson masses 
we get a value for $\Delta r^{e-\mu}_{SUSY}\leq 10^{-6}$ much below the actual experimental resolution. In addition, in the large $\tan\beta$ 
limit, $\alpha\rightarrow 0$ and $m_{A^0}\sim m_{H^0}\sim m_{H^{\pm}}$ and $\Delta r^{e-\mu}_{SUSY}$ tends to vanish.
The charginos/neutralinos sleptons ($\tilde l_{e,\mu}$) contributions to $\Delta r^{e-\mu}_{SUSY}$ are of the form
$$
\Delta r^{e-\mu}_{SUSY}\sim \frac{\alpha_{2}}{4\pi}
\left(\frac{\tilde{m}^{2}_{\mu}-\tilde{m}^{2}_{e}}{\tilde{m}^{2}_{\mu}+
\tilde{m}^{2}_{e}}\right)\frac{m^{2}_{W}}{M^{2}_{SUSY}},
$$
The degeneracy of slepton masses (in particular those of the first two generations) severely suppresses these contributions.
Even if we assume a quite large mass splitting among slepton masses (at the $10\%$ level for instance) we end up with 
$\Delta r^{e-\mu}_{SUSY}\leq 10^{-4}$. For the box-type non-universal contributions, we find similar or even 
more suppressed effects compared to those we have studied. So, finally, it turns out that all these LFC contributions yield values of
$\Delta r^{e-\mu}_{\!K\,SUSY}$ which are much smaller than the percent level required by the achieved experimental sensitivity.

On the other hand, one could wonder whether the quantity $\Delta r^{e-\mu}_{\!\,SUSY}$ can be constrained by the pion physics.
In principle, the sensitivity could be even higher: from 
$$
R^{exp.}_{\pi}=
(1.230\pm 0.004)\cdot 10^{-4}\,\,\,\,\,\,\,\rm{PDG},
$$ 
and by making a comparison with the SM prediction
$$
R^{SM}_{\pi}=(1.2354\pm 0.0002)\cdot 10^{-4},
$$
one obtains (at the $2\sigma$ level) 
$$
-0.0107\leq\Delta r^{e-\mu}_{\!NP}\leq 0.0022.
$$
Unfortunately, even in the most favorable cases, $\Delta r^{e-\mu}_{\!\,SUSY}$ remains much below its actual experimental upper bound.

In conclusion, SUSY effects  with flavor conservation in the leptonic sector can differently contribute to the $K\rightarrow e\nu_e$ and 
$K\rightarrow \mu\nu_{\mu}$ decays, hence inducing a $\mu-e$ non-universality in $R_K$, however such effects are still orders of magnitude below 
the level of the present experimental sensitivity  on $R_K$. The same conclusions hold for $R_{\pi}$.

\subsubsubsection{The lepton flavour violating case}\label{sec:RKtheory}

It is well known that models containing at least two Higgs doublets generally allow flavour violating couplings of the Higgs bosons with
the fermions \cite{Gunion:1989we}. In the MSSM such LFV couplings are absent at tree level. However, once non holomorphic terms are generated by
loop effects (so called HRS corrections \cite{Hall:1993gn}) and given a source of LFV among the sleptons, Higgs-mediated (radiatively induced)
$H\ell_i\ell_j$ LFV couplings are unavoidable \cite{Babu:2002et,Dedes:2002rh}. These effects have been widely discussed through the study of several processes, 
namely $\tau\!\rightarrow\!\ell_j\ell_k\ell_k$ \cite{Babu:2002et,Dedes:2002rh}, $\tau\!\rightarrow\!\mu\eta$
\cite{Sher:2002ew}, $\mu-e$ conversion in nuclei \cite{Kitano:2003wn}, $B\!\rightarrow\!\ell_j\tau$ \cite{Dedes:2002rh}, $H\!\rightarrow\!\ell_j\ell_k$ 
\cite{Brignole:2003iv} and $\ell_i\!\rightarrow\!\ell_j\gamma$ \cite{Paradisi:2005tk}.\\

Moreover, it has been shown \cite{Masiero:2005wr} that Higgs-mediated LFV couplings generate a breaking of the $\mu-e$ universality in the purely leptonic 
$\pi^{\pm}$ and $K^{\pm}$ decays.

One could naively think that SUSY effects in the LFV channels $M\rightarrow \ell_i\nu_k$ are further suppressed with respect to 
the LFC ones. On the contrary, charged Higgs mediated SUSY LFV contributions, in particular in the kaon decays into an electron or 
a muon and a tau neutrino, can be strongly enhanced. The quantity which now accounts for the deviation from the $\mu-e$ universality reads:
$$
R^{LFV}_{\pi,K}=
\frac{\sum_i\Gamma(\pi(K)\rightarrow e\nu_i)}
{\sum_i\Gamma(\pi(K)\rightarrow \mu\nu_{i})}
\,\,\,\,\,\,\,\,\,\,\,\,\,\,\,\,\,\,\,\,\,\,\,\,\,\,\,\,i= e,\mu,\tau.
$$
with the sum extended over all (anti)neutrino flavors (experimentally one determines only the charged lepton flavor in the decay products).

\begin{figure}[t]
\centering
\includegraphics[scale=1.0]{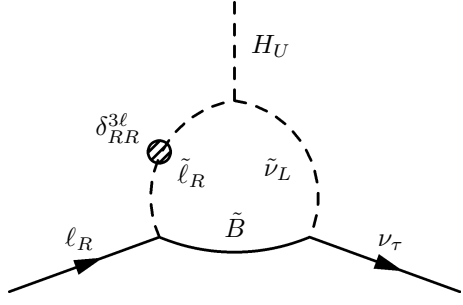}
\caption{\label{fig} 
Contribution to the effective  $\bar{\nu}_{\tau} \ell_R H^+$ coupling.}
\end{figure}

The dominant SUSY contributions to $R^{LFV}_{\pi,K}$ arise from the charged Higgs exchange. The effective LFV Yukawa couplings we consider are (see Fig.~\ref{fig}):
\begin{equation}
\label{coupl1}
\ell H^{\pm}{\nu_{\tau}}\rightarrow \frac{g_2}{\sqrt2}\frac{{m_{\tau}}}{M_W}\Delta^{3l}_{R}\tan^{2}\!\beta \,\,\,\,\,\,\,\,\,\,\,\,\ell=e,\mu.
\end{equation}
A crucial ingredient for the effects we are going to discuss is the quadratic dependence on $\tan\!\beta$ in the above coupling: one power of $\tan\!\beta$ 
comes from the trilinear scalar coupling in Fig.~\ref{fig}, while the second one is a specific feature of the above HRS mechanism.

The $\Delta^{3\ell}_{R}$ terms are induced at one loop level by the exchange of Bino (see Fig.~\ref{fig}) or Bino-Higgsino and sleptons.
Since the Yukawa operator is of dimension four, the quantities $\Delta^{3\ell}_{R}$ depend only on ratios of SUSY masses, 
hence avoiding SUSY decoupling. In the so called MI approximation the expression of $\Delta^{3\ell}_{R}$ is given by:
\begin{equation}
\Delta^{3\ell}_{R}\!\simeq\! \frac{\alpha_{1}}{4\pi}\mu M_1 m^{2}_{R} \delta^{3\ell}_{RR}
\left[I^{'}\!(M^{2}_{1},\mu^2,m^{2}_{R})\!-\!(\mu\!\leftrightarrow\! m_{L}) \right],
\end{equation}
where $\mu$ is the Higgs mixing parameter, $M_1$ is the Bino ($\tilde{B}$) mass and $m^{2}_{L(R)}$ stands for the left-left (right-right) slepton mass matrix entry.
The LFV MIs, i.e. $\delta^{3\ell}_{XX}\!=\!({\tilde m}^2_{\ell})^{3\ell}_{XX}/m^{2}_{X}$ $(X=L,R)$,
are the off-diagonal flavor changing entries of the slepton mass matrix. The loop function $I^{'}(x,y,z)$ is such that $I^{'}(x,y,z)= dI(x,y,z)/d z$,
where $I(x,y,z)$ refers to the standard three point one-loop integral which has mass dimension -2.

Making use of the LFV Yukawa coupling in Eq.~(\ref{coupl1}), it turns out that the dominant contribution to $\Delta r^{e-\mu}_{NP}$ reads \cite{Masiero:2005wr}:
\begin{equation}
\label{lfv}
R^{LFV}_{K}\simeq R^{SM}_{K} \left[1+\left(\frac{m^{4}_{K}}{M^{4}_{H}}\right) \!\left(\frac{m^{2}_{\tau}}{m^{2}_{e}}\right)|\Delta^{31}_{R}|^2\, \tan^{\!6}\!\beta\right].
\end{equation}
In Eq.~(\ref{lfv}) terms proportional to $\Delta^{32}_{R}$ are neglected given that they are suppressed by a factor $m^{2}_{e}/m^{2}_{\mu}$ with respect to the term
proportional to $\Delta^{31}_{R}$.

Taking $\Delta^{31}_{R}\!\simeq\!5\cdot 10^{-4}$ (by means of a numerical analysis, it turns out that $\Delta^{3\ell}_{R}\leq 10^{-3}$ \cite{Brignole:2003iv}),
$\tan\beta\!=\!40$ and $M_{H}\!=\!500 GeV$ we end up with $R^{LFV}_{K}\!\simeq\!R^{SM}_{K}(1+0.013)$.
We see that in the large (but not extreme) $\rm\tan\beta$ regime and with a relatively heavy $H^{\pm}$, it is possible to reach contributions
to $\Delta r^{e-\mu}_{\!K\,SUSY}$ at the percent level thanks to the possible LFV enhancements arising in SUSY models.

Turning to pion physics, one could wonder whether the analogous quantity $\Delta r^{e-\mu}_{\!\pi\,SUSY}$ is able to constrain SUSY LFV.
However,  the correlation between $\Delta r^{e-\mu}_{\!\pi\,SUSY}$ and $\Delta r^{e-\mu}_{\!K\,SUSY}$:
\begin{equation}
\label{lfvpi}
\Delta r^{e-\mu}_{\pi\,SUSY}\simeq\left(\frac{m_d}{m_u+m_d}\right)^{2} \left(\frac{m^{4}_{\pi}}{m^{4}_{k}}\right) \Delta r^{e-\mu}_{\!K\,SUSY},
\end{equation}
clearly shows that the constraints on $\Delta r^{e-\mu}_{\!K\,SUSY}$ force $\Delta r^{e-\mu}_{\pi\,SUSY}$ to be much below its actual experimental upper bound.

\subsubsubsection{On the sign of $\Delta r^{e-\mu}_{\!SUSY}$}

The above SUSY dominant contribution to $\Delta r^{e-\mu}_{\!NP}$ arises from LFV channels in the $K\!\rightarrow\!e\nu$ mode,
hence without any interference effect with the SM contribution. Thus, it can only increase the value of $R_{K}$ with respect to the SM
expectation. On the other hand, the recent NA48/2 result exhibits a central value lower than $R_{K}^{SM}$
(and, indeed, also lower than the previous PDG central value). One may wonder whether SUSY could account for such a lower $R_{K}$.
Obviously, the only way it can is through terms which, contributing to the LFC $K\!\rightarrow\!l\nu_{l}$ channels, can interfere (destructively)
with the SM contribution. We already commented that SUSY LFC contributions are subdominant. However, one can envisage the possibility of making use
of the large LFV contributions to give rise to LFC ones through double LFV MI that, as a final effect, preserves the flavour.

To see this point explicitly, we report the corrections to the LFC $H^{\pm}\ell\nu_{\ell}$ vertices induced by LFV effects
\begin{equation}
\label{coupl2}
\ell H^{\pm}{\nu_{\ell}}\!\rightarrow\! \frac{g_2}{\sqrt 2}\frac{{m_{\ell}}}{M_W}\tan\!\beta \left(\!1\!+\!\frac{m_{\tau}}{m_{\ell}}\Delta^{\ell\ell}_{RL}
\tan\!\beta\!\right)\,,
\end{equation}
where $\Delta^{\ell\ell}_{RL}$ is generated by the same diagram as in Fig.~\ref{fig} but with an additional $\delta^{3\ell}_{LL}$ MI
in the sneutrino propagator. In the MI approximation, $\Delta^{\ell\ell}_{RL}$ is given by
\begin{equation}
\Delta^{\ell\ell}_{RL}\!\simeq\! -\frac{\alpha_{1}}{4\pi}\mu M_1 m^{2}_{L} m^{2}_{R}
\,\delta^{\ell 3}_{RR}\delta^{3\ell}_{LL}\,I^{''}\!(M^{2}_{1},m^{2}_{L},m^{2}_{R})\,,
\end{equation}
where $I^{''}(x,y,z)= d^2I(x,y,z)/dydz$. In the large slepton mixing case, $\Delta^{\ell\ell}_{RL}$ terms are of the
same order of $\Delta^{3\ell}_{R}$ \footnote{Im($\delta^{13}_{RR}\delta^{31}_{LL}$)
is strongly constrained by the electron electric dipole moment \cite{Masina:2002mv}.
However, sizable contributions to $R^{LFV}_{K}$ can still be induced by Re($\delta^{13}_{RR}\delta^{31}_{LL}$).}.
These new effects modify the previous $R^{LFV}_{K}$ expression in the following way \cite{Masiero:2005wr}
\begin{eqnarray}
\label{lfclfv}
R^{LFV}_{K}\simeq R^{SM}_{K} \bigg[\,\bigg|1\!-\!\frac{m^{2}_{K}}{M^{2}_{H}} \frac{m_{\tau}}{m_{e}}\Delta^{11}_{RL}\,t_{\beta}^{3}\,\bigg|^{2}+
\bigg(\frac{m^{4}_{K}}{M^{4}_{H}}\bigg) \!\bigg(\frac{m^{2}_{\tau}}{m^{2}_{e}}\bigg) |\Delta^{31}_{R}|^2\,\tan^{6}\beta  \bigg].
\end{eqnarray}
In the above expression, besides the contributions reported in Eq.~(\ref{lfv}), we also included the interference between SM and SUSY LFC terms
(arising from a double LFV source). Setting the parameters as in the example of the above section and if  
$\Delta^{11}_{RL}\!=\!10^{-4}$ we get $R^{LFV}_{K}\!\simeq\! R^{SM}_{K}(1-0.032)$,
that is just within the expected experimental resolution reachable by NA48/2 once all the available data will be analyzed.
Finally, we remark that the above effects do not spoil the pion physics constraints.

The extension of the above results to $B \to \ell \nu$ \cite{Isidori:2006pk} is obtained with the replacement $m_K \to m_B$, while for 
the $D \to \ell \nu$ case $m^{2}_{K} \to (m_s/m_c) m^{2}_{D}$. In the most favorable scenarios, taking into account the constraints from LFV 
$\tau$ decays \cite{Paradisi:2005tk}, spectacular order-of-magnitude enhancements 
for $R_B^{e/\tau}$ and  $\mathcal O(50\%)$ deviations from the SM in $R_B^{\mu/\tau}$ are allowed \cite{Isidori:2006pk}.
There exists a stringent correlation between $R_B^{e/\tau}$ and $R_K^{e/\mu}$ so that:
\begin{equation}
R_B^{e/\tau} \simeq \left[r_H+\frac{m^{4}_{B}}{m^{4}_{K}}\Delta r^{e-\mu}_{\!K\,SUSY}\right] \leq 2\cdot 10^{2}\,.
\end{equation}
In particular, it turns out that $\Delta r^{e-\mu}_{\!K\,SUSY}$ is much more effective to constrain $R_B^{e/\tau}$ $\Gamma(B\rightarrow e\nu_{\tau})$ than LFV tau 
decay processes.

\subsubsubsection{Lepton universality in $M\rightarrow \ell \nu$ vs LFV $\tau$ decays}

Obviously, a legitimate worry when witnessing such a huge SUSY contribution through LFV terms is whether the bounds on LFV  tau decays, like
$\tau\rightarrow eX$ (with $X=\gamma,\eta,\mu\mu$), are respected \cite{Paradisi:2005tk}.
Higgs mediated $B(\tau\rightarrow \ell_j X)$ and $\Delta r^{e-\mu}_{\!K\,SUSY}$ have exactly the same SUSY dependence; hence, we can compute the upper bounds of
the relevant LFV tau decays which are obtained for those values of the SUSY parameters yielding $\Delta r^{e-\mu}_{\!K\,SUSY}$ at the percent level.

The most sensitive processes to Higgs mediated LFV in the $\tau$ lepton decay channels are $\tau\rightarrow \mu(e)\eta$, $\tau\rightarrow \mu(e)\mu\mu$ and 
$\tau\rightarrow \mu(e)\gamma$. The related branching ratios are \cite{Paradisi:2005tk}:
\begin{equation}
\frac{B(\tau\rightarrow l_j\eta)}{B(\tau\rightarrow l_j\bar{\nu_j}\nu_{\tau})} \simeq 18\pi^2\!\left(\frac{f^{8}_{\eta} m^{2}_{\eta}}{m_{\tau}}\right)^{\!2}\!
\!\left(1\!-\!\frac{m^{2}_{\eta}}{m^{2}_{\tau}}\right)^{\!2} \left(\frac{|\Delta^{3j}|^2\tan^{6}\beta }{m^{4}_{A}}\right),
\end{equation}
where $m^{2}_{\eta}/m^{2}_{\tau}\simeq 9.5\times10^{-2}$ and the relevant decay constant is $f^{8}_{\eta}\sim 110 MeV$,
\begin{equation}
\frac{B(\tau\rightarrow l_j\gamma)}{B(\tau\rightarrow l_j\bar{\nu_j}\nu_{\tau})} \simeq 10\left(\frac{\alpha_{el}}{\pi}\right)^3
\tan^{4}\beta |\Delta_{\tau j}|^{2} \bigg[\,\frac{m_{W}}{m_{A}} \log\left(\frac{m^{2}_{W}}{m^{2}_{A}}\right)\,\bigg]^4,
\end{equation}
\begin{eqnarray}
\frac{B(\tau\rightarrow l_j\mu\mu)}{B(\tau\rightarrow l_j\bar{\nu_j}\nu_{\tau})} &\simeq& \frac{m^2_{\tau}m^2_{\mu}}{32}
\left(\frac{|\Delta^{3j}|^2\tan^{6}\beta }{m^{4}_{A}}\right) \bigg(3+5\delta_{j\mu}\bigg),
\end{eqnarray}
where $|\Delta^{3j}|^2\!=\!|\Delta^{3j}_{L}|^2+|\Delta^{3j}_{R}|^2$. It is straightforward to check that, in the large $\tan\beta$ regime,
$B(\tau\rightarrow l_j\eta)$ and $B(\tau\rightarrow l_j\gamma)$ are of the same order of magnitude \cite{Paradisi:2005tk} and they are 
dominant compared to $B(\tau\rightarrow l_j\mu\mu)$. \footnote{It is remarkable that $\Delta r^{e-\mu}_{\!K\,SUSY}\sim |\Delta^{31}_{R}|^2$
while $B(\tau\rightarrow eX)\sim |\Delta^{31}_{L}|^2+|\Delta^{31}_{R}|^2$ (with $X=\eta, \gamma$ or $\mu\mu$). In practice, $\Delta r^{e-\mu}_{\!K\,SUSY}$ 
is sensitive only to RR-type LFV terms in the slepton mass matrix while $B(\tau\rightarrow eX)$ does not distinguish between left and right sectors.}.

Given that $\Delta r^{e-\mu}_{\!K\,SUSY}$ and $B(\tau\rightarrow l_j X)$ have the same SUSY dependence, once we saturate the $\Delta r^{e-\mu}_{\!K\,SUSY}$ 
value (at the \% level), the upper bounds on $B(\tau\rightarrow l_j X)$ (allowed by $|\Delta^{31}_{R}|^2$) are automatically predicted. We find that
\begin{equation}
B(\tau\rightarrow l_j\gamma)\sim B(\tau\rightarrow l_j\eta) \simeq 10^{-2}\left(\frac{|\Delta^{3j}|^2\tan^{6}\beta }{m^{4}_{A}}\right)
\simeq 10^{-8}\Delta r^{e-\mu}_{\!K\,SUSY}.
\end{equation}

So, employing the constraints for $\Delta r^{e-\mu}_{\!K\,SUSY}$ at the $\%$ level, we obtain the desired upper bounds:
$B(\tau\rightarrow e\eta), B(\tau\rightarrow e\gamma)\leq 10^{-10}$. Given the experimental upper bounds on the LFV $\tau$ lepton decays \cite{Abe:2003sx},
we conclude that it is possible to saturate the upper bound on $\Delta r^{e-\mu}_{\!K\,SUSY}$ while remaining much below the present and expected
future upper bounds on such LFV decays. There exist other SUSY contributions to LFV $\tau$ decays, like the one-loop neutralino-charged slepton exchanges,
for instance, where there is a direct dependence on the quantities $\delta^{3j}_{RR}$. Given that the existing bounds on the leptonic $\delta_{RR}$ involving
transitions to the third generation are rather loose \cite{Paradisi:2005fk}, it turns out that also these
contributions fail to reach the level of experimental sensitivity for LFV $\tau$ decays.

\subsubsubsection{$e$--$\mu$ universality in $\tau$ decays}

Studying the $\tau - \mu - e$ universality in the leptonic $\tau$ decays is an interesting laboratory for search for physics beyond the SM.
In the SM the $\tau$ decay partial width for the leptonic modes is:
\begin{eqnarray}
\Gamma\left(\tau\rightarrow l\bar\nu_l\nu_{\tau}(\gamma)\right) &=& \frac{G^2_{F} m^5_{\tau}}{192\pi^3} f(m^2_l/m^2_{\tau})\times\\
 &&\left[1 + \frac{3}{5}\frac{m^2_{\tau}}{M^2_W}\right] \left[1 + \frac{\alpha(m_{\tau})}{\pi}\left(\frac{25}{4}-\pi^2\right)\right],
\nonumber
\end{eqnarray}
where $f(x) = 1 - 8x + 8x^3 - x^4 -12x^2\log x$ is the lepton mass correction and the last two factors are corrections from the nonlocal
structure of the intermediate $W^{\pm}$ boson propagator and QED radiative corrections respectively. The Fermi constant $G_F$ is
determined by the muon life-time
\begin{eqnarray}
 G_F\equiv G_{\mu}= (1.16637\pm 0.00002)\times10^{-5}{\rm GeV}^{-2},
\end{eqnarray}
and absorbs all the remaining electroweak radiative (loop) corrections.

The main source of non$-$universal contributions would be the tree level contribution from the charged Higgs boson (mass dependent couplings) and
different slepton masses of the $\tilde\mu$, $\tilde\tau$ and $\tilde e$ sleptons exchanged in the one loop induced $\ell-W-\nu_{\ell}$ vertex.
On the other hand, as discussed in previous sections, the last contribution can provide a correction that can be at most as large as $10^{-4}$ 
(in the limiting case of very light sleptons and gauginos $\sim M_W$), very far for the actual and forthcoming experimental resolutions.
However, differently from the $M\rightarrow \ell\nu$ case, a tree level charged Higgs exchange breaks the Lepton Universality and it provides
a contribution that we are going to discuss.

The deviations from the $\tau-\mu-e$  universality can be conveniently discussed by studying the ratios $G_{\tau,e}/G_{\mu,e}$,
$G_{\tau,\mu}/G_{\mu,e}$ and $G_{\tau,\mu}/G_{\tau,e}$, given by the ratios of the corresponding branching fractions.
With the highly accurate experimental result for the $G_{\mu,e}$, the first two ratios are essentially a direct measure of
non-universality in the corresponding tau decays. When the statistical error of future experiments will become
negligible, the main problem for achieving maximum precision will be to reduce the systematic errors. One may
expect that certain systematic errors will be canceled in the ratio $G_{\tau,\mu}/G_{\tau,e}$.

The '04 world averaged data for the leptonic $\tau$ decay modes and $\tau$ lifetime 
are \cite{Eidelman:2004wy,Krawczyk:2004na}
\begin{equation}
B^e|_{exp} = (17.84 \pm 0.06) \%, \hspace{0.5cm}\,\,\,B^{\mu}|_{exp} = (17.37 \pm 0.06) \% ,\nonumber 
\end{equation}
\begin{equation}
\tau_{\tau} = (290.6 \pm 1.1) \times 10^{-15} s.
\end{equation}
Note that the  relative errors of the above measured  quantities are of the 0.34-0.38 \%, the biggest being for the lifetime.
One can parameterize a possible beyond the SM contribution by a quantity $\Delta^l$ ($l=e,\mu$),  defined as 
\begin{equation}
B^l = B^l |_{SM} (1 + \Delta^l).
\end{equation}
Including the W-propagator effect and QED radiative corrections, the following results for the branching ratios in the SM are obtained \cite{Krawczyk:2004na}: 
\begin{equation}
B^e|_{SM} = (17.80 \pm 0.07)\% ,\,\, \, B^\mu|_{SM} = (17.32 \pm 0.07)\%.
\end{equation}
Together with the experimental  data this leads  to the following $95\%$ C.L. bounds on $\Delta^l$, for the electron and muon decay mode, respectively \cite{Krawczyk:2004na}:
\begin{equation}
(-0.80 \leq \Delta^e \leq 1.21) \% ,\,\,\,\,\, (-0.76 \leq \Delta^{\mu} \leq 1.27) \%. 
\label{eq.explim}
\end{equation}
One can see that the negative contributions are constrained more strongly that the positive ones.
A tree level charged Higgs exchange leads to the following contribution \cite{Krawczyk:1987zj}
\begin{eqnarray}
\Gamma^{W^\pm+H^\pm} &=&\Gamma^{W^\pm} \left[1-2 \frac{m_l m_\tau \tan^2\beta}{M_{H^\pm}^2} \frac{m_l}{m_\tau}\kappa \left(\frac{m_l^2}{m_{\tau}^2} \right)
+\frac{m_{\tau}^2 m_l^2 \tan^4\beta}{4 M_{H^\pm}^4}\right]\nonumber \\
&\simeq& \Gamma^{W^\pm} \left[1-1.15\times 10^{-3}\left(\frac{200 \rm{GeV}}{M_{H^\pm}}\right)^2 \left(\frac{\tan\beta}{50}\right)^2\right],
\label{eq.treeH}
\end{eqnarray}
where $\kappa(x)=\frac{g(x)}{f(x)}\simeq 0.94$ with $g(x)=1+9x-9x^2-x^3+6x(1+x)\ln(x)$. In the above expression, the second term comes from the interference with the SM 
amplitude and it is much more important than the last one, that is suppressed by a factor $m_{\tau}^2 \tan^2\beta/8 M_{H^\pm}^2$.

For the future  precision of $G_{\tau,\mu}$ and $G_{\tau,e}$ measurements of order $0.1\%$ ($G_{\mu,e}$ is known with $0.002\%$ precision) 
the only effect that eventually can be observed is the slightly smaller value of $G_{\tau,\mu}$ as compared to $G_{\tau,e}$ and $G_{\mu,e}$
If measured, such effect would mean a rather precise information about MSSM: large $\tan\beta\geq 40$ and small $M_{H^{\pm}}\sim 200-300 \rm{GeV}$.

\subsubsubsection{Conclusions}

High precision electroweak tests, such as deviations from the SM expectations of the lepton universality 
breaking, represent a powerful tool to probe the SM and, hence, to constrain or obtain indirect hints of new physics beyond it.
Kaon and pion physics are obvious grounds where to perform such tests, for instance in the well studied $\pi\rightarrow \ell\nu_{\ell}$
and $K\rightarrow \ell\nu_{\ell}$ decays, where $l= e$ or $\mu$. In particular, a precise measurement of the flavor conserving
$K\rightarrow \ell\nu_{\ell}$ decays may shed light on the size of LFV in New Physics.
$\mu-e$ non-universality  in $K_{\ell2}$ is quite effective in constraining relevant regions of SUSY models with LFV.
A comparison with analogous bounds coming from $\tau$ LFV decays shows the relevance of the measurement of $R_{K}$ to probe LFV in SUSY.
Moreover, the $\tau - \mu - e$ universality in the leptonic $\tau$ decays is an additional interesting laboratory for searching for physics beyond the SM.


\subsection{EDMs from RGE effects in theories with low-energy supersymmetry}
\label{sec:EDM-RGE}
EDMs probe new physics in general and in particular low energy supersymmetry. For definiteness and for simplicity, we focus here on lepton EDMs,
as they are free from the theoretical uncertainties associated to the calculation of hadronic matrix elements. 
After a brief review of the constraints on slepton masses we discuss here a specific kind of sources of CPV, 
those induced radiatively by the Yukawa interactions of the heavy particles present in seesaw and/or grand-unified models.
It has been emphasized that these interactions could lead to LFV decays, 
in particular $\mu\to e\gamma$, at an observable rate; it is then natural to wonder whether this is also the case for EDMs.   

As shown in Section~\ref{sec:observables}, 
LFV decays, EDMs and additional contribution to MDMs all have a common origin,
the dimension 5 dipole operator possibly induced by some new physics beyond the SM:
\begin{eqnarray}
{\cal L}_{d=5} =\frac{1}{2} ~\bar \psi_{Ri} ~A_{ij}~ \psi_{Lj} ~\sigma^{\mu\nu} F_{\mu\nu} +h.c. ~~~~~~~~~~~~~~~~~\\
B(\ell_i\to \ell_j\gamma) \propto |A_{ij}|^2~~~~,~~~~~\delta a_{\ell_i} =\frac{2 m_{\ell_i}}{e}Re A_{ii}~~~~~,~~~~d_{\ell_i} =Im A_{ii}~~.
\end{eqnarray}
If induced at 1-loop, this amplitude displays a quadratic suppression with respect to the new physics mass scale, $M_{NP}$, 
and a linear dependence on the adimensional coupling $\Gamma^{NP}$ encoding the pattern of F and CP-violations 
(in the basis where the charged lepton mass matrix is diagonal): 
\begin{equation}
A_{ij} \approx \frac{e~ m_{\ell_i}}{(4\pi)^2} \frac{\Gamma^{NP}_{ij}}{M^2_{NP}}~~.
\end{equation}

For low energy supersymmetry, the loops involve exchange of gauginos and sleptons, so that 
$\Gamma^{NP}$ is proportional to the misalignment between leptons and sleptons, 
conveniently described by the flavor violating (FV) $\delta$'s of the mass insertion approximation.
It is well known that the flavor conserving (FC) $\mu$ and $a$ terms are potentially a very important source of CPV. 
In the expansion in powers of the FV $\delta$'s, they indeed contribute to $d_{\ell_i}$ at zero order:
\begin{equation}
Im (A_{ii}) ~=~ f_\mu ~m_{\ell_i}~  Arg(\mu)   ~+ f_a m_{\ell_i} Im (a_{i}) 
+ f_{LLRR} Im(\delta^{LL} m_\ell \delta^{RR} )_{ii}   +  ...
\label{deexpans}
\end{equation}
where the various $f$ represent supersymmetric loop functions and can
be found for instance in \cite{Masina:2002mv}.  Notice that the
contribution arising at second order in the FV $\delta$'s could be
even more important than the FC one, as happens for instance if CPV is
always associated to FV.

Assuming no cancellations between the amplitudes, we first review
briefly some limits considering for definiteness the mSUGRA scenario
with $\tan\beta =10$ and slepton masses in the range suggested by
$g_\mu$.  The strong impact of $\mu \to e \gamma$ on $\delta^{LL}$ has
been emphasized previously, where it was stressed that the impact of
$d_e$ is also remarkable in constraining the FC sources of CPV:
$\arg\mu\le 2\times 10^{-3}$, $Im a_e/m_R\le 0.2$.  As for the other
FV source in Eq.~(\ref{deexpans}), one obtains $Im(\delta^{LL}m_\ell
\delta^{RR})_{ee}/m_\tau\le 10^{-5}$.  The planned sensitivity
$d_\mu\le10^{-23}$ e cm would also give interesting bounds:
$\arg\mu\le 10^{-1}$, $Im(\delta^{LL}m_\ell
\delta^{RR})_{\mu\mu}/m_\tau\le 10^{-1}$.  Notice that, due to the
lepton mass scaling law of the $\mu$-term contribution, the present
bound on $\arg \mu$ from $d_e$ implies that the $\mu$-term
contribution to $d_\mu$ cannot exceed $2\times 10^{-25}$ e cm, below
the planned projects.  A positive measure of $d_\mu$ would thus signal
a different source of CPV, i.e. the $a_\mu$-term or the FV
contribution.  In the following we take real $\mu$.

The $a_i$-terms and the FV $\delta$'s at low energy can be thought as
the sum of two contributions.  The first is already present at the
Planck scale where soft masses are defined; we assume that this
contribution is absent because of some inhibition mechanism, as could
happen in supergravity.  The second contribution is induced
radiatively running from high to low energies by the Yukawa couplings
of heavy particles\footnote{The SM fermion Yukawa couplings induce
negligible effects.}  that potentially violate F and CP.  Since LFV
experiments are testing this radiatively induced misalignment, in the
following we will consider what happens for EDMs, beginning with the
pure seesaw model and then adding a grand-unification scenario, where
heavy colored Higgs triplets are present to complete the Higgs
doublets representations (in SU(5) for instance they complete the $5$
and $\bar 5$).  Notice that these triplets are important as in
supersymmetric theories proton decays mainly through their exchange.

Consider first the case of degenerate right handed neutrinos with mass $M$.
One can solve approximately the RGE by expanding in powers of $\ln( \Lambda/M)/(4\pi)^2$,
i.e. the log of the ratio of the two scales between which the neutrino Yukawa couplings $Y_\nu$ are present
times the corresponding loop factor suppression.
For LFV decays $\delta^{LL}_{ij}$ is induced at 1st order and is proportional to the combination
$(Y_\nu^\dagger Y_\nu)_{ij}$.
In particular, $\mu \to e \gamma$ constrains $(Y_\nu^\dagger Y_\nu)_{21}$ to be small and 
this has a strong impact on seesaw models.
To obtain an imaginary part for EDMs, one needs a non hermitian combination of Yukawa couplings,
which can be found only at 4th order: $Im( Y_\nu^\dagger Y_\nu [ Y_\nu^\dagger Y_\nu , Y_\ell^\dagger Y_\ell]  Y_\nu^\dagger Y_\nu)_{ii}$.
Such a contribution is negligibly small with respect to the present and planned experimental sensitivities.

Allowing for a non degenerate spectrum of right handed neutrinos, EDMs get strongly enhanced while LFV decays not. 
The latter are simply modified by taking into account the different mass thresholds:
\begin{equation}
\delta^{LL}_{ij} \propto \sum_{ k} C^k ~~~~,~~~~C^k\equiv {Y_\nu^\dagger}_{ik}  \ln \frac{\Lambda}{M_k}  {Y_\nu}_{kj} ~~.
\end{equation}
On the contrary for EDMs the seesaw-induced FC and FV contributions - coming respectively from 
$\mathrm{Im}a_i$ \cite{Ellis:2001xt,Ellis:2001yz,Ellis:2002fe,Masina:2003wt} 
and $\mathrm{Im}(\delta^{RR} m_{\ell} \delta^{LL} )_{ii}$ \cite{Masina:2003wt, Farzan:2004qu} - 
arise at 2nd and 3rd order and are proportional to the combinations \cite{Masina:2005am}:
\begin{eqnarray*}{}
\mathrm{Im} (a_i) 				&\propto&  \sum_{ k > k'} \frac{\ln (M_{k}/M_{k'})}{\ln (\Lambda/M_{k'})}~\mathrm{Im} (  C^{ k } C^{ k'} )_{ii}	\\ 
\mathrm{Im}(\delta^{RR}m_{\ell}\delta^{LL})_{ii}&\propto&  \sum_{ k > k'} \widetilde{\ln}^{k}_{k'} ~\mathrm{Im} (C^{k} m^2 _\ell C^{k'})_{ii}  ~~,\\
\end{eqnarray*}
where $\widetilde{\ln}^{k}_{k'}$ is a logarithmic function. 
The FV contribution generically dominates for $\tan\beta \gtrsim 10$. 
Without going in the details of this formulae, 
we just display some representative upper estimates for the seesaw-induced EDMs, 
considering for definiteness the $g_\mu$ region with $\tan\beta= 20$.
The seesaw induced $d_\mu$ is below the planned sensitivity, 
$d^{SS}_\mu \lesssim 10^{-25}$ e cm. On the contrary for $d_e$ it could be at hand,
$d_e^{SS}\lesssim 0.5\times 10^{-27}$ e cm; 
the expectation is however strongly model dependent and usually seesaw models that satisfy 
the bound from $\mu\to e\gamma$ predict a much smaller value \cite{Masina:2005am}. 
The possibility of large $d_e$ and its correlation with leptogenesis is discussed in \cite{Ellis:2002xg}.

Perspectives are much more interesting if there is also a stage of grand-unification.
In minimal supersymmetric $SU(5)$, the Higgs triplet Yukawa couplings contribute to the RGE-running 
for energies larger than their mass scale $M_T\sim M_{GUT}$.  
For LFV, $\delta^{RR}$ is generated at 1st order  
and is proportional to a combination of the up quark Yukawa couplings \cite{Barbieri:1995tw}:
\begin{equation}
\delta^{RR}_{ij} \propto  (Y_u^T Y_u^*)_{ij}  \ln \frac{\Lambda}{M_T} ~~. 
\end{equation}
Due to the weaker experimental bounds on $\delta^{RR}$, this contribution is not very significant.
On the other hand $\delta^{LL}$ is not changed, as also happens to the FC contribution to EDMs \cite{Masina:2003wt}.
The FV contribution to EDMs is on the contrary strongly enhanced: it arises at 2nd order (also for degenerate
right-handed neutrinos) and is proportional to:
\begin{equation}
\mathrm{Im}(\delta^{RR} m_{\ell} \delta^{LL} )_{ii} \propto  
\mathrm{Im} ( C  m_\ell  Y_u^T Y_u^* )_{ii}  \ln \frac{\Lambda}{M_T}  ~~.
\end{equation}
As a result, considering for definiteness the $g_\mu$ region with $\tan\beta= 20$ and the 
representative values for triplet and right-handed neutrino masses $M_T=2\times 10^{16}$ GeV 
and $M_3=10^{15}$ GeV,
the induced $d_\mu$ is still below planned, $d_\mu^{SS5}\lesssim 5\times 10^{-25}$ e cm,
but the induced $d_e$ could exceed by much the present limit: $d_e^{SS5} \lesssim 10^{-25}$ e cm. 
In turn this means that $Im(e^{-i\beta}C_{13})\lesssim 0.1$ ($\beta$ being the angle of the unitarity
triangle), which has of course an impact on seesaw models. Further details can be found in \cite{Masina:2003wt, Masina:2003iz}.
Notice however that, in addition to the problems with light fermion masses, 
minimal supersymmetric $SU(5)$ is generically considered to be ruled out 
by proton decay induced by Higgs triplet exchange. 

\begin{figure}[!h]
\centerline{
\includegraphics[width=\linewidth]{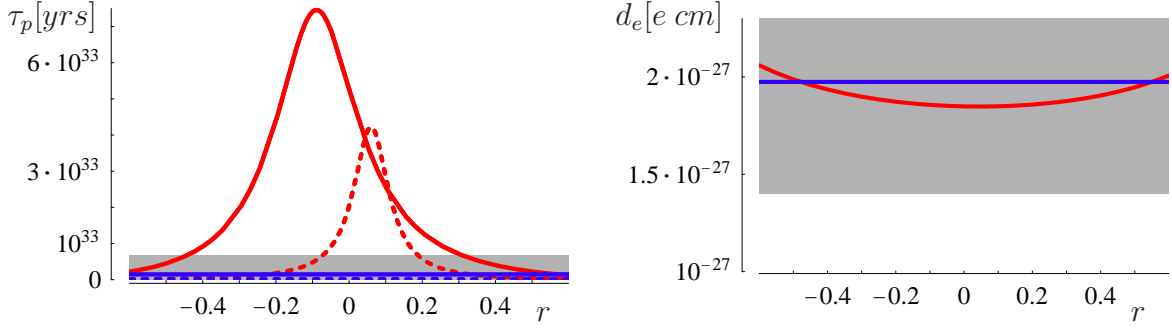} 
\put(-250, 8){\large $r$}  \put(-25, 8){\large $r$} 
\put(-450, 120){\large $\tau_p [yrs]$} \put(-220, 120){\large $d_e [e ~cm]$} 
}
\caption{ The predictions for $\tau_{p\to K\bar\nu}$ and $d_e$ are
shown as a function of $r$ for the degenerate (flat blue) and close to
pseudo Dirac (red) cases by taking: $m_T=10^{17}$ GeV,
$\Lambda=2\times10^{18}$ GeV and maximal CPV phase for $d_e$.  The
supersymmetric parameters $\tan\beta=3$, $\tilde M_1=200$ GeV, $\bar
m_R =400$ GeV, have been selected.  The shaded (grey) regions are
excluded experimentally. See \cite{Masina:2003iz} for more details.}
\label{fig-tpde}
\vskip 0 cm
\end{figure}

More realistic GUTs like $SO(10)$ succeed in suppressing the proton decay rate by introducing more Higgs triplets
and enforcing a peculiar structure for their mass matrix. What are the expectation for $d_e$ in this case?
Consider what happens in a semi-realistic $SO(10)$ model \cite{Masina:2003iz},
where in addition to the three $16$ fermion representations we introduce a couple of $10_H$'s
containing the Higgs doublets and triplets, $10_H^u=(H_D^u ,H_T^u )+(\bar H_D^u, \bar H_T^u )$, 
$10_H^d=(H_D^d, H_T^d )+( \bar H_D^d, \bar H_T^d )$.
Up and down quark fermion masses arise when the doublets $H_D^u$ and $\bar H_D^d$ acquire a VEV; 
in particular $Y_\nu=Y_u$, $Y_\ell=Y_d$, and also the triplet Yukawa couplings are fully determined
in terms of $Y_u$ and $Y_d$.
As for the mass matrices of the Higgses, the doublets are diagonal in this basis, while the triplets are a priori undetermined: 
\begin{equation}
(\begin{matrix} \bar H_D^d &  \bar H_D^u \end{matrix}) \left( \begin{matrix}  e.w. &  0 \\   0 & M_{GUT} \end{matrix} \right)
\left(\begin{matrix} H_D^u \\ H_D^d \end{matrix}\right)~~~~,~~~~~
(\begin{matrix} \bar H_T^d &  \bar H_T^u \end{matrix})~ M_T \left(\begin{matrix} H_T^u \\ H_T^d \end{matrix}\right)~~.
\end{equation}
Let consider two limiting cases for the pattern of the triplet mass matrix, diagonal degenerate and close to pseudo-Dirac:
\begin{equation}
M_T^{deg}=\left(  \begin{matrix} 1 & 0 \\ 0 & 1 \end{matrix}  \right) m_T~~~~~~~~M_T^{cpD}=\left(  \begin{matrix} 0 & 1 \\ 1 & r \end{matrix}  \right) m_T~~,
\end{equation}
where $r$ is a small real parameter, $r<1$, and the exact pseudo-Dirac form corresponds to the limit $r\to 0$. 
Notice that the close to pseudo-Dirac form is naturally obtained in the Dimopoulos-Wilczek mechanism to solve the doublet-triplet 
splitting problem.  
The prediction for proton lifetime displays a strong dependence on the structure of $M_T$, and only the pseudo-Dirac form is allowed,
as can be seen in Fig.~\ref{fig-tpde}
(there is an intrinsic ambiguity due to GUT phases, so that the prediction is in between the dotted and solid curves).
For EDMs on the contrary the Higgs triplets contribution to RGE is cumulative and, due to the log, mildly sensitive 
to the triplet mass matrix structure.
In the case of $O(1)$ CPV phase (a small phase would be unnatural in this context), 
$d_e$ would exceed the present bound for the values of supersymmetric parameters selected in Fig.~\ref{fig-tpde}. 
Planned searches will be a fortiori more constraining.
The impact of these results go beyond the essential model described above. 
Indeed, the week points of the model, like the fermion mass spectra, could be addressed without changing
by much the expectations for $d_e$.  
It is remarkable that {\it EDMs turns out to be complementary to proton decay in constraining supersymmetric GUTs}. 

In the above model one obtains the relation $d_\mu/d_e \sim \left| V_{ts}/V_{td} \right|^2 \approx 25$, so that the prediction for $d_\mu$ is below the
planned sensitivity. However, there are GUT models where this is not the case. For instance a significant $d_\mu$ is obtained in
L-R symmetric guts \cite{Babu:2000dq}.

\subsubsection{Electron--neutron EDM correlations in SUSY}
\begin{figure}[b]
\centerline{{\includegraphics[width=5.5cm]{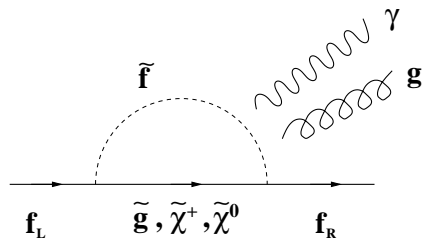}}}
\caption{One loop EDM contributions.}
\label{f0} 
\end{figure}

One of the questions we would like to address is whether non--zero EDM signals can constitute indirect evidence for supersymmetry. Supersymmetric models contain
additional sources of CP-violation compared to the SM, which induce considerable and usually too large  EDMs (Fig.\ref{f0}). In typical  (but not all)  SUSY models, the same
CP--violating source induces both hadronic and leptonic EDMs such that these are correlated. The most important source is usually the  CP--phase  of the $\mu$--term
and, in certain non--universal scenarios, the gaugino phases. The  CP--phases of the $A$--terms generally  lead to smaller contributions.

Typical SUSY models lead to  $\vert d_n \vert / \vert d_e \vert \sim {\cal O}(10)-{\cal O}(100)$. Thus, if both the  neutron and the electron EDMs are observed,
and this relation is found, it can be viewed as a clue pointing towards supersymmetry.

Since generic SUSY models suffer from the ``SUSY  CP problem'', EDMs should be analyzed in classes of models which allow for their  suppression. These include
models with either small CP phases or heavy spectra. $d_n $-$d_e$ EDM correlations have been analyzed in mSUGRA, the decoupling scenario with 2 heavy sfermion
generations, and split SUSY \cite{Abel:2005er}. Assuming that the SUSY CP phases are all of the same order of magnitude at the GUT scale, one finds
\begin{eqnarray}
{\rm mSUGRA}~: ~~ && d_e \sim 10^{-1} d_n \nonumber \\
{\rm split~ SUSY }~: ~~ && d_e \sim 10^{-1} d_n \nonumber \\
{\rm decoupling }~: ~~ && d_e \sim (10^{-1}- 10^{-2}) d_n \nonumber \;.
\end{eqnarray}
These results are insensitive to $\tan\beta$ and order one variations in the mass parameters. The
$d_e/d_n$ ratio is dominated by the factor  $m_e /m_q \sim 10^{-1} $, although different diagrams contribute to $d_e$ and $d_n$.

\begin{figure}[t]
\parbox{0.51\linewidth}{\includegraphics[width=\linewidth]{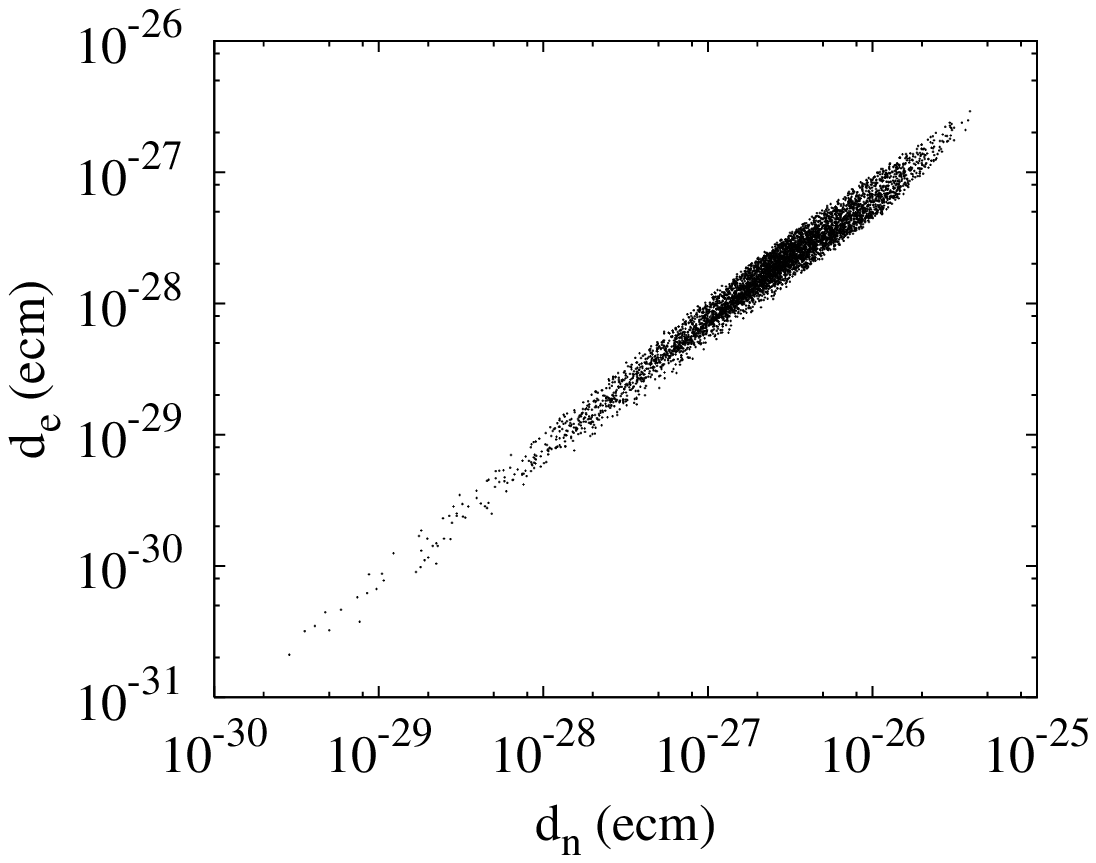}
\caption{EDM correlation in mSUGRA.\newline~\label{f1}}}\hspace*\fill 
\parbox{0.45\linewidth}{\includegraphics[width=\linewidth]{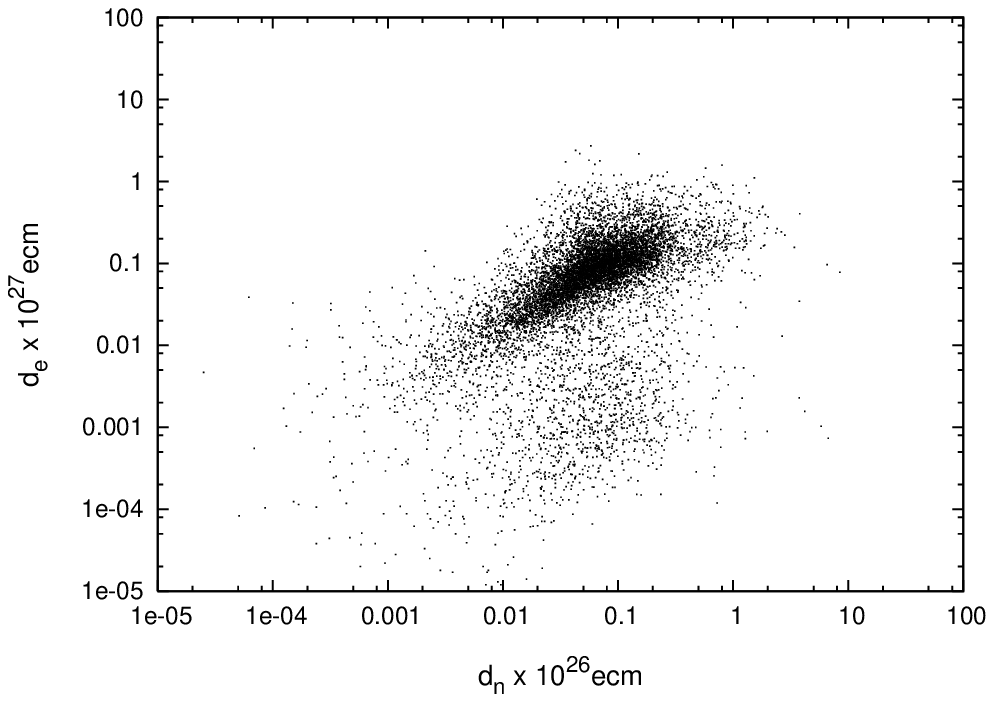}
\caption{EDM correlation in non--universal SUSY models.\label{f2}}}
\end{figure}

An example of the $d_n $-$d_e$ correlation in mSUGRA is presented in Fig.\ref{f1}. There $m_0, m_{1/2}, \vert A\vert$ are varied randomly
in the range [200 GeV, 1 TeV], $\tan\beta=5$ and the phase of the $\mu$-term $\phi_\mu$ is taken to be in the range
[-$\pi$/500, $\pi$/500]. The effect of the phase of the A--terms, $\phi_A$, is negligible as long is it is of the same order of magnitude as
$\phi_\mu$ at the GUT scale. Clearly, the relation $d_n / d_e \sim 10$ holds for essentially all parameter values.

As the next step, we would like to see how stable these correlations are. One might expect that  breaking universality at the GUT scale
would completely invalidate the above results. To answer this question, we study a non--universal MSSM parameterized   by
\begin{eqnarray}
&& m_{\rm squark}~, ~ m_{\rm slepton}~, ~M_3~,~M_1=M_2~,~\vert A \vert \nonumber\\ && \phi_\mu ~,~ \phi_A ~,~ \phi_{M_3}
\end{eqnarray}
at the GUT scale. The mass parameters are varied randomly in the range [200 GeV, 2 TeV] and the phases in the range [-$\pi$/300, $\pi$/300],
$\tan\beta=5$. We find  that although the correlation is not as precise as in the mSUGRA case, about 90\% of the points satisfy
the relation $d_n / d_e \sim 10-100$ (Fig.\ref{f2}). In most of the remaining 10\%, $  10^4 > d_n/d_e > 10^2 $,
which arise when the gluino phase dominates. The reason for the correlation is that
in most cases $\phi_\mu$ is significant  and induces both $d_n$ and $d_e$. Apart from the factor $m_q/m_e$, the SUSY EDM diagrams are comparable
as long as there are no large mass hierarchies in the SUSY spectrum. This means that the EDM correlation survives to a large extent, although
it is possible to violate it in certain cases.

It is instructive to compare the SUSY EDM `prediction' to those of other models. Start with the standard model.  The SM background due to the CKM phase is very small,
probably  beyond the experimental reach. The neutron EDM can also be induced by the QCD $\theta$--term,
\begin{equation}
d_n \sim 3 \times 10^{-16} ~\theta ~ e\; {\rm cm} \;,
\end{equation}
which does not affect  the electron EDM. Thus, one has $d_n >\! > \! > d_e$.

In extra dimensional models, usually there are no extra sources of CP-violation and the EDM predictions are similar to the SM values.
Two Higgs doublet models have additional sources of CP-violation, however, the leading EDM contributions appear at 2 or 3 loops such that the typical EDM values
are significantly smaller than those in SUSY models.

To conclude, we find that typical SUSY models predict   $\vert d_n \vert / \vert d_e \vert \sim {\cal O}(10)-{\cal O}(100)$. Thus, if
\begin{eqnarray}
&& d_e > d_n
\end{eqnarray}
or
\begin{eqnarray}
&& d_e < \!< d_n
\end{eqnarray}
is found, common SUSY scenarios would be disfavoured, although  such relations could still be obtained in baroque SUSY models.

It is interesting to consider SUSY GUT model, where CP phases in the neutrino Yukawa couplings contributes to hadronic EDMs.
For instance, in SU(5) SUSY GUT with right-handed neutrinos, not only large mixing but also CP-violating phases in neutrino
sector give significant contribution to the mixing and CP phases in the right-handed scalar down sector.
Though 1--2 mixing in the neutrino Yukawa coupling is strongly restricted by the $B(\mu\to e\gamma)$, 2--3 mixing in the neutrino
Yukawa couplings can be significantly large and this case is interesting in $B$ physics.
Large 2--3 mixing with CP-violation in neutrino sector may give a significant contribution not only to the $B(\tau\to \mu\gamma)$
but also to color EDM of $s$ quark which may affect \cite{Ibrahim:1998je,Hisano:2004tf} neutron and Hg EDM.

\newcounter{mysubequation}[equation]

\subsubsection{EDMs in split supersymmetry}

Supersymmetry-breaking terms involve many new sources of CP-violation. Particularly worrisome are the phases associated with the invariants
${\rm arg}(A^* M_{\tilde g})$ and ${\rm arg}(A^* B)$, where $A$ and $B$ represent the usual trilinear and bilinear soft terms and $M_{\tilde g}$
the gaugino masses. Such phases survive in the universal limit in which all the flavour structure originates from the SM Yukawa's. If these phases
are of order one, the electron and neutron EDMs induced at one-loop by gaugino-sfermion exchange are typically (barring accidental
cancellations~\cite{Ibrahim:1998je,Brhlik:1999ub,Abel:2001mc}) 
a couple of orders of magnitude above the 
limits~\cite{Ellis:1982tk,Abel:2001vy,Demir:2003js,Olive:2005ru},
a difficulty which is known as the supersymmetric CP problem.

Different remedies are available to this problem making the one loop sfermion contribution to the EDMs small enough, each with its pros and cons. 
One remedy is to have heavy enough sfermions (say heavier than $50$--$100\TeV$ 
to be on the safe side). Gauginos and Higgsinos are not required to be heavy, and can be closer to the electroweak scale, thus preserving the
supersymmetric solution to the dark matter problem and gauge coupling unification. This is the ``Split'' limit of the
MSSM~\cite{Arkani-Hamed:2004fb,Giudice:2004tc,Arkani-Hamed:2004yi}. In this limit, the heavy sfermions suppress the dangerous one-loop contributions
to a negligible level. Nevertheless, some phases survive below the sfermion mass scale and, if they do not vanish for an accidental or a symmetry
reason, they give rise to EDMs that are safely below the experimental limits, but sizeable enough to be well within the sensitivity of the next
generation of experiments~\cite{Arkani-Hamed:2004yi,Giudice:2005rz,Chang:2005ac,Abel:2005er}. Such contributions only arise at the two-loop level,
since the new phases appear in the gaugino-Higgsino sector, which is not directly coupled to the SM fermions. 

Besides the large EDMs, a number of additional unsatisfactory issues, all related to the presence of TeV scalars, plague the MSSM. The number of
parameters exceeds 100; flavour changing neutral current processes are also one or two orders of magnitudes above the experimental limits in most of
the wide parameter space; in the context of a grand unified theory, the proton decay rate associated to
sfermion-mediated dimension 5 operators is ruled out by the SuperKamiokande limit, at least in the minimal version of the supersymmetric SU(5) model;
in the supergravity context, another potential problem comes from the gravitino decay, whose rate is slow enough to interfere with primordial
nucleosynthesis. While none of those issues is of course deadly --- remedies are well known for each of them  --- it should be noted that the split
solution of the supersymmetric CP problem also solves all of those issues at once. At the same time, it gives rise to a predictive framework,
characterized by a rich, new phenomenology, mostly determined in terms of only 4 relevant parameters. Of course the price to be paid to make the
sfermions heavy is the large fine-tuning (FT) necessary to reproduce the Higgs mass, which exacerbates the FT problem already present in the MSSM.
This could be hard to accept, or not, depending on the interpretation of the FT problem, the two extreme attitudes being i) ignoring the problem, as
long as the tuning is not much worse than permille and ii) accepting a tuning in the Higgs mass as we accept the tuning of the cosmological constant,
as in Split Supersymmetry. The second possibility can in turn be considered as a manifestation of an anthropic selection
principle~\cite{Linde:1983gd,Weinberg:1987dv,Agrawal:1997gf,Bousso:2000xa,Arkani-Hamed:2005yv}. 

Before moving the quantitative discussion of the effect, let us note that the pure gaugino-Higgsino contribution to the EDMs, dominant in Split
Supersymmetry and possibly near the experimental limit, might also be important in the non-split case, depending on the mechanism invoked to push the
one-loop sfermion contribution below the experimental limit. 

\subsubsubsection{Sources of CP-violation in the split limit}

Below the heavy sfermion mass scale, denoted generically by $\tilde m$, the MSSM gauginos and Higgsinos, together with the SM fields constitute the
field content of the model. The only interactions of gauginos and Higgsinos besides the gauge ones are 
\begin{equation}
\label{eq:couplings}
-\mathcal{L} = 
\sqrt{2}\left( {\tilde g}_u H^\dagger \tilde W^a T_a \tilde H_u +  {\tilde g}^\prime_u
Y_{H_u}H^\dagger \tilde B \tilde H_u +  {\tilde g}_d H_c^\dagger
\tilde W^a T_a \tilde H_d +  {\tilde g}^\prime_d 
Y_{H_d}H_c^\dagger \tilde B \tilde H_d \right) +{\rm h.c.},
\end{equation}
where the Higgs-Higgsino-gaugino couplings ${\tilde g}_u$, ${\tilde g}_d$, ${\tilde g}^\prime_u$, ${\tilde g}^\prime_d$ can be expressed in terms of the gauge couplings and $\tan\beta$
through the matching with the supersymmetric gauge interactions at the scale $\tilde m$, $H_c = i\sigma_2 H^*$, $T_a$ are the $SU(2)$ generators, and
$Y_{H_u}=-Y_{H_d}=1/2$. CP-violating phases can enter the effective Lagrangian below the sfermion mass scale $\tilde m$ through the $\mu$-parameter,
the gaugino masses $M_i$, $i=1,2,3$, or the couplings ${\tilde g}_u$, ${\tilde g}_d$, ${\tilde g}^\prime_u$, ${\tilde g}^\prime_d$ (besides of course the Yukawa couplings, not relevant
here). Only three combinations of the phases of the above parameters are physical, in a basis in which the Higgs VEV is in its usual form,
$\left\langle H\right\rangle = (0,v)^T$, with $v$ positive. The three combinations are $\phi_1 = \arg({\tilde g}^{\prime *}_u {\tilde g}^{\prime *}_d M_1\mu)$, $\phi_2 = \arg({\tilde g}^*_u {\tilde g}^*_d M_2\mu)$,
$\xi = \arg({\tilde g}_u {\tilde g}^*_d {\tilde g}^\prime_d {\tilde g}^{\prime *}_u )$. Actually, the parameters above are not independent themselves. The tree-level matching with the full
theory above $\tilde m$ gives in fact $\arg({\tilde g}_u) = \arg({\tilde g}^\prime_u)$, $\arg({\tilde g}_d) = \arg({\tilde g}^\prime_d)$. As a consequence, the phase $\xi$ vanishes, thus
leaving only two independent phases. Moreover, if the phases of $M_1$ and $M_2$ are equal, as in most models of supersymmetry breaking, there is
actually only one CP-invariant: $\phi_2 = \arg({\tilde g}^*_u {\tilde g}^*_d M_2\mu)$. 

In terms of mass eigenstates, the relevant interactions are 
\begin{multline}
\label{eq:interactions}
-\mathcal{L} = \frac{g}{c_W} \overline{\chi^+_i} \gamma^\mu (G^R_{ij}P_R + G^L_{ij}P_L)\chi^+_j Z_\mu \\
+ \left[ g \overline{\chi^+_i} \gamma^\mu (C^R_{ij}P_R + C^L_{ij}P_L) \chi^0_j W^+_\mu +\frac{g}{\sqrt{2}} \overline{\chi^+_i}
  (D^R_{ij}P_R +D^L_{ij}P_L) \chi^+_j h +\text{h.c.}  \right] ,
\end{multline}
where
\refstepcounter{equation}\label{eq:mixings}
\begin{align}
&G^L_{ij} = V^{\phantom{\dagger}}_{iW^+}c_{W^+}V^\dagger_{W^+j} + V^{\phantom{\dagger}}_{ih_u^+}c_{h_u^+}V^\dagger_{h_u^+j} &
-&{G^{R}_{ij}}^* = U^{\phantom{\dagger}}_{iW^-}c_{W^-}U^\dagger_{W^-j} + U^{\phantom{\dagger}}_{ih^-_d}c_{h_d^-}U^\dagger_{h_d^-j} \stepcounter{mysubequation} \\
&C^L_{ij} = -V^{\phantom{*}}_{iW^+} N^*_{jW_3} +\frac{1}{\sqrt{2}} V^{\phantom{*}}_{ih^+_u}N^*_{jh^0_u} &
&C^R_{ij} = -U^*_{iW^-} N^{\phantom{*}}_{jW_3} -\frac{1}{\sqrt{2}} U^*_{ih^-_d}N^{\phantom{*}}_{jh^0_d} \stepcounter{mysubequation} \\
g&D^R_{ij} = {\tilde g}_u^* V_{ih^+_u}U_{jW^-} +{\tilde g}_d^* V_{iW^+}U_{jh^-_d} & &D^L=(D^R)^\dagger . \stepcounter{mysubequation}
\end{align}
In Eq.~(\ref{eq:mixings}a), $c_f = T_{3f} -s_W^2 Q_f$ ($s_W^2\equiv \sin^2\theta_W$) is the neutral current coupling coefficient of the fermion $\tilde
f$ and, accordingly, $c_{W^\pm}=\pm \cos^2\theta_W$, $c_{h_u^+,h_d^-}= \pm(1/2-s^2_W)$. The matrices $U$, $V$, $N$ diagonalize the complex chargino and
neutralino mass matrices, $M_{+} = U^TM^D_{+} V$, $M_0 = N^T N^D_0 N$, where $M^D_+ = \operatorname{Diag}(M^+_1,M^+_2)\geq 0$, $M^D_0 =
\operatorname{Diag}(M^0_1,\ldots,M^0_4)\geq 0$ and 
\begin{equation} 
 \label{eq:charginos}
 M_+ = \begin{pmatrix} M_2 & {\tilde g}_u v \\ {\tilde g}_d v & \mu \end{pmatrix}, \quad M_0 =
 \begin{pmatrix}
   M_1 & 0 & -{\tilde g}^\prime_d v/\sqrt{2} & {\tilde g}^\prime_u v/\sqrt{2} \\
   0 & M_2 & {\tilde g}_d v/\sqrt{2} & -{\tilde g}_u v/\sqrt{2} \\
   -{\tilde g}^\prime_d v/\sqrt{2} & {\tilde g}_d v/\sqrt{2} & 0 & -\mu \\
   {\tilde g}^\prime_u v/\sqrt{2} & -{\tilde g}_u v/\sqrt{2} & -\mu & 0
 \end{pmatrix}.
\end{equation}

\subsubsubsection{Two loop contributions to EDMs}

\begin{figure}
\begin{center}
\includegraphics[width=\linewidth]{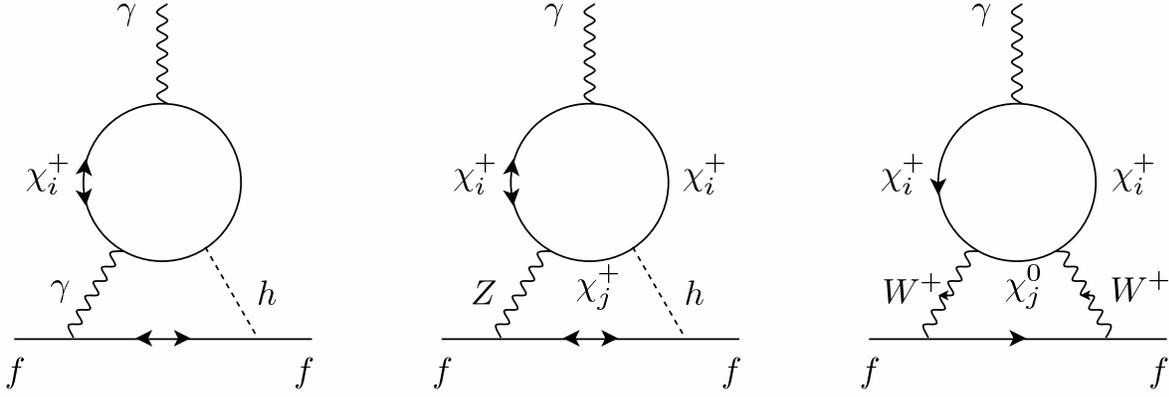}
\end{center}
\caption{Two loop contributions to the light SM fermion EDMs. The
third diagram is for a down-type fermion $f$.}
\label{fig:feynman}
\end{figure}

Fermion EDMs are generated only at two loops, since charginos and
neutralinos, which carry the information on CP-violation, are only
coupled to gauge and Higgs bosons. Three diagrams contribute to the
EDM of the light SM fermion $f$ at the two-loop level. They are
induced by the effective $\gamma\gamma h$, $\gamma Z h$, and $\gamma
WW$ effective couplings and are shown in Fig.~\ref{fig:feynman}. The
EDM $d_f$ of the fermion $f$ is then given by~\cite{Giudice:2005rz}
\begin{equation}
 \label{eq:d}
\phantom{\text{where}} \qquad d_f = d^{\gamma H}_f + d^{ZH}_f +
d^{WW}_f , \qquad\text{where}
\end{equation}
\refstepcounter{equation}\label{eq:dipoles}
\begin{align}
 d^{\gamma H}_f & = \frac{e Q_f\alpha^2}{4\sqrt{2}\pi^2s^2_W}
 \operatorname{Im}(D^R_{ii}) \frac{m_f M^+_i}{M_W m^2_H} f_{\gamma
 H}(r^+_{iH}) \stepcounter{mysubequation} \\ d^{ZH}_f &= \frac{e
 \left(T_{3f_L} -2 s^2_W Q_f \right)
 \alpha^2}{16\sqrt{2}\pi^2c^2_Ws^4_W} \operatorname{Im} \left(
 D^R_{ij}G^R_{ji} -D^L_{ij}G^L_{ji} \right) \frac{m_f M^+_i}{M_W
 m^2_H} f_{ZH}(r^{\phantom{\dagger}}_{ZH},r^+_{iH},r^+_{jH})
 \stepcounter{mysubequation} \\ d^{WW}_f & = \frac{eT_{3f_L}
 \alpha^2}{8\pi^2s^4_W} \operatorname{Im} \left( C^L_{ij}C^{R*}_{ij}
 \right) \frac{m_f M^+_i M^0_j}{M^4_W} f_{WW}(r^+_{iW},r^0_{jW})
 . \stepcounter{mysubequation}
\end{align}
In Eq.~(\ref{eq:dipoles}) a sum over indices $i,j$ is understood,
$Q_f$ is the charge of the fermion $f$, $T_{3f_L}$ is the third
component of the weak isospin of the fermion's left-handed
component. Also, $r_{ZH} = (M_Z/m_H)^2$, $r^+_{iH} = (M^+_i/m_H)^2$,
$r^+_{iW} = (M^+_i/M_W)^2$, $r^0_{iW} = (M^0_i/M_W)^2$, where $m_H$ is
the Higgs mass, and the loop functions are given by
\refstepcounter{equation}\label{eq:loopfunctions}
\begin{align}
 f_{\gamma H}(r) &= \int^1_0\frac{dx}{1-x}\,
 j\left(0,\frac{r}{x(1-x)}\right) \stepcounter{mysubequation} \\ f_{Z
 H}(r, r_1,r_2) &= \frac{1}{2}\int^1_0 \frac{dx}{x(1-x)}\, j\left(r,
 \frac{x r_1+(1-x) r_2}{x(1-x)}\right) \stepcounter{mysubequation} \\
 f_{WW}(r_1,r_2) &= \int^1_0\frac{dx}{1-x}\, j\left(0, \frac{x r_1 +
 (1-x) r_2}{x(1-x)}\right) . \stepcounter{mysubequation}
\end{align}
Their analytic expressions can be found in
Ref.~\cite{Giudice:2005rz}. The symmetric loop function $j(r,s)$ is
defined recursively by
\begin{equation}
 \label{eq:basicfunctions}
 j(r) = \frac{r\log r}{r-1}, \quad j(r,s) = \frac{j(r)-j(s)}{r-s}.
\end{equation}

Eq.~(\ref{eq:dipoles}) hold at the chargino mass scale $M^+$. The
neutron EDM is determined as a function of the down and up quark
dipoles at a much lower 
scale $\mu$, at which 
\begin{equation}
 \label{eq:eta}
d_q(\mu)=\eta_{\text{QCD}}~ d_q(M^+),\quad \eta_{\text{QCD}}=\left[
\frac{\alpha_s(M^+)}{\alpha_s(\mu)}\right]^{\gamma /2b},
\end{equation}
where the $\beta$-function coefficient is $b=11-2n_q/3$ and $n_q$ is
the number of effective light quarks. The anomalous-dimension
coefficient is $\gamma =8/3$. For $\alpha_s (M_Z)=0.118\pm 0.004$ and
$\mu=1\, {\rm GeV}$ (the scale of the neutron mass), the value of
$\eta_{\text{QCD}}$ is 0.75 for $M^+ =1\TeV$ and $0.77$ for $M^+
=200\, {\rm GeV}$. We expect an uncertainty of about 5\% from
next-to-leading order effects. This result~\cite{Giudice:2005rz} gives
a QCD renormalization coefficient about a factor of 2 smaller than
usually considered~\cite{Arnowitt:1990eh}, and it agrees with the
recent findings of Ref.~\cite{Degrassi:2005zd}.

The neutron EDM can be expressed in terms of the quark EDMs using QCD
sum rules~\cite{Pospelov:1999ha,Pospelov:2000bw}:
\begin{equation}
 \label{eq:neutron}
 d_n = (1\pm 0.5) \frac{f_\pi^2 m_\pi^2}{(m_u+m_d) (225\MeV)^3}
 \left(\frac{4}{3} d_d -\frac{1}{3} d_u \right) ,
\end{equation}
where $f_\pi \approx 92\MeV$ and we have neglected the contribution of
the quark chromo-electric dipoles, which does not arise at the
two-loop level in the heavy-squark mass limit. Since $d_d$ and $d_u$
are proportional to the corresponding quark masses, $d_n$ depends on
the light quark masses only through the ratio $m_u/m_d$, for which we
take the value $m_u/m_d = 0.553\pm 0.043$.

It is instructive to consider the limit $M_i, \mu \gg M_Z, m_H$ which
 simplifies the EDM dependence on the CP-violating invariants
 $|{\tilde g}_u {\tilde g}_d / M_2 \mu|\sin \phi_2$ and $|{\tilde
 g}^\prime_u {\tilde g}^\prime_d / M_1 \mu| \sin \phi_1$. The terms
 depending on the second invariant are actually suppressed, so that
 both the electron and neutron EDM are mostly characterized by a
 single invariant even in the case in which the phases of $M_1$ and
 $M_2$ are different. The relative importance of the three
 contributions to $d_f$ in Eq.~(\ref{eq:d}) can be estimated to
 leading order in $\log(M_2\mu/m^2_H)$ from
\begin{equation}
\label{eq:largelog2}
\frac{d_f^{ZH}}{d_f^{\gamma H}} \approx
\frac{(T_{3f_L}-2s^2_WQ_f)(3-4s_W^2)}{8c^2_WQ_f} \quad ; \quad
\frac{d_f^{WW}}{d_f^{\gamma H}} \approx -\frac{T_{3f_L}}{8s^2_WQ_f}
\quad \text{($M_2=\mu$)}.
\end{equation}
Numerically, Eq.~(\ref{eq:largelog2}) gives $d_e^{ZH} \approx
0.05\,d_e^{\gamma H}$, $d_e^{WW} \approx -0.3 \,d_e^{\gamma H}$ and
$d_n^{ZH} \approx d_n^{\gamma H}$, $d_n^{WW} \approx -0.7
\,d_n^{\gamma H}$. These simple estimates show the importance of the
$ZH$ contribution to the neutron EDM.

\subsubsubsection{Numerical results}

\begin{figure}
\begin{center}
\includegraphics[width=0.8\textwidth]{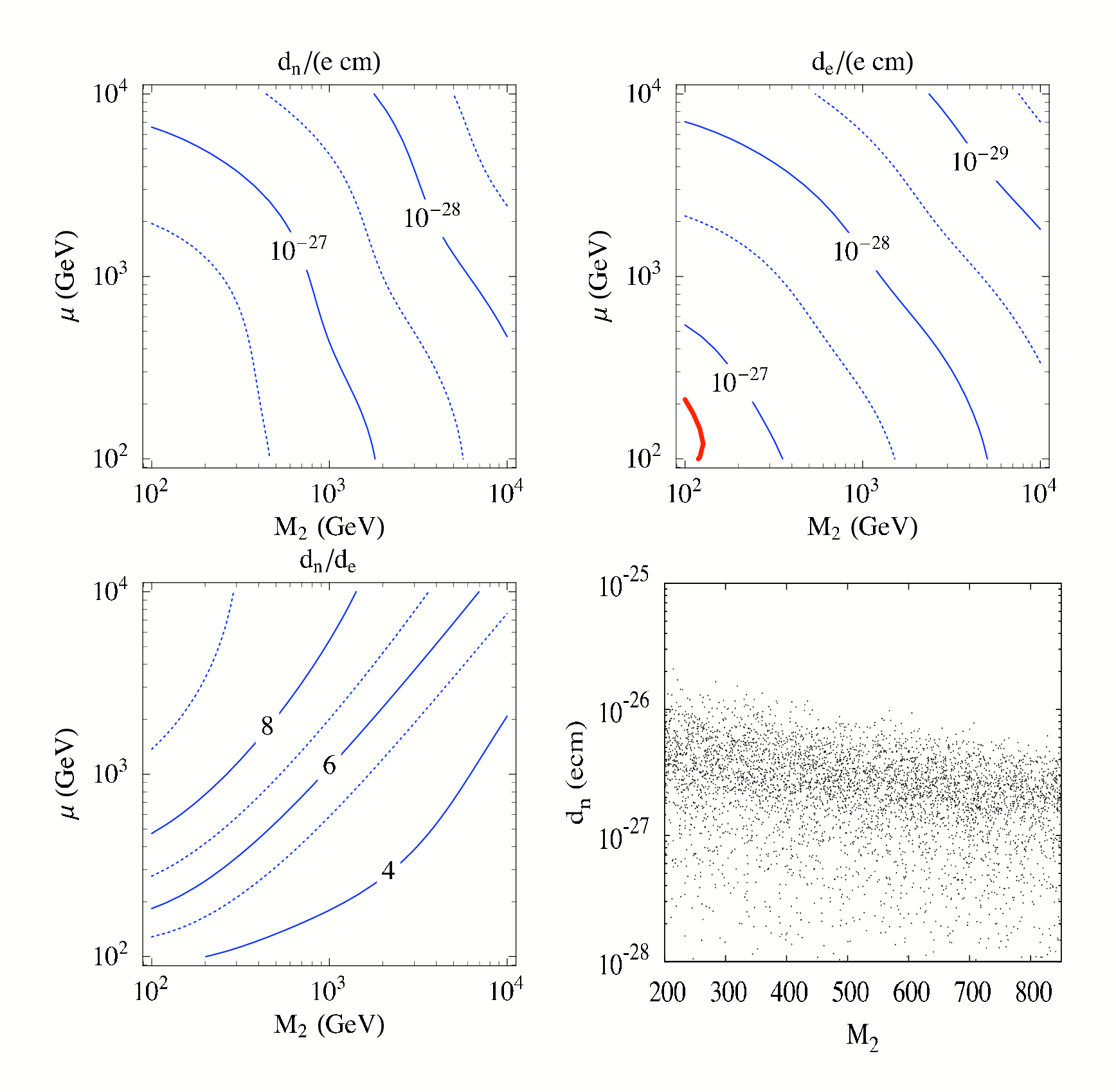}
\end{center}
\caption{Prediction for $d_n$, $d_e$, and their ratio $d_n/d_e$. In the contour plots we have chosen $\tan\beta = 1$, $\sin\phi = 1$, and $\tilde m =
10^9\, {\rm GeV}$. The results for $d_n$ and $d_e$ scale approximately linearly with $\sin 2\beta \sin\phi$, while the ratio is fairly independent of
$\tan\beta$, $\sin\phi$ and $\tilde m$. The red thick line corresponds to the present experimental limit
$d_e< 1.6\times 10^{-27}e\,\text{cm}$~\cite{Regan:2002ta}. Note that the uncertainty in $d_n$ is a factor of a few. The scatter plot shows $d_n$ values
when $M_{1,3}$ and $\mu$ are varied in the range $[200\, {\rm GeV}, 1\TeV]$, $m_h$ in $[100\, {\rm GeV}, 300\, {\rm GeV}]$
and the CP phase in the range $[-\pi,\pi]$.}
\label{fig:edm}
\end{figure}

Let us consider a standard unified framework for the gaugino masses at
the GUT scale. Using the RGEs given in
Refs.~\cite{Giudice:2004tc,Arvanitaki:2004eu}, the parameters in
Eq.~(\ref{eq:dipoles}) can be expressed in terms of the (single) phase
$\phi\equiv\phi_2$ and four positive parameters $M_2$, $\mu$
(evaluated at the low-energy scale), $\tan\beta$, and the sfermion
mass scale $\tilde m$. In first approximation, the dipoles depend on
$\beta$ and $\phi$ through an overall factor $\sin2\beta\sin\phi$.
The overall sfermion scale $\tilde m$ enters only logarithmically
through the RGE equations for ${\tilde g}_{u,d}$, ${\tilde
g}_{u,d}^\prime$. The numerical results for the electron and neutrino
EDMs can then conveniently be presented in the $M_2$--$\mu$ plane by
setting $\sin2\beta\sin\phi =1$ (it is then sufficient to multiply the
results by $\sin2\beta\sin\phi$) and, for example, $\tilde m = 10^9\,
{\rm GeV}$. Figure~\ref{fig:edm} shows the prediction for the electron
EDM, the neutron EDM, and their ratio $d_n/d_e$.  The red thick line
corresponds to the present experimental limits $d_e< 1.6\times
10^{-27}e\,\text{cm}$~\cite{Regan:2002ta}, while the limit
$d_n<3\times 10^{-26}e\,\text{cm}$~\cite{Baker:2006ts} does not impose
a constraint on the parameters shown in Fig.~\ref{fig:edm}.

An interesting test of Split Supersymmetry can be provided by a
measurement of both the electron and the neutron EDMs.  Indeed, in the
ratio $d_n/d_e$ the dependence on $\sin\phi$, $\tan\beta$ and $\tilde
m$ approximately cancels out. Nevertheless, because of the different
loop functions associated with the different contributions, the ratio
$d_n/d_e$ varies by $\mathcal{O}\left(100\%\right)$ when $M_2$ and
$\mu$ are varied in the range spanned in the figures. Still, the
variation of $d_n/d_e$ is comparable to the theoretical uncertainty in
Eq.~(\ref{eq:neutron}), and is significantly smaller than the
variation in the ordinary MSSM prediction, even in the case of
universal phases~\cite{Abel:2005er}\footnote{Note that the $ZH$
contribution is missing in the analysis of the Split Supersymmetry
case in Ref.~\cite{Abel:2005er}, which leads to a somewhat stronger
correlation between $d_e$ and $d_n$.}. On the other hand, the usual
tight correlation between the electron and muon EDMs, $d_\mu/d_e =
m_\mu/m_e$ persists.

\section{Experimental tests of charged lepton
  universality}
\label{sec:exp:universality}
Lepton universality postulates that lepton interactions do not
depend explicitly on lepton family number other than through their
different masses and mixings. The concept can be generalized to
include the quarks. Whereas there is little doubt about the
universality of electric charge there are scenarios outside the
Standard Model in which lepton universality is violated in the
interactions with $W$ and $Z$ bosons. Violations may also have their
origin in non-SM contributions to the transition amplitudes.  Such
apparent violations of lepton universality can be expected in various
particle decays:
\begin{list}{-}{}
\item
in $W$, $Z$ and $\pi$ decay resulting from R-parity violating extensions to the MSSM \cite{Lebedev:1999vc,Ramsey-Musolf:2000qn},
\item
in $W$ decay resulting from charged Higgs bosons \cite{Lebedev:2000ix,Park:2006gk},
\item
in $\pi$ decay resulting from box diagrams involving non-degenerate sleptons \cite{Ramsey-Musolf:2006vr},
\item
in $K$ decay resulting from LFV contributions in SUSY \cite{Masiero:2005wr} (see Sec.\ref{sec:unitarity}),
\item
in $\Upsilon$ decay resulting from a light Higgs boson \cite{Sanchis-Lozano:2006gx},
\item
in $\pi$ and K decay from scalar interactions \cite{Campbell:2003ir}, enhanced by the strong chiral suppression of the SM amplitude for decays into $e \overline{\nu}_e$ .
Since these contributions result in interference terms with the SM amplitude the deviations scale with the mass M of the exchange particle like $1/M^2$ rather than $1/M^4$
as may be expected naively.
\end{list}

Allowing for universality violations one can generalize the $V-A$ charged current weak interaction of leptons to
\footnote{Still more general violations lead to deviations from the $1-\gamma_5$ structure of the weak interaction.}:
\begin{equation}
{\mathcal L}=\sum_{l=e,\mu,\tau} \frac{g_l}{\sqrt{2}}W_{\mu} \overline{\nu}_l \gamma^{\mu} (\frac{1-\gamma_5}{2})l + {\rm h.c.}
\label{eq:nonuniversality}
\end{equation}

Experimental limits have recently been compiled by Loinaz {\it et al.}~\cite{Loinaz:2004qc}. Results are shown in Table~\ref{Tab:universalitylimits}.
\begin{table}[bht]
\begin{minipage}{\textwidth}
\caption{
Limits on lepton universality from various processes. One should keep in mind that violations may affect the various tests differently so which
constraint is best depends on the mechanism. Hypothetical non V-A contributions, for example, would lead to larger effects in decay modes with stronger
helicity suppression such as $\pi \to e \nu$ and $K \to e
\nu$. Adapted from Ref.~\cite{Loinaz:2004qc}. The ratios estimated
from tau decays are re-calculated using PDG averages, as described in
the text. 
\label{Tab:universalitylimits}}
\begin{tabular*}{\textwidth}{@{\extracolsep{\fill}}clrrl}
\hline\hline
\hspace*{1cm}&decay mode\hspace*{7cm}			&\multicolumn{3}{c}{\hspace*{3cm}constraint\hspace*{4cm}}\\
\hline
&$W   \to e   \; \overline{\nu}_e          	  $&$(g_{\mu} / g_e)_W =        $&0.999 	&$\pm$ 0.011	\\	
&$W   \to \mu \; \overline{\nu}_{\mu}      	  $&$(g_{\tau} / g_e)_W =       $&1.029 	&$\pm$ 0.014	\\
&$W   \to \tau\; \overline{\nu}_{\tau}     	  $&			         &		&		\\[3mm]
&$\mu \to e   \; \overline{\nu}_e     \; \nu_{\mu} $&$(g_{\mu} / g_e)_{\tau} =   $&1.0002 	&$\pm$ 0.0020	\\
&$\tau\to e   \; \overline{\nu}_e     \; \nu_{\tau}$&$(g_{\tau} / g_e)_{\tau\mu}=$&1.0012 	&$\pm$ 0.0023	\\
&$\tau\to \mu \; \overline{\nu}_{\mu} \; \nu_{\tau}$&				 &		&		\\[3mm]
&$\pi \to e   \; \overline{\nu}_e          	  $&$(g_{\mu} / g_e)_{\pi} =	$&1.0021	&$\pm$ 0.0016	\\	
&$\pi \to \mu \; \overline{\nu}_{\mu}      	  $&$(g_{\tau} / g_e)_{\tau\pi}=$&1.0030 	&$\pm$ 0.0034	\\
&$\tau\to \pi \; \overline{\nu}_{\tau}     	  $&				&		&		\\[3mm]
&$K   \to e   \; \overline{\nu}_e          	  $&$(g_{\mu} / g_e)_K =	$&1.024		&$\pm$ 0.020	\\	
&$K   \to \mu \; \overline{\nu}_{\mu}      	  $&$(g_{\tau}/g_{\mu})_{K\tau} =$&0.979 	&$\pm$ 0.017	\\
&$\tau\to K   \; \overline{\nu}_{\tau}     	  $&				&		&		\\
\hline\hline
\end{tabular*}
\end{minipage}
\end{table}

\begin{figure}
\includegraphics[height=0.48\linewidth, angle=90]{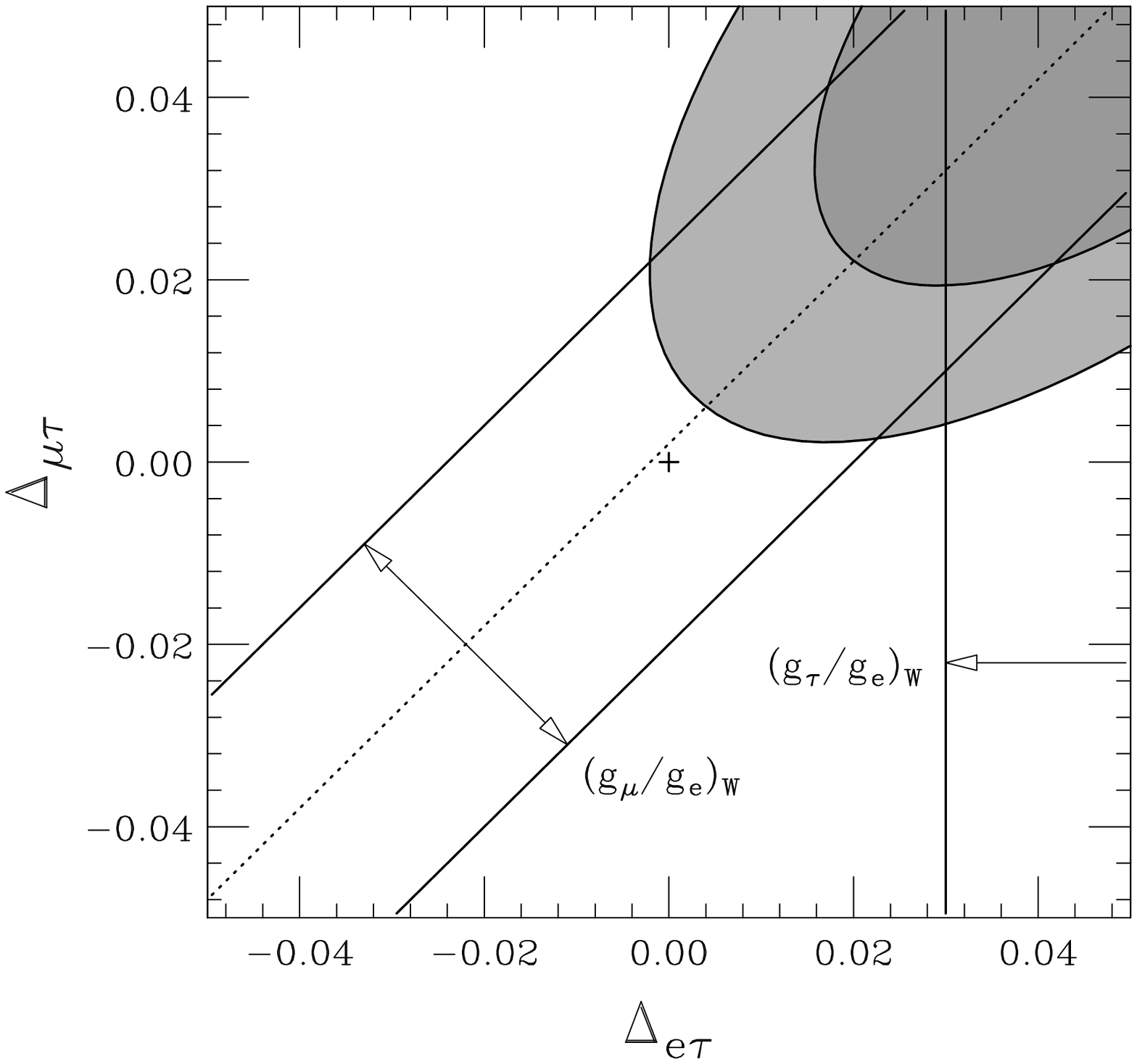}\begin{picture}(0,0) \put(-160,170){\LARGE a)} \end{picture}\hspace*{\fill}
\includegraphics[height=0.48\linewidth, angle=90]{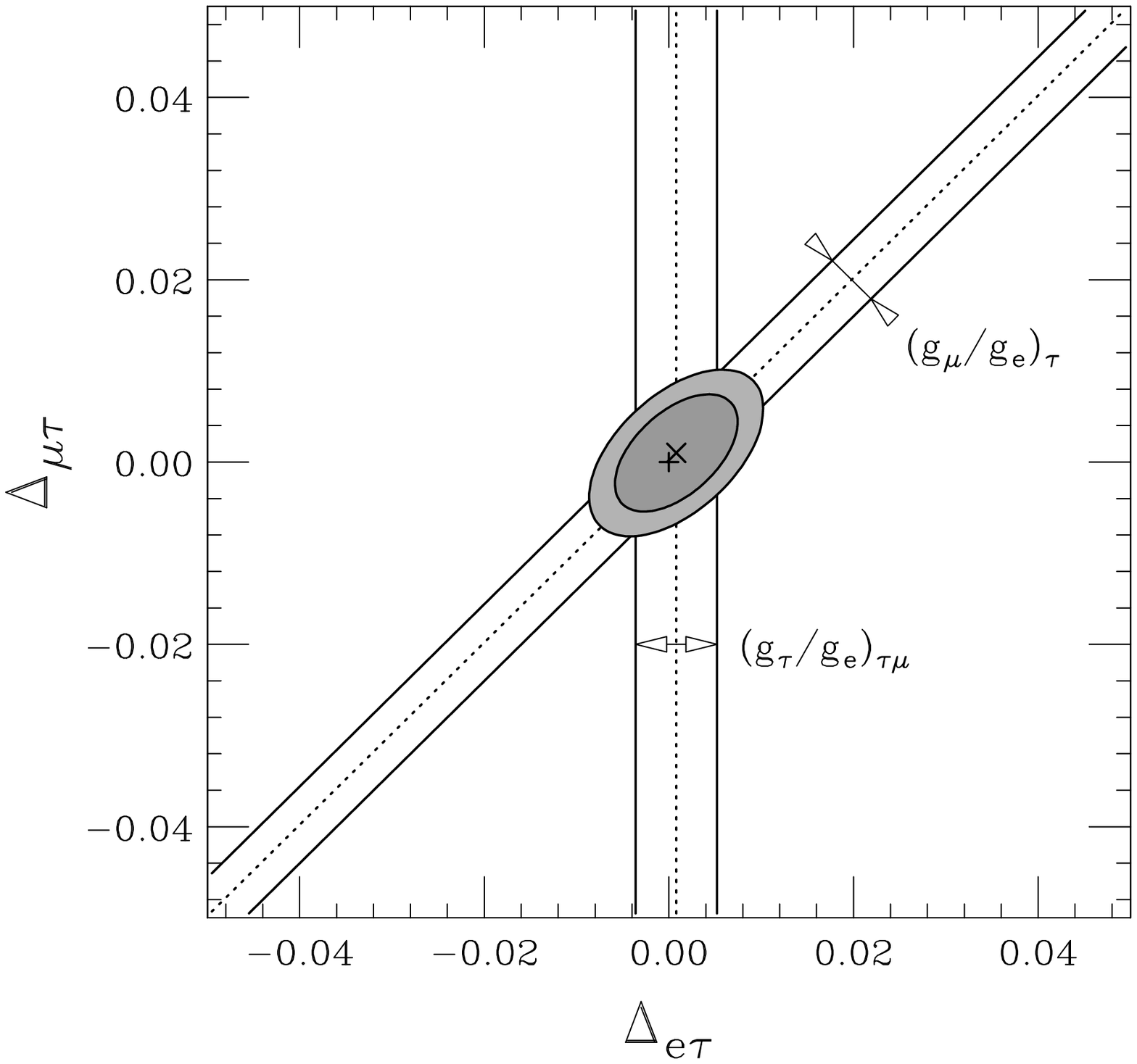}\begin{picture}(0,0) \put(-160,170){\LARGE b)} \end{picture}
\includegraphics[height=0.48\linewidth, angle=90]{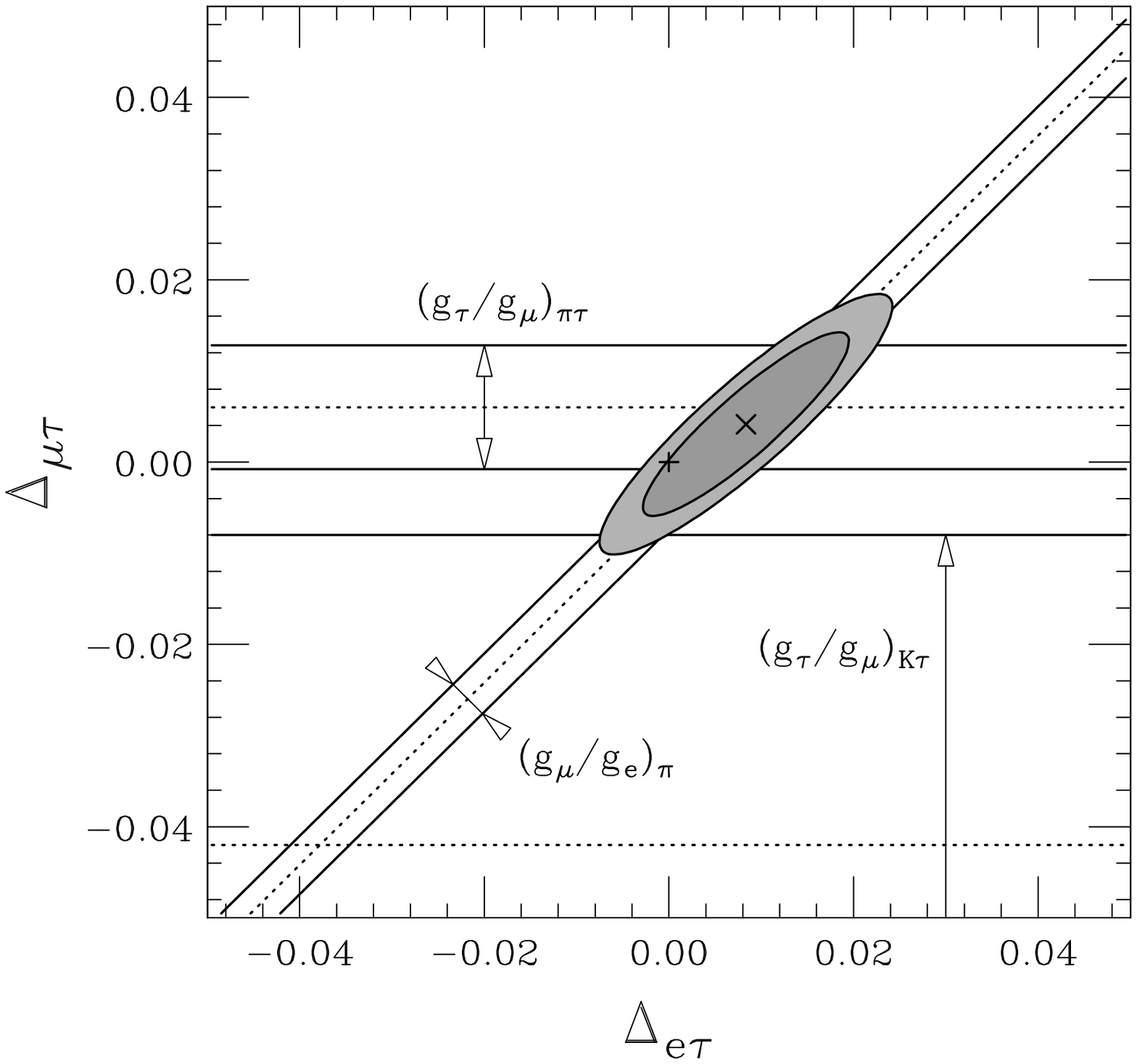}\begin{picture}(0,0) \put(-160,170){\LARGE c)} \end{picture}\hspace*{\fill}
\includegraphics[height=0.48\linewidth, angle=90]{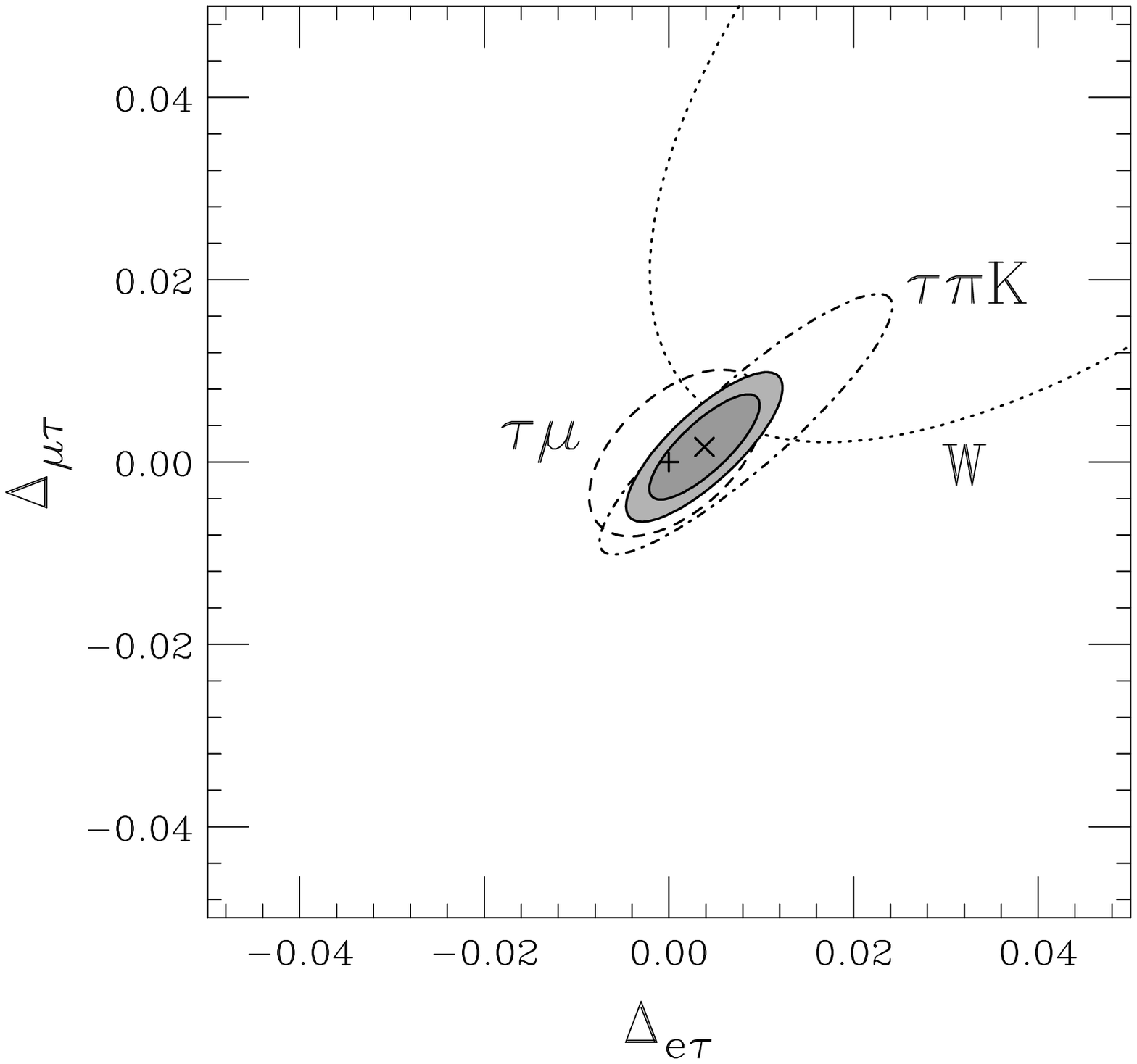}\begin{picture}(0,0) \put(-160,170){\LARGE d)} \end{picture}
\caption{Experimental limits on violations of lepton universality from a) $W$ decay, b) $\tau$ decay, c) $\pi$ and $K$ decay and d) the combination of a) - c). Parameters are
defined in the text. The $\pm 1 \sigma$ bands are indicated. The shaded areas correspond to 68\% and 90\% confidence levels. Results from the analysis in
Ref.~\cite{Loinaz:2004qc}. \label{fig:universality}}
\end{figure}

Following the notation of Ref.~\cite{Loinaz:2004qc} one may parametrize the violations by $g_l \equiv g (1 - \epsilon_l/2)$.
After introducing $\Delta_{ll^{\prime}} \equiv \epsilon_l-\epsilon_{l^{\prime}}$ the various experimental limits on deviations from lepton universality can be compared (see
Fig.~\ref{fig:universality}).

It is very fortunate that for most decay modes new dedicated experiments are being prepared. In the following subsections the status and prospects of these experimental tests of
lepton universality are presented.

\subsection{$\pi$ decay }\label{sec:exp:universality:pi}

In lowest order the decay width of $\pi \to l \overline{\nu}_l$ ($l = e, \mu$) is given by:

\begin{equation}
\Gamma^{\rm tree}_{\pi \to l \overline {\nu}_l} = \frac{g_l^2 g_{ud}^2 V_{ud}^2}{256\pi}\frac{f_{\pi}^2}{M_W^4}m_l^2m_{\pi}(1-\frac{m_l^2}{m_{\pi}^2})^2  \;\; .
\end{equation}
By taking the branching ratio the factors affected by hadronic uncertainties cancel:
\begin{equation}
R_{e/ \mu}^{\rm tree} \equiv \frac{\Gamma^{\rm tree}_{\pi \to e \overline {\nu}}}{\Gamma^{\rm tree}_{\pi \to \mu \overline {\nu}}}=
(\frac{g_e}{g_{\mu}}\times \frac{m_e}{m_{\mu}}\times \frac{1-m_e^2/m_{\pi}^2}{1-m_{\mu}^2/m_{\pi}^2})^2 \nonumber
\end{equation}
Radiative corrections lower this result by 3.74(3)\% \cite{Finkemeier:1994ev} when assuming that final states with additional photons are included.
Within the SM ({\it i.e.} $g_e=g_{\mu}$) this leads to:
\begin{equation}
R_{e/ \mu}^{\rm SM}=1.2354(2) \times 10^{-4} \;\; .
\end{equation}
Two experiments \cite{Czapek:1993kc,Britton:1992pg} contribute to the present world average for the measured value:
\begin{equation}
R_{e/ \mu}^{\rm exp}=1.231(4) \times 10^{-4} \;\; .
\end{equation}
As a result $\mu e$ universality has been tested at the level: $(g_{\mu}/g_e)_{\pi}$=1.0021(16).

Measurements of $R_{e/ \mu}$ are based on the analysis of $e^+$ energy
and time delay with respect to the stopping $\pi^+$. The decay $\pi
\to e \nu$ is characterized by $E_{e^+}=0.5 m_{\pi} c^2 = 69.3$~MeV
and an exponential time distribution following the pion life time
$\tau_{\pi}$=26~ns. In the case of the $\pi \to \mu \nu$ decay the
4~MeV muons, which have a range of about 1.4~mm in plastic
scintillator, can be kept inside the target and are monitored by the
observation of the subsequent decay $\mu \to e \nu \overline{\nu}$,
which is characterized by $E_{e^+} < 0.5 m_{\mu} c^2 = 52.3$~MeV, and a
time distribution which first grows according to the pion life time
and then falls with the muon life time. A major systematic error is
introduced by uncertainties in the low-energy tail of the $\pi \to e
\nu (\gamma)$ energy spectrum in the region below $0.5 m_{\mu}
c^2$. This tail fraction typically amounts to $\approx$1~\% .  The
low-energy tail can be studied by suppressing the $\pi \to \mu \to e$
chain by the selection of early decays and by vetoing events in which
the muon is observed in the target signal. Suppression factors of
typically $10^{-5}$ have been obtained. A study of this region is also
interesting since it might reveal the signal from a heavy sterile
neutrino \cite{Smirnov:2006bu}.

Although the two experiments contributing to the present world average of $R_{e/ \mu}$ reached very similar statistical and systematic errors there were
some significant differences. The TRIUMF experiment \cite{Britton:1992pg} made use of a single large NaI(Tl) crystal as main positron detector, with an energy
resolution of 5\% (fwhm) and a solid-angle acceptance of 2.9~\% of 4$\pi$~sr. The PSI experiment \cite{Czapek:1993kc} used a setup of 132 identical BGO
crystals with 99.8\% of 4$\pi$~sr acceptance and an energy resolution of 4.4\% (fwhm). A large solid angle reduces the low-energy tail of $\pi \to e \nu (\gamma)$
events but may also introduce a high energy tail for $\mu \to e \nu \overline{\nu} \gamma$.

Two new experiments have been approved recently aiming at a reduction of the experimental uncertainty by an order of magnitude. First results may be expected
in the year 2009.

\begin{list}{-}{}
\item
At PSI \cite{PSI_R05_01} the $3\pi$~sr CsI calorimeter built for a determination of the $\pi^+ \to \pi^0 e^+\nu$ branching ratio will be used. Large samples
of $\pi \to e \nu$ decays have been recorded parasitically in the past which were used as normalization for $\pi^+ \to \pi^0 e^+\nu$ with an accuracy of
$<0.3\%$, {\it i.e.} the level of the present experimental uncertainty of $R_{e/ \mu}$. The setup was also used for the most complete studies of the radiative
decays $\pi \to e \nu \gamma$ \cite{Bychkov:2005} and $\mu \to e \nu \overline{\nu}\gamma$ \cite{vanDevender:2005} done so far. Based on this experience an improvement
in precision for $R_{e/ \mu}$ by almost an order of magnitude is expected. 
\item
At TRIUMF \cite{TRIUMF_E1072} a single large NaI(Tl) detector will be
used again. The detector is similar in size to the one used in the
previous experiment but has significantly better energy
resolution. The crystal will be surrounded by CsI detectors to reduce
the low-energy tail of the $\pi \to e \nu$ response function.  By
reducing the distance between target and positron detector the
geometric acceptance will be increased by an order of magnitude.
\end{list}

\subsection{$K$ decay }\label{sec:exp:universality:K}
Despite the poor theoretical control over the meson decay constants,
ratios of leptonic decay widths of pseudoscalar mesons such as $R_K
\equiv \Gamma(K \to e \nu)/\Gamma(K \to \mu \nu)$ can be predicted
with high accuracy, and have been traditionally
considered as tests of the $V-A$ structure of weak interactions
through their helicity suppression and of $\mu-e$ universality. The
Standard Model predicts \cite{Finkemeier:1994ev}:
\begin{equation}
R_K(SM) = (2.472 \pm 0.001) \times 10^{-5}
\end{equation}
to be compared with the world average \cite{Yao:2006px} of published $R_K$ measurements:
\begin{equation}
  R_K(exp) = (2.44 \pm 0.11) \times 10^ {-5} \; .
\end{equation}
As mentioned above the strong helicity suppression of $\Gamma(K \to e
\nu)$ makes $R_K$ sensitive to physics beyond the SM. As discussed in
detail in section~\ref{sec:RKtheory} lepton
flavour violating contributions predicted in SUSY models may lead to a
deviation of $R_K$ from the SM value in the percent range.  Such
contributions, arising mainly from charged Higgs exchange with large
lepton flavour violating Yukawa couplings, do not decouple if SUSY
masses are large and exhibit a strong dependence on $\tan \beta$. For
large (but not extreme) values of this parameter, not excluded by
other measurements, the interference between the SM amplitude and a
double lepton-flavour violating contribution could produce a $-3\%$
effect. Other experimental constraints such as those from $R_{\pi}$ or
lepton flavour-violating $\tau$ decays were shown in
\cite{Masiero:2005wr} not to be competitive with those from $R_K$ in
this scenario.

\subsubsection{Preliminary NA48 results for $R_K$}

In the original NA48/2 proposal \cite{Proposal} the measurement of $K$ leptonic decays was not considered interesting enough to be mentioned. Nevertheless, triggers
for such decays were implemented during the 2003 run. Since these were not very selective they had to be highly down-scaled. The data still contain about 4000 $K_{e2}$
decays which is more than four times the previous world sample. In the analysis of these data \cite{FioriniThesis} $\sim$15\% background due to misidentified $K_{\mu2}$
decays was observed (see below). The preliminary result was presented at the HEP2005 Europhysics conference in Lisbon \cite{FioriniLisbon}:
\begin{equation}
  R_K(exp) = (2.416 \pm 0.043_{\mathrm stat} \pm 0.024_{\mathrm syst})   \times 10^ {-5} \; ,
\end{equation}
marginally consistent with the SM value. While the uncertainty in this result is dominated by the statistical error, the unoptimized $K_{e2}$ trigger and the lack of a
sufficiently large control sample resulted in a $\pm 0.8\%$ uncertainty.

During 2004 a 56 hours special run with simplified trigger logic at $\sim 1/4$ nominal beam intensity was performed, dedicated to the collection of semi-leptonic
$K^{\pm}$ decays for a measurement of $|V_{us}|$. About 4000 $K_{e2}$ decays were extracted from these data. The preliminary result for $R_K$ is consistent
with the 2003 value with similar uncertainty although the trigger efficiencies were better known.

The NA48 apparatus includes the following subsystems relevant for the  $R_K$ measurement
\begin{itemize}
\item  
a magnetic spectrometer, composed of four drift chambers and a dipole magnet (MNP33)
\item 
a scintillator hodoscope consisting of two planes segmented into vertical and horizontal strips, providing a fast level-1 (L1) trigger for charged particles
\item 
a liquid krypton electromagnetic calorimeter (LKr) with an L1 trigger system. 
\end{itemize}

In the analysis of the 2003-04 data $K_{e2}$ decays were selected using two main criteria:
\begin{itemize}
\item
$0.95 < E/pc < 1.05$ where $E$ is the energy deposited in LKr and $p$ is the momentum measured with the magnetic spectrometer.
\item
the missing mass $M_X$ must be zero within errors, as expected for a neutrino.
\end{itemize}
The main background resulted from misidentified $K_{\mu2}$ decays.
The $E/pc$ distribution of muons has a tail which extends to $E/pc \sim 1$ and the observed fraction of muons with $0.95 < E/pc < 1.05$
is $\sim5 \times 10^{-6}$. $K_{\mu2}$ background was present for $p > 25$ GeV/{\it c} where the $M_X$ resolution
provided by the magnetic spectrometer was insufficient to separate $K_{e2}$ from $K_{\mu2}$ decays.

\subsubsection{A new measurement of $\mathbf{\Gamma(K \to e\nu)/ \Gamma(K \to \mu \nu)}$ at the SPS }
During the Summer of 2007 NA62, the evolution of the NA48 experiment,
has accumulated more than 100K $K_{e2}$ decays. For this run the spectrometer
momentum resolution was improved by increasing the MNP33 momentum
kick from 120 MeV/{\it c} to
263 MeV/{\it c}. 

$K_{e2}$ decays are selected by requiring signals from the two
hodoscope planes (denoted by $Q_1$) and an energy deposition of at
least 10 GeV in the LKr calorimeter.  This trigger has an efficiency
$> 0.99$ for electron momenta $p > 15$ GeV/{\it c}. The same
down-scaled $Q_1$ trigger was used to collect $K_{\mu2}$ decays.
The beam intensity was adjusted to obtain a total trigger rate of
$10^4$ Hz, which saturates the data acquisition system.

\begin{figure}[htbp]
\parbox{0.52\linewidth}
{\includegraphics[width=\linewidth]{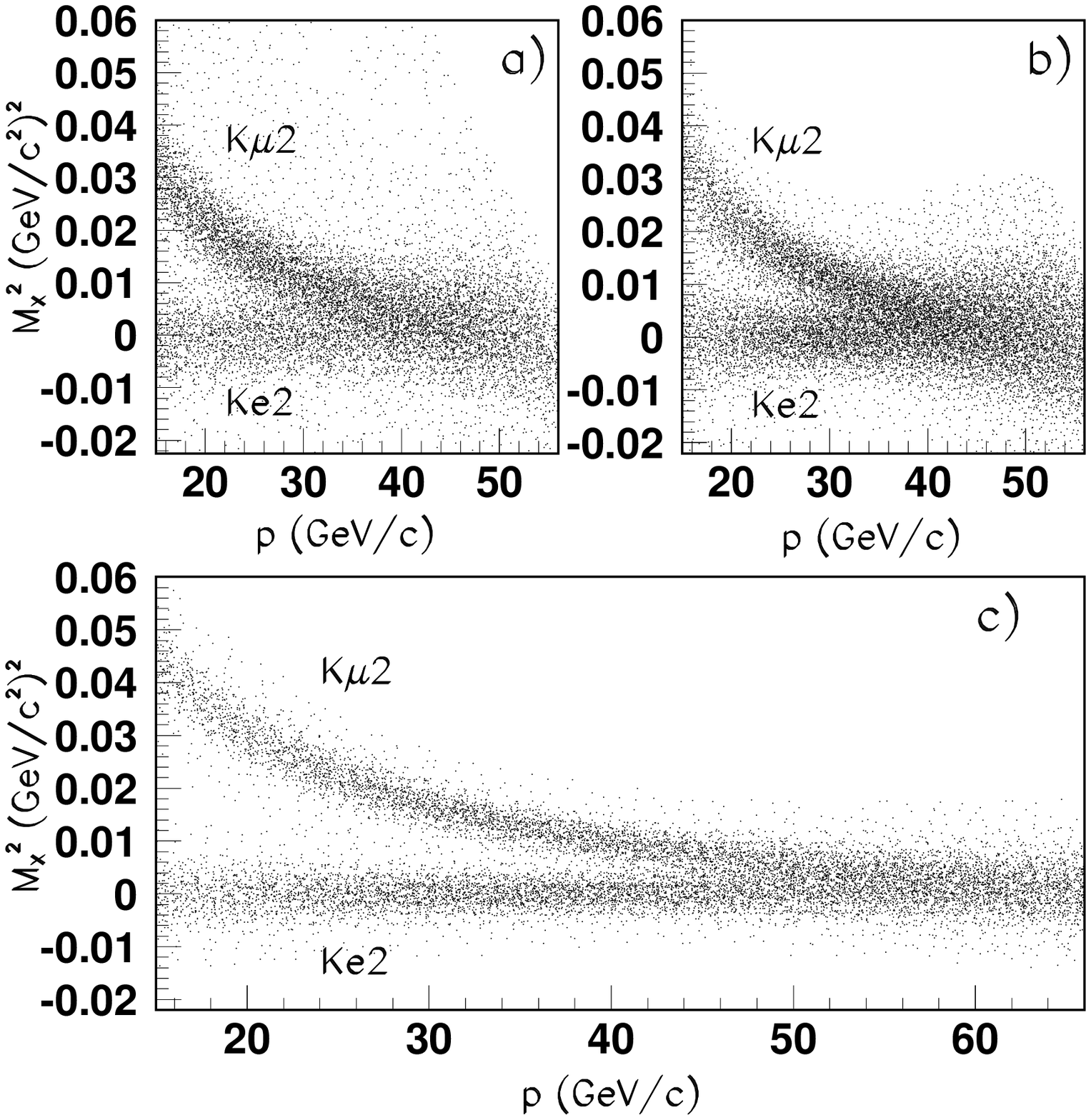}}   
\hspace*{\fill}
\parbox{0.46\linewidth}
{\includegraphics[width=1.1\linewidth]{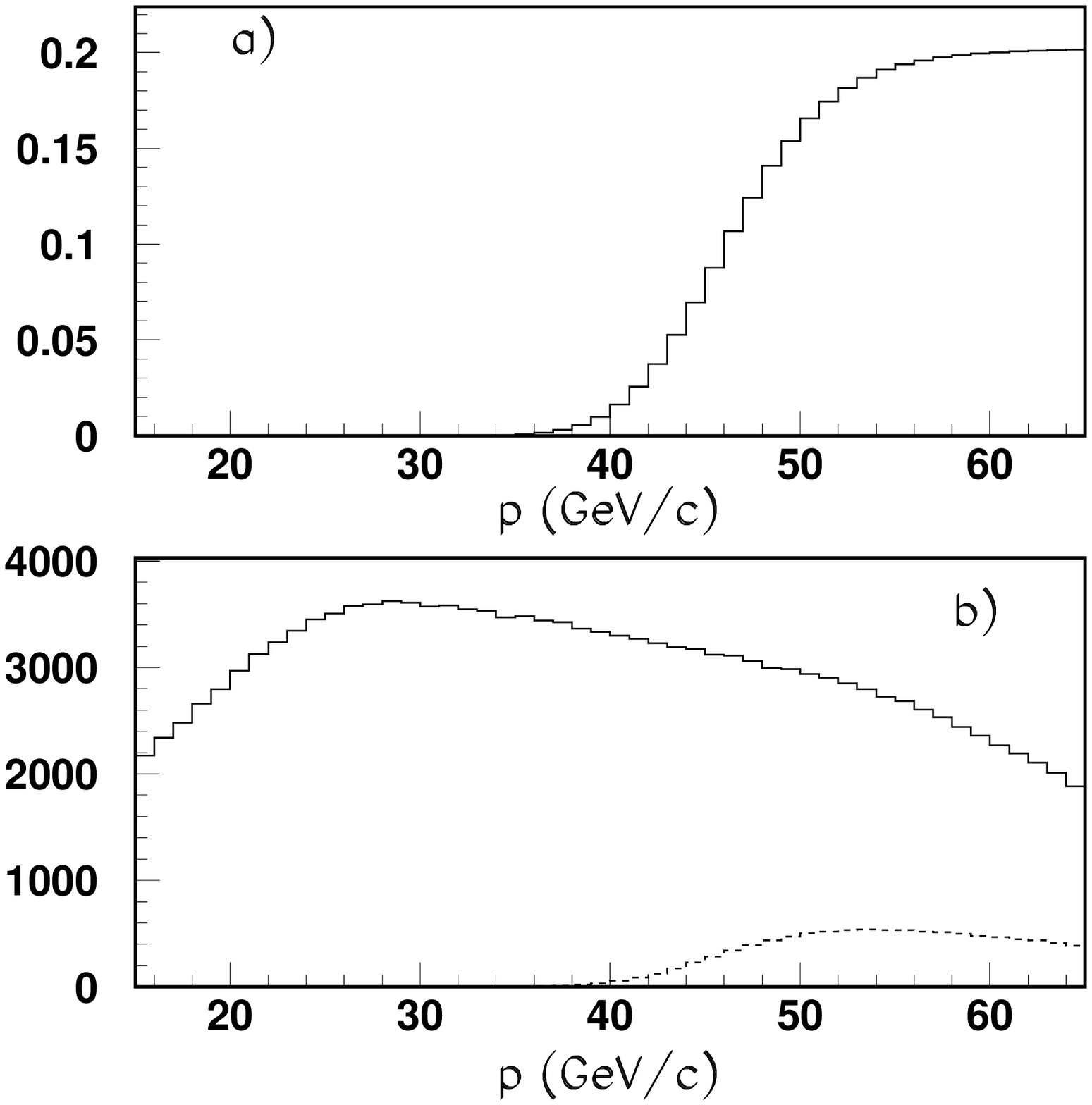}}
\parbox[t]{0.48\linewidth}{
\caption{Distributions of $M_{X}^2$ versus $p$ for $K_{e2}$ and
$K_{\mu2}$ decays. In the $M_{X}$ calculation the electron mass is
assumed for both processes: a) measured data from the 2004
run, b) Monte Carlo predictions for 2004 conditions,
c) Monte Carlo predictions for the conditions expected in
2007.\label{fig:mm-vs-p}}}
\hspace*{\fill}\parbox[t]{0.48\linewidth}{
\caption{a) $K_{\mu2}$ contamination in the $K_{e2}$ sample. b)
Simulated momentum distributions of genuine electrons from $K_{e2}$
decay (full histogram), and of fake electrons from $K_{\mu2}$ decays
(dashed histogram).
\label{fig:pe_distr}}}
\end{figure}

Figure~\ref{fig:mm-vs-p} shows the $M_{X}^2$ versus momentum
distribution for $K_{e2}$ and $K_{\mu2}$ decays for the 2004 data,
together with the predicted distributions for the 2004 run and for the
 2007 run, as obtained from a Monte Carlo simulation (for
$K_{\mu2}$ decays the electron mass is assigned to the muon). In the
2007 run, for electron momenta up to 35 GeV/{\it c} the $K_{\mu2}$
contamination to the $K_{e2}$ signal is reduced to a negligible level
thanks to the improved spectrometer momentum resolution (see
Fig.~\ref{fig:pe_distr}a). Using a lower limit of 15 GeV/{\it c} for
the electron momentum, and taking into account the detector
acceptance, this means that $\sim43\%$ of the $K_{e2}$ events will be
kinematically background free (see Fig.~\ref{fig:pe_distr}b).


The fraction of $K_{\mu2}$ faking $K_{e2}$ decays was measured at
all momenta in parallel with data taking. For this purpose a
$\sim$5~cm thick lead plate was inserted between the two hodoscope
planes covering six 6.5 cm wide vertical hodoscope counters. The
requirement that charged particles traverse the lead without
interacting helps to select a pure sample of $K_{\mu2}$ decay for
which the muon $E/pc$ distribution can be directly measured for the
evaluation of the $K_{\mu2}$ contamination to the $K_{e2}$
signal. Table~\ref{tab:beam} lists the relevant parameters describing
the running conditions both for the 2004 and 2007 runs.

\begin{table}[bht]
\begin{minipage}{\textwidth}
\caption{Comparison of the 2004 and 2007 running conditions.\label{tab:beam}}
\begin{tabular*}{\textwidth}{@{\extracolsep{\fill}}l|cc||l|cc}
\hline\hline
				&2004			&2007			&		 	&2004			&2007			\\
\hline
Acceptance (mr$^2$) 		&$0.36 \times 0.36$ 	&$0.18 \times
				0.18$ 	&SPS duty cycle (s/s) 	& 4.8
				/ 16.8 		& 4.8 / 16.8 		\\
$\quad \Delta \Omega$ (sr) 	&$4 \times 10^{-7}$ 	&$1 \times
				10^{-7}$	&live time (days)
				& 2.1 			& 100 \\ 
$\Delta p/p$ effective (\%) 	&$\pm 3$ 		&$\pm 2.5$
				&nr. of pulses 		& $1.08 \times
				10^4$ 	& $3 \times 10^5$  \\
$\quad$ RMS (\%) 		&$\sim 3.0$ 		&$\sim 1.8$ 		&Protons per pulse 	& $2.5 \times 10^{11}$ 	& $1.5 \times 10^{12}$ 	\\
TRIM3 $x'$ (mr) 		&0 			&$\pm 0.3$ 		&beam momentum (GeV/$c$)& $\approx 60$ 	& $\approx 75$ 			\\
$\quad p_T$ (MeV/c) 		&0 			&$\pm 22.5$
				&Triggers/pulse 	& 45,000
				& 48,000  \\
MNP33 $x'$ (mr) 		&$\pm 2.0$ 		&$\pm 3.5$
				&Good $K_{e2}$/pulse 	& $\sim$0.37
				& $\sim$ 0.5  \\
$\quad p_T$ (MeV/c) 		&$\pm 120$ 		&$\pm 263$ 		&Good $K_{e2}$ (total) 	& 4000 			& $>$100,000 \\
\hline\hline
\end{tabular*}
\end{minipage}
\end{table}

The overall statistical error, which includes the statistical
uncertainty on the background measurement, is expected to be $0.3\%$.
The uncertainty in the trigger efficiency will be reduced to less than
$\pm0.2\%$. The data collected in 2007 will provide a measurement of
$R_K$ with a total uncertainty (statistical and systematic errors
combined in quadrature) of less than $\pm0.5\%$.

\subsection{$\tau$ decay }\label{sec:exp:universality:tau}
There are two ways to test lepton universality in charged weak
interactions using $\tau$ decays :
\begin{itemize}
\item the universality of all three couplings can be tested by
comparing the rates of the decays $\tau\to \mu\nu\overline{\nu}$,
$\tau\to e\nu\overline{\nu}$ and $\mu \to e\nu\overline{\nu}$, and
\item $g_{\tau}/g_{\mu}$ can be extracted by comparing $\tau\to
\pi\nu$ and $\pi \to\mu\nu$.
\end{itemize}
When comparing the experimental constraints one should keep in mind
the complementarity of these two tests. Whereas the purely leptonic
decay modes are mediated by a transversely polarized $W$, the
semileptonic modes involve longitudinal polarization.

\subsubsection{Leptonic $\tau$ decays}
The decay width of $\ell_i\to\ell_f\nu\nu $ including radiative
corrections is given by \cite{Marciano:1988vm} :
\begin{equation}
\label{eq:tauellnunu}
\Gamma (\ell_i\to\ell_f\nu\nu ) = \frac{g^2_{\ell_i} g^2_{\ell_f}}{32
m_W^2} \frac{m_{\ell_i}^5}{192 \pi^3} (1+C_{\ell_i \ell_f}),
\end{equation}
where $(1+C_{\ell_i \ell_f}) = f(x) (1
+\frac{3}{5}\frac{m_{\ell_i}^2}{M^2_{W}})
(1+\frac{\alpha(m_{\ell_i})}{2\pi}(\frac{25}{4} - \pi^2))$ combines
weak and radiative corrections and $f(x) = 1-8x + 8x^3 - x^4 -
12x^2\ln x$ with $x \equiv m^2_{\ell_f}/m^2_{\ell_i}$.

Electron - muon universality could thus be tested at the 0.2\% level using:
\begin{equation}
\frac{g_{\mu}}{g_e} = \sqrt{ \frac{\ensuremath{B}(\tau\to \mu\nu\nu
)}{\ensuremath{B}(\tau\to e\nu\nu)} \frac{(1+C_{\tau e})}{(1+C_{\tau
\mu})} }= 1.0002 \pm 0.0020,
\end{equation}
where $C_{\tau e} = -0.004$ and $C_{\tau \mu} = -0.0313$ are the
corrections from Eq.~(\ref{eq:tauellnunu}). The values of the
branching ratios of leptonic $\tau$ decays are taken from
\cite{Yao:2006px} and are based mostly on measurements from LEP
experiments. $e-\tau$ universality has been verified with
similar precision:
\begin{equation}
\frac{g_{\tau}}{g_{e}} = \sqrt{\frac{(1+C_{\mu e})}{(1+C_{\tau
\mu})}\frac{\tau_{\mu}}{\tau_{\tau}}(\frac{m_{\mu}}{m_{\tau}})^5
\ensuremath{B}(\tau\to \mu\nu\nu) }= 1.0012 \pm 0.0023\; ,
\end{equation}
where $ \Gamma (\ell_i\to\ell_f\nu\nu ) =
\ensuremath{B}(\ell_i\to\ell_f\nu\nu ) / \tau_{\ell_i}$ has been used
and $C_{\mu e} = -0.0044 $. The measurement of 
$\mu-\tau$ universality can then be derived
from $g_{\tau}/g_{e}$ and $g_{\mu}/g_e$, giving $g_{\tau}/g_{\mu} =
1.0010\pm 0.0023$.

The measurements used in above formulas are relatively old
\cite{Yao:2006px}, and no input from BaBar or Belle is used.  The
measurements of leptonic branching fractions were done by the LEP
experiments in the course of the runs at or near the $Z^0$
resonance\cite{Yao:2006px}. The $\tau^+\tau^-$ events were selected via
their topology, and the $\tau$ decay products were required to
pass particle identification, using infomation from the calorimetry, tracking
devices, time projection chambers and muon systems. The largest
uncertainty on the measurement of tau branching ratios was statistical,
with systematics limitations arising from the simulation and from particle
identification.

The measurement of the $\tau$ lifetime\cite{Yao:2006px} comes from LEP experiments as
well. Due to the large $\sqrt{s}$, each $\tau$ in the event has a large boost
and travels 90 $\mu$m in average. However, as there is nothing but
$\tau$s produced in each event, their production vertex is unknown
and has to be estimated averaging over other events or by minimizing the sum of
impact parameters of both $\tau$'s decay products.

The most accurate published measurement of the $\tau$
mass~\cite{Bai:1995hf} was done by the BES experiment, through an
energy scan of the $\tau^+\tau^-$ production cross section in $e^+e^-$
collisions around the threshold region.  The collision energy scale
was calibrated with $J/\psi$ and $\psi(2S)$ resonances, with a
precision of 0.25~MeV.

Therefore the major contributions to the uncertainties on the ratios 
$g_{\tau}/g_{e}$ and $g_{\mu}/g_e$ are:
\begin{itemize}
\item the $\tau$ leptonic branching fractions (0.3\%), and
\item the $\tau$ lifetime (0.4\%).
\end{itemize}

In the calculation above, the measurements of leptonic $\tau$ decays are
taken as independent. However, there are common sources of systematic
uncertainties such as uncertainties on track reconstruction, number of
$\tau$ decays registered by an experiment and so on. If one
measured the branching ratio $B(\tau\to e\nu\nu)/B(\tau\to\mu\nu\nu)$
directly in one experiment, as was done by ARGUS and CLEO and as is
done for pion decays as well, most uncertainties would cancel. Taking
the PDG average on the branching ratio \cite{Yao:2006px} one obtains
$g_{\mu}/g_{e} = 1.0028 \pm 0.0055$.

The following improvements can be expected in the future. The KEDR
experiment is working, like BES, at the $\tau$-pair production
threshold. They plan to measure $m_{\tau}$ with a 0.15~MeV accuracy. A
preliminary result, with accuracy comparable to BES's measurement, is
available \cite{Anashin:2006xp}. Both BaBar and Belle have accumulated
large statistics of $\tau^+\tau^-$ events and should be able to
perform measurements of leptonic and semi-leptonic $\tau$ decays, as well
as to improve the measurement of the $\tau$ lifetime. While the collected
$\tau$ sample is much larger than at LEP, there are still significant
uncertainties remaining on luminosity, tracking and particle
identification. If the ratio of decay fractions is measured, then only
the particle identification uncertainties will remain. Currently the
electron and muon identification uncertainties for both BaBar and
Belle are around 1-2\%. At the $B$-factories the $\tau$ boost in the c.m
frame is much smaller than at LEP, and in addition the energies of the
$e^+$ and $e^-$ beams are not the same. This leads to significant differences in the technique of the lifetime
measurement. In particular the 3-dimensional reconstruction of the
trajectories of the decay products is poor and only the impact parameter
in the plane transverse to the beams, multiplied by the polar angle of the
total momentum vector of 3-prong $\tau$ decay products,  can be
used\cite{Lusiani:2005sy}.
While the statistics allows a very accurate measurement, the work
focuses on understanding the alignment of the vertex detector and the
systematics in the reconstruction of the impact parameter. The
measurement of the $\tau$ mass can also be done at the
$B$-factories. Belle has presented a mass measurement analyzing the
kinematic limit of the invariant mass of 3-prong $\tau$ decays
\cite{Abe:2006vf}. This measurement is however less precise than those
of BES or KEDR.

If one takes into account recent preliminary measurements of the
$\tau$ mass from the KEDR experiment\cite{Anashin:2006xp} and of the lifetime
from BaBar\cite{Lusiani:2005sy}, the determination of $\tau-e$
universality  changes slightly to $g_{\tau}/g_e =
1.0021 \pm 0.0020$.

\subsubsection{Hadronic $\tau$ decays}

Another way to test $\tau-\mu$ universality is to compare the
decay rates for $\tau\to\pi\nu$ and $\pi\to\mu\nu$:
\begin{equation}
\frac{g^2_{\tau}}{g^2_{\mu}} =
\frac{\ensuremath{B}(\tau\to\pi\nu)}{\ensuremath{B}(\pi\to\mu\nu)}\frac{\tau_{\pi}}{\tau_{\tau}}
\frac{2m_{\tau}m^2_{\mu}}{m_{\pi}^3}(\frac{m_{\pi}^2-m_{\mu}^2}{m_{\tau}^2-m_{\pi}^2})^2(1+C_{\tau\pi}),
\end{equation}
where $C_{\tau\pi} = -(1.6^{+0.9}_{-1.4}) 10^{-3}$
\cite{Marciano:1993sh,Decker:1994dd}.

Taking measurements from Ref.~\cite{Yao:2006px} one obtains $
g_{\tau}/g_{\mu} = 0.9996 \pm 0.037$. Here the main uncertainties come
from
\begin{itemize}
\item $\tau\to\pi\nu$ decay (1\%), where the dominant contribution is
due to $\tau\to\pi\pi^0\nu$ contamination and $\pi^0$ reconstruction,
\item the $\tau$ lifetime (0.3\%), and
\item the hadronic correction (0.1\%).
\end{itemize}

Again, no results from the $B$ factories are available yet, and one
should expect that the large $\tau$ samples collected by BaBar and
Belle will allow a significant improvement, in case the understanding
of particle identification will be improved.

\newcommand{\ps}{\ensuremath{\mathrm{Ps}}}
\newcommand{\ops}{\ensuremath{\mathrm{o\text{-}Ps}}}
\newcommand{\oPs}{\ensuremath{\mathrm{o\text{-}Ps}}}
\newcommand{\pps}{\ensuremath{\mathrm{p\text{-}Ps}}}
\newcommand{\nplus}{\ensuremath{N_+}}
\newcommand{\nminu}{\ensuremath{N_-}}
\newcommand{\kone}{\ensuremath{\hat{k}_1}}
\newcommand{\ktwo}{\ensuremath{\hat{k}_2}}
\newcommand{\kthr}{\ensuremath{\hat{k}_3}}
\newcommand{\shat}{\ensuremath{\hat{S}}}
\newcommand{\sopvec}{\ensuremath{\vec{S}_{OP}}}
\newcommand{\Bvec}{\ensuremath{\vec{B}}}
\newcommand{\ccp}{\ensuremath{{C_{CP}}}}
\newcommand{\au}{\ensuremath{A_{u}}}
\newcommand{\ap}{\ensuremath{A_{p}}}

\section{CP Violation with charged leptons}\label{sec:exp:CP}

There are two powerful motivations for probing CP~symmetry in lepton decays: 
\begin{itemize}
\item 
The discovery of CP~asymmetries in $B$ decays that are close to 100 \%
in a sense `de-mystifies' CP~violation. For it established that
complex CP~phases are not intrinsically small and can be close to 90
degrees even. This de-mystification would be completed, if
CP~violation were found in the decays of leptons as well.
\item 
We know that CKM dynamics, which is so successful in describing quark
flavour transitions, is not relevant to baryogenesis. There are
actually intriguing arguments for baryogenesis being merely a
secondary effect driven by primary leptogenesis
\cite{Buchmuller:2005eh}. To make the latter less speculative, one has
to find CP~violation in dynamics of the leptonic sector.
\end{itemize}
The strength of these motivations has been well recognized in the
community, as can be seen from the planned experiments to measure
CP~violation in neutrino oscillations and the ongoing heroic efforts
to find an electron EDM. Yet there are other avenues to this goal as
well that certainly are at least as difficult, namely to probe
CP~symmetry in muon and $\tau$ decays. Those two topics are addressed
below in Sections~\ref{sec:exp:CP:muon} and \ref{sec:exp:CP:tau}.  There
are also less orthodox probes, namely attempts (i) to extract an EDM
for $\tau$ leptons from $e^+e^- \to \tau^+\tau^-$, (ii) to search for
a T-odd correlation in polarized ortho-positronium decays and
(iii) to measure the muon transverse polarization in $K^+ \to \mu^+
\nu \pi^0$ decays. It is understood that the Standard Model does not
produce an observable effect in any of these three cases or the other
ones listed above (except for $\tau ^{\pm} \to \nu K_S \pi^{\pm}$, as
described below).

Concerning topic (i), one has to understand that one is searching for a
CP-odd effect in an {\em electromagnetic} production process unlike in
$\tau$ decays, which are controlled by weak forces.

In $[e^+e^-]_{OP} \to 3 \gamma$, topic (ii), one can construct various T-odd correlations or 
integrated moments between the spin vector $\vec S_{OP}$ of polarized 
ortho-positronium and the momenta $\vec k_i$ of two of the photons that define the decay plane: 
$$
A_{T\, odd} = \langle \vec S_{OP} \cdot (\vec k_1 \times \vec k_2)\rangle 
$$
\begin{equation}
A_{CP} = \langle (\vec S_{OP} \cdot \vec k_1)(\vec S_{OP} \cdot (\vec k_1 \times \vec k_2))\rangle 
\label{ACP}
\end{equation}
\begin{itemize}
\item 
The moment $A_{T\, odd}$ is P~and CP~{\em even}, yet T~{\em
odd}. Rather than by CP~or T~violation in the underlying
dynamics it is generated by higher order QED processes  .It has been
conjectured~\cite{BERN} 
that the leading effect is formally of order
$\alpha$ relative to the decay width due to the exchange of a photon
between the two initial lepton lines. From it one has to remove the
numerically leading contribution, which has to be absorbed into the
bound state wavefunction. The remaining contribution is presumably at
the sub-permille level. Alternatively $A_{T\, odd}$ can be generated at
order $\alpha ^2$ -- or at roughly the $10^{-5}$ level -- through the
interference of the lowest-order decay amplitude with one where a
fermion loop connects two of the photon lines.
\item 
On the other hand the moment $A_{CP}$ is odd under T~as well as
under P and in particular CP. Final state interactions can{\em not}
generate a CP-odd moment with CP~invariant dynamics. Observing
$A_{CP}\neq 0$ thus unambiguously establishes CP~violation.  The
present experimental upper bound is around few percent; it seems
feasible, see Section~\ref{sec:exp:CP:posi}, to improve the
sensitivity by more than three orders of magnitude, i.e. down to the
$10^{-5}$ level! The caveat arises at the theoretical level: with the
`natural' scale for {\em weak} interference effects in positronium
given by $G_F \; m_e^2 \sim 10^{-11}$, one needs a dramatic enhancement
to obtain an observable effect.
\end{itemize}

Discussing topic (iii) -- the muon transverse polarization in $K_{\mu
3}$ decays -- under the heading of CP~violation in the leptonic sector will
seem surprising at first. Yet a general, though hand waving argument,
suggests that the highly suppressed direct CP~violation in nonleptonic
$\Delta S=1$ -- as expressed through $\epsilon ^{\prime}$ -- rules
against an observable signal even in the presence of New Physics --
unless the latter has a special affinity for leptons.  The present
status of the data and future plans are discussed in
Section~\ref{sec:exp:CP:K}.

\subsection{$\mu$ decays }\label{sec:exp:CP:muon}

The muon decay $\mu^{-} \to e^{-} \overline{\nu}_{e} \nu_{\mu}$ and its
`inverse' $\nu_{\mu} e^{-} \to
\mu^{-} \nu_{e}$ are successfully described by the `V--A' interaction,
which is a particular case of the local, derivative-free,
lepton-number-conserving, four-fermion interaction
\cite{Michel_50}. The `V--A' form and the nature of the neutrinos
($\overline{\nu}_{e}$ and $\nu_{e}$) have been determined by
experiment \cite{FGJ_86,Langacker_89}.

The observables -- energy spectra, polarizations and angular distributions -- may be parameterized in terms of the dimensionless coupling constants 
$g_{\varepsilon \mu}^{\gamma}$ and the Fermi coupling constant $G_{\mathrm{F}}$. The matrix element is
\begin{equation}
	{\mathcal M} = \frac{4 G_{\mathrm{F}}}{\sqrt{2}}\,
	\sum_{\substack{\gamma = \mathrm{S, V, T} \\ \varepsilon, \mu =
	\mathrm{R, L} }}g^{\gamma}_{\varepsilon \mu} \langle
	\overline{e}_{\varepsilon} |\Gamma^{\gamma}|(\nu_{e})_{n}\rangle
	\langle (\overline{\nu}_{\mu})_{m}|
	\Gamma_{\gamma}|\mu_{\mu}\rangle \;.\label{eq:wfetscher:1}
\end{equation}
We use here the notation of Fetscher \etal, \cite{FGJ_86,Fetscher_00}
who in turn use the sign conventions and definitions of Scheck
\cite{Scheck_83}. Here $\gamma = \mathrm{S,V,T}$ indicate a (Lorentz)
scalar, vector, or tensor interaction, and the chirality of the 
electron or muon (right- or left-handed) is labeled by $\varepsilon,
\mu = \mathrm{R, L}$. The chiralities $n$ and $m$ of the $\nu_{e}$ and
the $\overline{\nu}_{\mu}$ are determined by given values of $\gamma$,
$\varepsilon$ and $\mu$. The 10 complex amplitudes $g_{\varepsilon
\mu}^{\gamma}$ and $G_{\mathrm{F}}$ constitute 19 independent
parameters to be determined by experiment. The `V--A' interaction
corresponds to $g_{\mathrm{LL}}^{\mathrm{V}} = 1$, with all other
amplitudes being 0.

Experiments show the interaction to be predominantly of the
vector type and left-handed [$g_{\mathrm{LL}}^{\mathrm{V}} >
0.96~(90~\%\mathrm{CL}$)] with no evidence for other couplings.  The
measurement of the muon lifetime yields the most precise determination
of the Fermi coupling constant $G_{\mathrm{F}}$, which is presently
known with a relative precision of $8 \times
10^{-6}$~\cite{FAST,Barczyk:2007hp}. Continued improvement of
this measurement is certainly an important goal~\cite{Marciano_00},
since $G_{\mathrm{F}}$ is one of the fundamental parameters of the
Standard Model.

\subsubsection{T invariance in $\mu$ decays}\label{MUONTINV}

$P_{\mathrm{T}_{2}}$ -- the component of the decay positron
polarization which is transverse to the positron momentum and the muon
polarisation -- is T odd and due to the practical absence of a
strong or electromagnetic final state interaction it probes T
invariance.  A second-generation experiment has been performed at PSI
by the ETH Z\"urich--Cracow--PSI Collaboration \cite{Barnett_00}. They
obtained, for the energy averaged
transverse polarization component:
\begin{equation}
\langle P_{\mathrm{T}_{2}} \rangle = 
(-3.7 \pm 7.7_{\mathrm{stat.}} \pm 3.4_{\mathrm{syst.}}) \times 10^{-3}
\label{PT2}
\end{equation}

\subsubsection{Future prospects}
\label{FUTPROSP}

The precision on the muon lifetime can presumably be increased over
the ongoing measurements by one order of magnitude~\cite{FAST}.
Improvement in measurements of the decay parameters seems more
difficult. The limits there are not given by the muon rates which
usually are high enough already ($\approx 3 \times 10^{8}$~s$^{-1}$ at
the $\mu$E1 beam at PSI, for example), but rather by effects like
positron depolarisation in matter or by the small available
polarisation ($< 7 \%$) of the electron targets used as analysers. The
measurement of the transverse positron polarisation might be improved
with a smaller phase space (lateral beam dimension of a few
millimetres or better).  This experiment needs a \emph{pulsed} beam
with high polarisation.

\subsection{CP violation in $\tau$ decays}\label{sec:exp:CP:tau}
\label{TAUCP}

The betting line is that $\tau$ decays -- next to the electron EDM and
$\nu$ oscillations -- provide the best stage to search for
manifestations of CP~breaking in the leptonic sector. There exists a
considerable literature on the subject started by discussions on a
tau-charm factory more than a decade ago
\cite{Tsai:1989ez,Tsai:1994rc,Kuhn:1992nz,Kuhn:1996dv} and attracting
renewed interest recently  
\cite{Bigi:2005ts,Datta:2006kd,Bernabeu:2006wf,Delepine:2006fv} 
stressing the following points:
\begin{itemize}
\item 
There are many more channels than in muon decays making the
constraints imposed by CPT symmetry much less restrictive.
\item 
The $\tau$ lepton has sizable rates into multi-body final states. Due
to their nontrivial kinematics asymmetries can emerge also in the
final state distributions, where they are likely to be significantly
larger than in the integrated widths. The channel
$$
K_L \to \pi^+\pi^-e^+e^-
$$
can illustrate this point. It commands only the tiny branching ratio
of $3\times 10^{-7}$.  The forward-backward asymmetry $\langle A
\rangle$ in the angle between the $\pi^+\pi^-$ and $e^+e^-$ planes
constitutes a CP {\em odd} observable. It has been measured by KTeV
and NA48 to be truly large, namely about 13\%, although it is driven
by the small value of $|\epsilon _K| \sim 0.002$. I.e., one can trade
branching ratio for the size an CP asymmetry.

\item 
New Physics in the form of multi-Higgs models can contribute on the tree-level like the SM W exchange. 
\item 
Some of the channels should exhibit enhanced sensitivity to New Physics.  
\item 
Having polarized $\tau$ leptons provides a powerful handle on CP~asymmetries and control over systematics. 
\end{itemize}
These features will be explained in more detail below. It seems clear
that such measurements can be performed only in $e^+e^-$ annihilation,
i.e. at the $B$ factories running now or better still at a
Super-Flavour factory, as discussed in the Working Group 2
report. There one has the added advantage that one can realistically
obtain highly polarized $\tau$ leptons: This can be achieved directly
by having the electron beam longitudinally polarized or more
indirectly even with unpolarized beams by using the spin alignment of
the produced $\tau$ pair to `tag' the spin of the $\tau$ under study
by the decay of the other $\tau$ like $\tau \to \nu \rho$.

\subsubsection{$\tau \to \nu K \pi$}

The most promising channels for exhibiting CP~asymmetries are $\tau^- \to \nu K_S \pi^-$, $\nu K^-\pi^0$ \cite{Kuhn:1996dv}: 
\begin{itemize}
\item 
Due to the heaviness of the lepton and quark flavours they are most sensitive to non-minimal Higgs dynamics while being Cabibbo suppressed in the SM. 
\item 
They can show asymmetries in the final state distributions. 
\end{itemize}
The SM does generate a CP~asymmetry in $\tau$ decays that should be observable. Based on known physics one can reliably predict a 
CP~asymmetry \cite{Bigi:2005ts}: 
\begin{equation} 
\frac{\Gamma(\tau^+\to K_S \pi^+ \overline \nu)-\Gamma(\tau^-\to K_S \pi^- \nu)}{\Gamma(\tau^+\to K_S \pi^+ \overline \nu)+\Gamma(\tau^-\to K_S \pi^- \nu)}= 
(3.27 \pm 0.12)\times 10^{-3}
\label{CPKS}
\end{equation}
due to $K_S$'s preference for antimatter over matter. Strictly speaking, this prediction is more general than the SM: no matter what produces the CP~impurity in
the $K_S$ wave function, the effect underlying Eq.~(\ref{CPKS}) has to be present, while of course not affecting $\tau^{\mp} \to \nu K^{\mp}\pi^0$. 

To generate a CP~asymmetry, one needs two different amplitudes contribute coherently. This requirement is satisfied, since the $K\pi$ system can be produced from
the (QCD) vacuum in a vector and scalar configuration with form factors $F_V$ and $F_S$, respectively. Both are present in the data, with the vector component (mainly
in the form of the $K^*$) dominant as expected \cite{Pich:2006nt}. Within the SM, there does not arise a weak phase between them on an observable level, yet it can 
readily be provided by a charged Higgs exchange in non-minimal Higgs models, which contributes to $F_S$. 

A few general remarks on the phenomenology might be helpful to set the stage. For a CP~violation in the underlying weak dynamics to generate an observable asymmetry
in partial widths or energy distributions one needs also a relative strong phase between the two amplitudes: 
\begin{eqnarray}
\Gamma (\tau ^- \to \nu K^- \pi^0) - \Gamma (\tau ^+ \to \bar \nu K^+
\pi^0)
&\propto& {\rm Im}( F_HF_V^*) {\rm Im}g_Hg_W^* \; , 
\\
\frac{d}{dE_K} \Gamma (\tau ^- \to \nu K^- \pi^0) - 
\frac{d}{dE_K} \Gamma (\tau ^+ \to \bar \nu K^+ \pi^0)
&\propto& {\rm Im}( F_HF_V^*) {\rm Im}g_Hg_W^* \; , 
\end{eqnarray}
where $F_H$ denotes the Higgs contribution to $F_S$ and $g_H$ its weak coupling. This should not represent a serious restriction, since the $K\pi$ system is produced 
in a mass range with several resonances. If on the other hand one is searching for a T-odd correlation like 
\begin{equation} 
O_T \equiv \langle \vec \sigma _{\tau} \cdot (\vec p_K \times \vec p_{\pi})\rangle \; , 
\end{equation} 
then CP~violation can surface even with{\em out} a relative strong phase 
\begin{equation} 
O_T \propto {\rm Re}( F_HF_V^*) {\rm Im}g_Hg_W^* \; . 
\label{OT1}
\end{equation}
Yet there is a caveat: final state interactions can generate T-odd moments even from T~invariant dynamics, when one has 
\begin{equation} 
O_T \propto {\rm Im}( F_HF_V^*) {\rm Re}g_Hg_W^* \; . 
\label{OT2}
\end{equation}
Fortunately one can differentiate between the two scenarios of Eqs.~(\ref{OT1},\ref{OT2}) at a $B$ or a Super-Flavour factory, where one can compare directly the
T-odd moments for the CP~conjugate pair $\tau^+$ and $\tau ^-$: 
\begin{equation} 
O_T(\tau^+) \neq O_T(\tau^-) \; \; \; \Longrightarrow \; \; \; {\rm CP~violation!} 
\end{equation}

A few numerical scenarios might illuminate the situation: a Higgs
amplitude 1\% or 0.1\% the strength of the SM $W$-exchange amplitude
-- the former [latter] contributing [mainly] to $F_S$ [$F_V$] -- is
safely in the `noise' of present measurements of partial widths; yet
it could conceivably create a CP~asymmetry as large 1\% or 0.1\%,
respectively. More generally a CP-odd observable in a SM allowed
process is merely {\em linear} in a New Physics amplitude, since the
SM provides the other amplitude. On the other hand SM forbidden
transitions -- say lepton flavour violation as in $\tau \to \mu
\gamma$ -- have to be {\em quadratic} in the New Physics amplitude.
\begin{equation} 
CP~{\rm odd} \propto |T^*_{SM}T_{NP}| \; \; {\rm vs.} \; \; {\rm LFV}
\propto |T_{NP}|^2
\end{equation}
Probing CP~symmetry at the 0.1\% level in $\tau \to \nu K\pi$ thus has
roughly the same sensitivity for a New Physics amplitude as searching
for B$(\tau \to \mu \gamma)$ at the $10^{-8}$ level.

CLEO has undertaken a pioneering search for a CP~asymmetry in the
angular distribution of $\tau \to \nu K_S \pi$ placing an upper bound
of a few percent
\cite{Anderson:1998ke}. 

\subsubsection{Other $\tau$ decay modes}

It appears unlikely that analogous asymmetries could be observed in
the Cabibbo allowed channel $\tau \to \nu \pi \pi$, yet detailed
studies of $\tau \nu 3\pi/4\pi$ look promising, also because the more
complex final state allows to form T-odd correlations with unpolarized
$\tau$ leptons; yet the decays of polarized $\tau$ might exhibit much
larger CP~asymmetries \cite{Datta:2006kd}.

Particular attention should be paid to $\tau \to \nu K 2\pi$, which
has potentially very significant additional advantages:

\begin{itemize}
\item One can interfere {\em vector} with {\em axial vector} $K2\pi$
configurations.  
\item The larger number of kinematical variables and of specific
channels should allow more internal cross checks of systematic
uncertainties like detection efficiencies for positive vs. negative
particles.
\end{itemize}

\subsection{Search for T violation in $K^+ \to \pi^0 \mu^+ \nu$ decay}\label{sec:exp:CP:K}
\label{KMU3}

The transverse muon polarization in $K^+\to\pi^0 \mu^+ \nu$ decay,
$P_T$, is an excellent probe of T violation, and thus of physics
beyond the Standard Model. Most recently the E246 experiment at the
KEK proton synchrotron has set an upper-bound of $\vert P_T \vert \le
0.0050$ (90\% C.L.). A next generation experiment is now being planned
for the high intensity accelerator J-PARC which is aiming at more than
one order of magnitude improvement in the sensitivity with
$\sigma(P_T)\sim 10^{-4}$.

\subsubsection{Transverse muon polarization}

A non-zero value for the transverse muon polarization ($P_T$) in the
three body decay $K\to\pi \mu \nu$ ($K_{\mu 3}$) violates T
conservation with its T-odd correlation\cite{saku}. Over the last
three decades dedicated experiments have been carried out in search
for a non-zero $P_T$.  Unlike other T-odd channels in e.g. nuclear
beta decays, $P_T$ in $K_{\mu 3}$ has the advantage that final state
interactions (FSI), which may induce a spurious T-odd effect, are very
small. With only one charged particle in the final state the FSI
contribution originates only in higher loop effects and has been shown
to be small. The single photon exchange contribution from two-loop
diagrams was estimated more than twenty years ago as $P_T^{FSI} \le
10^{-6}$ \cite{zhit}. Quite recently two-photon exchange contributions
have been studied \cite{efro}. The average value of $P_T^{FSI}$ over
the Dalitz plot was calculated to be less than $10^{-5}$.

An important feature of a $P_T$ study is the fact that the
contribution from the Standard Model (SM) is practically zero. Since
only a single element of the CKM matrix $V_{us}$ is involved for the
semileptonic $K_{\mu 3}$ decay in the SM, no CP violation appears in
first order. The lowest-order contribution comes from radiative
corrections to the $\bar{u}\gamma_{\mu}(1-\gamma_{5})sW^{\mu}$ vertex,
and this was estimated to be less than $10^{-7}$ \cite{bigi}.
Therefore, non-zero $P_T$ in the range of $10^{-3}$-$10^{-4}$ would
unambiguously imply the existence of a new physics contribution
\cite{bigi}.

\begin{figure}
\begin{center}
\includegraphics[width=0.65\linewidth]{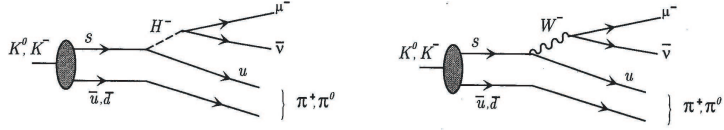} 
\caption{Two interfering diagrams inducing $P_T$ in the multi-Higgs
model (from Ref.\cite{bigi}). \label{fig:hex}} 
\end{center}
\end{figure}

Sizable $P_T$ can be accommodated in multi-Higgs doublet models
through CP violation in the Higgs sector \cite{mhdm}. $P_T$ can be
induced due to interference between charged Higgs exchange ($F_S$,
$F_P$) and $W$ exchange ($F_V$, $F_A$) as shown in
Fig.~\ref{fig:hex}. It is conceivable that the coupling of charged
Higgs fields to leptons is strongly enhanced relative to the coupling
to the up-type quarks \cite{gari} which would lead to an
experimentally detectable $P_T$ of $O(10^{-3})$. Thus, $P_T$ could
reveal a source of CP violation that escapes detection in $K\to 2\pi,
3\pi$ \cite{bigi}.

A number of other models also allow $P_T$ at an observable level without conflicting with other experimental constraints, and experimental limits on $P_T$ could
thus constrain those models. Among them SUSY models with $R$-parity breaking \cite{rpar} and a SUSY model with squark family mixing \cite{wung} should be mentioned.
A recent paper \cite{chan} discusses a generic effective operator leading to a $P_T$ expression in terms of a cut-off scale $\Lambda$ and the Wilson coefficients
$C_S$ and $C_T$.

\subsubsection{KEK E246 experiment}
\begin{figure}[h]
\begin{center}
\includegraphics[width=15cm]{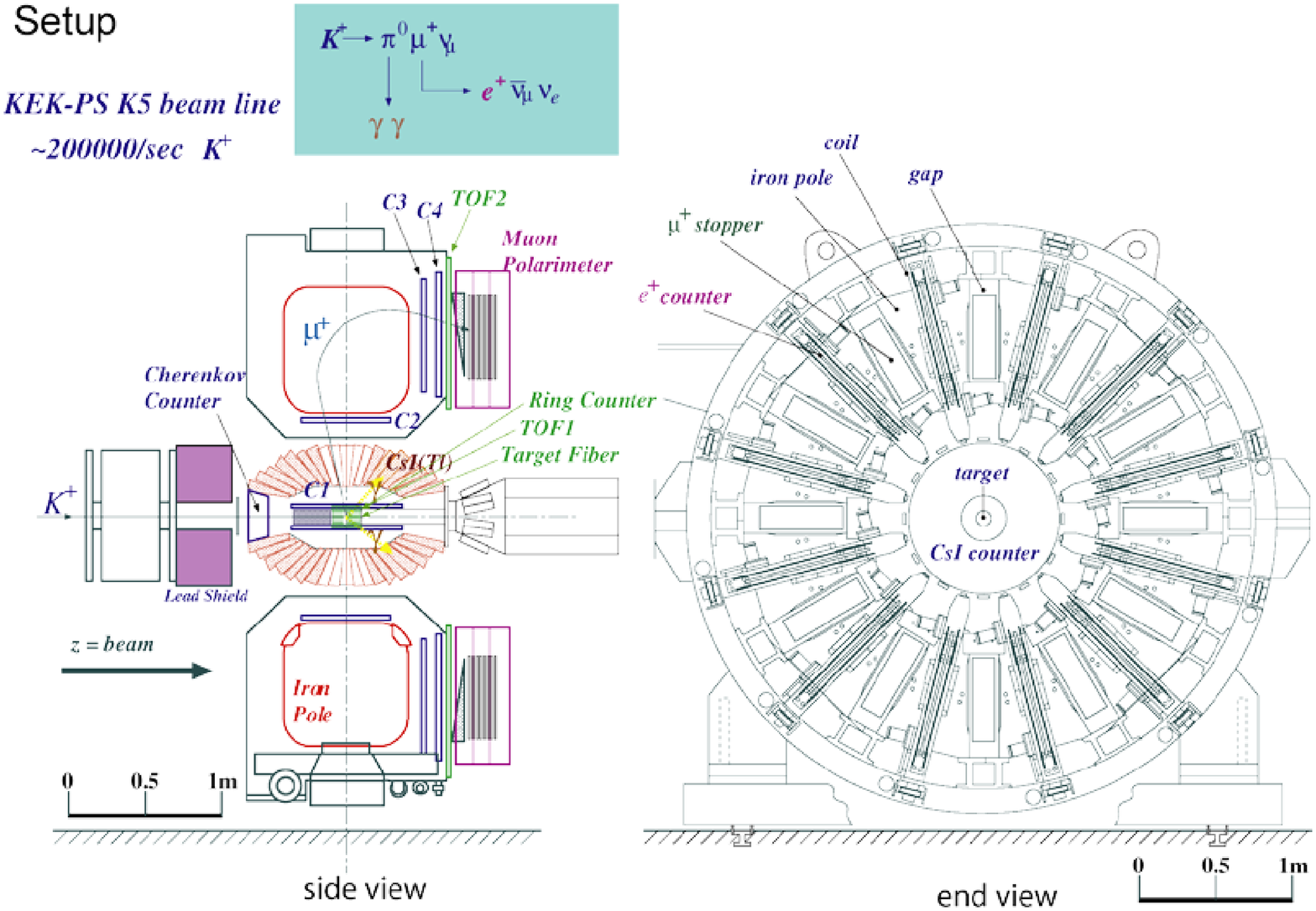} 
\caption{E246 setup using the superconducting toroidal spectrometer. The elaborate detector system \cite{nimp} consists of an active target (to monitor stopping $K^+$),
a large-acceptance CsI(Tl) barrel (to detect $\pi^0$), tracking chambers (to track $\mu^+$), and muon polarimeters (to measure $P_T$). \label{fig:E246}}
\end{center}
\end{figure}
The most recent and highest precision $P_T$ experiment was performed at the KEK proton synchrotron. The experiment used a stopped $K^+$ beam with an intensity of
$\sim10^5$/s and a setup with a superconducting toroidal spectrometer (Fig.~\ref{fig:E246}). 
Data were taken between 1996 and 2000 for a total of 5200 hours of beam time. The determination of the muon polarization was based on a measurement of the decay positron
azimuthal asymmetry in a longitudinal magnetic field using ``passive polarimeters''. Thanks to (i) the stopped beam method which enabled total coverage of the decay phase 
space and hence a forward/backward symmetric measurement with respect to the $\pi^0$ direction and (ii) the rotational-symmetric structure of the toroidal system,
systematic errors could be substantially suppressed.

The T-odd asymmetry was deduced using a double ratio scheme as
\begin{equation}
A_T=(A_{\mathrm{fwd}}-A_{\mathrm{bwd}})/2 \;\;,
\end{equation}
where the fwd(bwd) asymmetry was calculated using the ``clockwise'' and ``counter-clockwise'' positron emission rates $N_{cw}$ and $N_{ccw}$ as
\begin{equation}
 A_{\mathrm{fwd(bwd)}}=\frac{ N^{cw}_{\mathrm{fwd(bwd)}} -
 N^{ccw}_{\mathrm{fwd(bwd)} } }
 { N^{cw}_{\mathrm{fwd(bwd)}} + N^{ccw}_{\mathrm{fwd(bwd)}} } \;.
\end{equation}
$P_T$ was then deduced using
\begin{equation}
P_T=A_T/\{\alpha \langle\cos\theta_T \rangle\} \;\; '
\end{equation}
with $\alpha$ the analyzing power and $\langle\cos\theta_T\rangle$ 
the average kinematic attenuation factor. The final result was \cite{prdp}
\begin{eqnarray}
 P_T= -0.0017 \pm 0.0023 (stat) \pm 0.0011 (syst) \\ 
 {\rm Im}\xi = -0.0053 \pm 0.0071 (stat) \pm 0.0036 (syst) \;\; ,
\end{eqnarray}
corresponding to the upper limits of $\vert P_T \vert < 0.0050$ (90\%
C.L.) and $\vert {\rm Im}\xi
\vert <0.016$ (90\% C.L.), respectively. Here Im$\xi$ is the physics
parameter proportional to $P_T$ after removal of the kinematic factor.
This result constrained the three-Higgs doublet model parameter in the
way of $\vert {\rm Im}(\alpha_1 \gamma_1^*)
\vert < 544 (M_{H_{1}}/{\rm GeV})^2$, as the most stringent constraint
on this parameter. Systematic errors were investigated thoroughly,
although the total size was smaller than half of the statistical
error.  There were two items that could not be canceled out by any of
the two cancellation mechanisms of the 12-fold azimuthal rotation and
$\pi^0$-$fwd/bwd$: the effect from the decay plane rotation,
$\theta_z$ and the misalignment of the muon magnetic field,
$\delta_z$, which should both be eliminated in the next generation
J-PARC experiment.


\subsubsection{The proposed J-PARC E06 (TREK) experiment}
A new possible $P_T$ experiment, E06 (TREK), at J-PARC is aiming at a
sensitivity of $\sigma(P_T)\sim 10^{-4}$.  J-PARC is a high-intensity
proton accelerator research complex now under construction in Japan
with the first beam expected in 2008. In the initial phase of the
machine, the main synchrotron will deliver a 9$\mu$A proton beam at
30 GeV. A low momentum beam of $3\times 10^6$ $K^+$ per second will
be available for stopped $K^+$; this is about 30 times the beam
intensity used for E246. Essentially the same detector concept will be
adopted; namely the combination of a stopped $K^+$ beam and the
toroidal spectrometer, because this system has the advantage of
suppressing systematic errors by means of the double ratio measurement
scheme. However, the E246 setup will be upgraded significantly. 
The E246 detector will be upgraded in several parts so as to
accommodate the higher counting rate and to better control the
systematics. The major planned upgrades are the following:
\begin{itemize}
\item The muon polarimeter will become an active polarimeter,
providing the muon-decay vertex and the positron track, 
leading to an essentially background-free muon decay measurement, with
an increased positorn acceptance and analyzing power.
\item New dipole magnets will be added, improving the field uniformity and
the alignment accuracy.
\item The electronics and readout of the CsI(Tl) E246 calorimeter
 will be replaced to maximize the counting rate, fully exploiting the
 intrinsic crystal speed.
\item The tracking system and the active target will be improved for
 higher resolution and higher decay-in-flight background rejection. 
\end{itemize}
As a result, 20 times higher sensitivity to $P_T$ will be obtained
after a one year run.  The systematic errors will be controlled with
sufficient accuracy and a final experimental error of $\sim10^{-4}$
will be attained. A full description of the experiment can be found in
the proposal
\cite{e06p}.

It is now proposed to run for net $10^7$s corresponding to roughly one
year of J-PARC beam-time under the above mentioned beam
condition. This would yield $2.4\times 10^9$ good $K^+_{\mu3}$ events
in the $\pi^0$-fwd/bwd regions, providing an estimate of
$\sigma(P_T)_{stat}=1.35 \times 10^{-4}$. The inclusion of other
$\pi^0$ regions, enabled by the adoption of the active polarimeter,
would bring  the statistical sensitivity further down to the $10^{-4}$
level. The dominant  systematic errors is expected to arise from the
misalignment of the polarimeter and the muon magnetic field; this will be
determined from data, and  Monte
Carlo studies indicate a residual systematics at the $10^{-4}$
level.

It is proposed to run TREK in the early stage of J-PARC operation. 
The experimental group has already started relevant R\&D for the
upgrades after obtaining scientific approval, and the  exact
schedule will be determined after funding is granted.

\begin{table}[bht]
\begin{minipage}{\textwidth}
\caption{Goal of the J-PARC TREK experiment compared with the E246 result.}
\begin{tabular*}{\textwidth}{@{\extracolsep{\fill}}l|cc}
\hline\hline
	      		& E246 @ KEK-PS   			& TREK @ J-PARC   		\\
\hline
Detector       		& SC toroidal spectrometer  		& E246-upgraded   		\\
Proton beam energy		& 12 GeV			 	& 30 GeV		 	\\
Proton intensity		& $1.0\times 10^{12}$/s 		& $6\times 10^{13}$/s 		\\
$K^+$ intensity     	& $1.0\times 10^{5}$/s   		& $3\times 10^{6}$/s  	\\ 
Run time       		& $\sim 2.0\times 10^7$s   		& $1.0\times 10^7$s  		\\
$\sigma(P_T)_{\rm stat}$    	& $2.3\times 10^{-3}$  		& $\sim 1.0\times 10^{-4}$	\\
$\sigma(P_T)_{\rm syst}$    	& $1.1\times 10^{-3}$  		& $ < 1.0\times 10^{-4}$ 	\\
\hline\hline
\end{tabular*}
\end{minipage}
\end{table}

\subsection{Measurement of CP violation in ortho-positronium decay}\label{sec:exp:CP:posi}
\label{FEL}
CP violation in the \ops\ decay can be detected by an accurate
measurement of the angular correlation between the \ops\ spin \sopvec\
and the momenta of the photons from the \ops\
decay~\cite{Bernreuther:tt}, as shown in Eq.~(\ref{ACP}). It is useful
then to write the measurable quantity:
\begin{equation}
\ensuremath{N(\cos{\theta})=N_0(1+\ccp\cos{\theta})\ ,}
\end{equation}
with the CP violation amplitude parameter, \ccp, different from zero,
if CP violating interactions take part in the \ops\ decay.  In this
equation, $N(\cos{\theta})$ is the number of events with a measured
value $\cos{\theta}\pm|\Delta(cos(\theta))|$ (hereafter, for the sake
of simplicity, it will be refered to as the $\cos{\theta}$ value,
intending that this is measured with an uncertainty, depending on the
spatial resolution of the detector).  Here $\cos{\theta}$ is defined
as the product of $\cos{\theta_{1}}$, the cosine of the angle between
the \sopvec\ and the unit vector in the direction of highest energy
photon \kone, and $\cos{\theta_{n}}$, the cosine of the angle between
the
\sopvec\ and the unit vector in the direction perpendicular to the \ops\ decay plane, $\hat{n}\def\kone\times\ktwo$.
Within the Standard Model, \ccp\ in the positronium (Ps) system is expected to be very small, of the order of $10^{-9}$\cite{Skalsey:vt}.

The measured distribution $N(\cos{\theta})$ should show an asymmetry
given by
$\ensuremath{N(\cos{\theta_+})-N(\cos{\theta_-})}=2N_0\ccp\cos{\theta}$,
for $\cos{\theta_+}=-\cos{\theta_-}=\cos{\theta}$. The quantity \ccp\
can be determined by measuring the rate of events \nplus\ for a given
$\cos{\theta_+}=\cos{\theta}$ and \nminu\ for
$\cos{\theta_-}=-\cos{\theta}$. In practice, $\nplus\def\
N(\cos{\theta_+}$ is the number of events in which \ktwo\ forms an
angle with \kone\ smaller than $\pi$, and the \ops\ spin forms an
angle $\theta_{n}$ smaller than $\pi/2$ with the perpendicular to the
\ops\ decay plane. In the $\nminu\def\ N(\cos{\theta_-}$ events,
\ktwo\ forms an angle $2\pi-\theta_{12}$ with \kone\ and the \sopvec\
forms an angle $\pi-\theta_{n}$ with the normal to the \ops\ decay
plane. In other terms, in the \nminu\ events the perpendicular to the
\ops\ decay plane is reversed with respect to the \nplus\ events, by
flipping the direction of \ktwo\ specularly with respect to \kone.
Then the measurement of the asymmetry
\begin{equation}
A=\frac{(\nplus-\nminu)}{(\nplus+\nminu)}=\ccp\cos{\theta}
\label{equasy}
\end{equation}
allows to derive the experimental value of \ccp.

The measurement of the asymmetry $A$ implies that $\cos{\theta}$ in
Eq.~(\ref{equasy}) is a well defined quantity in the experiment. In
turn, this implies that the \ops\ spin direction is defined. This
direction can be selected using an external magnetic field \Bvec ,
which aligns the \ops\ spin parallel (m=1), perpendicular (m=0) or
antiparallel (m=$-1$) to the field direction. The magnetic field, in
addition, perturbs and mixes the two m=0 states (one for the para-Ps
and the other for the \ops).  Thus, two new states are possible for
the Ps system: the perturbed singlet and the perturbed triplet states,
both with m=0.  Their lifetimes depend on the \Bvec\ field intensity.
The perturbed singlet state has a lifetime shorter than 1 ns (as for
the unperturbed singlet state of the para-Ps), which is not relevant
in the measurement described here, because too short compared to the
typical detector time resolution of 1 ns.  For values of $\vert
\Bvec\vert$ of few kGauss,the perturbed \ops\ lifetime can be
substantially reduced~\cite{Felcini:2004yn} with respect to the
unperturbed value of about 142 ns~\cite{Jinnouchi:2003hr}.  Thanks to
this effect, it is possible to separate the m=0 from the m=$\pm 1$
states, by measuring the \ops\ decay time. This is the time between
the positron emission (by \eg\ a $^{22}$Na positron source) and the
detection of the \ops\ decay photons. The Ps is formed in a target
region, where SiO$_2$ powder is used as target material.  The value of
the $|\Bvec|$ field that maximizes the decay time separation between
m=0 and m=$\pm 1$ states is found to be B=4 kGauss, corresponding to a
m=0 perturbed \ops\ lifetime of 30 ns.

The measurement of the asymmetry $A$ is performed in the
following way. The direction and intensity of the $\Bvec$ field are
fixed. The \kone\ and \ktwo\ detectors are also fixed. In this way
$\cos{\theta}$ has a well defined value. For each event, the Ps decay
time and the energies of the three photons from the \oPs\ decay are
measured. The off-line analysis requires the highest energy photon in
the \kone\ detector to be within an energy range $\Delta
E_1=E_1^{max}-E_1^{min}$. The second highest energy photon must be
recorded in the \ktwo\ detector within an energy range $\Delta
E_2=E_2^{max}-E_2^{min}$. Then the \nplus\ and \nminu\ events are
counted to determine the asymmetry in Eq.~(\ref{equasy}). The
measurement of the asymmetry $A$ in both the perturbed states
(selected imposing short decay time, \eg\ between 10 and 60 ns) and
unperturbed states (selected imposing long decay time, between 60 and
170 ns) allows to eliminate the time-independent
systematics~\cite{Skalsey:vt}. Other systematics, which are
time-dependent, do not cancel out with this method and determine the
final uncertainty on the \ccp\ measurement.

\begin{figure}[tb]
\parbox{0.58\linewidth}{
\includegraphics[width=\linewidth]{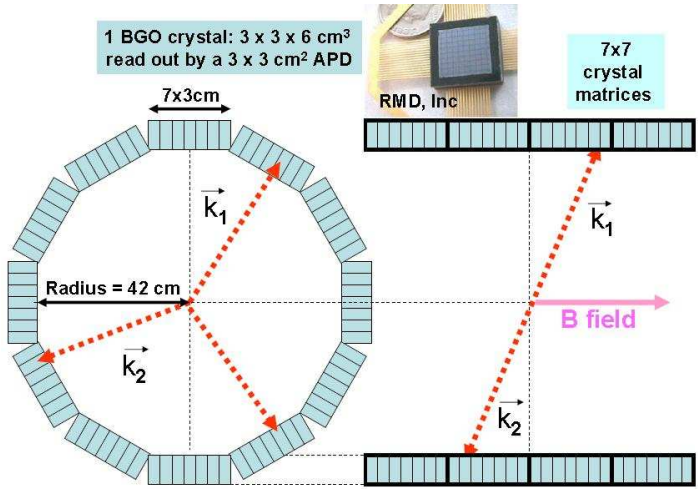}}\hspace*{\fill}
\parbox{0.38\linewidth}{
\caption{Schematic view of the BGO crystal barrel calorimeter used to detect the photons from the \ops\ decay: left, detector front view and definition of the \kone\ and
\ktwo\ vectors; right, detector side view, showing also the direction of the magnetic field \Bvec .\label{newdetec}}}
\end{figure}
An improved detector with superior spatial and energy resolution, as
compared to \cite{Skalsey:vt}, is sketched in Fig.~\ref{newdetec}.  It
consists of a barrel of BGO crystals with the \ops\ forming region at
its centre.  The crystal signals are read out by avalanche photodiodes
(APD), as the detector must work in the magnetic field. Improved
spatial and angular resolution is obtained thanks to the smaller size
of the crystal face exposed to the photons, $3\times 3$ cm$^2$, and
the larger barrel radius, 42 cm.  Note that such a detector could also
be used efficiently for PET scanning, combined with NMR diagnostic.
This possibility makes this detector a valuable investment also for
applications in nuclear medical imaging.

With this detector configuration and a simulation of the detector
response, the precision to be reached in the measurement of the \ccp\
parameter has been evaluated. A similar analysis was used for the
event selection as described in \cite{Felcini:2004yn}, except that no
veto is needed in the present configuration, thanks to the good
spatial resolution of the proposed crystals. Various uncertainties
affect the \ccp\ measurement.  The time-dependent uncertainties on the
asymmetry $A$ are induced mainly by the two-photon background, which
affect more strongly the events with shorter decay time, as well as by
the inhomogeneity of the \ops\ formation region, which affect the
measurement of the \ops\ decay time.  For high event statistics (at
least $10^{12}$ selected three photon events) the following
contributions to the asymmetry measurement were found: $\Delta
A_{stat} \sim 10^{-6}$, $\Delta A_{syst}(2\gamma\ bkgd)\sim 10^{-6}$,
$\Delta A_{syst}(\ops\ formation)\sim 2\times 10^{-6}$ resulting into
a total uncertainty: $\Delta A_{stat+syst}\sim 2.5\times
10^{-6}$. Being $\Delta\ccp$ related to the asymmetry total
uncertainty by the relation $\Delta\ccp=\Delta
A_{stat+syst}/Q$~\cite{Skalsey:vt} with Q, the analyzing power,
evaluated to be $\sim 0.5$ for this detector configuration, the total
uncertainty on the \ccp\ parameter is $\Delta\ccp\sim 5\times
10^{-6}$.

Although this precision is not sufficient to measure the expected Standard Model \ccp\ value of order of $10^{-9}$ , it is suitable to 
discover CP violating terms in the order of $10^{-5}$, which if detected would be signal of unexpected new physics beyond the Standard Model.

\section{LFV experiments}\label{sec:exp:LFV}
Mixing of leptonic states with different family number as observed in
neutrino oscillations does not necessarily imply measurable branching
ratios for LFV processes involving
the charged leptons. In the Standard Model the rates of LFV decays are
suppressed relative to the dominant family-number conserving modes by
a factor $(\delta m_{\nu}/m_W)^4$ which results in branching ratios
which are out of reach experimentally. Note that a similar family
changing quark decay such as $b \to s \gamma$ does obtain a very
significant branching ratio of $O(10^{-4})$ due to the large top mass.

As has been discussed in great detail in this report, in almost any
further extension to the Standard Model such as Supersymmetry, Grand
Unification or Extra Dimensions additional sources of LFV appear. For
each scenario a large number of model calculations can be found in the
literature and have been reviewed in previous sections, with
predictions that may well be accessible experimentally. Improved
searches for charged LFV thus may either reveal physics beyond the SM
or at least lead to a significant reduction in parameter space allowed
for such exotic contributions.

Charged LFV processes, i.e. transitions between $e$, $\mu$, and
$\tau$, might be found in the decay of almost any weakly decaying
particle. Although theoretical predictions generally depend on
numerous unknown parameters these uncertainties tend to cancel in the
relative strengths of these modes. Once LFV in the charged lepton
sector were found, the combined information from many different
experiments would allow us to discriminate between the various
interpretations.  Searches have been performed in $\mu$, $\tau$,
$\pi$, $K$, $B$, $D$, $W$ and $Z$ decay. Whereas highest experimental
sensitivities were reached in dedicated $\mu$ and $K$ experiments,
$\tau$ decay starts to become competitive as well.

\subsection{Rare $\mu$ decays}\label{sec:exp:LFV:mu}
LFV muon decays include the purely leptonic modes $\mu^+ \to e^+ \gamma$ and $\mu^+ \to e^+ e^+ e^-$, as well as the semi-leptonic $\mu - e$ conversion in muonic
atoms and the muonium - antimuonium oscillation. The present experimental limits are listed in Table~\ref{tab:mulim}.
\begin{table}[bht]
\begin{minipage}{\textwidth}
\caption{Present limits on rare $\mu$ decays. \label{tab:mulim}}
\begin{tabular*}{\textwidth}{@{\extracolsep{\fill}}lrccc}
\hline\hline
mode					&upper limit (90\% C.L.)\hspace*{-8mm}	&year	&Exp./Lab.		&Ref.				\\
\hline
$\mu^+\to e^+ \gamma$			&$1.2 \times 10^{-11}$			&2002	&MEGA / LAMPF		&\cite{Brooks:1999pu,Ahmed:2001eh}	\\
$\mu^+\to e^+ e^+ e^-$			&$1.0 \times 10^{-12}$			&1988	&SINDRUM I / PSI	&\cite{Bellgardt:1987du}		\\
$\mu^+ e^- \leftrightarrow \mu^- e^+$ 	&$8.3 \times 10^{-11}$			&1999	&PSI			&\cite{Willmann:1998gd}			\\
$\mu^-\;$Ti$\;\to e^-$Ti		&$6.1 \times 10^{-13}$			&1998	&SINDRUM II / PSI	&\cite{Wintz:1998rp}			\\
$\mu^-\;$Ti$\;\to e^+$Ca$^*$	 	&$3.6 \times 10^{-11}$			&1998	&SINDRUM II / PSI	&\cite{Kaulard:1998rb}			\\
$\mu^-\;$Pb$\;\to e^-$Pb	 	&$4.6 \times 10^{-11}$			&1996	&SINDRUM II / PSI	&\cite{Honecker:1996zf}			\\
$\mu^-\;$Au$\;\to e^-$Au	 	&$7 \times 10^{-13}$			&2006	&SINDRUM II / PSI	&\cite{Bertl:2006up}			\\
\hline\hline
\end{tabular*}
\end{minipage}
\end{table}

Whereas most theoretical models favor $\mu^+ \to e^+ \gamma$, this
mode has a  disadvantage from an experimental point of view
since the sensitivity is limited by accidental $e^+ \gamma$
coincidences and muon beam intensities have to be reduced now already.
Searches for $\mu - e$ conversion, on the other hand, are limited by
the available beam intensities and large improvements in sensitivity
may still be achieved.

All recent results for $\mu^+$ decays were obtained with ``surface'' muon beams containing muons originating in the decay of $\pi^+$'s that stopped
very close to the surface of the pion production target, or ``subsurface'' beams from pion decays just below that region. Such beams
are superior to conventional pion decay channels in terms of muon stop density and permit the use of relatively thin (typically 10~mg/cm$^2$) foils to stop the beam.
Such low-mass stopping targets are required for the ultimate resolution in positron momentum and emission angle, minimal photon yield, or the efficient production of
muonium in vacuum.

\subsubsection{$\mu \to e \gamma$}
Neglecting the positron mass the 2-body decay $\mu^+ \to e^+ \gamma$ of muons at rest is characterized by:
\begin{eqnarray*}
E_{\gamma} 	&=&	E_e = m_{\mu}c^2/2 = 52.8~{\rm MeV}	\\
\Theta_{e\gamma}&=&	180^{\circ}				\\
t_{\gamma} 	&=& 	t_e					
\end{eqnarray*}
All $\mu \to e \gamma$ searches performed during the past three decades were limited by accidental coincidences between a positron from normal muon decay,
$\mu \to e \nu \overline{\nu}$, and a photon produced in the decay of another muon, either by bremsstrahlung or by $e^+ e^-$ annihilation in flight. This background
dominates by far the intrinsic background from radiative muon decay $\mu \to e \nu \overline{\nu} \gamma$.  Accidental $e\gamma$ coincidences can be suppressed by testing
the three conditions listed above. The vertex constraint resulting from the ability to trace back positrons and photons to an extended stopping target
can further reduce background. Attempts have been made to suppress accidental coincidences by observing the low-energy positron associated with the photon, but 
with minimal success. High muon polarisation ($P_{\mu}$) could help if one would limit the solid angle to accept only positrons and photons (anti-)parallel to the muon
spin since their rate is suppressed by the factor $1-P_{\mu}$ for anti-parallel emission at $E = m_{\mu}c^2$/2 but the reduced solid angle would have to be compensated
by increased beam intensity which would raise the background again.

The most sensitive search to date was performed by the MEGA Collaboration at the Los Alamos Meson Physics Facility (LAMPF) which established
an upper limit  (90\% C.L.) on $B(\mu\to e \gamma)$ of $1.2 \times 10^{-11}$~\cite{Brooks:1999pu,Ahmed:2001eh}.
The MEG experiment~\cite{MEGprop} at PSI, aims at a single-event sensitivity of $\sim 10^{-13} - 10^{-14}$, and  began commissioning in early 2007.  
A straightforward improvement factor of more than an order of magnitude in suppression of accidental background results from the DC muon beam at PSI, as opposed
to the pulsed LAMPF beam which had a macro duty cycle of 7.7\% . Another order of magnitude improvement is achieved by superb time resolution ($\approx 0.15$~ns FWHM
on $t_{\gamma}- t_e$).
\begin{figure}[b!]
\includegraphics*[clip, trim=-35 111 -35 0, width=\linewidth]{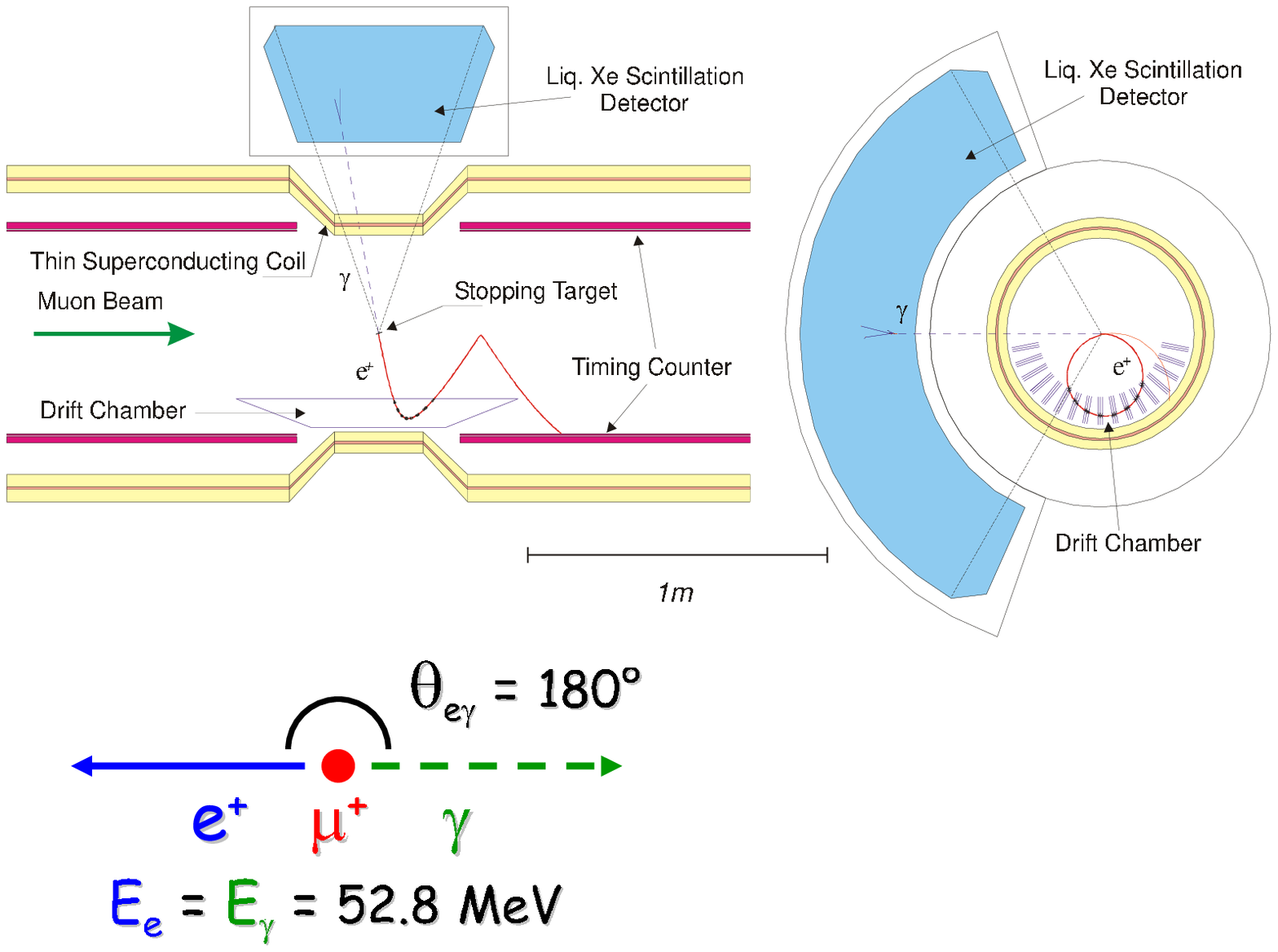}
\caption{Side and end views of the MEG setup. The magnetic field is shaped such that positrons are quickly swept out of the tracking region thus minimizing the load on
the detectors. The cylindrical 0.8~m$^3$ single-cell LXe detector is viewed from all sides by 846 PMTs immersed in the LXe allowing the reconstruction of photon energy,
time, conversion point and direction and the efficient rejection of pile-up signals.}
\label{fg:meg}
\end{figure}

\begin{figure}[t!]
\parbox{0.58\linewidth}{
\includegraphics*[clip, trim=0 0 0 0, width=\linewidth]{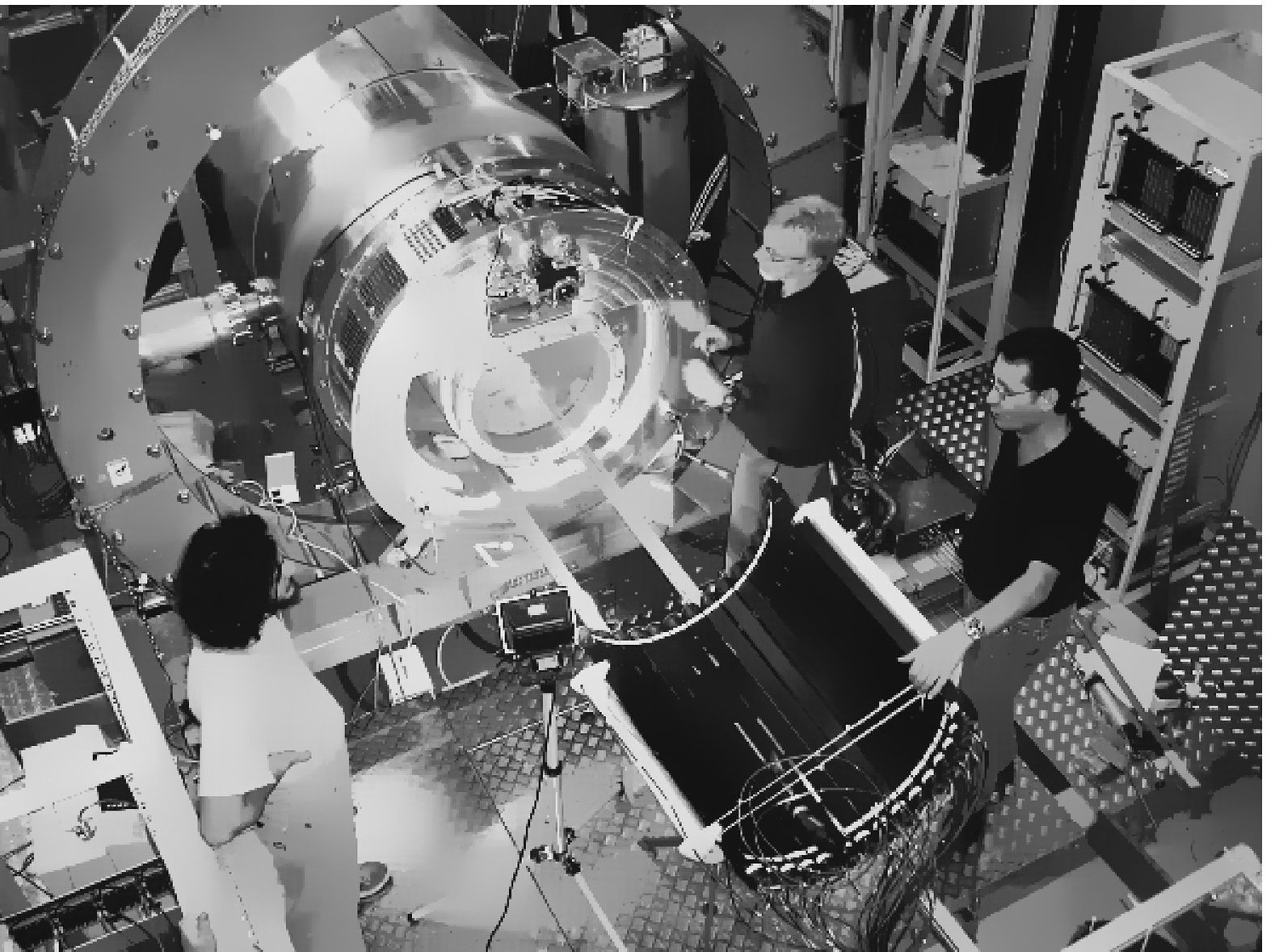}}\hspace*{\fill}
\parbox{0.39\linewidth}{\vspace*{3cm}
\caption{\newline Installing one of the timing counters into the COBRA magnet during the pilot run with the positron spectrometer at the end of 2006.
The large ring is one of two Helmholtz coils used to compensate the COBRA stray field at the locations of the photomultipliers of the LXe detector.\label{fig:MEG2}}}
\end{figure}
The MEG setup is shown in Fig.~\ref{fg:meg}. The spectrometer magnet makes use of a novel ``COBRA'' (COnstant Bending RAdius) design which results in a graded
magnetic field varying from 1.27~T at the centre to 0.49~T at both ends. This field distribution not only results in a constant projected bending radius for the
52.8~MeV positron, for polar emission angles $\theta$ with $|\cos\theta| < 0.35$ , but also sweeps away positrons with low longitudinal momenta much faster than a
constant field as used by MEGA. This design significantly reduces the instantaneous rates in the drift chambers.

The drift chambers are made of 12.5~$\mu$m thin foils supported by C-shaped carbon fibre
frames which are out of the way of the positrons. The foils have ``vernier" cathode pads which permit the measurement of the trajectory coordinate along the anode wires
with an accuracy of about 500~$\mu$m.  

There are two timing counters at both ends of the magnet (see Fig.~\ref{fig:MEG2}), each of which consists of a layer of plastic scintillator fibers and 15 plastic
scintillator bars of dimensions $4\times 4\times 90$~cm$^3$. The fibers give hit positions along the beam axis and the bars measure positron timings with a precision of
$\sigma = 40$~ps. The counters are placed at large radii so only high energy positrons reach them, giving a total rate of a few $10^4$/s for each bar.

High-strength Al-stabilized conductor for the magnet coil makes the
magnet as thin as 0.20 $X_0$ radially, so that 85\% of 52.8~MeV gamma
rays traverse the magnet without interaction before entering the gamma
detector placed outside the magnet. Whereas MEGA used rather
inefficient pair spectrometers to detect the photon, MEG developed a
novel liquid Xe scintillation detector, shown in Fig.~\ref{fg:meg}. By
viewing the scintillation light from all sides the electromagnetic
shower induced by the photon can be reconstructed which allows a
precise measurement of the photon conversion
point~\cite{Mihara:2004dj}. Special PMTs that work at LXe temperature
(-110$^\circ$C), persist under high pressures and are sensitive to the
VUV scintillation light of LXe ($\lambda\approx 178$~nm) have been
developed in collaboration with Hamamatsu Photonics. To identify and
separate pile-up efficiently, fast waveform digitizing is used for all
the PMT outputs.

The performance of the detector was measured with a prototype. The
results are shown in Table~\ref{tab:MEG}.  First data taking with the
complete setup took place during the second half of 2007. A
sensitivity of ${\cal O}(10^{-13})$ for the 90\% C.L. upper limit in
case no candidates are found should be reached after two years.

\begin{table}[b]
\begin{minipage}{\textwidth}
\caption{Performance of a prototype of the MEG LXe detector at $E_{\gamma}$=53~MeV. \label{tab:MEG}}
\begin{tabular*}{\textwidth}{@{\extracolsep{\fill}}llll}
\hline\hline
&observable		&resolution ($\sigma$)	&\\
\hline
&energy			&1.2\%			&\\
&time			&65~ps			&\\
&conversion point	&$\approx$4~mm		&\\
\hline\hline
\end{tabular*}
\end{minipage}
\end{table}

\subsubsubsection{Beyond MEG}
Ten times larger surface muon rates than used by MEG can be achieved at PSI today already but the background suppression would have to be improved by two orders of
magnitude.
Accidental background $N_{\rm acc}$ scales with the detector resolutions as:
\begin{equation*}
N_{\rm acc} \propto \Delta E_e \cdot \Delta t  \cdot (\Delta E_{\gamma} \cdot \Delta \Theta_{e\gamma} \cdot \Delta x_{\gamma})^2 \cdot A^{-1}_T \;\; ,
\end{equation*}
with $x_{\gamma}$ the coordinate of the photon trajectory at the target and $A_T$ the target area. Here it is assumed that the photon can be traced back to the
target with an uncertainty which is small compared to $A_T$. Since the angular resolution is dictated by the positron multiple scattering in the target this can
be written:
\begin{equation*}
N_{\rm acc} \propto \Delta E_e \cdot \Delta t  \cdot (\Delta E_{\gamma}  \cdot \Delta x_{\gamma})^2 \cdot \frac {d_T}{A_T}\;\; ,
\end{equation*}
with $d_T$ the target thickness. When using a series of $n$ target foils each of them could have a thickness of $d_T/n$ and the beam would still be stopped.
Since the area would increase like $n \cdot A_T$ the background could be reduced in proportion with $1/n^2$:
\begin{equation*}
N_{\rm acc} \propto \Delta E_e \cdot \Delta t  \cdot (\Delta E_{\gamma} \cdot \Delta x_{\gamma})^2 \cdot \frac {d_T/n }{n \cdot A_T}\; ,
\end{equation*}
so a geometry with ten targets, 1~mg/cm$^2$ each, would lead to the required background suppression.

\subsubsection{$\mu \to 3e$}
As has been discussed above the sensitivity of $\mu \to e \gamma$
searches is limited by background from accidental coincidences between
a positron and a photon originating in the independent decays of two
muons. Similarly, searches for the decay $\mu\to 3e$ suffer from
accidental coincidences between positrons from normal muon decay and
$e^+e^-$ pairs originating from photon conversions or scattering of
positrons off atomic electrons (Bhabha scattering). For this reason
the muon beam should be continuous on the time scale of the muon
lifetime and longer. In addition to the obvious constraints on
relative timing and total energy and momentum, which can be applied in
$\mu \to e \gamma$ searches as well, there are powerful constraints on
vertex quality and location to suppress the accidental
background. Since the final state contains only charged particles the
setup may consist of a magnetic spectrometer without the need for an
electromagnetic calorimeter with its limited performance in terms of
energy and directional resolution, rate capability, and event
definition in general. On the other hand, of major concern are the
high rates in the tracking system of a $\mu \to 3e$ setup which has to
stand the load of the full muon decay spectrum.

\subsubsubsection{The SINDRUM I experiment}

The present experimental limit, $B(\mu \to 3e) < 1 \times 10^{-12}$~\cite{Bellgardt:1987du}, was published way back in 1988. Since no new proposals
exist for this decay mode we shall analyse the prospects of an improved experiment with this SINDRUM experiment as a point of reference. A detailed description
of the experiment may be found in Ref.~\cite{Bertl:1985mw}.

Data were taken during six months using a 25~MeV/$c$ subsurface
beam. The beam was brought to rest with a rate of $6 \times 10^6\,
\mu^+$~s$^{-1}$ in a hollow double-cone foam target (length 220~mm,
diameter 58~mm, total mass 2.4~g). SINDRUM I is a solenoidal
spectrometer with a relatively low magnetic field of 0.33~T
corresponding to a transverse-momentum threshold around 18~MeV/$c$ for
particles crossing the tracking system. This system consisted of five
cylindrical MWPCs concentric with the beam axis. Three-dimensional
space points were found by measuring the charges induced on cathode
strips oriented $\pm 45^\circ$ relative to the sense wires. Gating
times were typically 50~ns.  The spectrometer acceptance for $\mu \to
3e$ was 24\% of 4$\pi$ sr (for a constant transition-matrix element)
so the only place for a significant improvement in sensitivity would
be the beam intensity.

\begin {figure}[htb]
\parbox{0.4\linewidth}{\includegraphics[width=\linewidth]{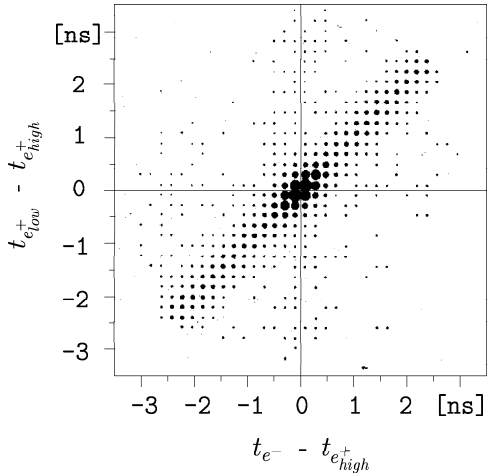}}\hspace*{\fill}
\parbox{0.56\linewidth}{
\caption{Relative timing of $e^+e^+e^-$ events. The two positrons are labelled {\em low} and {\em high} according to the invariant mass
when combined with the electron. One notices a contribution of correlated triples in the centre of the distribution. These events
are mainly $\mu \to 3e \nu \overline{\nu}$ decays. The concentration of events along the diagonal is due to low-invariant-mass $e^+e^-$
pairs in accidental coincidence with a positron originating in the decay of a second muon. The $e^+e^-$ pairs are predominantly due to
Bhabha scattering in the target.\label{SNDR_dt}}}
\end {figure}
Figure~\ref{SNDR_dt} shows the time distribution of the recorded $e^+e^+e^-$ triples. Apart from a prompt contribution of correlated
triples one notices a dominant contribution from accidental coincidences involving low-invariant-mass $e^+e^-$ pairs. Most of these are explained by
Bhabha scattering of positrons from normal muon decay $\mu \to e \nu \overline{\nu}$. The accidental
background thus scales with the target mass, but it is not obvious how to reduce this mass  significantly below the 11~mg/cm$^2$
achieved in this search.

\begin {figure}[htb]
\centerline{\includegraphics[height=60mm]{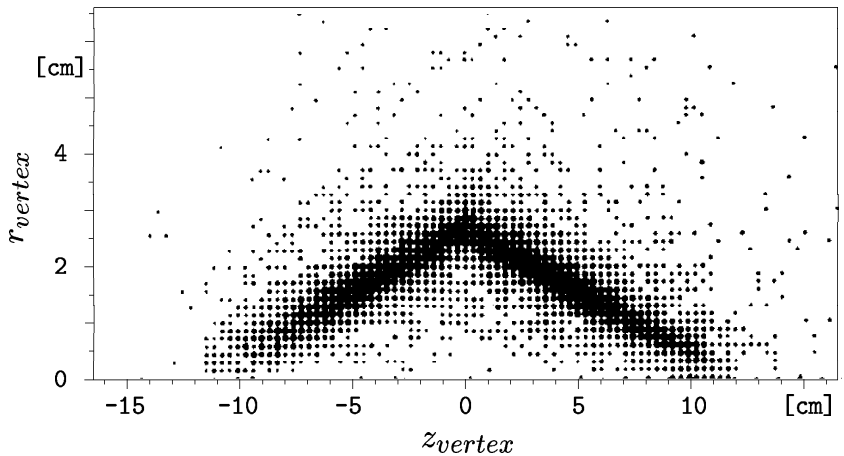}}
\caption{Spatial distribution of the vertex fitted to
prompt $e^+e^+e^-$ triples. One clearly notices the double-cone target.
\label{SNDR_vtx}}
\end {figure}

Figure~\ref{SNDR_vtx} shows the vertex distribution of prompt events. One should keep in mind that most of the uncorrelated triples contain
$e^+e^-$ pairs coming from the target and their vertex distribution will thus follow the target contour as well. This 1-fold accidental
background is
suppressed by the ratio of the vertex resolution (couple of mm$^2$) and the target area. There is no reason, other than the cost of  the
detection system, not to choose a much larger target. Such an  increase might also help to reduce the load on the tracking detectors.
Better vertex resolution would help as well. At these low energies tracking errors are dominated by multiple scattering in the first
detector layer but it should be possible to gain by bringing it closer to the target.

\begin {figure}[htb]
\begin{center}
\includegraphics[width=0.8\linewidth]{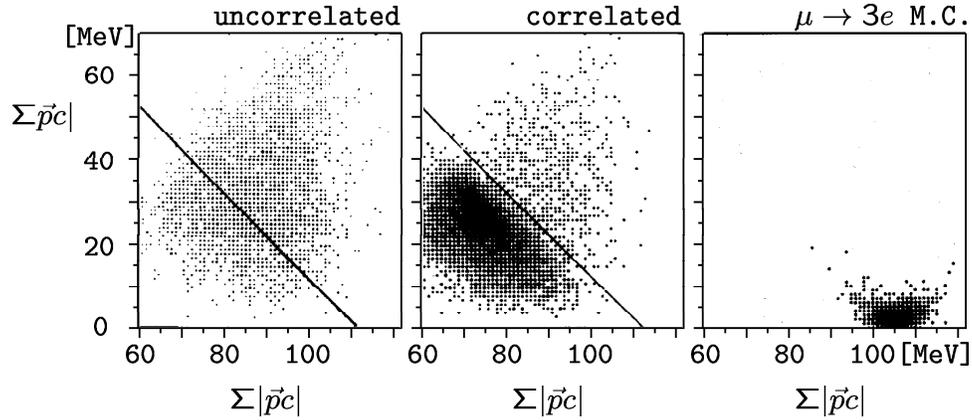}
\end{center}
\caption{Total momentum versus total energy for three event classes discussed in the text. The line shows the kinematic limit
(within resolution) defined by $\Sigma |\vec{p}c| + |\Sigma \vec{p}c| \leq m_{\mu}c^2$ for any muon decay. The enhancement in the distribution of correlated triples
below this limit is due to the decay $\mu \to 3e \nu \overline{\nu}$.
\label{SNDR_ep}}
\end {figure}

Finally, Fig.~\ref{SNDR_ep} shows the distribution of total momentum versus total energy for three classes of events, (i)
uncorrelated $e^+e^+e^-$  triples, (ii) correlated  $e^+e^+e^-$ triples, and (iii) simulated $\mu \to 3e$ decays. The distinction
between uncorrelated and correlated triples has been made on the basis of relative timing and vertex as discussed above.

\subsubsubsection{How to improve?}
What would  a $\mu \to 3e$ set-up look like that would aim at a single-event sensitivity around $10^{-16}$, i.e., would make use of a beam rate around
$10^{10}$~$\mu^+$/s? The SINDRUM I measurement was background-free at the level of $10^{-12}$ with a beam of $0.6  \times 10^{7}$~$\mu^+$/s. Taking into account that
background would have set in at $10^{-13}$, the increased stop rate would raise the background level to $\approx 10^{-10}$, so six orders of magnitude
in background reduction would have to be achieved. Increasing the target size and improving the tracking resolution should bring
two orders of magnitude from the vertex requirement alone. Since the dominant sources of background are accidental coincidences
between two decay positrons (one of which undergoes Bhabha scattering) the background rate scales with the momentum resolution
squared. Assuming an improvement by one order of magnitude, i.e., from the $\approx 10\%$ FWHM obtained by SINDRUM I to $\approx 1\%$
for a new search, one would gain two orders of magnitude from the constraint on total energy alone. The remaining factor 100 would result
from the test on the collinearity of the $e^+$ and the $e^+e^-$ pair.

As mentioned in Ref.~\cite{Bertl:1985mw} a dramatic suppression of background could be achieved by requiring a minimal opening angle (typically
30$^\circ$) for both $e^+e^-$ combinations. Depending on the mechanism for $\mu \to 3e$, such a cut might, however, lead to a strong loss in
$\mu \to 3e$ sensitivity as well.

Whereas background levels may be under control, the question remains whether detector concepts can be developed that work at the high beam
rates proposed. A large modularity will be required to solve problems of pattern recognition.

\subsubsection{$\mu - e$ conversion}
When negatively charged muons stop in matter they quickly form muonic
atoms which reach their ground states in a time much shorter than the
lifetime of the atom.  Muonic atoms decay mostly through {\it muon
decay in orbit} (MIO) $\mu^- (A,Z) \to e^- \nu_{\mu} \overline{\nu}_e
(A,Z)$ and {\it nuclear muon capture} (MC) $\mu^-(A,Z) \to\nu_{\mu}
(A,Z-1)^*$ which in lowest order may be interpreted as the incoherent
sum of elementary $\mu^-p \to n \nu_{\mu}$ captures. The MIO rate
decreases slightly for increasing values of $Z$ (down to 85\% of the
free muon rate in the case of muonic gold) due to the increasing muon
binding energy. The MC rate at the other hand increases roughly
proportional to $Z^4$. The two processes have about equal rates around
$Z=12$.

When the hypothetical $\mu - e$ conversion leaves the nucleus in its
ground state the nucleons act coherently, boosting the process
relative to the incoherent processes with exited final states. The
resulting $Z$ dependence has been studied by several
authors~\cite{Czarnecki:2002ac,Kosmas:2002if,Kitano:2002mt,Kosmas:1994ti}. For
$Z \lesssim 40$ all calculations predict a conversion probability
relative to the MC rate which follows the linear rise with $Z$
expected naively. The predictions may, however, deviate by factors 2-3
at higher $Z$ values.

As a result of the two-body final state the electrons produced in $\mu - e$ conversion are mono-energetic and their energy is given by:
\begin{equation}
E_{\mu e}=m_{\mu}c^2-B_{\mu}(Z)-R(A)\;,
\end{equation}
where $B_{\mu}(Z)$ is the atomic binding energy of the muon and R is the atomic recoil energy for a muonic atom with atomic number $Z$ and mass number $A$.
In first approximation $B_{\mu}(Z)\propto Z^2$ and $R(A)\propto A^{-1}$.

\subsubsubsection{Background}
Muon decay in orbit (MIO) constitutes an intrinsic background source which can only be suppressed with sufficient electron energy resolution. The process
predominantly results in electrons with energy $E_{\rm MIO}$ below $m_{\mu}c^2/2$, the kinematic endpoint in free muon decay, with a steeply falling high-energy
component reaching up to $E_{\mu e}$. In the endpoint region the MIO rate varies as $(E_{\mu e}-E_{\rm MIO})^5$ and a resolution of $1-2\,$MeV (FWHM) is sufficient
to keep MIO background under control. Since the MIO endpoint rises at lower $Z$ great care has to be taken to avoid low-$Z$ contaminations in and around the target.

Another background source is due to radiative muon capture (RMC) $\mu^-(A,Z) \to \gamma(A,Z-1)^*\nu_{\mu}$ after which the photon creates an $e^+e^-$ pair either
internally (Dalitz pair) or through $\gamma \to e^+e^-$ pair production in the target. The RMC endpoint can be kept below $E_{\mu e}$ for selected isotopes.

Most low-energy muon beams have large pion contaminations. Pions may produce background when stopping in the target through radiative pion capture (RPC) which
takes place with a probability of $O(10^{-2})$. Most RPC photons have energies above $E_{\mu e}$. As in the case of RMC these photons may produce background
through $\gamma \to e^+e^-$ pair production. There are various strategies to cope with RPC background:
\begin{itemize}
\item
One option is to keep the total number of $\pi^-$ stopping in the target during the live time of the experiment below $10^{4-5}$. This can be achieved with the
help of a moderator in the beam exploiting the range difference between pions and muons of given momentum or with a muon storage ring exploiting the difference
in lifetime.
\item
Another option is to exploit the fact that pion capture takes place at a time scale far below a nanosecond. The background can thus be suppressed with a beam
counter in front of the experimental target or by using a pulsed beam selecting only delayed events.
\end{itemize}

Cosmic rays (electrons, muons, photons) are a copious source of electrons with energies around $\approx100\,$MeV. With the exception of $\gamma \to e^+e^-$ pair
production in the target these events can be recognized by an incoming particle. In addition, passive shielding and veto counters above the detection system help
to suppress this background.

\begin{figure}[!th]
\begin{center}
\includegraphics[height=\linewidth, angle=90]{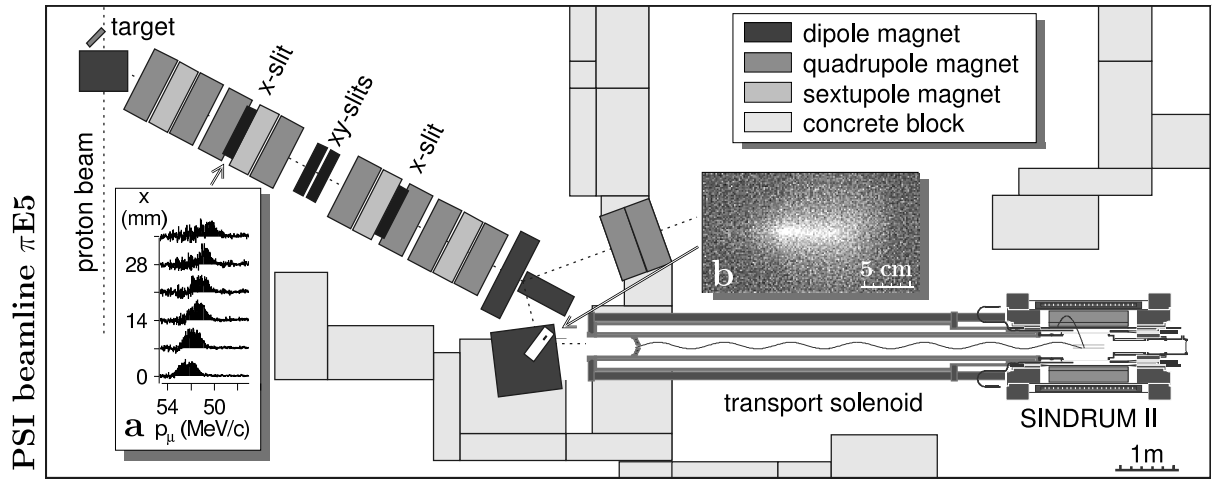}
\caption{Plan view of the SINDRUM II experiment. The $1\,$MW $590\,$MeV proton beam hits the $40\,$mm carbon production target (top left of the figure). The $\pi$E5
beam line transports secondary particles ($\pi, \mu, e$) emitted in the backward direction to a degrader situated at the entrance of a solenoid connected axially to
the SINDRUM~II spectrometer. Inset a) shows the momentum dispersion at the position of the first slit system.
Inset b) shows a cross section of the beam at the position of the beam focus.\label{fig:sindrum2}}
\end{center}
\end{figure}
\begin{figure}[bh!]
\parbox{0.48\linewidth}{\hspace*{0.05\linewidth}\includegraphics[clip, trim=20 210 0 0, width=0.95\linewidth]{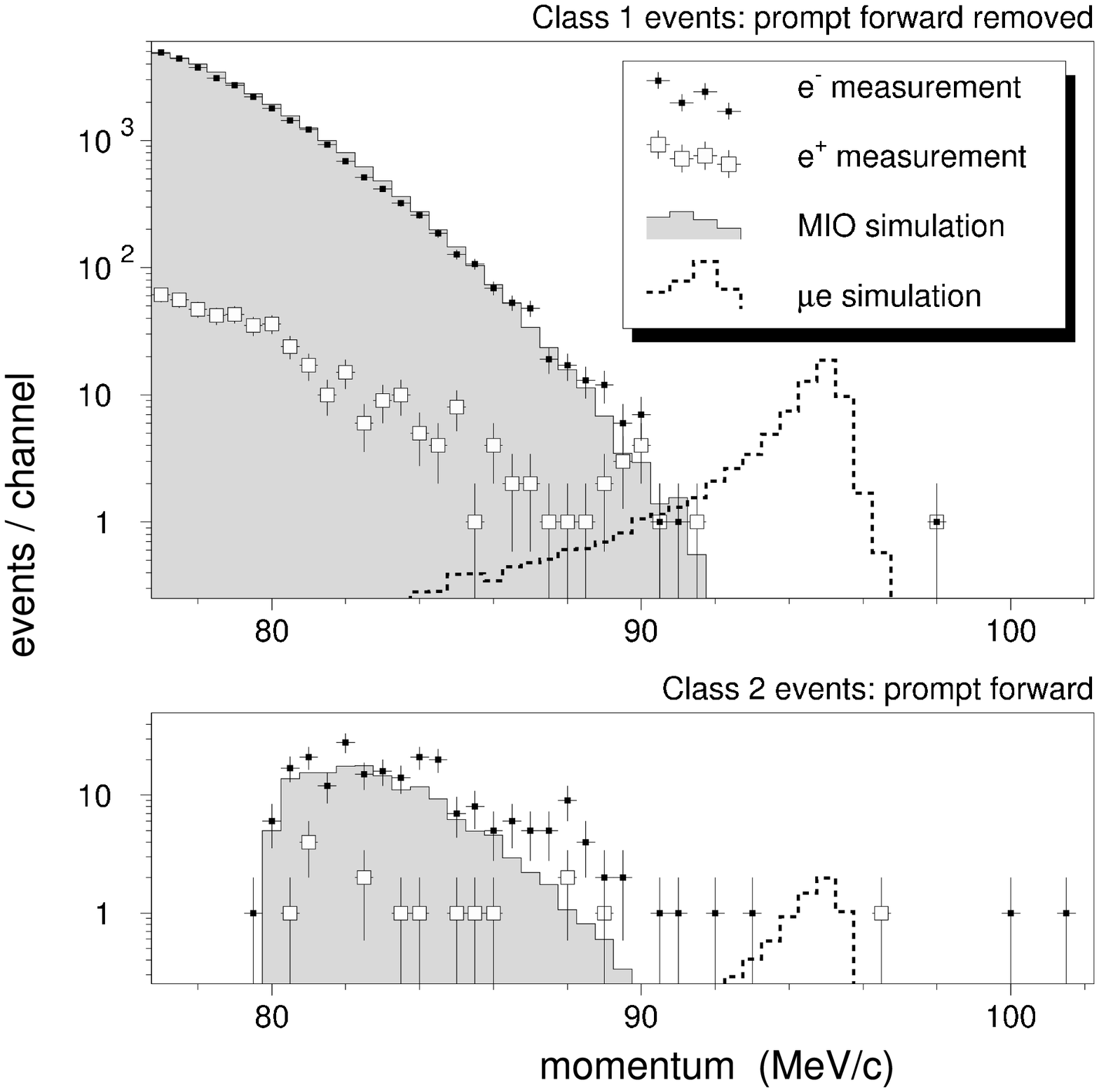}
\begin{picture}(0,0)(0,0) 
\put(0,50){\rotatebox{89.99999}{events/channel}}
\put(100,0){momentum (MeV/c)}
\end{picture}}\hspace*{\fill}
\parbox{0.48\linewidth}{\hspace*{0.05\linewidth}\includegraphics[clip, trim=20 20 0 320, width=0.95\linewidth]{figures/section81/etot_all.ps}
\begin{picture}(0,0)(0,0) 
\put(0,50){\rotatebox[origin=c]{89.99999}{events/channel}}
\put(100,0){momentum (MeV/c)}
\end{picture}
\caption{\newline Momentum distributions of electrons and positrons for two event classes described in the text. Measured distributions are compared with the results of
simulations of muon decay in orbit and $\mu - e$ conversion.\label{fig:sindrumgold}}}
\end{figure}

\subsubsubsection{SINDRUM II}
The present best limits (see Table~\ref{tab:mulim}) have been measured with the SINDRUM~II spectrometer at PSI. Most recently a search was performed on a gold
target~\cite{Bertl:2006up}. In this experiment (see Fig.~\ref{fig:sindrum2}) the pion suppression is based on the factor of two shorter range of pions as compared
to muons at the selected momentum of 52~MeV/c. A simulation using the measured range distribution shows that about one in $10^6$ pions cross an 8~mm thick CH$_2$
moderator. Since
these pions are relatively slow 99.9\% of them decay before reaching the gold target which is situated some $10\,$m further downstream. As a result pion stops in the
target have been reduced to a negligible level. What remains are radiative pion capture in the degrader and $\pi^- \to e^- \overline{\nu}_e$ decay in flight shortly
before entering the degrader. The resulting electrons may reach the target where they can scatter into the solid angle acceptance of the spectrometer. ${\cal O}$(10)
events are expected
with a flat energy distribution between 80 and $100\,$MeV. These events are peaked in forward direction and show a time correlation with the cyclotron rf signal. To cope
with this background two event classes have been introduced based on the values of polar angle and rf phase. Fig.~\ref{fig:sindrumgold} shows the corresponding momentum
distributions.

The spectra show no indication for $\mu - e$ conversion. The corresponding upper limit on
\begin{equation}
B_{\mu e} \equiv \Gamma(\mu^-{\rm Au}\to e^-{\rm Au_{g.s.}}) / \Gamma_{\rm capture}(\mu^-{\rm Au}) < 7 \times 10^{-13} \;\;\; 90\%\; C.L.
\label{eq:limit2}
\end{equation}
has been obtained with the help of a likelihood analysis of the momentum distributions shown in Fig.~\ref{fig:sindrumgold} taking into account muon decay in
orbit, $\mu - e$ conversion, a contribution taken
from the observed positron distribution describing processes with intermediate photons such as radiative
muon capture and a flat component from pion decay in flight or cosmic ray background.

\subsubsubsection{New initiatives} 
Based on a scheme originally
developed during the eighties for the Moscow Meson
Factory~\cite{Dzhilkibaev:1989zb} $\mu e$-conversion experiments are
being considered both in the USA and in Japan. The key elements are:
\begin{itemize}
\item
A pulsed proton beam allows to remove pion background by selecting events in a delayed time window. Proton extinction factors below $10^{-9}$ are needed.
\item
A large acceptance capture solenoid surrounding the pion production target leads to a major increase in muon flux.
\item
A bent solenoid transporting the muons to the experimental target results in a significant increase in momentum transmission compared to a conventional
quadrupole channel. A bent solenoid not only removes neutral particles and photons but also separates electric charges. 
\end{itemize}

Unfortunately, the MECO proposal at BNL~\cite{MECOprop} designed along
these lines was stopped because of the high costs. Presently the
possibilities are studied to perform a MECO-type of experiment at
Fermilab (mu2e). There is good hope that a proton beam with the
required characteristics can be produced with minor modifications to
the existing accelerator complex which will become available after the
Tevatron stops operation in 2009. A letter of intent is in
preparation.

\begin{figure}[bht]
\begin{center}
\includegraphics[width=0.8\linewidth]{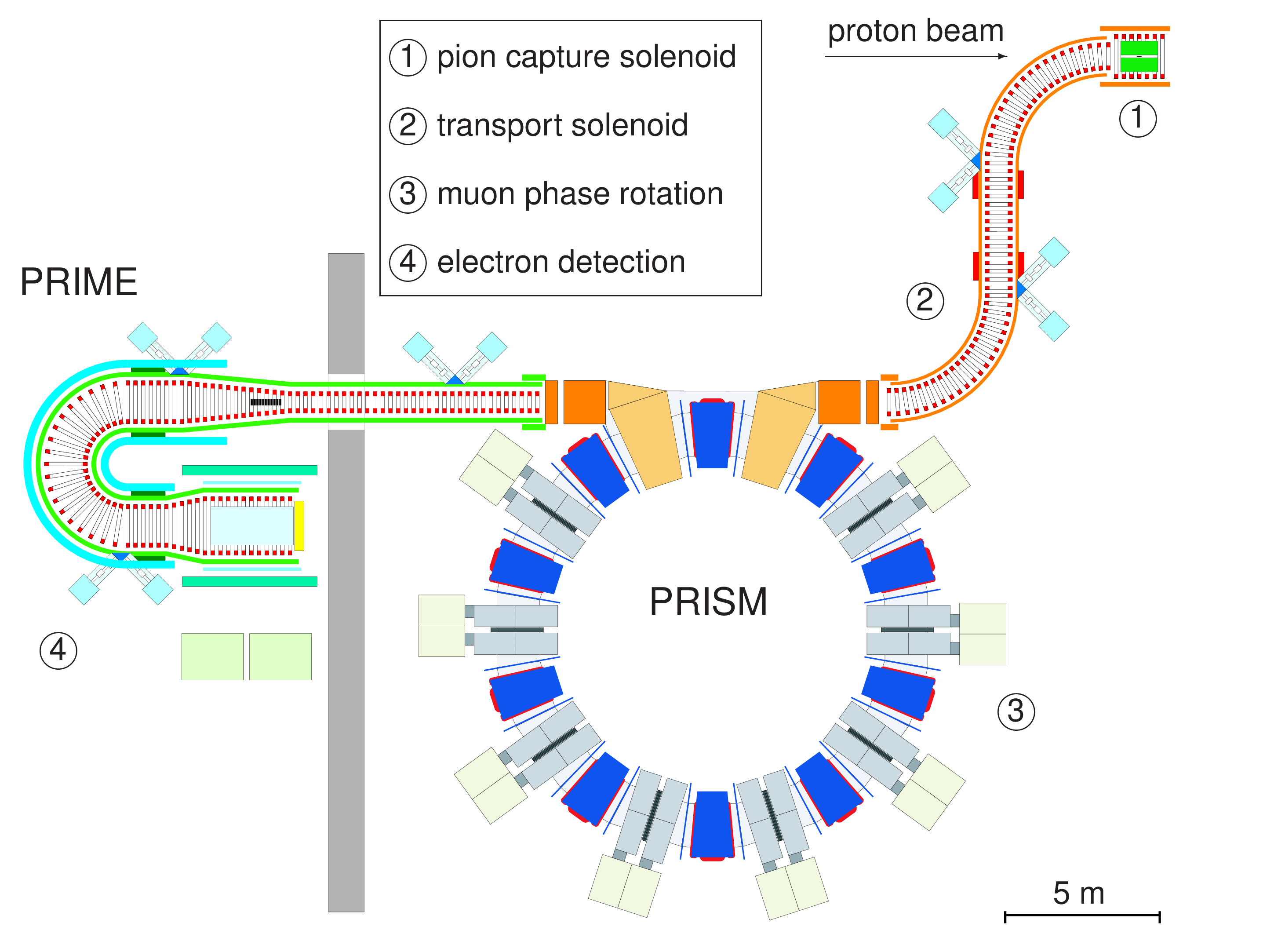}
\caption{Layout of PRISM/PRIME. The experimental target is situated at the entrance of the 180$^{\circ}$ bent solenoid that transports decay electrons to the
detection system. See text for further explanations.\label{fig:PRISM}}
\end{center}
\end{figure}

Further improvements are being considered for an experiment at J-PARC. To fully exploit the lifetime difference to suppress pion induced background the separation has to
occur in the beam line rather than after the muon has stopped since the lifetime of the muonic atom may be significantly shorter than the 2.2~$\mu$s of the free muon. For
this purpose a muon storage ring PRISM (Phase Rotated Intense Slow Muon source, see Fig.~\ref{fig:PRISM}) is being considered~\cite{PRISM} which makes use of
large-acceptance fixed-field alternating-gradient (FFAG) magnets. A portion of the PRISM-FFAG ring is presently under construction as r\&d project.
As the name suggests the ring is also used to reduce the momentum spread of the beam 
\begin{table}[thb]
\begin{minipage}{\textwidth}
\caption{$\mu-e$  conversion searches. \label{tab:mue_table}}
\begin{tabular*}{\textwidth}{@{\extracolsep{\fill}}lllcccc}
\hline\hline
project		&Lab		&status		&$E_p$ [GeV]	&$p_{\mu}$ [MeV/c]	&$\mu^-$ stops [s$^{-1}$]	&$S$
\footnote{single-event sensitivity: value of $B_{\mu e}$ corresponding to an expectation of one observed event}	\\
\hline
SINDRUM II	&PSI		&finished	&0.6		&52$\pm$1	&~~~~$10^7$		&$2\times10^{-13}$	\\
MECO		&BNL		&cancelled	&8~~		&45$\pm$25	&~~~~$10^{11}$		&$2\times 10^{-17}$	\\
mu2e		&FNAL		&under study	&8~~		&45$\pm$25	&0.6$\times 10^{10}$	&$4\times 10^{-17}$	\\
PRISM/PRIME	&J-PARC		&LOI		&40~~		&68$\pm$3	&~~~~$10^{12}$ 		&$5\times 10^{-19}$	\\
\hline\hline
\end{tabular*}
\end{minipage}
\end{table}
(from $\approx$30~\% to $\approx$3~\%) which is achieved by accelerating late muons
and decelerating early muons in RF electric fields. The scheme requires the construction of a pulsed proton beam~\cite{LOI26} a decision about which has not been made
yet. The low momentum spread of the muons allows the use of a relatively thin target which is an essential ingredient for high resolution in the momentum measurement 
with the PRIME detector~\cite{PRIME}.

Table~\ref{tab:mue_table} lists the $\mu ^-$ stop rates and single-event sensitivities for the various projects discussed above.

\newcommand{\BR}{\ensuremath{ B}}

\subsection{Searches for lepton flavour violation in $\tau$ decays}
Highest sensitivities to date are achieved at the $B$-factories and
further improvements are to be expected. At the LHC the modes with
three charged leptons in the final state such as $\tau \to 3\mu$ could
be sufficiently clean to reach even higher sensitivity. Studies have
been performed for LHCb~\cite{Shapkin:2007zz} and CMS (see below).

\subsubsection{$B$-factories}
Present generation $B$-factories operating around the $\Upsilon$(4S)
resonance also serve as $\tau$-factories, because the production cross
sections
$\sigma_{\bbbar} = 1.1 \nb$ and $\sigma_{\tautau} = 0.9 \nb$ are quite similar at center-of-mass energy near 10.58 \gev.
 \babar\ and BELLE have thus been able to reach the highest sensitivity to lepton flavour violating tau decays.

Many theories beyond the Standard Model allow for \taulg and \taulll decays, where $\ell^- = e^-, \mu^-$, at the level of $\sim {\cal{O}}
(10^{-10}-10^{-7})$. Examples are:
\begin{itemize}
\item
SM with additional heavy right-handed Majorana neutrinos or with left-handed and right-handed neutral isosinglets~\cite{Cvetic:2002jy};
\item
mSUGRA models with right handed neutrinos introduced via the seesaw mechanism~\cite{Ellis:1999uq,Ellis:2002fe};
\item
supersymmetric models with Higgs exchange~\cite{Dedes:2002rh,Brignole:2003iv} or SO(10) symmetry~\cite{Masiero:2002jn,Fukuyama:2003hn};
\item
technicolour models with non-universal Z$^\prime$ exchange~\cite{Yue:2002ja}.
\end{itemize}

Large neutrino mixing could induce large mixing between the supersymmetric partners of the leptons.
While some scenario's predict higher rates for $\taumg$ decays, others, for example with inverted mass hierarchy
for the sleptons~\cite{Ellis:2002fe}, predict higher rates for $\taueg$ decays.

Semi-leptonic neutrino-less decays involving pseudo-scalar mesons like $\tautolpz$, where $\pzero = \pi^0  , \eta, \eta^{\prime}$ may be enhanced over
\taulll decays in supersymmetric models, for example, arising out of exchange of CP-odd pseudo-scalar neutral Higgs boson,
which are further enhanced by colour factors associated with these decays. The large coupling of Higgs at the $s\bar{s}$ vertex enhances final states
containing the $\eta$ meson, giving a prediction of $\BRtaumeta : \BRtaummm: \BRtaumg = 8.4 : 1: 1.5$~\cite{Sher:2002ew}.
Some models with heavy Dirac neutrinos~\cite{Gonzalez-Garcia:1991be,Ilakovac:1999md}, two Higgs doublet models, R-parity violating supersymmetric models,
and flavour changing $Z^\prime$ models with non-universal couplings~\cite{Li:2005rr} allow for observable parameter space of new physics~\cite{Black:2002wh},
while respecting the existing experimental bounds at the level of $\sim {\cal{O}}(10^{-7})$.

\begin{figure}[!hbtp]
\begin{center}
\includegraphics[height=.42\linewidth]{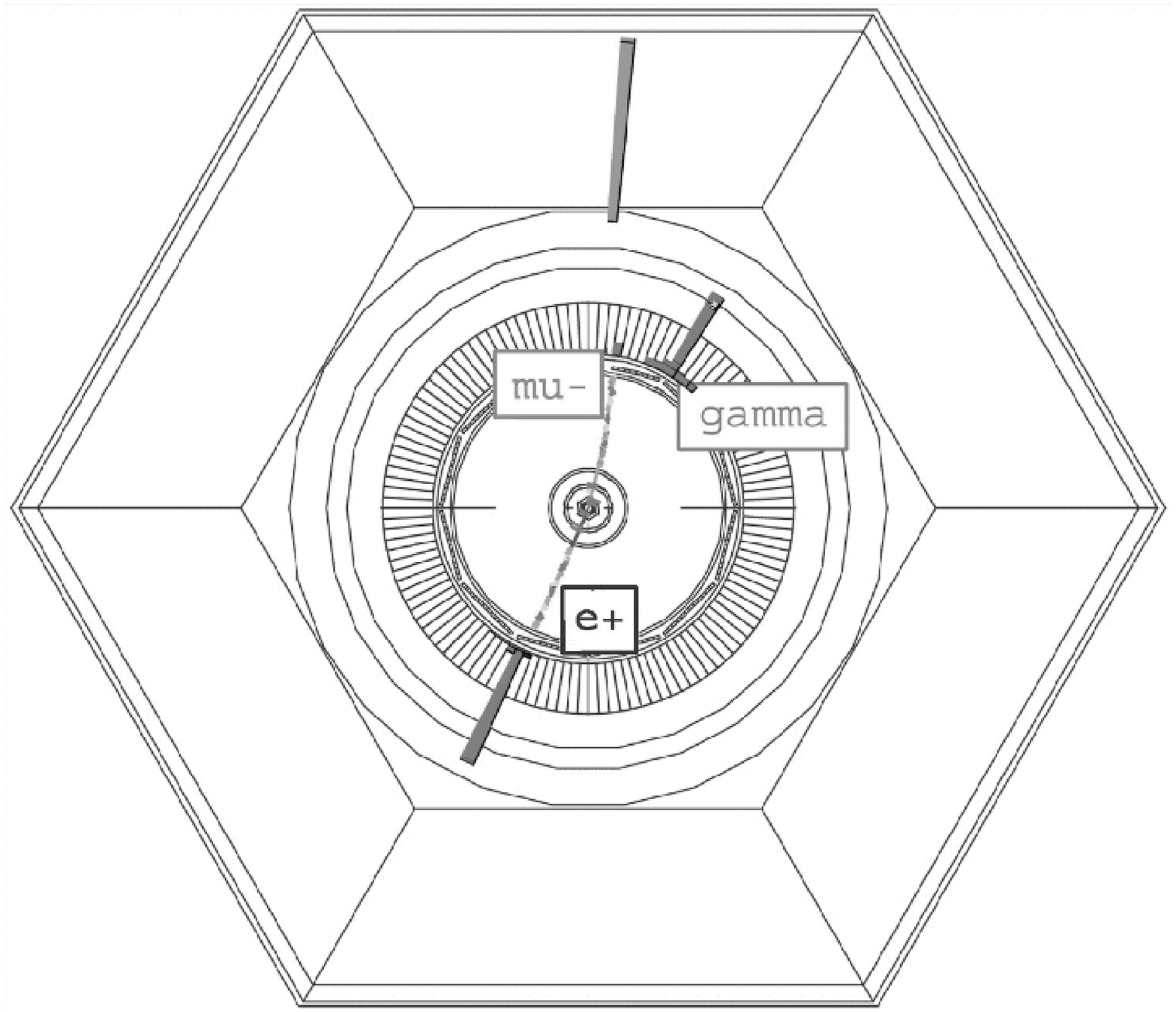}
\includegraphics[height=.42\linewidth]{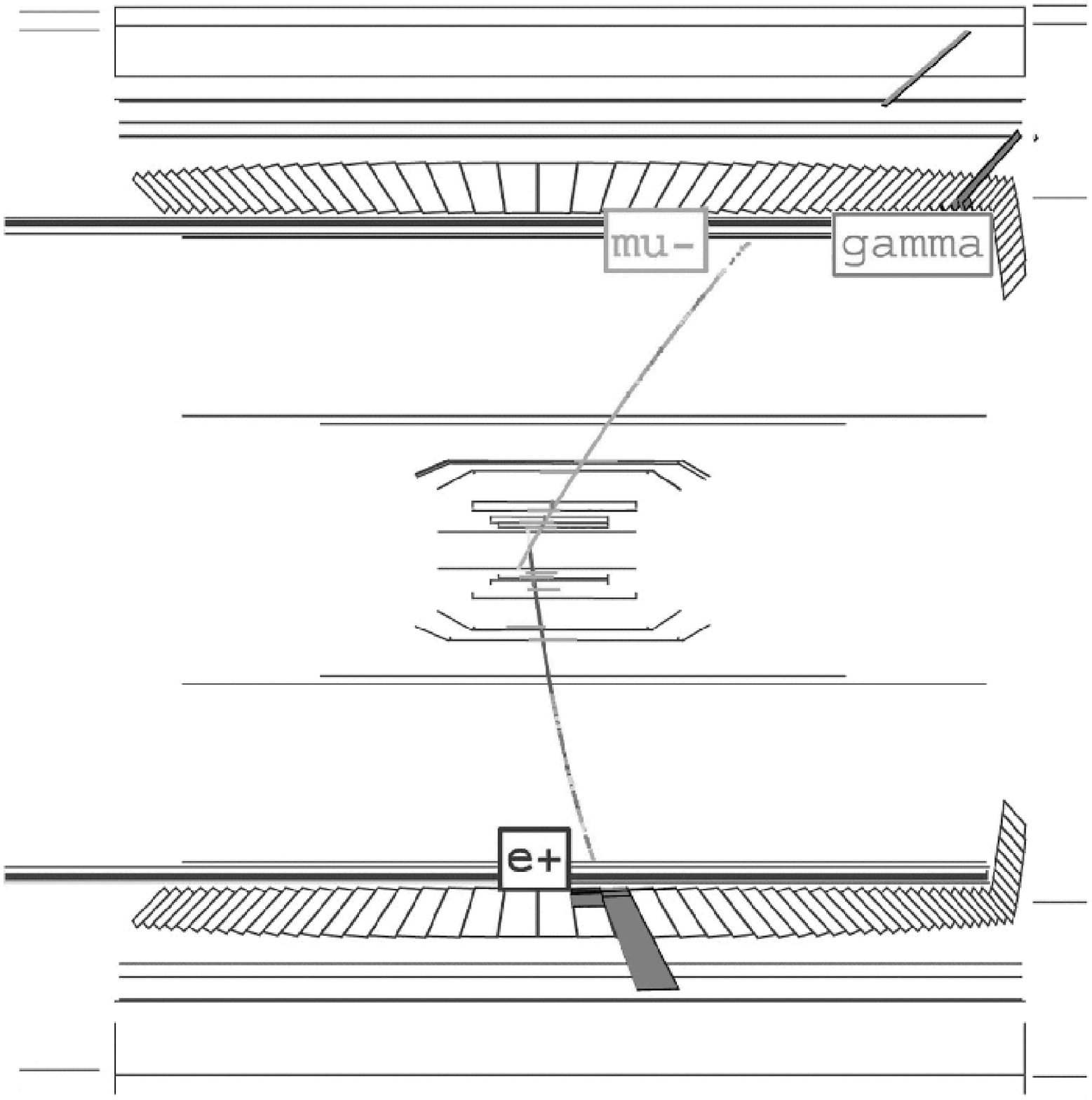}
\end{center}
\caption{Transverse and longitudinal views of a simulated $\tau\to\mu\gamma$ event in the \babar\ detector. The second tau decays to $e\nu\overline{\nu}$.
\label{explfv:taumugamma_topology}}
\end{figure}

\subsubsubsection{Search Strategy}
In the clean $\epem$ annihilation environment, the decay products of two taus produced are well separated in space as illustrated in
Fig.~\ref{explfv:taumugamma_topology}.

\begin{figure}[!hbtp]
\parbox{0.55\linewidth}{
\includegraphics[clip, trim=0 -10 0 -20, width=\linewidth]{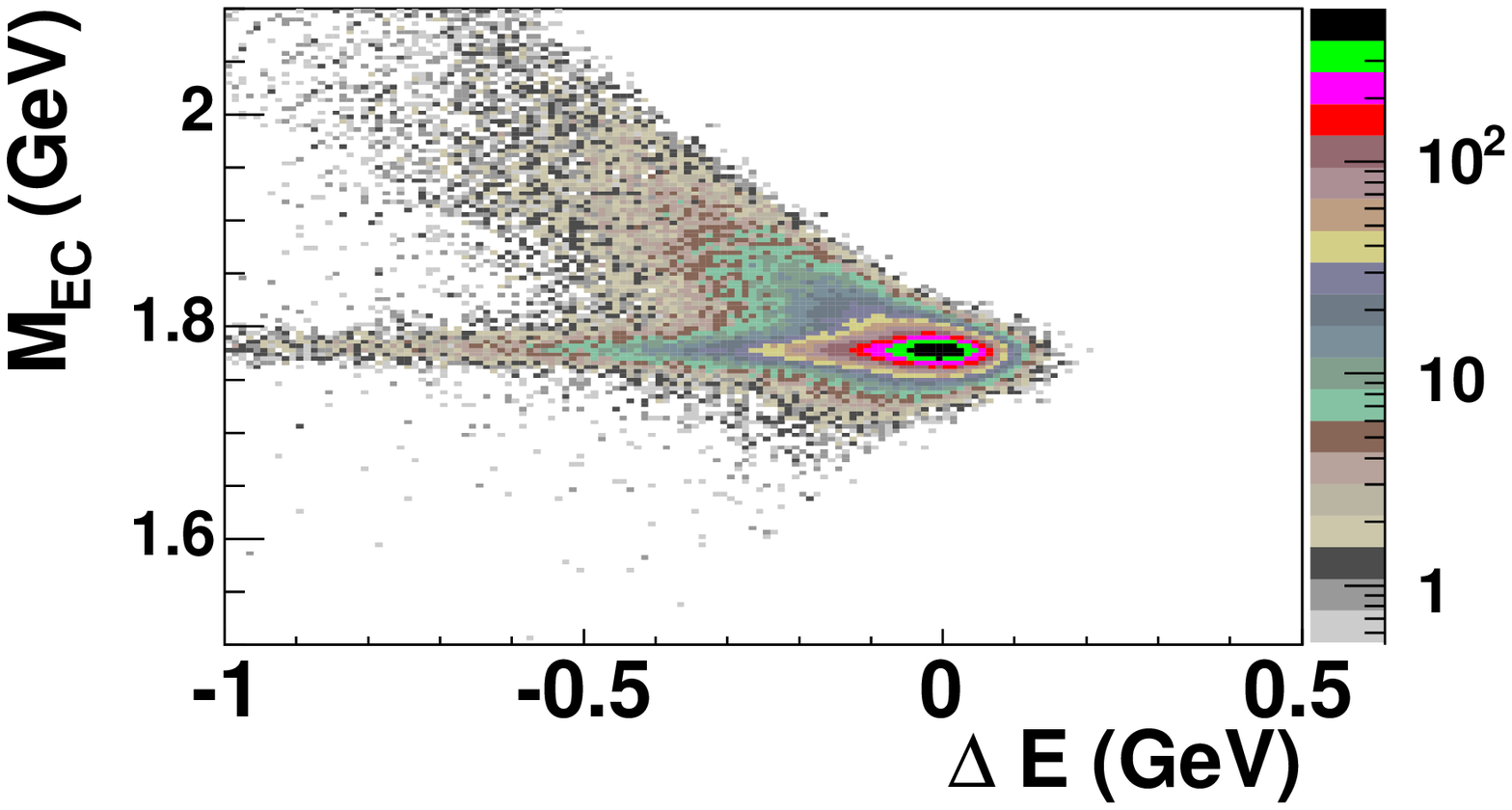}
\caption{ \mec{} vs. $\Delta E$ for simulated $\tau\to\mu\gamma$ events as reconstructed in the \babar\ detector.  The tails of distributions are due to initial
state radiation and photon energy reconstruction effects. Latter causes also the shift in $\langle \Delta E \rangle$.\label{explfv:taumugamma_mecdeltaea}}}
\hspace*{\fill}
\parbox{0.4\linewidth}{
\begin{center}
\includegraphics[clip, trim=0 0 0 10, width=0.8\linewidth]{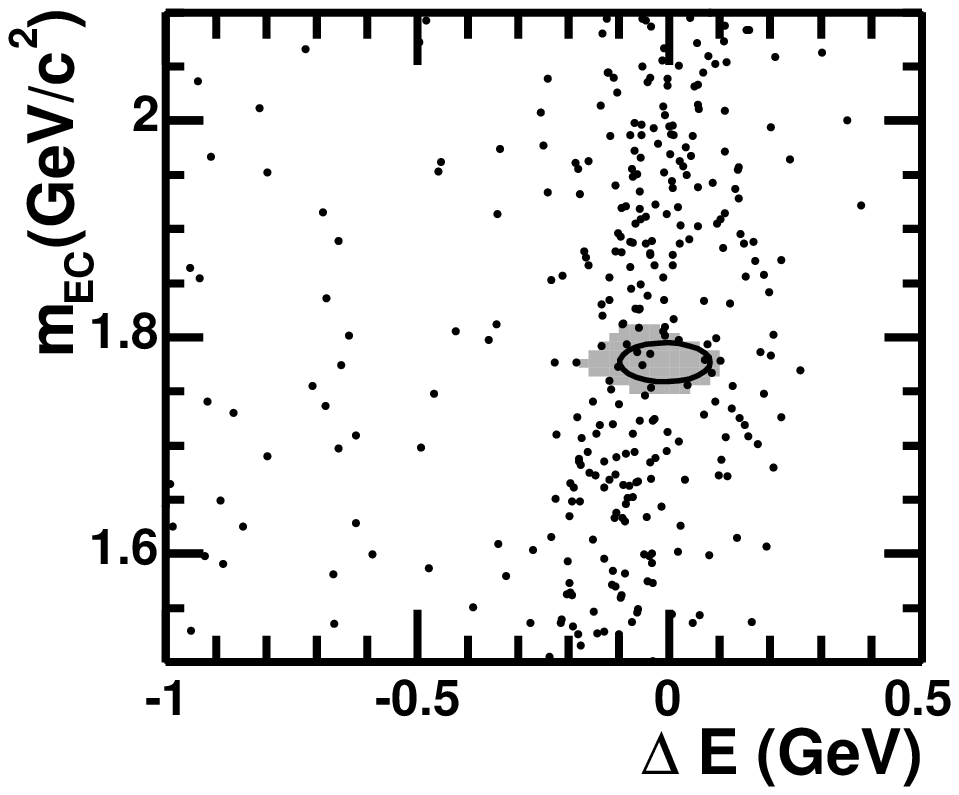}
\end{center}
\caption{Measured distribution of \mec{} vs. $\Delta E$ for $\tau\to\mu\gamma$ reconstructed by \babar~\cite{Aubert:2005ye}. The shaded region taken from
Fig.~\ref{explfv:taumugamma_mecdeltaea} contains 68\% of the hypothetical signal events.\label{explfv:taumugamma_mecdeltaeb}}}
\end{figure}
As shown in Fig.~\ref{explfv:taumugamma_mecdeltaea} neutrino-less $\tau$-decays have two characteristic features:
\begin{itemize}
\item the measured energy of $\tau$ daughters is close to half the center-of-mass energy,
\item the total invariant mass of the daughters is centered around the mass of the $\tau$ lepton.
\end{itemize}
While for $\ell\ell\ell$ modes the achieved mass resolution is excellent, the resolution ($\sigma$) of the $\ell\gamma$ final state improves from $\sim 20$~MeV
to 9~MeV by assigning the point of closest approach of the muon trajectory to the $\epem$ collision axis as the decay vertex and by using a kinematic fit
with the $\mu\gamma$ CM energy constrained to $\sqrt{s}/2$~\cite{Aubert:2005ye}. The energy resolution is typically 45~MeV with a long tail due to radiation.

The principal sources of background are radiative QED (di-muon or Bhabha) and continuum ($q\overline q$) events as well as $\tautau$ events
with a mis-identified standard model decay mode. There is also  some {\it irreducible} contribution from $\tautau$ events with hard
initial state radiation in which one of the $\tau$'s decays into a mode with the same charged particle as the signal.  For example,
$\tau\to\mu\nu\bar{\nu}$ decays accompanied by a hard $\gamma$ is an irreducible background in the $\tau\to\mu\gamma$ search.

The general strategy to search for the neutrino-less decays is to define a signal region, typically of size $\sim$ $2\sigma$, in the energy-mass plane of the
$\tau$ daughters and to reduce the background expectation from well-known CM processes inside the signal region by optimizing a set of selection criteria:
\begin{itemize}
\item
the missing momentum is consistent with the zero-mass hypothesis
\item
the missing momentum points inside the acceptance of the detector
\item
the second tau is found with the correct invariant mass
\item
minimal opening angle between two tau decay products
\item
minimal value for the highest momentum of any reconstructed track
\item
particle identification
\end{itemize}
The analyses are performed in a {\it blind} fashion by excluding events in the region of the signal box until all optimisations and systematic studies of the
selection criteria have been completed. The cut values are optimized using control samples, data sidebands and Monte Carlo extrapolation to the signal region
to yield the lowest expected upper limit under the no-signal hypothesis. The measured  \mec{} vs. $\Delta E$ distribution for the $\tau\to\mu\gamma$ search
after applying the constraints listed above is shown in Figure~\ref{explfv:taumugamma_mecdeltaeb}.

For the $\tautolpz$ searches, the pseudo-scalar mesons ($P^0$) are reconstructed in the following decay modes: $\pi^0\to\gamma\gamma$ for
$\tau^\pm\to\ell^\pm\pi^0$, $\etogg$ and $\etoppp$ ($\pi^0\to\gamma\gamma$) for $\tautoleta$, and $\eptoppe (\etogg)$ and
$\eptorg$ for $\tautoletap$.

\subsubsubsection{Experimental results from \babar\ and BELLE}

By the beginning of 2007 \babar\ and BELLE had recorded integrated luminosities of $\cal{L} \sim$400 and 700~\invfb{}, respectively,
which corresponds to a total of $\sim 10^9 \tau$-decays. Analysis of these data samples is still ongoing and published results include only part of the data
analysed. No signal has yet been observed in any of the probed channels and some limits and the corresponding integrated luminosities are summarized in
Table~\ref{expLFV:combtable}. Frequentist upper limits have been calculated for the combination of the two experiments~\cite{Banerjee:2007rj} using the technique
of Cousins and Highland~\cite{Cousins:1992qz} following the implementation of Barlow~\cite{Barlow:2002bk}.

\begin{table}[bht]
\begin{minipage}{\textwidth}
\caption{Integrated luminosities and observed upper limits on the branching fractions at 90\% C.L. for selected LFV tau decays by \babar\ and BELLE.
\label{expLFV:combtable}}
\begin{tabular*}{\textwidth}{@{\extracolsep{\fill}}l|rccrcc}
\hline\hline
		&\multicolumn{3}{c}{\babar}	 			&\multicolumn{3}{c}{BELLE}    				\\	
Channel		&$\cal{L}$	&$\BRul$ 	&Ref.			&$\cal{L}$	&$\BRul$      &Ref.			\\	
		&$(\invfb)$	&$(10^{-8})$ 	&			&$(\invfb)$	&$(10^{-8})$  &				\\	
\hline
\taueg		&232		&11		&\cite{Aubert:2005wa}	&535		&12           &\cite{Hayasaka:2007vc}	\\			
\taumg		&232		&6.8		&\cite{Aubert:2005ye}	&535		&4.5          &\cite{Hayasaka:2007vc} 	\\
\taulll         &92             &11 - 33        &\cite{Aubert:2003pc}	&535		&2 - 4	      &\cite{Abe:2007ev}	\\		
\tautoepiz	&339		&13		&\cite{Aubert:2006cz}	&401		&8.0          &\cite{Miyazaki:2007jp}	\\			
\tautompiz	&339		&11		&\cite{Aubert:2006cz}	&401		&12           &\cite{Miyazaki:2007jp}	\\			
\tautoeeta	&339		&16		&\cite{Aubert:2006cz}	&401		&9.2          &\cite{Miyazaki:2007jp}	\\			
\tautometa	&339		&15		&\cite{Aubert:2006cz}	&401		&6.5          &\cite{Miyazaki:2007jp}	\\			
\tautoeetap	&339		&24		&\cite{Aubert:2006cz}	&401		&16           &\cite{Miyazaki:2007jp}	\\			
\tautometap  	&339		&14		&\cite{Aubert:2006cz}	&401		&13           &\cite{Miyazaki:2007jp}	\\			
\hline\hline
\end{tabular*}
\end{minipage}
\end{table}

\subsubsubsection{Projection of limits to higher luminosities}

\BRtaumg and \BRtaummm have been lowered by five orders of magnitude over the past twenty-five years. 
Further significant improvements in sensitivity are expected during the next five years. Depending upon the nature of backgrounds contributing to a
given search, two extreme scenarios can be envisioned in extrapolating to higher luminosities:
\begin{itemize}
\item
If the expected background is kept below ${\cal{O}}(1)$ events, while maintaining the same efficiency $\BRulninety \propto 1/{\cal{L}}$ if no signal
events would be observed. In $\taummm$ searches, for example, the backgrounds are still quite low and the irreducible backgrounds are negligible even
for projected Super $B$-factories.
\item
If there is background now already and no reduction could be achieved in the future measurements $\BRulninety \propto 1/\sqrt{\cal{L}}$.
\end{itemize}
The $\sqrt{\cal{L}}$ scaling is, however, unduly pessimistic since the analyses improve steadily. Better understanding of the nature of the backgrounds
will lead to a more effective separation of signal and background.

The $\taumg$ searches suffer from significant background from both $\mu^+\mu^-$ and $\tautau$ events and to a lesser extend from $q\overline q$ production.
While one can expect to reduce these backgrounds with continued optimization with more luminosity at the present day $B$-factories, much of the background
from $\tautau$ events is irreducible coming from $\tau\to\mu\nu\bar{\nu}$ decays accompanied by initial state radiation. This source represents about 20\%
of the total background in the searches performed by the \babar\ experiment~\cite{Aubert:2005ye} and it is conceivable that an analysis can be developed
that reduces all but this background with minimal impact on the efficiency. One could also include new selection criteria such as a cut on the polar angle
of the photon which could reduce the radiative ``irreducible'' background by 85\% with a 40\% loss of signal efficiency. Table~\ref{LFVExptSensitivities}
summarizes the future sensitivities for various LFV decay modes.

\begin{table}[bht]
\begin{minipage}{\textwidth}
\caption{Expected 90\% CL upper limits on LFV $\tau$ decays with $75 \ {\rm ab}^{-1}$ assuming no signal is found and reducible
backgrounds are small ($\sim {\cal{O}}(1)$ events) and the irreducible backgrounds scale as $1/{\cal{L}}$.\label{LFVExptSensitivities}}
\begin{tabular*}{\textwidth}{@{\extracolsep{\fill}}lcll}
\hline\hline
~&Decay mode 				& Sensitivity 			&~\\
\hline
&$\tau \to \mu\,\gamma$   		& $2 \times 10^{-9}$  		&\\
&$\tau \to e\,\gamma$   		& $2 \times 10^{-9}$  		&\\
&$\tau \to \mu\, \mu\, \mu$  		& $2 \times 10^{-10}$ 		&\\
&$\tau \to e e e $ 			& $2 \times 10^{-10}$ 		&\\
&$\tau \to \mu \eta$ 			& $4 \times 10^{-10}$ 		&\\
&$\tau \to e \eta$ 			& $7 \times 10^{-10}$ 		&\\
\hline\hline
\end{tabular*}
\end{minipage}
\end{table}

In order to further reduce the impact of irreducible backgrounds at a
future Super B-factory experiment, one can consider what is necessary
to improve the mass resolution of the, e.g., $\mu-\gamma$
system. Currently, this resolution is limited by the $\gamma$ angular
resolution. Therefore improvements might be expected if the
granularity of the electromagnetic calorimeter is increased.

\subsubsection{CMS}
So far, only $\tau \to \mu$ transitions have been studied since muons are more easily identified and the CMS trigger thresholds for muons are generally lower
than for electrons. The $\tau\to\mu\gamma$ channel was studied in the past\cite{Unel:2005fj} both for CMS and for ATLAS but found not to be competitive with the
prospects at the B-factories. The $\tau\to3\mu$ channel looks more promising and will be discussed below.

\subsubsubsection{$\tau$ production at the LHC}\label{subsubsec:TauProduction}
It is planned to operate the LHC in three different phases. After a commissioning phase the LHC will be ramped up to an initial luminosity of
$\lumi = 10^{32}\percms$ followed by a low luminosity phase ($\lowlumi$). A high luminosity phase with $\hilumi$ will start in 2010 and last for a period of
several years. The integrated luminosity per year will be $10-30\fbinv$ and $100-300\fbinv$ for the low and high luminosity phases,
respectively\cite{DellaNegra:922757}.

The rate of $\tau$ leptons produced was estimated with the help of \PYTHIA 6.227 using the parton distribution function CTEQ5L. The results are shown in
Tab.~\ref{tab:TauPerYear}.
\begin{table}[thb]
\begin{minipage}{\textwidth}
\caption{Number of $\tau$ leptons per year produced during the low-luminosity phase of the LHC.\label{tab:TauPerYear}}
\begin{tabular*}{\textwidth}{@{\extracolsep{\fill}}l|llllll}
\hline\hline
production channel      & $W\to\tau\nu_{\tau}$	& $\gamma/Z\to\tau\tau$	&$B^0\to\tau X$ 	&$B^{\pm}\to\tau X$	&$B_s\to\tau X$  	&$D_s\to\tau X$ \\
\hline
$N_{\tau}$/$10\fbinv$ 	& $1.7\times 10^{8}$	& $3.2\times 10^{7}$	& $4.0\times 10^{11}$ 	&$3.8\times 10^{11}$	&$7.9\times 10^{10}$ 	&$1.5\times 10^{12}$\\
\hline\hline
\end{tabular*}
\end{minipage}
\end{table}
During the low luminosity phase assuming an integrated luminosity of only $10\fbinv$ per year about $10^{12}$ $\tau$ leptons will be produced within the CMS
detector. The dominant production sources of $\tau$ leptons at the LHC are the $D_s$ and various $B$ mesons. The $W$ and the $Z$ production sources will provide
considerably less $\tau$ leptons per year, but at higher energies which is an advantage for the efficient detection of their decay products (see below).

\subsubsubsection{$\tau\to3\mu$ detection} A key feature of CMS is a
$4 \mathrm{T}$ magnetic field, which ensures the measurement of
charged-particle momenta with a resolution of $\sigma_{\pt} /
\pt=1.5\%$ for $10\GeV$ muons\cite{DellaNegra:922757} using a
four-station muon system. A silicon pixel detector and tracker allow
to reconstruct secondary vertices with a resolution of about
$50\micron$\cite{NoteSpeer2006/032} and help to improve the muon
reconstruction. Furthermore, CMS has an electromagnetic calorimeter
(ECAL) composed of $\mathrm{PbWO_4}$ and a copper scintillator
hadronic calorimeter (HCAL). As a result of the high magnetic field
and the amount of material that has to be crossed only muons with
$\pt>3$~GeV/c are accepted. The reconstruction efficiency varies
between $\approx$70\% at $5\GeV$\cite{NoteJames2006/010} and
$\approx$98\% at 100~GeV/c\cite{DellaNegra:922757}.

The two levels of the CMS trigger system are called ``level 1'' (L1) and ``high level'' (HLT). The triggers relevant for this analysis are the dedicated single
and di-muon triggers. For the low luminosity phase it is planned to use as $\pt$ thresholds for single muons 14~GeV/c at L1 and 19~GeV/c for the HLT. The
thresholds for the di-muon trigger will be 3~GeV/c at L1 and 7~GeV/c for the HLT.

\begin{figure}[htbp]
\parbox{0.56\linewidth}{\includegraphics[width=\linewidth]{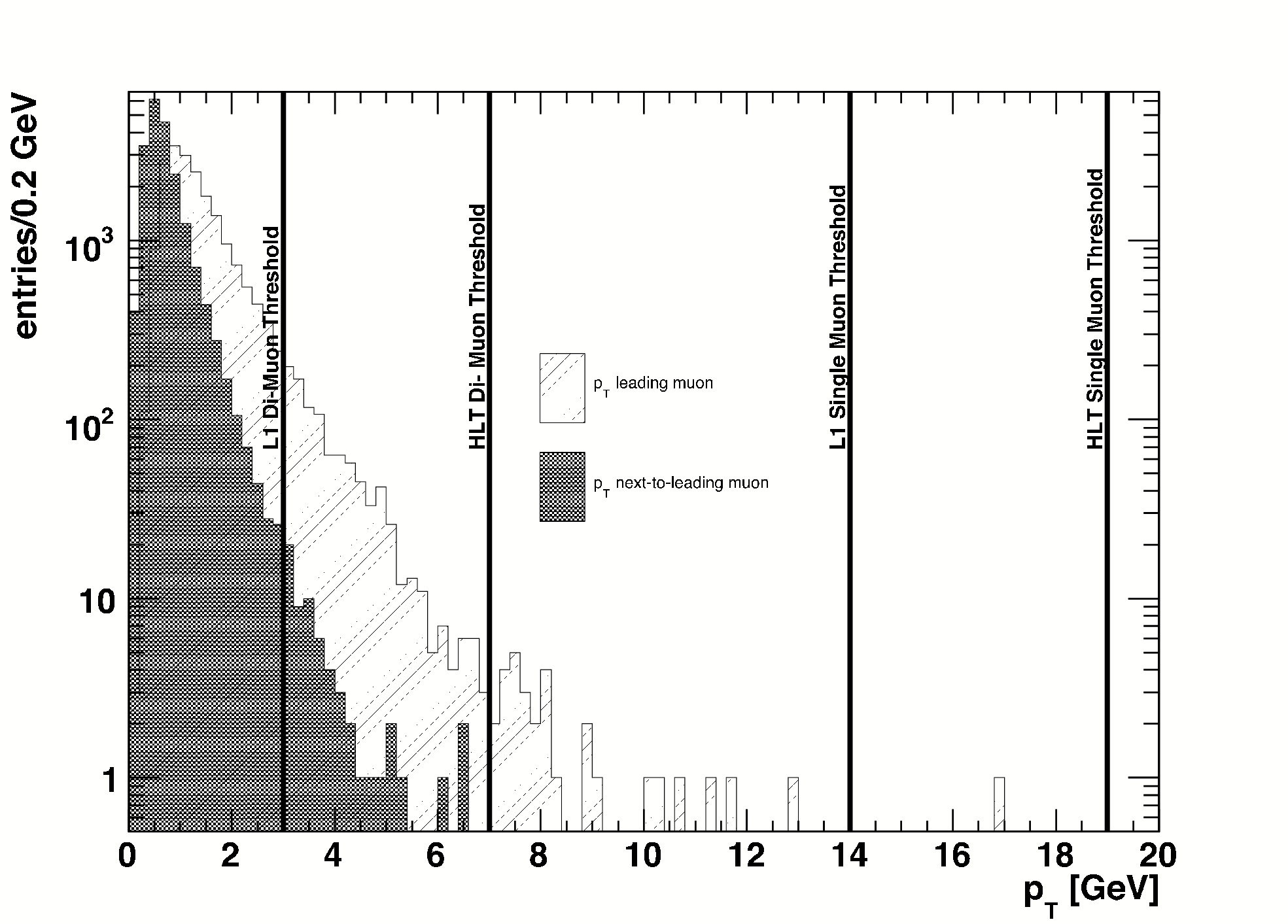}}
\hspace*{\fill}\parbox{0.4\linewidth}{\includegraphics[width=1.1\linewidth]{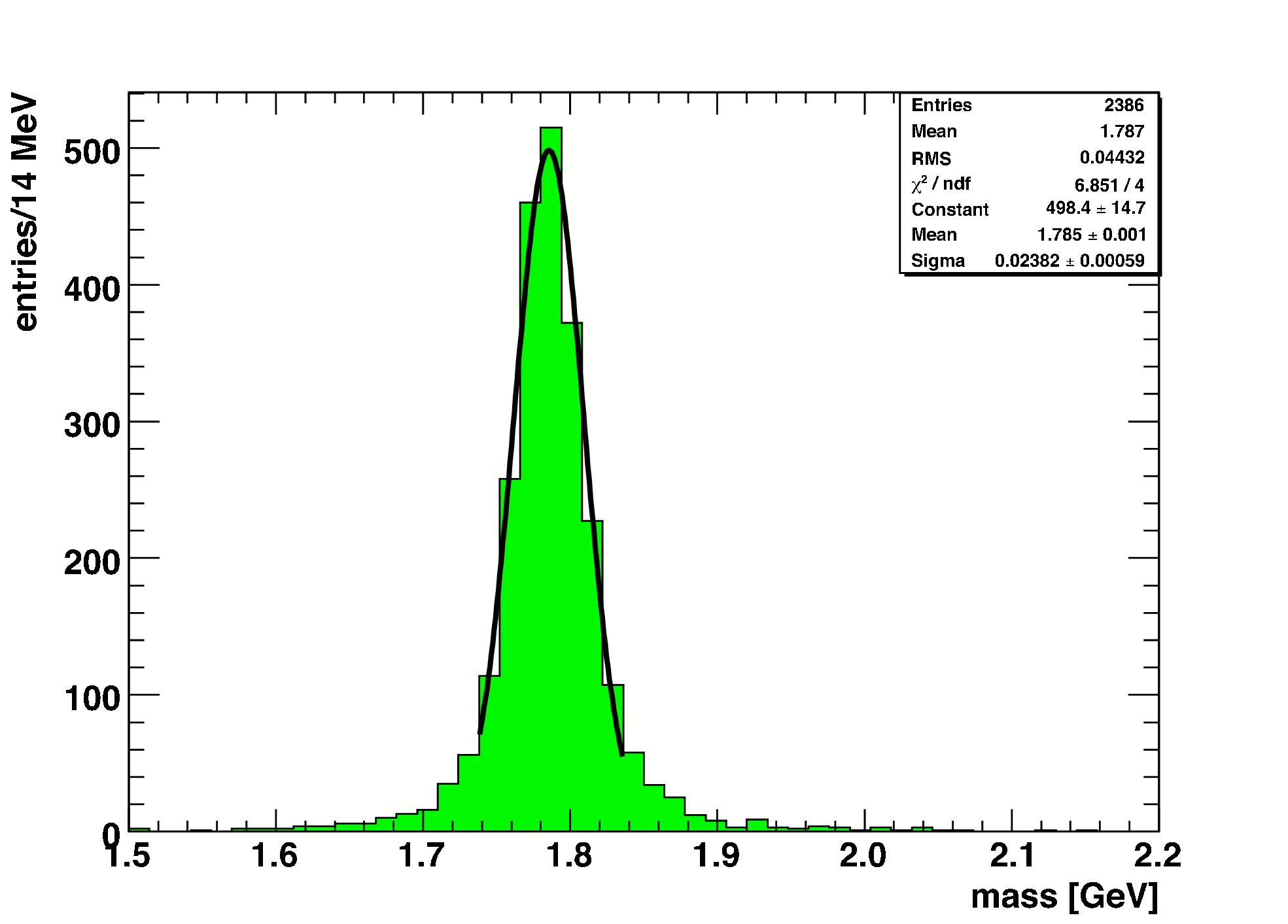}}
\parbox[t]{0.56\linewidth}{
\caption{$\pt$ distributions of the leading and next-to-leading muon from the decay $\tau \to 3\mu$ at CMS. The indicated trigger thresholds for the low
luminosity phase are clearly too high for the efficient detection of these events.\label{fig:PtLeadingMuon}}}
\hspace*{\fill}\parbox[t]{0.4\linewidth}{
\caption{Invariant mass distribution from the simulation of $\tau\to3\mu$ events.\label{fig:MassResolution}}}
\end{figure}

Most $\tau\to 3\mu$ events produced via $W$ and $Z$ decays will be accepted by the present triggers. Unfortunately, the low $\pt$ of the muons from the decays
of $\tau$'s originating in $D_s$ or $B$ decay result in a very low trigger efficiency (Fig.\ref{fig:PtLeadingMuon}). 
Dedicated trigger algorithms with improved efficiency are presently being studied.

To improve the identification of low $\pt$ muons a new method is currently under development combining the energy deposit in the ECAL, HCAL and the number of
reconstructed muon track segments in the muon systems.
The invariant mass distribution of reconstructed $\tau\to3\mu$ events is shown in Fig.\ref{fig:MassResolution}. The resolution is about 24~MeV/$c^2$, which
ensures a good capability to reduce background events. 

\subsubsubsection{Background and expected sensitivity} The main
sources of muons are decays of $D$ and $B$ mesons which are copiously
produced at LHC energies. A previous study \cite{SantinelliCMSNote}
suggested that these background events can be suppressed by
appropriate selection criteria. The probability to misidentify an
event from pile-up is small and cosmic rays can be rejected by
timing. Due to the high momentum of the muons from direct $W$ and $Z$
decays, the contribution to the background is
negligible\cite{Santinelli:2002ea}.

One rare decay that can mimic the signal is $D_s\to\phi\mu\nu_{\mu}$ followed by a decay $\phi\to\mu\mu$. This background can be reduced by an invariant mass
cut around the $\phi$ mass. Radiative $\phi$ decay $\phi\to\mu\mu\gamma$ survives this cut since the photon usually remains undetected. These radiative decays
and any other heavy meson decays may be suppressed using secondary vertex properties and isolation criteria and by exploring the three-muon angular distributions.
These studies are in progress. 

Predictions of the achieveable sensitivity are available in an older CMS Note \cite{SantinelliCMSNote}. In case no signal is observed the expected upper limit
on the $\tau\to3\mu$ branching ratio at $95\%$ CL for the $W$ source is $7.0\ten{-8}$ ($3.8\ten{-8}$) for $10\fbinv$ ($30\fbinv$) of collected data. For the $Z$
source a limit of $3.4\ten{-7}$ and for the $B$ meson source a limit of $2.1\ten{-7}$ was derived assuming an integrated luminosity of $30\fbinv$. The $D_s$
source was not studied in this early paper.

Potentially including the muons from $D$ and $B$ meson decays may lead to significant improvements of the sensitivity. Further studies are necessary to make
firm predictions.

\subsection{$B^0_{d,s}\to e^\pm\mu^\mp$}\label{sec:exp:LFV:B}
The present limits B$(B^0_d \to e\mu) < 1.7 \times
10^{-7}$~\cite{Chang:2003yy} determined by Belle and B$(B^0_s \to
e\mu) < 6.1 \times 10^{-6}$~\cite{Abe:1998bc} from CDF are of interest
since they place bounds on the masses of two Pati-Salam
leptoquarks~\cite{Pati:1974yy} (see below). Both measurements are
almost background free so significant improvements should be expected
from these experiments. These decay modes have similarities with the
$K^0_L \to \mu e$ decay for which an upper limit of $4.7 \times
10^{-12}$ exists~\cite{Ambrose:1998us}.

\begin{figure}[htb]
\includegraphics*[clip, trim=0 0 0 0, width=0.5\linewidth]{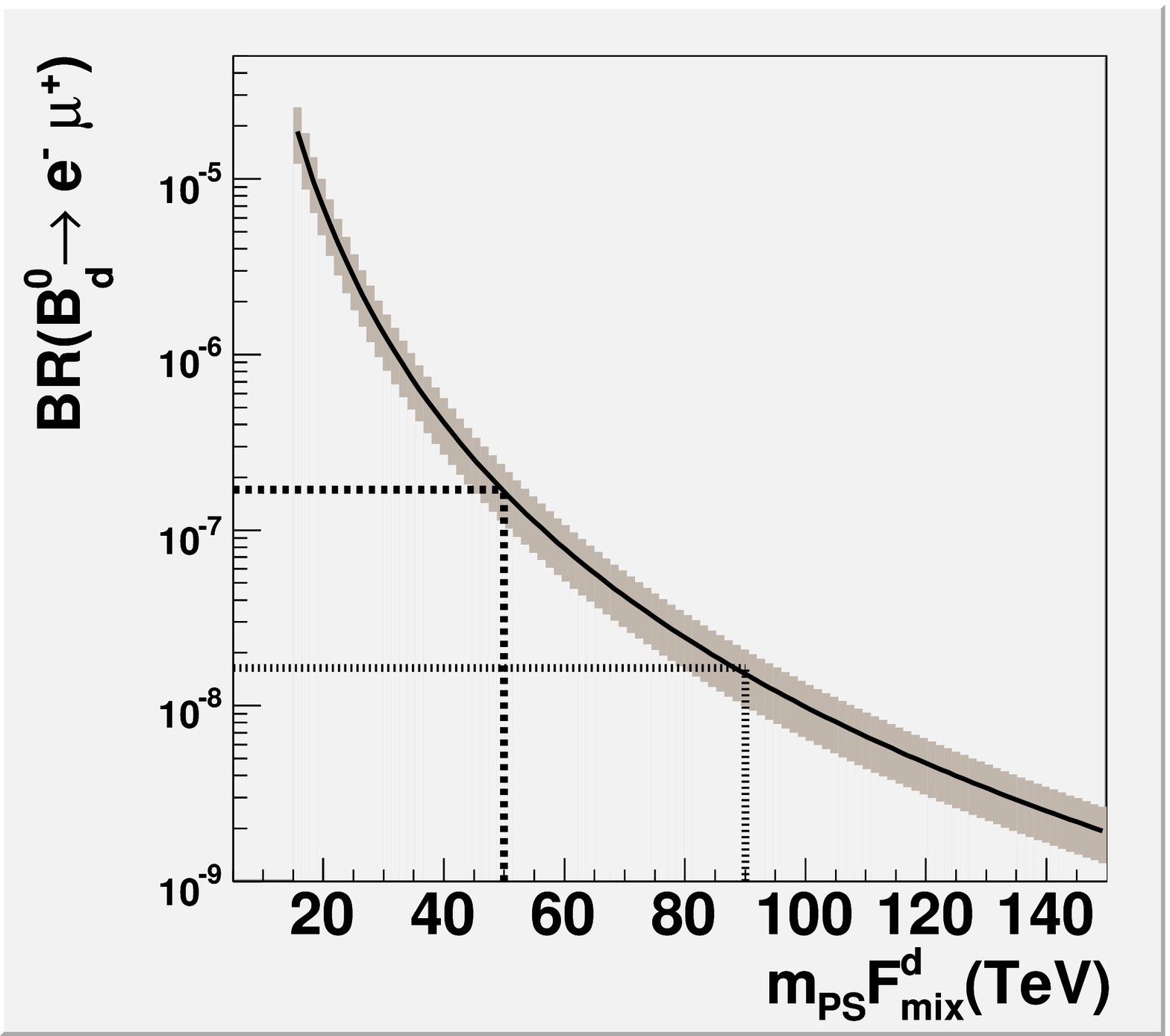}
\includegraphics*[clip, trim=0 0 0 0, width=0.5\linewidth]{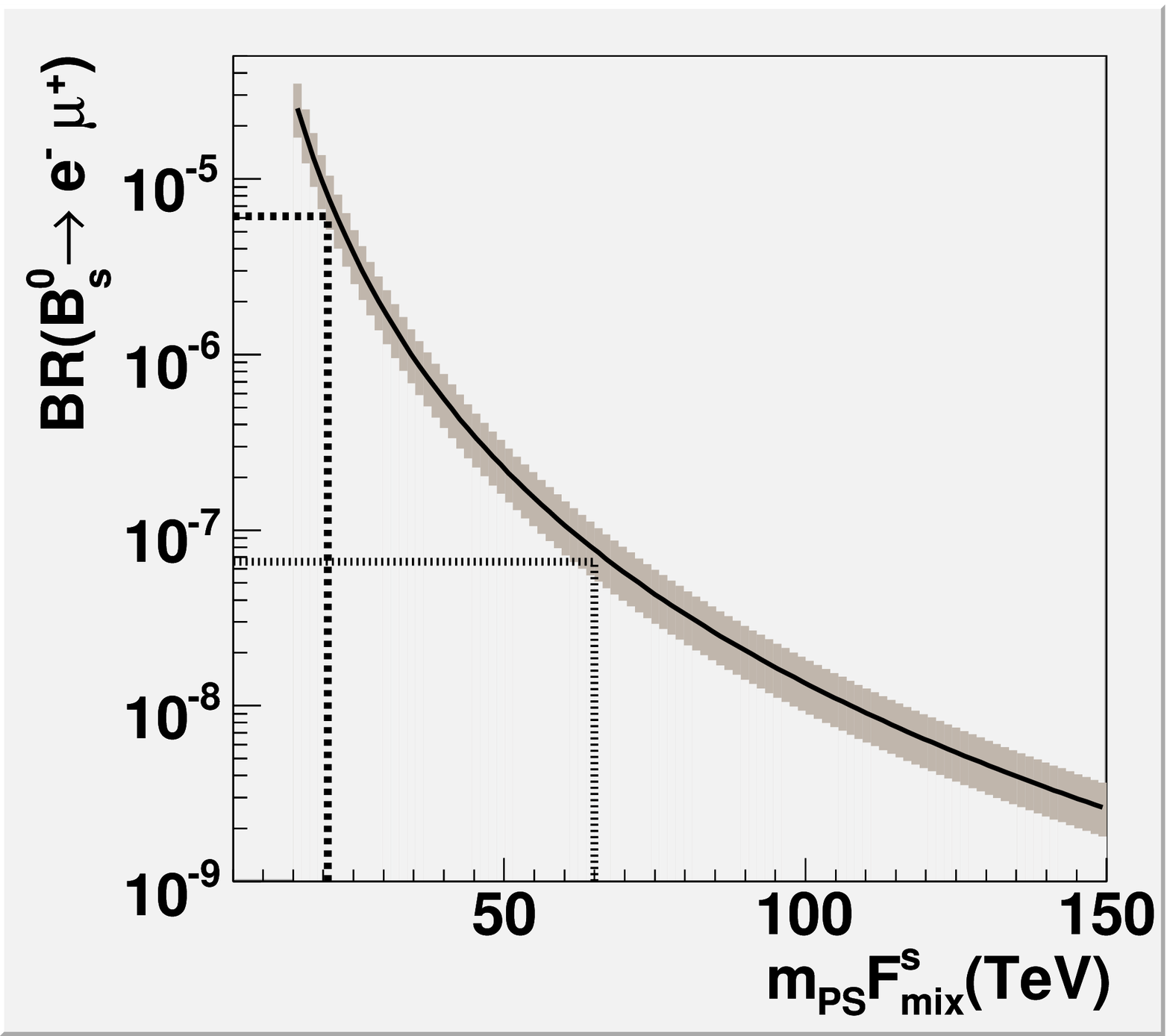}
\caption{90\% C.L. limits on B$(B^0_d \to e\mu)$ (left panel) and B$(B^0_s \to e\mu)$ (right panel) and the corresponding lower limits on the products of
Pati-Salam leptoquark mass and mixing. Present results are compared with results projected for LHCb for an integrated luminosity of 2fb$^{-1}$ in case no signal would be
observed. Dashed regions indicate the theoretical uncertainties in the relation between the variables.\label{fg:bmue}}
\end{figure}

The prospects of a more sensitive search have been studied for the
LHCb experiment~\cite{LHCb-2007-028}. Although background levels are
higher, this is more than compensated by the improved single-event
sensitivity. The event selection closely follows that of the
$B^0_{s}\to \mu^+\mu^-$ decay.
The
dominant backgrounds come from (i) events in which two b hadrons decay
into leptons combining to a fake vertex and (ii) from two-body
charmless hadronic decays when the two hadrons are misidentified as
leptons.  Signal and background are separated on the basis of particle
identification, invariant mass ($\sigma(m_B)$=50~MeV/c$^2$),
transverse momenta, proper distance and the isolation of the $B^0$
candidate from the other decay products. See Ref.~\cite{LHCb-2007-028}
for details. Simulation shows that for an integrated luminosity of
2fb$^{-1}$ the total background can be reduced to $\approx$80 events
with a selection efficiency of 1.4\%. Assuming no signal would be
found the 90\% C.L. upper limits would be $1.6\times 10^{-8}$ and
$6.5\times 10^{-8}$ for B($B^0_{d}\to e^\pm\mu^\mp)$ and B($B^0_{s}\to
e^\pm\mu^\mp$), respectively.  These values correspond to 90\%
C.L. lower limits on the Leptoquark mass and mixings of 90$\times
F^{d}_{mix}$~TeV and 65$\times F^{s}_{mix}$~TeV, where $F^{d,s}_{mix}$
are factors taking into account generation mixing within the
model. Present limits are 50~TeV and 21~TeV, respectively (see
Fig.~\ref{fg:bmue}).

\subsection{In flight conversions}\label{sec:exp:LFV:flight}

Lepton Flavour Violation could manifest itself in the conversion of
high-energy muons into tau leptons when scattering on nucleons in a
fixed target configuration~\cite{Gninenko:2004wq}.  Muons can be
produced much more copiously than tau leptons so $\mu \to \tau$
conversions could be more sensitive than neutrinoless $\tau \to \mu X$
decays.  When considering the effective lepton-flavour-violating
four-fermion couplings, tau decays mainly involve light quarks, so
heavy quark couplings are only loosely
constrained~\cite{Sher:2003vi}. In SUSY models, muon to tau conversion
could be greatly enhanced by Higgs mediation at energies where heavy
quarks contribute~\cite{Kanemura:2004jt}.

Within the context of this workshop the experimental feasibility of such experiment has been investigated.
The cross section for mu to tau lepton conversion on target has been estimated to be at most $550\,\text{ab}$~\cite{Sher:2003vi} for
$50\,\text{GeV}$ muons, using an effective model independent interpretation of the tau decay LFV constraints~\cite{Black:2002wh}
based on the 2000 data~\cite{Groom:2000in}. By rescaling the upper limit on $\text{B}(\tau\to\mu\pi^+\pi^-)$ to the current
value~\cite{Aubert:2005tp,Yusa:2006qq}, one obtains an upper limit at 90\% CL on the cross section of $4.7\,\text{ab}$. This value scales
roughly linearly with the muon energy.  In the context of the MSSM, the experimental
data available in 2004 constrained the cross section in the range from $0.1\,\text{ab}$ to $1\,\text{ab}$ for muon energies from
$100\,\text{GeV}$ to $300\,\text{GeV}~$\cite{Kanemura:2004jt}.\\

The following assumptions were made to assess the experimental feasibility:
\begin{itemize}
\item the goal is a sensitivity to the cross section corresponding to 1/10 of the present limits from tau decay, collecting at least thousand events
 per year;
\item the active target consists of 330 planes of $300\,\mu\text{m}$ thick silicon, with either strip or pixel readout;
\item the target has transverse dimensions corresponding to an area of $1\,\text{m}^2$ and the beam is distributed homogeneously over the target.
\end{itemize}

As a consequence, $3.75 \times 10^{19}$~muons/yr are needed which, assuming a 10\% duty cycle and an effective data-taking year of
$10^7$~s, corresponds to $3.75 \times 10^{13}$~muons/s (peak) and $3.75 \times 10^{12}$~muons/s (average).

Using the LEPTO 6.5.1 generator \cite{Ingelman:1996mq} deep inelastic muon scattering off nucleons was studied. The amount of power dissipated in the target is
sustainable, and the interaction rate is 0.6 interactions per 25\,ns, which is comparable to LHC experiments. Radiation levels and occupancy in the silicon
active target appear to be tractable, provided pixel readout is used.

When requiring momentum transfer above 2\,GeV and invariant mass of
the hadronic final state above 3\,GeV an effective interaction cross section of 47\,nb
was found. This value reduces to 15\,nb when applying the level 0 trigger requirement of at least 60\,GeV of hadronic energy which results in a rate of 7.7\,MHz.
The amount of data that needs to be extracted from the tracker for further event selection can probably be handled at such rate.

Unfortunately it appears that the required muon flux is incompatible with the operation of calorimeters as triggering and detecting
devices.  Assuming an LHCb-like electro-magnetic calorimeter with a 2.6\,cm thick lead absorber and an integration time of 25\,ns, and
assuming that electrons from muon decay travel unscreened for 4~m before encountering the electro-magnetic calorimeter, three high
energy electrons per 25\,ns integration time reach the calorimeter, preventing any effective way of triggering on electrons. Assuming an
LHCb-like hadronic calorimeter structured in towers consisting of 75 layers including $13\times 13\,\text{cm}^2$ scintillating pads and
16\,mm of iron each, each tower will detect 25\,TeV of equivalent hadronic energy for each 25\,ns of integration time just because of
the muon flux energy loss. The Poisson fluctuation of the number of muons will induce a fluctuation in the detected hadronic energy per
tower of about 200\,GeV, preventing the use of the hadronic calorimeter as a trigger for $\mu N \to \tau X$.

In conclusion, the idea of using an intense but transversely spread muon flux to produce and detect LFV muon conversions to tau leptons
does not appear feasible in this preliminary study, mainly because it does not appear possible to operate calorimeters at these rates.

\section{Experimental studies of electric and magnetic dipole
  moments}\label{sec:exp:dipoles} 

\subsection{Electric dipole moments}\label{sec:exp:dipoles:edm}
We review here the current status and prospects of the searches for
fundamental EDMs, a flavour-diagonal signal of CP
violation. At the non-relativistic level, the EDM $d$ determines a
coupling of the spin to an external electric field, ${\cal H} \sim d
\vec{E} \cdot \vec{S}$. Searches for intrinsic EDMs have a long
history, stretching back to the prescient work of Purcell and Ramsey
who used the neutron EDM as a test of parity in nuclear physics. At
the present time, there are two primary motivations for anticipating a
nonzero EDM at or near current sensitivity levels. Firstly, a viable
mechanism for baryogenesis requires a new $CP$-odd source, which if
tied to the electroweak scale necessarily has important implications
for EDMs. The second is that $CP$-odd phases appear quite generically
in models of new physics introduced for other reasons, e.g.  in
supersymmetric models. Indeed, it is only the limited field content of
the SM which limits the appearance of $CP$-violation to the CKM phase
and $\theta_{QCD}$. The lack of any observation of a nonzero EDM has,
on the flip-side, provided an impressive source of constraints on new
physics, and there is now a lengthy body of literature on the
constraints imposed, for example, on the soft-breaking sector of the
MSSM. Generically, the EDMs ensure that new $CP$-odd phases in this
sector are at most of ${\cal O}(10^{-1}-10^{-2})$, a tuning that
appears rather unwarranted given the ${\cal O}(1)$ value of the CKM
phase.

The strongest current EDM constraints are shown for three characteristic classes of observables in
Table~\ref{explimit}, and will be discussed in detail in the following.
\begin{table}[bht]
\begin{minipage}{\textwidth}
\caption{Current constraints within three representative classes of EDMs. \label{explimit}}
\begin{tabular*}{\textwidth}{@{\extracolsep{\fill}}lcrr}
\hline\hline 
Class 		& EDM 		& Current Bound\hspace*{8mm} 				&Ref.  			\\
\hline
Paramagnetic 	& $^{205}$Tl 	& $|d_{\rm Tl}| < 9 \times 10^{-25} e\, {\rm cm}$ 	&\cite{Regan:2002ta} 	\\ 
Diamagnetic 	& $^{199}$Hg 	& $|d_{\rm Hg}| < 2 \times  10^{-28} e\, {\rm cm}$ 	&\cite{Hg} 		\\ 
Nucleon 	& $n$ 		& $|d_n| < 3\times 10^{-26} e\, {\rm cm}$ 		&\cite{n} 		\\ 
\hline\hline
\end{tabular*}
\end{minipage}
\end{table}

We summarize first the details of the EDM constraints, and the induced bounds on a generic class of $CP$-odd operators normalized at 1 GeV,
commenting on how the next generation of experiments will impact significantly on the level of sensitivity in all sectors. We then turn
to a brief discussion of some of the constraints on new physics that ensue from these bounds.  More detailed discussions of phenomenology
of EDMs is given in the first half of this report (see e.g. Section~\ref{sec:EDM-RGE}). 

\subsubsection{ $CP$-odd operators and electric dipole moments}
We will briefly review the relevant formulae for the observable EDMs in terms of $CP$-odd operators normalized at 1 GeV. Including the most
significant flavour-diagonal $CP$-odd operators (see e.g. \cite{PRrev}) up to dimension six, the corresponding effective
Lagrangian takes the form,
\begin{eqnarray}
{\cal{L}}^{\rm 1\; GeV}_{eff} &=& \frac{g_s^2}{32\pi^{2}}\ \bar\theta\
G^{a}_{\mu\nu} \widetilde{G}^{\mu\nu , a} -
\frac{i}{2} \sum_{i=e,u,d,s} d_i\ \overline{\psi}_i (F\sigma)\gamma_5 \psi_i  -
\frac{i}{2} \sum_{i=u,d,s} 
\widetilde{d}_i\ \overline{\psi}_i g_s (G\sigma)\gamma_5\psi_i \nonumber\\
  && + \frac{1}{3} w\  f^{a b c} G^{a}_{\mu\nu} \widetilde{G}^{\nu \beta , b}
G^{~~ \mu , c}_{\beta}+ \sum_{i,j=e,q} C_{ij} (\bar{\psi}_i \psi_i) (\psi_j i\gamma_5 \psi_j)+ \cdots \label{leff}
\end{eqnarray}
The $\theta$-term, as it has a dimensionless coefficient, is particularly dangerous leading to the strong $CP$ problem and in what
follows we will invoke the axion mechanism \cite{PQ} to remove this term.

\begin{figure}[t]
\includegraphics[width=0.48\linewidth]{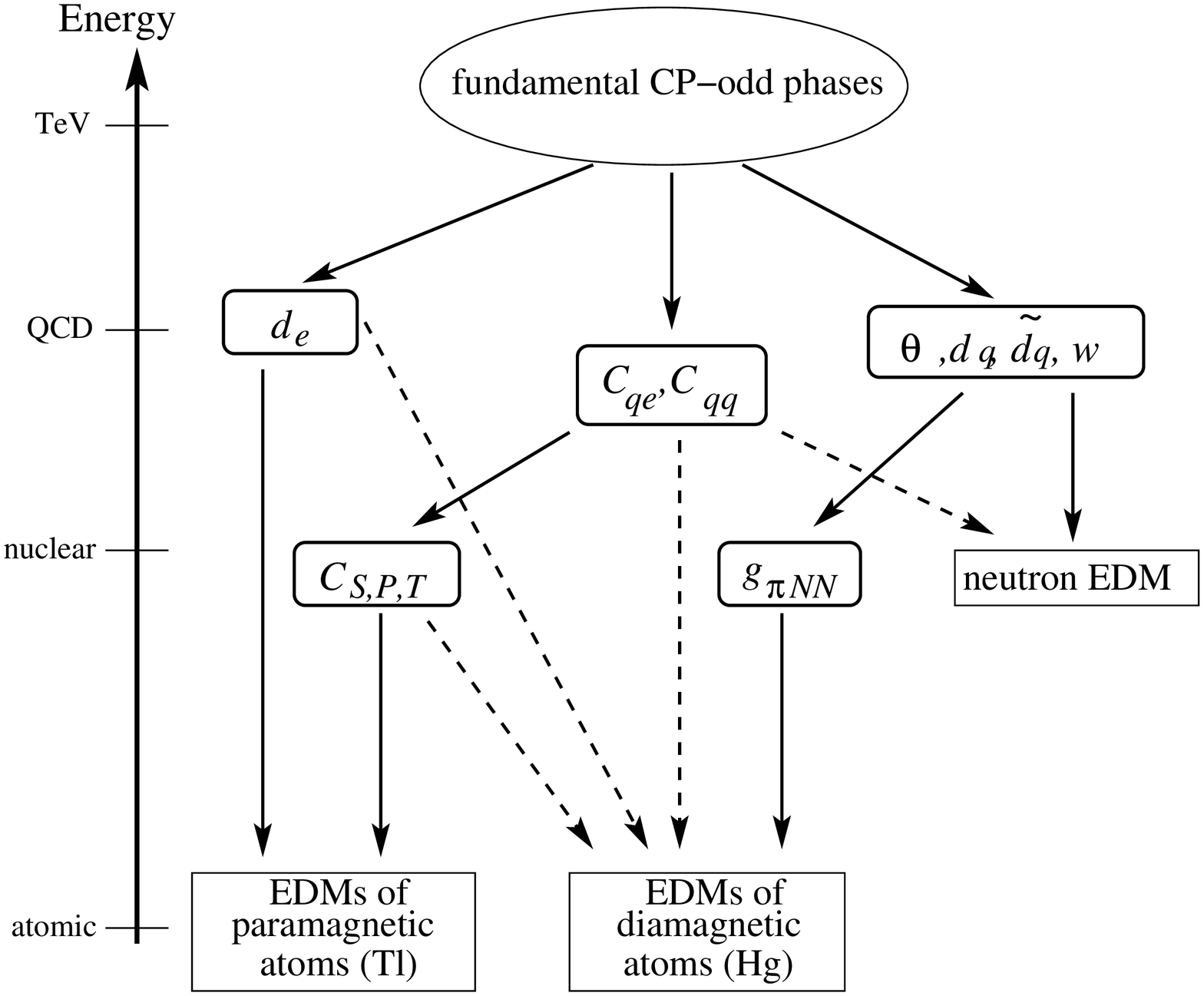}\hspace*{\fill}
\includegraphics[width=0.48\linewidth]{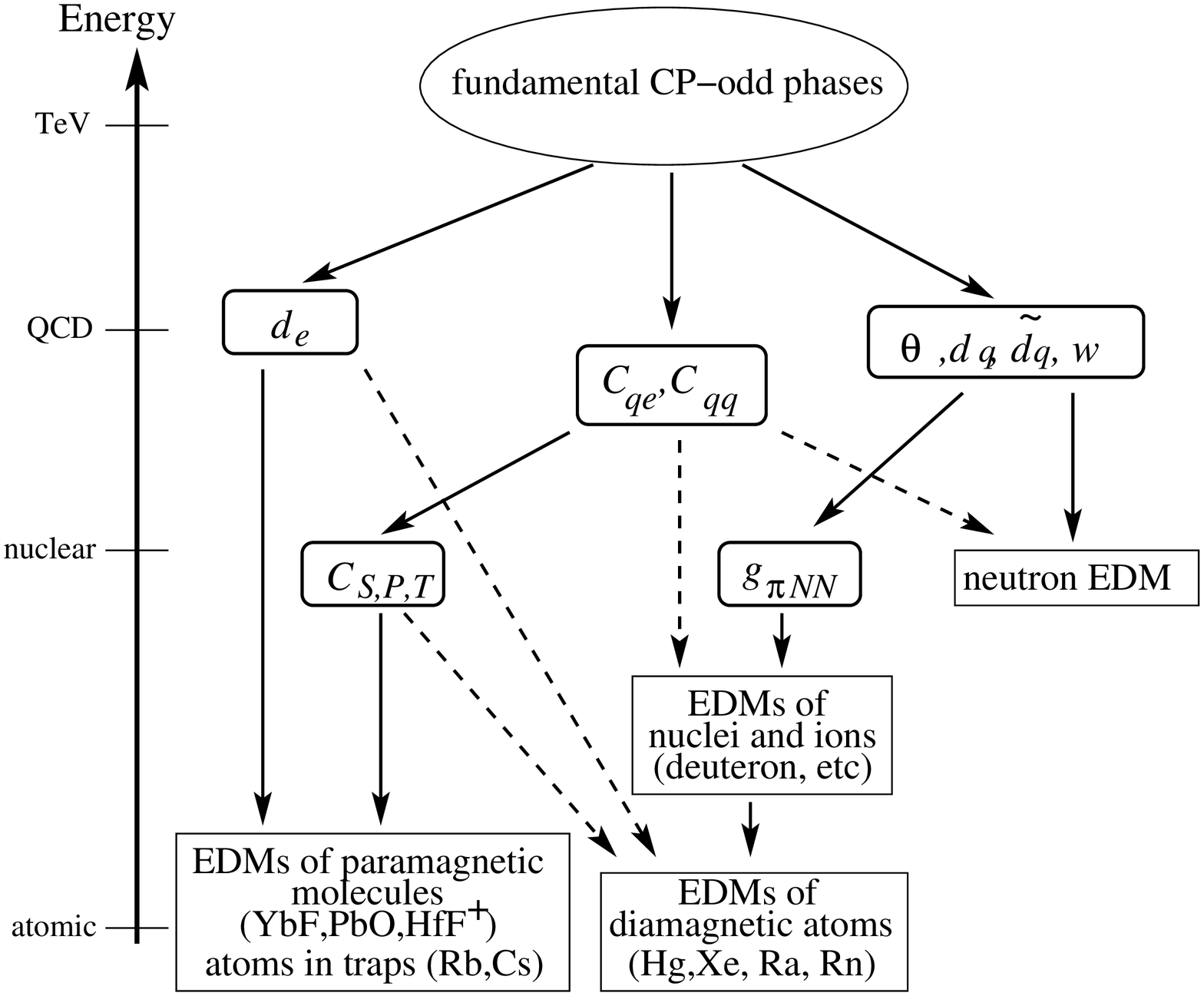}
\caption{\label{schemetb}A schematic plot of the hierarchy of scales between the leptonic and hadronic CP-odd
sources and three generic classes of observable EDMs. The dashed lines indicate generically
weaker dependencies in SUSY models. The current situation is given on the left, while on the right we show the 
dependencies of several classes of next-generation experiments.}
\end{figure}

The physical observables can be conveniently separated into three main categories, depending on the physical mechanisms via which an EDM can
be generated: EDMs of paramagnetic atoms and molecules; EDMs of diamagnetic atoms; and the neutron EDM. The inheritance pattern for
these three classes is represented schematically in Fig.~\ref{schemetb} and, while the experimental constraints on the
three classes of EDMs differ by several orders of magnitude, it is important that the actual sensitivity to the operators in (\ref{leff})
turns out to be quite comparable in all cases. This is due to various enhancements or suppression factors which are relevant in each case,
primarily associated with various violations of ``Schiff shielding'' -- the non-relativistic statement that an electric field applied to a
neutral atom must necessarily be screened and thus remove any sensitivity to the EDM.

\subsubsection{EDMs of paramagnetic atoms}
For paramagnetic atoms, Schiff shielding is violated by relativistic effects which can in fact be very large. One has roughly \cite{fgreview,kozlov},
\begin{equation}
 d_{\rm para}(d_e) \sim 10 \alpha^2 Z^3 d_e, 
\end{equation}
which for large atoms such as Thallium amounts to a huge enhancement of the field seen by the electron EDM (see e.g. \cite{fgreview,KL}),
which counteracts the apparently lower sensitivity of the Tl EDM bound,
\begin{equation} 
d_{\rm Tl} = -585 d_e -   43 ~{\rm GeV} \times e \, C_S^{\rm singlet}.
\label{dtl}
\end{equation}
We have also included here the most relevant $CP$-odd electron-nucleon interaction, namely $C_S\bar e i \gamma_5 e \bar NN$, which
in turn is related to the semileptonic 4-fermion operators in (\ref{leff}).

\subsubsection{EDMs of diamagnetic atoms}
For diamagnetic atoms, Schiff shielding is instead violated by the finite size of the nucleus and differences in the distribution of the
charge and the EDM. However, this is a rather subtle effect,
\begin{equation}
 d_{\rm dia} \sim 10 Z^2 (R_N/R_A)^2 \tilde{d}_q,
\end{equation}
and the suppression by the ratio of nuclear to atomic radii, $R_N/R_A$, generally leads to a suppression of the sensitivity to the
nuclear EDM, parameterized to leading order by the Schiff moment $S$, by a factor of $10^3$ (see e.g. \cite{fgreview, KL}). Thus, although
the apparent sensitivity to the Hg EDM is orders of magnitude stronger than for the Tl EDM, both experiments currently have comparable
sensitivity to various $CP$-odd operators and thus play a very complementary role.  Combining the atomic $d_{\rm Hg}( S)$, nuclear
$S(\bar g_{\pi NN})$, and QCD $\bar{g}_{\pi NN}^{(1)}(\tilde{d}_q)$, components of the calculation \cite{KL,PRrev}, we have
\begin{equation}
d_{\rm Hg} = 7\times 10^{-3}\,e\,(\tilde d_u - \tilde d_d)  + 10^{-2}\, d_e + {\mathcal O}(C_S,C_{qq})
\label{Hgmaster}
\end{equation}
where the overall uncertainty is rather large, a factor of 2-3, due to significant cancellations between various contributions.
A valuable feature of $d_{\rm Hg}$ is its sensitivity to the triplet combination of colour EDM operators $\tilde d_q$.

\subsubsection{Neutron EDM}
The neutron EDM measurement is of course not sensitive to the above atomic enhancement/suppression factors and, using 
the results obtained using QCD sum rule techniques \cite{PRrev} (see also \cite{chiral} for alternative chiral approaches), wherein under Peccei-Quinn 
relaxation of the axion the contribution of sea-quarks is also suppressed at leading order:
\begin{eqnarray}
 d_{n} &=& (1.4 \pm 0.6)(d_d-0.25d_u) + (1.1 \pm 0.5)e(\tilde d_d + 0.5\tilde d_u) 
 + 20\,{\rm MeV}\times e~ w +{\mathcal O}(C_{qq}).
\label{dn1}
\end{eqnarray}
Note that the proportionality to $d_q\langle \bar qq\rangle \sim  m_q\langle \bar qq\rangle \sim f_\pi^2m_\pi^2$ removes 
any sensitivity to the poorly known absolute value of the light quark masses. 

\subsubsection{Future developments}
The experimental situation is currently very active, and a number of new EDM experiments, as detailed in this report,
promise to improve the level of sensitivity in all three classes by one-two orders of magnitude in the coming years. These include: new searches for EDMs
of polarizable paramagnetic molecules, which aim to exploit additional polarization effects enhancing the effective field seen by the unpaired electron by
a remarkable factor of up to $10^5$, and are therefore primarily sensitive to the electron EDM; new searches for the EDM of the neutron in cryogenic systems;
and also proposed searches for EDMs of charged nuclei and ions using storage rings. This latter technique clearly aims to avoid the effect of Schiff
shielding and enhance sensitivity to the nuclear EDM and its hadronic constituents. A schematic summary of how a number of these new experiments will be
sensitive to the set of $CP$-odd operators is exhibited in Fig.~\ref{schemetb}.

\subsubsection{Constraints on new physics}
Taking the existing bounds, and the formulae above, we obtain the following set of constraints on the $CP$-odd sources at 1~GeV
(assuming an axion removes the dependence on $\bar{\theta}$),
\begin{eqnarray}
  \left| d_e + e (26\,{\rm MeV})^2 \left(3\frac{C_{ed}}{m_d} + 11  \frac{C_{es}}{m_s} + 5 \frac{C_{eb}}{m_b}\right)\right| < 1.6 \times
  10^{-27}\, e \; {\rm cm}\;\;\; {\rm from}\; d_{Tl},\\
 \left|  (\tilde{d}_d - \tilde{d}_u)+{\cal O}(\tilde{d}_s, d_e,
  C_{qq},C_{qe}) \right| < 2 \times 10^{-26} \, e\; {\rm cm}\;\;\; {\rm from}\; d_{Hg},\\
 \left| e(\tilde{d}_d + 0.56\tilde{d}_u) + 1.3(d_d - 0.25d_u) + {\cal O}(\tilde{d}_s,w,C_{qq})\right| < 2 \times 10^{-26}\, e\;{\rm cm}\;\;\; {\rm from}\; d_n,
\end{eqnarray}
where the additional ${\cal O}(\cdots)$ dependencies are known less precisely, but may not always be sub-leading in particular models. The
precision of these results varies from 10-15\% for the Tl bound, to around 50\% for the neutron bound, and to a factor of a few for Hg. It
is remarkable to note that, accounting for the naive mass-dependence $d_f \propto m_f$, all these constraints are of essentially the same
order of magnitude and thus highly complementary. Constraints obtained in the hadronic sector using other calculational techniques differ
somewhat but generally give results consistent with these within the quoted precision.

The application of these constraints to models of new physics has many facets and is discussed in several specific cases elsewhere in this
report. We will limit our attention here to just a few simple examples relevant to the motivations noted above.

\subsubsection{The SUSY $CP$-problem}
\begin{figure}[t]
\centerline{\includegraphics[width=5cm]{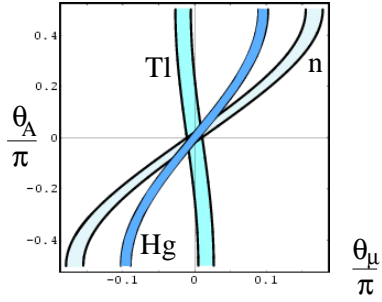}}
\caption{\label{fig:edm-phase}
Constraints on the CMSSM phases $\theta_A$ and $\theta_\mu$ from a combination of the three most sensitive EDM constraints, 
$d_n$, $d_{\rm Tl}$ and $d_{\rm Hg}$, 
for $M_{\rm SUSY} = 500$ GeV, and $\tan\beta=3$ (from \cite{PRrev}). The region allowed by EDM constraints is at the intersection
of all three bands around $\theta_A=\theta_\mu=0$. }
\end{figure}

It is now rather well-known that a generic spectrum of soft SUSY-breaking parameters in the MSSM will generate EDMs via 1-loop
diagrams~\cite{Ellis:1982tk} that violate the existing bounds by one-to-two orders of magnitude leading to the SUSY $CP$ problem.  The situation is
summarized rather schematically in Fig.~\ref{fig:edm-phase}.

In many respects the situation is better described by the amount of
fine tuning of the MSSM spectrum that is required to avoid these
leading-order contributions, and by how much the ability to avoid the
EDM constraints is limited by secondary constraints from numerous, and
more robust, 2-loop contributions~\cite{Chang:1998uc} and four-fermion
sources~\cite{Lebedev:2002ne}. 
Indeed, if we consider two extreme cases: (i) the 2HDM, where all SUSY fermions and sfermions are very heavy; and (ii) split
SUSY, where all SUSY scalars are very heavy, one finds that while 1-loop EDMs are suppressed, 2-loop contributions are already very
close to the current bounds~\cite{Giudice:2005rz,Chang:2005ac,Lebedev:2002ne}. This bodes well for the ability of
next-generation experiments to provide a comprehensive test of large SUSY phases at the electroweak scale, regardless of the detailed form
of the SUSY spectrum.

\subsubsection{Constraints on new SUSY thresholds}
If SUSY is indeed discovered at the LHC, but with no sign of phases in the soft-sector, one may instead consider the ability of EDMs to
detect new supersymmetric $CP$-odd thresholds. At dimension-five there are several $R$-parity--conserving operators, besides those well-known
examples associated with neutrino masses and baryon and lepton number violation \cite{WeinbergSY}.  Writing the relevant dimension-five
superpotential as \cite{prs}
\begin{equation}
\Delta {\cal W} = \frac{y_h}{\Lambda_{h}}H_dH_uH_dH_u +\frac{Y^{qe}_{ijkl}}{\Lambda_{qe}}(U_i Q_j )E_k L_l +\frac{Y^{qq}_{ijkl}}{\Lambda_{qq}}(U_iQ_{j}) (D_k Q_{l} )+
\frac{\tilde Y^{qq}_{ijkl}}{\Lambda_{qq}}(U_it^AQ_{j}) (D_kt^AQ_{l}),\label{qule}
\end{equation}
one finds that order-one $CP$-odd coefficients with a generic flavor structure, particularly for the semi-leptonic operators, are probed by the sensitivity of 
$d_{Tl}$ and $d_{Hg}$ at the remarkable
level of $\Lambda \sim 10^8$~GeV \cite{prs}. This is comparable to, or better than, the corresponding sensitivity of lepton-flavor violating observables. 

\subsubsection{Constraints on minimal electroweak baryogenesis}
As noted above, one of the primary motivations for anticipating nonzero EDMs at or near the current level of sensitivity is through
the need for a viable mechanism of baryogenesis. This is clear in essentially all baryogenesis mechanisms that are tied to the
electroweak scale. As a simple illustration, one can consider a {\it minimal} extension of the SM Higgs sector \cite{gsw,bfhs,Huber:2006ri},
\begin{equation}
 {\cal L}_{\rm dim\; 6} = \frac{1}{\Lambda^2} (H^\dagger H)^3 + \frac{Z^u_{ij}}{\Lambda_{\rm CP}^2} (H^{\dagger} H) U_i^c H Q_j 
 + \frac{Z^d_{ij}}{\Lambda_{\rm CP}^2} (H^{\dagger} H) D_i^c H^\dagger Q_j  + \frac{Z^e_{ij}}{\Lambda_{\rm CP}^2} (H^{\dagger} H) E_i^c H^\dagger L_j. \label{ops}
 \end{equation}
The first term is required to induce a sufficiently strong first-order electroweak phase transition, while the remaining operators provide the additional 
source (or sources) of $CP$-violation, where we have assumed consistency with the principle of minimal flavour violation. Modified Higgs couplings 
of this kind, including $CP$-violating effects, are currently the subject of significant research within collider physics, relevant to the LHC in 
particular \cite{Accomando:2006ga}, making EDM probes of models of this kind quite complementary.

As discussed in \cite{Huber:2006ri}, such a scenario can reproduce the required baryon-to-entropy ratio, $\eta_b = 8.9 \times 10^{-11}$,
while remaining consistent with the EDM bounds, provided the thresholds and the Higgs mass are quite low, e.g. $400\;{\rm GeV} <
\Lambda,\; \Lambda_{\rm CP} < 800\;{\rm GeV}$. The EDMs in this case are generated at the 2-loop level, and it is clear that an improvement
in EDM sensitivity by an order of magnitude would provide a conclusive test of minimal mechanisms of this form.

\subsection{Neutron EDM}
The neutron electric dipole moment is sensitive to many sources of CP violation.  Most famously, it constrains QCD (the ``strong CP problem"),
but it also puts tight constraints on Supersymmetry and other physics models beyond the Standard Model. The Standard Model
prediction of $\sim10^{-32}$~$e\,$cm is a factor of 10$^6$ below existing limits, so any convincing signal within current or foreseen
sensitivity ranges will be a clear indication of physics beyond the SM.

All current nEDM experiments use NMR techniques to search for electric-field induced changes in the Larmor precession frequency of
bottled ultracold neutrons.  Recent results from a room-temperature apparatus at ILL yielded a new limit of $\left| d_{n}\right| <2.9
\times 10^{-26}$~$e\,$cm (90\% CL) which rules out many ``natural" varieties of SUSY. Several new experiments hope to improve on this
limit: two of these involve new cryogenic techniques that promise an eventual increase in sensitivity by two orders of
magnitude.  First results, at the level of $\sim 10^{-27}$~$e\,$cm, are to be expected within about four years.

\subsubsection{ILL}
A measurement of the neutron EDM was carried out at the ILL between 1996 and 2002, by a collaboration from the University of Sussex, the
Rutherford Appleton Laboratory, and the ILL itself. The final published result provided a limit of
$\left| d_{n}\right| <2.9 \times 10^{-26}$~$e\,$cm (90\% CL) \cite{baker06}.  This represents a factor of two improvement beyond
the intermediate result \cite{harris99} and almost a factor of four
beyond the results existing prior to this experiment
\cite{smith90,altarev96}.  The collaboration, which has now expanded to include Oxford University and the University of Kure, has designed
and developed ``CryoEDM'', a cryogenic version of the experiment that is expected to achieve two orders of magnitude improvement in
sensitivity.  Construction and initial testing are underway at the time of writing.

{\noindent\it Experimental technique}\\
The room-temperature measurement was carried out using stored ultracold neutrons (i.e.\  having energies $\stackrel{\scriptstyle <}{\scriptstyle\sim}$ 200 neV)
from the ILL reactor.  The Ramsey technique of separated oscillatory fields was used to determine the Larmor precession frequency of the neutrons within $\vec{B}$
and $\vec{E}$ fields.  The signature of an EDM is a frequency shift proportional to any change in the applied electric field.

The innovative feature of this experiment was the use of a cohabiting atomic-mercury magnetometer \cite{green98}.  Spin-polarized Hg atoms shared the same volume
as the neutrons, and the measurement of their precession frequency provided a continuous high-resolution monitoring of the magnetic field drift: prior to this, such
drift entirely dominated the tiny $\vec{E}$-field induced frequency changes that were sought.  

{\noindent\it Systematics}\\
Analysis of the data revealed a new source of systematic error, which, as the problem of B-field drift had been virtually eliminated, became potentially the
dominant error.  Its origins lay in a geometric-phase (GP) effect  \cite{pendlebury04} - an unfortunate collusion between any small applied axial $\vec{B}$-field
gradient and the component of $\vec{B}$ induced in the particle's rest frame by the Lorentz transformation of the electric field.  This GP effect induced a frequency
shift proportional to $\vec{E}$, and hence a false EDM signal.  In fact, the Hg magnetometer itself was some 50 times more susceptible to this effect than were the
neutrons, so the introduction of the magnetometer brought the GP systematic with it.

This effect was overcome by careful measurement of the neutron-to-Hg frequency ratios for both polarities of magnetic field, in order to determine the point nominally
corresponding to zero applied axial B-field gradient, as well as by a series of auxiliary measurements to pin down small corrections due to local dipole \cite{harris06}
and quadrupole fields (as well as the Earth's rotation).  The final result therefore remained statistically limited.

The experiment is now complete and, as will be discussed below, the equipment will be used for further studies by another collaboration based largely at the PSI.

Still another collaboration, led by the PNPI in Russia, is developing a new room-temperature nEDM apparatus, which they plan to run at ILL.  It is also intended to
reach a sensitivity of $\sim10^{-27}$~$e\,$cm, to be achieved in part by the use of multiple back-to-back measurement chambers with opposing electric fields to cancel
some systematic errors.

{\noindent\it Cryogenic experiments overview}\\
It has been known for several decades \cite{golub_pendlebury77} that 8.9 \AA\ neutrons incident on superfluid $^4$He at 0.5 K will down-scatter, transferring their
energy and momentum to the helium and becoming ultracold neutrons (UCN) in the process.  This so-called super-thermal UCN source provides a much higher flux than is
available simply from the low-energy tail of the Maxwell distribution.  
In addition, 
the immersion of the apparatus in a bath of liquid helium should allow for the provision
of stronger electric fields than could be sustained {\it in vacuo}.  
The other two variables that contribute to the figure of merit for this experiment, namely
the polarization and the NMR coherence time, should also be improved: the incident cold neutron beam can be very highly polarized, and the polarization remains intact
during the down-scattering process; and the improved uniformity of magnetic field attainable with superconducting shields and coil will reduce depolarization during
storage, while losses from up-scattering will be much reduced due to the cryogenic temperatures of the walls of the neutron storage vessels.  

{\noindent\it ILL CryoEDM experiment status}\\ The majority of the
apparatus for the cryoEDM experiment has been installed at ILL, and
testing is underway.  UCN production via this superthermal mechanism
has been demonstrated \cite{baker03b}, and the solid-state UCN
detectors developed by the collaboration have also been shown to work
well \cite{baker03a}.  At the time of writing, there are still some
hardware problems to be resolved, in particular with components in and
around the Ramsey measurement chamber.  A high-precision scan of the
magnetic field was carried out in 2007, and measurements were made of
the neutron polarization.  An initial HV system will be installed in
spring 2008.  By the end of 2008, the system is expected to have a
statistical sensitivity of $\sim 10^{-27}$ $e\,$cm.

{\noindent\it Future plans}\\
In order to achieve optimum sensitivity, a number of improvements will need to be made:
\begin{list}{-}{}
\item
The superconducting magnetic shielding requires additional protective ``end caps" to shield fully the ends of the superconducting solenoid.
\item
The current measurement chamber only has two cells: one with HV applied, and one at ground as a control.  It is planned to upgrade to
a four-cell chamber, with the HV applied to the central electrode, in order to be able to carry out simultaneous measurements with electric
fields in opposite directions.  As well as canceling several potential systematic errors, this will reduce the statistical
uncertainty by doubling the number of neutrons counted.
\item
The ILL is preparing a new beam line with six times the currently available intensity of 8.9 \AA\ neutrons, and wishes to transfer the
experiment to that beam line in 2009. Funding for these improvements is expected to be contingent on successful running of the existing apparatus.
\end{list}
A sensitivity of $\sim2\times10^{-28}$~$e\,$cm should be achievable within two to three years of running at the new beam line.

\subsubsection{PSI}
The present best limit for the neutron electric dipole moment (EDM), $|d_n| < 2.9\times 10^{-26}\,e$\,cm~\cite{baker06}, was obtained by
the Sussex/Rutherford/ILL collaboration from measurements at the ILL
source for ultracold neutrons~\cite{Ste86}, as discussed in the
previous section.  The experiment is at this
point statistically limited and also facing systematic challenges not far away~\cite{baker06,pendlebury04,harris06}.  In order to make
further progress, both, statistical sensitivity and control of systematics, have to be improved.  Gaining in statistics requires new
sources for ultracold neutrons (UCN). These can be integrated into the experiment as for the new cryogenic EDM searches (see~\cite{Har06b}),
delivering UCN in superfluid helium, or a multipurpose UCN source, delivering UCN in vacuum.  This high
intensity UCN source is presently under construction at the Paul Scherrer Institut in Villigen, Switzerland~\cite{psi_ucn}. It is
expected to become operational towards the end of 2008 and to deliver UCN densities of more than 1000\,cm$^{-3}$ to typical experiments,
i.e. almost two orders of magnitude more than presently available.

The in-vacuum technique will be pushed to its limits, delivering first results in about 4 years. The following steps are planned by a sizeable international
collaboration\cite{psi_coll}:
\begin{list}{-}{}
\item
While the new UCN source is under construction the collaboration operates and improves the apparatus of the former Sussex/RAL/ILL collaboration 
at ILL Grenoble. In order to better control the systematic issues, the magnetic field and its gradients will be monitored and stabilized using an array of
laser optically pumped Cs-magnetometers~\cite{Gro05a,Gro05b}. An order of magnitude improvement compared to todays field fluctuations
over the typical measurement times of 100-1000\,s is certainly feasible. It is also necessary to improve the sensitivity of the Hg co-magnetometer~\cite{Gre98}. 
Other improvements of the system are with regard to UCN polarization and detection as well as upgrading the data acquisition system.
The hardware efforts are accompanied by a full simulation of the system.
\item
It is planned to move the apparatus from ILL to PSI towards the end of 2008 in order to be ready for data taking for about two years, 2009 and 2010. In addition to
the improvements of phase I, an external magnetic field stabilization system and a temperature stabilization are envisaged. Furthermore, work on developing a second
co-magnetometer using a hyper-polarized noble gas species is ongoing and might further improve the systematics control. In case of a successful development,
also the replacement of the Hg system together with an increase of the electric field strength may become possible.
In any case, a factor of 5 gain in sensitivity is expected from the higher UCN intensity, corresponding to a limit of about $5-6\times10^{-27}\,e\,$cm 
in case the EDM is not found. In parallel to the described activities, the design of a new experimental apparatus will  start in 2007. 
After a major design effort in 2008,  set-up of the new apparatus will start in 2009.
\item
The new experiment will be an optimized version of the room-temperature in-vacuum approach. Another order of magnitude gain in sensitivity will be obtained by a
considerable increase of the statistics due to a larger experimental volume ($\times \sqrt{5}$), a better adaption to the UCN source ($\times \sqrt{2}$),
longer running time ($\times \sqrt{3}$) and by an improvement of the electric field strength ($\times 2$). Completion of the new experimental apparatus is
anticipated for end of 2010, and data taking planned for 2011-2014. 
\end{list}
The features of the experiment include 
\begin{list}{-}{}
\item
continued use of the successful Ramsey-technique with UCN in vacuum and the apparatus at room-temperature, 
\item
increased sensitivity due to much larger UCN statistics at the new PSI source, larger experimental volume,
better polarization product and possibly larger electric field strength,
\item
application of a double neutron chamber system,
\item
improved magnetic field control and stabilization with multiple laser optically pumped Cs-mag\-neto\-meters, and 
\item
an improved co-magnetometry system.
\end{list}

As another very strong source for UCN is currently under construction at the FRMII in Munich, in the 
long run and for the optimum conditions for the experiment, the collaboration will have the opportunity to 
choose between PSI and FRMII.

\subsubsection{SNS}
A sizeable US collaboration \cite{sns_edm} is planning to develop a cryogenic experiment, following an early concept by Golub and
Lamoreaux \cite{golub94}.  It will be based at the SNS 1.4~MW spallation source at Oak Ridge. A fundamental neutron physics beam line
is under construction, which will include a double monochromator to select 8.9 \AA\ neutrons for UCN production in liquid helium.

In this experiment, spin-polarized $^3$He will be used both as a magnetometer and as a neutron detector.  The precession of the $^3$He
can, in principle, be detected with SQUID magnetometers.  Meanwhile, the cross section for the absorption reaction n $+ ^3$He $\rightarrow$
p + $^3$H + 764 keV is negligible for a total spin $J=1$, but very large ($\sim 5$ Mb) for $J=0$.  In consequence, a scintillation signal
from this reaction will be detected with a beat frequency corresponding to the difference between the Larmor precession
frequencies of the neutrons and the $^3$He.

An application for funding to construct this experiment is currently under review. Extensive tests are
underway to study, for example, the electric fields attainable in liquid helium, the $^3$He spin relaxation time and the diffusion of
$^3$He in $^4$He.  If construction goes according to plan, commissioning will be in approximately 2013, with results following
probably four or five years later. The ultimate sensitivity will be below 10$^{-28}$~$e\,$cm.

\subsection{The deuteron EDM}
A new concept of investigating the EDM of bare nuclei in magnetic storage rings has been developed by the storage ring EDM collaboration
(SREC) over the past several years.  The latest version of the methods analyzed turns out to be very sensitive for light (bare) nuclei and
promises the best EDM experiment for $\theta_{QCD}$, quark and quark-color EDMs.

The search for hadronic EDMs has been dominated by the search for a neutron EDM and nuclear Schiff moments in heavy diamagnetic
atoms, such as $^{199}$Hg.  The latter depend on nuclear theory to relate the measured Schiff moment to the underlying CP violating interaction. 

The sensitive `traditional' EDM experiments are, so far, all performed on electrically neutral systems, such as the neutron, atoms, or
molecules.  A strong electric field is imposed, together with a weak magnetic field, and using NMR techniques, a change of the Larmor
precession frequency is looked for.  The application of strong electric fields precludes a straightforward use of this technique on
charged particles.  These particles would accelerate out of the setup, leaving little time to make an accurate measurement.

Attempt to search for an EDM on simple nuclear systems, such as the proton or deuteron, when part of an atom, are severely hindered by
shielding.  This so-called Schiff-screening precludes an external electric field to penetrate to the nucleus.  Due to rearrangement of
the atomic electrons, the net effect of the electric field on the nucleus is essentially zero.  Known loop-holes include relativistic
effects, non-electric components in the binding of the electrons, and an extended size of the nucleus.  None of these loopholes are
sufficiently strong to allow a sensitive measurement on a light atom. For hydrogen atoms, the atomic EDM resulting from a nuclear EDM is
down by some seven orders of magnitude.

Nevertheless, light nuclei, and the deuteron in particular, are attractive to search for hadronic EDMs because of their relatively
simple structure.  Moreover, a novel experimental technique, using the motional electric field experienced by a relativistic particle when
traversing a magnetic field, make it possible to directly search for EDM on charged systems, such as the (bare) deuteron. 

\subsubsection{Theoretical considerations}

The deuteron is the simplest nucleus.  It consists of a weakly bound proton and neutron in a predominantly $^3$S$_1$ state, with a
small admixture of the D-state.  From a theoretical point of view, the deuteron is especially attractive, because it is the simplest system
in which the P-odd, T-odd nucleon-nucleon (NN) interaction contributes to an EDM.  Moreover, the deuteron properties are well understood, so
reliable and precise calculations are possible.

In \cite{Liu:2004tq}, a framework is presented that could serve as a starting point for the microscopic calculation of complex systems.
The most general form of the interaction, based only on symmetry considerations, contains ten P- and T-odd meson-nucleon coupling
constants for the lightest pseudo-scalar and vector mesons ($\pi$, $\rho$, $\eta$ and $\omega$).

This P-odd, T-odd interaction induces a $P$-wave admixture to the deuteron wave function.  It is this admixture that leads to an EDM.
Since the proton and neutron that make up the deuteron may also have an EDM, a disentanglement of one- and two-body contributions,
\begin{equation}
  d_{\mathcal{D}} \simeq d_{\mathcal{D}}^{(1)} + d_{\mathcal{D}}^{(2)}
\end{equation}
to the EDM is necessary to uncover the underlying structure of the P-odd T-odd physics.

The two-body component is predominantly due to the polarization effect, and shows little model dependence for all leading high-quality
potentials.  Additional contributions arrive from meson exchange.

The one body contribution is simply the sum of the proton and neutron EDMs.  The nucleon EDM has a wide variety of sources, as already
discussed for the neutron.  There exists no good model to describe the non-perturbative dynamics of bound quarks.  A commonly used method is
to evaluate hadronic loop diagrams, containing mesonic and baryonic degrees of freedom.  Within the framework presented in
\cite{Liu:2004tq}, the EDMs for the proton, neutron and deuteron are found (reproducing only the pion dependence),
\begin{equation}
  \begin{array}{cccccc}
 d_p 		& = & -0.05 \,\bar{g}_\pi^{(0)} &+0.03\,\bar{g}_\pi^{(1)} 	&  +0.14\,\bar{g}_\pi^{(2)} 	&+ \cdots\\
 d_n 		& = & +0.14 \,\bar{g}_\pi^{(0)} & 				& -0.14\,\bar{g}_\pi^{(2)} 	&+ \cdots\\
 d_{\mathcal{D}}& = & +0.09 \,\bar{g}_\pi^{(0)} &+0.23\,\bar{g}_\pi^{(1)} 	&				& + \cdots\\
\end{array}
\end{equation}
These dependences clearly show the complementarity of these three particles.

The magnitudes of the coupling constants can be calculated for several viable sources of CP-violation.  In the Standard Model, there is room
for CP-violation via the so-called $\bar{\theta}$ parameter.  In the case of the nucleons, one has the relation
\begin{equation}
  d_n \simeq -d_p \simeq 3\times 10^{-16}\,\bar{\theta}\;\; e \, {\rm cm}   , 
\end{equation}
which yields the severe constraint $\bar{\theta} < 1\times 10^{-10}$. For the deuteron, one finds
\begin{equation}
  d_{\mathcal{D}} \simeq -10^{-16}\,\bar{\theta}\;\; e \, {\rm cm}.
\end{equation}
At the level of $d_{\mathcal{D}} \simeq 10^{-29}\;\; e \, {\rm cm}$, one probes $\bar{\theta}$ at the level of $10^{-13}$.  Since $\bar{\theta}$
contributes differently to the neutron and the deuteron, it is clear that both experiments are complementary.  Indeed.  the prediction
\begin{equation}
  d_{\mathcal{D}} / d_n = -1/3
\end{equation}
provides a beautiful check as to whether $\bar{\theta}$ is the source of the observed EDMs, should both be measured.  In fact, measurement
of the EDMs of the proton, deuteron and $^3$He would allow to verify if they satisfy the relation
\begin{equation}
  d_{\mathcal{D}} : d_p : d_{^3\text{He}} \simeq 1:3: -3
\end{equation}
Here, it was assumed that $^3$He has properties very similar to the neutron, which provides most of the spin.

Generic supersymmetric models contain a plethora of new particles, which may be discovered at LHC, and new CP-violating phases. 
Following the work by Lebedev {\em et al.}\cite{Lebedev:2004va} and the review by Pospelov and Ritz\cite{Pospelov:2005pr}, we find that SUSY
loops give rise to ordinary quark EDMs, $d_q$, as well as quark-color EDMs, $\tilde{d}_q$.  For the neutron and deuteron one finds (with the
color EDM part divided in isoscalar and isovector parts)
\begin{equation}
  \begin{array}{rcccc}
d_n 		& \simeq & 1.4 \left(d_d - 0.25 d_u \right) 	& +~0.83 e \left(\tilde{d}_d + \tilde{d}_u\right)& +~0.27e\left(\tilde{d}_d - \tilde{d}_u \right) \\
d_\mathcal{D} 	& \simeq & \left(d_d + d_u \right) 		&-~0.2e \left(\tilde{d}_d + \tilde{d}_u \right) 	&+~6e \left(\tilde{d}_d - \tilde{d}_u \right),
  \end{array}
\end{equation}
and similar relations for {\em e.g.} the mercury EDM.  The isovector
part is limited to $| ec(\tilde{d}_d-\tilde{d}_u )|
<2\times 10^{-26}\;\; e \, {\rm cm}$ by the present limit on
the $^{199}$Hg atom.  The experimental bound on the neutron suggests
that $| e(\tilde{d}_d+\tilde{d}_u) |<4\times
10^{-26}\;\; e \, {\rm cm}$, assuming the isoscalar contribution to be
dominant.  Also in this case, the deuteron and neutron show
complementarity.  This is in particular in their sensitivity to the
isovector contribution, which is 20 times larger for the deuteron.

The large sensitivity to new physics (see {\em e.g.}\cite{Lebedev:2004va}) and the relative simplicity of calculating
the nuclear wavefunction, make it clear that small nuclei hold great discovery potential and should therefore be vigorously pursued.

\subsubsection{Experimental approach}

All sensitive EDM searches are performed on neutral systems, which are (essentially) at rest.  The only exception is the proposed use of
molecular ions (HfF$^+$ and ThF$^+$)\cite{MolecIon}, but also for this experiment, the motion of the molecules is not crucial.

In the recent past, several novel techniques have been proposed to use the motional electric field sensed by a particle moving through a
magnetic field at relativistic velocities.  The evolution of the spin orientation for a spin-1/2 particle in an electromagnetic field
($\vec{E},\vec{B}$) is described by the so-called Thomas or BMT equation\cite{Jackson}.  The spin precession vector $\vec{\Omega}$,
{\em relative} to the momentum of the particle, is given by \cite{Silenko:2006fm}
\begin{equation}
  \vec{\Omega} = \frac{e}{m}
    \left[ a\vec{B}
            + \left( a-\frac{1}{\gamma^2-1}\right)\vec{\beta}\times\vec{E}
            + \frac{\eta}{2}\,\left(\vec{E} + \vec{\beta}\times\vec{B}
            - \frac{\gamma}{\gamma+1}\vec{\beta}(\vec{\beta}\cdot\vec{E}) \right)
    \right]
  \label{eq:gm2andEDM}
\end{equation}
with $\vec{\mu} = 2(1+a)\,(e/m)\,\vec{S}$ and $\vec{d} = \eta/2\,(e/m)\,\vec{S}$.  It was assumed that
$\vec{\beta}\cdot\vec{B}=0$.  The first two terms between brackets
will be referred to as $\omega_a$, whereas the last one will be referred
to as $\omega_\eta$.

For fast particles, the electric field in the rest frame of the particle is dominated by $\vec{\beta} \times \vec{B}$.  For
commonplace storage rings, this field can exceed the size of a static electric field made in the laboratory by more than an order of
magnitude, thus giving the storage ring method a distinct advantage.

In a homogeneous magnetic field, $\vec{\omega}_a \propto \vec{B}$ and $\vec{\omega}_\eta\propto\vec{\beta}\times\vec{B}$ are orthogonal,
leading to a small tilt in the precession plane and an second order increase in the precession frequency.  Although this was used to set a
limit on the muon EDM\cite{Bailey:1977sw,McNabb:2004tj}, it does not allow for a sensitive search.

The application of a radially oriented electric field $E_r$ to slow down $\omega_a$ and thus to increase the tilt, was proposed in
\cite{Farley:2003wt}.  For a field strength
\begin{equation}
  E_r = \frac{a\beta}{1-(1+a)\beta^2} B_z
\end{equation}
the spin of an originally longitudinally polarized beam remains aligned with the momentum at all times.  In this case $\vec{\beta}\cdot\vec{E}=0$, and thus
\begin{equation}
  \vec{\Omega} = \cfrac{e}{m}\, \cfrac{\eta}{2}\,\left(\vec{E} + \vec{\beta}\times\vec{B} \right) \label{eq:frozenspin}
\end{equation}
The EDM thus manifests itself as a precession of the spin around the motional electric field $\vec{E}^* = \gamma\left[\vec{E} +
\vec{\beta}\times\vec{B}\right]$, {\em i.e.} as a growing vertical polarization component parallel to $\vec{B}$.  This approach can be
used for all particles with a small magnetic anomaly, so that the necessary electric field strength remains feasible.  Concept
experiments, employing this technique, have been worked out for the muon\cite{Semertzidis:1999kv,Miller:2004nw,Adelmann:2006ab} and the
deuteron \cite{Semertzidis:2003iq}.  Other candidate particles have been identified as well (see {\em e.g.} \cite{Khriplovich:1998zq}).

A third, most sensitive approach is reminiscent of the magnetic
resonance technique introduced by Rabi\cite{Rabi:1938aa}.  The spin is
allowed to precess under the influence of a dipole magnetic field. In
the original application, an oscillating magnetic field oriented
perpendicular to the driving field is applied.  By scanning the
oscillation frequency, a resonance will be observed when the frequency
of the oscillating field matches the spin precession frequency.

In this application, the oscillating magnetic field are replaced by an
oscillating electric field\cite{Orlov:2006su}.  When at resonance, the
electric field coherently interacts with the electric dipole
moment. As a consequence, the polarization component along the
magnetic field oscillates in the case of a sizeable EDM.  In practice,
only the onset of the first oscillation cycle will be visible in the
form of a slow growth of the vertical polarization, proportional to
the EDM.

The oscillating electric field is obtained by modulating the velocity of the deuterons circulating in a magnetic field, setting up a
so-called synchrotron oscillation.  For a time dependent velocity $\beta(t) = \beta_0 + \delta\beta(t)$ generated by an oscillating
longitudinal electric field $E_{RF}(t)$ and a constant magnetic field $B$, the spin evolution follows from
\begin{equation}
  \vec{\Omega} = \cfrac{e}{m}\,\left[  \left\{a\,B + \frac{\eta}{2}\vec{\beta_0}\times\vec{B}\right\} + \frac{\eta}{2}\left\{\delta\vec{\beta}(t)\times\vec{B}
 -  \frac{\beta^2\gamma}{\gamma+1}\vec{E}_{RF}(t) \right\} \right] \equiv \vec{\Omega}_0 + \vec{\delta\Omega}(t)
\end{equation}
The first term yields spin precession about $\vec{\Omega}_0$, without affecting the polarization parallel to it.  For $\delta\beta(t) =
\delta\beta\cos(\omega\,t+\psi)$, and $B\delta\beta\gg\beta^2\gamma/(\gamma+1)E_{RF},$ the parallel polarization component is given by
\begin{equation}
  dP_\parallel/dt \simeq \frac{e}{m}\,P_\circ\,\eta\,\delta\beta\,B\,  \cos\left(\Delta\omega\,t + \Delta\phi\right),
  \label{eq:resMetEssential}
\end{equation}
with $\Delta\omega \equiv \Omega_0-\omega$ and $\Delta\phi \equiv \phi-\psi$.  The beam is assumed to have a longitudinal polarization
$P_0$ at injection time.  For $\Delta\omega=0$ the vertical polarization will grow continuously at a rate proportional to the EDM.
Maximum sensitivity is obtained for $\Delta\phi=0$ or $\pi$, whereas for $\Delta\phi = \pi/2$ or $3\pi/2$ there is no sensitivity to the
EDM.  The latter will prove useful in controlling systematic errors. At the same time, the radial polarization component is given by
\begin{equation}
  P_\perp \simeq P_0 \sin (\Omega_0 t + \phi).  \label{eq:resMetPerp}
\end{equation}
This polarization component can be incorporated in a feedback cycle, to phase-lock the velocity modulation to the spin precession, {\em
i.e.} to guarantee $\Delta\omega=0$ and $\Delta\phi$ constant.  In addition, observation of $\Omega_0$ allows to measure or stabilize the magnetic field.

From Eq.~(\ref{eq:resMetEssential}) and (\ref{eq:resMetPerp}), the main design criteria are easily derived, several of which are common
to all other EDM experiments.  They include
\begin{itemize}
\item high initial polarization $P_0$;
\item large field strength $E^{eff} \propto (\delta\beta\,B)$;
\item close control over the resonance conditions $\Delta\omega$ and phase $\Delta\phi$;
\item long spin coherence time $P_\circ(t)$;
\item long synchrotron coherence time $\delta\beta(t)$;
\item sensitive method for independent observation of $P_\parallel$ and $P_\perp$.
\end{itemize}

The parameters of the current concept deuteron EDM ring are presented in Tab.~\ref{tab:EDMringparameters}.
\begin{table}[bht]
\begin{minipage}{\textwidth}
\caption{Parameters of the concept deuteron EDM storage ring.\label{tab:EDMringparameters}}
\begin{tabular*}{\textwidth}{@{\extracolsep{\fill}}lllrl}
\hline\hline
\hspace*{5mm}&\hspace*{5mm}Parameter&\hspace*{-5mm}symbol \hspace*{10mm}	&\multicolumn{2}{l}{design value}	\\
\hline
&Deuteron momentum 		&$p_\mathcal{D}$	& 1500 			& MeV/$c$ 	\\
&Magnetic field strength 	&$B$			& 2 			& T 		\\
&Bending radius 			&$\rho$			& 2.5 			& m 		\\
&Length of each straight section &$l$			& 5 			& m 		\\
&Orbit length 			&$L$			& 26 			& m 		\\
&Momentum compaction 		&$\alpha_p$		& 1 			& 		\\
&Cyclotron period 		&$t_{c}$		& 137 			& ns 		\\
&Spin precession period 		&$t_{s}$		& 660 			& ns 		\\
&Spin coherence time 		&$\tau_{s}$		& 1000 			& s 		\\
&Motional electric field 	&$E^*/\gamma$		& 375			& MV/m 		\\
&Synchrotron amplitude 		&$\delta\beta/\beta$	& 1 			& \% 		\\
&Synchrotron harmonic 		&$h$			& 40 			& \hspace*{34mm}\\
&Particles per fill 		&$N$			& $10^{12}$ 		& 		\\
&Initial polarization 		&$P_\circ$		& 0.9 			& 		\\
&EDM precession rate @ $d = 10^{-26}\;\; e \, {\rm cm}$  \hspace*{20mm}
				&$\omega_\eta$		& 1 			& $\mu$rad/s 	\\
\hline\hline
\end{tabular*}
\end{minipage}
\end{table}
Coherent synchrotron oscillation can be obtained by a set of two RF cavities, one operating at a harmonic of the revolution frequency to
bring the beam close to the spin-synchrotron resonance, and a second operating at the resonance frequency to create a forced oscillation.

The statistical reach of the experiment is determined by the number of particles used to determine the polarization, as well as the analyzing
power of the polarimeter.  The most efficient way to probe the deuteron polarization at the energy considered is by nuclear
scattering.  To obtain high efficiency, conventional techniques, in which a target is inserted into the beam are unsuitable.  Instead,
slow extraction of the beam onto a thick analyzer target is necessary. Slow extraction could be realized by exciting a weak beam
resonance, or alternatively, by Coulomb scattering off a thin gas jet. The thickness of the analyzer target is optimized to yield maximum
efficiency, which may reach the percent level.

The EDM will create a left-right asymmetry in the scattered particle rate, whose initial rate of growth is proportional to the EDM.  False
signals from, {\em e.g.}, oscillating radial magnetic fields in the ring will be mitigated by varying the lattice parameters.  This will
change the systematic error amplitude, while leaving the EDM signal unchanged.  Various features of the ring design and bunches with
opposite EDM signals will be used to reduce the impact of other systematic effects.

The expected very high observability of most of the field imperfections in the experiment comes from the combination of gross
amplification of the original perturbations in the control bunches, and observation and correction of the amplified parasitic growth of
the vertical polarization component.  This growth is many orders of magnitude more sensitive to ring imperfections than any other beam
parameter.  Preliminary studies shows no unmanageable sources of systematic errors at the level of the expected statistical uncertainty
of $10^{-29}\;\; e \,$cm.

There is currently great interest in EDM experiments because of their potential to find new physics complementary to and even reaching
beyond that which can be found at future accelerators (LHC and beyond). The new approach described here would be the most sensitive
experiment for the measurement of several possible sources of EDMs in nucleons and nuclei for the foreseeable future, if systematic
uncertainties can be controlled.

\subsection{EDM of deformed nuclei: ${225}$Ra}

In the nuclear sector, the strongest EDM limits have been set by cell measurements which restrict the EDM of $^{199}$Hg to $< 2.1\times
10^{-28}e\,$cm. A promising avenue for extending these searches is to take advantage of the large enhancements in the
atomic EDM predicted for octupole-deformed nuclei. One such case is $^{225}$Ra, which is predicted to be two to three orders of magnitude
more sensitive to T-violating interactions than $^{199}$Hg. The next generation EDM search around laser-cooled and trapped $^{225}$Ra is
being developed by the Argon group. They have demonstrated transverse cooling, Zeeman slowing, and capturing of $^{225}$Ra and $^{226}$Ra
atoms in a magneto-optical trap (MOT). They have measured many of the transition frequencies, lifetimes, hyperfine splittings and isotope
shifts of the critical transitions. This new development should enable them to launch a new generation of nuclear EDM searches. The
combination of optical trapping and the use of octupole deformed nuclei should extend the reach of a new EDM search by two orders of
magnitude. A non-zero EDM in diamagnetic atoms is expected to be most sensitive to a chromo-electric induced EDM effect.

Radium-225 is an especially good case for the search of the EDM
because it has a relatively long lifetime ($t_{1/2}$~=~14.9~d), has
spin 1/2 which eliminates systematic effects due to electric
quadrupole coupling, is available in relatively large quantities from
the decay of the long-lived $^{229}$Th ($t_{1/2}$~=~7300~yr), and has
a well-established octupole nature. The octupole deformation enhances
 parity doubling of the energy levels. For example, the sensitivity to
T-odd, P-odd effects in $^{225}$Ra is expected to be a factor of
approximately 400 larger than in $^{199}$Hg, which has been used by
previous searches to set the lowest limit ($< 2 \times 10^{-28}~e\,$cm)
so far on the atomic EDM. The 14.9-day half-life for $^{225}$Ra is
sufficiently long that measurements can be performed and systematics
can be checked without resorting to an accelerator-based
experiment. Nevertheless, if a $^{225}$Ra beam facility were available
for this experiment, approximately a hundred times more atoms could be
produced which could have the impact of improving the sensitivity by
yet another order of magnitude.

Laser cooling and trapping of $^{225}$Ra atoms was developed in preparation of an EDM search. The laser trap allows one to collect and
store the radioactive $^{225}$Ra atoms that are otherwise too rare to be used for the search with conventional atomic-beam or vapor-cell
type methods. Moreover, an EDM measurement on atoms in a laser trap would benefit from the advantages of high electric field, long
coherence time, and a negligible so-called ``$v \times E$" systematic effect.

\begin{figure}
\parbox{0.65\linewidth}{
\includegraphics[width=\linewidth]{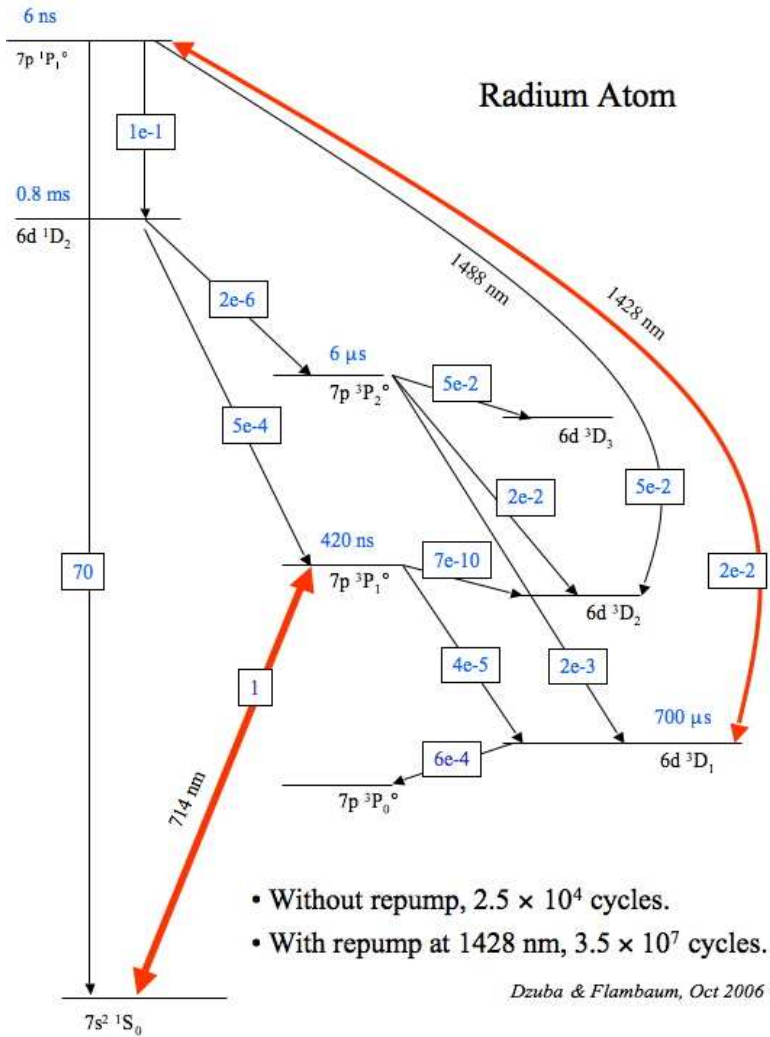}}
\parbox{0.35\linewidth}{
\caption{Atomic level structure of radium-225 indicating the cycling transition at 714 nm and the re-pump transition at 1428 nm. The values in boxes
indicate the relative transition probabilities.\label{fig:rad}}}
\end{figure}

The Argon group has demonstrated a magneto optical trap (MOT) of Ra atoms by using the $7s^2$ $^1S_0 \to$ $7s7p$ $^3P_1$ transition as the
primary trapping transition, and $7s6d$ $^3D_1 \to$ $7s7p$ $^1P_1$ as the re-pump transition (see Fig.~\ref{fig:rad}). They used a
Ti:Sapphire ring laser system to generate the 714~nm light to excite the $7s^2$ $^1S_0 \to$ $7s7p$ $^3P_1$ transition.
The primary leak channel from this two-level quasi-cycling system is the decay from $7s7p$ $^3P_1$ to
$7s6d$ $^3D_1$, from which the atoms were pumped back to the ground-level via the $7s6d$ $^3D_1 \to$ $7s7p$ $^1P_1$ transition
followed by a spontaneous decay from $7s7p$ $^1P_1$ back to the ground-level. The re-pump was induced by laser light at 1428.6~nm
generated by a diode laser. This re-pump transition can be excited for an average of 1400 times before the atom leaks to other metastable
levels. Therefore, with the re-pump in place, an atom can cycle for an average of $3.5 \times 10^7$ times and stay in the MOT for at least
30~s before it leaks to dark levels. Here the MOT is used only to capture the atoms; the trapped atoms would then be transferred to an
optical dipole trap for storage and measurement. They plan to achieve a lifetime of 300~s in the dipole trap.

The ultimate goal of the present series of measurements is to provide a measurement that is comparable in sensitivity to the atomic EDM
experiment for $^{199}$Hg. Because of the enhancement from the octupole deformation of $^{225}$Ra, the measurement would then be more
than two orders of magnitude more sensitive to T-violating effects in the nucleus than that of the $^{199}$Hg experiment. The immediate goal
over the next two years is to provide an initial atomic EDM limit of $\sim 1 \times 10^{-26}\;e\,$cm. Thereafter, the plan is to improve the
experiment until the ultimate goal is achieved.

\subsection{Electrons bound in atoms and molecules}
\subsubsection{Theoretical aspects}
We discuss here permanent EDMs of diatomic molecules induced by the EDM of the electron and by $P$- and $T$-odd
$e$-$N$ neutral currents. In heavy molecules the effective electric field $E_\mathrm{eff}$ on unpaired electron(s) is many orders of
magnitude higher than the external laboratory field required to polarize the molecule. As a result, the EDM of such molecules is strongly
enhanced. The exact value of the enhancement factor is very sensitive to relativistic effects and to electronic correlations. In recent
years several methods to calculate $E_\mathrm{eff}$ were suggested and reliable results were obtained for a number of molecules.

The study of a non-relativistic electron in a stationary state immediately leads to the zero energy shift $\delta \varepsilon$ for an
atom in the external field $\bm{E}_0$ induced by the electron EDM $\bm{d}_e = d_e \bm{\sigma}$. Indeed, the average acceleration
$\langle \bm{a}\rangle = 0$, so the average force $-e \langle\bm{E}\rangle=0$. Therefore, $\delta \varepsilon = -\bm{d}_e
\cdot\langle \bm{E}\rangle =0$. This statement is known as Schiff theorem. In the relativistic case, the position-dependence of the
Lorentz contraction of the electron EDM leads instead to a net overall atomic EDM~\cite{jackson07}. Even though $\left\langle\mathbf{E}\right\rangle = 0$, it still 
can be (and indeed is) the case that $\left\langle\mathbf{d}_e \cdot \mathbf{E}\right\rangle \neq 0$, if $\mathbf{d}_e$ is not spatially 
uniform. Taking account of the fact that the length-contracted value of $\mathbf{d}_e$ is NOT spatially uniform for an electron inside the 
Coulomb field of an atom exactly reproduces the form of the enhancement factor. 

Reliable calculations of atomic energy shifts are easier with the relativistic EDM Hamiltonian for the Dirac electron, which
automatically turns to zero in the non-relativistic approximation~\cite{Commins91}: 

\begin{equation}\label{Eq:H_d}
H_d = 2d_e \left(\begin{array}{cc} 0 & 0 \\ 0 & \bm{\sigma} \end{array}\right)\cdot\bm{E}
\,\cong\, 2d_e \left(\begin{array}{cc}0 & 0 \\0 & \bm{\sigma}\end{array}\right)\cdot\bm{E}_\mathrm{int}~.
\end{equation}

This Hamiltonian is singular at the origin and we neglected the external field $E_0$. Using Eq.~\eqref{Eq:H_d} it is straightforward
to show that the induced EDM of the heavy atom $d_\mathrm{at}$ is of the order of $10\alpha^2 Z^3 d_e$, where $Z$ is the number of protons
in the nucleus. If $Z\sim 10^2$ the atomic enhancement factor $k_\mathrm{at}\equiv d_\mathrm{at}/d_e \sim 10^3$. This estimate holds
for atoms with an unpaired electron with $j=\tfrac12$. For higher angular momentum $j$ the centrifugal barrier strongly suppresses $d_\mathrm{at}$.

Atomic EDM can be also induced by a scalar $P,T$-odd $e$-$N$ neutral current~\cite{Commins91}:
\begin{equation}\label{Eq:eNcurrents}
H_S = i\frac{G\alpha}{2^{1/2}} Zk_S \gamma_0\gamma_5 n(r),
\end{equation}
where $G$ is Fermi constant, $\gamma_i$ are Dirac matrices, $n(r)$ is the nuclear density normalized to unity, and $Zk_S=Zk_{S,p}+Nk_{S,n}$
is the dimensionless coupling constant for a nucleus with $Z$ protons and $N$ neutrons. Atomic EDMs induced by the interactions
(\ref{Eq:H_d},\ref{Eq:eNcurrents}) are obviously sensitive to relativistic corrections to the wave function. Numerical calculations
also show their sensitivity to correlation effects. For example, the Dirac-Fock calculation for Tl gives $d_\mathrm{Tl}=-1910 d_e$ while
the final answer within all-order many-body perturbation theory is $d_\mathrm{Tl}= -585 d_e$ (see Ref.~\cite{Commins91} for details). Note
that the present limit on the electron EDM follows from the experiment with Tl \cite{Regan:2002ta}.

The internal electric field in a polar molecule, $E_\mathrm{mol}\sim \frac{e}{R_o^2} \sim 10^9\,\mathrm{V/cm}$, is 4~--~5 orders of
magnitude larger than the typical laboratory field in an atomic EDM experiment. This field is directed along the molecular axis and is
averaged to zero by the rotation of the molecule. The molecular axis can be polarized in the direction of the external electric field
$\bm{E}_0$. One usually needs the field $E_0\sim 10^4\mathrm{~V/cm}$ to fully polarize the heavy diatomic molecule.  The corresponding molecular
enhancement factor is $k_\mathrm{mol} \sim k_\mathrm{at}\times \frac{E_\mathrm{mol}}{E_0} \sim 10^4 k_\mathrm{at}$.

For closed-shell molecules all electrons are coupled and the net EDM is zero. Therefore one needs a molecule with at least one unpaired
electron. Such molecules have nonzero projection $\Omega$ of electronic angular momentum on the molecular axis. Again, as in the
case of atoms, for the molecules with one unpaired electron the largest enhancement corresponds to $\Omega=\tfrac12$. The centrifugal
barrier leads to strong suppression of the factor $k_\mathrm{mol}$ for higher values of $\Omega$. On the other hand, such molecules can be
polarized in a much weaker external field.

For strong external field $E_0$ the factor $k_\mathrm{mol}$ depends on $E_0$ and it is more practical to define an effective electric field
on the electron $E_\mathrm{eff}$ so, that the $P,T$-odd energy shift for a fully polarized molecule is equal to:
\begin{equation}\label{Eq:Heff}
\delta\varepsilon_{P,T}=E_\mathrm{eff}d_e+\tfrac12 W_S k_S\, ,
\end{equation}
where two terms correspond to interactions \eqref{Eq:H_d} and \eqref{Eq:eNcurrents}. Calculated values of $E_\mathrm{eff}$ and $W_S$
for a number of molecules are listed in Table~\ref{tab:Eeff}.
\begin{table}[!tbh]
\begin{minipage}{\textwidth}
\caption{ Calculated values of parameters $E_\mathrm{eff}$ and $W_S$ from Eq.~\eqref{Eq:Heff}
for diatomic molecules.  The question marks reflect the uncertainty in the knowledge of the ground state.\label{tab:Eeff}}
\begin{tabular*}{\textwidth}{@{\extracolsep{\fill}}llcccl}
\hline \hline
Molecule & State & $\Omega$ &\multicolumn{1}{r}{$E_\mathrm{eff}\, \left(10^{9}\frac{\mathrm{V}}{\mathrm{cm}}\right)$}&$W_S$\,(kHz)&\multicolumn{1}{c}{Ref.}\\
\hline
BaF    & ground     &$ 1/2 $&$ -7.5\pm 0.8 $&$ -12\pm 1   $& \cite{KTMS97,NCD07}    \\
YbF    & ground     &$ 1/2 $&$ -25\pm 3    $&$ -44\pm 5   $& \cite{Mosyagin98,NCD07}    \\
HgF    & ground     &$ 1/2 $&$ -100\pm 15  $&$ -190\pm 30 $& \cite{KL95}    \\
HgH    & ground     &$ 1/2 $&$  -79        $&$ -144       $& \cite{KL95}    \\
PbF    & ground     &$ 1/2 $&$  +29        $&$  +55       $& \cite{KL95}    \\
PbO    & metastable &$ 1   $&$  -26        $&$            $& \cite{Petrov05}    \\
HI$^+$ & ground     &$ 3/2 $&$   -4        $&$            $& \cite{Isaev05}    \\
PtH$^+$& ground (?) &$   3 $&$   20        $&$            $& \cite{Meyer06}    \\
HfF$^+$& metastable (?)&$ 1$&$   24        $&$            $& \cite{PMI07}    \\
\hline\hline
\end{tabular*}
\end{minipage}
\end{table}
An EDM experiment is currently going on with YbF molecules. This molecule has a ground state with $\Omega=\tfrac12$. The $P,T$-odd parameters \eqref{Eq:Heff} were
calculated with different methods by several groups, and estimates of the systematic uncertainty are available. 
Several other molecules and molecular ions have been suggested for the search for electron EDM including PbO, PbF, HgH, and
PtH$^+$. PbF and HgH have $\Omega=\tfrac12$ and calculations are similar to the YbF case. The ground state of PbO has closed shells and the
experiment is done on the metastable state with two unpaired electrons and $\Omega=1$. Here electronic correlations are much stronger and
calculations are more difficult.

Finally, molecular ions like PtH$^+$ are less studied and even their ground states are not known exactly. It is anticipated that such ions can be trapped and
a long coherence time for the EDM experiment can be achieved. Recently the first estimates of the effective field for PtH$^+$ and several other
molecular ions were reported \cite{Meyer06}. These estimates are based on non-relativistic molecular calculations. Proper relativistic
molecular calculations for these ions may be extremely challenging.

\subsubsection{Experimental aspects}

Over a dozen different experiments searching for the electron electric dipole moment that are under way or planned will be reviewed here.  At
present the experimental upper limit on \(d_{e}\) is \cite{Regan:2002ta}: $|d_{e}| \leq 1.6 \times 10^{-27} e\,$cm, where \(e\) is the
unit of electronic charge.  

Most of this work is being done in small groups on university campuses. These experiments employ a wide range
of technologies and conceptual approaches. Many of the latest generation of experiments promise two or more orders of magnitude
improvement in statistical sensitivity, and most have means to suppress systematic errors well beyond those obtained in the previous generation.

To detect $d_e$, most experiments rely on the energy shift \(\Delta E = -\bm{d}_{e} \cdot \vec{\bf E}\) upon application of \(\vec{\bf E}\) to an
electron.  Until recently, most experimental searches for \(d_{e}\) used gas-phase paramagnetic atoms or molecules and employed the
standard methods of atomic, molecular, and optical physics (laser and rf spectroscopy, optical pumping, atomic and molecular beams or vapor
cells, etc.) in order to directly measure the energy shift $\Delta E$. Recently, another class of experiments has been actively pursued,
in which paramagnetic atoms bound in a solid are studied.  Here the principles are rather different than for the gas-phase experiments,
and techniques are more similar to those used in condensed matter physics (magnetization and electric polarization of macroscopic
samples, cryogenic methods, etc.).  We discuss these two classes of experiments separately.

\paragraph{A simple model experiment using gas-phase atoms or molecules}

Experimental searches for \(d_{e}\) using gas-phase atoms or molecules share many broad features. Each consists of a state selector, where the initial
spin state of the system is prepared; an interaction interval in which the system evolves for a time \(\tau\) in an electric field \(\vec{\bf E}\) (and often
a magnetic field \(\bm{B} \parallel \vec{\bf E}\) as well); and a detector to determine the final state of the spin. To understand the essential features, we
consider a simple model that is readily adapted to describe most realistic experimental conditions.  In this model, an ``atom'' of spin 1/2 with
enhancement factor $R$, containing an unpaired electron with spin magnetic moment \(\mu\) and EDM \(d_{e}\). The spin is initially prepared to lie along
$\hat{x}$, i.e., is in the eigenstate $\left|\chi^{x}_{+}\right\rangle$ of spin along $\hat{x}$:
\(\left|\psi_{0}\right\rangle = \left|\chi^{x}_{+}\right\rangle \equiv \frac{1}{\sqrt{2}} \left(\begin{array}{c} 1\\ 1\end{array}\right)\).
During the interaction interval the spin precesses about \(\vec{\bf E} = \Esca \hat{z}\) and \(\bm{B} = B\hat{z}\), in the $xy$ plane, by angle
$2\phi = -(d_{e}R\Esca + \mu B)\tau/\hbar$. At time \(\tau\) the quantum state has then evolved to
$\left|\psi\right\rangle = \frac{1}{\sqrt{2}} \left(\begin{array}{c} e^{-i\phi}\\ e^{i\phi}\end{array}\right)$.
Finally, the detector measures the probability that the resulting spin state lies along $\hat{y}$.  This is determined by the overlap of the wavefunction
$\left|\psi\right\rangle$ with $\left|\chi^{y}_{+}\right\rangle \equiv \frac{1}{\sqrt{2}} \left(\begin{array}{c} 1\\ i\end{array}\right)$. 
Hence the signal $S$ from $N$ detected atoms observed in time \(\tau\) is
$S = N \left|\left\langle \chi^{y}_{+} | \psi \right\rangle \right|^2 =  N\cos^{2}{\phi}$.  
  
The angle \(\phi\) is the sum of a large term $\phi_{1} = -\mu B\tau/(2\hbar)$ and an extremely small term $\phi_{2} = -d_{e}R\EuScript{E}\tau/(2\hbar)$.
To isolate \(\phi_{2}\) one observes $S$ for \(\vec{\bf E}\) and \(\bm{B}\) both parallel and anti-parallel. Reversing \(\vec{\bf E} \cdot \bm{B}\) changes the
relative sign of \(\phi_{1}\) and \(\phi_{2}\) and thus changes $S$; the largest change in $S$ occurs by choosing $B$ such that \(\phi_{1} = \pm\pi/4\).
With this choice, we have $S_{\pm} \equiv S(\vec{\bf E} \cdot \bm{B}~{ }^{>}_{<}~0) = \frac{N}{2} (1 \pm 2\phi_{2})$. The minimum uncertainty in determination
of the phase $\phi_2$ in time \(\tau\), due to shot noise, is \(\delta\phi_{2} = \sqrt{\frac{1}{N}}\). If the experiment is repeated \(T/\tau\) times for
a total time of observation \textit{T}, the statistical uncertainty in \(d_{e}\) is 
$\delta d_{e} = \sqrt{\frac{1}{N_{0}}}\sqrt{\frac{1}{T\tau}} \left|\frac{\hbar}{\EuScript{E}_{\mathrm{eff}}}\right|$, where we used
\(R\EuScript{E} = \EuScript{E}_{\mathrm{eff}}\). In practice, other ``technical'' noise sources can significantly increase this uncertainty, particularly fluctuations in the magnetic field.  Hence, careful magnetic shielding is required in all EDM experiments. 

\paragraph{Systematic errors}
\label{subsec:01.02}

The EDM is revealed by a term in the signal proportional to a P, T-odd pseudoscalar such as \(\vec{\bf E} \cdot \bm{B}\). False terms of the same apparent form
can appear even without P, T violation through a variety of experimental imperfections.  The most dangerous effects appear when \(\bm{B}\) depends on the
sign of \(\vec{\bf E}\), which can occur in several ways. For example, leakage currents flowing through insulators separating the electric field electrodes can
generate an undesired magnetic field $\bm{B}_L$. Also, if the atoms or molecules have a non-zero velocity \(\bm{v}\), a motional magnetic field
\(\bm{B}_{\textrm{mot}} = \frac{1}{c} \vec{\bf E} \times \bm{v}\) exists in addition to the applied magnetic field \(\bm{B}\); along with various other
imperfections in the system, this effect can lead to systematic errors.  A related systematic effect involves geometric phases, which appear if the
direction of the quantization axis (often determined by $\bm{B}_{\textrm{total}} = \bm{B} + \bm{B}_{\textrm{mot}}$) varies between the state selector
and the analyzer \cite{Commins91}. 

A variety of approaches are employed to deal with these and other systematics.  Aside from leakage currents, most systematics depend on a combination of
two or more imperfections in the experiment (i.e. misaligned or stray fields); these can be isolated by deliberately enhancing one imperfection and
looking for a change in the EDM signal. Some experiments utilize, in addition to the atoms of interest, additional species as so-called
``co-magnetometers''. These co-magnetometer species (e.g., paramagnetic atoms with low $R$) are chosen to have negligible or small enhancement factors,
but retain sensitivity to magnetic systematics such as those mentioned above. 

In paramagnetic molecule experiments, issues with systematic effects are somewhat different.  Here the ratio
\( \EuScript{E}_{\mathrm{eff}} / \EuScript{E}_{\mathrm{ext}} \) is enhanced, and relative sensitivity to magnetic systematics is correspondingly reduced.
The \(\vec{\bf E} \times \bm{v}\) effect is effectively eliminated by the large tensor Stark effect \cite{Player70} typically found in molecular states. The
saturation of the molecular polarization $|P|$ (and hence \(\EuScript{E}_{\mathrm{eff}}\)) leads to a well-understood non-linear dependence of the EDM
signal on \(\EuScript{E}_{\mathrm{ext}}\) that can discriminate against certain systematics.  Conversely, the extreme electric polarizability leads to a
variety of new effects, such as a dependence of the magnetic moment $\mu$ on \(\EuScript{E}_{\mathrm{ext}}\), and geometric phase-induced systematic
errors related to variations in the direction of \(\vec{\bf E}_{\mathrm{ext}}\).  

\paragraph{Experiments with gas-phase atoms and molecules}

~\\
\noindent
- {\bf The Berkeley thallium atomic beam experiment}

\begin{figure}[ht]
\begin{center} 
\includegraphics[width=0.5\linewidth]{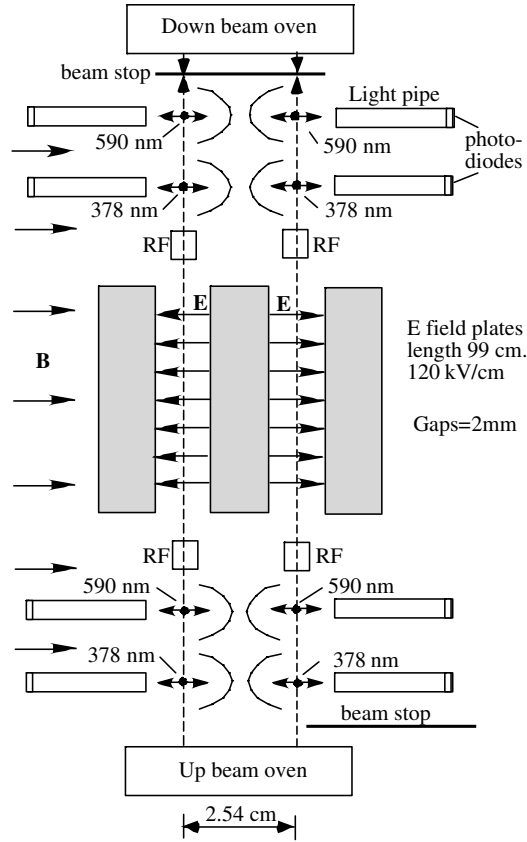} 
\caption{\label{fig:04} Schematic diagram of the Berkeley thallium experiment \cite{Regan:2002ta}, not to scale. Laser beams for state selection and analysis
at 590 nm (for Na) and 378 nm (for Tl) are perpendicular to the page, with indicated linear polarizations. The diagram shows the up-going atomic beams
active. }
\end{center}
\end{figure}

This experiment gives the best current limit on $d_e$.  In its final version \cite{Regan:2002ta}, two pairs of vertical counter-propagating atomic beams,
each consisting of Tl ($Z = 81,~R_{\mathrm{Tl}} = -585$ \cite{Liu92}) and Na $(Z = 11,~R_{\mathrm{Na}}=0.32)$, were employed (See Fig.~\ref{fig:04}).
Spin alignment and rotation of the \(6^{2}P_{1/2} (F = 1)\) state of Tl and the \(3^{2}S_{1/2} (F = 2, F = 1)\) states of Na were accomplished,
respectively, by laser optical pumping and by atomic beam magnetic resonance with separated oscillating rf fields of the Ramsey type. Detection was
achieved via alignment-sensitive laser-induced fluorescence. In the interaction region, with length \(\approx 1\) meter, the side-by-side atomic beams
were exposed to nominally identical \(\bm{B}\) fields, but opposite \(\vec{\bf E}\) fields of \(\approx 120\)~kV/cm. This provided common-mode rejection of
magnetic noise and control of some systematic effects.  Average thermal velocities corresponded to an interaction time $\tau \approx 2.3$ ms (1 ms) for
Tl (Na) atoms.  Use of counter-propagating atomic beams served to cancel all but a very small remnant of the \(\vec{\bf E} \times \bm{v}\) effect.  Various
auxiliary measurements, including use of Na as a co-magnetometer, further reduced this remnant and isolated the geometric phase effect. \(\Esca\) and
leakage currents were measured using auxiliary measurements based on the observable quadratic Stark effect in Tl. About \(5.2 \times 10^{13}\)
photo-electrons of signal per up/down beam pair were collected by the fluorescence detectors. The final result is:
$d_{e} = (6.9 \pm 7.4) \times 10^{-28}e\,$cm, which yields the limit $|d_{e}| \leq 1.6 \times 10^{-27} e\,$cm (90\% conf.).

~\\
\noindent
- {\bf Cesium vapor cell experiments}\label{subsec:02.02}

An experiment to search for \(d_{e}\) in a vapor cell of Cs ($Z=55$; $R_{\mathrm{Cs}} = 115$ \cite{Johnson86}) was reported by L. Hunter and co-workers
\cite{Murthy89} at Amherst in 1989. The method is being revisited in a present-day search by led by M. Romalis at Princeton \cite{RomalisPriv}. The
Amherst experiment was carried out with two glass cells, one stacked on the other in the $z$ direction. Nominally equal and opposite \(\vec{\bf E}\) fields
were applied in the two cells. The cells were filled with Cs, as well as N$_2$ buffer gas to minimize Cs spin relaxation. Circularly polarized laser
beams, directed along $x$, were used for spin polarization via optical pumping. Magnetic field components in all three directions were reduced to less
than \(10^{-7}\)~G.  Thus precession of the atomic polarization in the $xy$ plane was nominally due to \(\vec{\bf E}\) alone.  The final spin orientation was
monitored by a probe laser beam directed along $y$. The effective interaction time was the spin relaxation time \(\tau \approx 15\)~ms. The signals were
the intensities of the probe beams transmitted through each cell. A non-zero EDM would have been indicated by a dependence of these signals on the
rotational invariant \( \bm{J} \cdot ( \bm{\sigma} \times \vec{\bf E} ) \tau \), where \(\bm{\sigma},\bm{J}\) were the pump and probe circular
polarizations, respectively. The most important sources of possible systematic error were leakage currents and imperfect reversal of \(\vec{\bf E}\).  The
result was $d_{e} = (-1.5 \pm 5.5 \pm 1.5) \times 10^{-26}~e\,$cm.

In the new experiment at Princeton, each cell also contains \({}^{129}\mathrm{Xe}\) at high pressure. Cs polarization is transferred to the
\({}^{129}\mathrm{Xe}\) nuclei by spin-exchange collisions.  Under certain conditions this coupling can also give rise to a self-compensation mechanism,
where slow changes in components of magnetic field transverse to the initial polarization axis are nearly canceled by interaction between the alkali
electron spin and the noble gas nuclear spin.  This leaves only a signal proportional to an anomalous interaction that does not scale with the magnetic
moments--for example, interaction of $d_e$ with \(\Esca_{\mathrm{eff}}\). This mechanism (which is understood in some detail \cite{Kornack02}) has the
potential to reduce both the effect of magnetic noise, and some systematic errors. 

~\\
\noindent
- {\bf Experiments with laser-cooled atoms}\label{subsec:02.03}

Laser-cooled atoms offer significant advantages for electron EDM searches.  The low velocities of cold atoms yield long interaction times, and also
suppress \(\vec{\bf E} \times \bm{v}\) effects.  However, these techniques typically yield small numbers of detectable atoms, and magnetic noise must be
controlled at unprecedented levels.  New systematics due to, e.g., electric forces on atoms and/or perturbations due to trapping fields (see e.g.
\cite{Romalis99}) can appear.

Experiments based on atoms trapped in an optical lattice have been proposed by a number of investigators \cite{Chin01,HeinzenPriv,Weiss03}. Two such
experiments, similar in their design, are currently being developed: one led by D.S. Weiss at Pennsylvania State University and another led by D. Heinzen
at the University of Texas. Both plan to use Cs atoms to detect $d_e$, along with Rb atoms ($Z=37$, $R_{\mathrm{Rb}} = 25$) as a co-magnetometer.  The
Texas apparatus consists of two side-by-side far-off-resonance optical dipole traps, each in a vertical 1-D lattice configuration. These traps are placed
in nominally equal and opposite \(\vec{\bf E}\) fields and a common \(\bm{B}\) field of several mG parallel to $\vec{\bf E}$. To load the atoms into the optical
lattice, cold atomic beams from 2D magneto-optical traps exterior to the shields will be captured with optical molasses between the \(\vec{\bf E}\)-field
plates.  The electric field plates will be constructed from glass coated with a transparent, conductive indium tin oxide layer. 

We are aware of two other EDM experiments based on laser-cooled atoms.  One employing a slow ``fountain'', in which Cs atoms are launched upwards and
then fall back down due to gravity, has been proposed and developed by H. Gould and co-workers at the Lawrence Berkeley National Laboratory \cite{Amini06}.
Another, using ${}^{210}\mathrm{Fr} (\tau = 3.2~\mathrm{min}; Z=87, R_{\mathrm{Fr}}=1150)$, has been proposed and is being developed by a group at the
Research Center of Nuclear Physics (RCNP), Osaka University, Japan \cite{SakemiPriv}. 

~\\
\noindent
- {\bf The YbF experiment}\label{subsec:02.04}

\begin{figure}[h] 
\begin{center}
\includegraphics[width=0.5\linewidth]{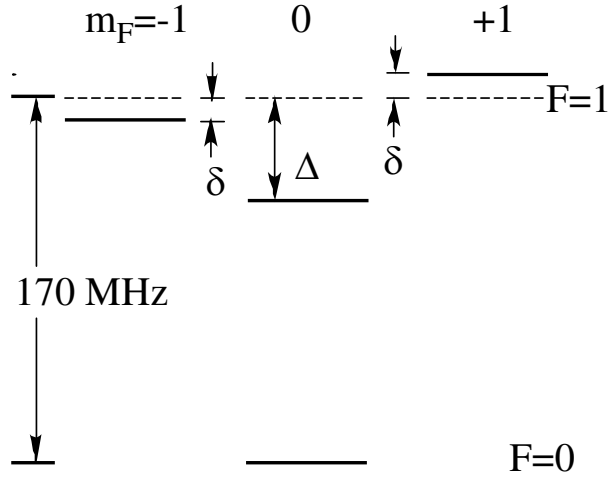} 
\caption{\label{fig:06} Schematic diagram, not to scale, of the hyperfine structure of the \(X^{2}\Sigma\) electronic state of $^{174}$YbF
in the lowest vibrational and rotational level. $\Delta$ is the tensor Stark shift.  \(\delta\) is the shift caused by the combination of the Zeeman
effect and the effect of \(d_{e}\) in \(\vec{\bf E}_{\mathrm{eff}}\).}
\end{center}
\end{figure}

E. A. Hinds and co-workers \cite{Hudson02} at Imperial College, London have developed a molecular beam experiment for investigation of \(d_{e}\) using
YbF. Figure~\ref{fig:06}
shows the relevant energy level structure of the \(X~ {}^{2}\Sigma_{1/2}^{+} (v = 0, N = 0)~ J = 1/2\) ground state of a \({}^{174}\mathrm{YbF}\) molecule.
\({}^{174}\mathrm{Yb}\) has nuclear spin $I_{\mathrm{Yb}}=0$, while $I_{\mathrm{F}} = 1/2$; hence the $J=1/2$ state has two hyperfine components,
\(F = 1\) and \(F = 0\), separated by 170 MHz. An external electric field $\vec{\bf E}_{\mathrm{ext}}$ along $\hat{z}$ with magnitude
$\Esca_{\mathrm{ext}} = 8.3$ kV/cm corresponds to \(\Esca_{\mathrm{eff}} \approx 13\) GV/cm \cite{Hudson02,Mosyagin98}, which splits the
$F=1, m_F = \pm 1$ levels by $2d_e\Esca_{\mathrm{eff}}$. In this external field, the level \(F = 1, m_{F} = 0\) is shifted downward relative to
\(m_{F} = \pm 1\) by an amount \(\Delta = 6.7~\mathrm{MHz}\) due to the large tensor Stark shift associated with the molecular electric dipole. 

In the experiment, a cold beam of YbF molecules is generated by chemical reactions within a supersonic expansion of Ar or Xe carrier gas. Laser optical
pumping removes all \(F = 1\) state molecules, leaving only \(F = 0\) remaining in the beam. Next, a 170 MHz rf magnetic field along $x$ excites molecules
from \(F = 0\) to the coherent superposition  \(|\psi\rangle = \frac{1}{\sqrt{2}} |F = 1, m_{F} = 1\rangle + \frac{1}{\sqrt{2}} |1, -1\rangle\). While
flying through the central interaction region of length 65~cm, the beam is exposed to parallel electric and magnetic fields
\((\pm \EuScript{E},\pm B)\hat{z}\) ($B \sim  0.1$ mGauss).  Next, an rf field drives each \(F = 1\) molecule back to \(F = 0\). Because of the phase
shift \(2\phi\) developed in the central region, the final population of \(F = 0\) molecules is proportional to \(\cos^{2}{\phi}\). These $F=0$ molecules
are detected by laser-induced fluorescence in the probe region.

The most significant systematic errors in this experiment are expected to arise from variation in the direction and magnitude of $\vec{\bf E}$ along the beam
axis. If the direction of $\vec{\bf E}$ changes in an absolute sense, a geometric phase could be generated, and if $\vec{\bf E}$ changes relative to $\bm{B}$, the
magnetic precession phase $\phi_1$, proportional to $\vec{\bf E}_{\mathrm{ext}} \cdot \bm{B}/|\vec{\bf E}_{\mathrm{ext}}|$, could be affected. A preliminary result of
the YbF experiment \cite{Hudson02}, published in 2002, is: $d_{e} = (-0.2 \pm 3.2) \times 10^{-26} e\,$cm. Many significant improvements have been
made since 2002, and it is likely that this experiment will yield a much more precise result in the near future.

~\\
\noindent
- {\bf The PbO experiment}\label{subsec:02.05}

A search for \(d_{e}\) using the metastable \(a(1)^{3}\Sigma_{1}\) state of PbO is being carried out at Yale \cite{DeMille00}. The $a(1)$
state has a relatively long natural lifetime: \(\tau\left[a(1)\right] = 82(2)~\mu s\), and can be populated in large numbers using laser
excitation in a vapor cell.  In this state, the level of total (rotational + electronic) angular momentum $J=1$ contains two
closely-spaced ``\(\Omega\) doublet'' states of opposite parity, denoted as \(e^{-}\) and \(f^{+}\).  An external electric field
\(\vec{\bf E}_{\mathrm{ext}} = \EuScript{E}_{\mathrm{ext}}\hat{z}\) mixes \(e^{-}\) and \(f^{+}\) states with the same value of $M$,
yielding molecular states with equal but opposite electrical polarization $P$. The degree of polarization \(|P| \approx 1\) for
$\Esca_{\mathrm{ext}} \gtrsim 10$ V/cm.  When \(|P| = 1\) the effective molecular field is calculated to be
\(\EuScript{E}_{\mathrm{eff}} \cong 26~\mathrm{GV/cm}\) \cite{Petrov05}. The opposite molecular polarization in the two
$\Omega$-doublet levels leads to a sign difference in the EDM-induced energy shift between these two levels.  This difference provides an
excellent opportunity for effective control of systematic errors, since comparison of the energy shifts in the upper and lower states
acts as an ``internal co-magnetometer'' requiring only minor changes in experimental parameters to monitor.

The Yale experiment is carried out in a cell containing PbO vapor, consisting of an alumina body supporting top and bottom gold foil electrodes, and flat
sapphire windows on all 4 sides. The electric field \(\vec{\bf E}_{\mathrm{ext}} = \EuScript{E}_{\mathrm{ext}}\hat{z}\) is quite uniform over a large
cylindrical volume (diameter 5~cm, height 4~cm), and is chosen in the range 30-90 V/cm. The magnetic field $B_z$ is chosen in the range 50-200 mG. The cell
is enclosed in an oven mounted in a vacuum chamber. At the operating temperature 700 C, the PbO density is
\( n_{\mathrm{PbO}} \approx 4 \times 10^{13}~\mathrm{cm^{-3}} \).

A state with simultaneously well-defined spin and electrical polarization is populated as follows. A pulsed laser beam with $z$ linear polarization
excites the transition $X[J=0^{+}] \rightarrow a(1)[J=1^{-},M=0]$. ($X$ is the electronic ground state of PbO.) Following the laser pulse a Raman
transition is driven by two microwave beams. The first, with $x$ linear polarization, excites the upward 28.2 GHz transition
$a(1) [J=1^{-},M=0] \rightarrow a(1)[J=2^{+},M=\pm 1]$.
The second, with $z$ linear polarization and detuned to the red or blue with respect to the first by 20--60 MHz, drives the downward transition
$a(1) [J=2^{+},M=\pm 1] \rightarrow a(1) [J=1,M=\pm 1]$. The net result is that about 50\% of the \(J=1^{-}, M=0\) molecules are transferred to a coherent
superposition of \(M=\pm 1\) levels in a single desired \(\Omega\)-doublet component. The subsequent spin precession (due to to $\Esca$ and $B$) is
detected by observing the frequency of quantum beats in the fluorescence that accompanies spontaneous decay to the \(X\) state. The signature of a
non-zero EDM is a term in the quantum beat frequency that is proportional to \(\vec{\bf E}_{\mathrm{ext}} \cdot \bm{B}\) and that changes sign when one
switches from one \(\Omega\)-doublet component to the other. 

The present experimental configuration is sufficient to yield statistical uncertainty comparable to the present limit on \(d_{e}\) in a reasonable
integration time of a few weeks.  However, large improvements can be made in a next generation of the experiment.  In the new scheme, detection will be
accomplished via absorption of a resonant microwave probe beam tuned to the 28.2 GHz transition described above.  With this method, the signal-to-noise
ratio is linearly proportional to the path length of the probe beam in the PbO vapor. In a second generation experiment the cell can be made $\sim 10$
times longer than it is now, and the probe beam can pass through the cell multiple times by using suitable mirrors.  Improvements in sensitivity of up to
a factor of 3000 over the current generation are envisioned.

~\\
\noindent
- {\bf Other molecule experiments} \label{subsec:02.06}

E. Cornell and co-workers at the Joint Institute for Laboratory Astrophysics (Boulder, Colorado) have proposed an experiment \cite{CornellPriv} to
search for \(d_{e}\) in the \({}^{3}\Delta_{1}\) electronic state of the molecular ion \(\mathrm{HfF^{+}}\). The premise is to take advantage of the
long spin coherence times typical for trapped ion experiments with atoms, along with the large effective electric field acting on $d_e$ in a molecule.
Preliminary calculations \cite{Isaev05} suggest that the \({}^{3}\Delta_{1}\) state is a low-lying metastable state with very small \(\Omega\)-doublet
splittings; as in PbO, this state could thus be polarized by small external electric fields (\(\lesssim 10\)~V/cm) to yield
$\EuScript{E}_{\mathrm{eff}} \approx 18~\mathrm{GV/cm}$ \cite{Meyer06}. To search for \(d_{e}\), electron-spin-resonance spectroscopy, using the Ramsey
method, is to be performed in the presence of rotating electric and magnetic fields. The electric field polarizes the ions and its rotation prevents them
from being accelerated out of the trap. As in PbO, use of both upper and lower \(\Omega\)-doublet components will yield opposite signs of the EDM signal,
but nearly identical signals due to systematic effects. However, this experiment has the unique disadvantage that it is impossible to reverse the electric
field: in the laboratory frame it must always point inward toward the trap center.

N. Shafer-Ray and co-workers at Oklahoma University have proposed an experiment to search for \(d_{e}\) in the ground $^2\Pi_{1/2}$ electronic state of
PbF \cite{Shafer-Ray06}. The proposed scheme is similar to the YbF experiment, and the value of \(\EuScript{E}_{\mathrm{eff}}\) is also approximately the
same as for YbF.  The primary advantage of PbF is that its electric field-dependent magnetic moment should vanish when a suitable, large external electric
field \(\Esca_{0}\approx 67\) kV/cm is applied \cite{Shafer-Ray06}. This could dramatically reduce magnetic field-related systematic errors.

\paragraph{Experiments with solid-state samples}

Recently, S. Lamoreaux \cite{Lamoreaux02} revived an old idea of F. Shapiro \cite{Shapiro68} to search for $d_e$ by applying an electric field
\(\vec{\bf E}_{\mathrm{ext}}\) to a solid sample with unpaired electron spins. If \(d_{e} \neq 0\), at sufficiently low temperature the sample can acquire
significant spin-polarization and thus a detectable magnetization along the axis of \(\vec{\bf E}_{\mathrm{ext}}\). Lamoreaux pointed out that use of modern
magnetometric techniques and materials (such as \(\mathrm{Gd_{3}Ga_{5}O_{12}}\): gadolinum gallium garnet, or GGG) could yield impressive sensitivity to
$d_e$. GGG has a number of attractive properties.  Its resistivity is so high (\(> 10^{16}\) Ohm-cm for \(T < 77\) K) that it can support large applied
electric fields (\(\vec{\bf E}_{\mathrm{ext}} \approx 10\) kV/cm) with very small leakage currents. Moreover, the ion of interest in GGG,
\(\mathrm{Gd^{3+}}\) (\(Z=64\)) has a non-negligible enhancement factor \cite{Dzuba02}. A complementary experiment is being done by L. Hunter and
co-workers \cite{Heidenreich05} at Amherst College.  Here, a strong external magnetic field is applied to the ferrimagnetic solid
\(\mathrm{Gd_{3-x}Y_{x}Fe_{5}O_{12}}\)  (gadolinium yttrium iron garnet, or GdYIG), thus causing substantial polarization of the \(\mathrm{Gd^{3+}}\)
electron spins. If \(d_{e} \neq 0\), this results in electric charge polarization of the sample, and thus a voltage developed across the sample that
reverses with applied magnetic field. 

The basic theoretical considerations that must be taken into account to estimate the expected signals \cite{Mukhamedjanov03} in these solid-state
experiments include the same types of calculations needed for free atoms.  In addition, however, it is necessary to construct models for the modification
of atomic electron orbitals in the solid material, as well as the response of the material to the EDM-induced perturbation of the heavy paramagnetic atom.
The results of the calculations are as follows.  When all Gd spins are polarized in the GdIG sample, the resulting macroscopic electric field across the
sample is:
$\EuScript{E} = 0.7 \times 10^{-10} (d_e/10^{-27} e\,$cm) V/cm. A similar calculation can be used to determine the
degree of spin polarization of GGG upon application of an external electric field \cite{Lamoreaux02}.  An externally applied electric field of 10 kV/cm
yields an effective electric field $\EuScript{E}^* = -\Delta E/d_e = 3.6 \times 10^{5}$ V/cm acting on the EDM (\cite{Mukhamedjanov03}; see also
\cite{Lamoreaux07}). The resulting magnetization $M$ of the sample is simply related to its magnetic susceptibility $\chi$:
$M = \chi d \EuScript{E}^*/\mu_a$, where $\mu_a$ is the magnetic moment of a Gd$^{3+}$ ion.  Using the standard expression for $\chi(T)$ in a paramagnetic
sample, one finds $M \approx 8 n_{\mathrm{Gd}} (d_e \EuScript{E}^*)/(k_B T)$. Here $k_B$ is the Boltzmann constant and $T$ is the sample temperature.
This yields a magnetic flux $\Phi = 4 \pi M S$ over an area $S$ of an infinite flat sheet.  In a recent development \cite{Lamoreaux07}, Lamoreaux has
pointed out that this type of electrically-induced spin polarization can be amplified in a system that is super-paramagnetic, so that its magnetic
susceptibility $\chi$ is extremely large.  It appears that GdIG (GdYIG with $x=0$) has this property at sufficiently low magnetic field.  If so, the
sensitivity of a magnetization measurement in GdIG at $T = 4$K could be similar to that of GGG at much lower temperatures, greatly simplifying the
required experimental techniques.

~\\
\noindent
- {\bf The Indiana GGG experiment}\label{subsec:03.01}

C. Y. Liu of Indiana University has devised a prototype experiment \cite{Liu04,LiuPriv} in which two GGG disks, 4~cm in diameter and of thickness
\(\approx 1\)~cm, are sandwiched between three planar electrodes. High voltages are applied so that the electric fields in the top and bottom samples are
in the same direction. If \(d_{e} \neq 0\), a magnetic field similar to a dipole field should be generated, and this is to be detected by a flux pickup
coil located in the central ground plane. The latter is designed as a planar gradiometer with 3 concentric loops, arranged to sum up the returning flux
and to reject common-mode magnetic fluctuations. As the electric field polarization is modulated, the gradiometer detects the changing flux and feeds it
to a SQUID sensor. The entire assembly is immersed in a liquid helium bath.

The EDM sensitivity of the prototype experiment is estimated to be $\delta d_e \approx 4 \times 10^{-26} e\,$cm. Although this falls short of
the ultimate desired sensitivity of $10^{-30} e\,$cm, the prototype experiment is useful as a learning tool for solving some basic
technical problems. At Indiana, a second-generation experiment is also being planned, which will operate at much lower temperatures
(\(\approx 10\)--15 mK), and will employ lower-noise SQUID magnetometers.  However, questions remain as to the nature of the
magnetic susceptibility $\chi$ of GGG at such low temperatures.

Some thought has gone into possible systematic effects in this system.  Although crystals with inversion symmetry such as GGG and GdIG should not exhibit
a linear magnetoelectric effect \cite{Mercier74}, crystal defects and substitutional impurities can spoil this ideal. Furthermore a quadratic
magnetoelectric effect does exist, and to avoid systematic errors arising from it, good control of electric field reversal is required.

~\\
\noindent
- {\bf The Amherst GdYIG experiment}\label{subsec:03.03}

GdYIG is ferrimagnetic, and both Gd$^{3+}$ ions and Fe lattices contribute to its magnetization $M$. Their contributions are generally of opposite sign,
but at moderately low temperatures $T$ the Fe component is roughly constant while the Gd component changes rapidly with $T$. There exists a
``compensation'' temperature \(T_{C}\) where the Gd and Fe magnetizations cancel each other, and the net magnetization $M$ vanishes.  For
\(T > T_{C} (<T_{C})\), $M$ is dominated by Fe (Gd). The Gd contribution to $M$ can be reduced by replacing some \(\mathrm{Gd}^{3+}\) ions with
non-magnetic \(\mathrm{Y}^{3+}\). With $x$ the average number of Y ions per unit cell, (so that 3-$x$ is the average number of Gd ions per unit cell), the
compensation temperature becomes $T_{C} = \left[290 - 115(3 - x)\right] \mathrm{K}$. This dependence of \(T_{C}\) on $x$ is exploited in the Amherst GdYIG
experiment. A toroidal sample is employed, consisting of two half-toroids, each in the shape of the letter C. One ``C'' has \(x = 1.35\) with a
corresponding \(T_{C} = 103\)~K. The other ``C'' has \(x = 1.8\) with a corresponding \(T_{C} = 154\)~K. These are joined together with copper foil
electrodes at the interface. At \(T = 127\)~K, the magnetizations of the 2 ``C's'' are identical, but their Gd magnetizations are nominally opposite.
When a magnetic field $H$ is applied to the sample with a toroidal current coil, all Gd spins are nominally oriented toward the same copper electrode.
Thus EDM signals from \(C_{1}\) and \(C_{2}\) add constructively. However below 103 K (above 154 K) the Gd magnetization is parallel (antiparallel) to
$M$ in both C's, which results in cancellation of one EDM signal by the other. Data are acquired by observing the voltage difference $A$ $(B)$ between
the two foil electrodes for positive (negative) polarity of the applied magnetic field $H$. An EDM should be revealed by the appearance of an asymmetry
\(d = A - B\) that has a specific temperature dependence, as described above.

A large spurious effect has been seen that mimics an EDM signal when \(T < 180\)~K, but which deviates grossly from expectations for \(T > 180\)~K.
This effect, which is associated with a component of magnetization that does not reverse with $H$, has so far frustrated efforts to realize the full
potential of the GdYIG experiment.  The best limit that has been achieved so far is \cite{Heidenreich05}: $d_{e} < 5 \times 10^{-24}~e\,$cm.

\subsection{Muon EDM}
The best direct upper limits for an electric dipole moment (EDM) of the muon come from the experiments  measuring the muon anomalous
magnetic moment (g--2).  The CERN experiment obtained $1.1\times 10^{-18}\,e\,$cm (95\% C.L.)~\cite{Bailey:1977sw} and the preliminary
limit from Brookhaven is $2.8 \times 10^{-19}\,e$\,cm~\cite{McNabb:2004tj}. Assuming lepton universality,
the electron EDM limit of $d_e < 2.2 \times 10^{-27}\,e$\,cm~\cite{Regan:2002ta} can be scaled by the electron to
muon mass ratio, in order to obtain an indirect limit of $d_\mu < 5 \times 10^{-25}\,e$\,cm. However, viable models exist in which the
simple linear mass scaling does not apply and the value for the muon EDM could be pushed up to values in the $10^{-22}\,e$\,cm region (see,
e.g.,~\cite{Babu:2000dq,Feng:2001sq,Romanino:2001zf,Bartl:2003ju}). In order for experimental searches to become sufficiently sensitive,
dedicated efforts are needed. Several years ago, a letter of intent for a dedicated experiment at JPARC~\cite{Aok03} was presented,
proposing a new sensitive ``frozen spin'' method~\cite{Semertzidis:1999kv,Farley:2003wt}: The anomalous magnetic
moment precession of the muon spin in a storage ring can be compensated by the application of a radial electric field, thus
freezing the spin; a potential electric dipole moment would lead to a rotation of the spin out of the orbital plane and thus an observable
up-down asymmetry which increases with time. The projected sensitivity of the proposed experiment (0.5\,GeV/c muon momentum, 7\,m ring
radius) is $10^{-24}-10^{-25}\,e$\,cm. Recently it has been pointed out that there is no immediate advantage from working at high muon
momenta and a sensitive approach with a very compact setup (125\,MeV/c muon momentum, 0.42\,m ring radius) was
outlined~\cite{Adelmann:2006ab}. Already at an existing beam line, such as the $\mu$E1 beam at PSI, a measurement with a sensitivity of
better than $d_\mu \sim 5 \times 10^{-23}\,e$\,cm within one year of data taking appears feasible. The estimates for the sensitivity assume
an operation in a ``one-muon-per-time'' mode and the experiment would appear to be statistically limited. With an improved muon accumulation
and injection scheme, the sensitivity could be further increased~\cite{Ade07}. Thus the compact storage ring approach at an
existing facility could bring the proof of principle for the frozen spin technique and cover the next 3-4 orders of magnitude in
experimental sensitivity to a possible muon EDM.

\subsection{Muon g--2}\label{sec:exp:dipoles:g-2}

In his famous 1928 paper\cite{Dirac} Dirac pointed out that the interaction of an electron with external electric and magnetic fields may have
two extra terms where ``the two extra terms 
\begin{equation}
\frac{e h }{ c}\left({\mathbf \sigma} , {\mathbf H} \right) + i \frac{e h}{ c}\rho_1 \left( {\mathbf \sigma} , {\mathbf E} \right) ,
\label{eq:dirac-dpm}
\end{equation}
\dots when divided by the factor $2m$ can be regarded as the additional potential energy of the electron due to its new degree of freedom.''
These terms represent the magnetic dipole (Dirac) moment and electric dipole moment interactions with the external fields. 

In modern notation, for the magnetic dipole moment of the muon we have:
\begin{equation}
\bar u_{\mu}\left[ eF_1(q^2)\gamma_{\beta} +
\frac{i e}{ 2m_{\mu}}F_2(q^2)\sigma_{\beta \delta}q^{\delta}\right] u_{\mu}
\end{equation}
where $F_1(0) = 1,$ and $F_2(0) = a_{\mu}$.

The magnetic dipole moment of a charged lepton can differ from its Dirac value ($g = 2$) for several reasons. Recall that the proton's
$g$-value is 5.6 ($a_p =1.79$), a manifestation of its quark-gluon internal structure. On the other hand, the leptons appear to have no
internal structure, and the magnetic dipole moments are thought to deviate from 2 through radiative corrections, i.e. resulting from
virtual particles that couple to the lepton. We should emphasize that these radiative corrections need not be limited to the Standard Model
particles. While the current experimental uncertainty of $\pm0.5$~ppm on the muon anomaly is 770 times larger than that on the electron
anomaly~\cite{Odom:2006zz}, the former is far more sensitive to the effects of high mass scales. In the lowest-order diagram where mass
effects appear, the contribution of heavy virtual particles with mass $M$ scales as $(m_{\rm lepton}/M)^2$, giving the muon a factor of
$(m_\mu/m_e)^2\simeq 43000$ increase in sensitivity over the electron.

\subsubsection{The Standard Model value of the anomalous magnetic moment}

The standard model value of a lepton's anomalous magnetic moment ({\em the anomaly})
\begin{equation}
a_\ell \equiv  \frac{(g_s -2)}{2}\nonumber
\end{equation}
has contributions from three different sets of radiative
processes: quantum electrodynamics (QED) -- with loops containing leptons ($e,\mu,\tau$) and photons; hadronic -- with hadrons in vacuum polarization loops;
and weak -- with loops involving the bosons $W,Z,$ and Higgs:
\begin{equation}
a_\ell^{\rm SM} = a_\ell^{\rm QED} + a_\ell^{\rm hadronic} + a_\ell^{\rm weak}\, .
\end{equation}

\begin{figure}[h!]
\begin{center}
\includegraphics[width=0.8\textwidth,angle=0]{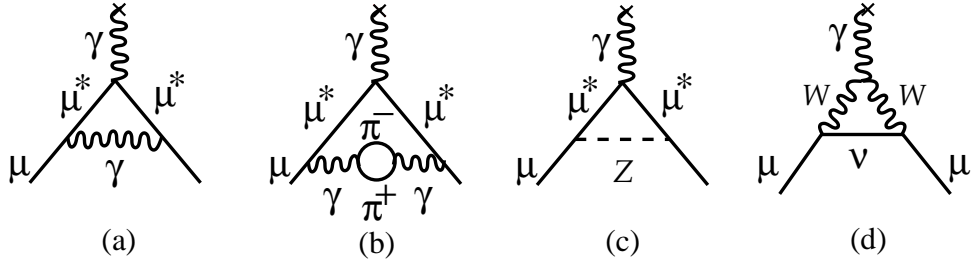}
\end{center}
\caption{The Feynman graphs for: (a) lowest-order QED (Schwinger) term; (b) lowest-order hadronic correction; (c) and (d) lowest order electroweak
terms. The * emphasizes that in the loop the muon is off-shell. With the known limits on the Higgs mass, the contribution from the single Higgs loop
is negligible.}\label{fg:radcor}
\end{figure}

The QED contribution has been calculated up to the leading five-loop corrections\cite{Kinoshita:2005sm}. The dominant ''Schwinger term''\cite{Sch48}
$a^{(2)} = \alpha/2 \pi$, is shown diagrammatically in Fig.~\ref{fg:radcor}(a). Examples of the hadronic and weak contributions are given in
Fig.~\ref{fg:radcor}(b)-(d).

The hadronic contribution cannot be calculated directly from QCD, since the energy scale ($m_{\mu} c^2$) is very low, although Blum has performed 
a proof of principle calculation on the lattice\cite{blum}. Fortunately, dispersion theory gives a relationship between the vacuum polarization loop
and the cross section for $e^+ e^- \to {\rm hadrons}$,
\begin{equation}
a_{\mu}({\rm Had;1})=(\frac{\alpha m_{\mu}}{ 3\pi})^2 \int^{\infty} _{4m_{\pi}^2} \frac{ds }{ s^2}K(s)R(s)\; ,
\end{equation}
where
\begin{equation}
R\equiv \sigma_{\rm tot}(e^+e^-\to{\rm hadrons}) / \sigma_{\rm tot}(e^+e^-\to\mu^+\mu^-)
\end{equation}
and experimental data are used as input\cite{Miller:2007kk,hadcom}

The Standard Model value of the muon anomaly has recently been
reviewed\cite{Miller:2007kk}, and the latest values of the
contributions are given in Table~\ref{tb:SM06}.
\begin{table}[bht]
\begin{minipage}{\textwidth}
\caption{Standard-model contributions to the muon anomalous magnetic dipole moment, $a_\mu$. All values are taken from Ref.~\cite{Miller:2007kk}.\label{tb:SM06}}
\begin{tabular*}{\textwidth}{@{\extracolsep{\fill}}lrclc}
\hline\hline
QED \hspace*{50mm}&\multicolumn{3}{l}{$116~584~718.09\pm 0.14_{5\rm loops}\pm 0.08_{\alpha} \pm 0.04_{\rm masses}$} 				&$\times 10^{-11}$	\\
Hadronic (lowest order) &$a_{\mu}[\rm HVP(06)]$	&=&$6901 \pm 42_{\mbox{\rm \tiny exp}} \pm 19_{\mbox{\rm \tiny rad}} \pm 7_{\mbox{\rm \tiny QCD}}$&$\times 10^{-11}$ \\
Hadronic (higher order) &$a_{\mu}[\rm HVP~h.o.]$&=&$-97.9\pm 0.9_{\mbox{\rm \tiny exp}} \pm 0.3_{\mbox{\rm \tiny rad}}$ 		&$\times 10^{-11}$		\\
Hadronic (light-by-light) &$a_{\mu}[\rm HLLS]$	&=&$110\pm 40$										&$\times 10^{-11}$		\\
Electroweak 		&$a_{\mu}[EW]$		&=&$154\pm 2\pm 1$ 									&$\times 10^{-11}$		\\
\hline\hline
\end{tabular*}
\end{minipage}
\end{table}
The sum of these contributions, adding experimental and theoretical errors in quadrature, gives 
\begin{equation}
\label{eq:sm06}
a_{\mu}^{\mbox{\rm\tiny SM(06)}}=11~659~1785~(61)\times 10^{-11}\,,
\end{equation}
which should be compared with the experimental world average\cite{Bennett:2006fi} 
\begin{equation}
a_{\mu}^{\mbox{\rm\tiny exp}}~=11~659~2080~(63)\times 10^{-11}\, \ . 
\label{eq:E821wa}
\end{equation}
One finds $\Delta a_{\mu}=295(88)\times 10^{-11}$, a $3.4~\sigma$ difference. It is clear that both the theoretical and the experimental uncertainty
should be reduced to clarify whether there is a true discrepancy or a statistical fluctuation. We will discuss potential improvements to the experiment below.

\subsubsection{Measurement of the magnetic dipole moment}

The measured value of the muon anomaly has a 0.46~ppm statistical uncertainty and a 0.28~ppm systematic uncertainty, which are combined in quadrature to
obtain the total error of 0.54~ppm. To significantly improve the measured value, both errors must be reduced. We first discuss the experimental technique,
and then the systematic errors.

In all but the first experiments by Garwin {\em et al.}~\cite{Garwin:1957hc} the measurement of the magnetic anomaly made use of the spin rotation in a magnetic field
relative to the momentum rotation:
\begin{eqnarray}
\vec \omega_S &=& - \frac{qg \vec B }{ 2m} - \frac{q \vec B }{ \gamma m }(1-\gamma)\nonumber\\
\vec \omega_C &=& - \frac{q \vec B }{ m \gamma}\nonumber\\[2mm]
\vec \omega_a &\equiv& \vec \omega_S - \vec \omega_C\nonumber\\
              &=& - \left( \frac{g-2 }{ 2} \right) \frac{q \vec B }{ m} = - a_\mu \frac{ q \vec B }{ m}.
\end{eqnarray}
A series of three beautiful experiments at CERN culminated in a 7.3~ppm measure of $a_\mu$\cite{Bailey:1978mn}.
In the third CERN experiment, a new technique was developed based on the observation that electrostatic quadrupoles
could be used for vertical focusing. With the velocity transverse to the magnetic field ($\vec \beta \cdot \vec B = 0$), the spin precession formula becomes
\begin{equation}
\vec \omega_a = - \ \frac{q }{ m} \left[ a_{\mu} \vec B - \left( a_{\mu}- \frac{1 }{ \gamma^2 - 1} \right)
\frac { {\vec \beta \times \vec E }}{ c } \right]\,.
\label{eq:omega}
\end{equation}
For $\gamma_{\rm magic} = 29.3$, ($p = 3.09$~GeV/$c$), the second term vanishes so $\omega_a$ becomes independent of the electric field
and the precise knowledge of the muon momentum. Also the knowledge of the muon trajectories to determine the average magnetic field becomes
less critical which reduces the uncertainty in $B$.

\begin{figure}[h]
\begin{center}
\includegraphics[clip, trim= 0 10 0 2, width=0.6\textwidth]{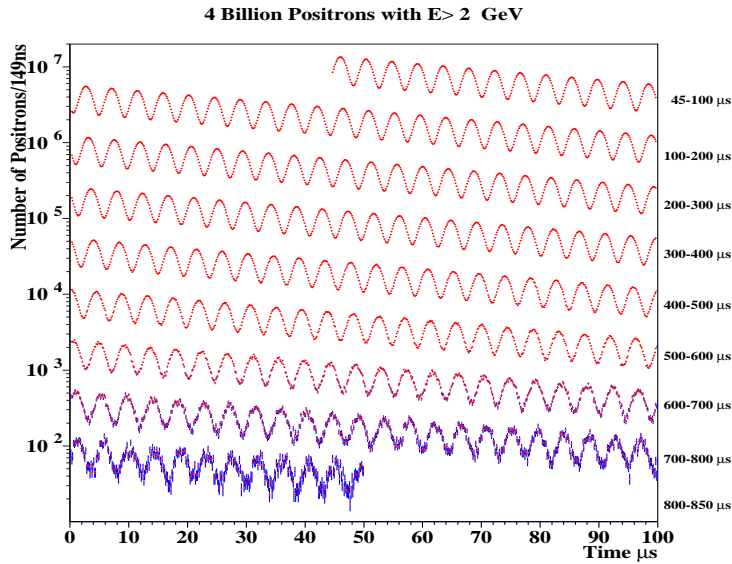}
\caption{The time spectrum of $10^9$ positrons with energy greater than 1.8 GeV from the Y2000 run. The endpoint energy is 3.1 GeV.
The time interval for each of the diagonal ``wiggles'' is given
on the right.\label{fg:wiggles}}
\end{center}
\end{figure}

This technique was used also in experiment E821 at the Brookhaven National Laboratory Alternating Gradient Synchrotron (AGS)~\cite{Miller:2007kk,Bennett:2006fi}.
The AGS proton beam is used to produce a beam of pions, that decay to muons in an 80~m pion decay channel. Muons with $p_{magic}$ are brought into the
storage ring and stored using a fast muon kicker. Calorimeters, placed on the inner-radius of the storage ring measure both the energy and arrival time
of the decay electrons. Since the highest energy electrons are emitted anti-parallel to the muon spin the rate of high-energy electrons is modulated by
the spin precession frequency:
\begin{equation}
N(t,E_{th}) = N_{0}(E_{th})e^{ {-t}/{\gamma\tau} } [1+A(E_{th})\cos(\omega_{a}t+\phi(E_{th}))]. \label{eq:fivep} 
\end{equation}
The time spectrum for electrons with $E > E_{th} = 1.8$~GeV is shown in\cite{Bennett:2006fi} Fig.~\ref{fg:wiggles}. The value of $\omega_a$ is obtained from
these data using the 5-parameter function (Eq.~(\ref{eq:fivep})) as a starting point, but many additional small effects must be taken into
account\cite{Bennett:2006fi,Miller:2007kk}.

The magnetic field is measured with nuclear magnetic resonance (NMR) probes, and tied through calibration to the Larmor frequency of the free 
proton\cite{Bennett:2006fi}. The anomaly is determined from
\begin{equation}
a_{\mu} = \frac{ {\tilde \omega_a / \omega_p} }{ \lambda - {\tilde \omega_a / \omega_p}}= \frac { {\mathcal R} }{ \lambda - {\mathcal R}} ,
\label{eq:lambda}
\end{equation}
where the tilde on $\tilde \omega_a$ indicates that the measured muon precession frequency has been adjusted for any necessary
(small) corrections, such as the pitch and radial electric field corrections\cite{Miller:2007kk}, and $\lambda= \mu_\mu / \mu_p$ 
is the ratio of the muon to proton magnetic moments. 

\subsubsection{An improved g--2 experiment}
One of the major features of an upgraded experiment would be a substantially increased flux of muons into the storage ring. The BNL
beam\cite{Bennett:2006fi} took forward muons from pion decays, and selected muons 1.7\% below the pion momentum. With this scheme,
approximately half of the injected beam consisted of pions.  An upgraded experiment would need to quadruple the quadrupoles in the
pion decay channel, to increase the beam-line acceptance. To decrease the hadron flash at injection one would need to go further away from
the pion momentum. Alternatively one could increase the pion momentum to 5.32~GeV/c so that backward decays would produce muons at the magic
momentum. Then the pion flash would be completely eliminated, which would significantly reduce the systematic error from gain
instabilities.

The inflector magnet that permits the beam to enter the storage ring undeflected would need to be replaced, since the present model loses
half of the beam through multiple scattering in material across the beam channel. The fast muon kicker would also need to be improved.
With the significant increase in beam, the detectors would have to be segmented, new readout electronics would be needed, and a better
measure of lost muons would also be needed.

To reduce the magnetic field systematic errors, significant effort will be needed to improve on the tracking of the field with time, and the
calibration procedure used to tie the NMR frequency in the probes to the free proton Larmor frequency\cite{Bennett:2006fi}. 

While there are technical issues to be resolved, the present technique-- magic $\gamma$, electrostatic focusing, uniform magnetic
field -- could be pushed to below 0.1~ppm. To go further would probably require a new technique. One possibility discussed by Francis
Farley \cite{Farley:2003mj} would be to use muons at much higher energy, say 15~GeV, which would increase the number of precessions
that can be observed. The storage ring would consist of a small number of discrete magnets with uniform field and edge focusing and the field
averaged over the orbit would be independent of orbit radius (particle momentum). The averaged field could be calibrated by injecting
polarized protons and observing the proton g--2 precession.

\section{Acknowledgements}
The authors of this report are grateful to all the additional Workshop
participants who contributed with their presentations during the
Working Group meetings: A. Baldini, W. Bertl, G. Colangelo, F. Farley,
L. Fiorini, G. Gabrielse, J.R. Guest, A. Hoecker, J. Hosek,
P. Iaydijev, Y. Kuno, S. Lavignac, D. Leone, A. Luccio, J. Miller,
W. Morse, H. Nishiguchi, Y. Orlov, E. Paoloni, A. Pilaftsis, W. Porod,
N. Ramsey, S. Redin, W. Rodejohann, M.-A. Sanchis-Lozano,
N. Shafer-Ray, A. Soni, A. Strumia, G. Venanzoni, T. Yamashita,
Z. Was, H. Wilschut.

This work was supported in part by the Marie Curie research training network
``HEPTOOLS'' (MRTN-CT-2006-035505).

\end{document}